\documentclass[backref=page,cornellheadings,smallerheadings]{fiu}
\pdfoutput=1

\PassOptionsToPackage{hyperfootnotes=true}{hyperref}

\renewcommand{\bibname}{LIST OF REFERENCES}
\usepackage{color}   
\usepackage{url}
\usepackage{appendix}
\usepackage{capt-of}
\usepackage{siunitx}
\usepackage{subfig}
\usepackage{caption}
\usepackage{graphicx}  
\usepackage{dcolumn}   
\usepackage{bm}        
\usepackage{amssymb}   
\usepackage{physics}
\usepackage{xcolor}
\usepackage[numberlinked=true]{footnotebackref}
\usepackage{amsmath}
\usepackage{cite}
\usepackage{float}
\usepackage{empheq}
\usepackage{notoccite}
\usepackage[bottom, flushmargin, hang, multiple]{footmisc} 
\usepackage{hyperref}

\hypersetup{
    colorlinks=false, 
    linktoc=all,     
    linkcolor=blue,  
}

\title{Cross Section Measurements of Deuteron Electro-Disintegration at Very High Recoil Momenta and Large 4-Momentum Transfers ($Q^{2}$)}
\author{Carlos Yero Perez}
\conferraldate{April}{2020}
\defensedate{July 02, 2020}
\advisor{Werner Boeglin}
\memberone{Misak Sargsian}
\membertwo{Joerg Reinhold}
\memberthree{Brian Raue}
\degreefield{Physics}
\college{College of Arts, Sciences and Education}
\collegedean{Dean Michael R. Heithaus}
\gradschooldean{Andr$\acute{e}$s G. Gil}

\begin{document} 

\setcounter{page}{1}
\pagenumbering{roman}
\pagestyle{plain}


\maketitle

\makeapproval{4}

\makecopyright

\begin{dedication}
To my late uncle and nuclear engineer Raul Yero Sosa.
\end{dedication}

\begin{acknowledgments}
\raggedright
First, I would like to thank my parents for supporting me during my years as an undergraduate student, specially when I decided to take
the important step of changing my major from Biology to Physics during the first semester at Florida International University.\\
\hspace{0.25in} I would like to thank my advisor, Professor Werner Boeglin, for giving me the opportunity to work on one of Jefferson Lab's experiments
at Hall C which allowed me to gain an unimaginable amount of hands-on experience on both hardware and software related tasks during the initial
phases of the 12 GeV experimental program. It is not often that a graduate student has the opportunity to work from the ground up on a nuclear physics
experiment and be able to have a global perspective on all the different aspects of what constitutes a nuclear/particle physics experiment. For this, I considered myself
very lucky to have been given this opportunity. I am also very thankful to all the Experimental Hall C Staff and Users for all the useful discussions I had with them
on different aspects of experimental nuclear physics which allowed me to gain a better perspective on some of the most difficult (but also the most fun!) topics
which I considered to be spectrometer optics and what constitutes the set-up of an electronics trigger. \\
\hspace{0.25in} I would also like to thank Dr. Mark Jones from Hall C for his constant support and guidance during the analysis of this experiment.
I am very grateful to Mark as a friend and colleague who has demonstrated infinite patience even when I ask the most stupid questions one could ever
imagine. I cannot really thank him enough for all the help I received. \\
\hspace{0.25in} Finally, my gratitude goes to theorists Misak Sargsian, Jean-Marc Laget, Sabine Jeschonnek and J.W. Van Orden for providing the theoretical calculations
as well as helpful discussions on this topic. And from the experimental side, a special thanks goes to Dr. Dave Mack from Hall C for diligently (and voluntarily)
revising my thesis and making sure to point out any inconsistencies, typos and silly mistakes. \\
\hspace{0.25in} This work was supported in part by the U.S. Department of Energy (DOE), Office of Science, Office of Nuclear Physics
under grant No. DE-SC0013620 and contract DE-AC05-06OR23177, the Nuclear Regulatory Commission (NRC) Fellowship
under grant No. NRC-HQ-84-14-G-0040 and the Doctoral Evidence Acquisition (DEA) Fellowship.

\end{acknowledgments}

\begin{abstract}
The $^{2}$H$(e,e'p)n$ cross sections have been measured at negative 4-momentum transfers of $Q^{2} = 4.5\pm0.5$ (GeV/c)$^{2}$
and $Q^{2} = 3.5\pm0.5$ (GeV/c)$^{2}$ reaching neutron recoil (missing) momenta up to $p_{\mathrm{r}}\sim$1.0 GeV/c. The data
have been obtained at fixed neutron recoil angles $5^{\circ}\leq\theta_{nq}\leq95^{\circ}$ with respect to the 3-momentum transfer $\vec{q}$.
The new data agree well with the previous data which reached $p_{\mathrm{r}}\sim550$ MeV/c. At $\theta_{nq}=35^{\circ}$ and $45^{\circ}$,
final state interactions (FSI), meson exchange currents (MEC) and isobar configurations (IC) are suppressed and the plane wave impulse
approximation (PWIA) provides the dominant cross section contribution. The new data are compared to recent theoretical calculations,
and a significant disagreement for recoil momenta $p_{\mathrm{r}}>700$ MeV/c is observed.\\
\indent The experiment was carried out in experimental Hall C at the Thomas Jefferson National Accelerator Facility (TJNAF) and formed part of
a group of four experiments that were used to commission the new Super High Momentum Spectrometer (SHMS). The experiment consisted of a 10.6 GeV
electron beam incident on a liquid deuterium target which resulted in the break-up of the deuteron into a proton and neutron. The scattered
electrons were detected by the SHMS in coincidence with the knocked-out protons detected in the previously existing High Momentum Spectrometer (HMS) and the recoiling neutrons
were reconstructed from energy-momentum conservation laws. To ensure that the $^{2}$H$(e,e'p)n$ reaction channel was selected, we required the
missing energy of the system to be the binding energy of the deuteron ($\sim$2.22 MeV). \\
\indent The spectrometers' central angles and momenta were set to measure three central missing momentum settings of the neutron corresponding to $p_{\mathrm{r}}=80, 580$
and 750 MeV/c, which required the SHMS central angle and momentum to be fixed and the HMS to be rotated from smaller to larger angles corresponding to the lower and higher missing momentum
settings, respectively. The experiment was carried out in a time period of six days with typical electron beam currents of 45-60 $\mu$A at about 50$\%$ beam efficiency.

\end{abstract}

\contentspage

\tablelistpage

\figurelistpage

\normalspacing
\setcounter{page}{1}
\pagenumbering{arabic}
\pagestyle{cornell}

\chapter{INTRODUCTION}\label{chap:chapter1}
In this introductory chapter I will first give a historical overview of the deuteron. Then I will briefly discuss
the transition from meson theory to phenomenology as well as historical electron-scattering experiments on the deuteron that
were done in an effort to understand how the nucleon-nucleon ($NN$) interaction works. Finally, I will give the motivation for doing this experiment.

\section{The Deuteron and the Beginning of Nuclear Forces}
The deuteron (originally called by various names such as ``deuton,'' ``diplon'' or ``diplogen'') was discovered in 1931 by
H. Urey\cite{d2_discovery_Urey} while spectroscopically examining the residue of a distillation of liquid hydrogen.
It was not until the discovery of the neutron by J. Chadwick\cite{neut_disc_chadwick1932} a few months later that the deuteron mass 
could be explained. Within a few months, the first attempt to describe the nuclear force between the proton and neutron using a quantum-mechanical 
approach was made by W. Heisenberg\cite{Heisenberg_1932_1, Heisenberg_1932_2, Heisenberg_1932_3} under the faulty assumption that the neutron was 
a bound system of a proton and an electron, as this was the existing view of the nucleus at the time. In 1934, H. Bethe and R. Peierls introduced 
for the first time the Hamiltonian of the deuteron\cite{Bethe_1935} (``diplon'' at the time) treating it as a two-body system with a nucleon-nucleon 
($NN$) interactive potential, even though the details of the interaction were unkown at the time. The approach to describe the $NN$ potential via a Hamiltonian
would become a basis for the successful description of nuclear systems and reactions in the future\cite{Orden_2001deuteron}. In that same year, the first semi-successful attempt at explaining how 
the nuclear force worked was presented by H. Yukawa using the idea of particle exchange introduced in the Quantum Field Theory (QFT) of electromagnetic interactions
known as Quantum Electrodynamics (QED) and developed by P.M. Dirac in the late 1920s. In a simplified version of Yukawa's theory, the $NN$ potential is expressed as:
\begin{equation}
V_{\mathrm{Yukawa}} = -C_{\mathrm{Y}}\frac{e^{-r/R}}{r}, \hspace{3mm} R\equiv\frac{\hbar c}{m_{\pi}c^{2}}
\label{eq:eq1.1}
\end{equation}
where the overall ``-'' sign means that the force is attractive, $C_{\mathrm{Y}}$ is related to the coupling strength between the nucleons, $r$ is the distance between the nucleons, 
$R$ is the range of interaction, $\hbar c=197.3$ MeV$\cdot$fm, and $m_{\pi}$ is the mass of the exchanged particle. The attractive force between two nucleons is mediated 
by the exchange of a single massive boson (meson) that Yukawa estimated to be $m_{\pi}\sim200$ times the mass of an electron. The particle was later 
discovered in a cosmic ray experiment in 1947\cite{pion_discovey_1947}, and became known as the charged pion, which earned Yukawa the Physics Nobel Prize in 1949. 
Since the exchanged particle has a finite mass (unlike the virtual photon in QED), the strong nuclear force operates at small distances where the mass of the exchanged 
mesons mediating the $NN$ interaction is inversely proportional to the interaction range. Using Heisenberg's uncertainty principle and the known pion mass, the $NN$ interaction range is estimated 
to be $R\sim1.4$ fm. In reality, the Yukawa potential is more complicated than presented in Eq. \ref{eq:eq1.1} and only describes the long-range part
of the $NN$ interaction. \\
\indent Before additional discussion of nuclear interactions, it is worth mentioning the implications that an important experimental discovery had on the nature of the
nuclear force. In 1939, Rabbi \textit{et al.}\cite{d2_quadMoment_discovery_1939, d2_quadMomentRep_1940} measured the deuteron's electric quadrupole moment ($Q_{zz}=+2.73$ $e$fm$^{2}$). 
The implications of a static quadrupole moment were that the nuclear potential did not only have a central (spherical symmetric) part, but also a complicated non-central component that
needed to be accounted for. To understand the non-central component, consider the multipole expansion of a charge distribution in the presence of an external electric field, $\vec{E}=-\vec{\nabla}V$, which can be expressed as:
\begin{equation}
E_{\mathrm{int}} = V(0)q + \frac{\partial V}{\partial z}\Big|_{0} p_{z} - \frac{1}{4}\frac{\partial^{2}V}{\partial z^{2}}\Big|_{0}Q_{zz} + . . .
\end{equation}
where $E_{\mathrm{int}}$ is the interaction energy, $V$ is the electric potential and $q, p_{z}$ and $Q_{zz}$ are the monopole ($L=0$), dipole ($L=1$) and quadrupole ($L=2$) terms, respectively. 
\begin{figure}[H]
\centering
\includegraphics[scale=0.2]{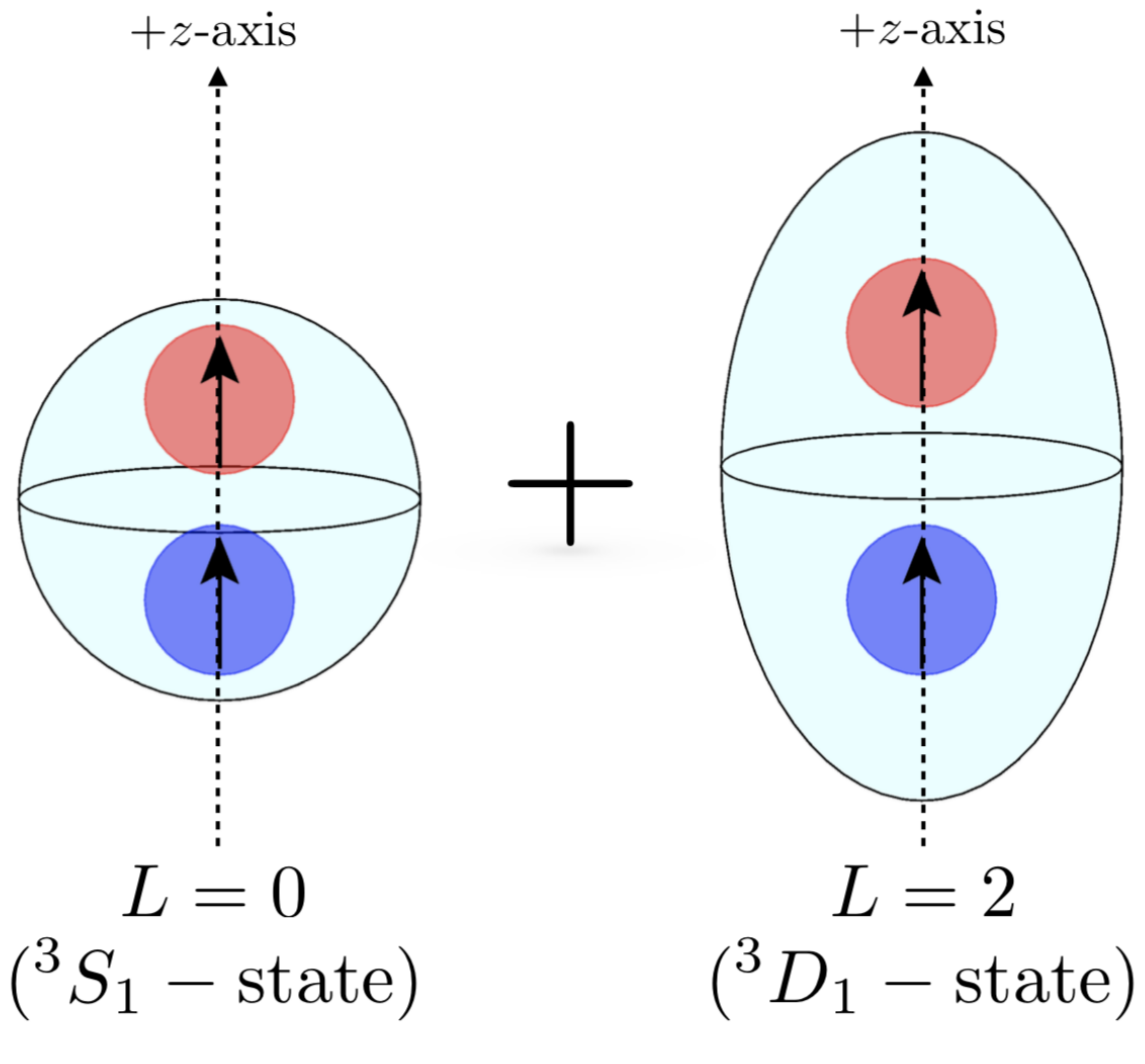}
\caption{The $L=0$ (spherically symmetric) and $L=2$ (prolate spheroid) are the charge distributions of the deuteron, where the spins of the proton (blue) and neutron (red) are aligned.
  The existence of a positive electric quadrupole moment indicates that the deuteron charge distribution is actually elongated about the axis of rotation ($z$-axis).}
\label{fig:fig1.1}
\end{figure}
\indent The monopole term conserves angular momentum ($L=0$), which is a property of central forces and corresponds to a spherically symmetric charge distribution with radius $\langle r \rangle^{2} = \langle x^{2} + y^{2} + z^{2} \rangle \equiv 3\langle z^{2} \rangle$
assuming that the expectation values of the square of the distance from the center to the surface, $\langle x^{2} \rangle = \langle y^{2} \rangle = \langle z^{2} \rangle$, are equal.
The existence of a quadrupole term ($Q_{zz}=e(3z^{2} - r^{2})$) in the deuteron, however, indicates that the nuclear force has a tensor component that arises from the spin-orbit interaction between the
angular momentum ($L=2$) and the intrinsic nuclear spins. There also exist an interaction between the intrinsic particle spins known as the spin-spin interaction that contributes to the
tensor component of the nuclear force, however, this interaction arises from a purely quantum-mechanical effect and has no classical analog. The quadrupole term measures the lowest order departure from a spherical charge distribution in a nucleus (see Fig. \ref{fig:fig1.1}).
To understand the additional tensor component, consider the
following:
\begin{align}
J = L + S,  \label{eq:1.3}
\end{align}   
where $S = s_{p} + s_{n}$, $(s_{p}, s_{n})$ are the proton and neutron intrinsic spins, $L$ is the relative angular momentum between the two nucleons and $J$ is the total angular momentum.
The range of possible angular momentum states is given by
\begin{align}
|L - S| \leq J \leq L + S. \label{eq:1.4}
\end{align}
From the experimental fact that $J=1\hbar$\cite{d2_spinMeas_Murphy1934} for the deuteron, the possible combinations are:
\begin{subequations}
\begin{align}
&L=0, \hspace{3mm} S=1 \hspace{5mm} \text {$(s_{p}, s_{n})$  parallel} \label{eq:1.5a}\\
&L=1, \hspace{3mm} S=0 \hspace{5mm} \text {$(s_{p}, s_{n})$  anti-parallel} \label{eq:1.5b}\\
&L=1, \hspace{3mm} S=1 \hspace{5mm} \text {$(s_{p}, s_{n})$  parallel} \label{eq:1.5c}\\
&L=2, \hspace{3mm} S=1 \hspace{5mm} \text {$(s_{p}, s_{n})$  parallel} \label{eq:1.5d} 
\end{align}
\end{subequations}
From the observation that the deuteron parity\footnote{\singlespacing Parity refers to the eigenvalue of the angular wave function under the trasnformation: $\hat{P}Y(\theta,\phi)=Y(\pi-\theta,\phi+\pi)=PY(\theta,\phi)$, 
where $P=(-1)^{L}$} is even or ``+,'' only even values of relative angular momentum are allowed, which implies that the deuteron wave function is not in a pure $L=0$ state,
but rather a superposition of $L=0$ and $L=2$ states. Using the spectroscopic notation ($^{2S+1}L_{J}$), the $L=0$ and $L=2$ are referred to as ``sharp'' (or $S$-wave) and ``diffuse'' (or $D$-wave) components, respectively.
The deuteron wave function can then be expressed as a linear combination of two possible states
\begin{equation}
\ket{\Psi_{\mathrm{d}}} = p_{S}\ket{^{3}S_{1}} + p_{D}\ket{^{3}D_{1}}, \label{eq:1.6}
\end{equation}
where  $p_{S}^{2} + p_{D}^{2} = 1$ and the normalization coefficiencts, $(p_{S}, p_{D})$, represent the probability of finding the deuteron
in either an $S$-state ($p_{S}$) or $D$-state ($p_{D}$). The relative contribution from the $S$- or $D$-state are sensitive to the radial part of the deuteron
wave function, which is determined phenomenologically. A summary of the $D$-state probability for different $NN$ potentials can be found in Ref.\cite{d2_properties_zhaba2017},
with typical ranges $p_{D}\sim3-7\%$. 

\section{From Meson Theory to Phenomenology} \label{sec:section1.2}
\begin{figure}
\centering
\includegraphics[scale=0.30]{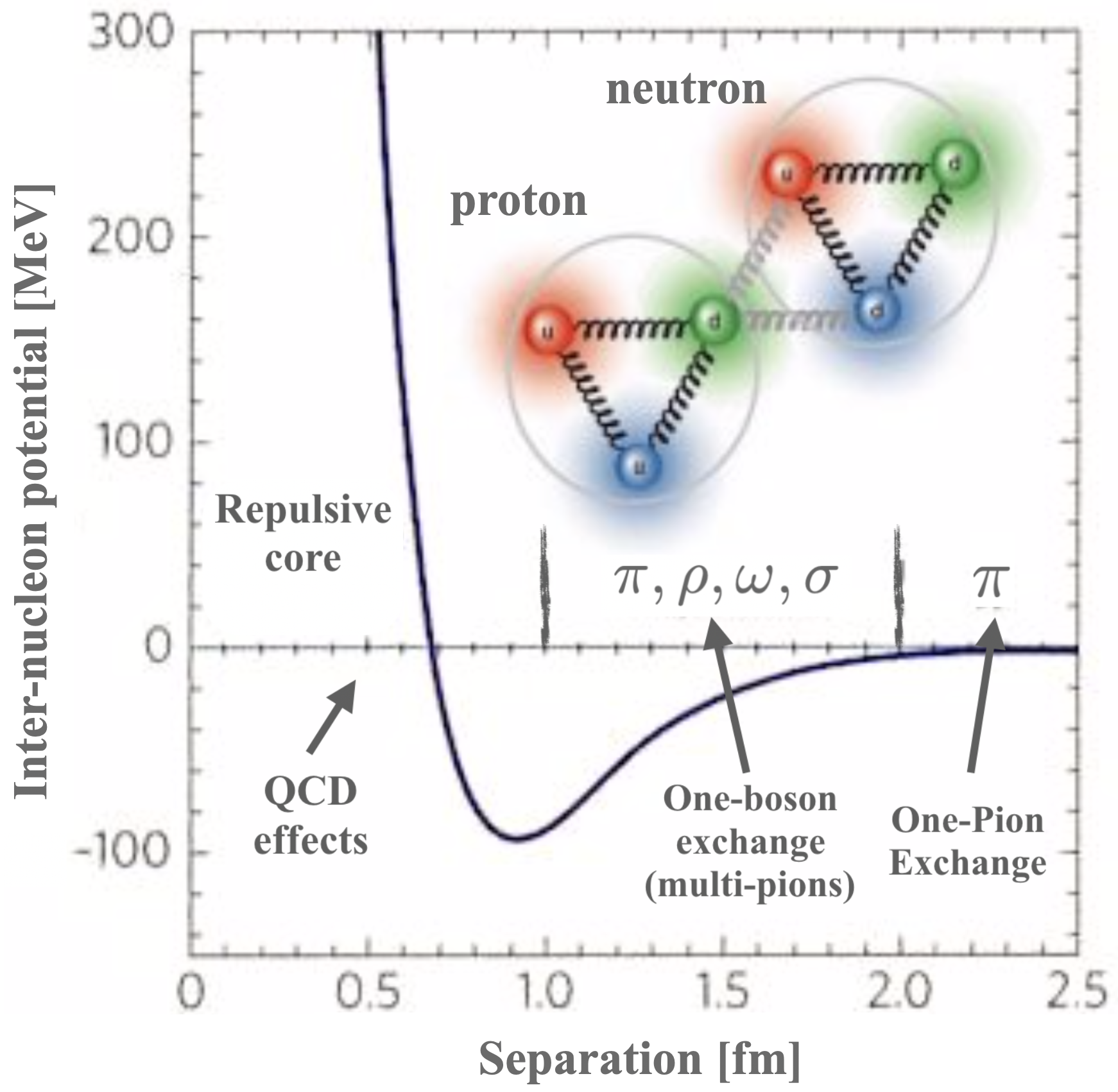}
\caption{Qualitative $NN$-potential versus inter-nucleon separation distance. Note: Reprinted from Ref.\cite{Naghdi_2014}.}
\label{fig:fig1.2}
\end{figure}

After the discovery of the pion in cosmic rays in 1947 and its artificial production in the lab at the Berkeley Cyclotron in 1948\cite{PionLab_1948},
a great deal of effort was devoted to the development of meson theory as the fundamental theory of nuclear forces. In 1951, Taketani, Nakamura and Sasaki\cite{Sasaki_1951}
proposed that the nuclear potential should be divided into different regions that should be treated separately (see Fig. \ref{fig:fig1.2}). \\
\indent It was suggested that the long-range part of the potential should be treated using meson theory, while the intermediate and short-range parts should be approached
phenomenologically as additional complications because of heavy mesons, higher order perturbations, coupling strengths and relativistic effects become difficult to solve. 
Nevertheless, in the early 1950s there were various attempts to develop a fundamental theory of strong interactions (meson theory)\cite{Taketani1952, Brueckner_1953, Brueckner2_1953, Taketani1954} 
that ultimately failed when multi-pion exchanges where included in the theory. Only the long-range part of the $NN$ potential---or the One-Pion Exchange Potential (OPEP)---was 
found to describe the $NN$ scattering data at the time. A more general overview of the development of pion theory can be found in Refs.\cite{Iwadare1956, Taketani_1956}. \\
\indent In the early 1960s, the possibility of the existence of heavier mesons started to emerge theoretically\cite{hvyMeson_thrPred} and experimentally\cite{hvyMeson_expPred}.
These ideas led to the development of the One-Boson Exchange Potential (OBEP)\cite{Hoshizaki_1961}, where the idea of a single pion exchange between two nucleons was generalized to
a single boson exchange, in which heavier mesons were also included in the model and would account for shorter distances in the $NN$ potential. Soon afterwards, several heavier mesons 
were discovered experimentally, most notably the $\rho(770)$\cite{rho_1961} and $\omega(783)$\cite{omega_1961} mesons. With the discovery of heavier mesons, increased efforts were 
devoted to the development of the OBEP\cite{Hoshizaki_1962, Ogawa_1967} and soon afterwards, the first $NN$ potential models emerged that seemed to describe the $NN$ scattering data better
than any previous models to date. Some of the best known potentials of the 1960s were by Hamada-Johnston (HJ)\cite{HJ_NN_1962} and Reid68\cite{Reid68_NN_1968}. \\
\indent In the 1970s and 1980s efforts continued towards the development of improved nuclear phenomenological models using the OBEP. Of particular importance was the
development of the relativistic OBEP\cite{relOBEP_1971,relOBEP_1972,relOBEP_1974,relOBEP_1975} in the 1970s where the full relativistic scattering amplitudes were used in
the calculations. The inclusion of relativistic amplitudes produced a significant improvement in the agreement between $NN$ scattering data and phenomenological models using the relativistic OBEP as compared to
previous non-relativistic models (see Fig. 2 of Ref.\cite{Machleidt_1993}). During the 1970s, there was also an effort devoted to derive the $2\pi-$exchange contributions 
to the nuclear potential, which accounted for the intermediate range of the nuclear force. The reason was that during its initial years, the OBEP models had to introduce the
$\sigma$ meson in order to describe the intermediate-range nuclear force, however, no experimental evidence for the $\sigma$ meson had been found. \\
\indent Well known examples of potentials that
included $2\pi$ exchange contributions were the Stony-Brook\cite{StonyBrook_NN_Jackson1975} and Paris\cite{Paris_NN_Lacombe1973,Paris_NN_Lacombe1975} group potentials.
In the 1980s, more sophisticated potentials based on the OBE approach were constructed, particularly by the Argonne and Bonn groups with $NN$ potentials that included
$2\pi$ exchange contributions such as the Argonne V14 and V28 (AV14 and AV28)\cite{AV14_1984} potentials and the full Bonn\cite{Bonn_NN_Machleidt1987} potential, which
included both $2\pi$ exchange contributions as well as relativistic effects on the OBEP. It is important to note that some of the potentials mentioned above were improved
further in later years. As an example, in this experiment we used the parametrized Paris\cite{Paris_NN_Lacombe1980}, AV18\cite{AV18_1995} and charge-dependent
Bonn (CD-Bonn)\cite{CDBonn_NN_Machleidt2001} potentials to compare with experimental data.
\section{Historical $^{2}$H$(e,e'p)n$ Experiments} \label{sec:section1.3}
In order to probe the internal structure and dynamics of nuclei, electron-nucleon scattering serves as the most valuable tool since the interaction is described by the well established theory of QED, which is
capable of making accurate predictions. Electron scattering experiments can be separated into inclusive or exclusive types. In the former, only
the electron is detected in the final state (single-arm experiments), and one studies the nucleus in question by integrating over all possible final states\cite{coin_exp_bowglin1995}.
In the latter, one or more particles are detected in coincidence with the scattered electron, which allows one to investigate properties unique to the specific reaction
in question. In deuteron electro-disintegration ($^{2}$H$(e,e'p)n$), for example, one detects the scattered electron in coincidence with the proton and the missing neutron is reconstructed
from momentum conservation laws. The $^{2}$H$(e,e'p)n$ reaction proves to be the most direct way of probing the internal structure of the deuteron since it is possible to deduce the internal
momentum of the nucleons from the neutron recoil (``missing'') momenta. \\
\indent Historical $^{2}$H$(e,e'p)n$ experiments were started in 1962 at the Stanford Mark III Linear Accelerator (linac) at a very low $Q^{2}=0.085$ (GeV/c)$^{2}$\cite{d2_1st_edisin_exp_1962}
and shortly after in 1965 at the Orsay linac at $Q^{2}=0.264$ (GeV/c)$^{2}$\cite{d2_edisin_exp_1965}. At the time, the smallest cross sections measured were limited 
by the duty factor\footnote{\singlespacing The duty factor is defined by the ratio $f_{\mathrm{duty}} = \delta T_{\mathrm{pulse}}/\delta T_{\mathrm{rep}}$, where $\delta T_{\mathrm{pulse}}$ is the pulse length and
$\delta T_{\mathrm{rep}}$ is the pulse repetition period of the electron beam. The small duty factor of the accelerators leads to high instantaneous particle rates
and therefore high accidental coincidence rates, or equivalently, low signal-to-noise ratio. As a result, the amount of beam time required to measure smaller cross sections is not feasible
due to the high accidentals rate\cite{coin_exp_bowglin1995}.} of the particle accelerators at the time ($f_{\mathrm{duty}}\sim10^{-5}$)\cite{sargsian_2015}. In the 1970s and
1980s, the duty factor of accelerators increased and it became possible to measure smaller $^{2}$H$(e,e'p)n$ cross sections, corresponding to larger missing momenta.
\begin{figure}[H]
\centering
\includegraphics[scale=0.42]{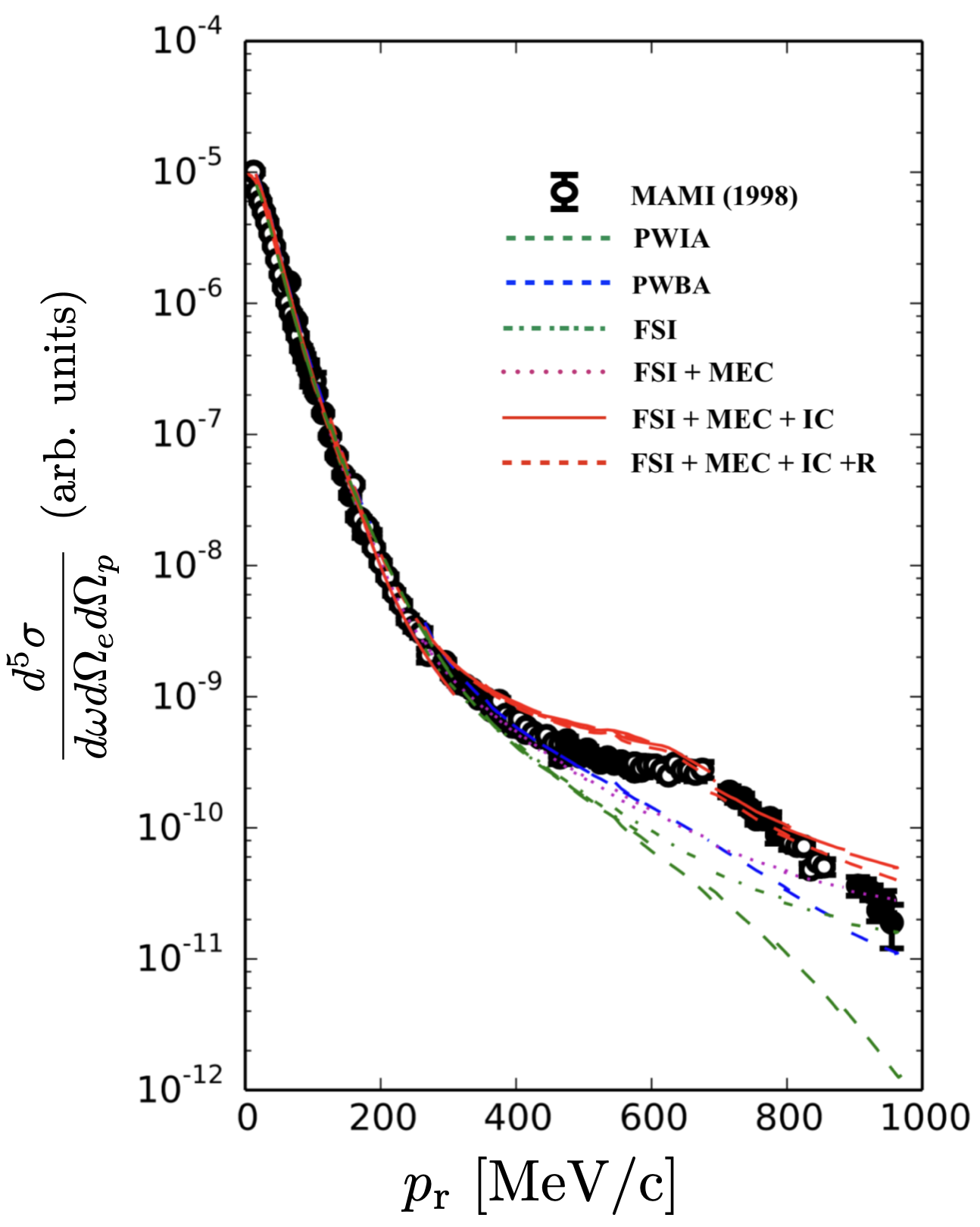}
\caption{$^{2}$H$(e,e'p)n$ cross section versus neutron recoil momentum from the MAMI (1998) experiment\cite{MAMI_1998}. Theoretical calculations were performed by H. Arenh\"{o}vel\cite{Arenhovel_1976}. Note: Reprinted from Ref.\cite{sargsian_2015}.}
\label{fig:fig1.3}
\end{figure}
For example, the Kharkov Institute\cite{d2_edisin_exp_1975}
2 GeV linac extended the missing momentum range to $\sim300$ MeV/c and experiments done at SACLAY\cite{d2_exp_1981,d2_exp_1984} measured $^{2}$H$(e,e'p)n$ cross sections up to missing momenta
$\sim500$ MeV/c. In the 1990s, the duty factor of electron accelerators increased further ($f_{\mathrm{duty}}\sim1$) such as the Amsterdan Pulse Stretcher (AmPS) at the National Institute for Nuclear 
and High Energy Physics (NIKHEF) in the Netherlands, the Mainz Microtron (MAMI) in Germany and Thomas Jefferson National Accelerator Facility (TJNAF) in the United States. The dramatic improvement
in the electron accelerator duty factor allowed for the first time measurements of very small cross sections in relatively short periods of time. For example, $^{2}$H$(e,e'p)n$ cross sections were
measured at missing momenta up to $\sim700$ MeV/c ($Q^{2}=0.28$ (GeV/c)$^{2}$) at NIKHEF\cite{Kasdorp1998} and $\sim928$ MeV/c ($Q^{2}=0.33$ (GeV/c)$^{2}$) at MAMI\cite{MAMI_1998}. While
at TJNAF\cite{PhysRevLett.89.062301}, the unique combination of high energy, duty factor and beam current allowed the measurements to be carried out for the first time at relatively high missing 
momnetum up to $\sim550$ MeV/c and $Q^{2}=0.665$ (GeV/c)$^{2}$.  \\
\indent The comparison of the results (see Fig. \ref{fig:fig1.3}) from the MAMI (1998) experiment with H. Arenh\"{o}vel's calculations\cite{Arenhovel_1976} demonstrated that only for very specific kinematics (e.g., SACLAY experiment in Ref.\cite{d2_exp_1981}), at missing momenta below $\sim200$ MeV/c
meson exchange currents (MEC), isobar configurations (IC) and final state interactions (FSI) are relatively small and cancel, leaving the plane wave born approximation (PWBA)\footnote{\singlespacing In the plane wave impulse approximation (PWIA), it is assumed that only the proton gets knocked out by the virtual photon
  whereas in the PWBA, the process in which the neutron is knock-out is also considered.} as the dominant contribution to the cross section. However, above $\sim300$ MeV/c, the PWBA (dashed blue), FSI (dashed-dotted green), MEC (dotted purple)
and IC (solid red) all contribute significantly to the $^{2}$H$(e,e'p)n$ cross section and obscure any possibility of extracting the momentum distributions (PWIA in dashed green). 

\section{First $^{2}$H$(e,e'p)n$ Experiments at Large $Q^{2}$ } \label{sec:section1.4}
\begin{figure}[H]
\centering
\includegraphics[scale=0.4]{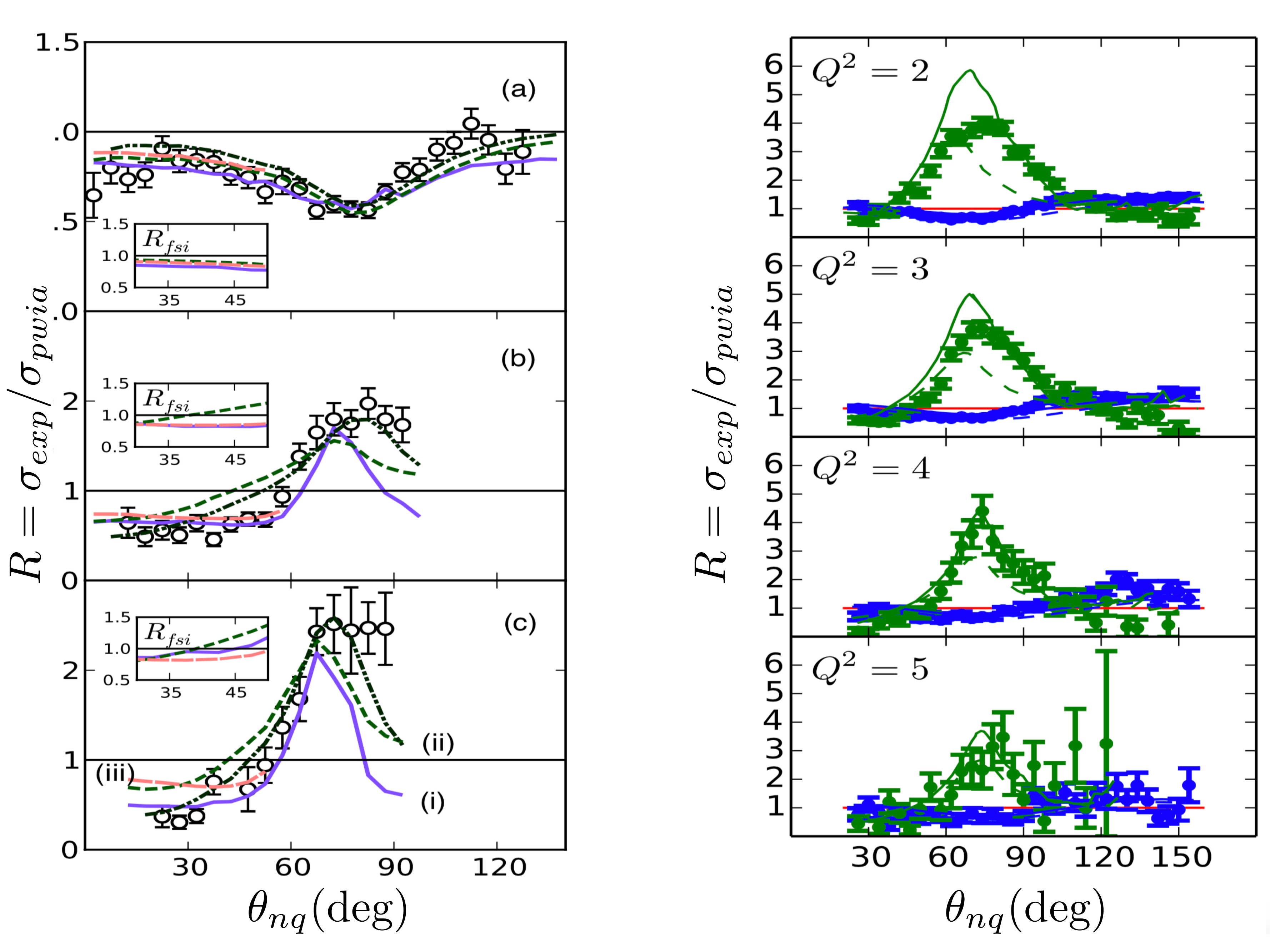}
\caption{$^{2}$H$(e,e'p)n$ angular distributions of the cross section ratio, $R = \sigma_{exp}/\sigma_{pwia}$. (Left) Hall A data at $Q^{2}=3.5\pm0.25$ (GeV/c)$^{2}$ and recoil momentum settings (a) $p_{\mathrm{r}}=0.2$ GeV/c,
  (b) $p_{\mathrm{r}}=0.4$ GeV/c and (c)  $p_{\mathrm{r}}=0.5$ GeV/c. Theoretical calculations for (i) solid (purple) curves using the CD-Bonn potential by M. Sargsian\cite{PhysRevC.82.014612},
  (ii) dashed (green) curves using FSI and dashed-double dotted (black) curves using FSI+MEC+IC by J.M. Laget\cite{LAGET2005} using the Paris potential and (iii) dashed
  (pink) curves denote calculations by J.W. Van Orden\cite{PhysRevC.78.014007}. (Right) Hall B data at various $Q^{2}$ settings. The green data (with FSI re-scattering peak) correspond to $400\leq p_{\mathrm{r}} \leq600$ MeV/c, and the blue
  data (no FSI re-scattering) correspond to $200\leq p_{\mathrm{r}}\leq 300$ MeV/c. The solid curves are calculations from J.M. Laget\cite{LAGET2005} and the dashed curves are from M. Sargsian\cite{PhysRevC.82.014612}. Note: Reprinted from Ref.\cite{sargsian_2015}.}
\label{fig:fig1.4}
\end{figure}

The first $^{2}$H$(e,e'p)n$ experiments at $Q^{2}>1$ (GeV/c)$^{2}$ were carried out at TJNAF in experimental Halls A\cite{PhysRevLett.107.262501} and B\cite{PhysRevLett.98.262502}.
Both experiments determined that the cross sections for fixed recoil momenta indeed exhibited a strong angular dependence with neutron recoil angles, peaking at $\theta_{nq}\sim70^{\circ}$ in
agreement with the generalized eikonal approximation (GEA) calculations\cite{Sargsian_2001,PhysRevC.56.1124} at high missing momentum and $Q^{2}>2$ (GeV/c)$^{2}$ (see Fig. \ref{fig:fig1.4}). \\
\indent In Hall B, the CEBAF Large Acceptance Spectrometer (CLAS) measured angular distributions for a range of $Q^{2}$ values as well as momentum distributions. However,
statistical limitations made it necessary to integrate over a wide angular range to determine momentum distributions that are therefore dominated by FSI, MEC and IC for recoil
momenta above $\sim300$ MeV/c.\\
\indent In Hall A, the pair of high resolution spectrometers (HRS) made it possible to measure the missing momentum dependence of the cross section for fixed neutron recoil angles ($\theta_{nq}$)
reaching missing momenta up to $p_{\mathrm{r}}=550$ MeV/c at $Q^{2}=3.5\pm0.25$ (GeV/c)$^{2}$. For the first time, very different momentum distributions were found for $\theta_{nq}=35\pm5^{\circ}$
and $45\pm5^{\circ}$ compared to $\theta_{nq}=75\pm5^{\circ}$. Theoretical models attributed this difference to the suppression of FSI at the smaller angles ($\theta_{nq}=35,45^{\circ}$) compared to
FSI dominance at $\theta_{nq}=75^{\circ}$\cite{PhysRevLett.107.262501}.

\section{Motivation} \label{sec:section1.5}
Being the most simple neutron-proton ($np$) bound state, the deuteron serves as a starting point to study the strong nuclear force (or $NN$ potential) without additional
complications that arise from $A>2$ nuclei. As mentioned before, the $NN$ potential is sub-divided into three regions with inter-nucleon distance $r$
as follows:
\begin{itemize}
\item the long range part (LR), where $r > 2$ fm
\item the intermediate or mid-range part (MR), where $1 < r < 2$ fm 
\item the short range part (SR), where $r < 1$ fm
\end{itemize}
The LR part is dominated by a single $\pi$ exchange, where usually the OPEP is used by most phenomenological models. The MR part is dominated by
2$\pi$ exchange or the exchange of a heavier mesons. Finally, the SR part is often modeled by a repulsive hard core and is determined
completely phenomenologically. It is this part that is least known from a theoretical point of view and the most difficult to access experimentally.
At such small inter-nucleon distances, from a Quantum Chromodynamics (QCD) perspective, a repulsive force is expected. For example, if one considers
the three quarks inside the nucleon, as the proton and neutron start to overlap, the quarks in each nucleon cannot be considered independent of the other.
Given that quarks are fermions, the Pauli exclusion principle prevents any two fermions from occupying the same quantum state. As a consequence, 
any three quarks must go to energy states above the lowest states occupied by the other three\cite{intro_nuclphys_wong1998}. This process requires a large
amount of energy that shows up as a resistance (repulsive hard core) to bring the two nucleons to sub-Fermi distances. \\
\indent From a nuclear physics perspective, the overlap between the nucleons in the deuteron is directly related to short-range correlations (SRCs) observed
in $A>2$ nuclei\cite{PhysRevC.68.014313,PhysRevLett.96.082501,PhysRevLett.99.072501,Fomin_2017}. Short-range studies of the deuteron are also important in determining
whether, or to what extent, the description of nuclei in terms of nucleon/meson degrees of freedom is still valid before having to include explicit quark effects, an issue
of fundamental importance in nuclear physics\cite{pr01-020}.\\
\indent Presently, there are only a few nuclear physics experiments from which a transition between nucleonic to quark degrees of freedom have been observed\cite{PhysRevLett.81.4576,PhysRevLett.87.102302,PhysRevC.66.042201,PhysRevC.70.014005}.
The experiment presented in this dissertation seeks to study the short range structure of the deuteron by extending the previous Hall A measurements\cite{PhysRevLett.107.262501} of the
$^{2}$H$(e,e'p)n$ cross section to $Q^{2}=4.5\pm0.5$ (GeV/c)$^{2}$ at Bjorken scale $x_{\mathrm{Bj}}>1$ and neutron recoil momenta up to $p_{\mathrm{r}}\sim1$ GeV/c, which is almost double
of the maximum recoil momentum previously measured in Hall A. Measurements at such large $Q^{2}$ and high missing momenta required a high beam energy and small electron scattering angles leading 
to the detection of electrons at $\sim$8.5 GeV/c, made possible with the newly commissioned Hall C Super High Momentum Spectrometer (SHMS). At the selected kinematic
settings with neutron recoil angles between 35$^{\circ}$ and 45$^{\circ}$, MEC, IC and FSI are mostly suppressed. This leaves the PWIA as the dominant
contribution to the $^{2}$H$(e,e'p)n$ cross section giving access to the high momentum components of the deuteron wave function.



\chapter{THEORETICAL BACKGROUND}\label{chap:chapter2}
In this chapter I will derive the basic formulas for the $^{2}$H$(e,e'p)$ reaction kinematics
using the Feynmann diagram of Fig. \ref{fig:fig2.1} (assume natural units for speed of light, $c = 1$). Then I will briefly
discuss the general reaction cross section and the various reaction mechanisms
that can occur. Finally, the theoretical models that are used to compare
with the experimental data will be discussed.

\section{The $^{2}$H$(e,e'p)n$ Reaction Kinematics} \label{sec:section2.1}
\begin{figure}[H]
\centering
\includegraphics[scale=0.338]{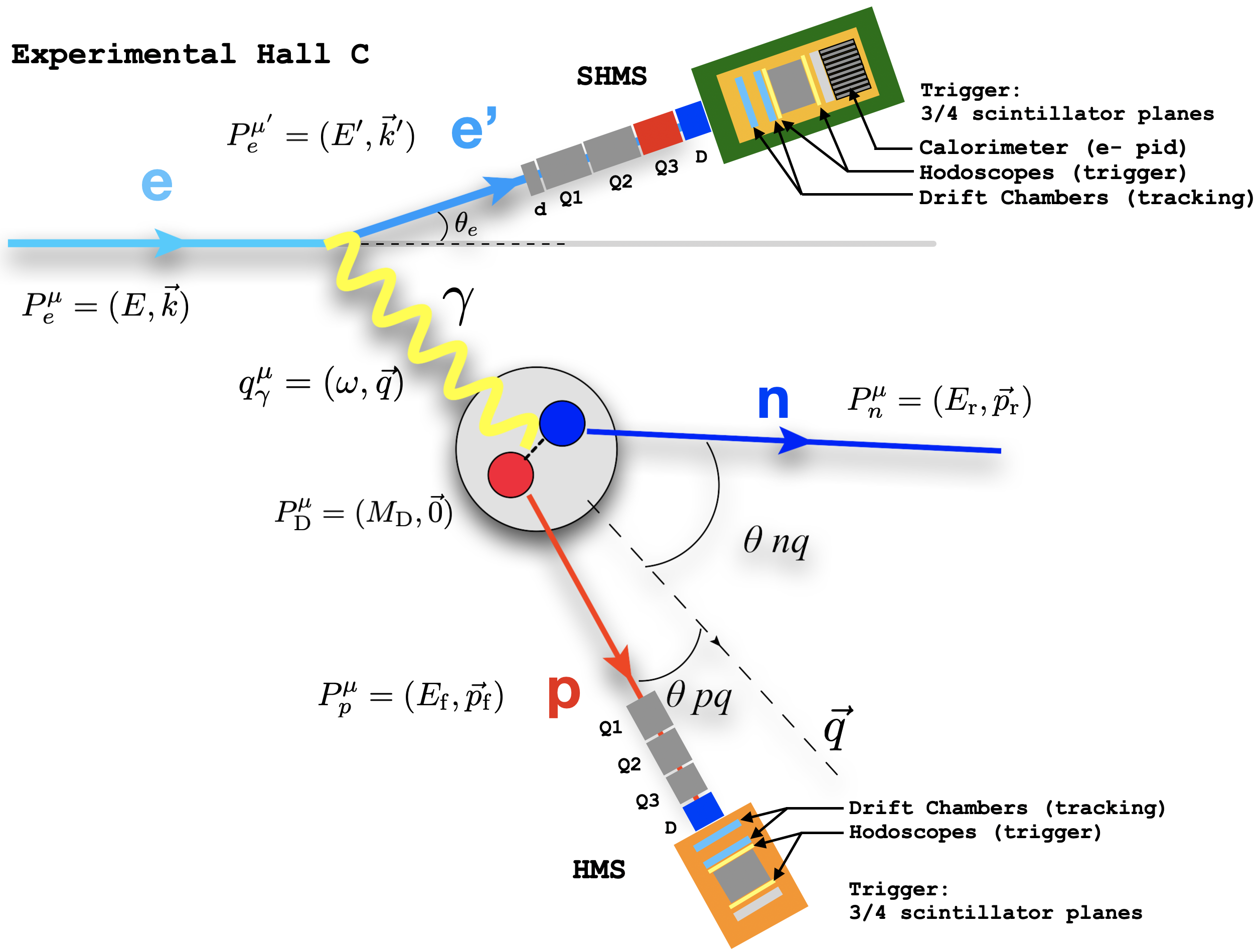}
\caption{Simplified Feynman diagram of the $^{2}$H$(e,e'p)n$ reaction kinematics and the respective four momenta of the interacting particles.}
\label{fig:fig2.1}
\end{figure} 
The deuteron electro-disintegration reaction kinematics can be described in QED by the exchange of a single virtual photon between the electron and deuteron
assuming a One-Photon Exchange Approximation (OPEA). The Feynman diagram in Fig. \ref{fig:fig2.1} describes a typical deuteron electro-disintegration
reaction, where the electron interacts with the deuteron via the exchange of a virtual photon that breaks the deuteron up into a proton and a neutron.
The scattered electron is detected by the Super High Momentum Spectrometer (SHMS) in coincidence with the knocked out proton in the High Momentum Spectrometer (HMS).
The ``missing'' (recoil) neutron is reconstructed from momentum conservation laws. \\
\begin{figure}[H]
\centering
\includegraphics[scale=0.3]{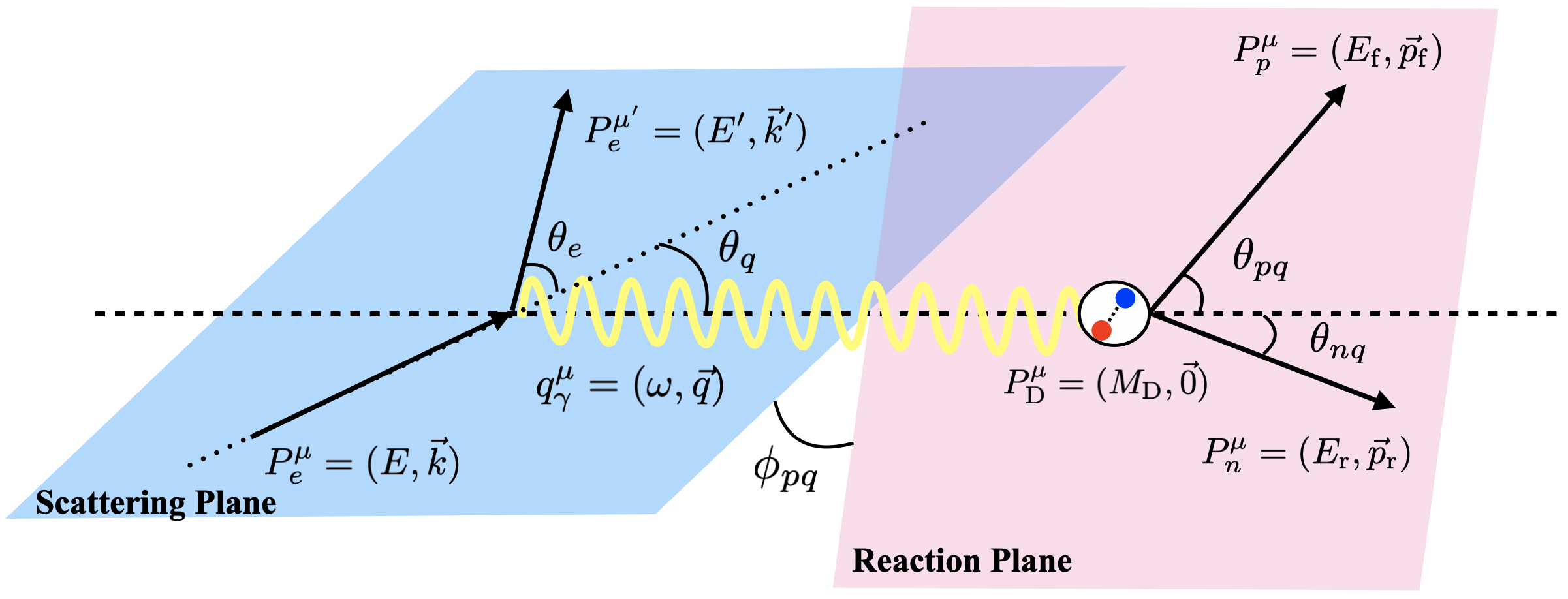}
\caption{General $^{2}$H$(e,e'p)n$ reaction kinematics.}
\label{fig:fig2.2}
\end{figure}
\indent Figure \ref{fig:fig2.2} shows a more detailed diagram of the $^{2}$H$(e,e'p)n$ reaction kinematics, where the scattering plane is defined by $\hat{y}_{\mathrm{sct}}=\hat{k}\cross\hat{k'}$
where $\hat{k}$ and $\hat{k'}$ are unit vectors in the direction of the incident and scattered electron, respectively, and $\hat{y}_{\mathrm{sct}}$ is a unit vector normal to the scattering plane which defines
its orientation. Similarly, the orientation of the reaction plane is defined by $\hat{y}_{\mathrm{react}}=\hat{q}\cross\hat{p_{\mathrm{f}}}$, where $\hat{q}$ and $\hat{p_{\mathrm{f}}}$ are unit vectors in
the direction of the virtual photon
and final state proton, respectively, and $\hat{y}_{\mathrm{react}}$ is a unit vector normal to the reaction plane.
The angle between the two planes is defined by $\hat{y}_{\mathrm{sct}}\cdot\hat{y}_{\mathrm{react}} = \cos(\phi_{pq})$. In Hall C, the angle
$\phi_{pq}$ is referred to as the out-of-plane angle between the two spectrometers, but since the spectrometers are always in the same plane, the two possible values are $\phi_{pq}=0^{\circ}$ or $180^{\circ}$, which
only apply to the central ray of the spectrometers. \\
\indent The relevant kinematic variables in the $^{2}$H$(e,e'p)n$ reaction can be obtained by applying energy and momentum conservation at the electron and hadron vertices in Fig. \ref{fig:fig2.2}.
At the electron vertex, the initial and final electron four momenta are $P^{\mu}_{e}=(E,\vec{k})$ and $P'^{\mu}_{e}=(E',\vec{k}')$, where the final electron scatters at angle $\theta_{e}$ relative to the incident
electron direction. The energy and momentum transfer carried by the virtual photon are defined as
\begin{align}
  q^{\mu}_{\gamma} \equiv P^{\mu}_{e} - P'^{\mu}_{e} = (E-E', \vec{k}-\vec{k}') = (\omega, \vec{q}).
  \label{eq:2.1}
\end{align}
By taking the negative square of Eq. \ref{eq:2.1} and assuming $E,E' \sim k, k'$ (electron mass $m^{2}_{e} \ll k^{2}, k'^{2}$), it is
convenient to define the four-momentum transfer of the virtual photon (also known as the virtuality) as
\begin{align}
  Q^{2} \equiv -q^{\mu}q_{\mu} &=  q^{2} - \omega^{2} \approx 4kk'\sin^{2}(\theta_{e}/2).
  \label{eq:2.2}
\end{align}
It is also convenient to define the Bjorken scale, $x_{\mathrm{Bj}}\equiv\frac{Q^{2}}{2M_{p}\omega}$, where $M_{p}$ is the proton mass.
At the hadron vertex, the deuteron nucleus with mass $M_{\mathrm{D}}$ is stationary with total internal momentum of the proton and neutron,
$\vec{p}_{p,\mathrm{i}} + \vec{p}_{n,\mathrm{i}} = \vec{0}$, and four-momentum, $P^{\mu}_{\mathrm{D}}=(M_{\mathrm{D}}, \vec{0})$. The final
state proton and neutron four-momenta are defined as $P^{\mu}_{p}=(E_{\mathrm{f}}, \vec{p}_{\mathrm{f}})$ and $P^{\mu}_{n}=(E_{\mathrm{r}}, \vec{p}_{\mathrm{r}})$, respectively.
Applying energy-momentum conservation at the hadron vertex,
\begin{align}
  q^{\mu}_{\gamma} + P^{\mu}_{\mathrm{D}} = P^{\mu}_{p} + P^{\mu}_{n} \implies (\omega, \vec{q}) + (M_{\mathrm{D}}, \vec{0}) = (E_{\mathrm{f}}, \vec{p}_{\mathrm{f}}) + (E_{\mathrm{r}}, \vec{p}_{\mathrm{r}}).
  \label{eq:2.3}
\end{align}
From energy conservation of Eq. \ref{eq:2.3},
\begin{align}
  &\omega + M_{\mathrm{D}} = E_{\mathrm{f}} + E_{\mathrm{r}} = T_{p} + M_{p} + T_{n} + M_{n} \nonumber \\
  &\implies E_{\mathrm{m}} \equiv E_{\mathrm{BE}} =  M_{p} +  M_{n} - M_{D} = \omega - T_{p} - T_{n},
  \label{eq:2.4}
\end{align}
where the missing energy ($E_{\mathrm{m}}$) is defined as the binding energy ($E_{\mathrm{BE}}$) of the deuteron and $(T_{p}, T_{n})$ are the final kinetic energies of the proton and neutron, respectively.
From momentum conservation of Eq. \ref{eq:2.3},
\begin{align}
  &\vec{p}_{\mathrm{f}} = \vec{q} - \vec{p}_{\mathrm{r}} \implies p_{\mathrm{f}}^{2} = q^{2} + p_{\mathrm{r}}^{2} - 2qp_{\mathrm{r}}\cos(\theta_{nq}).
  \label{eq:2.5}
\end{align}
Or equivalenlty, the neutron recoil momentum from Eq. \ref{eq:2.3} can be expressed as,
\begin{align}
  &\vec{p}_{\mathrm{r}} = \vec{q} - \vec{p}_{\mathrm{f}} \implies p_{\mathrm{r}}^{2} = q^{2} + p_{\mathrm{f}}^{2} - 2qp_{\mathrm{f}}\cos(\theta_{pq}).
  \label{eq:2.6}
\end{align}
Substituting Eq. \ref{eq:2.6} into Eq. \ref{eq:2.5} and solving for $\cos(\theta_{nq})$,
\begin{equation}
  \cos(\theta_{nq}) = \frac{q - p_{\mathrm{f}}\cos(\theta_{pq})}{\sqrt{q^{2} + p_{\mathrm{f}}^{2} - 2qp_{\mathrm{f}}\cos(\theta_{pq})}},
  \label{eq:2.7}
\end{equation}
where $(\theta_{nq}, \theta_{pq})$ refers to the angle between the virtual photon and the recoiling neutron ($\theta_{nq}$) or scattered proton ($\theta_{pq}$) direction.
From Eq. \ref{eq:2.7}, under the assumption $\vec{q}>\vec{p}_{p,\mathrm{i}}$ and that the proton is struck and the neutron is a spectator without further interaction, the
limiting cases are shown in Fig. \ref{fig:fig2.3}.\\
\indent From Fig. \ref{fig:fig2.3}, the proton (red) and neutron (blue) are initially inside the deuteron moving in opposite direction with total internal momentum ($\vec{p}_{p,\mathrm{i}} + \vec{p}_{n,\mathrm{i}} = \vec{0}$)
represented by the dashed vectors. The virtual photon (black solid vector) can interact with the proton as follows:
\begin{itemize}
  \item \underline{\textit{Anti-Parallel Kinematics:}} The virtual photon knocks out a proton initially moving along $\vec{q}$, transferring all its momentum to the proton
  in the final state (solid red vector) such that $\vec{q}<\vec{p}_{\mathrm{f}}$. The neutron recoils in opposite direction to $\vec{q}$ with missing momentum same as its internal
  momentum in the deuteron. 
  \item \underline{\textit{Parallel Kinematics:}} The virtual photon knocks out a proton initially moving opposite to $\vec{q}$, transferring all its momentum to the proton
  causing it to change direction in the final state such that $\vec{q}>\vec{p}_{\mathrm{f}}$. The neutron recoils along the direction of $\vec{q}$ with missing momentum same as its internal
  momentum in the deuteron.
  \item \underline{\textit{Perpendicular Kinematics:}} The virtual photon knocks out a proton initially moving perpendicular to $\vec{q}$, transferring all its momentum to the proton
    causing it to change direction in the final state such that $|\vec{q}|\sim|\vec{p}_{\mathrm{f}}|$ and $\theta_{pq}$ is at very small angles. The neutron recoils perpendicular to $\vec{q}$
    with missing momentum same as its internal momentum in the deuteron.
\end{itemize}
\begin{figure}[H]
\centering
\includegraphics[scale=0.34]{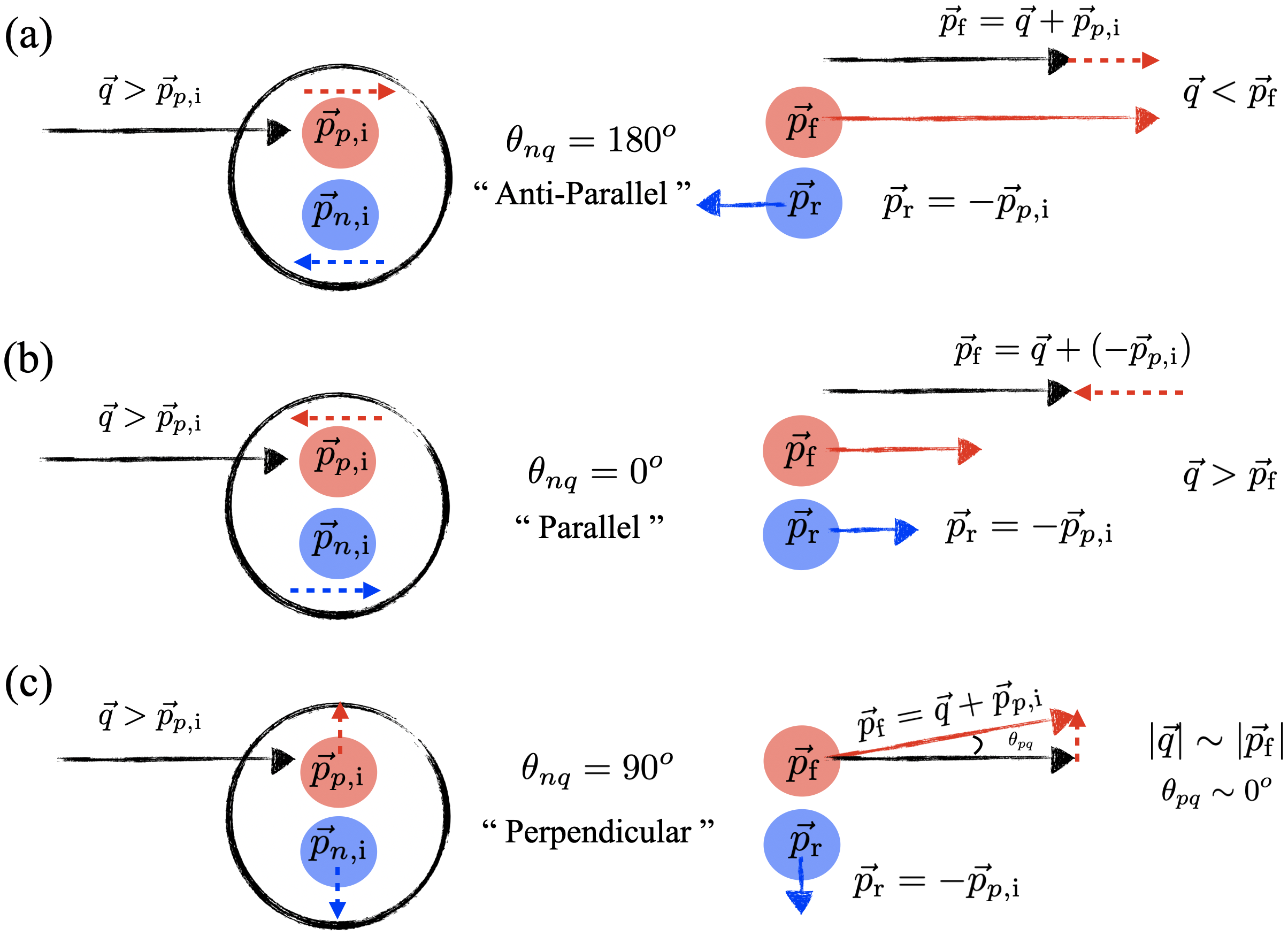}
\caption{(a) Anti-parallel kinematics. The neutron (blue) recoils in the opposite direction to $\vec{p_{\mathrm{f}}}$ (red) and $\vec{q}$ (black). (b) Parallel kinematics. The
neutron recoils in the same direction as $\vec{p_{\mathrm{f}}}$ and $\vec{q}$. (c) Perpendicular kinematics. The neutron recoils in a perpendicular direction relative to $\vec{q}$.}
\label{fig:fig2.3}
\end{figure}
These are limiting cases, but in general, the vectors do not have to be perfectly aligned with $\vec{q}$ when referring to these kinematics. It is sufficient if the final state vectors are approximately along $\vec{q}$.
The \textit{Parallel} and \textit{Anti-Parallel Kinematics} are more directly related to the short-range structure of the deuteron as FSI are expected to be reduced at these kinematics whereas
in the \textit{Perpendicular Kinematics}, FSI become dominant at higher missing momentum, which can lead to a larger inferred initial momentum than the true internal momentum of the proton\cite{pr01-020}.
This experiment (E12-10-003) has chosen kinematics ($\theta_{nq}$ at forward angles) that favor the \textit{Parallel Kinematics} for short-range structure studies of the deuteron.

\section{The $^{2}$H$(e,e'p)n$ Cross Section} \label{sec:theory_Xsec}
\begin{figure}[H]
\centering
\includegraphics[scale=0.38]{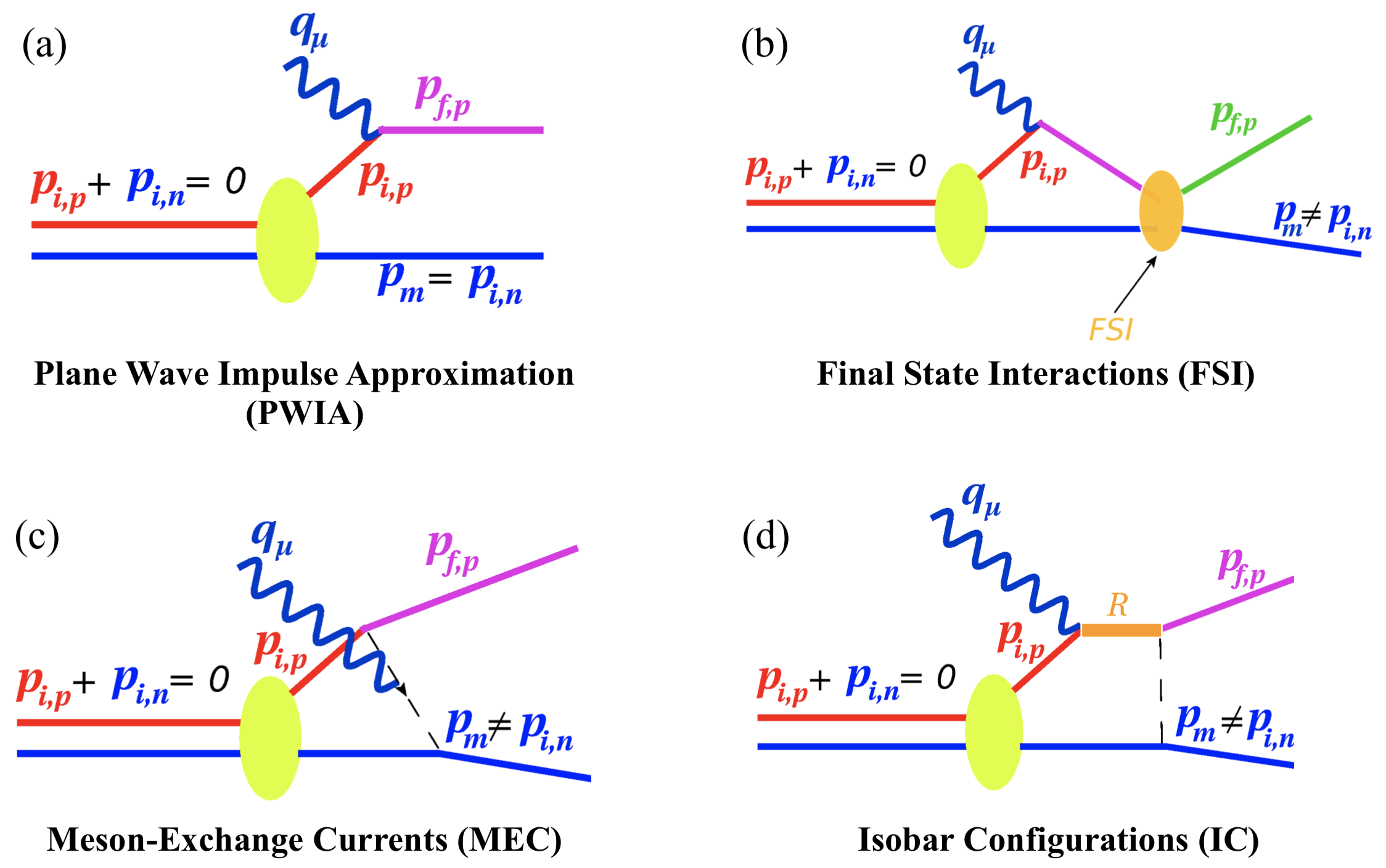}
\caption{$^{2}$H$(e,e'p)n$ Feynman Diagrams. The colors are: (yellow) deuteron target, (red) initial state proton, (blue solid) initial and final state neutron, 
  (magenta, green) final state proton, (orange oval) FSI re-scattering, (orange rectangle) intermediate resonance state, (navy blue) virtual photon and 
  (black dashed line) meson exchange.}
\label{fig:fig2.4}
\end{figure}

Assuming the OPEA, for the general $A(e,e'p)$ reaction where an electron is detected in coincidence with a knocked-out proton and the residual $(A-1)$ system recoils, the unpolarized 6-fold 
differential cross section can be expressed as (See Chapter 6 of Ref.\cite{Boffi_1993}):
\begin{equation}
\frac{d^{6}\sigma}{dE'd\Omega_{e}d\Omega_{p}dT_{p}} = \sigma_{Mott}(v_{\mathrm{L}}W_{\mathrm{L}} + v_{\mathrm{T}}W_{\mathrm{T}} + v_{\mathrm{LT}}W_{\mathrm{LT}}\cos\phi_{pq} + v_{\mathrm{TT}}W_{\mathrm{TT}}\cos2\phi_{pq}),
\label{eq:2.8}
\end{equation}
where the longitudinal ($W_{\mathrm{L}}$), transverse ($W_{\mathrm{T}}$) and interference ($W_{\mathrm{LT}}, W_{\mathrm{TT}}$) nuclear response functions are determined from matrix 
elements of the hadronic four-current operator and the leptonic kinematic factors $(v_{\mathrm{L}}, v_{\mathrm{T}}, v_{\mathrm{LT}}, v_{\mathrm{TT}})$ are determined from matrix elements of 
the leptonic four-current operator. The Mott cross section, $\sigma_{\mathrm{Mott}}$, describes the scattering of an electron off an infinitely 
massive and spinless point charge and is defined as
\begin{equation}
\sigma_{\mathrm{Mott}} = \Big(\frac{2\alpha k'\cos(\theta_{e}/2)}{Q^{2}}\Big)^{2},
\label{eq:2.9}
\end{equation}
where $\alpha\sim 1/137$ is referred to as the fine structure constant and characterizes the coupling strength of the electromagnetic interaction. \\
\newpage
\indent The leptonic kinematic factors and nuclear response functions are summarized in Tables 2 and 4 of Ref.\cite{Boffi_1993}. 
For a more detailed discussion of the formalism used to derive the leptonic and hadronic matrix elements from their respective
four-current operators refer to Chapter 2 and 6 of Ref.\cite{Boffi_1993}. \\
\indent The cross section in Eq. \ref{eq:2.8} can include the various nuclear processes such as MEC, IC and FSI,
which can significantly alter the nuclei momenta in the final state. For the deuteron in particular, the Feynman diagrams in Fig. \ref{fig:fig2.4} describe
possible reaction mechanisms, which are further discussed in the following sections.
\section{Plane Wave Impulse Approximation}
In the PWIA (Fig. \ref{fig:fig2.4}(a)), the virtual photon couples directly to the bound proton, which is subsequently ejected from the deuteron without any further interaction with the
recoiling neutron. The recoiling neutron carries a momentum equal in magnitude but opposite in direction to the initial momentum of the bound proton, 
$\vec{p}_{\mathrm{r}} = -\vec{p}_{p,\mathrm{i}}$, thus providing information on the momentum of the bound proton and its momentum distribution. Within
the PWIA, the general cross section in Eq. \ref{eq:2.8} can be factorized as follows:
\begin{equation}
\frac{d^{6}\sigma}{dE'd\Omega_{e}d\Omega_{p}dT_{p}} = K\sigma_{eN}S(\vec{p}_{p,\mathrm{i}}, E_{\mathrm{m}}),
\label{eq:2.10}
\end{equation}  
where $\sigma_{eN}$ describes the elementary cross section for an electron scattering off a bound (off-shell) nucleon where the deForest\cite{DEFOREST1983}
off-shell cross sections, $\sigma_{\mathrm{cc1}}$ or $\sigma_{\mathrm{cc2}}$, are commonly used. The kinematic factor that results from the factorization is defined
as $K\equiv E_{\mathrm{f}}p_{\mathrm{f}}$, and $S(\vec{p}_{p,\mathrm{i}}, E_{\mathrm{m}})$ is referred to as a spectral function, which describes the probability of
finding a bound proton with momentum $\vec{p}_{p,\mathrm{i}}$ and separation energy $E_{\mathrm{m}}$. The separation (binding) energy of the bound state can be 
integrated out of Eq. \ref{eq:2.10} to obtain,
\begin{equation}
\sigma_{\mathrm{theory}} \equiv \frac{d^{5}\sigma_{\mathrm{theory}}}{dE'd\Omega_{e}d\Omega_{p}} = Kf_{\mathrm{rec}}\sigma_{eN}S(\vec{p}_{p,\mathrm{i}}),
\label{eq:2.11}
\end{equation}
where $f_{\mathrm{rec}}$ is the recoil factor that arises from the integration in $E_{\mathrm{m}}$ and is defined as\cite{WB_privOct19}
\begin{equation}
  f_{\mathrm{rec}} \equiv \frac{1}{1 - \frac{1}{2}\frac{E_{\mathrm{f}}}{E_{\mathrm{r}}}\frac{q^{2} - (p^{2}_{\mathrm{f}}+p^{2}_{\mathrm{r}})}{p^{2}_{\mathrm{f}}}}.
  \label{eq:2.12}
\end{equation}
For the deuteron, the spectral function is interpreted as the momentum distribution of the proton inside a nucleus. Experimentally, the reduced cross section
is determined from the experimental cross section by
\begin{equation}
  \sigma_{\mathrm{red}} \equiv \frac{\sigma_{\mathrm{exp}}}{Kf_{\mathrm{rec}}\sigma_{eN}},
  \label{eq:2.13}
\end{equation}
where $\sigma_{\mathrm{exp}} \equiv \frac{d^{5}\sigma_{\mathrm{exp}}}{dE'd\Omega_{e}d\Omega_{p}}$.
If the PWIA were completely valid, $\sigma_{\mathrm{red}}$ would be the deuteron momentum distribution.  \\
\indent The inclusion of the process (not shown in Fig. \ref{fig:fig2.4}) in which the virtual photon couples to the neutron and the proton is a spectator is often defined as the Plane Wave Born Approximation (PWBA) and
can be suppressed by choosing the appropiate kinematics such that
the 3-momentum transfer ($\vec{q}$) is significantly greater than the largest missing momentum ($p_{m}$) studied and approximately on the order of
the momentum of the ejected proton. Both of these conditions are satisfied in this experiment. \\
\indent In reality, long-range processes such as FSI, MEC and IC always contribute to some extent to the total $^{2}$H$(e,e'p)n$ cross section,  
hence the word \textit{``Approximation''} in PWIA. As will be discussed next, these long-range contributions can significantly alter the recoiling 
neutron momentum, thereby obscuring the initial momentum distribution of the bound nucleon reducing the possibility of directly probing the high momentum component of
the deuteron wave function.
\section{Final State Interactions}
In direct FSI (Fig. \ref{fig:fig2.4}(b)), the ejected proton and recoiling neutron continue to interact further causing re-scattering of both nucleons.
This situation is unfavorable for the extraction a momentum distribution as during the interaction of the knocked-out proton with the recoiling neutron,
momentum is being exchanged leading to $\vec{p}_{\mathrm{r}}\neq -\vec{p}_{p,\mathrm{i}}$. As any possible momentum can be exchanged between the final state
particles, they are not considered plane waves but rather distorted waves and the factorization of the cross section breaks down. If the remaining conditions for the PWIA are still valid, the
spectral function in Eq. \ref{eq:2.11} can be replaced by a distorted spectral function, $S_{\mathrm{D}}(\vec{p}_{\mathrm{r}}, \vec{p}_{\mathrm{f}})$, and the approximation is
regarded as a Distorted Wave Impulse Approximation (DWIA). See Section 6.4 of Ref.\cite{Boffi_1993} for details. \\
\indent At large missing momenta ($p_{\mathrm{r}}>300$ MeV/c) and high $Q^{2}$, FSI exhibit a strong angular dependence on $\theta_{nq}$ with maximal FSI re-scattering
at $\theta_{nq}\sim70^{\circ}$ and a minimal re-scattering at $\theta_{nq}=40^{\circ}$ and $120^{\circ}$ as predicted by the GEA\cite{PhysRevC.56.1124,Sargsian_2001} and confirmed by the previous 
Halls A and B experiments\cite{PhysRevLett.98.262502,PhysRevLett.107.262501}. From these observations, it became clear that FSI dominates the deuteron cross section at 
the \textit{Perpendicular Kinematics} ($\theta_{nq}\sim70^{\circ}$) whereas in the \textit{Parallel/Anti-Parallel Kinematics} ($\theta_{nq}\sim40^{\circ},120^{\circ}$) it is significantly reduced (see Fig. \ref{fig:fig2.3}).
\section{Meson Exchange Currents and Isobar Configurations}\label{sec:MEC_IC_theory}
In the MEC diagram (Fig. \ref{fig:fig2.4}(c)), the virtual photon couples to the virtual meson being exchanged between the two nucleons, whereas in the
IC diagram (Fig. \ref{fig:fig2.4}(d)), the virtual photon excites a bound nucleon into an intermediate isobar resonance state ($\Delta$) that subsequently decays ($\Delta N\rightarrow NN$) via FSI to the ground state
causing further re-scattering between the final state nucleons via the exchange of a pion. \\
\indent Early $^{2}$H$(e,e'p)n$ experiments\cite{d2_exp_1984, MAMI_1998, Kasdorp1998, PhysRevLett.89.062301} 
showed that at low $Q^{2}$ and high missing momenta, MEC and IC contribute significantly to the deuteron cross section. At large $Q^{2}$, however, 
from a theoretical perspective, these contributions are expected to be significantly reduced. \\
\indent The suppression of MEC can be understood from the fact that the estimated MEC scattering amplitude ($A_{\mathrm{MEC}}$) is proportional to
the meson propagator in the electromagnetic current operator($J^{\mu}_{m}(Q^{2})$) and the $NN$-meson form factor ($\Gamma_{MNN}(Q^{2})$)
that have the following $Q^{2}$ dependence\cite{Sargsian_2001},
\begin{align}
  A_{\mathrm{MEC}} \propto J^{\mu}_{m}(Q^{2})\Gamma_{MNN}(Q^{2}) \propto  \frac{1}{(1 + Q^{2}/m^{2}_{\mathrm{meson}})} \frac{1}{(1 + Q^{2}/\Lambda)^{2}},
\end{align}
where $m_{\mathrm{meson}}\sim0.71$ GeV and $\Lambda\sim0.8-1$ GeV$^{2}$.
Therefore, at $Q^{2}>1$ (GeV/c)$^{2}$, MEC are expected to be suppressed by an
overall factor of $\sim1/Q^{6}$ as compared to the PWIA. \\
\begin{figure}[H]
\centering
\includegraphics[scale=0.46]{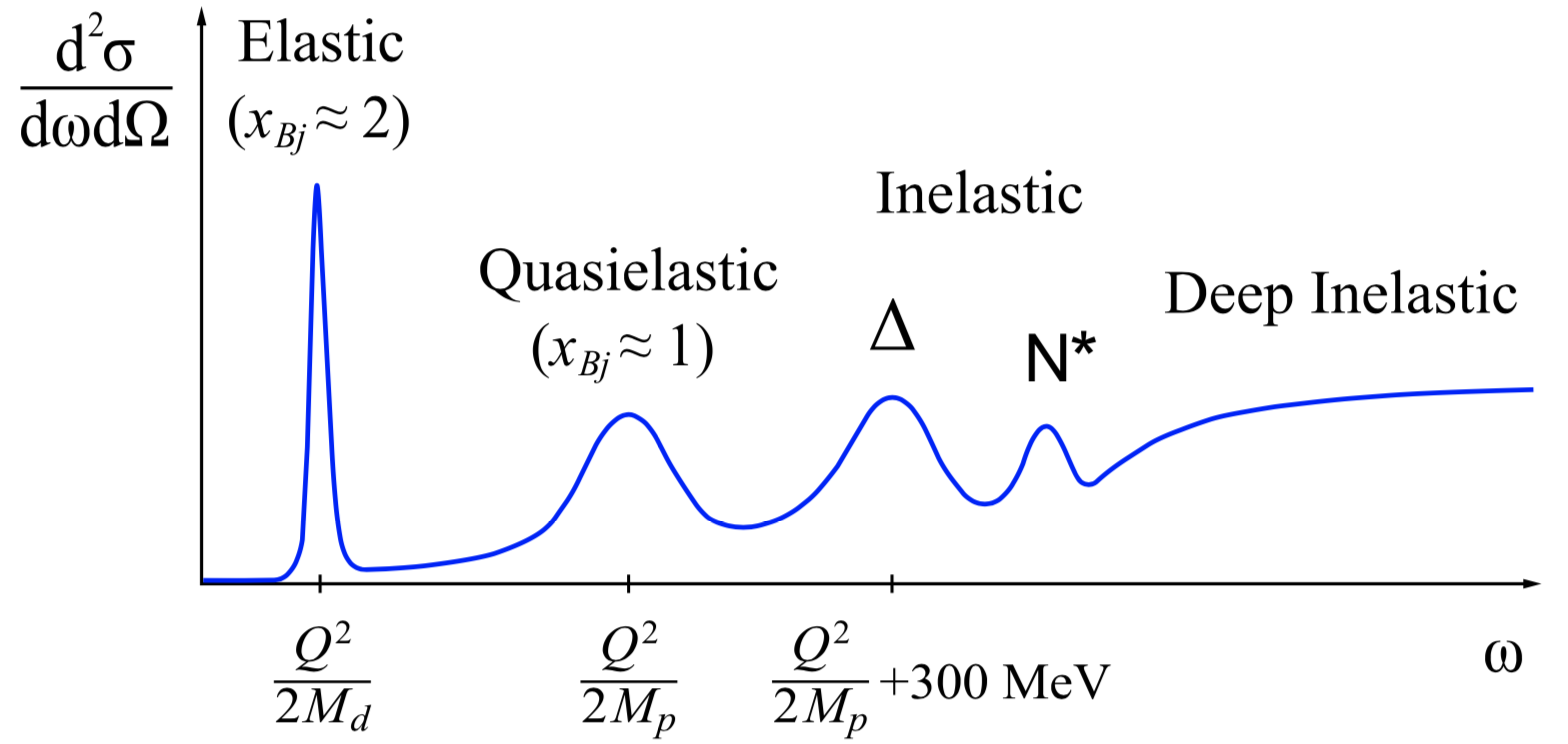}
\caption{Qualitative inclusive deuteron-electron scattering cross section. Note: Reprinted from Ref.\cite{HIbraham_phdthesis}.}
\label{fig:fig2.5}
\end{figure}
\indent The suppression of IC arises in part due to the kinematics chosen. At large $Q^{2}$, one is able to
select $x_{\mathrm{Bj}}>1$, which corresponds to probing the lower energy ($\omega$) part of the deuteron quasi-elastic peak, which is
maximally far away from the inelastic resonance electroproduction threshold. From Fig. \ref{fig:fig2.5}, the inclusive $^{2}$H$(e,e')$
shows qualitatively that at the left end of the quasi-elastic peak ($x_{\mathrm{Bj}}>1$) one is maximally away from the inelastic $\Delta$ and $N^{*}$ resonance
electroproduction region and corresponds to the kinematics where this experiment was done.
\section{From Theoretical Potentials to Cross Sections}
\begin{figure}[H]
\centering
\includegraphics[scale=0.40]{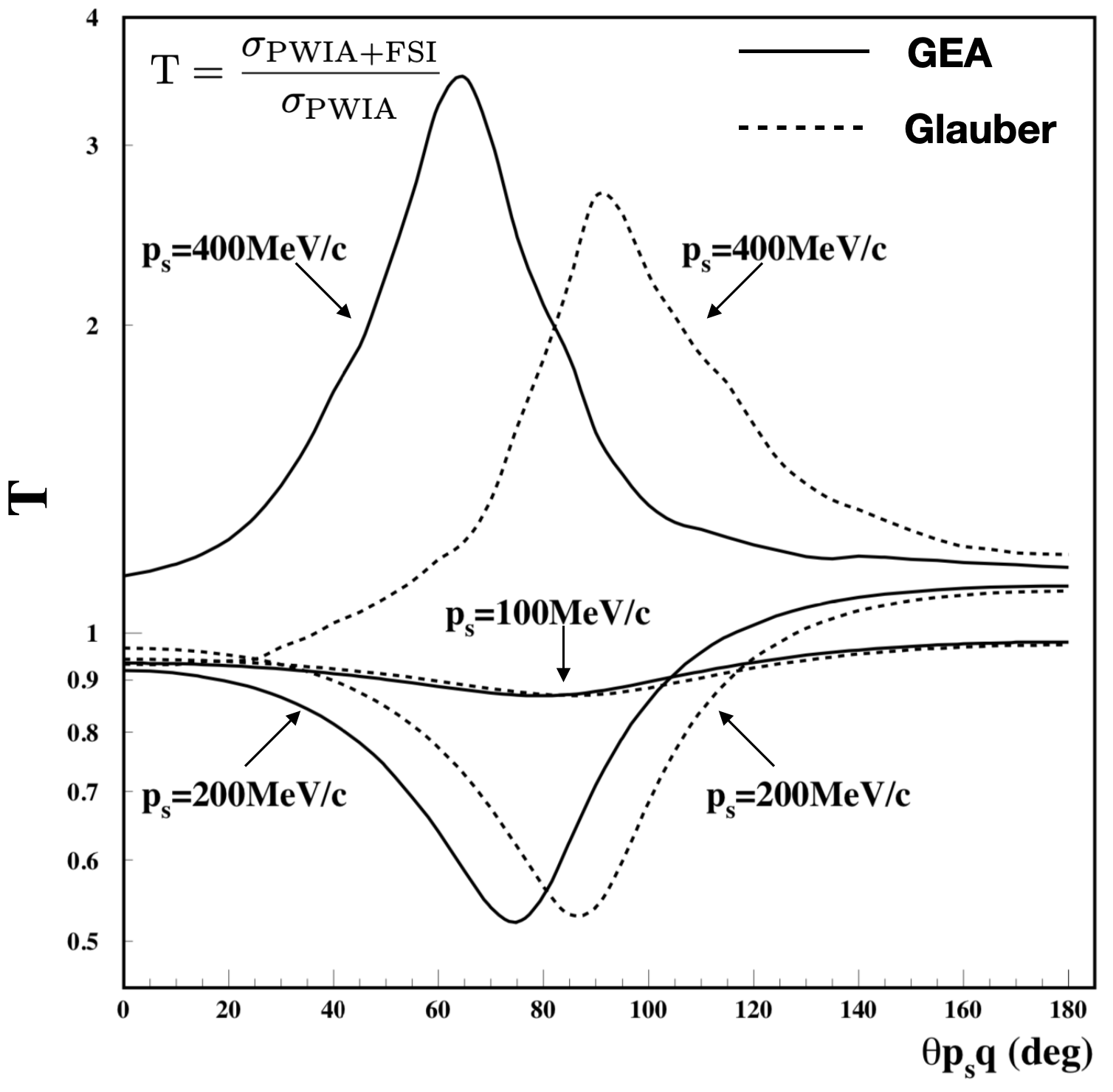}
\caption{Ratio of the theoretical cross section ratios versus neureon recoil angles $\theta_{nq}$ (denoted as $\theta_{\mathrm{p_{s}q}}$ in the figure) calculated within both the GEA (solid) and Glauber approximation (dashed) for
varios neutron recoil momenta $p_{\mathrm{r}}$ (denoted as $p_{\mathrm{s}}$ in the figure).  Note: Reprinted from Ref.\cite{Sargsian_2001}. }
\label{fig:fig2.6}
\end{figure}

In the E12-10-003 experiment, the theoretical cross sections used to compare to data were determined using the
following phenomenological $NN$-potentials:\\
\begin{itemize}
\item Parametrized Paris (1980)\cite{Paris_NN_Lacombe1980}
\item Argonne V18 (AV18) (1995)\cite{AV18_1995}
\item Charge-Dependent Bonn (CD-Bonn) (2001)\cite{CDBonn_NN_Machleidt2001}
\end{itemize}
Each of these potentials are improved versions of the original potentials and were developed by Paris, Argonne and Bonn theoretical groups, respectively. 
The groups have employed different techniques used in their approach to describe the intermediate and short range parts of the $NN$ potential, 
whereas for the long-range part, all have used the well-known One Pion Exchange Potential.\\ 
\indent In general, the construction of $NN$ potentials is largely based on parameters that the model must fit to either 
neutron-neutron ($nn$), proton-proton ($pp$) or neutron-proton ($np$) scattering data and the results are usually 
presented in texts as $\chi^{2}/$datum to determine the success of the model in describing the experimental data. \\
\indent The calculations to determine the theoretical cross sections from an $NN$ potential are based on solving the Schrodinger equation,
\begin{equation}
\hat{H_{\mathrm{D}}}\psi_{\mathrm{D}}(\vec{r}) = E_{\mathrm{D}}\psi_{\mathrm{D}}(\vec{r}) \implies (\hat{T_{p}} + \hat{T_{n}} + \hat{V}_{NN})\psi_{\mathrm{D}}(\vec{r}) = E_{\mathrm{D}}\psi_{\mathrm{D}}(\vec{r}),
\label{eq:2.15}
\end{equation}
where $\hat{H_{\mathrm{D}}}$ is the Hamiltonian operator that acts on the deuteron wave function ($\psi_{D}(\vec{r})$) and is expressed in terms of the 
proton and neutron kinetic energy operators ($\hat{T_{p}},\hat{T_{n}}$) and the interactive $NN$ potential ($\hat{V}_{NN}$), which is determined
by the theory groups. By solving Eq. \ref{eq:2.15}, the deuteron wave function as well as the scattering amplitude
(and theoretical cross section) can be determined. In reality, Eq. \ref{eq:2.15} is restricted to the non-relativistic description of the wave function as it uses a
classical definition of the kinetic energies in the Hamiltonian. In this situation, a generalized form of the Schrodinger wave equation can be used to describe the system relativistically.
Alternatively, the Bethe-Salpeter equation\cite{PhysRev.84.1232}, which uses a relativistically covariant formalism (Feynman S-matrix formalism),
can also used to describe a 2-body bound state including relativistic effects. \\
\indent Different theoretical calculations\cite{Arenhovel_1976,Bianconi_1995,PhysRevC.82.014612,PhysRevC.78.014007,LAGET2005,PhysRevC.90.064006} have been developed to describe the deuteron wave function
within the PWIA as well as to account for additional processes such as FSI, MEC or IC that are not described by theoretical potentials. In addition, some of the most recent
theoretical calculations also account for off-shell effects\footnote{\singlespacing The off-shell effects arise from the fact that for a bound system, the energy-momentum conservation applies to the nucleus as a whole, but the momentum of a pair
of nucleons within the nucleus is no longer restricted and the individual particles are considered to be ``off the energy shell'' (off-shell). Whereas for a pair of free interacting nucleons, the energy-momentum
conservation applies and the particles are said to be ``on the energy-shell'' (on-shell).}, which become important at higher missing momenta\cite{PhysRevC.78.014007,PhysRevC.82.014612}. Some theoretical potentials may also include off-shell effects in their
models, however, there is no way of knowing \textit{a priori} whether they are correct since these potentials were derived from $NN$ scattering data where the interacting particles are by definition on their energy shell (free interacting particles).
See Chapter 2 of Ref.\cite{intro_nuclphys_wong1998} for a detailed discussion. \\
\indent  In this experiment, the theoretical calculations used to determine the $^{2}$H$(e,e'p)n$ cross sections from the AV18 and CD-Bonn potentials were performed by M. Sargsian\cite{PhysRevC.82.014612}, while those for 
the Paris potential were by J.M. Laget\cite{LAGET2005}. In the former, an effective Feynman diagrammatic approach described in Ref.\cite{Sargsian_2001} is used to calculate the scattering amplitudes within the virtual
nucleon approximation. This approximation has three main assumptions described in Ref.\cite{PhysRevC.82.014612}, which also defines its range of validity. The first two assumptions are satisfied by requiring the neutron recoil momenta
to be $p_{\mathrm{r}}\leq 700$ MeV/c, while the third assumption made is that at large $Q^{2}$($>$1 GeV$^{2}$), MEC are considered to be a sufficiently small correction (see Section \ref{sec:MEC_IC_theory}) such that they can be ignored.
The assumptions of the virtual nucleon approximation restrict the Feynman diagrams to the PWIA (Fig. \ref{fig:fig2.4}(a)), direct FSI (Fig. \ref{fig:fig2.4}(b)), charge exchange FSI\footnote{\singlespacing This process corresponds to the
  scenario in which the virtual photon strikes a neutron that re-interacts with the spectator proton in the final state via $np\rightarrow pn$ charge-exchange re-scattering.} (not shown) and IC (Fig. \ref{fig:fig2.4}(d)) where the IC can be suppressed
kinematically by choosing $x_{\mathrm{Bj}}>1$ and was not considered in the calculations from Ref.\cite{PhysRevC.82.014612}. These Feynman diagrams constitute the basis for the theoretical framework of the generalized eikonal approximation (GEA)\cite{PhysRevC.56.1124,Sargsian_2001},
which uses the effective Feynman diagram rules described in Ref.\cite{Sargsian_2001} to determine the PWIA and FSI scattering amplitudes in covariant form that account for off-shell effects. \\
\indent The GEA predicts a strong angular anisotropy observed in FSI as a function of the neutron recoil angles peaking at $\theta_{nq}\sim 70^{\circ}$. This prediction was confirmed by the first high $Q^{2}$ deuteron electro-disintegration experiments
carried out at Halls A\cite{PhysRevLett.107.262501} and B\cite{PhysRevLett.98.262502} of Jefferson Lab (see Fig. \ref{fig:fig1.4}). Additionally, it was also found that at very forward and backward neutron recoil angles, FSI were significantly
reduced and comparable to the PWIA. The reduction in FSI can be understood from the fact in the high energy limit ($Q^{2}>1$ GeV$^{2}$) of the GEA, the $pn$ re-scattering amplitude is mostly imaginary:
\begin{align}
  A &= A_{\mathrm{PWIA}} + iA_{\mathrm{FSI}},
  \label{eq:2.16}
\end{align}
with $A_{\mathrm{FSI}} \approx i|A_{\mathrm{FSI}}|$, where the total scattering amplitude $A$ is expressed as the sum of the PWIA ($A_{\mathrm{PWIA}}$) and the imaginary part of the FSI ($A_{\mathrm{FSI}}$) scattering amplitudes. The total theoretical cross section can then be
obtained by taking the modulus square of the total scattering amplitude and can be expressed as
\begin{equation}
  \sigma_{\mathrm{PWIA+FSI}} \sim |A|^{2} = |A_{\mathrm{PWIA}}|^{2} - 2\underbrace{|A_{\mathrm{PWIA}}||A_{\mathrm{FSI}}|}_{\substack{ \text{``Screening'' or } \\ \text{interference term} }} + \underbrace{|A_{\mathrm{FSI}}|^{2}}_{\text{re-scattering term}}.
\end{equation}
Taking the ratio of the total to the PWIA part of the cross section,
\begin{equation}
  R = \frac{\sigma_{\mathrm{PWIA+FSI}}}{\sigma_{\mathrm{PWIA}}} = 1 - 2\frac{|A_{\mathrm{PWIA}}||A_{\mathrm{FSI}}|}{|A_{\mathrm{PWIA}}|^{2}} + \frac{|A_{\mathrm{FSI}}|^{2}}{|A_{\mathrm{PWIA}}|^{2}}.
\end{equation}
From the ratio of cross sections the interference term enters with an opposite sign as compared to the re-scattering term, which provides an opportunity for an approximate cancellation at certain
neutron recoil angles as shown in Fig. \ref{fig:fig1.4} of Ref.\cite{sargsian_2015}. This cancellation is also approximately independent of the neutron recoil momenta, which opens a kinematic
window at $\theta_{nq}\sim$40$^{\circ}$ where one can probe the short-range structure of the deuteron beyond $p_{\mathrm{r}}\sim500$ MeV/c and is the main focus of this experiment. \\
\indent In contrast to GEA approach used by M. Sargsian, J.M. Laget uses a diagrammatic approach described in Ref.\cite{LAGET2005}, which he first introduced in Ref.\cite{LAGET1981} and is used to calculate the
theoretical cross sections including the IC, MEC and FSI contributions. The kinematics of this experiment suppress IC and MEC contributions and therefore we only used the PWIA and FSI
contributions to the theoretical cross sections to comparare with the data. The PWIA and FSI scattering amplitudes for the $^{2}$H$(e,e'p)n$ reaction have been reproduced by Laget in Ref.\cite{LAGET2005}
and utilize the relativistic expressions of the proton and neutron on-shell current density operators, ($J_{p}(q)$, $J_{n}(q)$), in both amplitudes. The current densities use the conventional dipole expression
for the magnetic form factors of the proton and neutron, while for the neutron and proton electric form factor, the Galster parametrization\cite{Galster_param1971} and the results from the Hall A
experiment described in Ref.\cite{PhysRevLett.88.092301} were used, respectively.\\
\indent Similar to the predictions from the GEA, the FSI calculations from J.M. Laget also show that the FSI peak at $\theta_{nq}\sim$70$^{\circ}$ for $p_{\mathrm{r}}\sim$500 MeV/c, whereas for lower recoil
momenta, the peak shifts towards larger recoil angles with a dip at about $\theta_{nq}\sim90^{\circ}$ for the smallest recoil momenta. This can be understood from the fact that as the incident electron
scatters from a proton at rest, it transfers most of its energy and momentum to the struck proton while the neutron (also at rest) recoils at $\theta_{nq}\sim90^{\circ}$, which is predicted by
a non-relativistic eikonal approximation known as the Glauber approximation\cite{PhysRev.100.242}. \\
\indent The Glauber approximation assumes the bound nucleons are stationary within the nucleus and predicts
an FSI re-scattering peak at $\theta_{nq}\sim90^{\circ}$ corresponding to the transverse re-scattering of the stationary neutron relative to the exchanged virtual photon direction. For configurations
in which the internal momenta of the nucleons increases, this approximation is valid up to a certain extent for nucleon recoil momenta up to about
the fermi momentum, $k_{\mathrm{F}}\sim 250$ MeV/c. Beyond the fermi momentum, however, the relativistic effects within the nucleus cannot be ignored and must be accounted for in the theoretical calculations.
In the classical Glauber approximation, these relativistic effects are ignored since the nucleon propagator is linearized and the FSI peak stays at $\theta_{nq}\sim90^{\circ}$ for recoil momenta $p_{\mathrm{r}}>k_{\mathrm{F}}$.
When relativistic effects are accounted for, as it is done for both the GEA and Laget's diagrammatic approach, the FSI re-scattering peak shifts towards $\theta_{nq}\sim70^{\circ}$ with increasing recoil momenta. The
agreement of the FSI peak location between M. Sargsian's and J.M. Laget's approach can be understood from the fact that in the GEA, the higher order recoil terms in the nucleon propagator are accounted for while in the
calculations by Laget, the full kinematics of the reaction are taken into account from the beginning of the calculations as stated in Ref.\cite{LAGET2005}.\\
\indent To illustrate the results from this discussion, Fig. \ref{fig:fig2.6} shows the ratio of the theoretical cross sections using the PWIA+FSI calculations to cross sections calculated within the PWIA plotted versus neutron
recoil angles. At the lowest missing momenta ($p_{\mathrm{r}}=100$ MeV/c), the GEA and Glauber calculations are within almost a perfect agreement, which validates the GEA approach, which reduces to the Glauber approximation at very
small recoil momenta well within the fermi momentum of the nucleons. At $p_{\mathrm{r}}=200$ MeV/c, however, a shift in the FSI peak can already be observed towards $\sim80^{\circ}$ whereas at $p_{r}=400$ MeV/c, a significant shift of $\sim30^{\circ}$
from $\theta_{nq}\sim90^{\circ}$ to $\theta_{nq}\sim70^{\circ}$ can be observed. While for the Glauber approximation, the FSI peak stays ``fixed'' at $\theta_{nq}\sim90^{\circ}$.


\newcommand{\hhodthrs}{-44.5 mV }       
\newcommand{\hhodgate}{60 ns}
\newcommand{\hPrShLo}{-40 mV }
\newcommand{\hPrShHi}{-60 mV }
\newcommand{\hSHLo}{-45 mV }
\newcommand{\hPrShLogate}{30 ns}
\newcommand{\hPrShHigate}{30 ns}
\newcommand{\hSHLogate}{30 ns}
\newcommand{\hcerthrs}{-50 mV }
\newcommand{\hcergate}{30 ns}

\newcommand{\shodthrs}{-30 mV }
\newcommand{\quartzthrs}{-60 mV }
\newcommand{\shodgate}{60 ns} 
\newcommand{\sngcthrs}{-50 mV}
\newcommand{\sngcgate}{30 ns}
\newcommand{\shgcthrs}{-50 mV}
\newcommand{\shgcgate}{30 ns}
\newcommand{\saerthrs}{-50 mV}
\newcommand{\saergate}{30 ns}

\chapter{EXPERIMENTAL SETUP}	\label{chap:chapter3}
In this chapter I will discuss the experimental equipment used to carry out the
12 GeV Hall C commissioning experiments at the Continuous Electron Beam
Accelerator Facility (CEBAF). First I will give a brief overview of the accelerator and then discuss the Hall C 12 GeV upgrade
and components required for experiments: beamline, target, spectrometer systems (magnets and detectors)
and the trigger electronics setup used to collect data.
\section{CEBAF Accelerator Overview}
With the discovery of quarks inside the proton in a series of $ep$ scattering experiments at
Stanford Linear Accelerator (SLAC)\cite{PhysRevLett.23.930,PhysRevLett.23.935} in the late 1960s
and the development of a new theory of strong interactions (QCD) in the early 1970s, many questions
regarding the role of quarks in nuclear forces arose. For example, ``Why weren't the effects of the
underlying quark structure immediately visible?" or ``Could new phenomena be discovered that were a direct
consequence of QCD and our new understanding of nuclear theory?''\cite{Gross_2011}. To answer these questions, electron-hadron
coincidence experiments would have to be carried out at high energies in relatively short periods of time---a task
that could not be done by the accelerators at the time due to the low duty factors and high accidental rates (see Section \ref{sec:section1.3}).
As a possible solution to this issue it was recommended by both the Friedlander panel (1976) and the Livingston panel (1977)
that a new high energy, continuous wave (CW) beam, electron accelerator should be built for nuclear physics research\cite{Gross_2011}.\\
\indent In 1985, the United States Department of Energy (DOE) approved the concept of CEBAF based on superconducting radio-frequency (SRF)
technology that would allow for a high energy and high duty factor machine to be built and in February 1987, the construction project of CEBAF
along with three experimental end stations (Halls A, B and C) officially began\cite{Gross_2011}.
\begin{figure}[!h]
\centering
\includegraphics[scale=0.6]{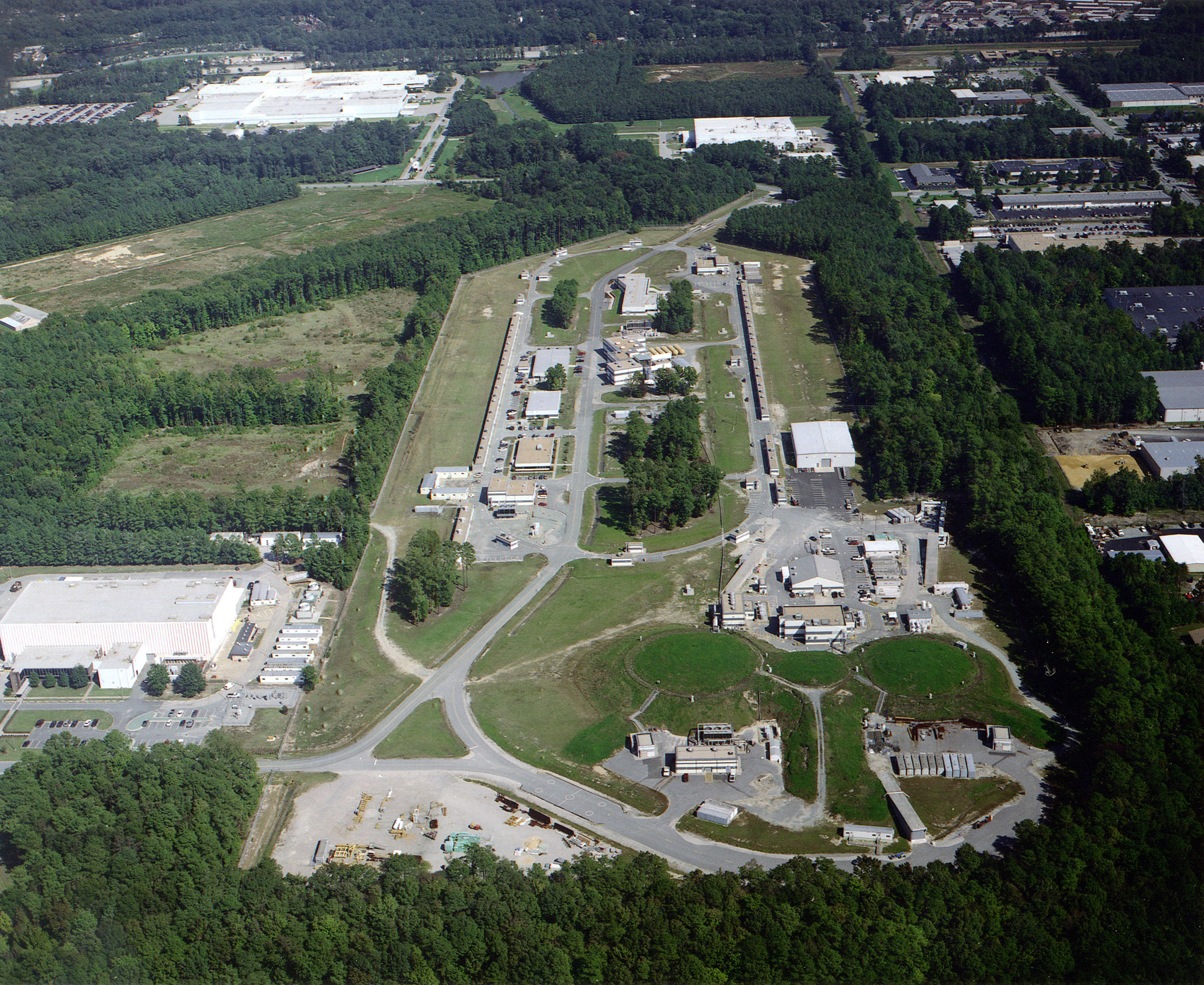}
\caption{Aerial view of CEBAF at Newport News, Virginia. The service buildings mark the 7/8-mile (1.4 km) racetrack-shaped 
  accelerator 30 feet (base of tunnel) below the surface.
  The dome-shaped terrain represent the accelerator end-stations (experimental halls), which are also underground. }
\label{fig:fig3.1}
\end{figure}
In 1994, the first beam was successfully delivered to experimental
Hall C and the following year CEBAF reached the design energy of 4 GeV. Finally, in June 1998, beam was successfully delivered simultaneously to all
three experimental halls\cite{CEBAF_Reece2016, CEBAF_Leeman2001}. \\
\indent Although the CEBAF was initially designed to operate at 4 GeV, the research and development work on SRF technology at Jefferson Lab
allowed the accelerator to be upgraded to beam energies of nearly 6 GeV and total beam currents of up to 200 $\mu$A combined for all 
experimental halls starting in the year 2000\cite{JLab_C50_Drury2011,CEBAF_Reece2016}. CEBAF operations at 6 GeV concluded 
in Spring 2012 by completing its 178th experiment since 1994.
\subsection{Accelerator Upgrade to 12 GeV}
The idea of a 12 GeV upgrade at CEBAF had already started in the late 1990s with the purpose of probing 
the nuclear structure at even smaller scales (larger $Q^{2}$) that would enable new insights into the structure of the
nucleon, the transition between hadronic and quark/gluons degrees of freedom and the nature of confinement. In 2004, the U.S. DOE approved the development of 
the 12 GeV conceptual design and approved start of construction in September 2008\cite{McKeown_2011}. \\
\begin{figure}[!h]
\centering
\includegraphics[scale=0.35]{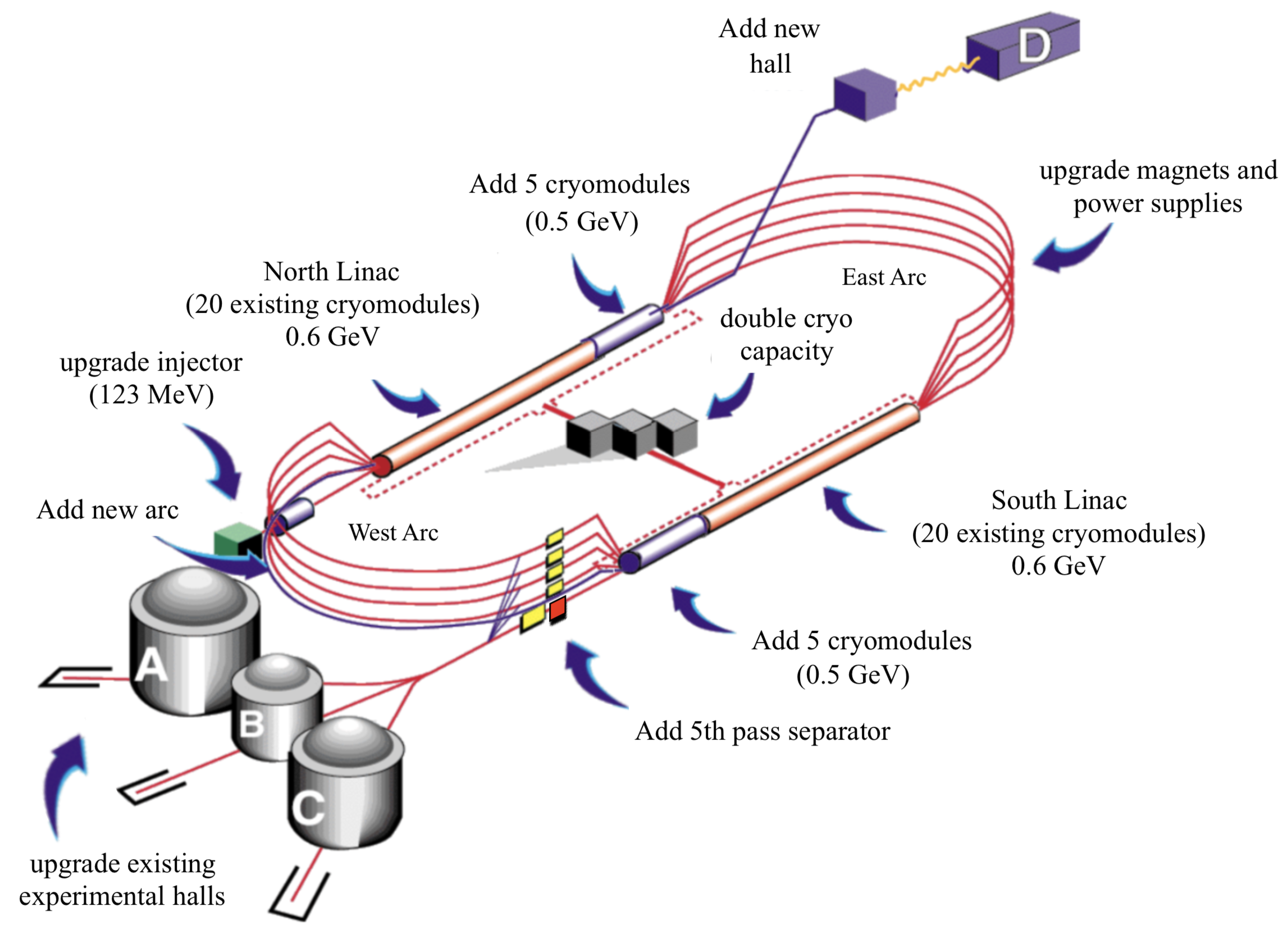}
\caption{Schematic of CEBAF 12 GeV Upgrade. Note: Reprinted from Ref.\cite{JLab12GeV_Pilat2012}.}
\label{fig:fig3.2}
\end{figure}\\
Figure \ref{fig:fig3.2} shows a schematic of CEBAF with the 12 GeV upgrade components. The racetrack-shaped accelerator
site consists of an injector, 2 ($\sim1/4$ mile each) anti-parallel SRF linear accelerators (linac), 2 recirculation arcs, a helium 
refrigerator (Central Helium Liquifer or CHL-1), and the end stations of each experimental hall. \\
\indent The main upgrades of the 12 GeV era were as follows:
\begin{itemize}
\item \textit{5 new cryomodules (C100) per linac}\cite{JLab12GeV_Pilat2012}: The C100 cryomodules are an improved design of the original C20 and refurbished C50 cryomodules of the 6 GeV era\cite{JLab_C50_Drury2011}.
A single C100 cryomodule (see Fig. \ref{fig:fig3.3}) consists of 8 7-cell 1497 MHz Niobium SRF cavities as compared to the previous 5-cell cavities and can accelerate electrons up to 100 MeV/cryomodule,
which yields 0.5 GeV acceleration per linac. The existing cryomodules accelerate electrons to 0.6 GeV/linac for a total acceleration of 1.1 GeV/linac or 2.2 GeV per pass. 
\begin{figure}[!h]
\centering
\includegraphics[scale=0.5]{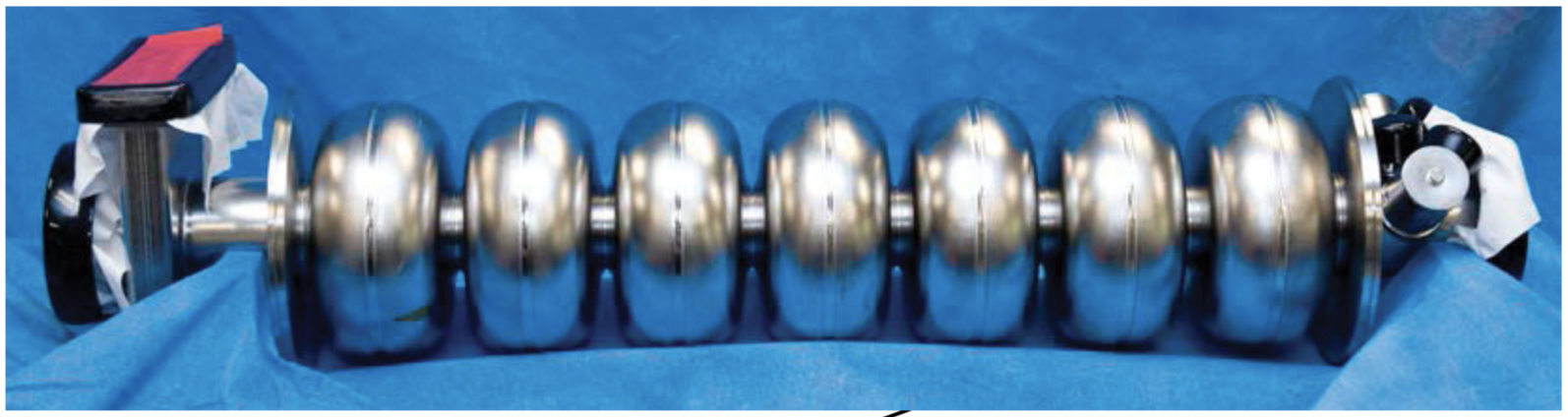}
\caption{A single (C100) 7-cell Niobium cavity.\cite{Rimmer_2020}. Note: Reprinted from Ref.\cite{JLab12GeV_Pilat2012}.}
\label{fig:fig3.3}
\end{figure}
\item \textit{upgrade recirculating arc magnets}\cite{JLab12GeV_Pilat2012,Rimmer_2020}: The arcs dipole magnets were upgraded in order to accomodate the higher beam energies. 
  In addition, a 5th pass separator and 10th arc were added in order to extract and steer the beam to the new experimental Hall D, which receives an extra half-pass for a total
  of 5.5 passes (12.1 GeV beam), whereas the other halls, at a maximum of 5 passes, receive beam energies only up to 11 GeV\footnote{\singlespacing Beam energies are actually smaller by a few 100 MeV mostly due to the inability of the cryomodules to maintain sufficiently high gradients at acceptable trip rates and partly due to energy loss due to synchrotron radiation in the arcs.}.
\item \textit{double cryogenic capacity}\cite{JLab12GeV_Pilat2012}: The upgraded SRF linacs (adding the new cryomodules) required doubling the cryogenics supply for the cryomodules to
  operate at 2 K temperatures. This was done by constructing a 2nd helium refrigerator building (CHL-2) to meet the demands.
\item \textit{upgrade injector to 123 MeV}\cite{JLab12GeV_Pilat2012}: The injector energy was upgraded by adding a new C100 cryomodule towards the final acceleration portion 
  of the injector to increase the electrons' acceleration from 67 to 123 MeV before entering the north linac. An additional 4th laser was added for the new Hall D operation.
\item \textit{upgrade experimental halls}: To meet the demands of the higher beam energies and the new experimental programs\cite{burkert2012jlab}, the three existing experimental halls
  were upgraded as well\cite{McKeown_2011}. In addition, a new experimental hall (Hall D) was built to carry out the \textit{GlueX}  physics program which requires a 9 GeV polarized photon beam
  from a 12 GeV electron beam.
\end{itemize}
\subsection{Particle Acceleration at CEBAF}
\begin{figure}[H]
\centering
\includegraphics[scale=0.25]{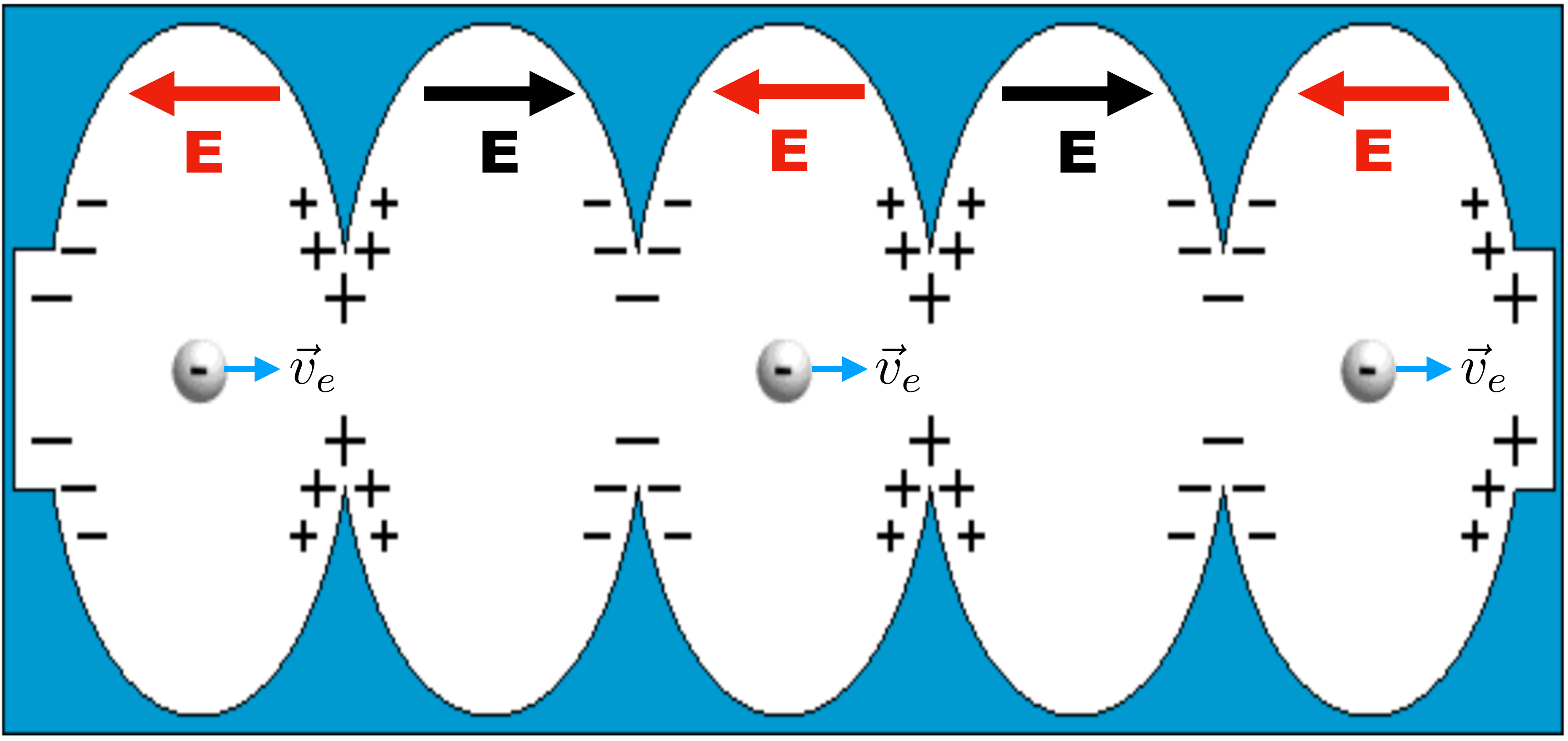}\caption{A cartoon of electrons being accelerated by a 5-cell cavity. The principle of operation is the same regardless of the SRF cavity design.}
\label{fig:fig3.4}
\end{figure}

To accelerate the electrons at CEBAF, the SRF resonant cavities (operating at 2K in a $^{4}$He bath) are excited at the fundamental frequency $f_{0}=1497$ MHz.
The resulting oscillating electric field causes the electrons to be accelerated (see Fig. \ref{fig:fig3.4}).
\indent As the accelerated electrons reach the cell boundary, the electric field is reversed (in periodic cycles of T = 1/$f_{0}$) so that when the electrons reach the adjacent cell
they are accelerated again. To achieve this synchronicity, the electrons injected into the north linac must have a frequency that is a sub-harmonic ($f_{0}/n$, where $n$ is an integer) 
of the fundamental machine frequency $f_{0}$. To achieve a sub-harmonic of the fundamental frequency it is important to consider how the electrons are generated at the
injector. \\
\indent The electron beam is generated in the injector\cite{Kazimi_2004} by shining a laser with frequency $f_{0}/n$ into a GaAs photocathode creating electron beam bunches of the same frequency
as the laser. For simultaneous hall operations, a laser is used for each hall\footnote{\singlespacing For simultaneous hall operation: In the 6 GeV era, three lasers each operating at frequencies $f_{0}/3$ (499 MHz) and out-of-phase by
  120$^{\circ}$ were incident on the same GaAs photocathode to produce 3 separate electron bunches for each hall. In the 12 GeV era, with the addition of a fourth hall, a fourth new laser as well as the other three lasers
  need to operate at the same frequency (249.5 MHz) for simultaneous four-hall operations to be possible. Details can be found in Refs.\cite{Kazimi_2016, Kazimi_2019}.}. 
The electron bunches for each hall are then sent into the injector beamline and are further accelerated (up to 123 MeV) before entering the north linac. Once in the north linac, the electrons are accelerated further
with a gain of 1.1 GeV before being steered by the east arc into the south linac for an additional 1.1 GeV gain in acceleration before completing a single pass. At this point, 
it depends on the physics demands of each hall whether to receive 1-pass (2.2 GeV), 2-pass (4.4 GeV), 3-pass (6.6), 4-pass (8.8 GeV), 5-pass (11 GeV) or in the case of Hall D,
5.5 pass (12.1 GeV). Towards the end of the south linac are devices called separators (see Fig. \ref{fig:fig3.2}) used to separate the interleaved electron beam bunches to be sent
to their respective experimental hall. In the case of Hall D (at 5.5-pass), the 5th pass separator (operating at 750 MHz) is used to separate Halls A, B and C from Hall D electron beam bunches. The Hall D beam
bunches have to travel an additional 1/2 pass through the 10th west arc and into the north linac for an additional 1.1 GeV boost to reach the 12.1 GeV beam as required by the Hall D physics program.

\section{Hall C 12 GeV Upgrade}
The Jefferson Lab 12 GeV project was successfully completed in Spring 2017. Both the accelerator and experimental halls passed the 
Key Performance Parameters (KPP) test that required them to meet the operational goals for the project\cite{CEBAF_Spata2018}.
\begin{figure}[H]
\centering
\includegraphics[scale=0.32]{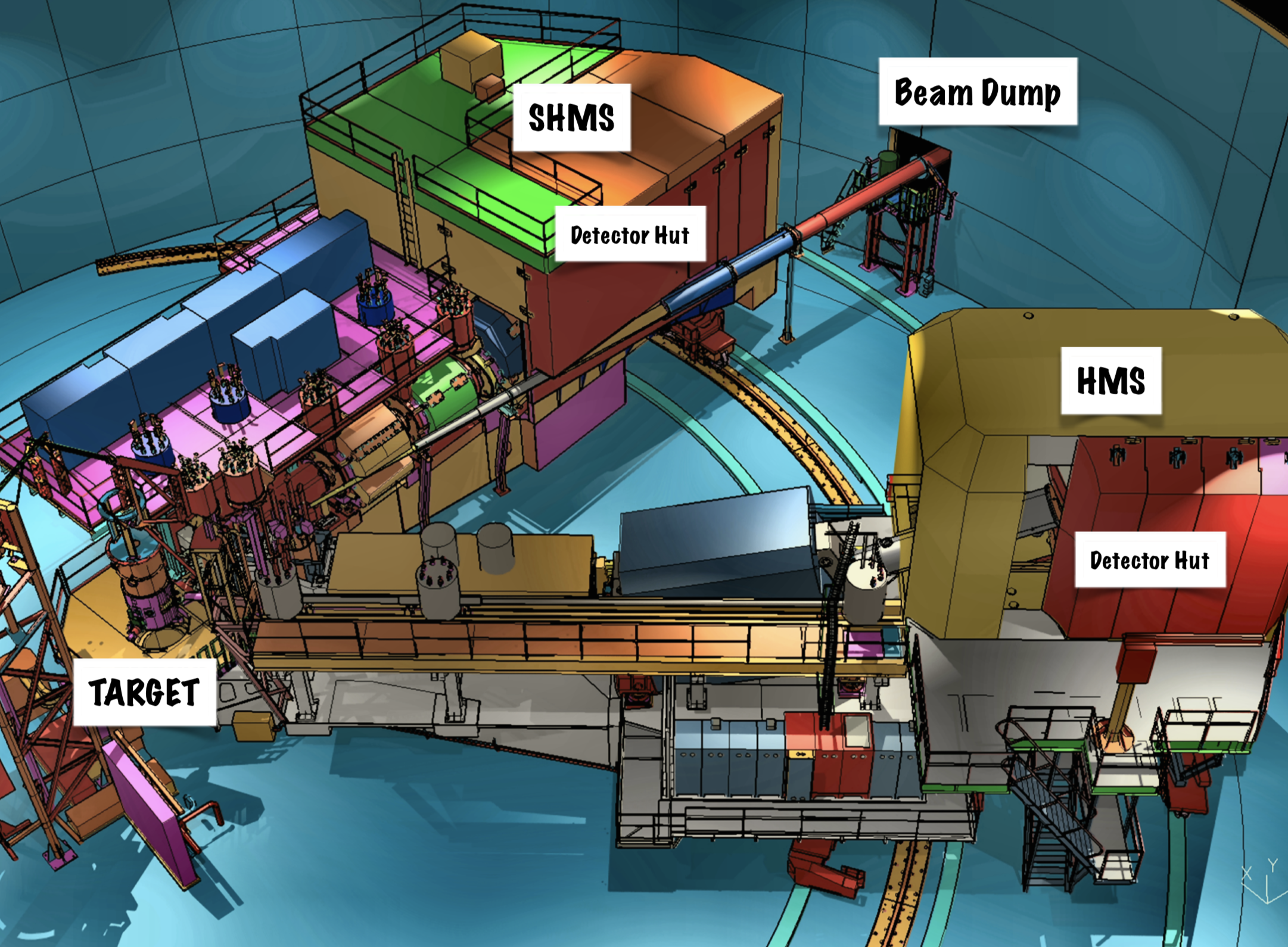}
\caption{Artist's rendering of Experimental Hall C after the 12 GeV upgrade. Note: Reprinted from Ref.\cite{HallC_SEM_saw2019}.}
\label{fig:fig3.5}
\end{figure}
Figure \ref{fig:fig3.5} shows a general view of Hall C with the new Super High Momentum Spectrometer (SHMS) alongside the
High Momentum Spectrometer (HMS) from the 6 GeV era. During the KPP, Hall C showed its capability to run continuously for 8 hours with
stable beam at 3-pass (beam currents $I_{\mathrm{b}}\sim5 \mu$A) and demonstrated a satisfactory detector performance and particle identification for the new spectrometer\cite{HallC_KPP}.
In Spring 2018, a group of experiments was used to commission the new spectrometer and test its full capabilities. The commissioning experiments
were a small part of the rich 12 GeV physics program developed by the Hall C collaboration\cite{HallC_12GeV_CDR} and included physics topics at kinematics that were
beyond the capabilities of CEBAF in the 6 GeV era such as nuclear dynamics at short distances using exclusive
reactions\cite{e12_10_003_proposal}, nucleon structure functions at high $x_{\mathrm{Bj}}$\cite{e12_10_002_proposal},
and nuclear effects from QCD\cite{e12_06_107,e12_10_008}. 

\section{E12-10-003 Experiment Overview} \label{sec:exp_overview}
This experiment was part of the group of experiments that commissioned the new Hall C spectrometer. The experiment ran for six days (April 3-9, 2018) 
and was divided into various groups of runs for specific studies (see Table \ref{tab:table3.1}).\\
\indent For the main part of the experiment, a 10.6 GeV electron beam was incident on a 10 centimeter-long liquid deuterium target (LD$_{2}$). The scattered electron was detected by the SHMS
in coincidence with the knocked-out proton detected in the HMS. The ``missing''(undetected) neutron was reconstructed from energy-momentum conservation laws (see Section \ref{sec:section2.1}).
The beam currents delivered by the accelerator ranged between 45-60 $\mu$A and the beam was rastered over a 2$\times$2 mm$^{2}$ area to reduce the effects of localized target density reduction
on the cryogenic targets (hydrogen and deuterium). \\
\indent The event trigger for each spectrometer was set to require at least three of four scintillator planes to fire, which is the
lowest level trigger definition at Hall C. In additon to the scintillators (hodoscope planes), each spectrometer also used a pair of drift chambers for the determination of particle tracks. Additionally in the SHMS,
a calorimeter was used to improve the identification of electrons. However, due to the extremely low coincidence rates and the absence of any significant pion background, the electron sample collected
in the SHMS calorimeter was very clean and the calorimeter cut was found to have little to no effect on the final electron yield.\\
\indent We measured three missing momentum settings for the deuteron centered at $p_{\mathrm{r}}=80, 580$ and 750 MeV/c. At each of these settings, 
the electron arm (SHMS) was ``fixed'' in momentum and angle and the proton arm (HMS) was rotated from smaller to larger angles corresponding to the lower 
and higher missing momentum settings, respectively. In reality, the spectrometers' angle and momentum changed back and forth multiple times at each setting
which made the reproducibility of the exact setting impossible. As a result, the data collected from the 580 and 750 MeV/c settings were separated into multiple datasets, each
corresponding to a change in either spectrometer. We analyzed separately 2 data sets for the 580 MeV/c setting, and 3 data sets for the 750 MeV/c setting.\\
\indent Hydrogen elastic $^{1}$H$(e,e')p$ data were acquired at kinematic settings close to the deuteron 80 MeV/c setting for cross-checks with the spectrometer acceptance 
modeled using the Hall C Monte Carlo simulation program, SIMC\cite{SIMC_overview_Gaskell2009}. Additional $^{1}$H$(e,e')p$ data were also taken at three other kinematic settings that covered the
SHMS momentum acceptance range for the deuteron and were used for spectrometer optics optimization, momentum calibration, and the determination of spectrometer offsets 
and kinematic uncertainties. In addition to elastic data, SHMS data were obtained using a 3-foil carbon target and a sieve slit to check and fix a problem encountered with one
of the spectrometer magnets. A complete list of the kinematic settings is given in Table \ref{tab:table3.1}.\\
\indent From Table \ref{tab:table3.1}, only the data taken after the SHMS Optics studies (with the exception of Proton Absorption measurements) were analyzed since during data taking
it was found by the experts that the SHMS Q3 and HB magnets had a saturation
\begin{table} [H]
\centering
\scalebox{0.8}{
\begin{tabular}[t]{llllccccc}
\hline
Date&Study&Runs&Target&$E_{\mathrm{b}}$&P$_{\mathrm{SHMS}}$&$\theta_{\mathrm{SHMS}}$&P$_{\mathrm{HMS}}$&$\theta_{\mathrm{HMS}}$\\
(April)&&&&[GeV]&[GeV]&[deg]&[GeV]&[deg]\\
\hline
\hline
03&Carbon Hole&3242&$^{12}$C&10.60314&8.7&12.2&2.938&37.29\\
        \\
        &$^{1}$H$(e,e')p$&3243-3248&$^{1}$H&10.60314&8.7&12.2&2.938&37.29\\
        \\
        &Proton Absorption$^{\ast}$&3249-3251&$^{1}$H&10.60314&8.7&12.2&2.938&37.29\\
        \\
04&Aluminum Dummy&3252-3258&Al&10.60314&8.7&12.2&2.938&37.29\\
        \\
        &Proton Absorption$^{\ast}$&3259-3263&$^{1}$H&10.60314&8.7&12.2&2.938&37.29\\
        \\
        &80 MeV/c (set0)&3264-3268&$^{2}$H&10.60314&8.7&12.2&2.8438&38.89\\
        \\
04-05&580 MeV/c (set0)&3269-3282&$^{2}$H&10.60314&8.7&12.2&2.194&54.945\\
        \\
05&SHMS Optics$^{\dagger}$&3283-3285&$^{12}$C&10.60314&8.7&8.938&2.194&54.945\\
  &           &3286     &$^{12}$C&10.60314&8.7&8.938&2.765&37.338\\
  &           &3287    &$^{12}$C&10.60314&8.7&12.06&2.938&37.338\\
  \\
  &$^{1}$H$(e,e')p$&3288&$^{1}$H&10.60314&8.7&12.194&2.938&37.338\\
  \\
  &80 MeV/c (set1)&3289&$^{2}$H&10.60314&8.7&12.194&2.843&38.896\\
  \\
05-06&580 MeV/c (set1)&3290-3305&$^{2}$H&10.60314&8.7&12.194&2.194&54.992\\
  \\
06-08&750 MeV/c (set1)&3306-3340&$^{2}$H&10.60314&8.7&12.194&2.091&58.391\\
  \\
08    &580 MeV/c (set2)&3341-3356&$^{2}$H&10.60314&8.7&12.194&2.194&55.0\\
    \\
08-09  &750 MeV/c (set2)&3357-3368&$^{2}$H&10.60314&8.7&12.194&2.091&58.394\\
    \\
09 &$^{1}$H$(e,e')p$&3371&$^{1}$H&10.60314&8.7&13.93&3.48&33.545\\
   &$^{1}$H$(e,e')p$$^{\ast}$&3372&$^{1}$H&10.60314&8.7&9.928&3.48&33.545\\
   &$^{1}$H$(e,e')p$$^{\ast}$&3373&$^{1}$H&10.60314&8.7&9.928&3.017&42.9\\
   &$^{1}$H$(e,e')p$&3374&$^{1}$H&10.60314&8.7&9.928&2.31&42.9\\
   &$^{1}$H$(e,e')p$$^{\ast}$&3375&$^{1}$H&10.60314&8.7&8.495&1.8904&42.9\\
   &$^{1}$H$(e,e')p$&3376&$^{1}$H&10.60314&8.7&8.495&1.8899&47.605\\
   &$^{1}$H$(e,e')p$&3377&$^{1}$H&10.60314&8.7&8.495&1.8899&47.605\\
   &$^{1}$H$(e,e')p$&3379&$^{1}$H&10.60314&8.7&8.495&1.8898&47.605\\
   \\
09  &750 MeV/c (set3)&3380-3387&$^{2}$H&10.60314&8.7&12.21&2.091&58.391\\
\hline
\end{tabular}
}
\caption{The E12-10-003 experiment comprehensive run list. The spectrometer central momentum and angle were determined based on the dipole NMR set value and spectrometer camera, respectively. 
The beam energy in this table is uncorrected for synchrotron radiation (see Section \ref{sec:Eb_meas}). In the column Study, the ($\ast$) are data taken with SHMS single-arm (electron trigger ONLY) 
and the ($\dagger$) represents data taken with SHMS single-arm and Centered Sieve Slit positioned. The remaining runs are taken with SHMS-HMS coincidence trigger only. }
\label{tab:table3.1}
\end{table}
\noindent correction that was not needed. This correction was removed from the magnet controls software\footnote{\singlespacing See HCLOG entries below \\1. \url{https://logbooks.jlab.org/entry/3555385}\\2. \url{https://logbooks.jlab.org/entry/3555428}\\3. \url{https://logbooks.jlab.org/entry/3555436}\\4. \url{https://logbooks.jlab.org/entry/3555447}}
and the experiment resumed data-taking starting at run 3288 without the HB/Q3 correction. The analysis of Proton Absorption studies was not impacted as it involved taking yield ratios.
\section{The Hall C Beamline}
When the electron beam exits the south linac, it is steered at the beam switchyard to any one of the three experimental hall beamlines (A, B or C). In Hall C,
the beam is sent through a transport line with an entrance channel of 63.5 mm inner diameter stainless steel tubing connected with conflat flanges which reduces the inner diameter to 25.4 mm
when passing through the steering magnets (dipoles, quadrupoles, hexapoles and beam correctors)\cite{HallC_SEM_saw2019}. To reach the hall entrance, the beam is bent by
a series of 8 dipole magnets located at the hall arc (see Fig. \ref{fig:fig3.6}). The beam is then transported through the Hall C alcove\footnote{\singlespacing Transport line between green shield 
wall and hall entrance where the Compton and M{\o}ller Polarimeters are located.} into the scattering (target) chamber and the beam dump at the other end of the hall. Several 
beam diagnostics components are placed throughout multiple locations in the accelerator and hall beamlines. The relevant ones used to monitor the beam in Hall C are the harps 
(wire scanners), beam position monitors (BPMs) and beam current monitors (BCMs). The beamline is also equipped with two permanent beam raster systems with the possibility to add a third raster.

\subsection{Beam Energy Measurement}\label{sec:Eb_meas}
The accurate determination of the absolute beam energy is important as its uncertainty is directly related to the uncertainty in the cross section. Various
methods have been proposed to determine the beam energy at CEBAF\cite{beamE_meas_PUlmer1993}. 
In Hall C, the beam energy is determined by using the beamline arc as a spectrometer (first proposed in Ref.\cite{beamE_meas_Carlini1993}) and is the method used
in this experiment. The method is based on the equations of motion of a charged particle in a magnetic field. For an electron the magnetic force is given by
\begin{equation}
|\vec{F}_{B}| = e|\vec{v}_{e} \cross \vec{B}| = ev_{e}B_{\perp} = \frac{\gamma m_{e}v^{2}_{e}}{\rho_{\mathrm{c}}},
\label{eq:3.1}
\end{equation}
where $e, v_{e}$, $B_{\perp}$ and $\rho_{\mathrm{c}}$ are the elementary charge, electron velocity, magnetic field (perpendicular to the velocity) and the local radius of curvature,
respectively, and $\gamma\equiv(1 - v^{2}_{e}/c^{2})$, is the Lorentz factor to account for relativistic effects.
From Eq. \ref{eq:3.1}, the electron momentum is given by $p_{e} = \gamma m_{e}v_{e}$ and the radius of curvature can be expressed as $\rho_{c} = dL_{\parallel} / d\theta_{\mathrm{arc}}$
where $dL_{\parallel}$ and $d\theta_{\mathrm{arc}}$ are the infinitesimal arc lenth and arc angle, respectively. Using these definitions, Eq. \ref{eq:3.1} may be expressed as
\begin{equation}
p_{e} = \frac{eB_{\perp}dL_{\parallel}}{d\theta_{\mathrm{arc}}}.
\label{eq:3.2}
\end{equation}
In reality, as the electron beam traverses through the Hall C arc, the dipole magnetic fields are not homogeneous and need to be integrated over infinitesimal
($dL_{\parallel}$) arc elements along the beam trajectory. Eq. \ref{eq:3.2} can then be expressed as
\begin{equation}
  p_{e} = C_{k}\frac{\int^{L}_{0}B_{\perp}dL_{\parallel}}{\int^{\theta_{\mathrm{arc}}}_{0}d\theta_{\mathrm{arc}}},
  \label{eq:3.3}
\end{equation}
where $C_{k}$ is a constant determined from dimensional analysis and using the conversion factor 1 [C][T][m] $ \equiv $ 1.87115736 $\cross$ 10$^{18}$ GeV/c
and $e = 1.602\cross10^{-19}$ C:
\begin{align*}
  &C_{k} \equiv 1.602\cross10^{-19} \mathrm{C} \cross \frac{1.871 \cross 10^{18} \mathrm{GeV/c}}{1 \mathrm{[C][T][m]}} = 0.29979 \mathrm{[GeV/c][T^{-1}][m^{-1}]},
\end{align*}
which gives Eq. \ref{eq:3.3} in units of [GeV/c] assuming the magnetic field and arc length are given in [T][m]. This is the final form (Eq. \ref{eq:3.3}) used
in the beam eneergy measurements.
\begin{figure}[H]
\centering
\includegraphics[scale=0.35]{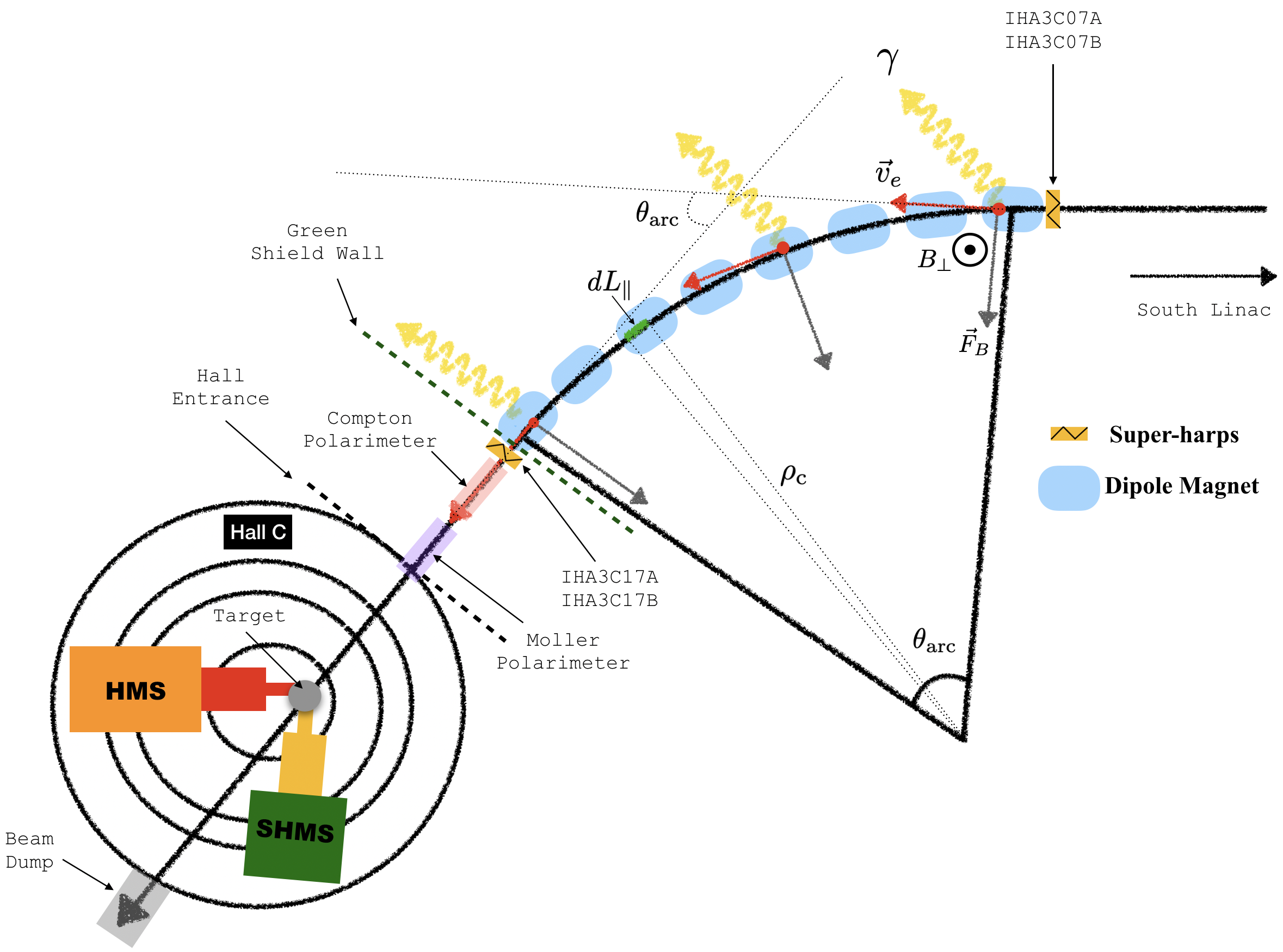}
\caption{The Hall C arc with the relevant beamline components for the beam energy measurements are shown. The electron (red vector) loses energy (synchrotron radiation shown as yellow wiggly arrows) as it
  traverses the arc under a perpendicular magnetic field $B_{\perp}$. Two superharps (wire-scanners) at each end of the arc are used to determine small variations in the beam direction. }
\label{fig:fig3.6}
\end{figure}
\indent The integrated field $\int B_{\perp}dL_{\parallel}$ is determined by carefully mapping the magnetic
fields of the arc dipoles at the corresponding dipole current. The bend angle $\theta_{\mathrm{arc}}$ is determined from a survey
by the relative orientation of the beam at the arc entrance and exit (see Fig. \ref{fig:fig3.6}).
The superharps\footnote{\singlespacing Compared to harps, superharps have been more accurately fiducialized and surveyed for absolute
  position measurements\cite{JRA_phdthesis}.} at both ends of the arc are used to determine the absolute beam position and direction. See
Ref.\cite{HallC_Superharp1995} for technical details of the Hall C superharps. \\
\indent During the beam energy measurement, Machine Control Center (MCC)\footnote{\singlespacing MCC operators control and steer the
  beam around the accelerator and into the experimental end stations.}
operators turn off all the arc quadrupole and beam corrector magnets, which would otherwise provide an achromatic beam\footnote{\singlespacing Quadrupole magnets function as an achromatic (or in this case, momentum independent) lens to the beam by providing the necessary restoring forces to focus the beam and minimize dispersion.}
and only use the dipoles to steer the beam. As a result, dispersion
(momentum dependent position) builds up across the arc, which provides a very sensitive energy measurement as the
beam will be spread out based on small energy differences. The negative side effect of dispersion is that it becomes
very difficult for operators to guide the beam across the arc due to this spread. If this is done successfully,
the operators use a lookup table determined from the field mapping
to convert the dipole current to $\int B_{\perp}dL_{\parallel}$ across the arc. \\
\indent To measure the beam direction, a pair of superharps located at the arc entrance and exit (see Fig. \ref{fig:fig3.6}) are used and controlled by MCC as they are
invasive to the beam. During the harp scans, the signals produced by two of the superharps were unexpectedly wide and it was decided not to use this information.
This was not a cause of concern as the variations in the beam direction allowed by the beamline diameter were sufficiently small and were expected to have a small effect
on the $\int B_{\perp}dL_{\parallel}$ measurements\cite{MKJ_privJan2020}. The measured beam energy at the arc entrance (uncorrected for synchrotron radiation) is shown in Table \ref{tab:table3.1}. 
A detailed table with beam energy measurements performed at different times can be found in Ref.\cite{HCwiki_Ebeam}.\\
\indent As the electron beam traverses the Hall C arc, it changes direction, which causes the beam to lose energy due to synchrotron radiation.
This loss is not accounted for in the field integral measurements and must be determined separately. The usual formula for energy loss due to
synchrotron radiation is given by (see Ref.\cite{SyncRad_2013})
\begin{equation}
  \delta E_{\mathrm{sync}}[\mathrm{keV}] = 88.46 \frac{E^{4}_{\mathrm{meas}}[\mathrm{GeV}]}{\rho_{\mathrm{c}}[\mathrm{m}]} \frac{\theta_{\mathrm{arc}}}{360^{\circ}},
  \label{eq:3.4}
\end{equation}
where $E_{\mathrm{meas}}$ is the measured beam energy at the arc entrance.
Since the original energy loss formula is per $360^{\circ}$, the fractional energy loss in the Hall C arc is
$\theta_{\mathrm{arc}}/360^{\circ}$, where $\theta_{\mathrm{arc}} = 34.3^{\circ}$ from the survey and the arc radius of curvature is $\rho_{\mathrm{c}}=40.09$ m.
Substituting the beam energy from Table \ref{tab:table3.1} and the geometrical values from the arc in Eq. \ref{eq:3.4} and converting keV to GeV, one obtains
\begin{align}
  \delta E_{\mathrm{sync}} = 0.00265 \text{ GeV}.
  \label{eq:3.5}
\end{align}
The corrected beam energy and its relative error at the target are then given by
\begin{align}
  &E_{\mathrm{tgt}} =  E_{\mathrm{meas}} - \delta E_{\mathrm{sync}} \label{eq:3.6},\\
  &\Big(\frac{\delta E_{\mathrm{tgt}}}{E_{\mathrm{tgt}}}\Big)^{2} = \Big(\frac{\delta E_{\mathrm{meas}}}{E_{\mathrm{meas}}}\Big)^{2} + \Big(\frac{\delta E_{\mathrm{sync}}}{E_{\mathrm{meas}}}\Big)^{2}, \label{eq:3.7}
\end{align}
where $\delta E_{\mathrm{meas}}/E_{\mathrm{meas}}$ is the relative error due to the field integral and $\delta E_{\mathrm{sync}}/E_{\mathrm{meas}}$ is the relative error due to the measured beam energy due to synchrotron radiation.
From the beam energy measurement on April 30, 2018\cite{HCwiki_Ebeam}:
\begin{align}
E_{\mathrm{meas}} \pm \delta E_{\mathrm{meas}}= 10.60314 \pm 0.00415 \text{ GeV}. \label{eq:3.8}
\end{align}
Substituting the numerical values of Eqs.\ref{eq:3.5} and \ref{eq:3.8} in Eqs. \ref{eq:3.6} and \ref{eq:3.7} one obtains the corrected beam energy and its relative uncertainty
\begin{align}
  &E_{\mathrm{tgt}} = 10.6005 \text{ GeV} \label{eq:3.9},\\
  &\frac{\delta E_{\mathrm{tgt}}}{E_{\mathrm{tgt}}} = 4.64 \times 10^{-4}.  \label{eq:3.10}
\end{align}
\subsection{Beamline Components}
\begin{figure}[H]
\centering
\includegraphics[scale=0.31]{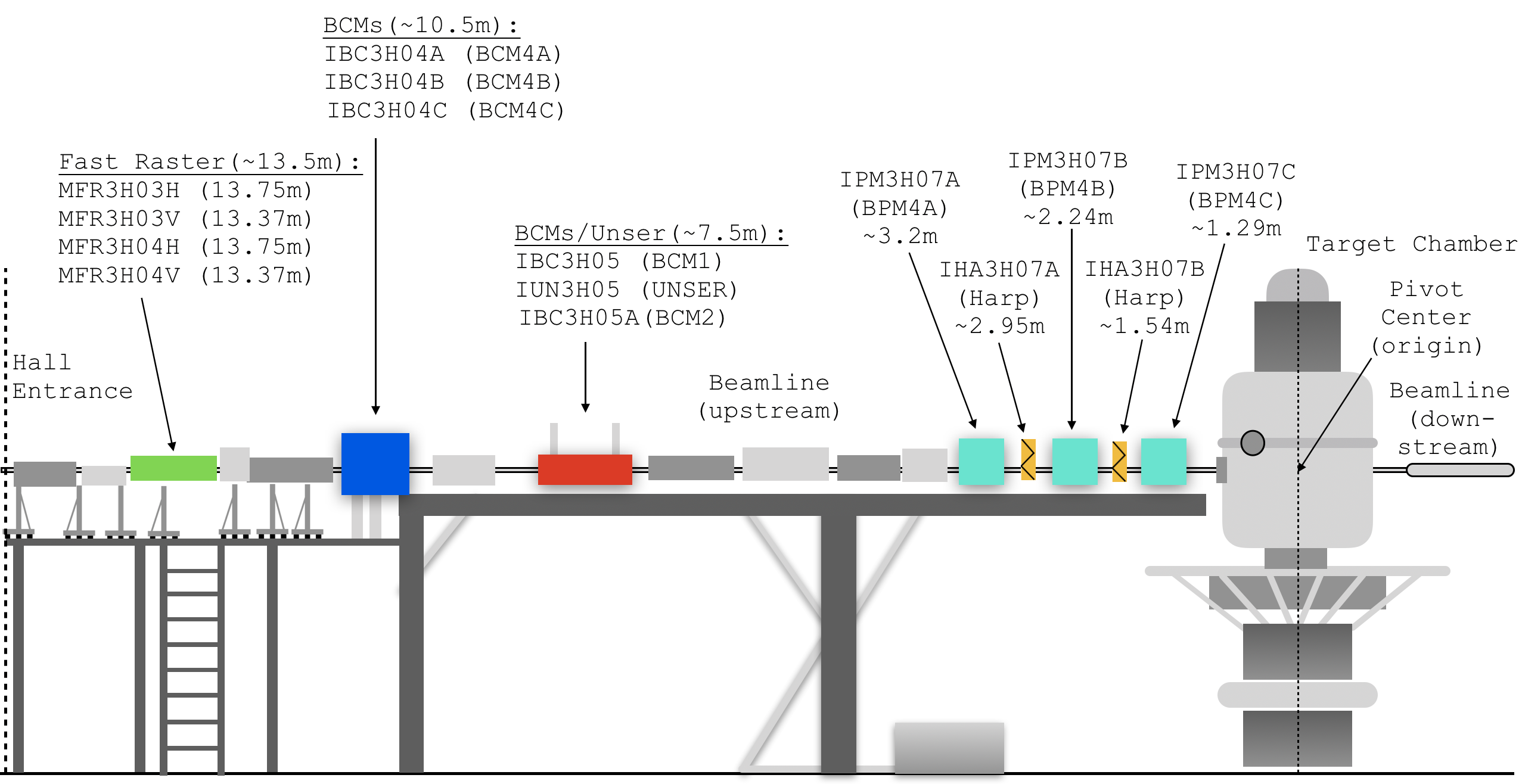}
\caption{Hall C beamline from hall entrance to target chamber. Distances to the relevant beamline components are measured from the origin (the pivot center) and given in meters.
  The first three colored boxes (green, blue and red) have multiple components with the relevant distances to the target origin. The codenames used in the Fast Raster
  magnets refer to the horizontally (H) and vertically (V) bending air-core magnets. The commonly used names of the other beamline components are indicated in parentheses. }
\label{fig:fig3.7}
\end{figure}

\begin{figure}[H]
\centering
\includegraphics[scale=0.3]{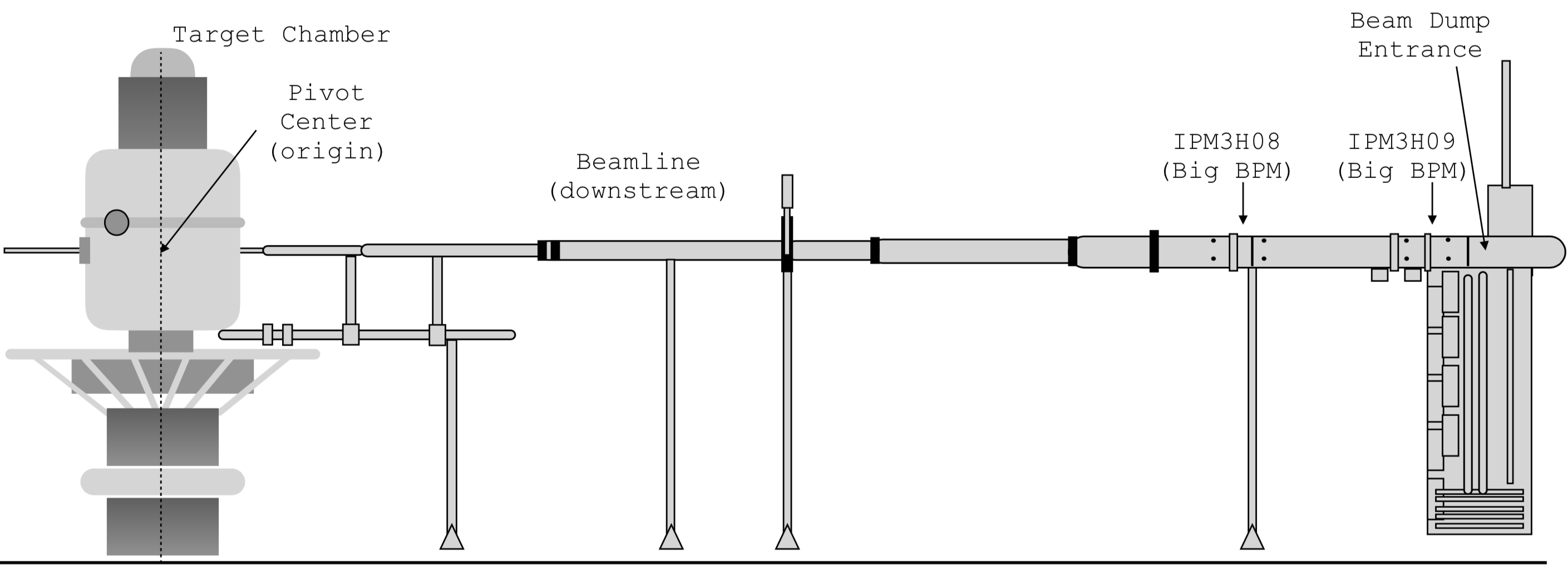}
\caption{Hall C beamline from target chamber to beam dump.}
\label{fig:fig3.8}
\end{figure}

As the beam enters Hall C, it passes through various beamline components (see Figs. \ref{fig:fig3.7} and \ref{fig:fig3.8}) as it is transported to the target chamber 
and into the beam dump. Upstream of the target chamber, are the fast raster (FR), beam position and current monitors (BPMs, BCMs),
and harps which were briefly mentioned in the previous section.
\indent Downstream of the target chamber, the entire beam pipe is $\sim27.4$ m long measured from the exit of the chamber to the entrance of the beam dump
with two 1.5 meter-long, removable sections of 24-inch diameter beam pipes that can be replaced each with Big BPMs towards the beam dump. These BPMs are used to measure the beam position downstream of the
target chamber. These are necessary because the fringe fields of the SHMS magnets can change the beam direction during experimental configurations where the SHMS is at small angles (typically $<$10$^{\circ}$).
\subsubsection{Harps}
\begin{figure}[!ht]
\centering
\includegraphics[scale=0.4]{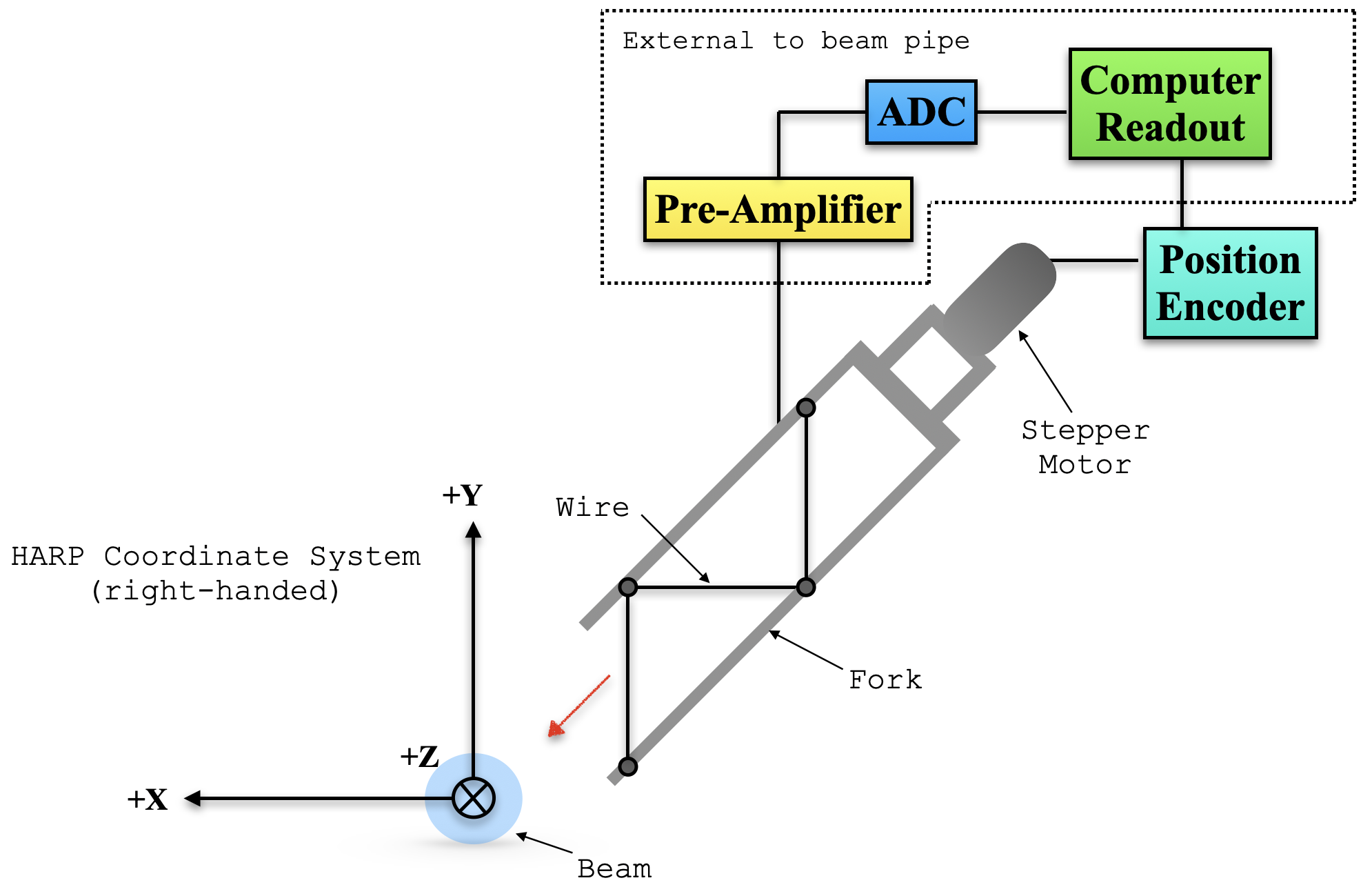}
\caption{Hall C beamline harp diagram. The harp enters (red arrow shows direction of motion) at a 45$^{\circ}$ angle. The two vertical wires measure the beam position along the $x$-axis and a vertical wire measures the position alng the $y$-axis.}
\label{fig:fig3.9}
\end{figure}

The harps consist of a fork with three wires (see Fig. \ref{fig:fig3.9}) and a stepper motor attached that enables the entire system to move invasively through the unrastered beam.
As each of the wires comes in contact with the beam, a current is produced in the wire due to secondary electron emission. This current signal is amplified before being sent to an Analog-to-Digital Converter
(ADC). The ADC spectrum formed from the digitized signals is fit to obtain the beam profile (size). To determine the absolute beam position, as each wire passes through
the beam a position encoder generates the number of pulses equivalent to the number of steps the motor has moved, which corresponds to an absolute beam position. 
\begin{figure}[!h]
\centering
\includegraphics[scale=0.28]{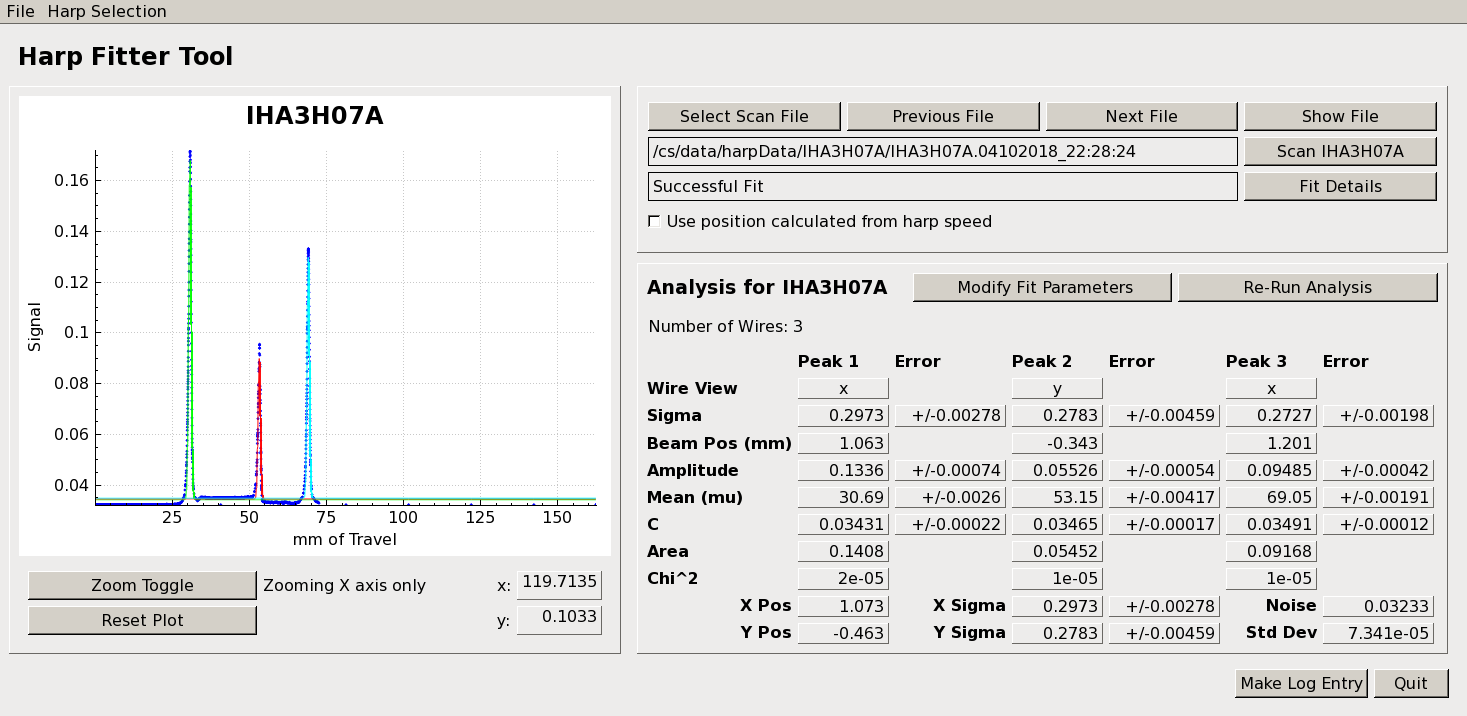}
\caption{Results from a harp scan of harp IHA3H07A taken at 5-pass on April 2018.}
\label{fig:fig3.10}
\end{figure}\\
\indent Figure \ref{fig:fig3.10} shows the results of a typical harp scan (with beam currents $\sim$5 $\mu$A CW) where the $y$-axis represents the ADC value plotted versus the distance the fork travelled (mm) shown in the $x$-axis.
The fit results for each wire (gaussian peak) are shown at the right of the plot. The overall results of the absolute beam position (X Pos (mm), Y Pos (mm)) and beam profile (X Sigma (mm), Y Sigma (mm)) 
are shown at the bottom. Since the harps are invasive to the beam, the scan is not performed during normal experiment operations. Therefore, to monitor the beam positions in real time during the experiment, 
the BPMs must be calibrated using the absolute beam positions from the harp scans.

\subsubsection{Beam Raster Systems}\label{sec:beam_raster}
The intrinsic electron beam size of CEBAF in the 12 GeV era\footnote{\singlespacing In the 6 GeV era, CEBAF delivered average beam spot sizes of 50-200 $\mu$m, which were comparatively
  smaller than in the 12 GeV era. The increase in the intrinsic beam size from the 6 to 12 GeV era is attributed to an increase in synchrotron radiation emitted by the beam.} is typically 200-700 $\mu$m in diameter
(approximately a gaussian full width, 2$\sigma$). From the harp scan results in Fig. \ref{fig:fig3.10} for example,
a typical gaussian has a full width of $2\sigma\sim$600 $\mu$m, which is a reasonably good approximation for the diameter of the unrastered beam, considering that
the beam is often asymmetric. The amount of power per unit area (intensity) deposited by such a small beam size on either the target, 
target chamber or the beam dump for extended periods of time can cause damage to these components by overheating. For cryogenic targets such as liquid hydrogen and
deuterium used in this experiment for example, there are two effects\cite{DMack_privJul2020}:
\begin{itemize}
\item At lower beam currents, the target density change is due to warming of the cryogenic fluid with a density variation of the order $\sim1\%$/K.
  Rastering the beam reduces this temperature rise and hence the systematic density change. 
\item At higher beam currents, bubbles also start to form and break-off at the target cell windows.
\end{itemize}
These effects have a direct impact on the high-precision cross section measurements required by the Hall C physics program as significant target density changes cause the data yield to
be significantly lower. To solve this issue, the intrinsic beam is smeared out (rastered) to reduce the temperature changes over a larger area.\\
\indent The Hall C raster system consists of two beam rasters permanently installed in the Hall C beamline. The
M{\o}ller Raster (not shown) is located upstream of the M{\o}ller target in the Hall C alcove and the fast raster is located $\sim$14 m upstream of the Target Chamber (see Fig. \ref{fig:fig3.7}).
A third raster (Slow Raster) can be added for experiments that require a polarized target but does not form part of the standard beamline components\cite{HallC_SEM_saw2019}.
For most experiments (including this experiment), the fast raster is used and is discussed in more detail in Appendix \hyperref[appendix:appB]{B}.\\

\begin{figure}[ht]
\centering
\includegraphics[scale=0.35]{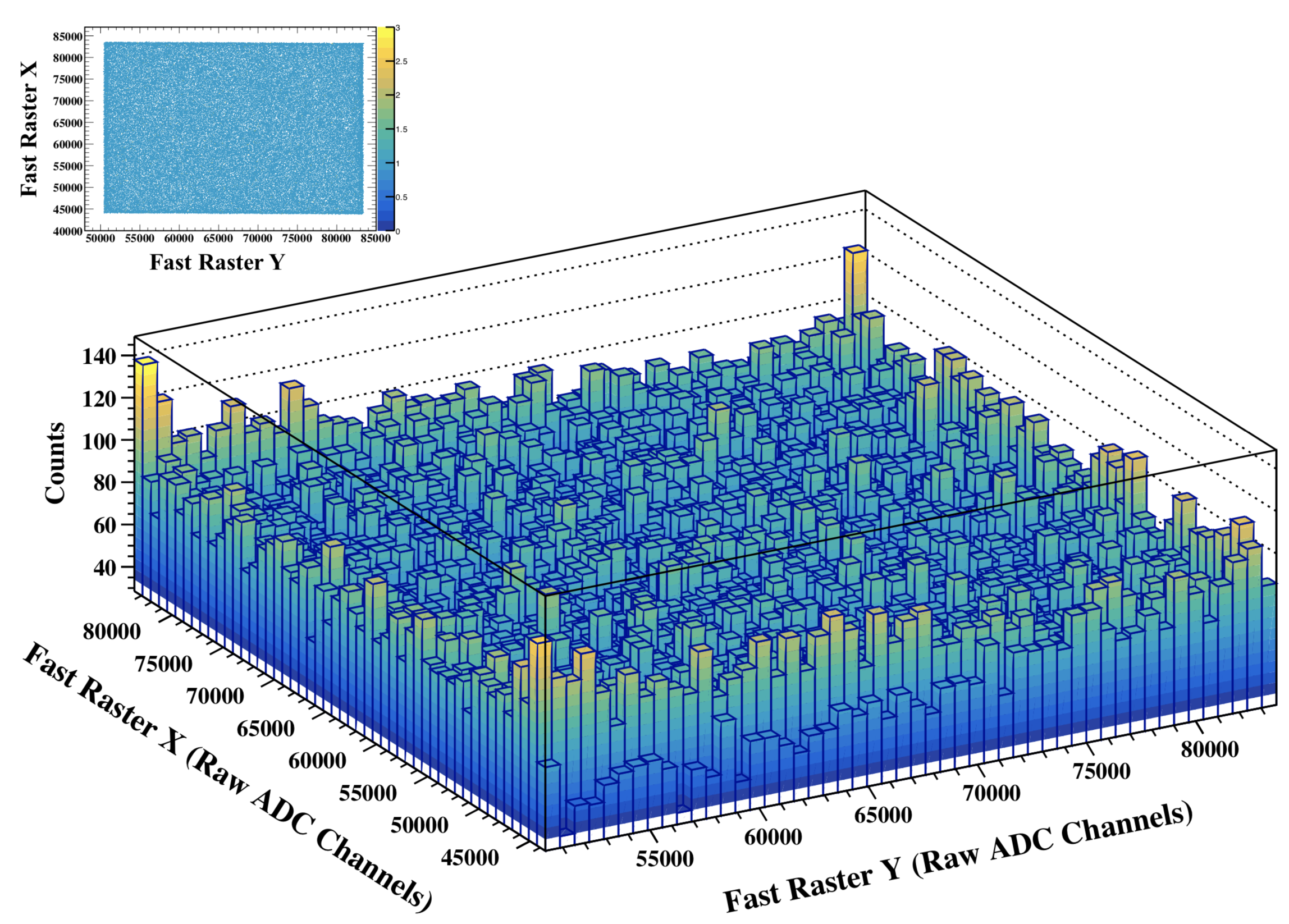}
\caption{Fast raster (X,Y) raw ADC signals measured by the pickup probe during run 3289 for the 80 MeV/c setting. The 3D plot (and inset 2D representation) show an approximately uniform XY raster distribution.}
\label{fig:fig3.12}
\end{figure}
\indent Figure \ref{fig:fig3.12} shows a 3D (and inset 2D projection) of the fast-raster raw ADC signal distribution measured during the E12-10-003 experiment.
The raster was set to $2\times 2$ mm$^{2}$ to minimize localized density changes in the 10-cm liquid deuterium cryogenic target. The raster distribution in Fig. \ref{fig:fig3.12} shows that the beam is uniformly distributed across the entire $2\times2$ mm$^{2}$ raster, especially at the boundaries, which was
a major issue with the original Hall C raster in use from 1996 to 2002\cite{Raster_Yan2005}. 
\subsubsection{Beam Position Monitors (BPM)}
The BPMs are cylindrical cavities that form part of the beamline and are used to make continuous, non-invasive measurements of the beam position during normal beam operations.
\begin{figure}[!ht]
\centering
\includegraphics[scale=0.32]{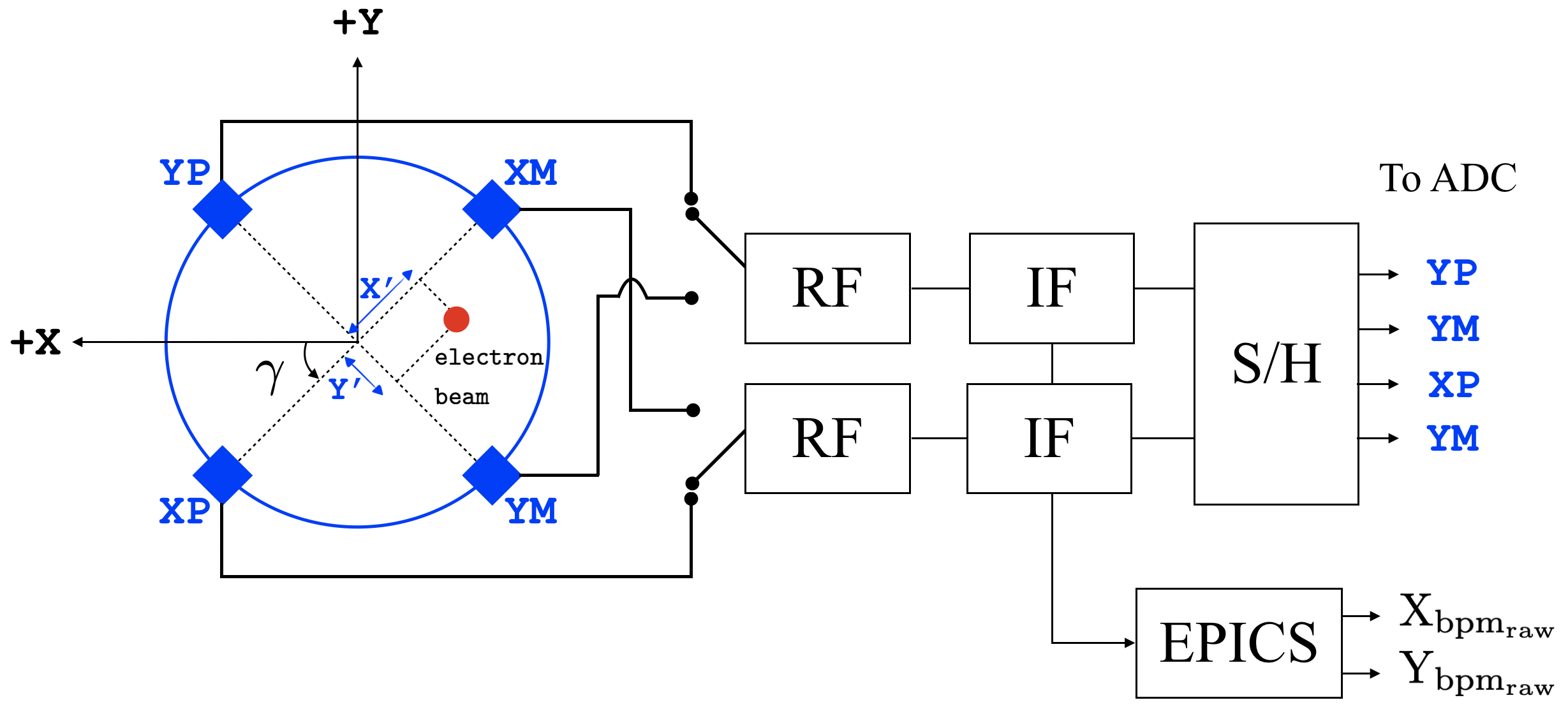}
\caption{Hall C BPM and electronics diagram. In EPICS coordinate system (left-handed), the beam is directed out of the page.
The antennae are located along the axes of a coordinate system (blue) that is oriented 45$^{\circ}$ relative to the EPICS coordinate system. Note: Reprinted from Ref.\cite{HallC_SEM_saw2019}.}
\label{fig:fig3.13}
\end{figure}
To measure the beam position in Hall C, three BPMs located upstream of the target are used (see Fig. \ref{fig:fig3.7}). Each BPM consists of an enclosure that 
forms part of the beam-pipe with four wire-antennae attached to feedthroughs on the interior wall of the pipe. 
The antennae (blue) are in a coordinate system oriented at $\gamma=$45$^{\circ}$ relative to the EPICS\cite{ANL_EPICS} coordinate system (see Fig. \ref{fig:fig3.13}).
When the beam (499 MHz sub-harmonic of $f_{0}$) passes through the
BPM cavity it induces a signal in the antennae with an amplitude inversely proportional to the distance between
the beam and each of the antennae. This RF (radio-frequency) signal is then converted to a more convenient lower frequency
known as IF (intermediate-frequency). This signal is subsequently detected by the S/H (sample-and-hold) section, which as its name indicates, samples the input signal 
and holds it until it can be further processed by the ADCs to be analyzed by software. This method, 
however, requires the user to retrieve the constants and perform a separate calculation to convert the raw ADC (processed antennae signals) to raw beam position values.
Alternatively, the antennae signals are interpreted and calibrated by the EPICS readout chain using the standard difference-over-sum method to determine the raw beam positions. 
The beam position is averaged over 0.3 seconds and is logged into the EPICS database with 1 Hz updating frequency and injected in the data-stream every few seconds, unsynchronized but
with a reference timestamp\cite{HallC_SEM_saw2019}. Using the Hall C analysis software, the raw BPM positions are retrieved from EPICS and calibrated relative to the
absolute beam position determined from the harp scans.

\subsubsection{Beam Current Monitors (BCM)}
The experimental cross section measurements at Hall C require the data yield to be normalized by the total charge at the target.
To achieve this, multiple BCMs are used for the continuous, non-invasive measurement of the beam current inside the hall. 
The primary system consists of two BCMs and an adjacent Unser Monitor (BCM1, Unser, BCM2) located $\sim$7.5 m upstream of the target. Three
supplementary BCMs (BCM4A, BCM4B, BCM4C) located $\sim$10.5 m upstream of the target are also used to measure the beam
current in parallel with the primary system, but must be removed during certain configurations of the beamline (e.g., during 
polarized target experiments that require the slow raster system)\cite{HallC_SEM_saw2019}.\\
\indent Each BCM consists of a stainless steel, cylindrically-shaped cavity. The beam excites the cavity at its resonant frequency
of 1497 MHz. An antenna inside the cavity couples some of this RF power into a heliax cable which goes to the counting house for further processing. 
Electrically, the resonant cavities act like a resistor through which the beam current flows, extracting $\lesssim$1 mW from 
a beam that usually has 0.1-1 MW of total power\cite{DMack_privFeb2020}. Although the extracted power (antennae signal) is a small fraction of the total beam power, 
it is still considered a large RF signal that has to be converted to a lower frequency by down-converters to be processed by the electronics.\\
\indent  The Unser monitor (Parametric Current Transformer or PCT)\cite{Unser_1992} is a beam current monitor that is toroidal in shape with circular
strips of an extremely permeable material\footnote{\singlespacing The circular strips used in the Unser toroid consist of (CoFe)$_{70}$(MoSiB)$_{30}$---an amorphous magnetic
  alloy that exhibits extreme magnetic permeability, which means the material internal dipoles become easily aligned in response to an applied external magnetic field\cite{Unser_1992}.}.
As the electron beam passes through the toroid axis of symmetry, its circular magnetic field magnetizes the strips of permeable material in the toroid. A modulator-demodulator
circuit senses this magnetization and sends a current to a compensating coil to cancel out the field established by the beam. This compensating current is proportional to the
beam current\cite{BCMs_Mack1993}.\\
\indent Both the resonant cavities (BCMs) and the Unser are sensitive to temperature variations in their surroundings.
The variations in temperature cause thermal expansion or contraction of the BCM cavities and the Unser toroid material
resulting in the detuning of the cavities and undesirable zero drifts\footnote{\singlespacing The ``zero drift'' refers to the effect where the zero reading of an instrument is
  modified by the ambient conditions.} in the output signal of the Unser monitor. To minimize the effects due to temperature variations, the Hall C BCMs 
and Unser are thermally insulated in a box and kept at a temperature of 110 $^{\circ}$C with a tolerance of $\pm$0.2 $^{\circ}$C, which is monitored 
periodically.\\
\begin{figure}[h!]
\centering
\includegraphics[scale=0.35]{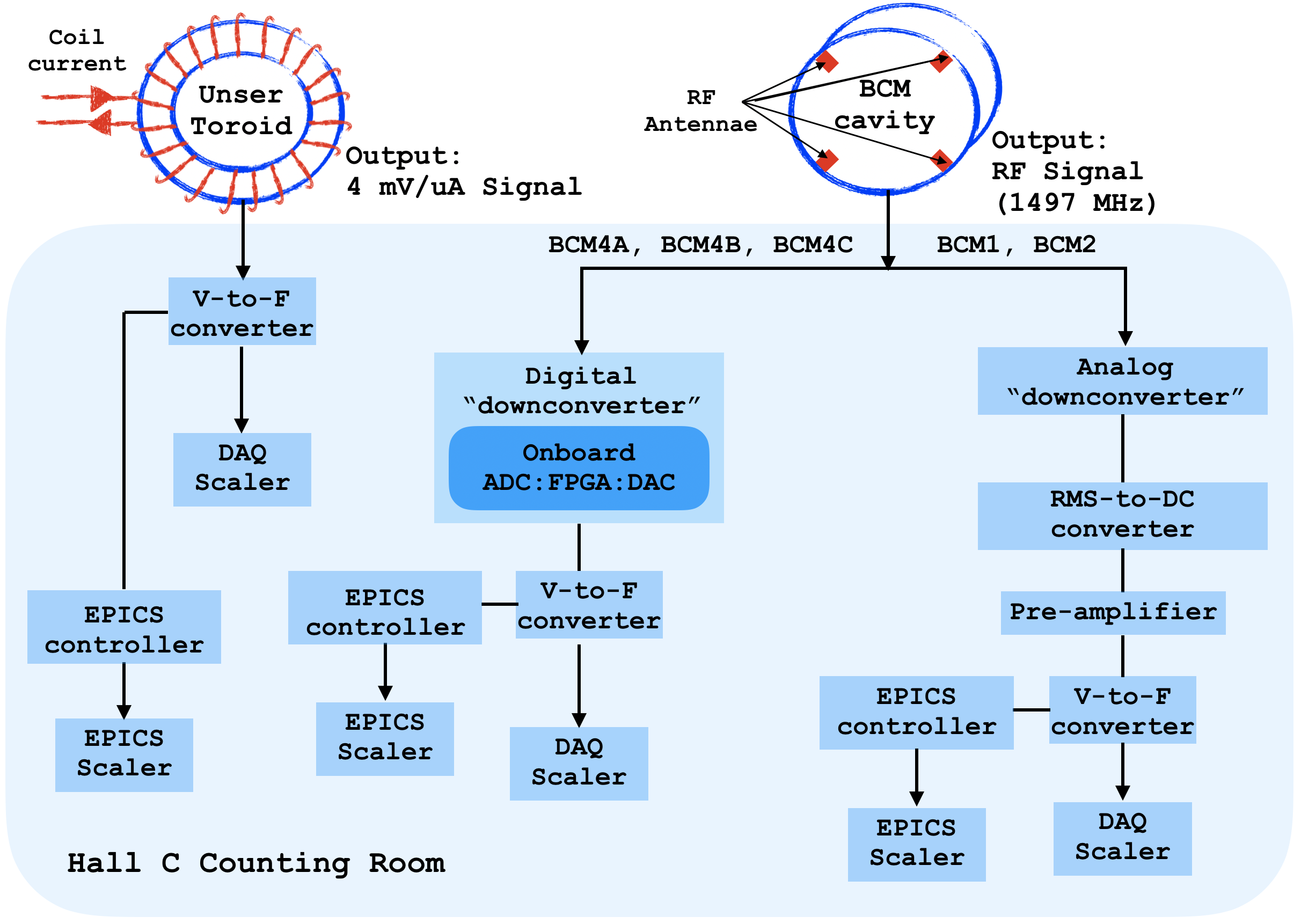}
\caption{Hall C BCM electronics diagram\cite{HallC_SEM_saw2019}.}
\label{fig:fig3.14}
\end{figure}\\
\indent Figure \ref{fig:fig3.14} shows a schematic diagram of the Hall C BCMs and Unser. The output RF signals from the BCMs are at 1497 MHz and must be
converted to a lower frequency signal by down-converters to be processed by the Hall C electronics. In the primary BCMs (BCM1 and BCM2), the RF signal is converted to a 
lower frequency ($\sim$100 kHz) by an analog down-converter and fed into an RMS-to-DC converter with a 20 kHz bandpass filter. 
The output signal is amplified and sent to a Voltage-to-Frequency converter (V-to-F) before being read out by the scalers. The supplementary BCMs (BCM4A, BCM4B, BCM4C) RF signals 
are processed by a digital down-converter with an onboard ADC, FPGA and DAC, which provides an analog signal that can be sent to a V-to-F and finally to the scalers\cite{HallC_SEM_saw2019}.
The Unser has a nominal output signal of 4 mV/$\mu$A and is also sent to a V-to-F before being read out by a scaler. The BCM/Unser signals from
the V-to-F are also read out by EPICS scalers directed by the EPICS controller located in the Hall C Counting Room. The Unser gain is verified
during downtimes by running a precision current through a wire which passes through the toroid head. \\
\indent Even though the Unser has an extremely stable gain, its output signal offset can drift significantly on a time scale of several minutes and cannot
be used to continuously monitor the beam current. On the other hand, the BCMs in general have a stable offset but their gain needs to be calibrated
and is not as stable as the Unser gain. Therefore, to use the BCMs as continuous current monitors they must be calibrated relative to the absolute beam
current determined by the Unser.  
\section{Target Chamber}
\begin{figure}
\centering
\includegraphics[scale=0.35]{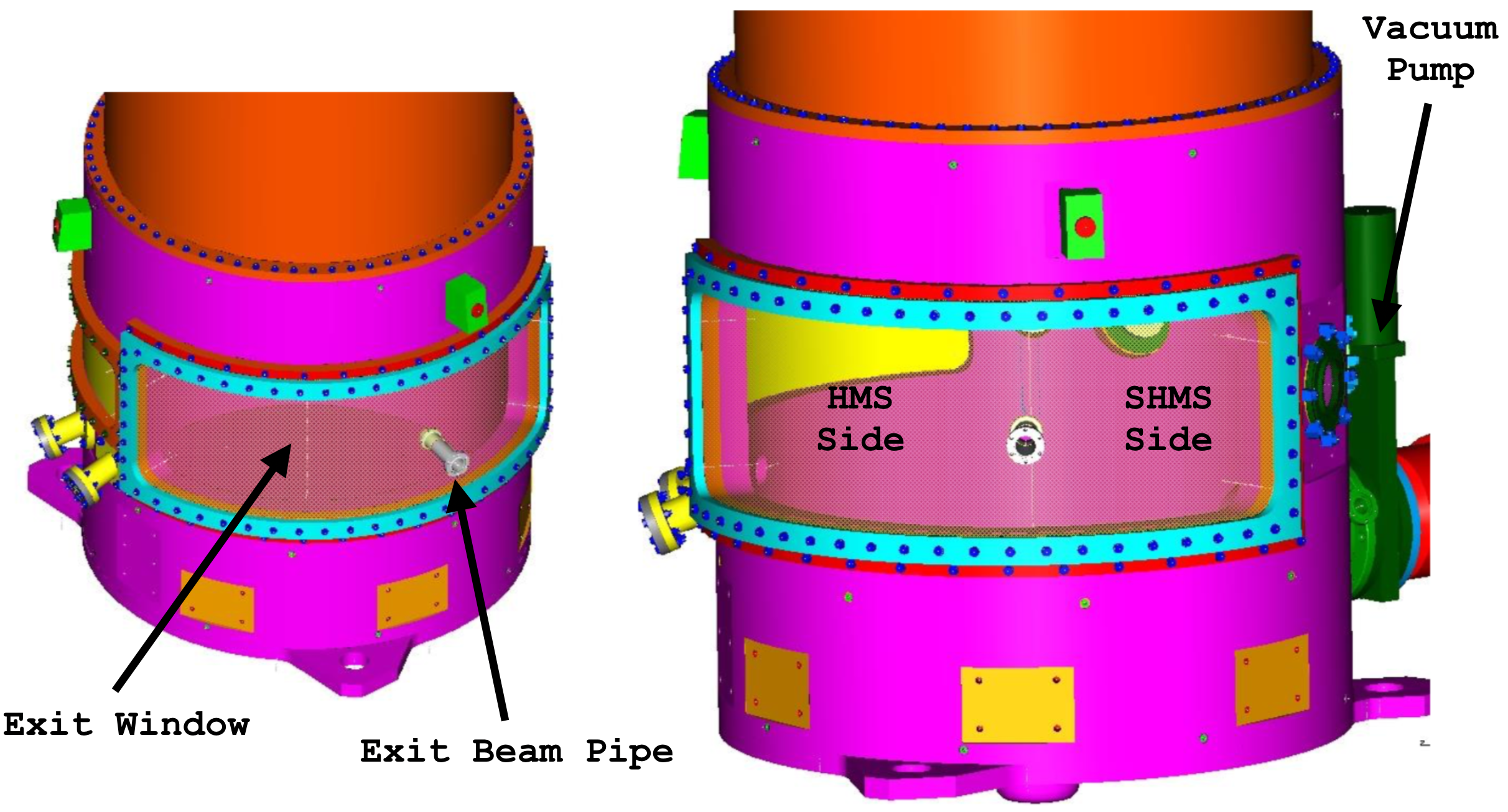}
\caption{A CAD (computer-aided design) drawing of the Hall C Target Chamber design. Note: Reprinted from Ref.\cite{HallC_SEM_saw2019}.}
\label{fig:fig3.15}
\end{figure}

The target chamber is a large evacuated cylindrical aluminum tank in two stacked sections (see Fig. \ref{fig:fig3.15}) that contains the solid and cryotargets in a target ladder. 
The aluminum chamber is nominally 2 inches thick with an inner diameter of 41 inches and outer diameter of 45 inches. The vacuum in the chamber is kept
at a pressure of a few 10$^{-6}$ torr by a turbomolecular vacuum pump connected through a gate valve in the lower cylinder. In the standard configuration, both spectrometers
share a single chamber exit window that covers a horizontal angular range from 3.2$^{\circ}$ to 77.0$^{\circ}$ on the HMS side and 3.2$^{\circ}$ to 47.0$^{\circ}$ on the SHMS side
with a vertical coverage of $\pm17.3^{\circ}$ for both sides.\\
\indent Both spectrometers' entrance window are actually very close to, but not in contact (or vacuum coupled) with the chamber window itself. So as the
electron beam scatters from the target, the outgoing particles have to exit the target cell, pass through
the exit window of the chamber into the air
and then through the spectrometer entrance window before entering the spectrometer vacuum. The unscattered beam exits the chamber through an opening in the exit window to which the exit beam pipe 
is connected via a threaded compression flange. There are also various openings in the target chamber through which the beam can enter, two pumping ports, several viewports 
and some spare ports. The viewports are used with a remote TV camera and light to observe the target motion and position in the counting room\cite{HallC_SEM_saw2019}.
\subsection{Target Ladder}
\begin{figure}[H]
\centering
\includegraphics[scale=0.31]{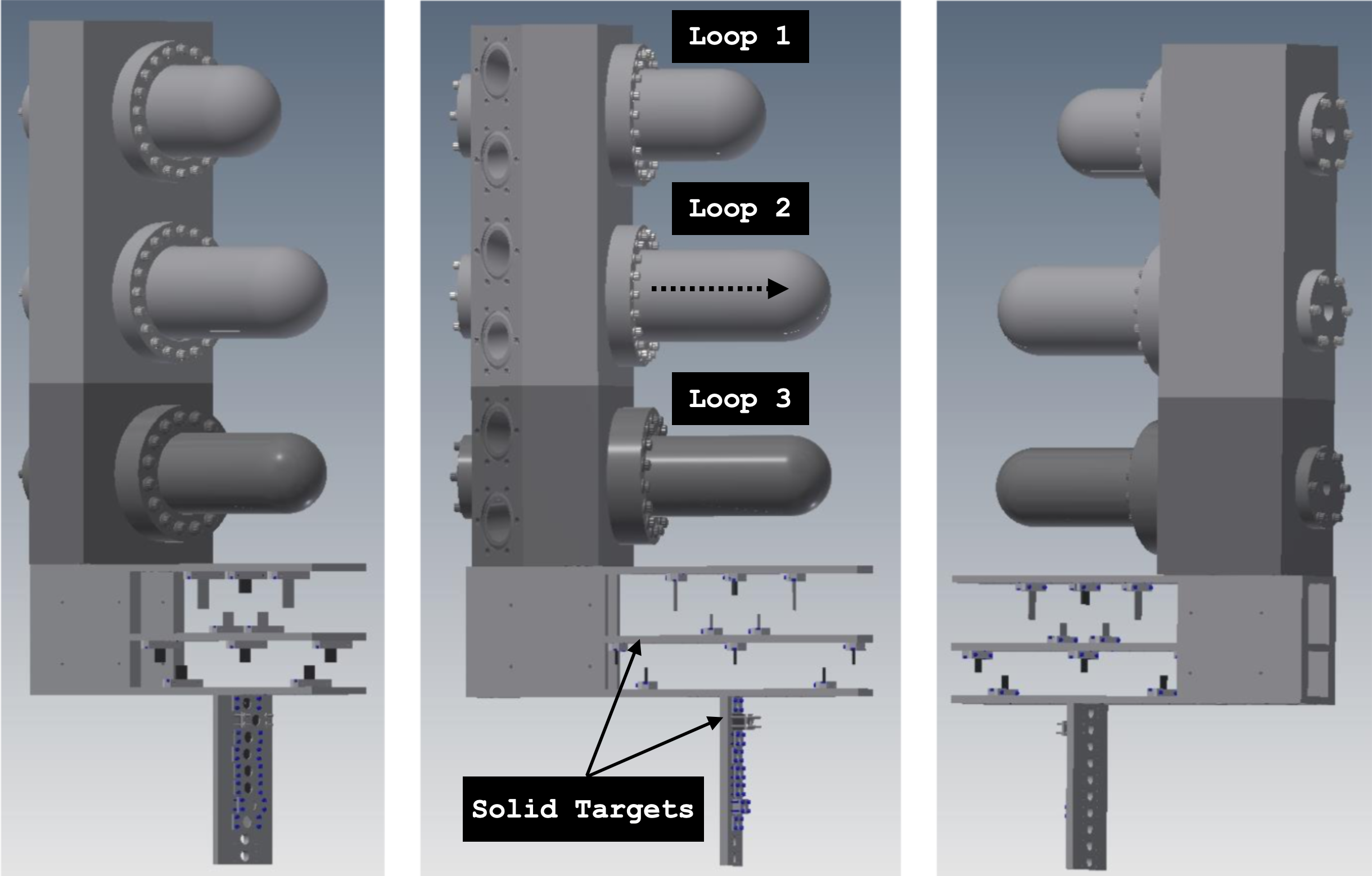}
\caption{A CAD drawing of the Hall C Target Ladder. The arrow shows the beam direction. Note: Reprinted from Ref.\cite{HallC_SEM_saw2019}.}
\label{fig:fig3.16}
\end{figure}

\begin{figure}[H]
\centering
\includegraphics[scale=0.3]{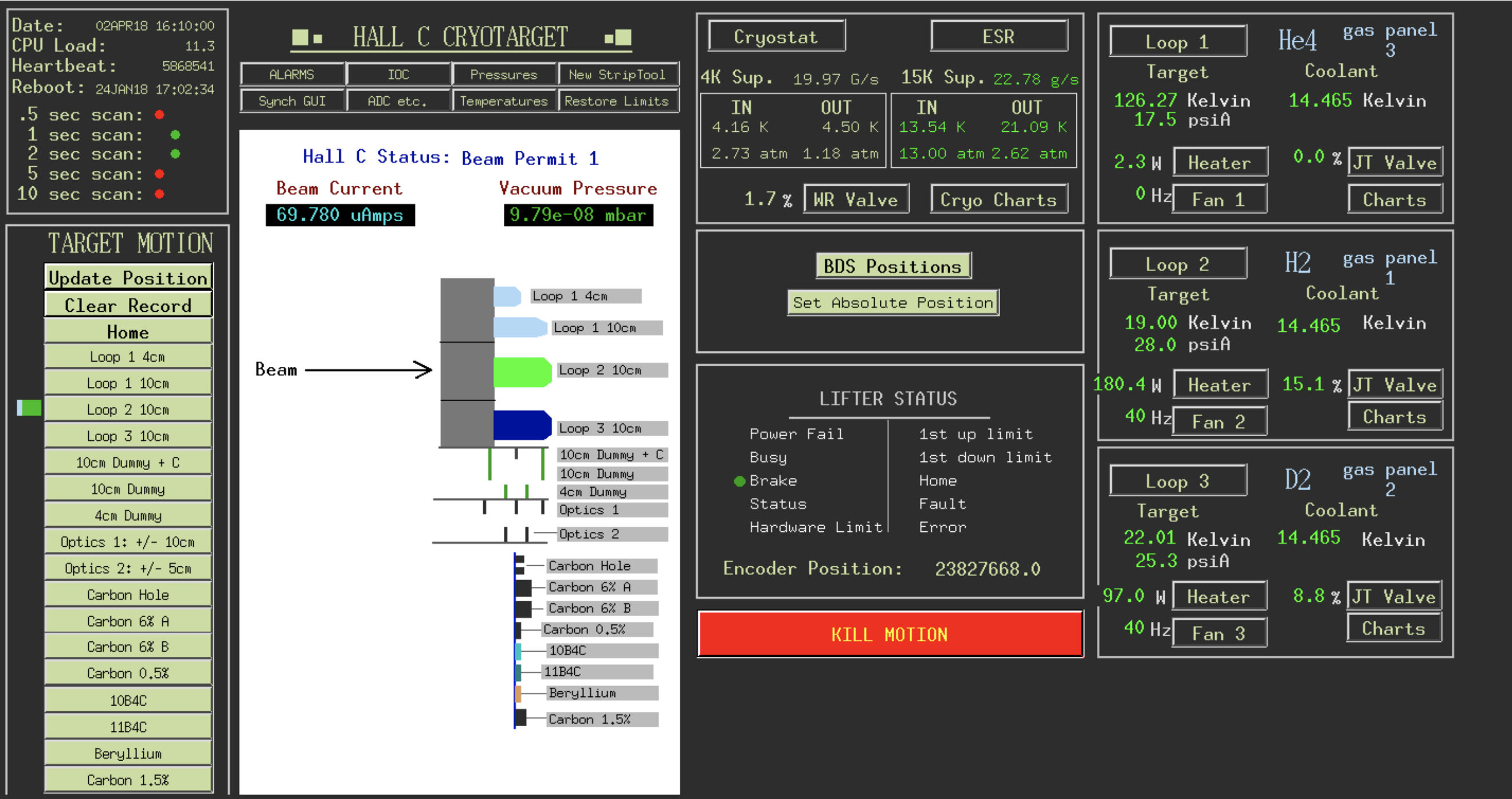}
\caption{Hall C Target GUI screen during the E12-10-003 experiment.}
\label{fig:fig3.17}
\end{figure}

The solid and cryogenic targets are accommodated in a target ladder with a single axis (vertical) motion system employed to select the desired target.
\noindent Figure \ref{fig:fig3.16} shows a CAD representation of the target ladder with the three cryogenic target cells above the solid targets. The target ladder motion is
controlled remotely via a target GUI in the Counting Room. During this experiment, the cryogenic target cells were filled with liquid helium (Loop 1), hydrogen (Loop 2)
and deuterium (Loop 3), respectively. \\
\indent Figure \ref{fig:fig3.17} shows a typical target GUI screen during this experiment. The central white screen shows a representation of the targets
and an arrow indicating which target was being hit by the beam. The live status of the beam current and the target chamber vacuum pressure are also visible at 
the top of that screen. To the extreme left of the GUI is the panel used to move the target ladder to a specific target. During this operation, it is extremely
important for the beam to be taken away to prevent any damage to the target ladder system. To the right of the white screen are two panels with live feedback of
the helium coolant supply from the End Station Refrigerator (ESR) as well as the cryogenic targets temperature and pressure and other relevant information which 
can be displayed on strip charts. See Refs.\cite{HallC_SEM_saw2019,12GeV_tgtTraining} for detailed information.\\
\indent The relevant targets\footnote{\singlespacing See HCLOG entry:\url{https://logbooks.jlab.org/entry/3555843} for detailed
  information and pictures on the Hall C Target Configuration for Spring 2018 run period.} used in this experiment were:
\begin{itemize}
  \item \textbf{Carbon Hole:} A carbon hole target consists of a thin carbon foil with a central hole of 2 mm in diameter. With a rastered beam of at least $2\times 2$ mm$^{2}$, this target can be
  used to check how well is the beam centered at the target. As the beam passes through the hole, the edges of the beam do interact with the carbon at the edges of the hole causing the
  electrons to re-scatter and be detected by either spectrometer. The $(x,y)$ raster values are plotted (when the spectrometer recorded a particle) forming a raster pattern with a hole (see Fig. \ref{fig:fig3.18}).
  \begin{figure}[!ht]
    \centering
    \includegraphics[scale=0.35]{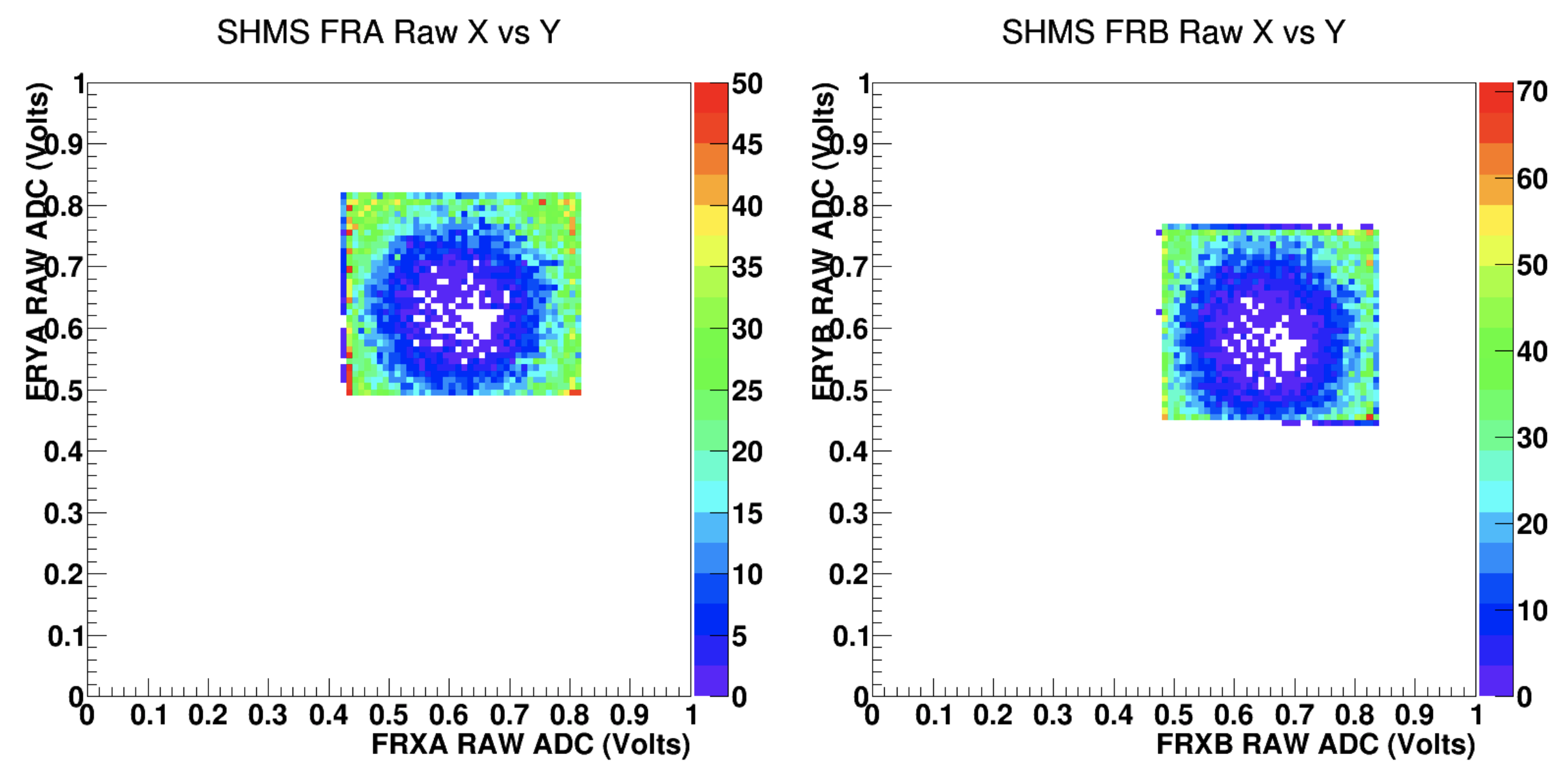}
    \caption{Carbon hole check during the E12-10-003 experiment shows the raster pattern for FR-A (left) and FR-B (right) raster magents.}
    \label{fig:fig3.18}
  \end{figure}
\item \textbf{Aluminum Dummy (10 cm):} The dummy target consists of aluminum foils mounted on separate frames at the locations Z = $\pm5$ cm corresponding to the
  cryogenic target entrance and exit windows. The dummy target runs are used for the subtraction of the background associated with the actual aluminum cryotarget windows.
  Note: dummy target windows are $\sim$ 10 times thicker than the actual cryotarget windows which must be accounted for.
\item \textbf{Optics-1:} The optics target consists of carbon foils located at Z = -10, 0, 10 cm along the beam axis and are used for spectrometer optics optimization studies.
  Even though these foils are beyond the target length used in Hall C, they are necessary for an optics reconstruction along the full target length.
\item \textbf{Liquid Hydrogen (10 cm):} The cryogenic liquid hydrogen (LH$_{2}$) is kept at a temperature of T$_{\mathrm{LH}_{2}} = 19\pm0.1$ K ($\sim25$ psia).
  LH$_{2}$ freezing and boiling points are T$_{F}$ = 13.8 K and T$_{B}$ = 22.1 K, respectively. A $2\times 2$ mm$^{2}$ raster is used to minimize density reduction at high beam currents. 
\item \textbf{Liquid Deuterium (10 cm):} The cryogenic liquid deuterium (LD$_{2}$) is kept at a temperature of T$_{\mathrm{LD}_{2}} = 22\pm0.1$ K ($\sim23$ psia).
  LD$_{2}$ freezing and boiling points are T$_{F}$ = 18.7 K and T$_{B}$ = 25.3 K, respectively. A $2\times 2$ mm$^{2}$ raster is used to minimize density reduction at high beam currents.
\end{itemize}
\subsection{Cryotarget Loop Anatomy}
\begin{figure}[H]
\centering
\includegraphics[scale=0.35]{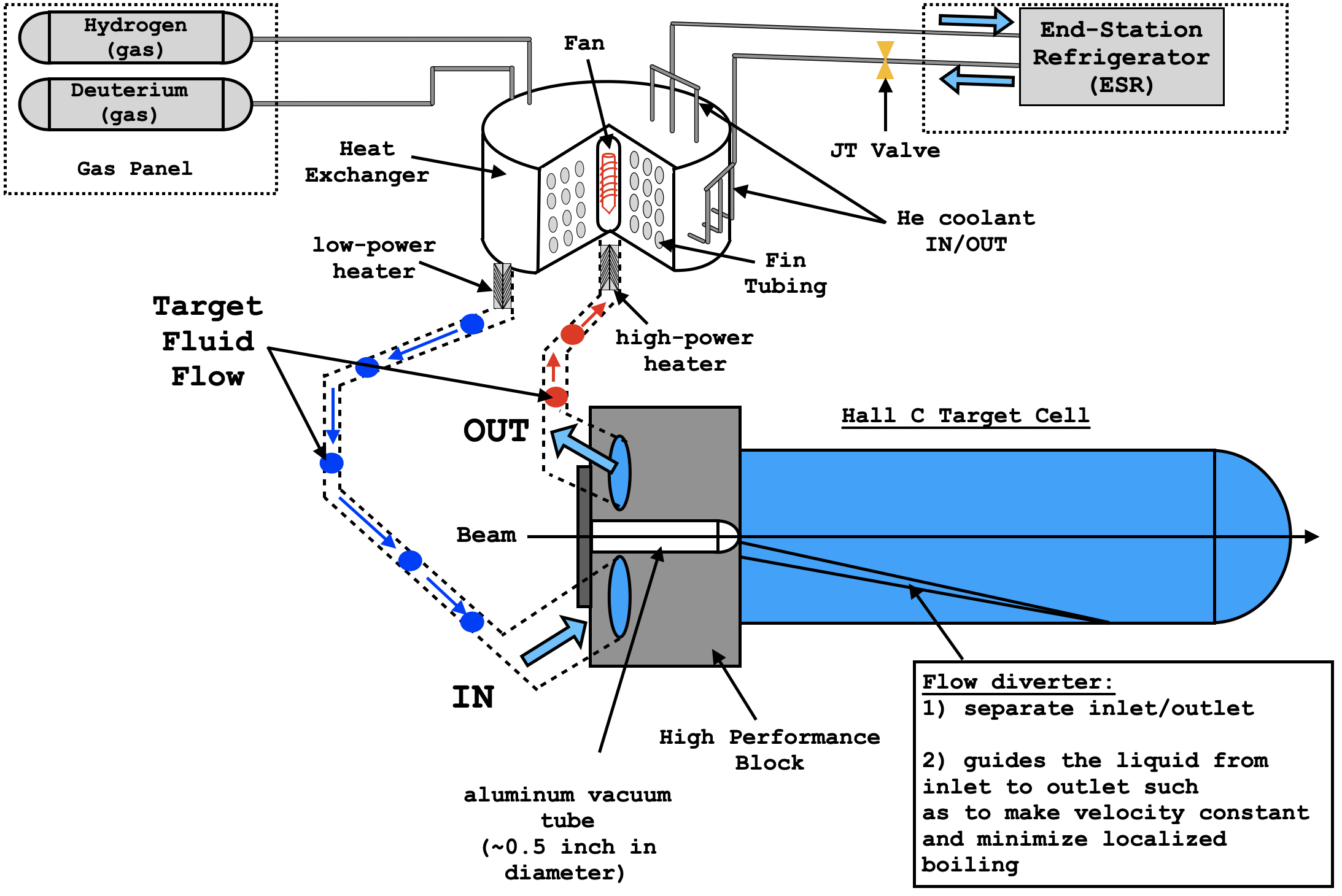}
\caption{Hall C cryotarget loop anatomy for the 12 GeV era (not to scale). Figure adaptation from Refs.\cite{12GeV_tgtTraining}\cite{SCovrig_privMarch2020}.}
\label{fig:fig3.19}
\end{figure}

Cryogenic targets are more complex than the solid targets, which are kept cool by thermal conduction (or direct contact) between
the cryotarget and solid target ladders. To keep the cryotargets at very low temperatures, the hydrogen and deuterium must be re-circulated constantly through a heat exchanger.
Figure \ref{fig:fig3.19} shows a simplified version of the loop anatomy for a typical target cell used in the Hall C 12 GeV era.
\indent A gas panel outside the building provides the heat exchanger with a constant supply of either hydrogen or deuterium, which is cooled to either 19 K (hydrogen) or 22 K (deuterium) by
the 15 K He coolant supply from the ESR. The amount of coolant sent to the heat exchanger is controlled via a Joule-Thompson (JT) valve. The target fluid is
then sent to the target cell and enters the high-performance block through an inlet. The flow diverter inside the target cell then guides the liquid from the inlet to the
outlet such as to make the flow velocity constant and minimize localized density changes caused by the beam. The liquid then leaves the target cell back to the heat exchanger at a higher
temperature than which it entered the cell, mostly due to the heat deposited by the electron beam. The liquid is then cooled again by the 15 K He supply from the ESR completing the loop.
This way, the target fluid is constantly recirculated between the target cell and heat exchanger to keep the fluid at operating temperatures.
\section{Hall C Spectrometers}
The main experimental equipment in Hall C consists of a pair of magnetic spectrometers designed to perform high-precision cross section measurements at
a relatively high luminosity\footnote{\singlespacing Luminosity in nuclear physics is defined as ($\#$ of beam particles/second) $\times$ ($\#$ of target particles/cm$^{2}$),
  typically expressed in units of cm$^{-2}$s$^{-1}$ (see Section \ref{sec:exp_Xsec}).}. The spectrometers have bearings at the pivot which permit rapid, remote spectrometer rotation.
Each spectrometer consists of a series of optical elements (quadrupoles and dipoles) followed by a series of particle detectors that are housed in a heavily shielded detector hut.
The optical elements are used to transport the scattered particles from the target chamber to the particle detectors. The tracks are then reconstructed at the focal
plane and translated back to the target.\\
\indent The entire spectrometer system rests on a rotatable support structure or carriage that keeps the optical elements and detectors aligned relative to each other and to the target. The support
structure rides on steel wheels mounted on steel rails that enable spectrometer rotation. The rotation
can be controlled remotely via the spectrometer control screens in the Counting Room. The central angles are determined from a camera attached at a surveyed
location towards the rear end of each spectrometer and is monitored remotely from a TV screen in the Counting Room. 
\begin{table}[H]
  \centering
  \scalebox{0.9}{
  \begin{tabular}[t]{lll}
    \hline
    \textbf{\textit{Parameter}} & \textbf{\textit{HMS}}         &\textbf{\textit{SHMS}}         \\
    & \textbf{\textit{Performance}} &\textbf{\textit{Specification}}\\
    \hline
    \hline
    Range of Central Momentum &  0.4 to 7.4 GeV/c &  2 to 11 GeV/c \\
    Momentum Acceptance  &  $\pm10\%$ &  $-10\%$ to $+22\%$ \\
    Momentum Resolution &  $0.1\%-0.15\%$ &  $0.03\%-0.08\%$ \\
    Scattering Angle Range &  10.5$^\circ$ to 90$^\circ$ & 5.5$^\circ$ to 40$^\circ$ \\
    \\
    Target Length Accepted at 90$^\circ$ &  10 cm &  50 cm \\
    Horizontal Angle Acceptance   &  $\pm32$ mrad &  $\pm18$ mrad \\
    Vertical Angle Acceptance   &  $\pm85$ mrad & $\pm50$ mrad \\
    Solid Angle Acceptance             &  8.1 msr  &  $>$4 msr \\
    \\
    Horizontal Angle Resolution  &  0.8 mrad &  0.5 -- 1.2 mrad \\
    Vertical Angle Resolution  &  1.0 mrad &  0.3 -- 1.1 mrad \\
    Target resolution (ytar)  &  0.3 cm &  0.1 - 0.3 cm \\
    \\
    Maximum Event Rate  &  2000 Hz &  10,000 Hz \\
    Max. Flux within Acceptance    &  $\sim 5$ MHz &  $\sim5$ MHz \\
    \\
    e/h Discrimination   &  $>$1000:1 at $98\%$ efficiency &   $>$1000:1 at $98\%$ efficiency \\
    $\pi$/K Discrimination  &  100:1 at $95\%$ efficiency &  100:1 at $95\%$ efficiency \\
    \hline
  \end{tabular}
  }
  \caption{Demonstrated performance of the HMS and design specifications for the SHMS.} 
  \label{tab:tab3.2}
\end{table}
\indent The major component of the Hall C 12 GeV upgrade is the new Super High Momentum Spectrometer (SHMS) that replaced the orginal companion of the HMS known as the Short Orbit
Spectrometer (SOS). The new SHMS-HMS pair makes Hall C the only facility in the
world capable of carrying out the rich nuclear physics program detailed in Refs.\cite{HallC_12GeV_CDR,12GeV_program_JRA2005}. To be able to carry out this program successfully,
the SHMS\footnote{\singlespacing The SHMS was built from
2009 to 2016, and was first commissioned in 2017 as part of the Hall C KPP.} was designed to achieve a maximum momentum of 11 GeV/c, which is well matched
with the maximum beam energy delivered to Hall C. The SHMS is also able to rotate to very small
forward central angles down to 5.5$^{\circ}$ as well as operate at an unprecedented high luminosity of 10$^{39}$ cm$^{-2}$s$^{-1}$ and has a 32$\%$ momentum acceptance which
measures the percent deviation of the particle momentum relative to the central momentum of the spectrometer. A detailed description of the HMS performance parameters
and SHMS design specifications is given in Table \ref{tab:tab3.2}.

\subsection{Spectrometer Slit System}
\begin{figure}[H]
  \centering
  \includegraphics[scale=0.4]{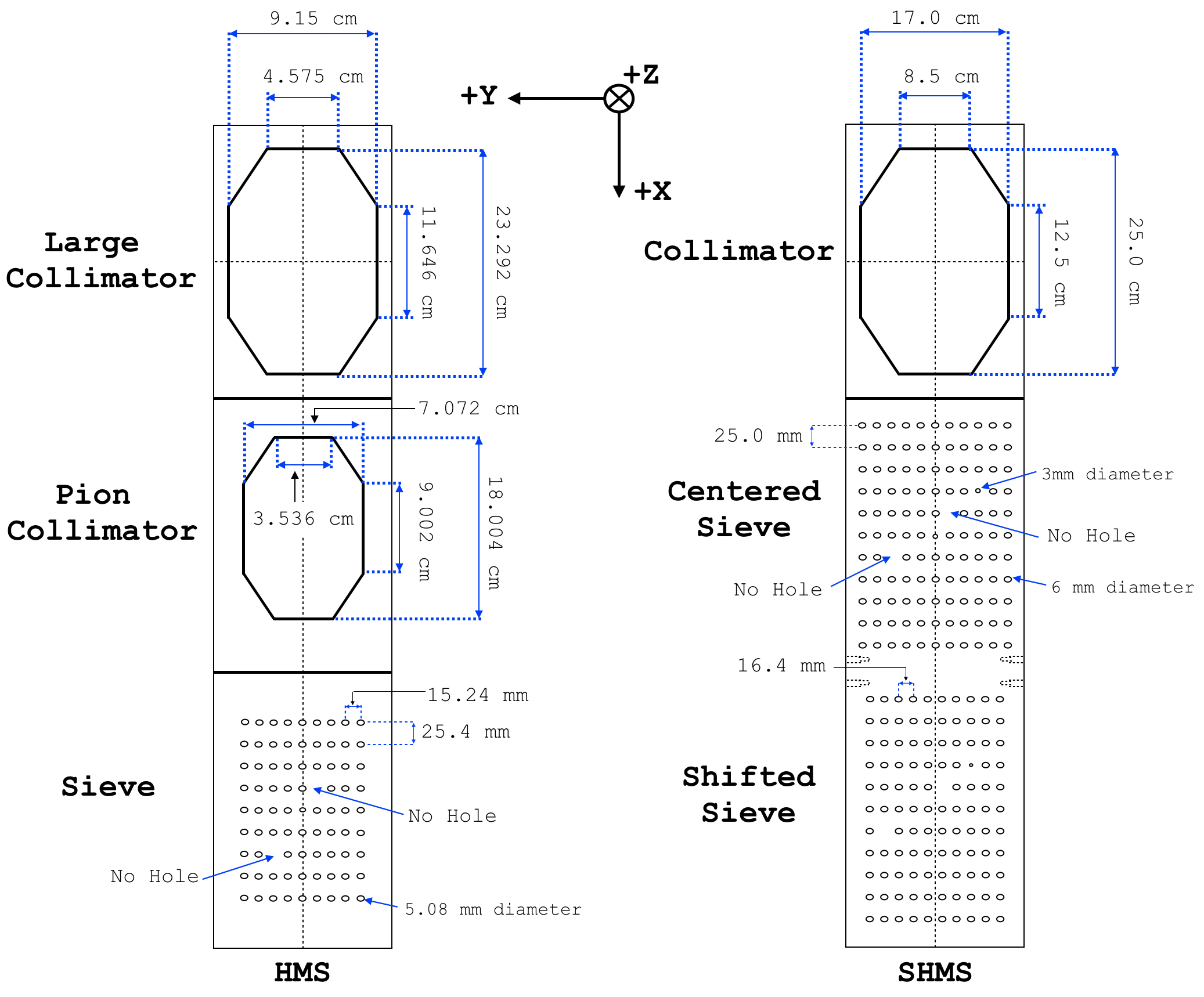}
  \caption{Spectrometer slit system.}
  \label{fig:fig3.20}
\end{figure}

As the electron beam interacts with the target atoms, the final-state particles scatter radially outwards in all possible directions, but only a small fraction of these fall within
the momentum and angular acceptance set by the spectrometer. Each spectrometer is equipped with a slit system containing collimators
defining the angular acceptance, and a sieve slit used for spectrometer optics studies. Table \ref{tab:tab3.3} gives a summary of the apertures
as well as the corresponding solid angles defined by the collimators in each spectrometer.\\
\indent The slit system in each spectrometer is housed in a vacuum box with a movable slit ladder that has space for three separate slits. Each ladder position can either hold
a collimator or an optics sieve slit in accordance to the experimental requirements. The ladder motion is controlled remotely from the Hall C Counting Room.
Figure \ref{fig:fig3.20} shows the slit configuration used for each spectrometer during the commissioning experiments.
\begin{table}[H]
  \centering
  \scalebox{0.83}{
  \begin{tabular}[t]{lllll}
    \hline
    & \textbf{Horizontal} (mr) & \textbf{Vertical} (mr) & \textbf{Solid Angle} (msr) & \textbf{Shape}\\
    \hline
    \hline
    HMS &  &  & \\
    \textit{Large Collimator} & $\pm$70  & $\pm$27.5  & 6.74 & Octagonal, Flared\\
    \textit{Pion Collimator} & $\pm$54  & $\pm$21.3  & 4.03 & Octagonal, Flared\\
    \\
    SHMS & & &\\
    \textit{Collimator} & $\pm$49.4 & $\pm$33.6 & 5.81 & Octagonal, Flared\\
    \hline
  \end{tabular}
  }
  \caption{Spectrometer apertures at the collimator entrance.} 
  \label{tab:tab3.3}
\end{table}
\subsubsection{The HMS Slits}
In the HMS, the slit system is installed at the entrance of the first quadrupole magnet (Q1) at a distance of Z$^{\mathrm{(HMS)}}_{\mathrm{coll}}$=166.37 cm from the center of the target to the collimator
entrance. The slit system consists of two collimators and a sieve slit. Each slit is a rectangular block of a heavy alloy metal (90$\%$ W, 10$\%$ Cu/Ni) with a density of 17 g/cm$^{3}$. The
collimators have a machined octagonal-shaped opening whereas the sieve slit has several rows of holes drilled into it with the exception of two specified locations that are used to determine
the orientation of the slit in the reconstruction analysis. During this experiment, there was no need to insert the sieve as the HMS optics is well understood and it was decided to only use
the new \textit{Large Collimator} and not the original \textit{Pion Collimator} (6 GeV era) to define the acceptance.
\subsubsection{The SHMS Slits}
In the SHMS, the slit system is installed between the horizontal bender (HB) and first quadrupole (Q1) magnets at a distance of Z$^{\mathrm{(SHMS)}}_{\mathrm{coll}}$=253 cm from the center of the target to
the collimator entrance. The slit system consists of a collimator and two sieve slits. The collimator and two sieves are made of Mi-Tech$^{\mathrm{TM}}$ Tungsten HD-17 (90$\%$ W, 6$\%$ Ni, 4$\%$ Cu)
with a density of 17 g/cm$^{3}$. Similar to the HMS, the SHMS collimator has an octagonal-shaped opening and the sieves have several rows of holes drilled into it (see Fig. \ref{fig:fig3.20}).
The centered sieve has 11 columns of holes with the sixth column at the center, whereas the shifted sieve has 10 columns and shifted from the central axis.
More details about the HMS/SHMS slit system can be found in Refs. \cite{THorn_Jan2009,HallC_SEM_saw2019}. 
\subsection{Spectrometer Magnets}
The spectrometers' optical elements consist of a series of superconducting magnets that guide the scattered particles towards a detector stack. To keep the magnet coils
at superconducting temperatures, a constant supply of liquid $^{4}$He at a temperature of 4.5 K is provided by ESR to Hall C. The cryogenic supply is distributed to each spectrometer
via flexible transfer lines emanating from a main distribution box located over the pivot into each spectrometer cryogenics network. The magnet power supplies are located on the
spectrometer support structure adjacent to the magnets. To remove the excess heat, the power supplies are all water-cooled by a constant flow rate that can be monitored by a water
flow meter located on the electronic boxes on the floor near the pivot\cite{HallC_SEM_saw2019}. The magnet cryogenics and power supplies, as with the spectrometer rotation controls,
are operated and monitored remotely via the magnet control screens in the Counting Room. \\
\begin{table}[H]
  \centering
  \scalebox{0.77}{
  \begin{tabular}[t]{llllllll}
    \hline
                    &               &                           &                   &                    &                   &                         & \textbf{Stored}  \\
    \textbf{Magnet} & \textbf{Type} & \textbf{EFL}\footnotemark & \textbf{Aperture} & \textbf{Momentum}  & \textbf{Current}  & \textbf{Field/Gradient} & \textbf{Energy} \\
                    &               &     (m)      &        (cm)       &     (GeV/c)        &     (A)           &                         &    (MJ)  \\
    \hline
    \hline
    HMS Q1  & Cold Fe & 1.867  & 40         & 7.4 & 1012  & 7.148 T/m  & 0.335\\
    HMS Q2  & Cold Fe & 2.104  & 60         & 7.4 & 1023  & 6.167 T/m  & 1.59\\
    HMS Q3  & Cold Fe & 2.104  & 60         & 7.4 & 1023  & 6.167 T/m  & 1.59\\
    HMS D   & Warm Fe & 5.122  & 40         & 7.4 & 3000  & 2.073 T    & 9.79  \\
    \\
    SHMS d & ``C'' Septum       & 0.752 & 14.5 x 18  & 11  & 3930  & 2.56 T     & 0.2\\
    SHMS Q1 & Cold Fe        & 1.86  & 40         & 11  & 2460  & 7.9 T/m    & 0.382\\
    SHMS Q2 & cos(2$\theta$) & 1.64  & 60         & 11  & 3630  & 11.8 T/m   & 7.6\\
    SHMS Q3 & cos(2$\theta$) & 1.64  & 60         & 11  & 2480  & 7.9 T/m    & 3.4\\
    SHMS D  & cos($\theta$)  & 2.85  & 60         & 11  & 3270  & 3.9 T      & 13.7\\
    \hline
  \end{tabular}
  }
  \caption{Spectrometer magnets design parameters\cite{SLassiter_privMarch2020}.} 
  \label{tab:tab3.4}
\end{table}
\footnotetext{\singlespacing EFL refers to the \textit{Effective Field Length} of the magnet.}
\indent Table \ref{tab:tab3.4} summarizes the design parameters of the spectrometer magnets.
Each spectrometer is designed to provide point-to-point focusing (Q1, Q2, Q3) and a vertical momentum dispersion (D) and can be configured to
transport either positive or negatively charged particles by setting the individual magnets to a ``+'' or ``-'' polarity in an alternating pattern. 
The central momentum is also set individually for each magnet via a field setting program that uses a current-to-field map associated with a central
momentum for each magnet. The detailed procedure of how to operate the spectrometer magnets from the magnet GUI is discussed in Ref.\cite{HCwiki_magnets}. 
\subsubsection{The HMS Magnets}
\begin{figure}[H]
\centering
\includegraphics[scale=0.35]{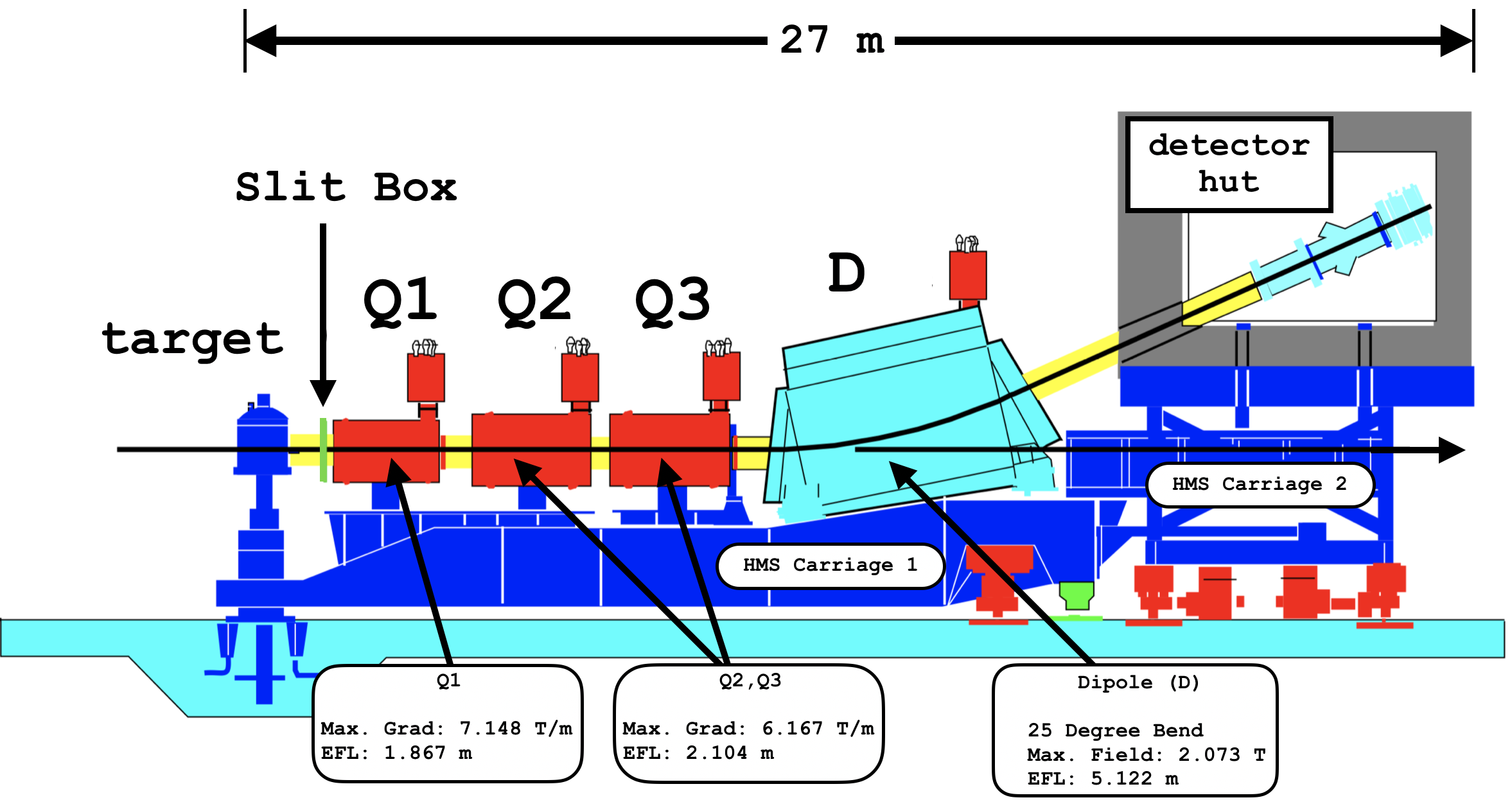}
\caption{High Momentum Spectrometer (HMS) side view.}
\label{fig:fig3.21}
\end{figure}

The HMS optics elements consist of three quadrupoles (Q1, Q2, Q3) and a dipole (D) magnet arranged in a (QQQD) configuration that are used to
transport the scattered particles into a series of particle detectors located in a detector hut (see Fig. \ref{fig:fig3.21}).
The quadrupoles focus the collimated particles into the dipole that bends the central momentum particles vertically by 25$^{\circ}$ into the detector stack.
\indent The HMS is capable of detecting particles with central momentum from 0.4 to 7.4 GeV/c and can be rotated from 10.5$^{\circ}$ to 85$^{\circ}$ where the
minimum/maximum angles are restricted by administrative, software, and hardware limits. These limits depend on the beamline configuration and obstructions
in the Hall at the time of the experiment\cite{HallC_SEM_saw2019}. The spectrometer
magnets and shield hut (with detectors) are actually supported by two separate, but firmly attached carriages that keep the detectors and magnets aligned to each other and to the target. \\
\indent Even though the HMS is a well understood spectrometer from the 6 GeV era, it underwent minor modifications in preparation for the experimental requirements
of the 12 GeV era and had to be re-commissioned. First, the NMR probe\footnote{\singlespacing The NMR probe is used to determine the spectrometer central momentum, which
  is mostly determined by the dipole. Ideally, the probe is placed at the center of the dipole and picks up a field reading from it and regulates this
  field by re-adjusting the dipole current to achieve a more precise dipole field and hence a more precise central momentum. In reality, more than one
probe is used in this procedure (see Ref.\cite{holly_HMS_Optics2017} for details).}
used for precise field regulation of the HMS dipole was replaced
and second, the old HMS drift chambers were replaced with a new design similar to the SHMS drift chambers.
With the replacement of the NMR probe, the precise mapping between the NMR probe reading and the dipole magnetic field had to be re-done 
(see Ref.\cite{holly_HMS_Optics2017}) and, with the installation of the new HMS drift chambers, the tracking and optics
reconstruction to the target had to be checked as well. Furthermore, some of the 12 GeV era experiments required the HMS central momentum to
operate above $\sim4$ GeV/c, where saturation effects in the dipole and quadrupole magnets were expected but had not been previously studied since
no experiment in the 6 GeV era required HMS central momenta $>4$ GeV/c. For this purpose, special runs using hydrogen elastic scattering and carbon data with
the HMS sieve inserted were taken and used for the re-optimization of the HMS reconstruction optics\cite{holly_HMS_Optics2017}. 
\subsubsection{The SHMS Magnets}
\begin{figure}[!h]
\centering
\includegraphics[scale=0.32]{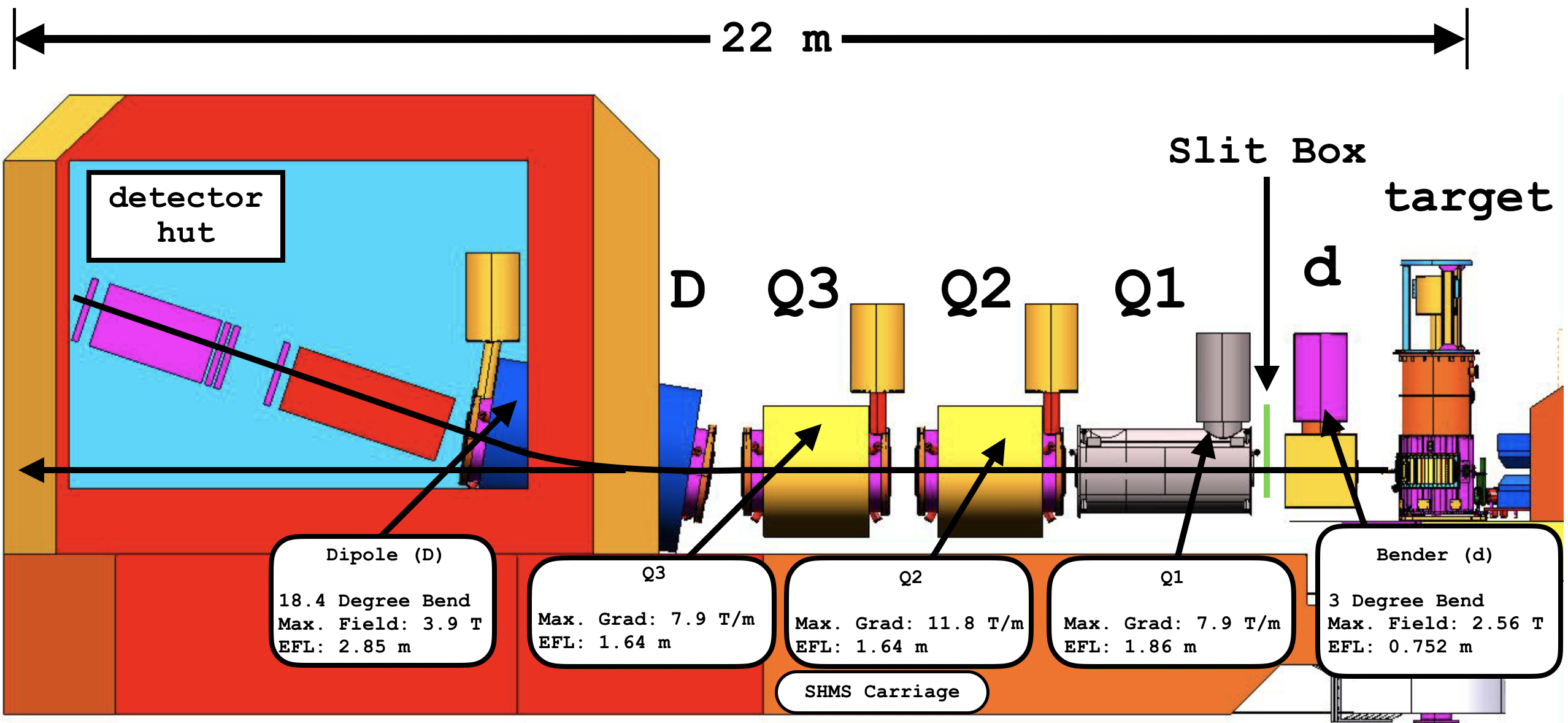}
\caption{Super High Momentum Spectrometer (SHMS) side view.}
\label{fig:fig3.22}
\end{figure}

Similar to the HMS, the SHMS optics elements consist of an array of three quadrupoles (Q1, Q2, Q3) and dipole (D) magnet used to guide the scattered particles into a series
of particle detectors in a shielded hut. The SHMS has an additional dipole magnet (d) located between the target chamber and the first quadrupole known as the Horizontal Bender (HB).
The magnets are arranged in a (dQQQD) optics configuration (see Fig. \ref{fig:fig3.22}). 
\indent The HB is used to horizontally bend the scattered particles that match the SHMS central momentum by 3$^{\circ}$ away from the beamline and towards the collimator before entering Q1. To achieve a horizontal bend,
the optical axis of the HB is oriented 3$^{\circ}$ lower (towards the beamline) relative to the optical axis of the rest of the spectrometer (collimator, Q1, Q2, Q3, D, detectors). With the HB equipped,
the SHMS is able to detect particles at angles that would have otherwise been impossible to reach due to the obstruction of the quadrupole magnets and shield house with the beamline.\\
\indent As the particles are bent horizontally towards the collimator and enter Q1, they are focused through the remaining quadrupoles (Q2, Q3) and into the dipole (D) where
the central-momentum particles are vertically bent by 18.4$^{\circ}$ into the detector stack. The SHMS can detect particles with central momentum from 0.2 to 11 GeV/c and can be rotated from a central angle of 8.5$^{\circ}$ up to 40$^{\circ}$. In reality, due to the $3^{\circ}$
bend by the HB, the central-ray particles detected at the hut actually scatter from the target 3$^{\circ}$ lower relative to the hut, therefore, the SHMS full angular coverage is from 5.5$^{\circ}$ to 37$^{\circ}$.\\
\indent Given that the SHMS is a new spectrometer, a significant amount of work during the Fall-2017 to Spring-2018 run period has been devoted towards understanding and optimizing the magnetic optics as well as
commissioing the particle detectors. See Ref.\cite{holly_SHMS_Optics2019} for details of the optics commissioning work for the SHMS. Details on the detector calibrations and SHMS optics
optimization for this experiment will be discussed in detail in Chapter \ref{chap:chapter4}. 
\subsection{Spectrometer Detectors}
\begin{figure}[H]
\centering
\includegraphics[scale=0.4]{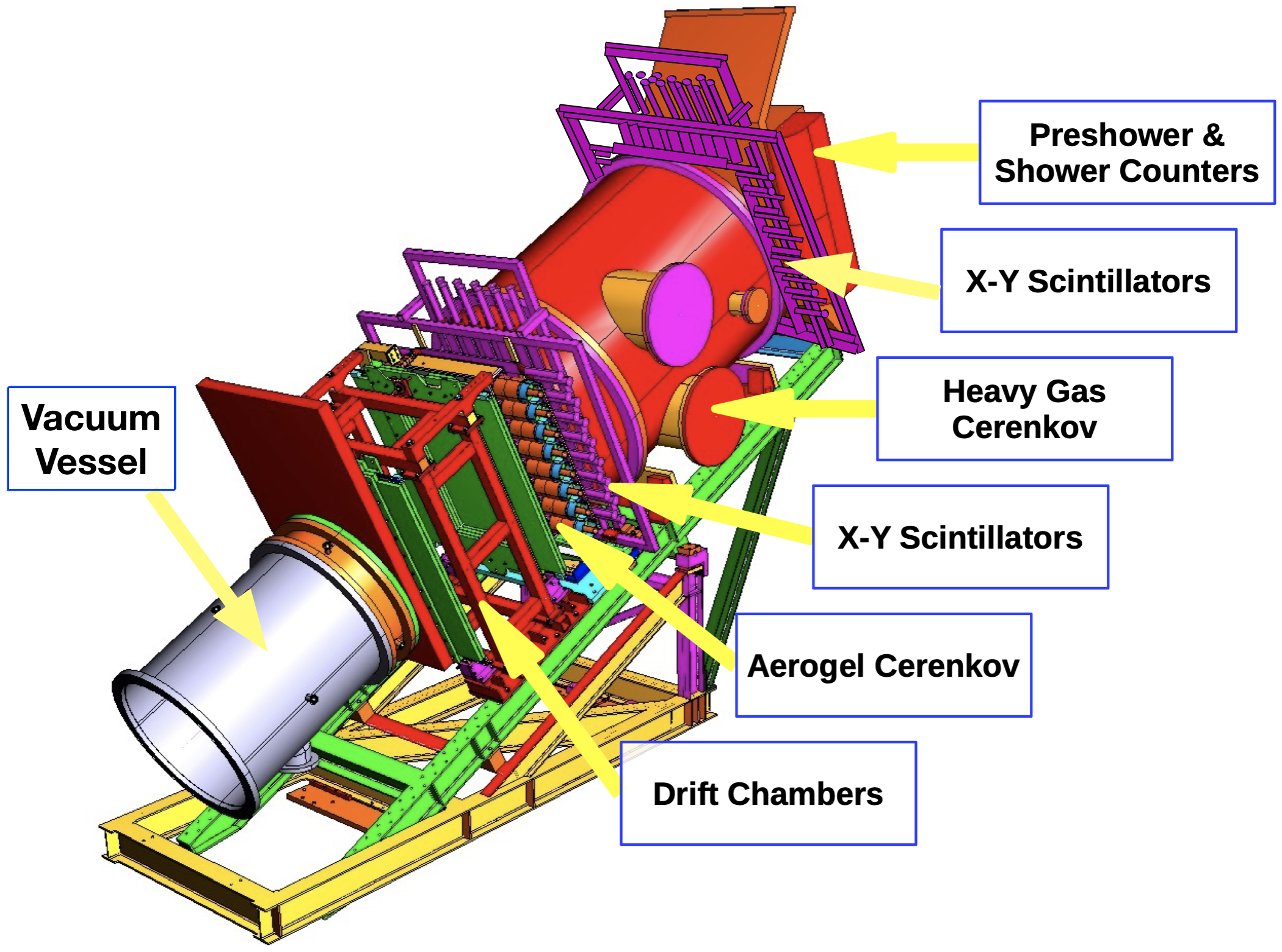}
\caption{High Momentum Spectrometer (HMS) detector stack.}
\label{fig:fig3.23}
\end{figure}

\begin{figure}[H]
\centering
\includegraphics[scale=0.32]{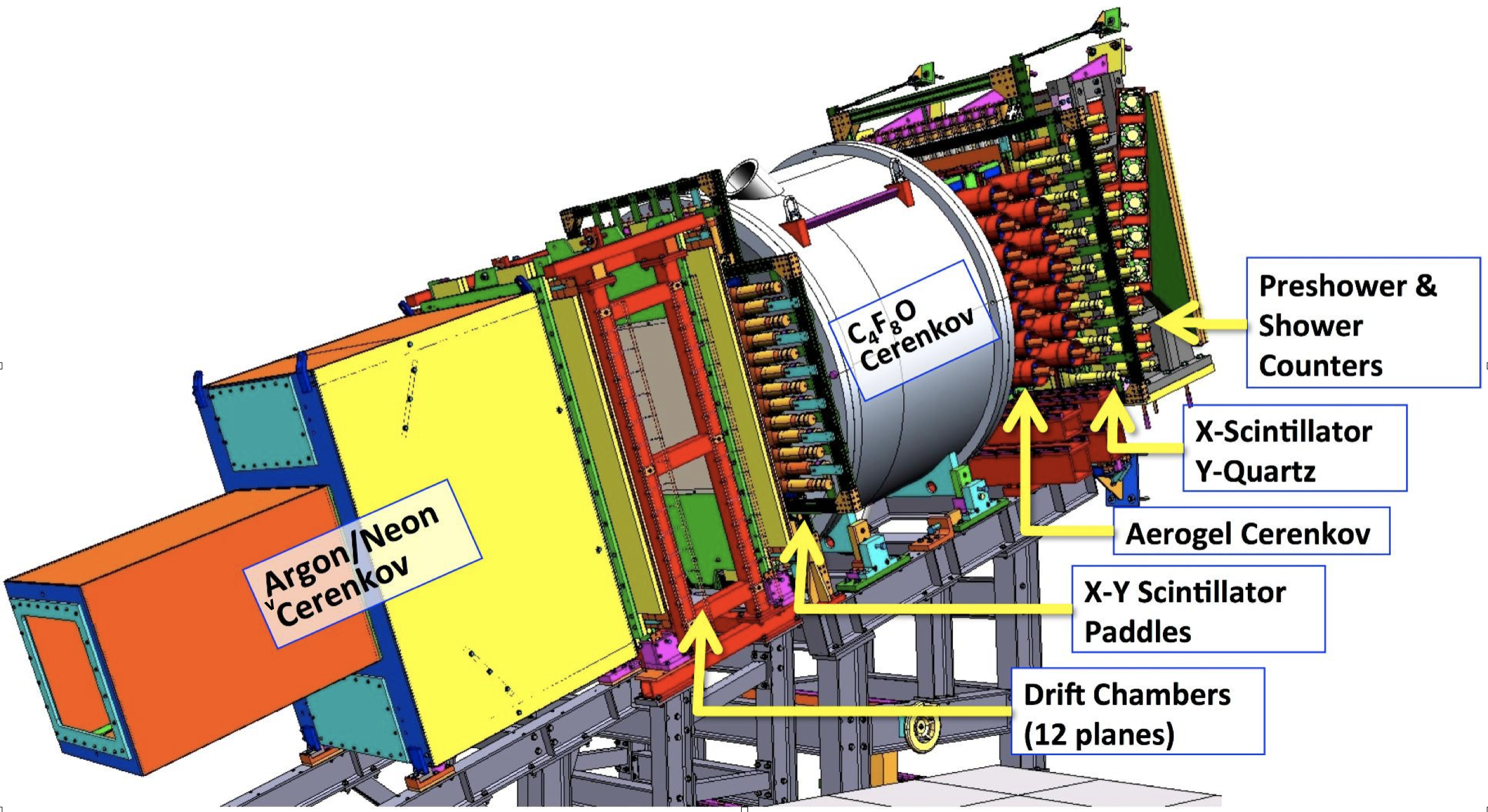}
\caption{Super High Momentum Spectrometer (SHMS) detector stack.}
\label{fig:fig3.24}
\end{figure}

Each spectrometer is equipped with a similar set of particle detectors housed in a heavily shielded hut. The detector package consists of a pair of drift chambers (DC1 and DC2) used for track reconstruction,
two pairs of hodoscope planes used for particle triggering, a calorimeter used for $e/\pi$ separation, and a gas and aerogel \v{C}erenkov used for additional particle identification.\\
\indent In the HMS (see Fig. \ref{fig:fig3.23}), the particles enter the detector hut through a cylindrical vacuum vessel that extends from the dipole exit window (outside the hut) to just upstream of DC1.
As the particles exit the vacuum vessel, they first pass through the drift chamber pair followed by an aerogel \v{C}erenkov detector, a first pair of XY hodoscope planes,
a Heavy Gas \v{C}erenkov (HGC), a second pair of XY hodoscope planes, and towards the end, a preshower and shower counters that make up the calorimeter detector. During the commissioning run period, the
aerogel detector was not installed in the HMS detector stack.\\
\indent In the SHMS (see Fig. \ref{fig:fig3.24}), as the particles enter the detector hut, they pass through a vacuum vessel (similar to HMS) coupled to the dipole, which
partially protrudes inside the hut due to space constraints. Depending on the experimental requirements for particle identification, the vacuum
extension pipe can be replaced with the Noble Gas \v{C}erenkov (NGC) at higher spectrometer momenta, where the effects of multiple scattering are minimized.\\
\indent As the particles exit the vacuum vessel (or NGC), they pass through a pair of drift chambers, followed by a pair of XY hodoscope planes, a Heavy Gas \v{C}erenkov, an aerogel \v{C}erenkov,
a second pair of XY hodoscope planes and finally, the preshower and shower counters, which constitute the calorimeter detector.\\
\indent During the commissioning run period, the NGC was installed as the spectrometer central momentum was relatively high in each of the experiments.
In this experiment (E12-10-003) in particular, both spectrometers used the standard hodoscopoe and drift chamber detectors for event triggering and tracking.
Due to the negligible background and low coincidence trigger rates the need of additional particle identification was minimal. Only the SHMS
calorimeter was used to select a clean sample of electrons.
\subsubsection{Drift Chambers}
\begin{figure}[H]
\centering
\includegraphics[scale=0.36]{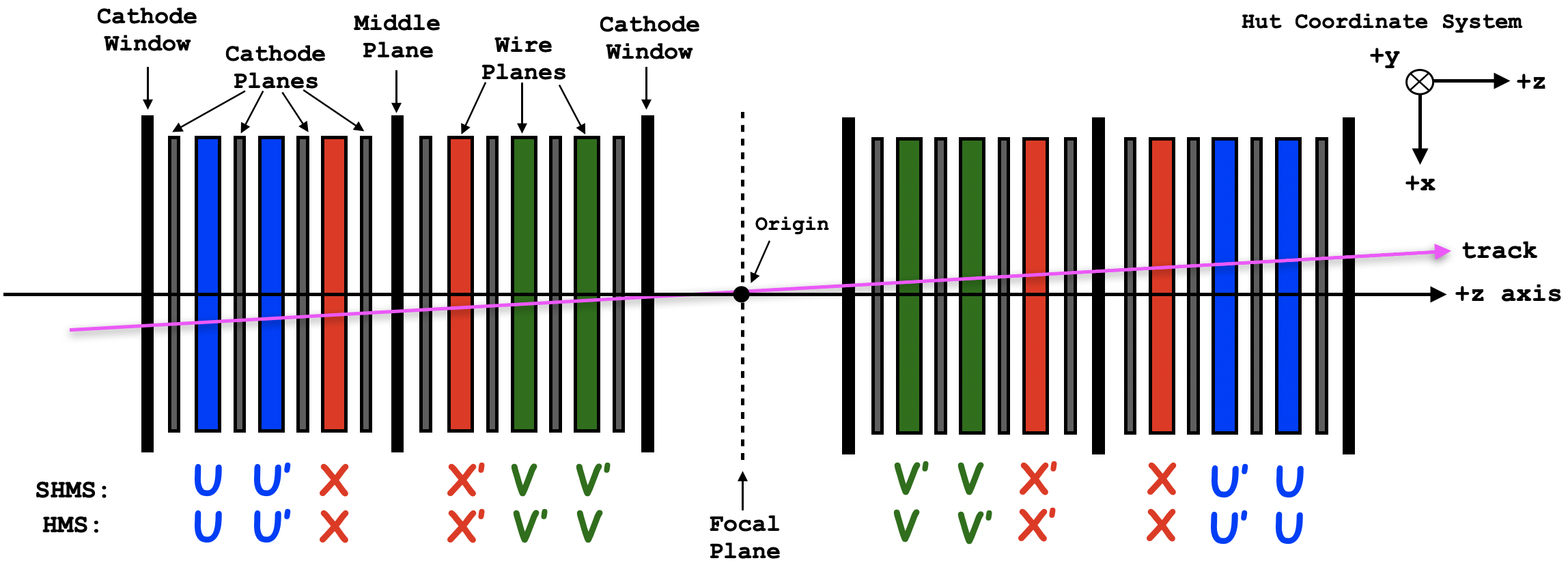}
\caption{Side view of the plane orientation for the DC1 (left) where the colored planes represent the wire planes, and
  DC2 (right) which is identical in design to DC1 rotated by 180$^{\circ}$
  about the $x$-axis (vertical) forming a mirror image along the $z$-axis.}
\label{fig:fig3.25}
\end{figure}

\begin{figure}[H]
\centering
\includegraphics[scale=0.4]{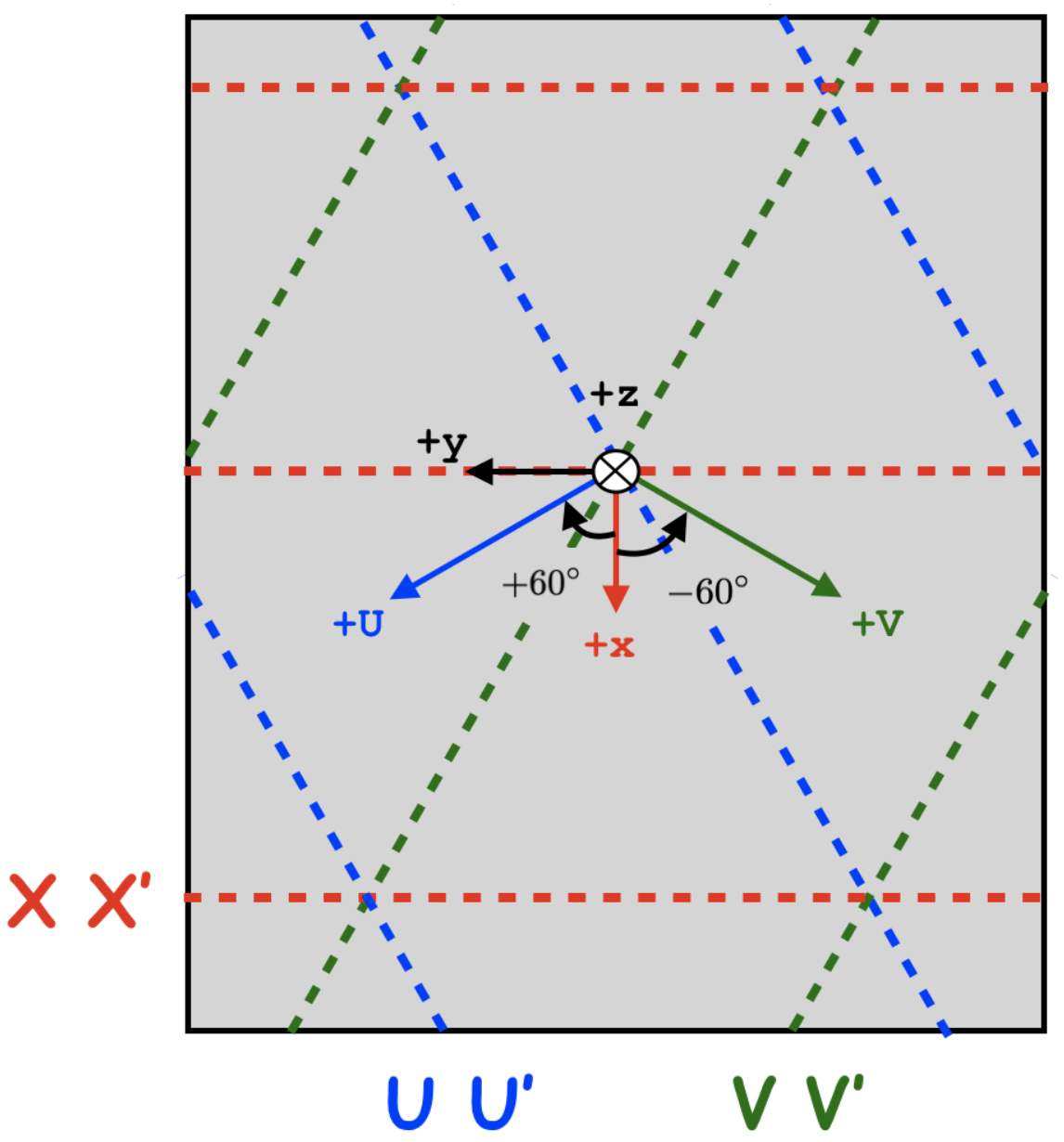}
\caption{Front view of the wire (dashed) orientations
  for each plane, indicated by representative sense wires of different colors, where the +$z$-axis (particle direction) is into the page. The wires in each plane are superimposed onto a single plane in this figure for
  convenience and their orientation is defined by the vector normal to the wire.}
\label{fig:fig3.26}
\end{figure}

The drift chambers in both the HMS and SHMS are of similar design (see Refs.\cite{hms_dc_slide,shms_dc_techreport}). In each spectrometer, the two drift chambers are mounted on an aluminum frame and are separated by about 80 cm
as measured from their middle plane. Each chamber consists of 6 anode (wire) planes and 8 cathode planes confined between two cathode windows. The middle plane is used for mounting (on both sides of the plane) the 16-channel amplifier
discriminator cards required for sense wire readout. The middle plane also divides the chamber in half, each consisting of three wire planes and four cathode planes (see Fig. \ref{fig:fig3.25}).\\
\indent DC1 and DC2 are separated by an ``imaginary'' plane referred to as the focal plane, which is chosen such that the focal point of the
spectrometer optics coincides with the origin of the focal plane. This means that the particles transported to the hut are focused at the focal plane and those
with a momentum equal to the central spectrometer momentum are focused at the origin. This assumes that the spectrometer is also positioned at the central angles
corresponding to the central momentum, otherwise, the focal point will be shifted. \\
\indent The wire planes for each chamber were designed such that a 180$^{\circ}$ rotation of the unprimed wire planes about the $z$-axis produce the primed planes with
wires at the same orientation, but slightly shifted, which allows the resolution of the left/right ambiguities\footnote{\singlespacing The left/right ambiguities refers to our ignorance of whether a particle passed to the left or
  right side of the sense wire that detected it}. For each wire plane, the wire orientation is defined by a vector perpendicular to the wire where (U,U') and (V,V')
are oriented at $\pm60^{\circ}$ relative to the (X,X') wires as illustrated in Fig. \ref{fig:fig3.26}.
\indent A wire plane consists of alternating field and sense wires. In both the HMS and SHMS, the sense wires are made of 20 $\mu$m gold-plated tungsten and the field
wires are made of copper plated beryllium with thickness of 100 $\mu$m in HMS and 80 $\mu$m in SHMS. The numbering scheme of the sense wires is determined by the direction
of the perpendicular vector to the wire (see Fig. \ref{fig:fig3.26}) which points towards increasing wire numbers. For the HMS drift chambers, the (U, U', V, V') consist of
96 sense wires per plane and the (X, X') consist of 102 sense wires per plane. In contrast, the SHMS (U, U', V, V') consist of 107 sense wires per plane and the (X, X')
consist of 79 sense wires per plane. \\
\indent During operations, each chamber was filled with a carefully chosen gas mixture (50:50 argon/ethane) such that the charged particles that pass through the chamber ionize the
surrounding argon gas atoms producing an avalanche of electrons where the ethane served as the quenching element. In addition, the cathode planes and field wires were kept at a negative potential ($\sim$ -1940 V) relative to the sense wires, which were kept grounded at zero potential.
The potential difference between the field wires and cathode planes relative to each sense wire established an electric field with field lines pointing away from the sense wires and towards the adjacent field wires and cathode planes.
The calibration procedures will be discussed in detail in Section \ref{sec:DC_Calibration}.
\subsubsection{Hodoscopes}
\begin{figure}[!h]
\centering
\includegraphics[scale=0.36]{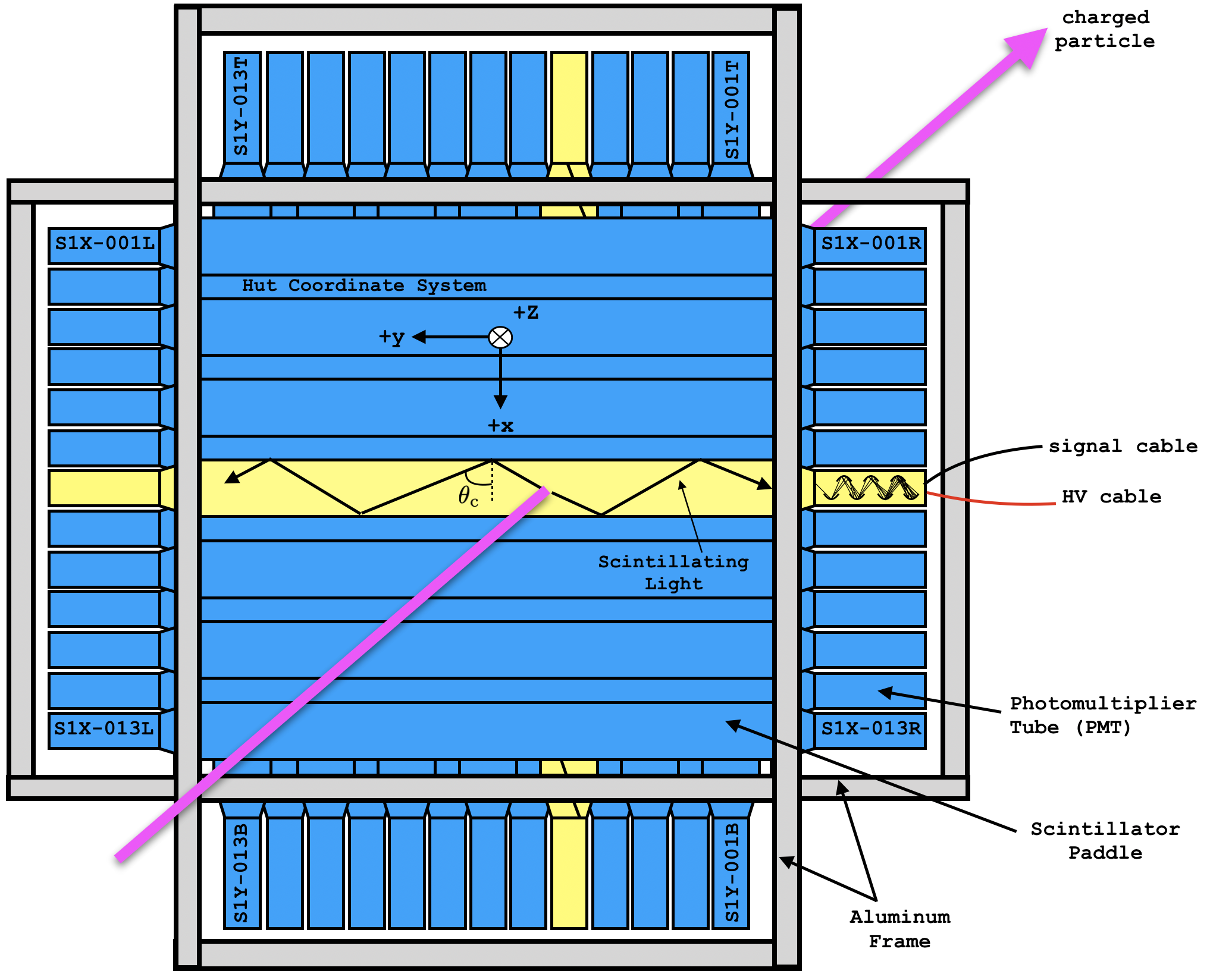}
\caption{Front view of the SHMS S1X (front) and S1Y (back) hodoscope planes.}
\label{fig:fig3.27}
\end{figure}

Each spectrometer is equipped with a series of four scintillator arrays (hodoscope planes) grouped into two pairs separated by a distance of about 2.2 m. Each pair is segmented along the
dispersive ($x$-axis) and non-dispersive ($y$-axis) direction by an array of long rectangular elements that can be either a plastic \textit{scintillator paddle} or \textit{quartz bar} with
a photomultiplier tube (PMT) coupled at each end. The plastic scintillating materials used in the HMS and the first three planes of the SHMS
are the BC-404\cite{hms_hodo_material} from Saint-Gobain Crystals and RP-408\cite{shms_hodo_material} from Rexon Corporarion, respectively. The last plane in the SHMS, known as the \textit{quartz plane}, is
composed of Corning HPFS 7980 Fused Silica (or quartz)\cite{shms_quartz_material} bars. To eliminate the possible gaps and avoid dead spots between adjacent elements where a particle
could pass undetected, the paddles/bars are slightly overlapped by a few millimeters in every plane. Table \ref{tab:tab3.4} summarizes the dimensions of each paddle for every hodoscope plane
in both spectrometers.
\begin{table}[H]
  \centering
  \scalebox{0.85}{
  \begin{tabular}[t]{lllll}
    \hline
    \textbf{Plane} & \textbf{Thickness} (mm) & \textbf{Width} (cm) & \textbf{Length} (cm) & \textbf{$\#$ of Elements}\\
    \hline
    \hline
    HMS  &  &  & \\
    1X & 2.12  & 8.0  & 75.5  & 16 \\
    1Y & 2.12  & 8.0  & 75.5  & 10 \\
    2X & 2.12  & 8.0  & 75.5  & 16 \\
    2Y & 2.12  & 8.0  & 75.5  & 10 \\
    \\
    SHMS    &   &  & \\
    1X & 5  & 8.0  & 100  & 13\\
    1Y & 5  & 8.0  & 100  & 13\\
    2X & 5  & 10.0 & 110  & 14\\
    2Y & 25 & 5.5  & 125  & 21\\
    \hline
  \end{tabular}
  }
  \caption{Summary of hodoscopes paddle dimensions for each plane.} 
  \label{tab:tab3.5}
\end{table}
\indent The fast timing properties of the plastic scintillators makes the hodoscope detector ideal for particle triggering specially after the 12 GeV energy upgrade where particle rates become significantly
higher. The addition of the quartz plane in the SHMS detector package provides a clean detection of charged particles while maintaining a high level of background rejection that optimizes the hodoscope
tracking efficiency at higher rates where the background is also expected to be larger.\\
\indent Even though the plastic scintillators and the quartz both emit light due to charged particle interactions, the process by which the light is produced is different.
In a scintillator, as a charged particle traverses the medium, it excites the molecules in the scintillator material that decay back into the ground state via the emission of
scintillation photons in (or near) the visible light range, a process known as fluorescence. The photons propagate towards the end of the scintillator paddles where they
are detected by the PMT. As the photons interact with the PMT photocathode, a certain number of photoelectrons will be produced via the photoelectric effect. These electrons are accelerated towards a series
of dynodes creating an electron avalanche towards the end of the PMT at the anode creating a measurable analog signal that is sent via a signal cable to the Counting Room for further signal processing
(see Fig. \ref{fig:fig3.27}).\\
\indent As the light propagates through the scintillator material, when it reaches the boundaries, the light may not be completely reflected and can be lost due to refraction. If the light ray incident on the
boundary exceeds the critical angle\footnote{\singlespacing The critical angle required for total internal reflection depends on the index of refraction between the two media at the boundary.}, it is reflected via total
internal reflection where no losses occur at the boundary. To maximize the scintillator light output in case of partial refraction, the scintillators are wrapped in a layer of a highly reflective
material (aluminum foil) and multiple layers of Tedlar (HMS) or electrical tape (SHMS) to ensure light tightness.\\
\indent In contrast to the plastic scintillators, the radiation produced in the quartz plane is based on the \v{C}erenkov effect, which will be discussed in more detail in the next detector section.
Additional technical information on the hodoscopes design and construction can be found in Refs.\cite{shms_hodo_techreport, shms_hodo_slides_2017, Quartz_mthesis2014}. 
\subsubsection{Threshold \v{C}erenkovs}
\begin{figure}[H]
\centering
\includegraphics[scale=0.3]{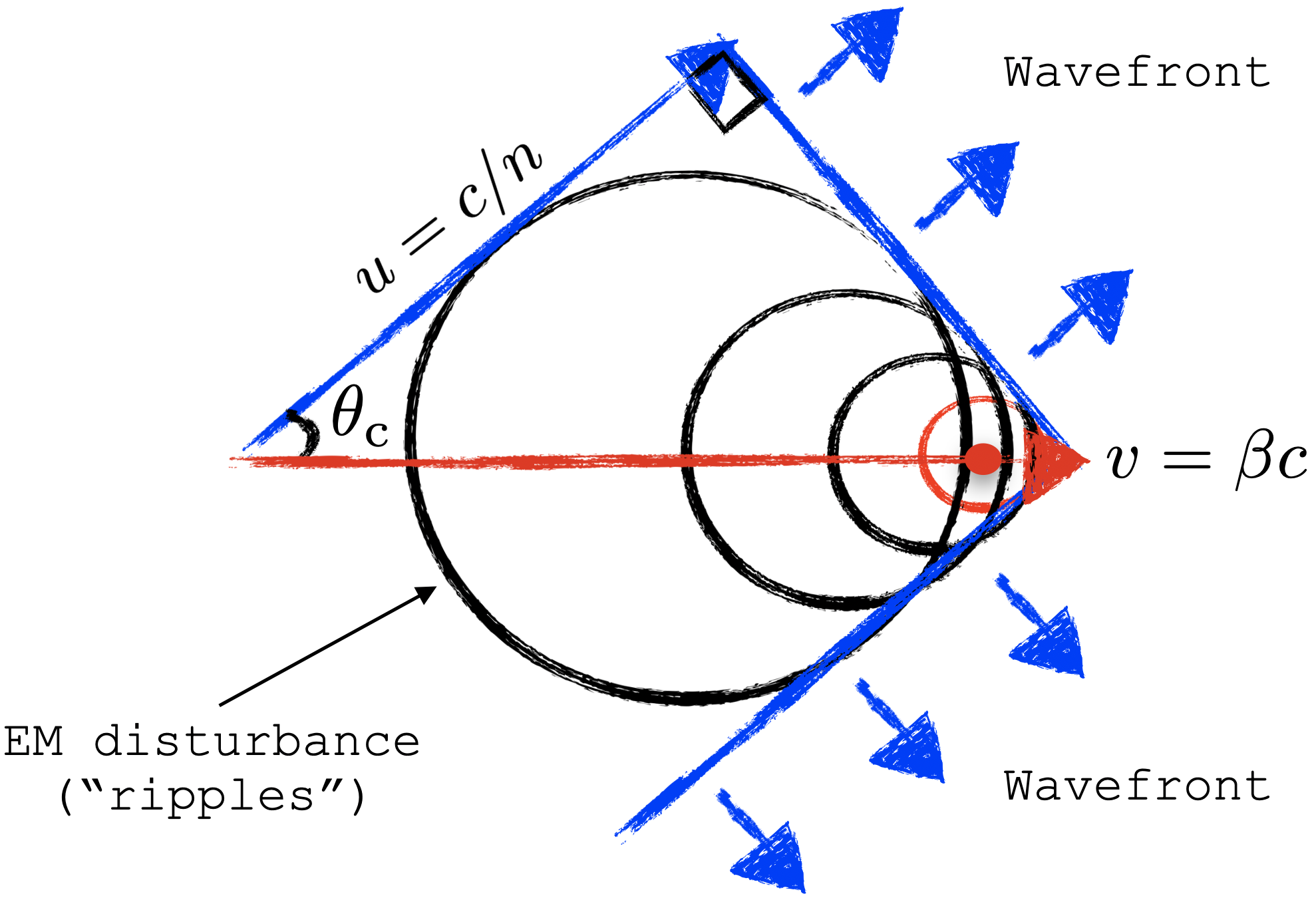}
\caption{Cartoon of the \v{C}erenkov effect. The charged particle (red) traverses a medium faster than the speed of light (blue) in that medium, producing a conical light wavefront.}
\label{fig:fig3.28}
\end{figure}

At high particle momenta ($\geq3$ GeV/c), the use of the hodoscope Time-of-Flight (TOF) method to identify different particles becomes practically useless, which can be understood from the relation,
$\Delta t \sim 1/p^{2}$, where $\Delta t$ is the hodoscope time difference between the first and second pair of planes and $p$ is the particle momentum. This means that at higher particle momenta, it becomes very difficult
to identify each particle due to the small and indistinguishable time difference between different particle masses. Therefore, the use of additional particle identification detectors becomes a necessity.\\
\indent The Hall C spectrometers are equipped with various threshold \v{C}erenkov particle detectors. These detectors
depend on \v{C}erenkov effect, which occurs when a charged particle traverses a transparent medium faster than the speed of light in the medium. As a result, the charged particle creates an electromagnetic disturbance
in the medium that causes \v{C}erenkov radiation to be emitted and distributed in a conical shape about the tracjectory of the particle (see Fig. \ref{fig:fig3.28}).\\
\indent From Fig. \ref{fig:fig3.28}, the electromagnetic disturbance (light) created by the passage of the charge particle is analogous to the sound waves
emitted by a supersonic jet. Since the charged particle moves faster than the spherical waves it emits, a conical shape is formed given by the relation
\begin{equation}
  \cos(\theta_{\mathrm{c}}) = \frac{1}{n\beta},
  \label{eq:3.22}
\end{equation}
where $\beta=v/c$ is the ratio of the velocity ($v$) of the charged particle to the speed of light
in vacuum ($c$) and $n=c/u$ is the index of refraction of the medium, where $u$ is the speed of light in the medium.
Alternatively,  $\beta=p/\sqrt{m^{2}+p^{2}}$ where $m$ is the particle's mass. Since the charged particle must travel faster that light in the medium for \v{C}erenkov light to be emitted,
one requires $v > c/n \implies n > 1/\beta$ which can be expressed in terms of momentum as
\begin{equation}
  n > \frac{\sqrt{m^{2} + p^{2}}}{p}.
  \label{eq:3.23}
\end{equation}
From this inequality, the index of refraction of the medium can be adjusted accordingly such that at a fixed momentum, the mass of the particle will determine whether or not \v{C}erenkov
radiation will be produced. \\
\indent Both the gas and aerogel threshold \v{C}erenkov detectors in Hall C utilize this basic inequality for particle identification. In addition, for the gas \v{C}erenkovs, the quantity $(n-1)$ is proportional
to the gas pressure, which can be adjusted to change the index of refraction and hence select the desired particle mass that will trigger the \v{C}erenkov effect.
Below is a brief description of each \v{C}erenkov detector.\\
\\
\textit{\textbf{Heavy Gas \v{C}erenkov (HGC)}} Each spectrometer is equipped with an HGC detector located between the front and rear hodoscope planes. The
detector is filled with a gas that is kept at a specific pressure depending on the experimental requirements. The detector is also equipped with several mirrors mounted and oriented so as to reflect and focus the
\v{C}erenkov light towards the PMTs.\\
\begin{figure}[H]
\centering
\includegraphics[scale=0.4]{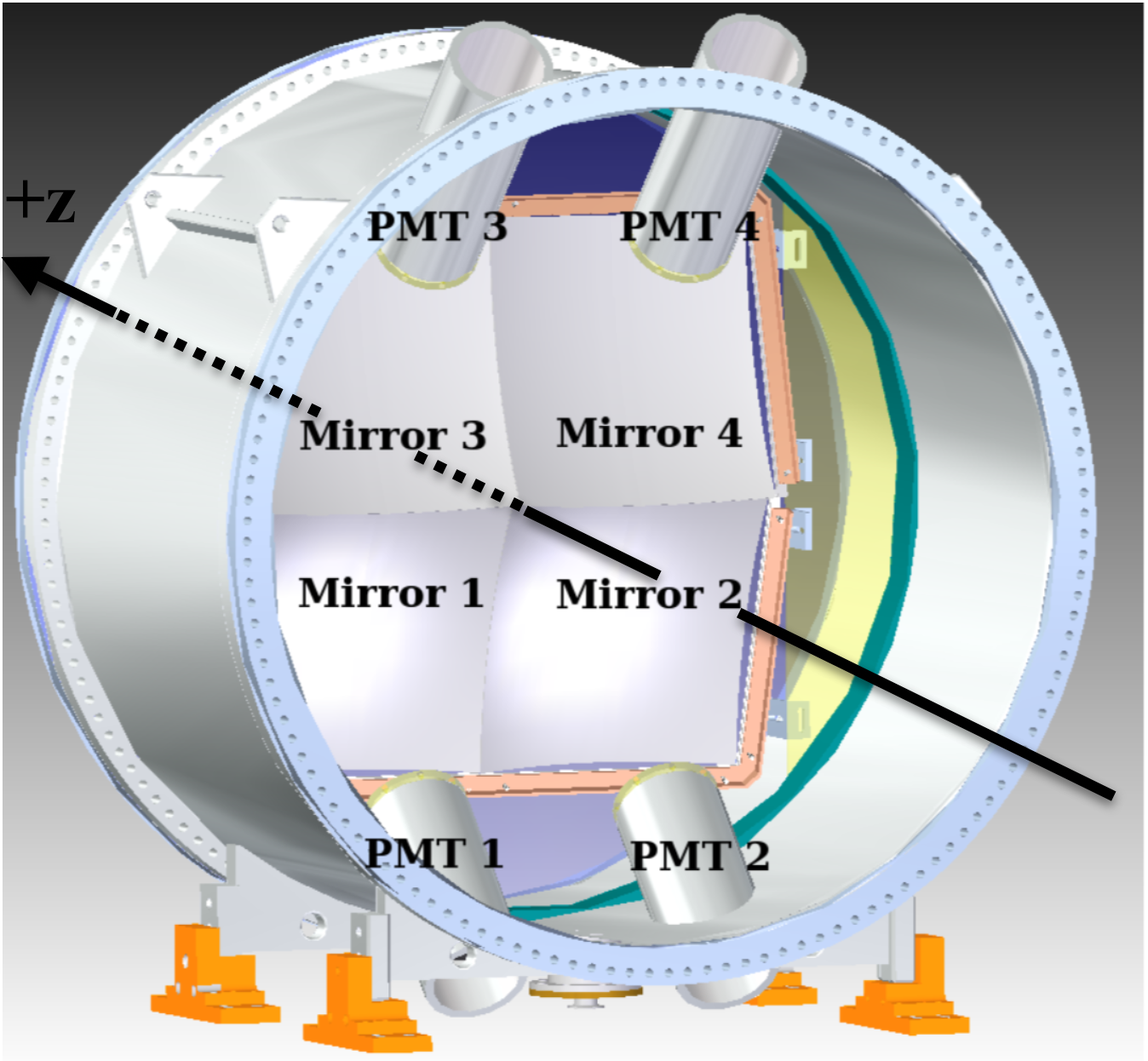}
\caption{CAD rendering of the SHMS Heavy Gas \v{C}erenkov detector. Note: Reprinted from Ref.\cite{shms_hgc}.}
\label{fig:fig3.29}
\end{figure}
The HMS \v{C}erenkov\cite{hms_cer,HallC_SEM_saw2019} consists of a 1.5 meter-long cylindrical tank with 2 spherical mirrors installed that focus the \v{C}erenkov light onto 2 PMTs.
The detector can operate as an $e/\pi$ or $\pi/p$ discriminator, depending on whether the tank is filled with C$_{4}$F$_{10}$ or N$_{2}$ gas for the former, or Freon-12 gas for the latter, at a range of operating
pressures depending on the gas used. In contrast, the SHMS \v{C}erenkov\cite{shms_hgc,HallC_SEM_saw2019} (see Fig. \ref{fig:fig3.29}) consists of a 1.3 meter-long cylindrical tank (1.88 m in diameter) and has 4 mirrors installed that focus the light
onto 4 PMTs. The detector is filled with C$_{4}$F$_{10}$ or C$_{4}$F$_{8}$O gas, which are functionally equivalent, and is able to operate as an $e/\pi$ or $\pi/K$ discriminator depending on the gas pressure used.\\
\\
\textit{\textbf{Noble Gas \v{C}erenkov (NGC)}} The SHMS is equipped with an additional
gas \v{C}erenkov detector\cite{shms_ngc_techreport,HallC_SEM_saw2019} (see Fig. \ref{fig:fig3.30}) located in
front of the drift chambers due to space constraints in the hut. The detector consists of a rectangular tank 2.5
\begin{figure}[H]
\centering
\includegraphics[scale=0.35]{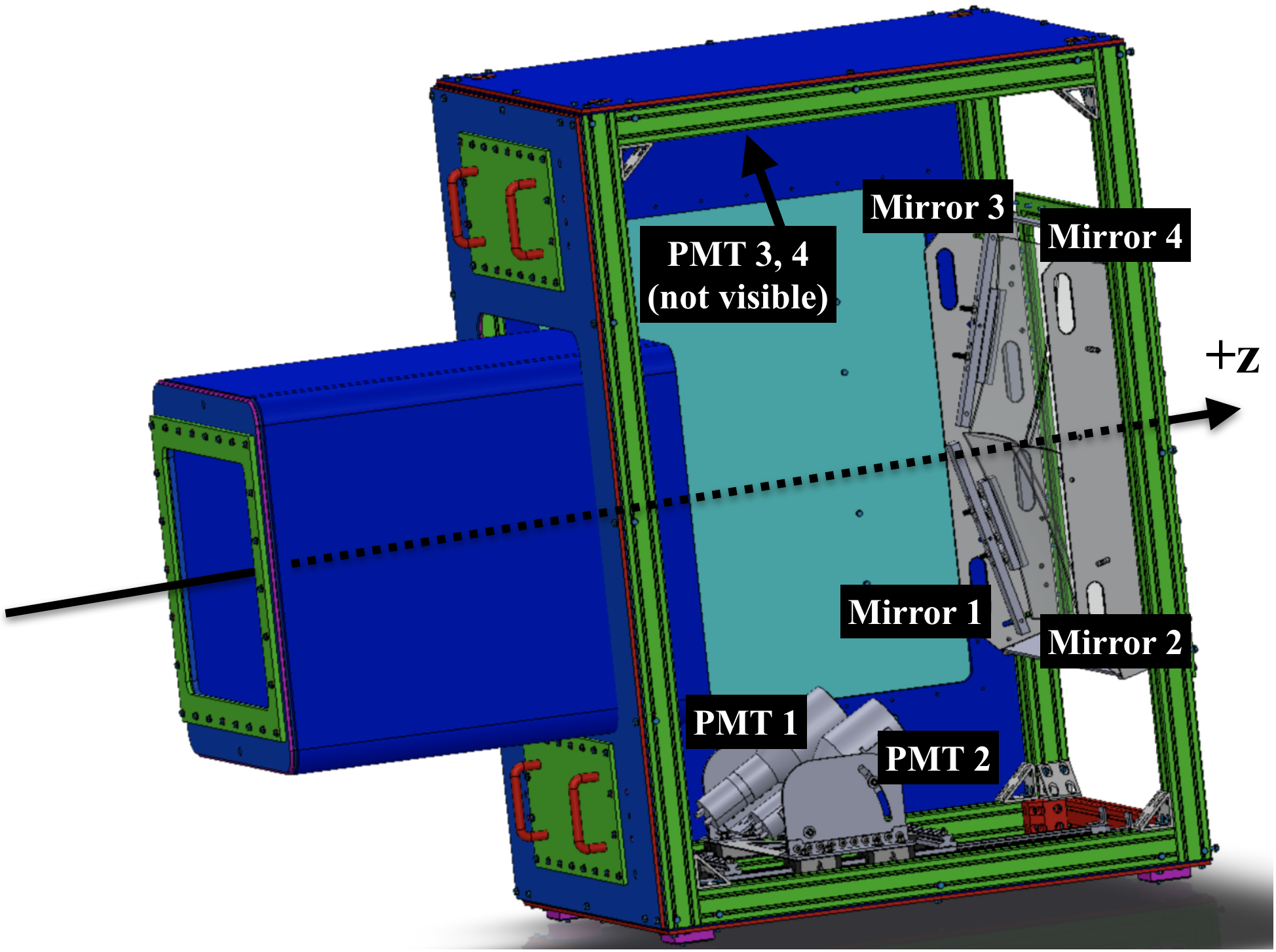}
\caption{CAD rendering of the SHMS Noble Gas \v{C}erenkov detector.}
\label{fig:fig3.30}
\end{figure}
\noindent meters in length (along $z$-axis) and 0.8 meters wide with 4 spherical mirrors used to reflect and focus the \v{C}erenkov light onto 4 PMTs. The tank is designed to operate at a gas pressure of 1 atm of either argon,
neon or a mixture of the two gases, which provides $e/\pi$ discrimination at momenta above 6 GeV/c. At lower central momenta, the NGC is replaced with an extended vacuum pipe of the same length to minimize the effects
of multiple scattering on the particle trajectory.\\
\\
\textit{\textbf{Aerogel \v{C}erenkov}} The strange physics part of the program at Hall C requires the capability to carry out a clean $p/\pi/K$ separation.
To achieve a clean particle identification, both spectrometers are equipped with an aerogel \v{C}erenkov detector.
\begin{figure}[H]
\centering
\includegraphics[scale=0.5]{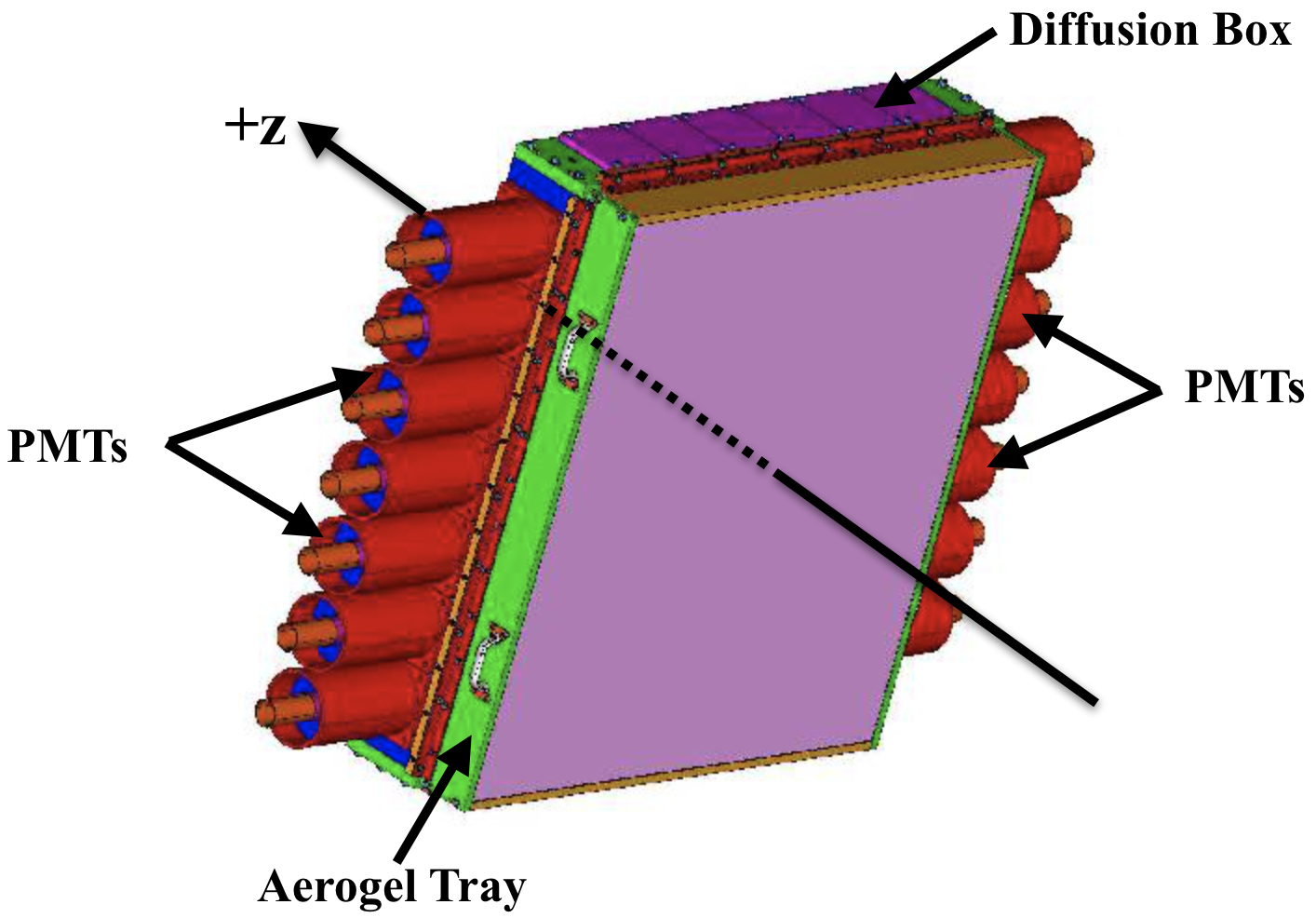}
\caption{CAD rendering of the SHMS aerogel \v{C}erenkov detector. The HMS design is very similar with slightly different dimensions and an additional PMT at both ends.}
\label{fig:fig3.31}
\end{figure}
In contrast to the gas \v{C}erenkovs, this detector
uses an aerogel, which is a transparent, highly porous material with a refractive index typically between those of gases and liquids. The aerogel detector in each spectrometer
are of similar design consisting of an aerogel tray followed by a light diffusion box with a highly reflective material at the inner boundaries. As the charged particle passes
through the aerogel material, depending on the refractive index, specific particles will emit \v{C}erenkov light that travels into the diffusion box where it is reflected at
the boundaries of the box and into an array of PMTs mounted on each side. The analog signal from the light collected by the PMTs is sent to the Counting Room for further signal
processing. A detailed description of the aerogel detector including the dimensions and material specifications can be found in Refs.\cite{hms_aero,shms_aero}.
\subsubsection{Calorimeters}
\begin{figure}[h]
\centering
\includegraphics[scale=0.5]{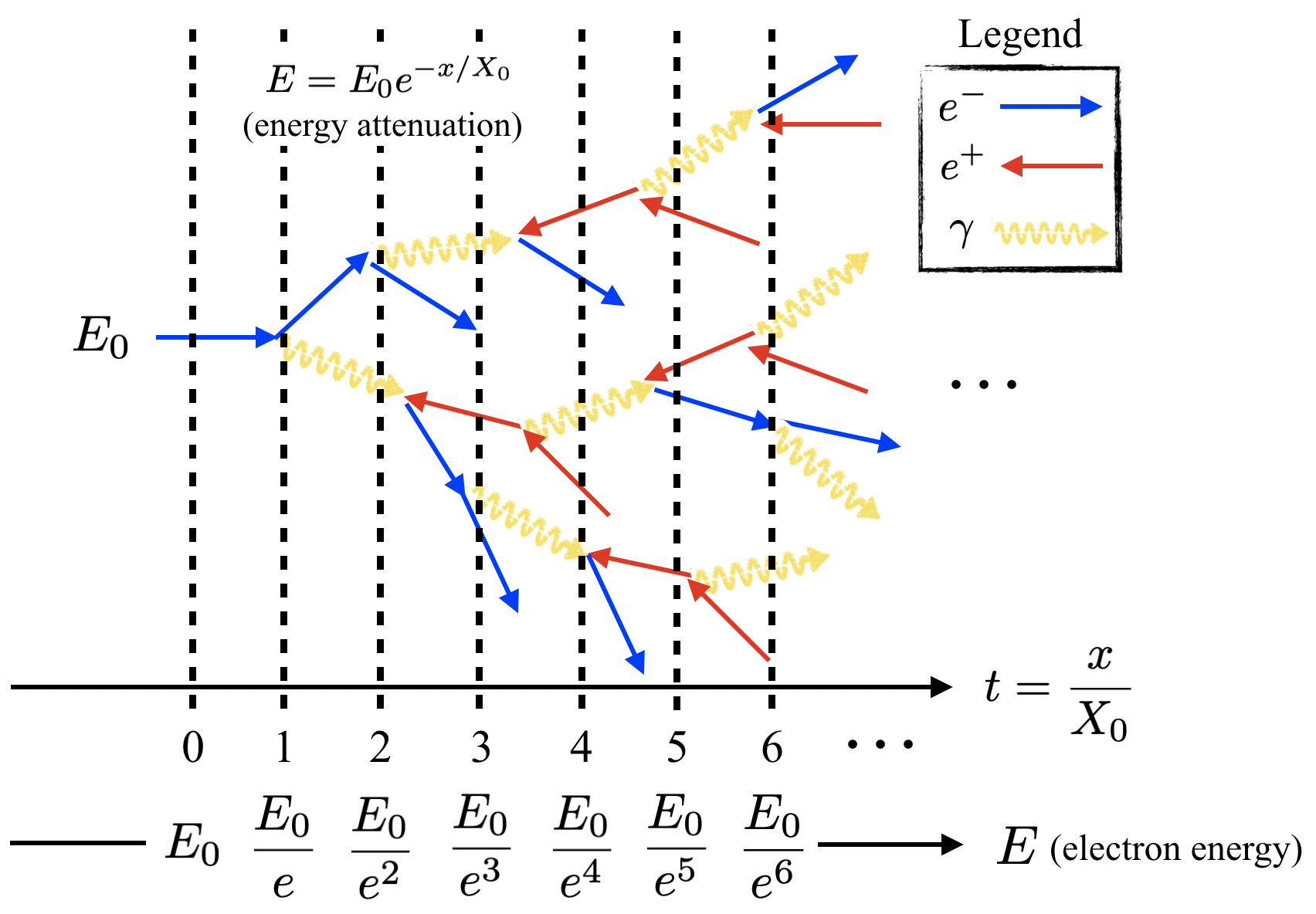}
\caption{Typical electromagnetic shower cascade in a calorimeter.}
\label{fig:fig3.32}
\end{figure}

\begin{figure}[!h]
\centering
\includegraphics[scale=0.38]{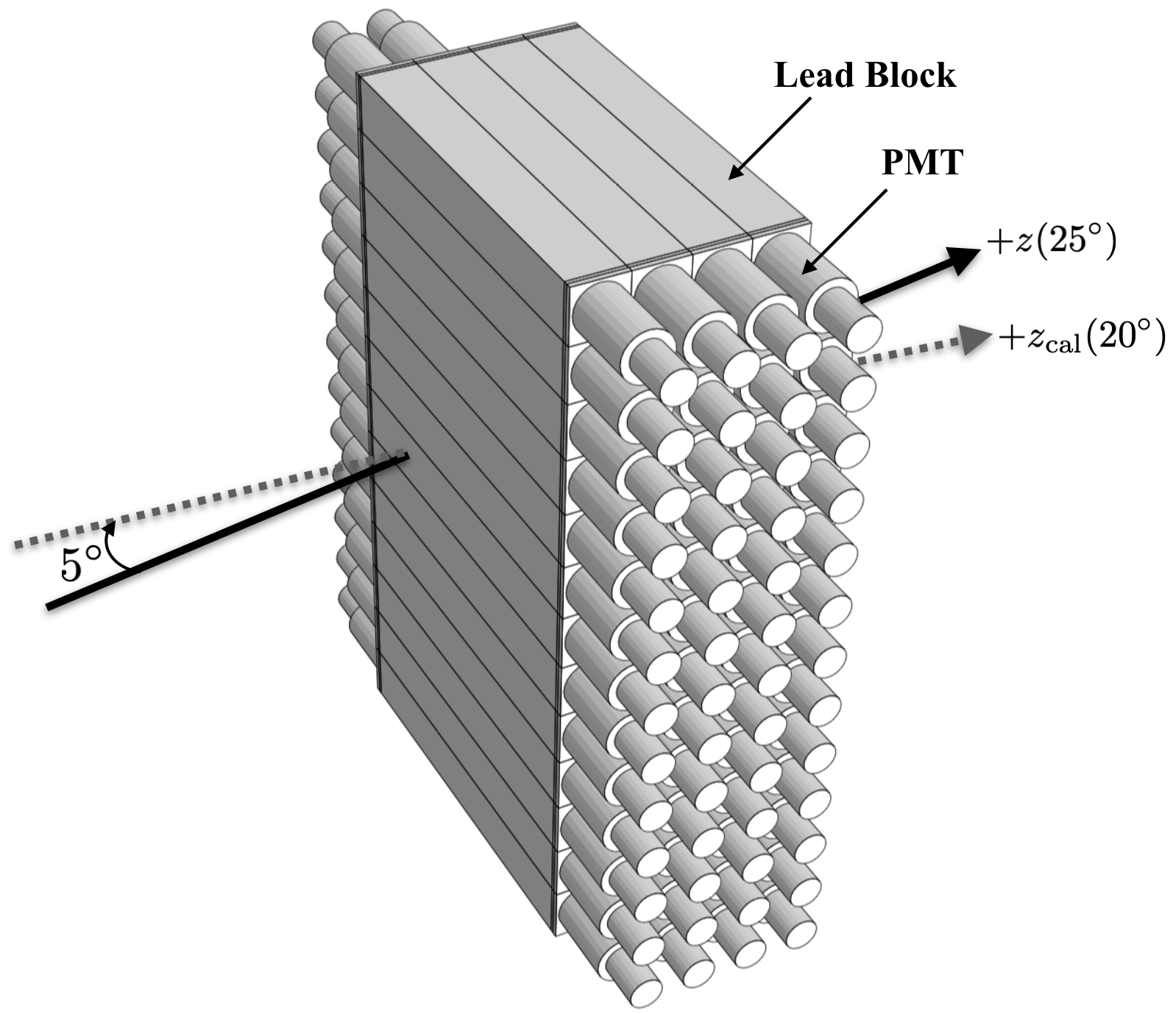}
\caption{HMS electromagnetic calorimeter. The entire detector is tilted vertically by 5$^{\circ}$ lower relative to the central ray of the spectrometer hut.}
\label{fig:fig3.33}
\end{figure}

\begin{figure}[!h]
\centering
\includegraphics[scale=0.48]{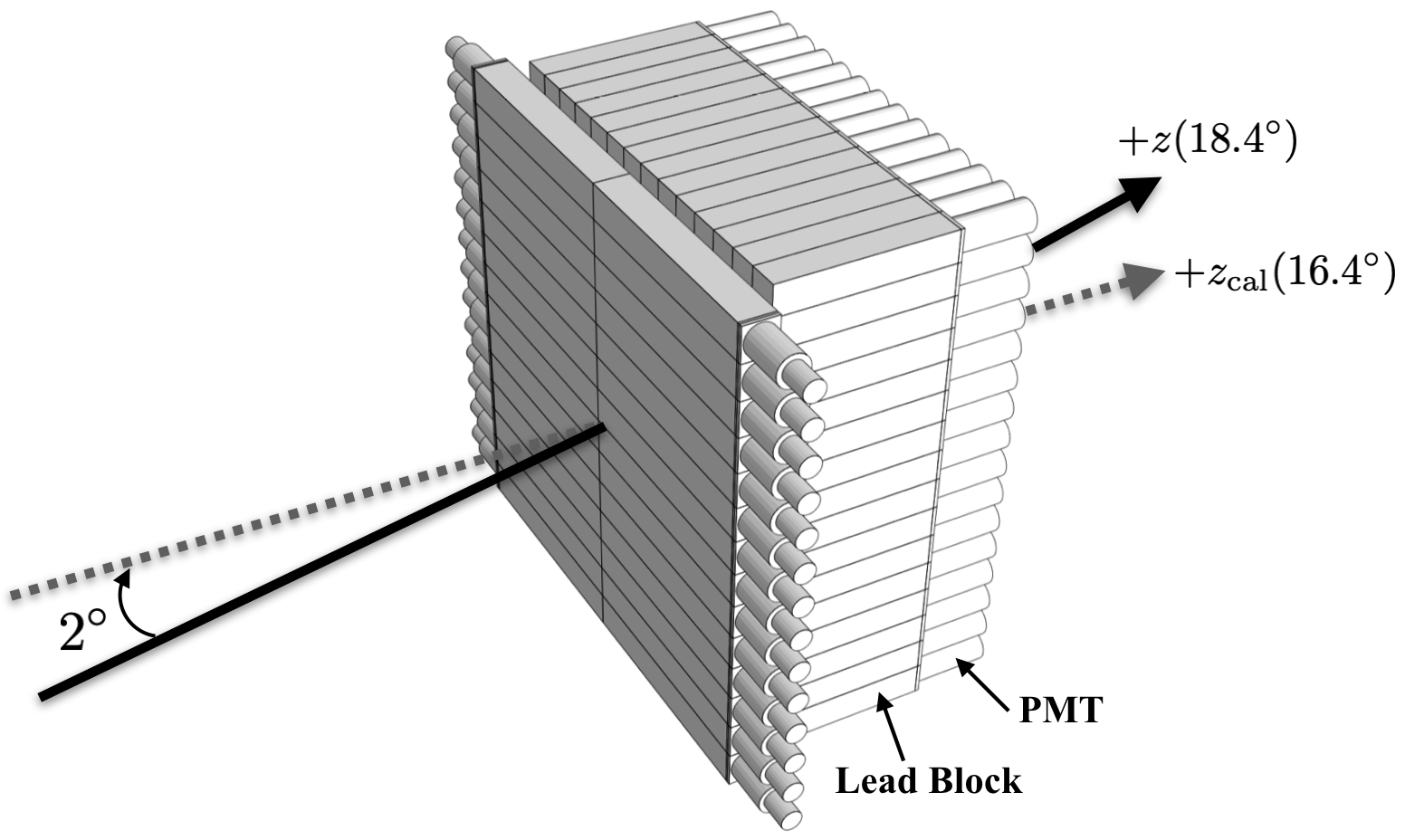}
\caption{SHMS electromagnetic calorimeter. The entire detector is tilted vertically by 2$^{\circ}$ lower relative to the central ray of the spectrometer hut.}
\label{fig:fig3.34}
\end{figure}

The electromagnetic (EM) calorimeter in each spectrometer is primarily used for $e/\pi$ discrimination and to complement the gas \v{C}erenkov detectors
for a more robust electron identification and pion suppression. The calorimeters provide a destructive measurement of the projectile particle energy
and are therefore located at the end of the detector stack. The projectile energy is measured by the detection of \v{C}erenkov radiation primarily from EM showers
produced mainly via bremsstrahlung and pair production processes. As the electrons traverse the calorimeter, they are slowed down (decelerated) by
the calorimeter radiator and emit bremsstrahlung photons that decay to $e^{+}e^{-}$ pairs via pair production. These pairs further radiate bremsstrahlung
photons triggering an EM shower cascade reaction until most or all the initial electron energy has been deposited in the calorimeter.\\
\indent From Fig. \ref{fig:fig3.32}, a single electron with initial energy, $E_{0}$, enters the calorimeter radiator material and produces
a particle cascade in a chain reaction that becomes less energetic as it traverses the radiator. The horizontal axis (along particle trajectory) is defined as a multiple of one
radiation length ($X_{0}$ [g/cm$^{2}$]), which is defined as the mean distance over which a high-energy electron loses all but $1/e$ of its initial energy ($E_{0}$)
due to bremsstrahlung radiation\cite{PDG_2018}. The radiation length can also be expressed in [cm] by dividing $X_{0}$ by the density of the material in [g/cm$^{3}$].\\
\indent To ensure that all (or most) of the incident projectile energy is deposited in the radiator material, the Hall C calorimeters are made of several stacked layers
of thick lead glass blocks (or modules) that are tilted a few degrees lower relative to the spectrometer central ray. The original HMS calorimeter\cite{Mkrtchyan_2013} (see Fig. \ref{fig:fig3.33})
was commissioned in 1994 as one of the first detectors to operate in Hall C and remains in the stack since no significant deterioration in performance has been observed. The detector
consists of 52 TF-1 lead glass modules (refractive index 1.65, density 3.86 g/cm$^{3}$) stacked in four layers of 13 blocks/layer with dimensions of $10 \times 10 \times 70$ cm$^{3}$ per block.
A single layer measures 3.65 radiation lengths along the particle trajectory ($+z$) for a total of $\sim$14.6 radiation lengths for the entire calorimeter, sufficient to
absorb most of the electron projectile energy. The total energy deposited is read out by PMTs coupled at both ends of the first two layers, and PMTs coupled at one end on the last
two layers of the calorimeter.\\
\indent The \textit{new} SHMS calorimeter\cite{Mkrtchyan_2013} (see Fig. \ref{fig:fig3.34}) consists of TF-1 and F-101 type lead glass modules (refractive index 1.65, density 3.86 g/cm$^{3}$) assembled separately
into a preshower (TF-1) and shower (F-101) counter. The preshower blocks used in the SHMS were removed from the decommissioned SOS calorimeter and consist of 28 modules stacked
in two adjacent columns. Each module has dimensions of $10 \times 10 \times 70$ cm$^{3}$ and is coupled to a PMT at one end. In contrast to the preshower, the shower counter blocks were obtained
from the decommissioned HERMES calorimeter detector and consist of 224 modules of dimensions $8.9 \times 8.9 \times 50$ cm$^{3}$ per module with a coupled PMT towards the long end of the block.
The modules were stacked in a \textit{fly's eye} configuration behind the preshower plane. This configuration, which is $\sim18$ radiation lengths deep, guarantees that the EM showers from
the highest energetic projectiles ($\sim10$ GeV) will be mostly absorbed. The preshower counter, in contrast, is only $3.6$ radiation lengths thick and is specifically positioned in front
of the shower to improve the particle identification capabilities by detecting the early onset of EM showers. A detailed report of the HMS/SHMS calorimeter detector performance and design
specifications can be found in Ref.\cite{Mkrtchyan_2013}.
\section{The Hall C Electronics Trigger Setup}\label{section3.7}
The majority of the detector electronics in the HMS/SHMS detector huts are read out by \textit{Read-Out Controllers} (ROCs) crates located in the Counting Room except for the HMS/SHMS drift
chambers and the SHMS shower counter signals, which are read in their respective ROCs in the detector huts. In the HMS/SHMS huts, the drift chamber
signals are transmitted by 20-25 foot-long ribbon cables that are read out in the hut electronics rack (see Figs. \ref{fig:fig3.35} and \ref{fig:fig3.38}).
On the SHMS side, the shower counter consists of 224 signal cables read directly in the hut electronics rack.
\subsection{HMS Detector Hut Electronics}
\begin{figure}[h!]
  \centering
  \includegraphics[scale=0.55]{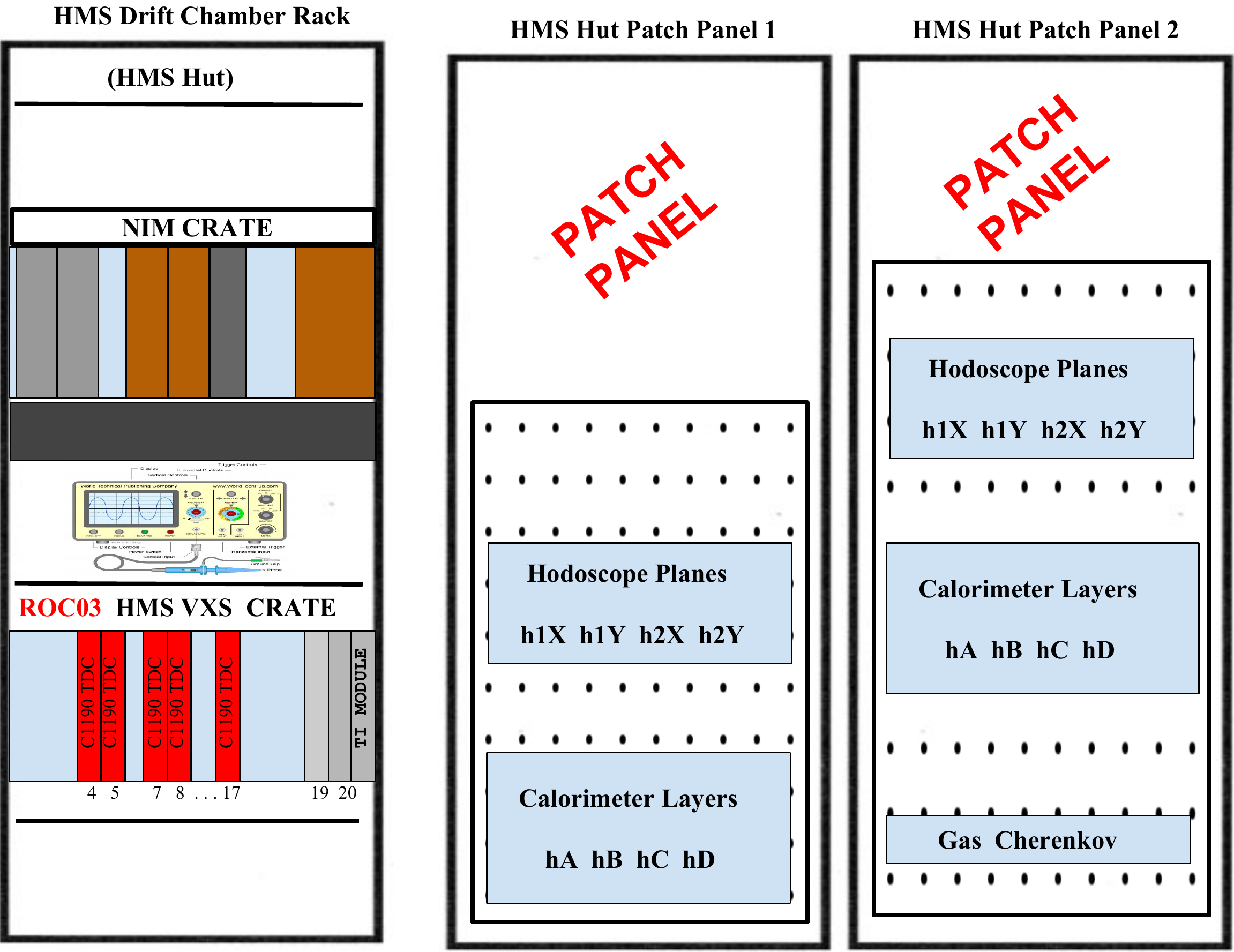}
  \caption{HMS detector hut electronics rack (left) and patch panels (right).}
  \label{fig:fig3.35}
\end{figure}

\begin{figure}[h!]
  \centering
  \includegraphics[scale=0.36]{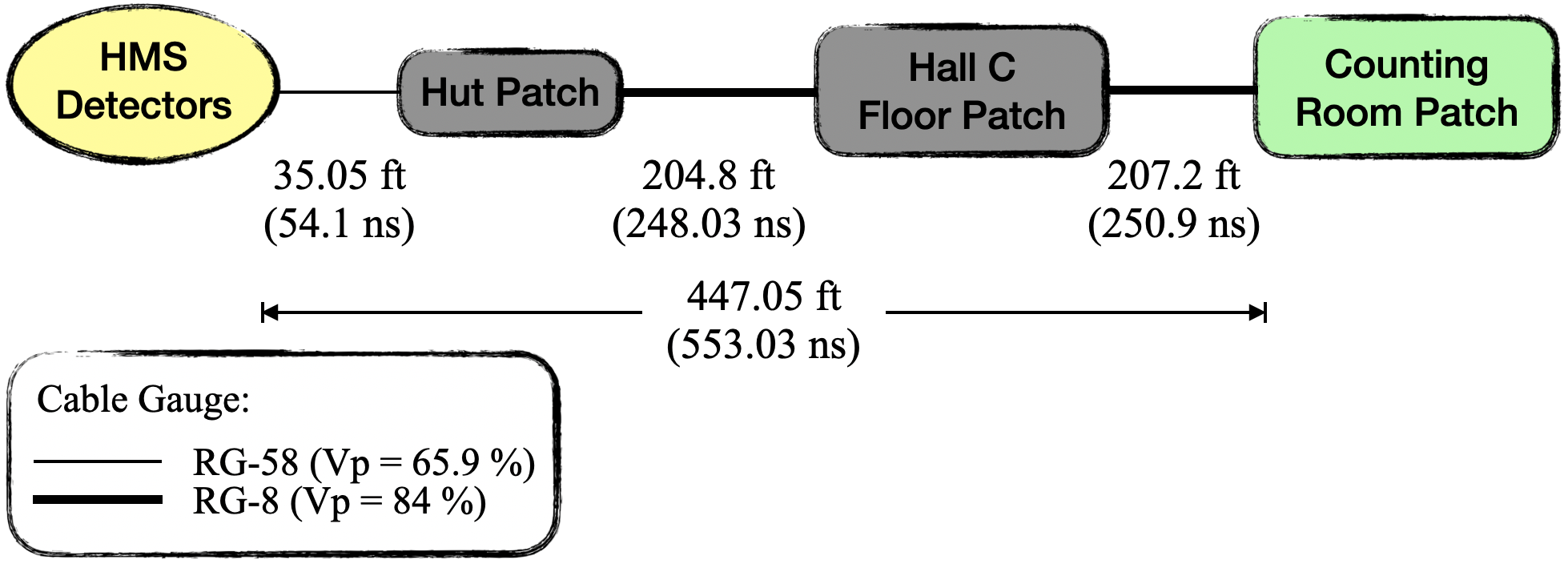}
  \caption{HMS patch diagram from detectors to Counting Room.}
  \label{fig:fig3.36}
\end{figure}

The HMS drift chambers are read out through a VXS Crate (\textbf{ROC 03}) in the detector hut electronics rack (see Fig. \ref{fig:fig3.35}). The signals are carried through 16-channel ribbon cables
fed into various CAEN1190 (C1190)\cite{C1190_TDC} TDC modules. The Trigger Interface (TI)\cite{Trigger_Interface} module at the front end of the crate distributes the readout
trigger throughout all modules in the crate and initiates data readout.\\
\indent The rest of the HMS detector signals (gas \v{C}erenkov, hodoscopes, calorimeter) are
sent to the Hall C floor patch panel via the hut patch, with the exception of the aerogel, which is sent directly from the detector to the floor patch.
All the signals are then sent to the Counting Room patch panel to be processed by the electronics (see Fig. \ref{fig:fig3.36}).
\subsection{SHMS Detector Hut Electronics}
\begin{figure}[H]
  \centering
  \includegraphics[scale=0.35]{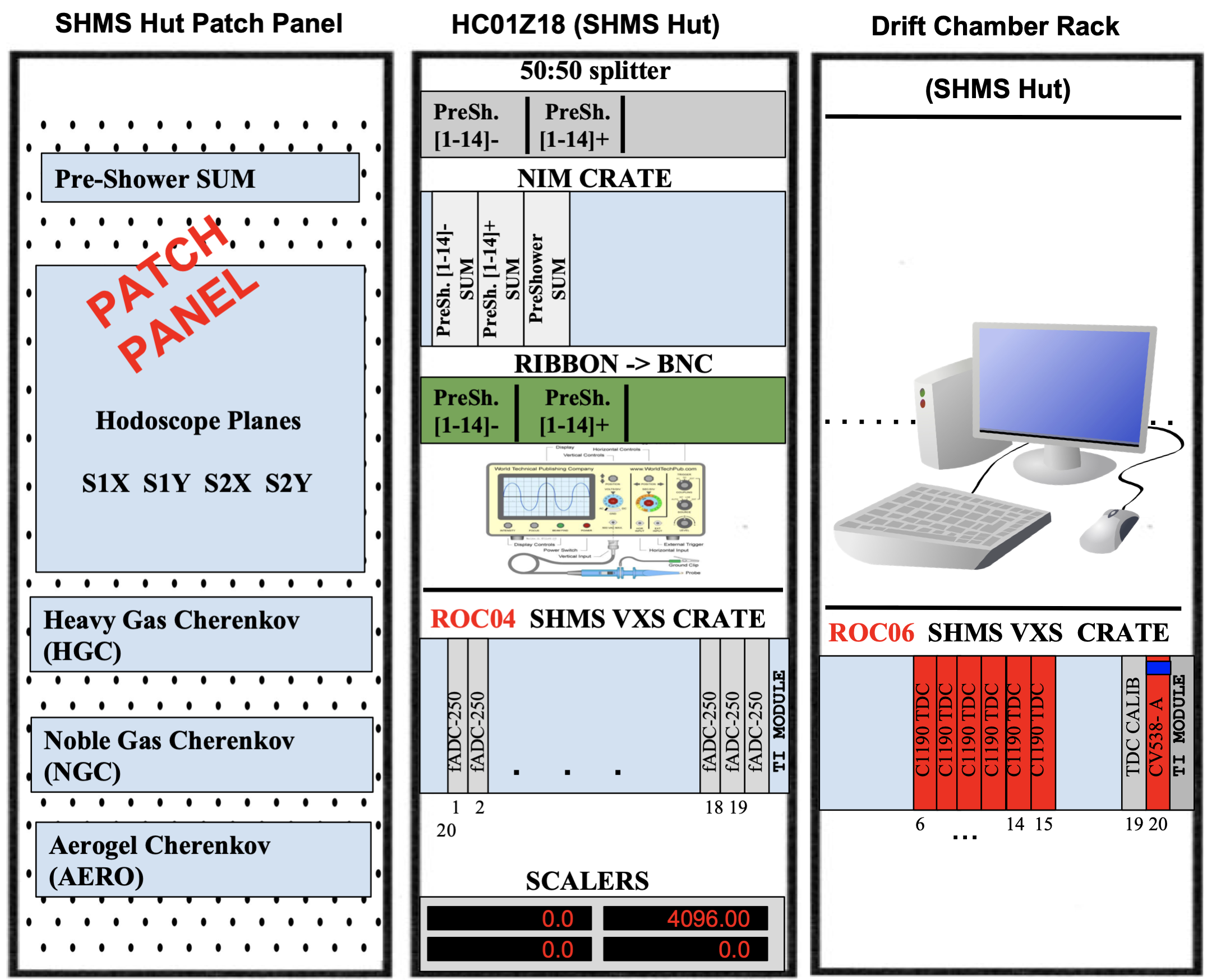}
  \caption{SHMS hut patch panel (left) and electronics racks (right).}
  \label{fig:fig3.37}
\end{figure}

\begin{figure}[H]
  \centering
  \includegraphics[scale=0.36]{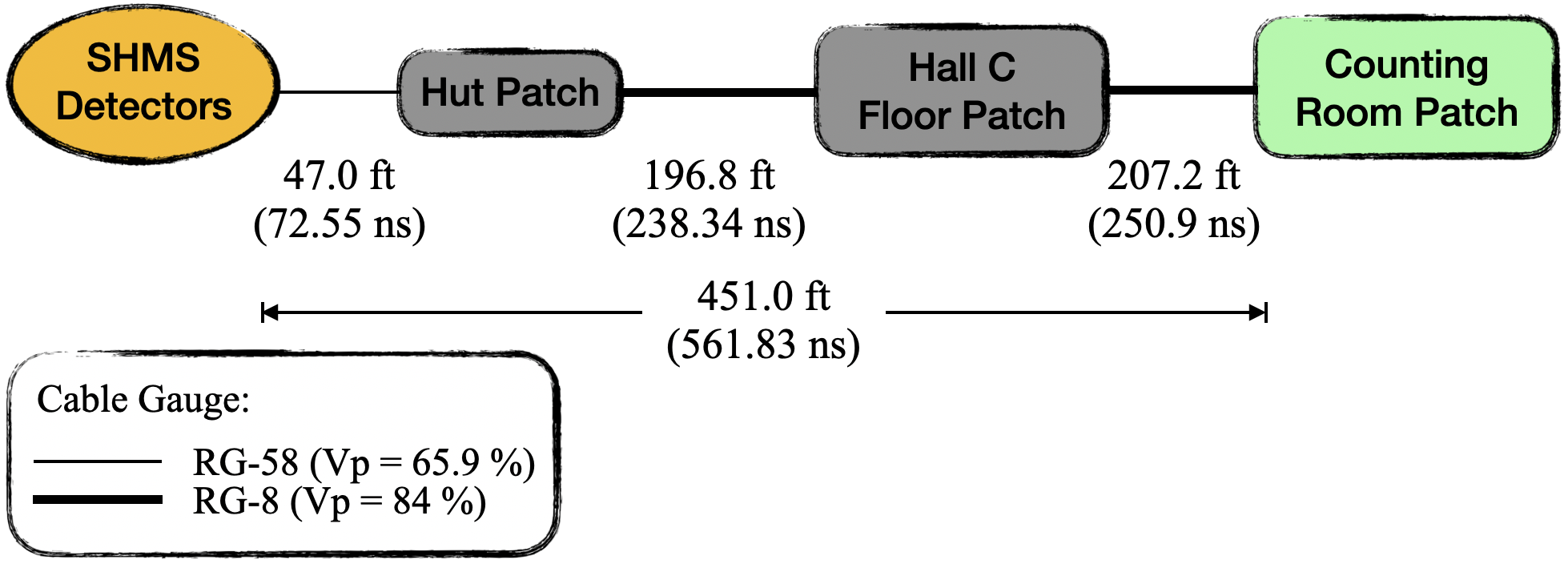}
  \caption{SHMS patch diagram from detectors to Counting Room.}
  \label{fig:fig3.38}
\end{figure}

Similar to the HMS drift chambers, the SHMS drift chambers are also read out by TDCs in a VXS Crate in the SHMS electronics hut (see Fig. \ref{fig:fig3.37}).
\indent The 224 shower counter signals are directly connected to 250 MHz flash ADCs\cite{F250_ADC}, hereafter referred to as fADCs, in a separate VXS Crate (\textbf{ROC 04}). The preshower signals (x14/side) pass
through a 50:50 splitter where a part of the signal is fed to an fADC and the other part is partially summed in the hut and sent via the hut patch panel to the
Counting Room patch. The rest of the SHMS detector signals (HGC/NGC, hodoscopes, aerogel) are sent to the Counting Room via the hut patch panel to be processed by the electronics (see Fig. \ref{fig:fig3.38}). \\
\subsection{Hall C Counting Room Electronics}
\begin{figure}[h!]
  \centering
  \includegraphics[scale=0.44]{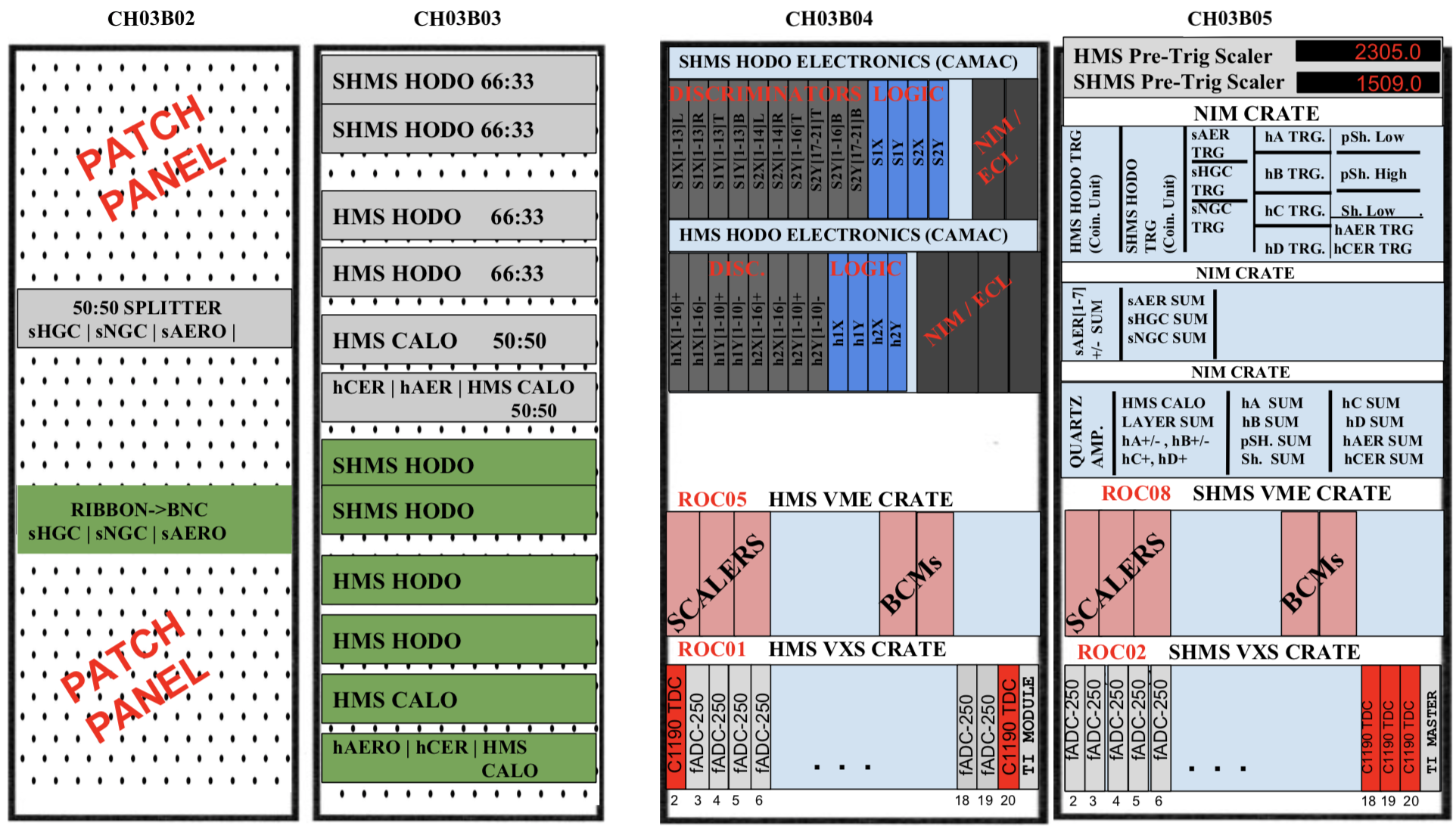}
  \caption{Counting Room patch panels (left 2 racks) and main electronic racks for HMS/SHMS detector signal processing (right 2 racks).}
  \label{fig:fig3.39}
\end{figure}

Once the detector signals arrive at the Counting Room patch (see Fig. \ref{fig:fig3.39}(left)), they are processed by
the NIM/CAMAC\footnote{\singlespacing NIM or \textit{Nuclear Instrumentation Modules} and CAMAC or
  \textit{Computer Automated Measurement and Control} define a set of standard modular-crate electronics
  commonly used in experimental nuclear/particle physics.}
electronics (see Fig. \ref{fig:fig3.39}(right)) to form the single-arm or coincidence triggers
for each spectrometer. The signals are also sent to fADCs/TDCs to determine energy and timing information for individual
detectors.
\subsection{HMS Trigger Setup}
\indent The XY scintillator arrays (hodoscope planes) form part of the standard HMS trigger configuration.  
Additional particle detectors may also be incorporated into the HMS trigger as required by different experiments. The gas \v{C}erenkov and calorimeter triggers are
used for $e/\pi$ discrimination, whereas the aerogel \v{C}erenkov trigger is used for $\pi/K/p$ discrimination. 
\subsubsection{Hodoscopes Pre-Trigger}
\begin{figure}[h!]
  \centering
  \includegraphics[scale=0.4]{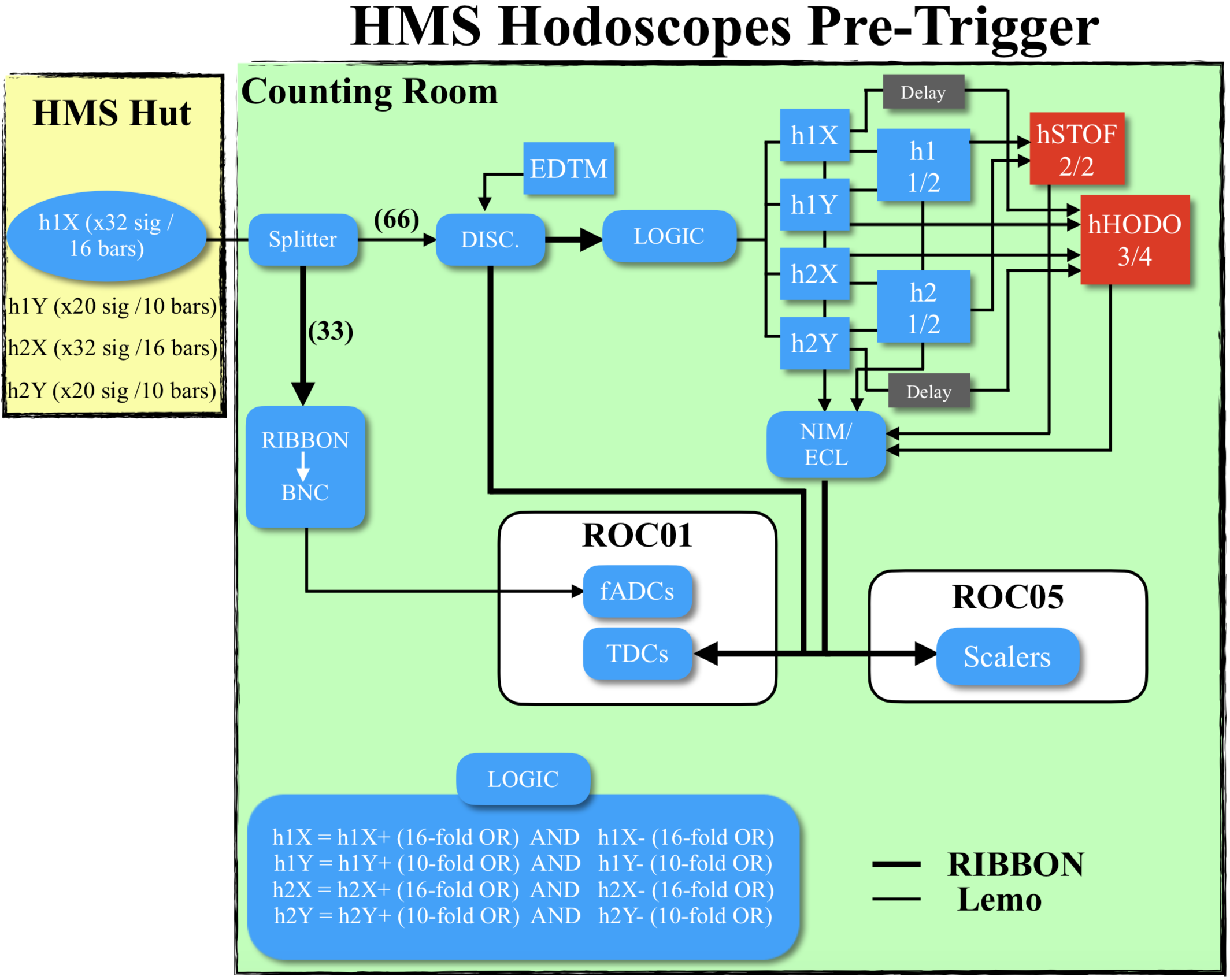}
  \caption{HMS hodoscopes electronics diagram.}
  \label{fig:hHODO_diagram}
\end{figure}
\noindent Each hodoscope plane consists of an array of scintillator paddles coupled to a PMT at each end (see Fig. \ref{fig:fig3.23}), so each paddle reads out two signals. In Fig. \ref{fig:hHODO_diagram},
for example, hodoscope plane h1X consists of 32 signals (16 paddles) read out in the Counting House (CH) patch. Each side of the plane (x16 signals/side) is fed into a 64-channel input passive splitter (16 Ch./set).
One-third of the signal amplitude is sent via a 16-channel ribbon cable to a 64-channel input Ribbon-to-BNC converter (16 Ch./set) fed into a 16-channel NIM input fADC.
The remaining two-thirds of the signal amplitude is sent to a 16-channel input CAMAC discriminator unit. The HMS discriminator thresholds and gate widths were set to \hhodthrs and \hhodgate, respectively,
but may be subject to change.\\
\indent The discriminated signals are sent via two ribbon cable outputs to C1190 TDCs and scalers (daisy-chained) and to a LeCroy 4564 CAMAC logic unit to form the plane pre-triggers.
The logic unit takes four sets of 16-channel input ribbon cables and forms a 16-fold OR for each set by default. Further boolean operations are done through the module backplane by connecting
a twisted pair cable to the pin corresponding to the desired boolean operation. For the HMS hodoscope plane pre-triggers, the boolean operations are as follows:
\begin{empheq}[box=\fbox]{align*}
& \text{\textbf{h1X} = h1X+ (16-fold OR) \text{AND} h1X- (16-fold OR)} \\ 
& \text{\textbf{h1Y} = h1Y+ (10-fold OR) \text{AND} h1Y- (10-fold OR)} \\
& \text{\textbf{h2X} = h2X+ (16-fold OR) \text{AND} h2X- (16-fold OR)} \\ 
& \text{\textbf{h2Y} = h2Y+ (10-fold OR) \text{AND} h2Y- (10-fold OR)} 
\end{empheq}
\indent Once a pre-trigger has been made for each plane, they are sent to a NIM/ECL converter (Level Translator - Phillips Scientific (or P/S) Model 7126) via twisted pair cables. The NIM output is then sent to individual
sets of a P/S Model 752 NIM logic unit to adjust the widths of each of the plane pre-triggers as necessary before making coincidence. An XY hodoscope plane
coincidence (\textbf{h1} = \textbf{h1X} OR \textbf{h1Y}, \textbf{h2} = \textbf{h2X} OR \textbf{h2Y}) is then made by feeding each hodoscope XY plane pair into a P/S Model 755 NIM logic unit\footnote{\singlespacing The output
  widths of the P/S Model 755 logic units were set to $\sim$ 50 ns for the HMS. See HCLOG entry \url{https://logbooks.jlab.org/entry/3501357}.}. A copy of each of
the four individual plane pre-triggers is also sent to another set of P/S Model 755 to make a 3/4 or 4/4 plane coincidence (via a front-panel knob), which defines the standard hodoscope pre-trigger (\textbf{hHODO 3/4}).
An additional pre-trigger (\textbf{hSTOF} = \textbf{h1} \text{AND} \textbf{h2}) is formed by requiring the coincidence between any two of the front (\textbf{h1}) and back (\textbf{h2}) scintillator plane pairs to measure the time-of-flight (TOF)
between any of the two front and back planes. A copy of all the pre-triggers discussed above are sent to TDCs and scalers via a NIM/ECL converter for timing and counting information (see Fig. \ref{fig:hHODO_diagram}). 
\subsubsection{Calorimeter Pre-Trigger}
\begin{figure}[H]
  \centering
  \includegraphics[scale=0.35]{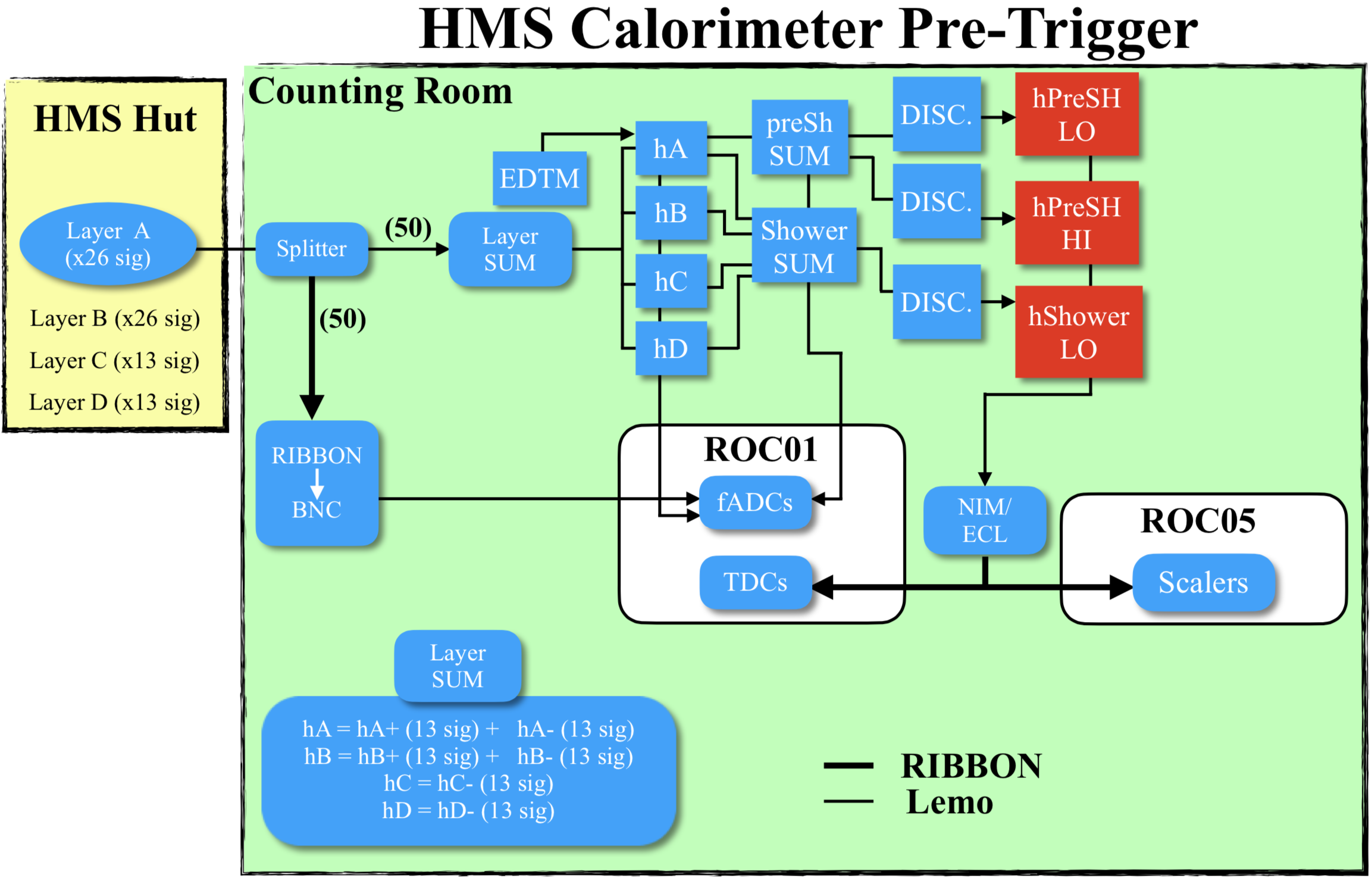}
  \caption{HMS calorimeter electronics diagram.}
  \label{fig:hCAL_diagram}
\end{figure}
\noindent The HMS calorimeter consists of four layers of lead blocks. Layers A and B read out 26 PMT signals per layer (13 signals/side) while layers C and D
read out 13 signals/layer on one side. The first layer forms the preshower counter while all four layers (A, B, C and D) form the shower counter.  
Each layer is read out in the Counting Room patch and fed into 50:50 splitters. One output of the splitter is connected to an fADC via a Ribbon-to-BNC converter (same as hodoscopes)
while the other output is sent to P/S Model 740 NIM Linear FI/FO summing modules. Each side of a layer is summed first (hA+, hA-, hB+, hB-, hC and hD sums). The sums are connected into
a LeCroy Model 428F summing module where layers hA+/- and hB+/- are summed to form hA and hB sums. A copy of each layer sum is then sent to an fADC. The preshower sum (\textbf{preSh SUM}) is made from
layer A, while the shower sum (\textbf{Shower SUM}) is made by summing all four layers. A copy of the preshower and shower sums is also sent to an fADC channel. The preshower and shower sums are also sent to a P/S Model 715
NIM discriminator unit to form the preshower Low/High (\textbf{hPreSH LO},\textbf{hPreSH HI}) and shower Low (\textbf{hShower LO}) pre-triggers with thresholds \hPrShLo, \hPrShHi and \hSHLo, respectively, with all
gate widths set to 30 ns. A copy of the pre-triggers is sent to TDC and scaler modules for trigger timing and counting information.
\subsubsection{Gas \v{C}erenkov Pre-Trigger} \label{sec:hms_cer_section}
\begin{figure}[!h]
  \centering
  \includegraphics[scale=0.35]{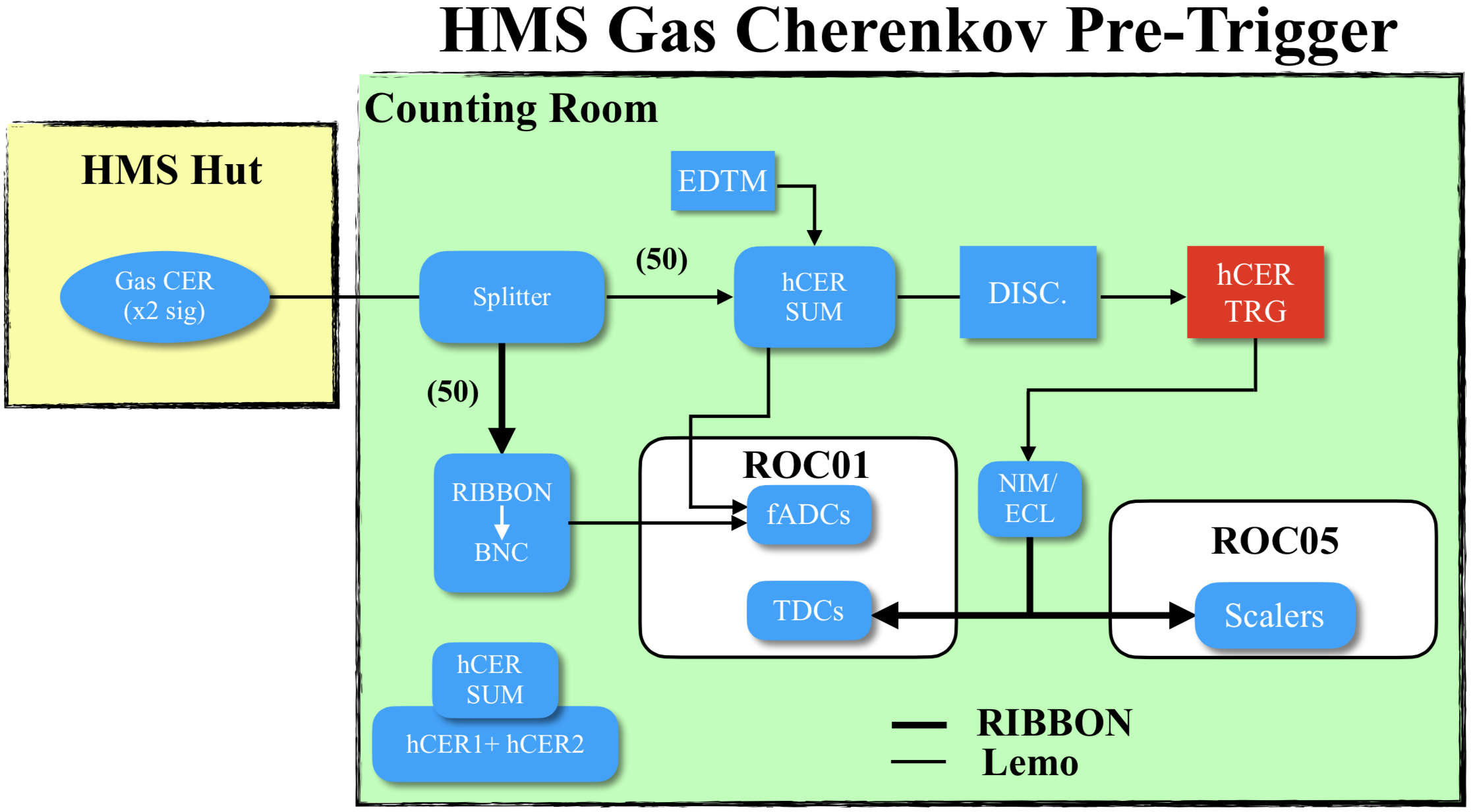}
  \caption{HMS gas \v{C}erenkov electronics diagram. Same electronics diagram applies for SHMS gas \v{C}erenkovs.}
  \label{fig:hCER_diagram}
\end{figure}
\noindent The HMS gas \v{C}erenkov detector consists of a 1.5 m long cylindrical tank between the first and second set of hodoscope planes (see Fig. \ref{fig:fig3.23}).
The tank is filled with a gas and has two spherical mirrors that focus the \v{C}erenkov photons towards two 5-inch PMTs. The
signals are read out in the Counting Room patch and pass through a 50:50 splitter. One output is fed into
an fADC module via a Ribbon-to-BNC converter. The other output is sent to a LeCroy Model 428F summing module, and a copy of the sum is fed to an fADC. The sum is
also sent to a P/S Model 715 NIM discriminator to form the \v{C}erenkov pre-trigger (\textbf{hCER TRG}) with a threshold and gate width set to \hcerthrs and \hcergate.
A copy of the discriminated signal is also sent to TDCs and scalers via a NIM/ECL converter for trigger and counting rate information.
\subsubsection{Aerogel \v{C}erenkov Pre-Trigger}
\indent The HMS aerogel \v{C}erenkov detector signals are sent directly to the Hall C floor patch panel and then
sent to the Counting Room patch and connected to a 50:50 splitter. One output leads to an fADC module via a Ribbon-to-BNC converter. The other output is
sent to a summing module, and a copy of the sum is sent to an fADC. The sum is also sent to a NIM discriminator to form the aerogel pre-trigger (\textbf{hAERO TRG}).
A copy of the discriminated signal is registered by TDCs and scalers via a NIM/ECL converter for trigger and counting information purposes.
The electronics diagram is the same as in Fig. \ref{fig:hCER_diagram}.
\subsubsection{HMS Single Arm Pre-Trigger}\label{ssec:hms_single_arm_sec}
\begin{figure}[H]
  \centering
  \includegraphics[scale=0.35]{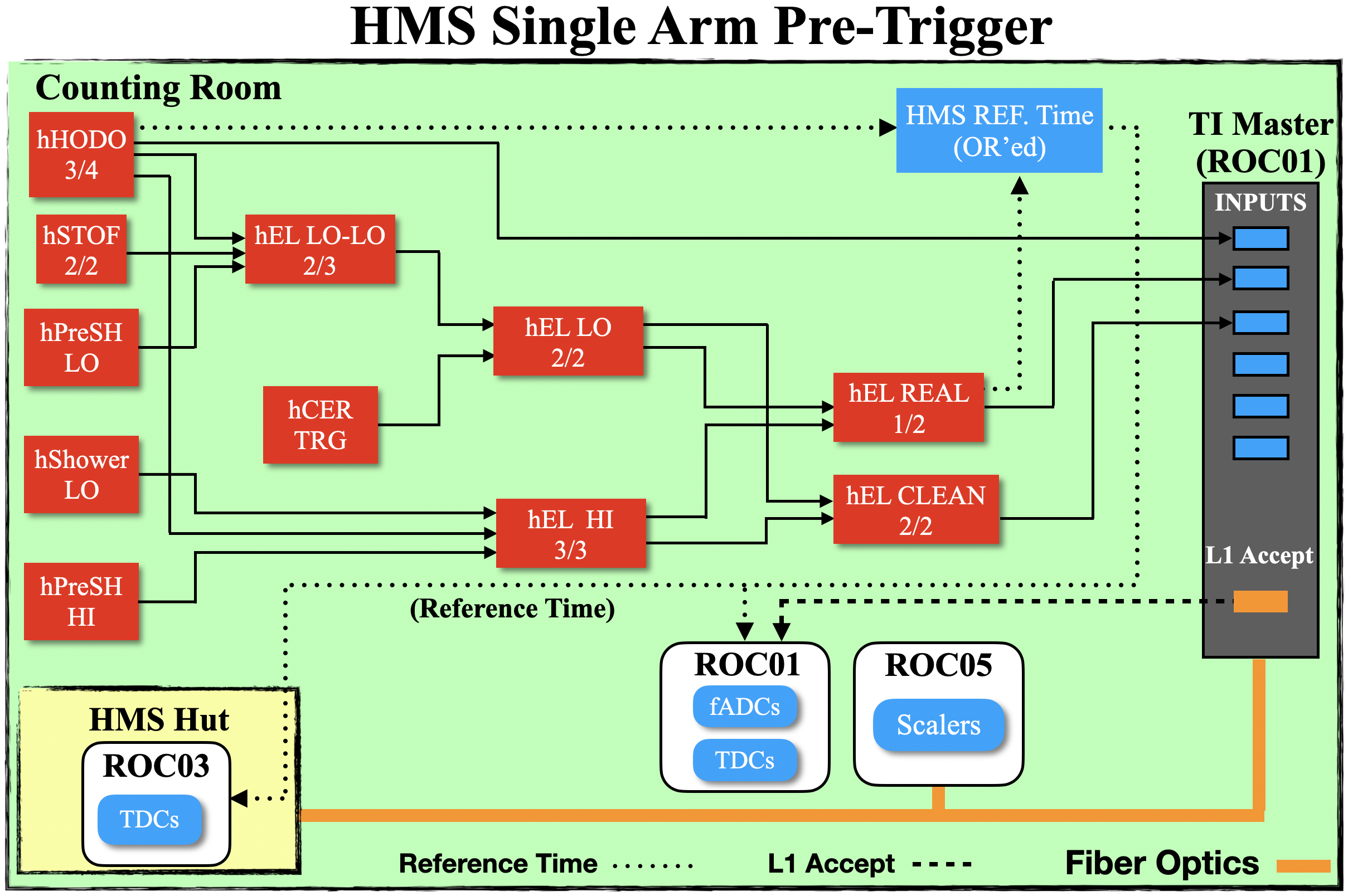}
  \caption{HMS single arm pre-trigger electronics diagram.}
  \label{fig:HMS_SingleArm_diagram}
\end{figure}

\noindent The HMS single arm pre-trigger is formed from the standard pre-trigger (hodoscopes) and a combination of other detector pre-triggers as required by the
experiment. The standard and other experiment-specific pre-triggers are sent to a P/S  Model 755 NIM logic unit to form a final single-arm pre-trigger (\textbf{hHODO 3/4}, \textbf{hEL REAL}, \textbf{hEL CLEAN}).
A copy of every pre-trigger (shown in red in Fig. \ref{fig:HMS_SingleArm_diagram}) is sent to scalers/TDCs (not shown). The final pre-triggers are sent to the front-end
of the TI module in \textbf{ROC 01}, which can receive up to 6 individual pre-triggers. A copy of the accepted trigger (Level 1 or L1 Accept) is sent via fiber optics cables to all the ROCs associated
with the HMS for data readout by all fADC/TDC modules. 
\noindent A copy of certain final pre-triggers are also OR'ed and are ultimately distributed to all ROCs with fADC/TDC modules
to function as a reference time associated with the L1 Accept.
The reference time is subtracted from every channel in every fADC/TDC module on an event-by-event basis to reduce intrinsic jitter
and achieve the design resolution of the module. To guarantee that every event has an associated reference time, the HMS standard pre-trigger (\textbf{hHODO 3/4}) is OR'ed with the \textbf{hEL-REAL}
pre-trigger to guarantee a reference time in the rare case where the \textbf{hHODO 3/4} fails due to trigger inefficiency which is very small ($\lesssim1 \%$). 
\subsection{SHMS Trigger Setup}
The three planes (X1, Y1, X2) of scintillator arrays and the quartz plane (Y2) form part of the standard SHMS trigger configuration (see Fig. \ref{fig:fig3.24}).  
Additional particle detectors may also be incorporated into the SHMS trigger as required by different experiments. The NGC and
calorimeter triggers are used for $e/\pi$ discrimination, whereas the HGC and aerogel \v{C}erenkov triggers are used for $e/\pi/p$ and $\pi/K/p$
discrimination, respectively, depending on the gas pressure and aerogel material used. 
\subsubsection{Hodoscopes Pre-Trigger}
\begin{figure}[h!]
  \centering
  \includegraphics[scale=0.36]{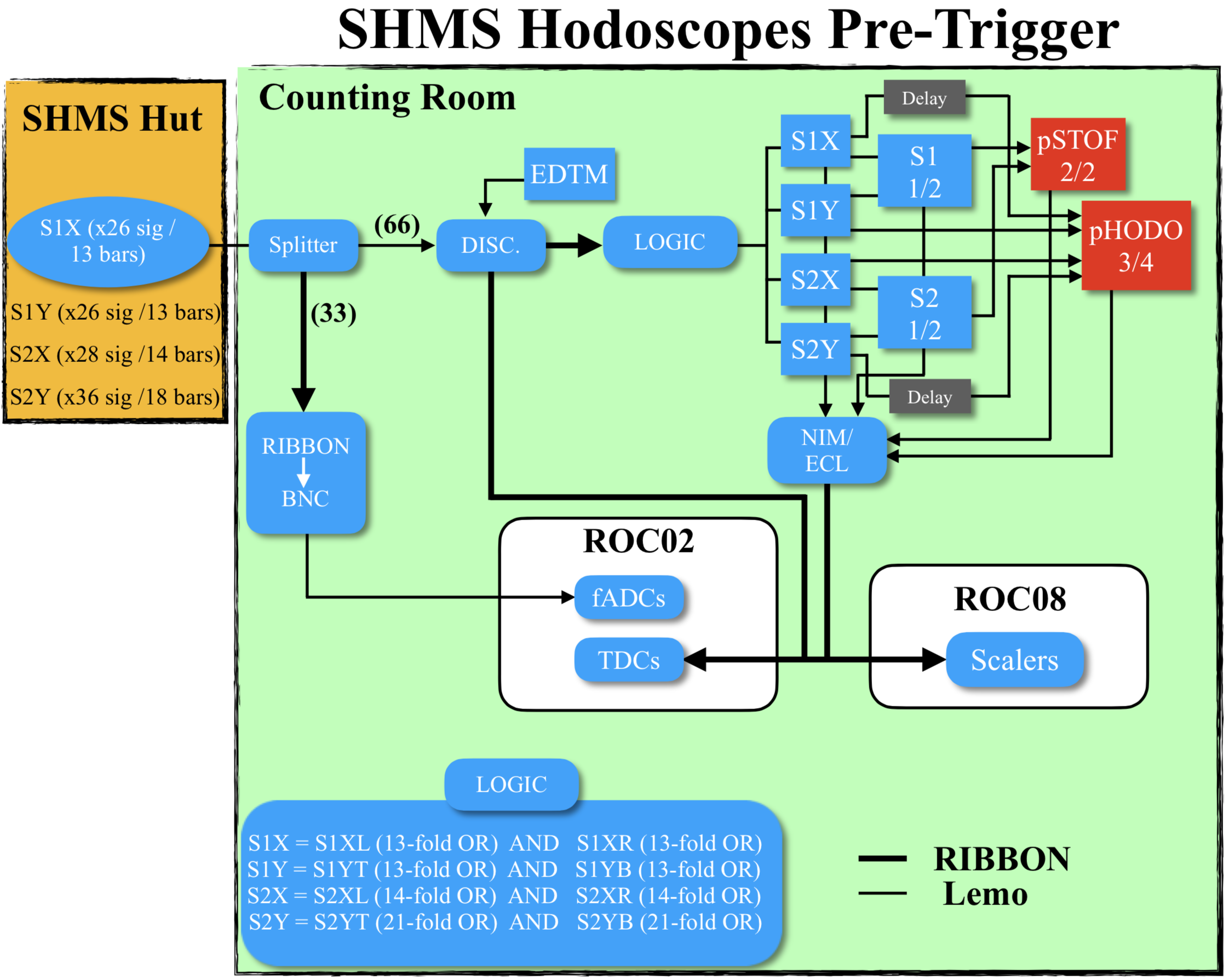}
  \caption{SHMS hodoscopes electronics diagram. It is important to note that only 18 of the 21 quartz bars are currently usable.}
  \label{fig:pHODO_diagram}
\end{figure}
\indent Each hodoscope plane consists of an array of scintillator paddles (or quartz bars) coupled to a PMT at each end (see Fig. \ref{fig:fig3.27}), so each bar reads out two signals. As shown in Fig. \ref{fig:pHODO_diagram},
for example, hodoscope plane S1X consists of 26 signals (16 paddles) read out in the Counting House (CH) patch. Each side of the plane (x13 signals/side) is connected to a 64-channel input passive splitter (16 Ch./set).
One-third of the signal amplitude is sent via a 16-channel ribbon cable to a 64-channel input Ribbon-to-BNC converter (16 Ch./set) and subsequently into a 16-channel NIM input fADC.
The remaining two-thirds of the signal amplitude is sent to a 16-channel input CAMAC discriminator unit. The SHMS scintillator discriminators thresholds and gate widths were set to \shodthrs and \shodgate, respectively,
whereas the quartz plane discriminators thresholds and gate widths were set to \quartzthrs and \shodgate, respectively.\\
\indent The discriminated signals are sent via two ribbon-cable outputs to C1190 TDCs and scalers (daisy-chained) and to a LeCroy 4564 CAMAC logic unit to form the plane pre-triggers.
The logic unit takes four sets of 16-Ch. input ribbon cable and forms a 16-fold OR for each set by default. Further boolean operations are done through the module backplane by connecting a twisted pair cable to
the pin corresponding to the desired boolean operation. For the SHMS hodoscope plane pre-triggers, the boolean operations are as follows:
\begin{empheq}[box=\fbox]{align*}
& \text{\textbf{S1X} = S1XL (13-fold OR) \text{AND} S1XR (13-fold OR)} \\ 
& \text{\textbf{S1Y} = S1YT (13-fold OR) \text{AND} S1YB (13-fold OR)} \\
& \text{\textbf{S2X} = S2XL (14-fold OR) \text{AND} S2XR (14-fold OR)} \\ 
& \text{\textbf{S2Y} = \big\{S2Y[1-16]T OR S2Y[17-21]T\big\} \text{AND} \big\{S2Y[1-16]B OR S2Y[17-21]B\big\}}
\end{empheq}
\indent Once a pre-trigger has been made for each plane, they are sent to a NIM/ECL converter (Level Translator P/S Model 7126) via twisted pair cables to convert the ECL signal
(twisted pair) to a NIM signal. The NIM output is then sent to individual sets of a P/S Model 752 NIM logic unit to adjust the widths of each of the plane pre-triggers as necessary before making a coincidence.
An XY hodoscope plane coincidence (\textbf{S1} = S1X \text{OR} S1Y, \textbf{S2} = S2X \text{OR} S2Y) is then made by connecting each hodoscope XY plane pair into a P/S Model 755 NIM logic unit\footnote{\singlespacing The output
  widths of the P/S Model 755 logic units were set to $\sim$ 100 ns for the SHMS. See HCLOG entry \url{https://logbooks.jlab.org/entry/3501354}.}. A copy of each of the four
individual plane pre-triggers is also sent to another set of P/S Model 755 to make a 3/4 or 4/4 plane coincidence (configured via a front-panel knob) which defines the standard hodoscope pre-trigger (\textbf{pHODO 3/4}).
An additional pre-trigger (\textbf{pSTOF} = \textbf{S1} \text{AND} \textbf{S2}) is formed by requiring the coincidence between any two of the front (\textbf{S1}) and back (\textbf{S2}) scintillator (or quartz) plane pair to measure the TOF
between any of the two front and back planes. A copy of the hodoscope pre-triggers are also sent to TDCs and scalers via a NIM/ECL converter for timing and counting information. 
\subsubsection{PreShower and Shower Calorimeter Pre-Trigger}
\noindent The SHMS preshower consists of two sets of fourteen PMT-coupled lead blocks oriented perpendicular to the shower counter blocks. The initial sum was done in the SHMS electronics hut. The PMT signals
in groups of four blocks were summed to form:
\begin{empheq}[box=\fbox]{align*}
  & \text{preSh SUM [1-4]: [1-4]L + [1-4]R} \\ 
  & \text{preSh SUM [5-8]: [5-8]L + [5-8]R} \\ 
  & \text{preSh SUM [9-12]: [9-12]L + [9-12]R} \\ 
  & \text{preSh SUM [13-14]: [13-14]L + [13-14]R}
\end{empheq}
The partial preshower signal sum was sent to the Counting Room patch where a final sum was made. Two copies of the final sum were sent to a discriminator to form two preshower pre-triggers
(\textbf{pPreSH HI}, \textbf{pPreSH LO}) with a lower and higher threshold, respectively. \\
\begin{figure}[H]
  \centering
  \includegraphics[scale=0.33]{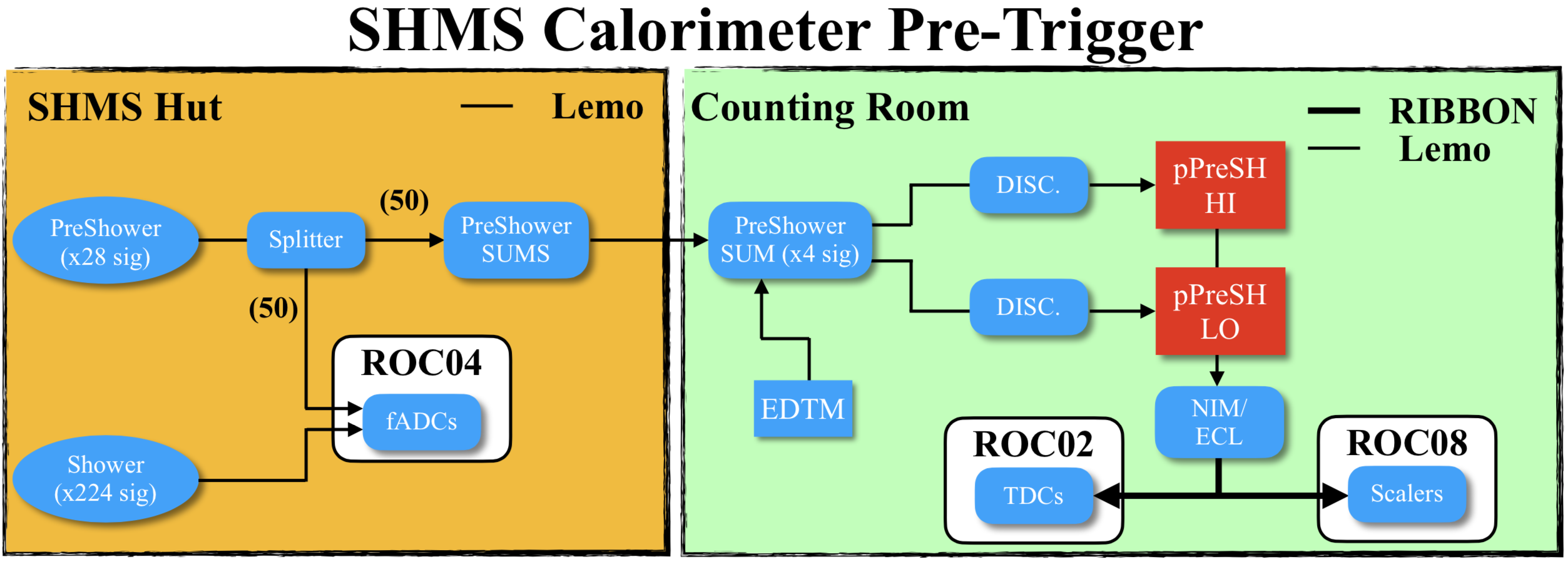}
  \caption{SHMS PreShower and Shower electronics diagram.}
  \label{fig:pCAL_diagram}
\end{figure}
A copy of the pre-triggers were sent to TDCs and scalers via a NIM/ECL converter for timing and counting information. 
The shower counter consists of 224 lead blocks, each coupled to a PMT at the end. Becasue of the high channel density of this detector, its signals do not form part of the trigger and are sent
directly to the \textbf{ROC 04} fADCs in the SHMS detector hut.
\subsubsection{Heavy and Noble Gas \v{C}erenkov Pre-Trigger}
\begin{figure}[h!]
  \centering
  \includegraphics[scale=0.35]{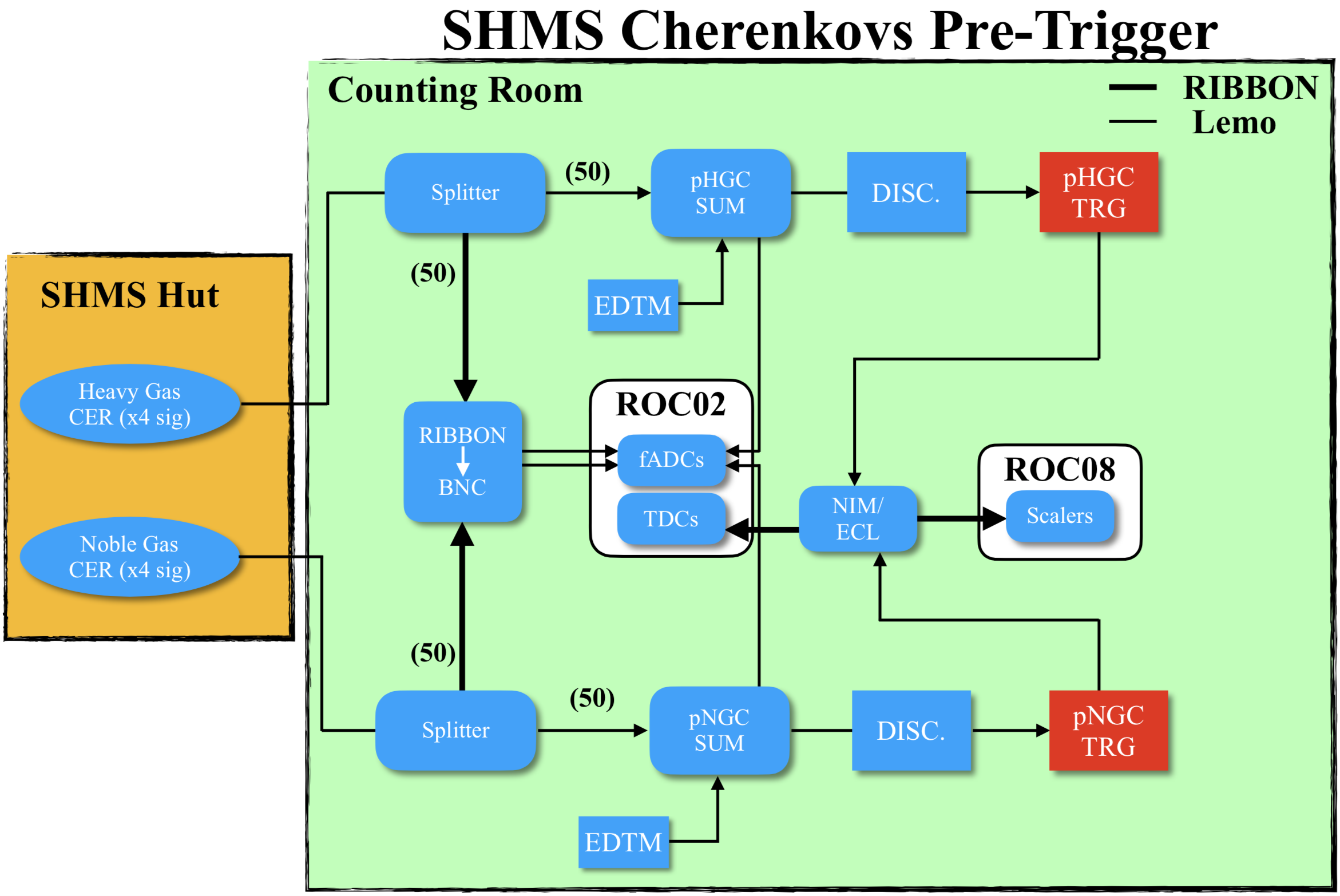}
  \caption{SHMS gas \v{C}erenkovs electronics diagram.}
  \label{fig:pCER_diagram}
\end{figure}
\noindent The SHMS HGC detector consists of a 1 meter-long, 1.6 meters in diameter cylindrical tank located between the front and back sets of hodoscope planes (see Fig. \ref{fig:fig3.24}).
The tank is filled with a gas and has four thin spherical mirrors that focus the \v{C}erenkov light towards four 5-inch PMTs. \\
\indent The SHMS NGC detector consists of a 2 meter-long active length of argon/neon gas tank located before the first drift chamber (see Fig. \ref{fig:fig3.24}).
The tank is filled with a gas and has four overlapping mirrors that focus the \v{C}erenkov photons towards four 5-inch PMTs. The electronics trigger
setup for the SHMS \v{C}erenkovs is shown in Fig. \ref{fig:pCER_diagram} and is very similar to the HMS \v{C}erenkovs. Refer to Fig. \ref{fig:hCER_diagram} and read the corresponding
section for a full description of the corresponding trigger setup.
\subsubsection{Aerogel \v{C}erenkov Pre-Trigger}
The SHMS aerogel \v{C}erenkov detector consists of a $110 \times 100 \times 24.5$ cm$^{3}$ rectangular aerogel tray coupled to a diffusion box. The diffusion box has seven 5-inch PMTs on each side
which detect \v{C}erenkov light produced by interactions with the aerogel material. The detector is located between HGC and second set of hodoscope planes
(see Fig. \ref{fig:fig3.24}). The electronics diagram is the same as in Fig. \ref{fig:pCER_diagram}.
\subsubsection{SHMS Single Arm Pre-Trigger}
\noindent The SHMS single arm trigger is formed exactly as the HMS single arm trigger, with the exception of the detectors pre-triggers involved which depend on the experiment.
Refer to Fig. \ref{fig:HMS_SingleArm_diagram} and read the corresponding section for a detailed description of the electronic diagrams.
\begin{figure}[H]
  \centering
  \includegraphics[scale=0.34]{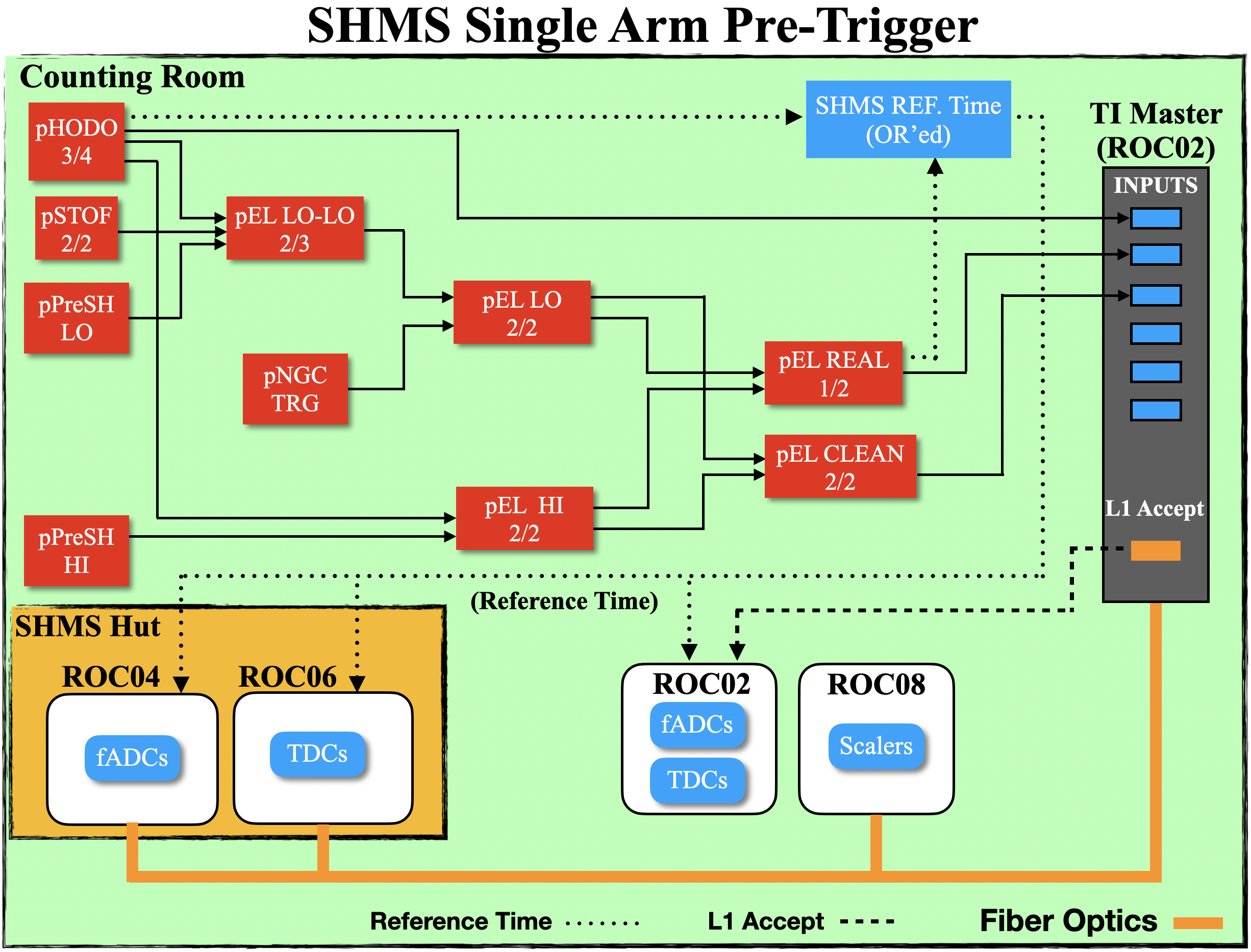}
  \caption{SHMS single arm trigger electronics diagram.}
  \label{fig:SHMS_SingleArm_diagram}
\end{figure}
\subsubsection{Coincidence Trigger Set-Up}
\begin{figure}[H]
  \centering
  \includegraphics[scale=0.35]{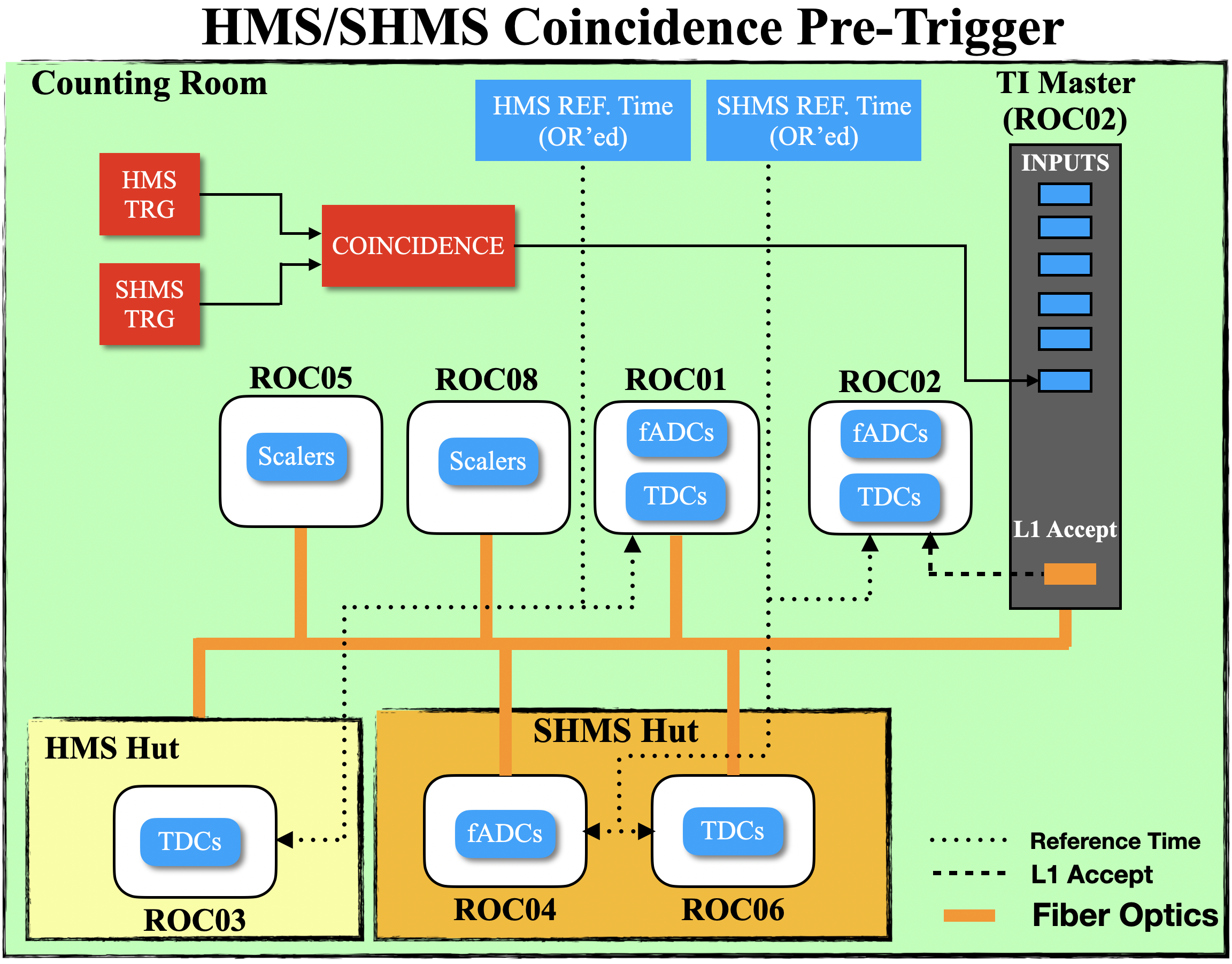}
  \caption{Coincidence trigger electronics diagram.}
  \label{fig:Coin_TRG}
\end{figure}

\noindent In coincidence mode (see Fig. \ref{fig:Coin_TRG}), the HMS and SHMS pre-triggers are sent to a NIM logic module where the first spectrometer pre-trigger that arrives will open
a coincidence time window during which the second spectrometer pre-trigger may or may not arrive in that time. This will determine whether two spectrometers pre-triggers
are correlated with the event originated at the target. If the coincidence pre-trigger is formed, a copy is sent to scalers/TDCs while another copy is sent to the fron-end
of the TI module in \textbf{ROC 02} which acts as the Trigger Master (TM) in coincidence mode. Once the TM accepts the coincidence trigger, multiple copies of the L1 Accept
are distributed to all HMS and SHMS ROCs (except \textbf{ROC 02}) via fiber obtics cables in all crates for data readout. An additional copy of the L1 Accept is also sent to
the front-end of the TDCs in \textbf{ROC 02}. Multiple copies of the HMS/SHMS pre-triggers (reference times) are also distributed to their respective spectrometer ROCs with
fADC/TDC modules to function as a reference time associated with the coincidence trigger.
\subsection{Electronic Dead Time Monitoring (EDTM)}\label{subsec:edtm}
\noindent The EDTM system is a new method used in Hall C to measure the total dead time of the data acquisition (DAQ) system. It consists of introducing a controlled (fixed frequency) pulse
as near as possible to the detectors that form part of the trigger. Ideally, one would send the EDTM pulses at the detector level in the hut such that both the real physics and EDTM signals
pass through the same electronics. Since this is not easy or practical to do, the EDMT logic pulses are injected at the trigger logic level in the Counting Room.\\
\begin{figure}[H]
  \centering
  \includegraphics[scale=0.54]{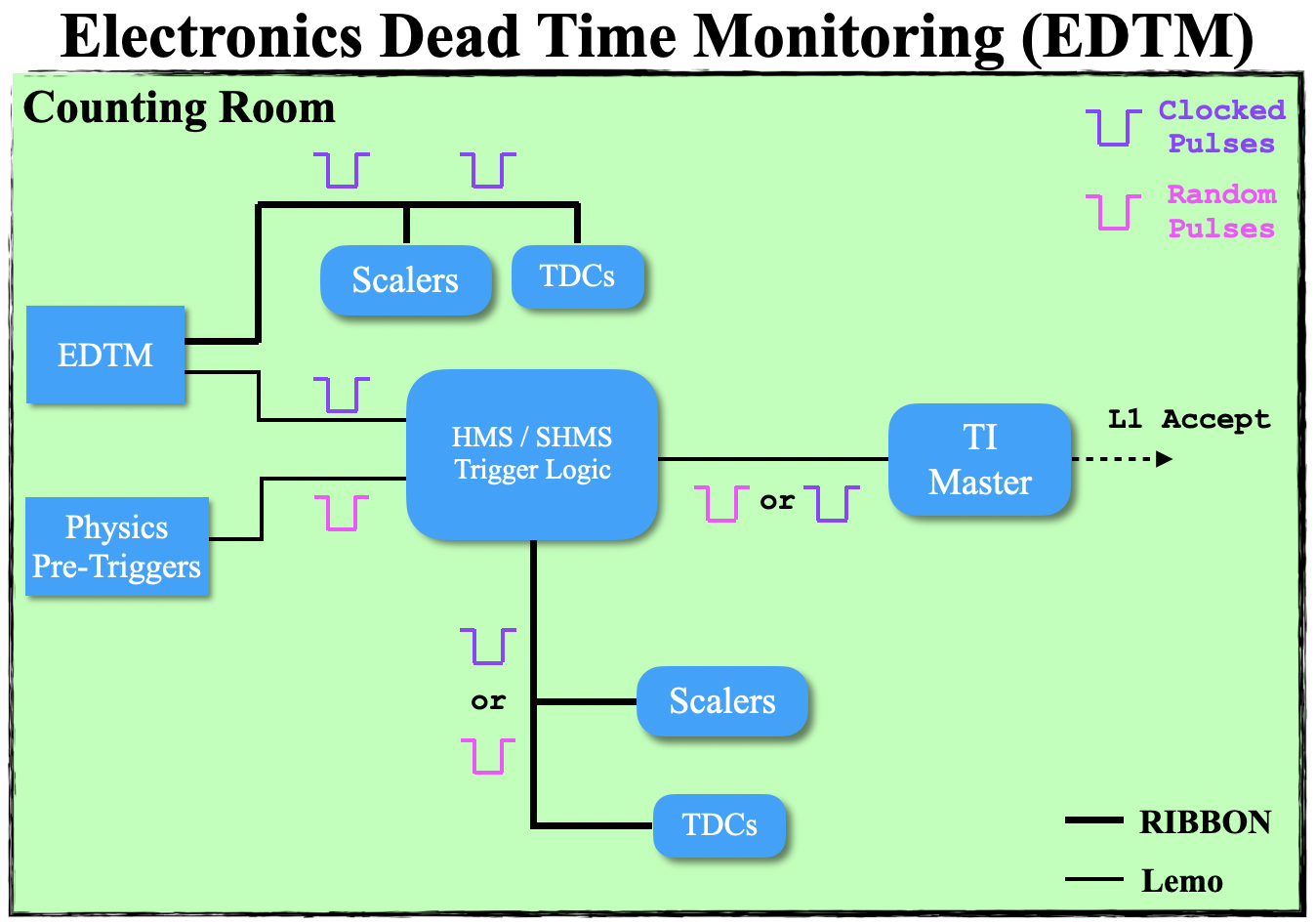}
  \caption{EDTM electronics diagram.}
  \label{fig:EDTM_diagram}
\end{figure}
\indent By design, the EDTM is a \textit{real} trigger as measured by the electronics and readout systems. Since the EDTM is invasive to the trigger electronics, its frequency should be small
enough to minimize the probability of blocking actual physics triggers, but sufficiently large to gather the necessary statistics for a precise \textit{dead time} measurement during the course of a run. \\
\indent Figure \ref{fig:EDTM_diagram} shows a simplified diagram of the EDTM signal distribution through the trigger electronics. The EDTM logic signals (purple) are injected into the trigger logic
where they mix with the physics pre-triggers (magenta). A separate copy of the EDTM is also sent to scalers/TDCs to be used in the dead time calculation. If the
EDTM makes it to the front-end of the Trigger Interface (TI) module and gets accepted (L1 Accept), it has esentially measured both the electronics and computer dead time.\\
\begin{figure}[H]
  \centering
  \includegraphics[scale=0.40]{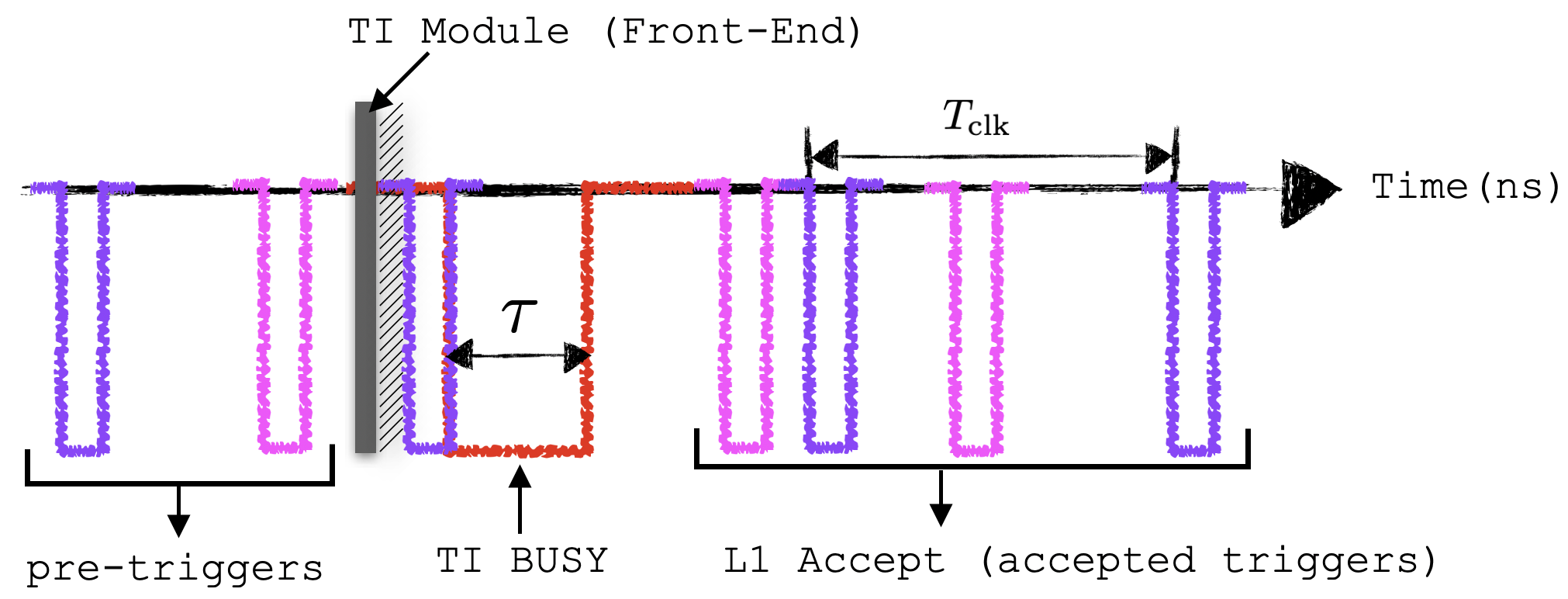}
  \caption{Cartoon representation of EDTM (purple) and physics (magenta) pre-triggers at the TI module front-end.}
  \label{fig:EDTM_TI}
\end{figure}
\indent Figure \ref{fig:EDTM_TI} shows the random physics (magenta) and clocked (purple) EDTM pulses at an input channel of the TI front-end where the EDTM has
been set to a sufficiently large frequency ($1/T_{\mathrm{clk}}$) to ensure that enough EDTM signals get accepted in order to make a statistically significant and
reliable dead time calculation. In this example, an EDTM signal has been accepted by the TI which triggered a \textit{BUSY} signal for a time $\tau$ during which all other incoming pre-triggers are blocked
contributing to the DAQ computer dead time. The accepted pre-triggers are distributed to all ROCs for data readout. \\
\indent Over the course of a run, the total dead time ($T_{\mathrm{TDT}}$), or alternatively, the total live time ($T_{\mathrm{TLT}}$) in terms of the EDTM is defined as
\begin{equation}
 T_{\mathrm{TDT}} \equiv 1 - T_{\mathrm{TLT}} = 1 - \frac{N_{\mathrm{edtm,acc}}}{N_{\mathrm{edtm,scl}}},
  \label{eq:3.24}
\end{equation}
where $N_{\mathrm{edtm,acc}}$ is the number of accepted EDTM counts obtained by requiring a non-zero hit on the EDTM TDC spectrum, and $N_{\mathrm{edtm,scl}}$ is the number of EDTM scaler counts regardless of whether or not
the EDTM was accepted. In reality, frequent beam trips occur during the course of a run which makes this calculation biased since one can measure live times of $\sim$100$\%$ during beam-off periods as only
the EDTM signal (and cosmic rays) are measured. To eliminate this bias, the live time calculation was done by making a software cut on the beam current above a certain threshold (refer to target density corrections in Section \ref{sec:tgt_boil_corr}).
Furthermore, since the EDTM events are generated by a clock, rather than a poisson source, it introduces an additional bias since the EDTM cannot block itself. To account for this bias, an additional correction
to the total live time was derived and can be found in Ref.\cite{DMack_livetime_2019}. This correction, however, is negligible provided that the EDTM rate is sufficiently low as was the case for this experiment ($\sim 2$ Hz). \\
\indent Even though only the total live time is required as a correction factor in the measured cross section, one may also calculate the computer live time defined as
\begin{equation}
  T_{\mathrm{CLT}} = \frac{N_{\mathrm{phy,acc}}}{N_{\mathrm{phy,scl}}},
  \label{eq:3.25}
\end{equation}
where $N_{\mathrm{phy,acc}}$ is the number of accepted physics triggers obtained by requiring a zero hit on the EDTM TDC spectrum (EDTM rejected by TI) and $N_{\mathrm{phy,scl}}$ is the number of physics trigger scaler counts after
having subtracted the EDTM scaler counts. The electronic live time can then be obtained from the following formula:
\begin{equation}
  T_{\mathrm{TLT}} = T_{\mathrm{CLT}} \cdot T_{\mathrm{ELT}},
  \label{eq:3.26}
\end{equation}
where $T_{\mathrm{ELT}}$ is the electroninc live time expressed as a fraction (not percent). \\
\indent In the E12-10-003 experiment, the data from the main analysis was read out by an unprescaled coincidence trigger which simplified the live time calculations discussed above,
since there was no need to divide by a prescale factor or account for simultaneous multiple input triggers. For a more detailed discussion on the live time calculations and its correction factors
see Refs.\cite{pooser_livetime_2019,DMack_livetime_2019}.

\chapter{GENERAL HALL C ANALYSIS OVERVIEW}	\label{chap:chapter4}
The general Hall C analysis procedure for experiments in the 12 GeV era is discussed. The procedure
outlines the first necessary steps in the data analysis regardless of the nature of the experiment. These include, but are not
limited to, setting reference time cuts, detector time window cuts, and performing detector calibrations.
Optics checks and optimization analysis of the SHMS reconstruction matrix using $^{1}$H$(e,e')p$ elastic data are discussed for this experiment. Finally,
the data-to-simulation comparisons of the spectrometer acceptance as well as the event selection criteria for $^{1}$H$(e,e')p$ elastics and
the $^{2}$H$(e,e'p)n$ reaction are shown.
\section{Reference Time Cuts}
\noindent The first step in Hall C data analysis is to make sure the reference time cuts are set properly,
as one needs to make sure the reference times correlated with the trigger are selected. The \textit{reference time}\footnote{\singlespacing As with any hardware electronics
signal, the reference time signal is arbitrary and is only useful when compared relative to any other arbitrary signal such as to
measure a relative time.} signal is defined as a copy of either one or multiple OR'ed pre-trigger logic signals described in the electronics diagram of
Figs. \ref{fig:HMS_SingleArm_diagram}, \ref{fig:SHMS_SingleArm_diagram} and \ref{fig:Coin_TRG}. The reference time is distributed to all
fADCs\footnote{\singlespacing Before being sent to an fADC, the reference time logic signal must be converted into an analog signal. This is done
with a passive circuit.}
and C1190 TDC modules of all Read-Out Controllers (ROCs). The fADC and TDC modules register either analog (fADC) or discriminated logic (TDC) signals from every detector
output as well as the corresponding reference time signal generated by the trigger electronics. The main objective of the reference time signal is 2-fold:
\begin{itemize}
  \item the reference time signal serves as a \textit{common stop} (initiates a look-back window) for all detector input channels in each fADC/TDC module 
  \item the reference time signal is used to determine time intervals from the raw detector signals sent to the TDC module
\end{itemize}
\begin{figure}[H]
  \centering
  \includegraphics[scale=0.50]{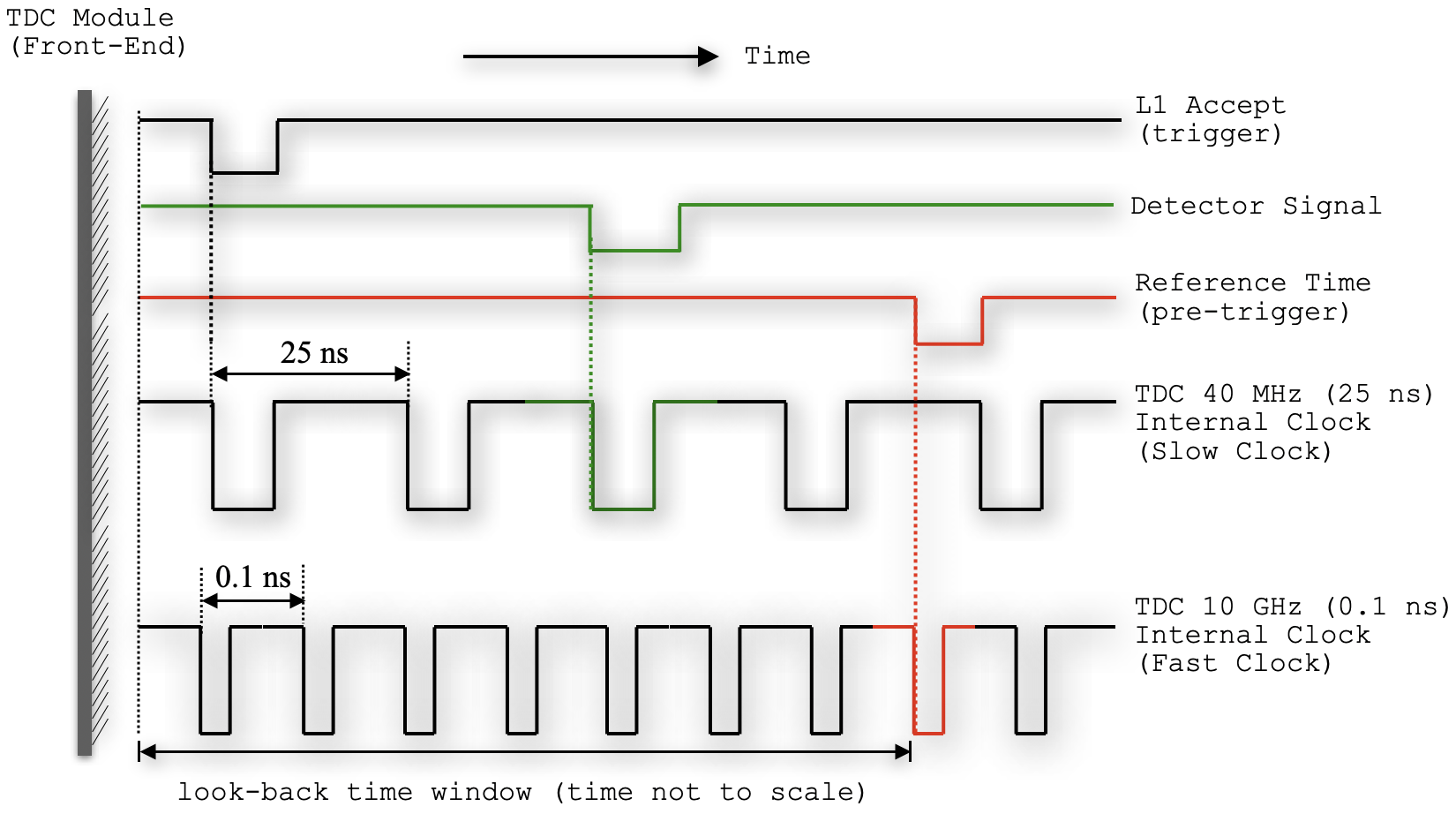}
  \caption{Cartoon illustrating the synchronization of the a detector signal with the internal clocks of a C1190 TDC Module.}
  \label{fig:TDC_clock}
\end{figure}
\indent Before modern TDCs such as the C1190 TDC\cite{C1190_TDC}, in the original fastbus TDC modules\cite{fastbus_1877_TDC} a L1 Accept (accepted trigger) was sent to the front-end of
the module and acted as a \textit{common start time} to all channels of the module as well as initiated data readout. The common start time was measured relative to the stop signal
which was provided by the individual input channels on the TDC module. This time difference was converted to a number and histogrammed to form a TDC spectrum of counts vs. channel
number. \\
\newpage
\indent On modern TDCs (see Fig. \ref{fig:TDC_clock}), the L1 accept, which has an intrinsic 4 ns jitter\footnote{\singlespacing In electronics terminology, jitter refers to a small,
  irregular variation or unsteadiness in an otherwise periodic signal.}, is sent to the front-end to initiate data readout. The detector signal (green) sent to the
front-end of the module serves as the \textit{TDC Start} and is synchronized with the internal 40 MHz clock of the TDC (slow clock). The detector signal (TDC Start) latches onto
the leading edge of the next 40 MHz clock cycle, which means the signal could have landed anywhere in a 25 ns range between the
previous and next clock cycle. As a result, a 25 ns jitter arises intrinsically when the raw TDC signal is measured relative to the L1 accept, which has an additional 4 ns jitter.
This means that the raw TDC detector signals cannot be determined better than 29 ns resolution, which is well above the module specifications of 0.1 ns resolution. To improve the timing
resolution, the TDC uses a second internal clock (fast clock) at 10 GHz or 0.1 ns periodicity. The reference time (copy of pre-trigger which is effectively a delayed L1 accept)
is sent to the TDC module at a time delay relative to all the detector signals and effectively serves as the \textit{TDC Stop}, which latches onto the leading edge of the next clock cycle of the 10 GHz high resolution clock, and initiates
a look-back time window (usually a few $\mu$s) corresponding to the full TDC spectrum (e.g., see Fig. \ref{fig:pDCref_time_prob}).
This internal reference time, which is known to approximately 0.1 ns, is subtracted from the raw detector signal TDC time, thereby, improving the timing resolution of the detector signals to $\sim$0.1 ns per channel. \\
\indent The reference time signal is common to all modules in a given ROC and is therefore subtracted from all input channels of every fADC/TDC module present in said ROC.
The reference time subtraction is performed by the Hall C analyzer \textit{hcana} during the analysis replay. When using the reference time,
\textit{hcana} chooses the first hit in the time window if multiple hits are present per event. In this scenario, the first hit may not
necessarily be the \textit{good hit} and the wrong reference time would be chosen resulting in the wrong time being subtracted in the fADC/TDC spectra.
By placing a reference time cut, the analyzer then considers the first hit after the cut, which is likely to be a \textit{good hit}. \\
\indent As an example, consider the $^{1}$H$(e,e')p$ elastic coincidence run 3377, which had the highest SHMS rate of elastics taken during the E12-10-003 experiment.
\begin{figure}[H]
  \centering
  \includegraphics[scale=0.36]{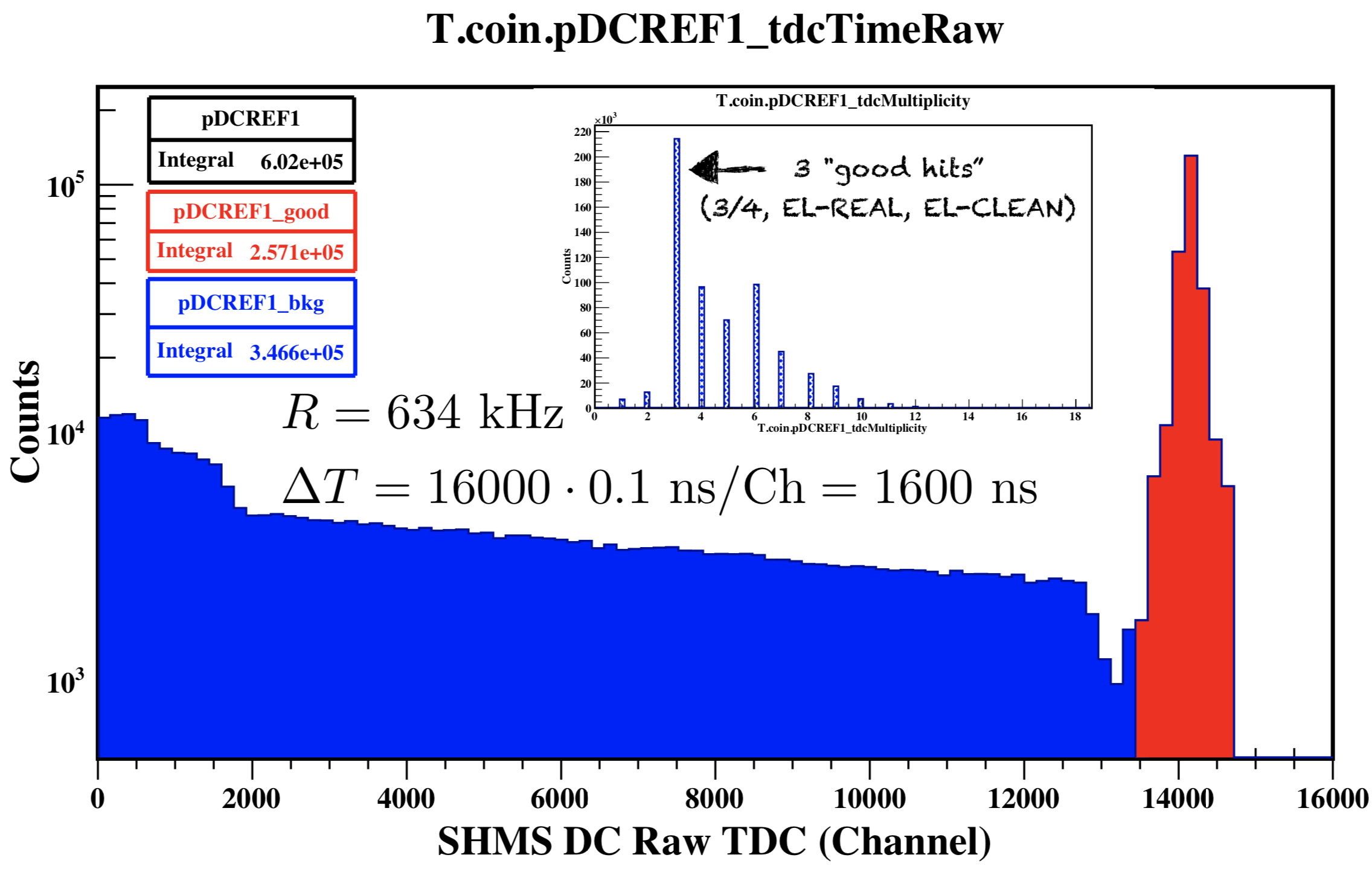}
  \caption{SHMS reference time spectrum for coincidence run 3377 of the E12-10-003 experiment. Background hits are shown in blue and good hits in red. Inset: Multiplicity histogram
  corresponding to the reference time spectrum.}
  \label{fig:pDCref_time_prob}
\end{figure}
\noindent Figure \ref{fig:pDCref_time_prob} shows an uncorrected reference time spectrum in the SHMS drift chambers crate (ROC 06) where the red spectrum
represents the prompt peak corresponding to the reference time signal events and the blue spectrum represents background events corresponding to signals other than
the reference time. The inset plot shows the multiplicity histogram corresponding to the TDC spectrum. The multiplicity histogram shows the total number of counts or L1 accept ($y$-axis) versus the
number of TDC hits (or multiplicity) corresponding to each event where the number of TDC hits are defined as the total hits in the TDC readout window for any given event. For example,
the inset of Fig. \ref{fig:pDCref_time_prob} shows that most of the events had a multiplicty of three, which means that for every event, the TDC readout window registered 3 hits. In this
case, the 3 hits correspond to three reference time signals, which were OR'ed and used as an effective reference time. The other multiplicities ($>3$ hits) represent additional background hits
that were present in the readout window and can potentially pass as a reference time, thereby blocking the true reference time, resulting in the wrong reference time being subtracted. The
multiplicities below 3 hits represent the very unlikely case (hodoscope trigger inefficiency) in which one or more of the reference time signals is not formed, in which case the other
reference times are used.\\
\indent Following the poisson behavior of physics triggers, the probability that 2 hits (one background hit, one good hit) fall within a certain
time window $\Delta T$ at a given physics rate $R$ is given by
\begin{equation}
  P(\lambda; k) = e^{-\lambda}\frac{\lambda^{k}}{k!},
  \label{eq:4.1}
\end{equation}
where $\lambda=R \Delta T$ and $k$ is the number of TDC hits.
From the multiplicity in Fig. \ref{fig:pDCref_time_prob}, most reference time events had 3 good hits (HODO 3/4, EL-REAL, EL-CLEAN).
For simplicity of the calculation, we redefine 3 good hits as a single good hit. Then, from Eq. \ref{eq:4.1} and Fig. \ref{fig:pDCref_time_prob},
the probability of finding 2 hits within the drift chamber time window is
\begin{align}
  P(\lambda;k) = e^{-R\Delta T}\frac{(R\Delta T)^{2}}{2!} = 0.1865
  \label{eq:4.2}
\end{align}
From Fig. \ref{fig:pDCref_time_prob}, this probability is given by taking the ratio of the background (blue) to the total number of events normalized to one good hit (divide by 3) to obtain
\begin{align}
  P_{\mathrm{data}} = \frac{1}{3}\frac{346600}{602000} = 0.1919
  \label{eq:4.3}
\end{align}
The two results agree to $\leq 1\% $. These results indicate that if the reference time had not been set for this run, then $\sim 19\%$ of the
events would have the incorrect reference time and a lower tracking efficiency by $\sim 19\%$, hence, the importance of setting the reference times.\\
\indent In Hall C, each spectrometer has multiple pre-triggers (HODO 3/4, STOF, EL-REAL, EL-CLEAN) that may change depending on the nature of the experiment.
The base pre-trigger is the HODO 3/4, which requires at least 3 of 4 hodoscope planes to fire. During the commissioning phase of the spectrometers, the
reference time logic was initially defined to be\footnote{\singlespacing See December 2017 HC-Log Entry \url{https://logbooks.jlab.org/entry/3501198}.}:
\begin{align}
  T^{\mathrm{logic}}_{\mathrm{reftime,init}} \equiv &\textbf{ p(h)HODO 3/4} \text{ OR } \textbf{p(h)STOF} \nonumber  \\
  &\text{ OR } \textbf{p(h)EL-REAL} \text{ OR } \textbf{p(h)EL-CLEAN}
  \label{eq:4.4}
\end{align}
where the logic pre-trigger signals described in Eq. \ref{eq:4.4} have been delayed in time relative to each other and the \textbf{p(h)} refers to prefix used in the software to denote the SHMS (HMS).  
Figure \ref{fig:scope_trace} shows a visual representation of Eq. \ref{eq:4.4} of how typical reference time logic signals might appear on an oscilloscope.
On January 2018, the STOF trigger was removed from this definition\footnote{\singlespacing See HC-Log Entry \url{https://logbooks.jlab.org/entry/3519686}.}.
Finally, on August 2018, EL-CLEAN was removed from the reference time definition\footnote{\singlespacing See HC-Log Entry \url{https://logbooks.jlab.org/entry/3585301}.} as well.
It was determined that any pre-trigger that required the HODO 3/4 was unnecessary and redundant to have in the reference time definition so they were removed.
\begin{figure}[H]
  \centering
  \includegraphics[scale=0.45]{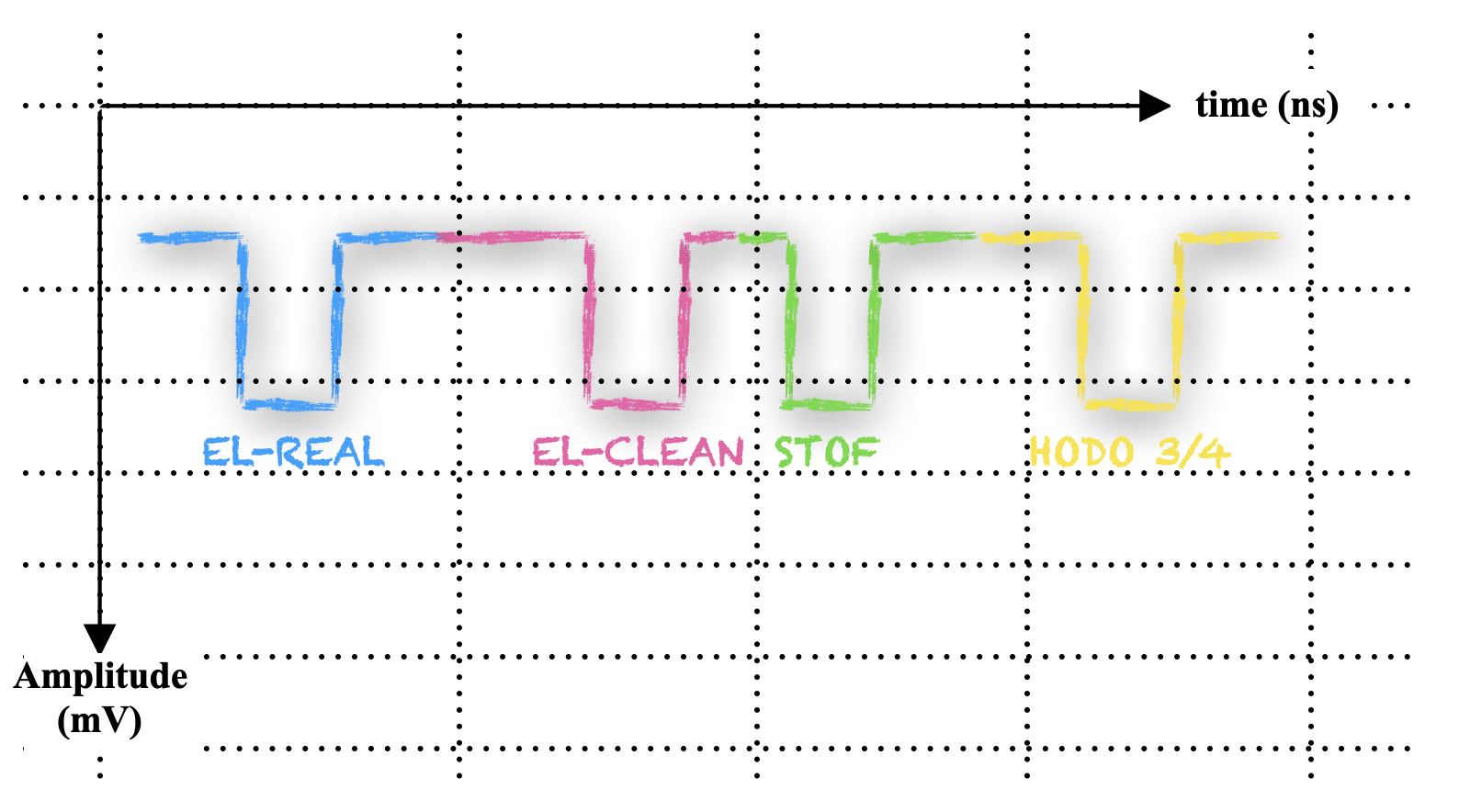}
  \caption{Cartoon illustrating how reference time logic definition of Eq. \ref{eq:4.4} might appear on an oscilloscope. This corresponds to a multiplicity of four,
  or equivalently, 4 good hits in the TDC readout window.}
  \label{fig:scope_trace}
\end{figure}
\indent As of the Fall 2019 run period, the reference time logic definition in Hall C was:
\begin{align}
  T^{\mathrm{logic}}_{\mathrm{reftime}} \equiv \textbf{p(h)HODO 3/4} \text{ OR } \textbf{p(h)EL-REAL} 
  \label{eq:4.5}
\end{align}
as the EL-REAL did not require a HODO 3/4, and in the rare instances the latter is missing, the former can be used as a reference time.
The STOF was also completely removed from the trigger definition and the reference time was
redefined as HODO 3/4. Since STOF was removed, the EL-REAL now requires a HODO 3/4 and it was determined that this reference time (EL-REAL) was no longer
needed so it was removed from the reference time definition as well. Therefore, as of the current run period (Spring 2020), for the A1n/d2n experiment, the reference time
is defined to be: $T^{\mathrm{logic}}_{\mathrm{reftime}} \equiv \textbf{p(h)HODO 3/4}$, which is the lowest level pre-trigger required to form all other pre-triggers.  \\
\indent Figure \ref{fig:shms_ref_times} shows the reference time histograms with the set reference time cuts for the 80 MeV/c setting of this experiment.
The same reference time cuts used for the higher missing momentum setting (580/750 MeV/c) as there should not be a significant
shift in the reference time signals provided that there is no change in the hardware (signal cable lengths, threhsolds, etc.)
or the DAQ ROCs readout window, which can only be modified by the experts and should not change during the course of an experiment.
See Ref.\cite{cyero_HC_Analysis_steps} for a detailed list of the reference times as well as instructions on how to set the reference time cuts
in the analysis.
\begin{figure}[H]
  \centering
  \includegraphics[scale=0.36]{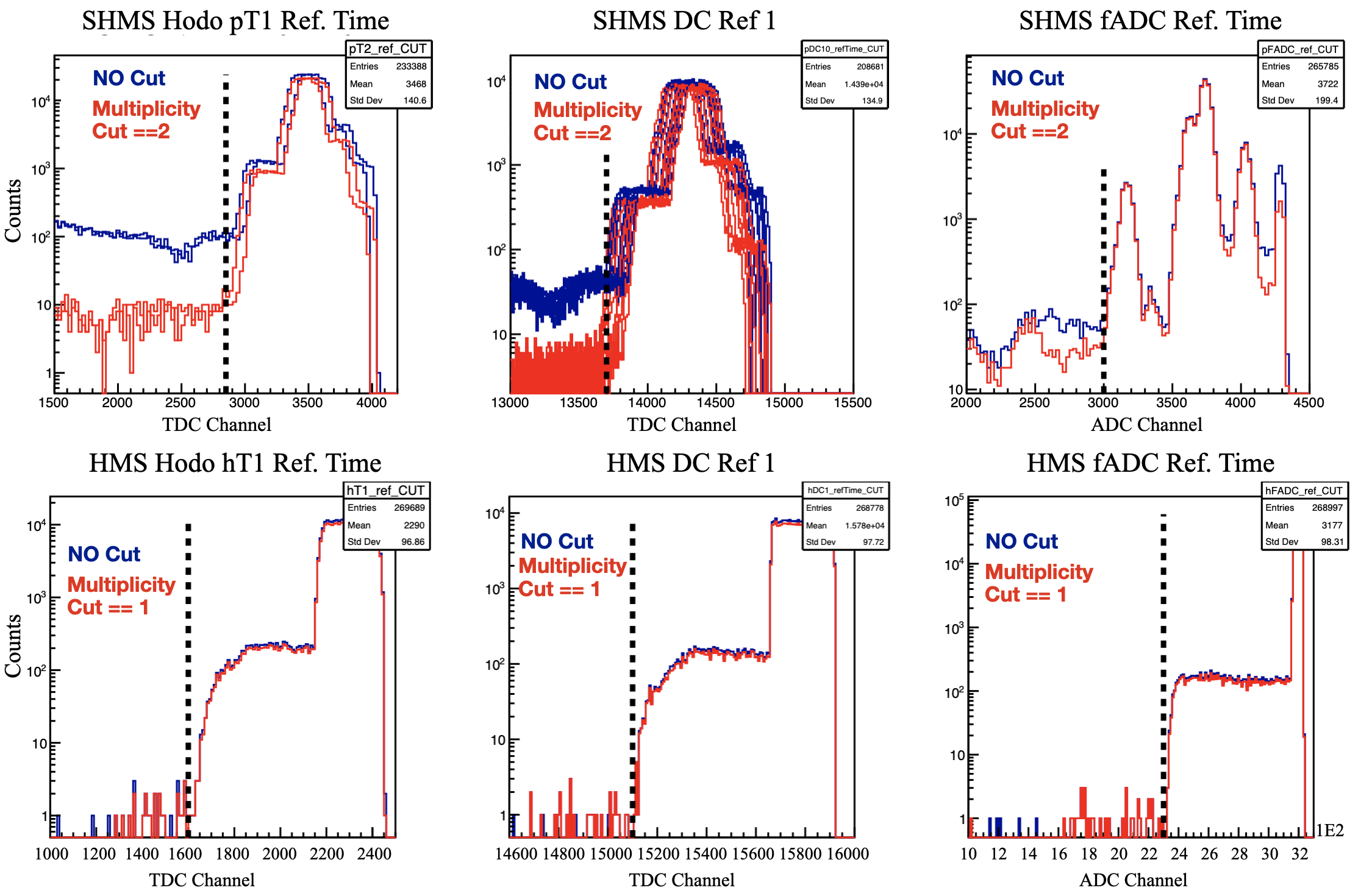}
  \caption{SHMS (top panel) and HMS (bottom panel) reference time cuts for coincidence run 3289 of the E12-10-003 experiment. The conversion from TDC channel to time is $\sim$ 0.1 ns/Ch. The conversion from fADC channel to time is 0.0625 ns/Ch. }
  \label{fig:shms_ref_times}
\end{figure}
\section{Detector Time Window Cuts}
\begin{figure}[H]
  \centering
  \includegraphics[scale=0.34]{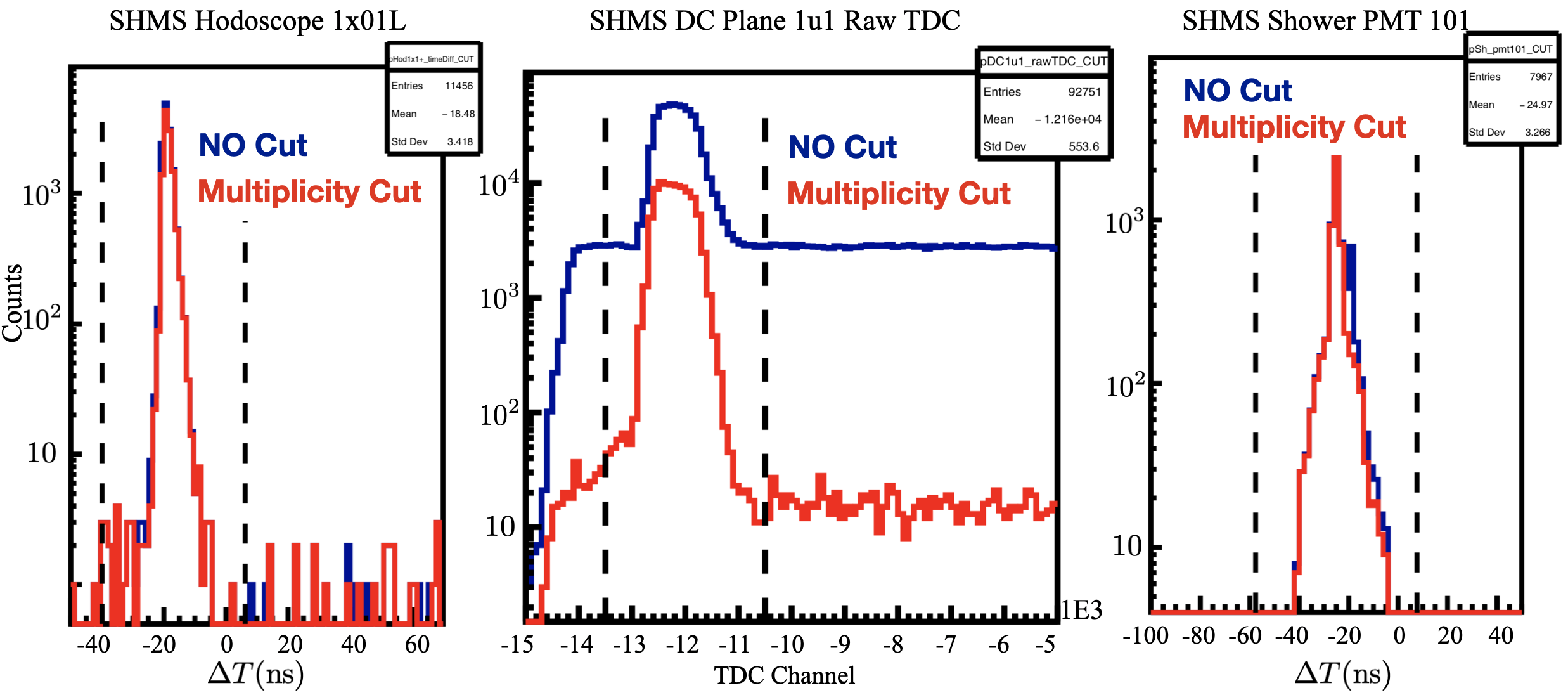}
  \caption{SHMS hodoscope (left), drift chamber (middle) and calorimeter (right) time window cuts for an individual PMT channel (or DC plane).
    The time difference between fADC and TDC pulse times is denoted by $\Delta T$.}
  \label{fig:detec_tWin}
\end{figure}

The next step in the analysis procedure is setting up the detector time window cuts. These are necessary to reduce sources of background that slip into the detector
time windows when detecting the physics signals of interest. The time window cut is made on a time difference between the fADC and TDC times on a PMT basis  for all the detectors
except the drift chamber, which cut on the raw drift times for each plane. The time difference is defined in \textit{hcana} as
\begin{align}
  &\text{AdcTdcDiffTime} = \text{TdcTime[ipmt][jhit] - AdcPulseTime[ipmt][jhit]} \nonumber\\
  &\text{AdcTdcDiffTime} = \text{HodoStartTime - AdcPulseTime[ipmt][jhit]}  \nonumber
\end{align}
where the HodoStartTime is the hodoscope time projected at the focal plane\footnote{\singlespacing The focal plane is an imaginary mid-plane in-between the first and second drift chambers.
  Its nominal origin coincides with the focus of the spectrometer for momentum acceptance $\delta=0 \%$, $Y_{\text{tar}}=0$ cm. (see Fig. \ref{fig:fig3.25}).\label{foot:focalplane}}, and the (TdcTime[ipmt][jhit], AdcPulseTime[ipmt][jhit]) are the TDC and fADC pulse time, respectively,
for any given i$^{\text{th}}$ PMT and the j$^{\text{th}}$ hit within the corresponding i$^{\text{th}}$ PMT fADC/TDC look-back time window. The pulse times are timing signals corresponding to a detector output that
have been measured relative to the reference time and are therefore considered corrected pulse times as opposed to the raw detector arbitrary pulse times that are sent to the front-end
of the module. If the event is truly a physics event originating from the target, then in principle, the time difference should be a $\delta$-function, however, due to the finite
detector/module timing resolution, it has a finite width and gaussian shape. Events that are far away from the main peak are clearly out-of-time indicating that the fADC pulse time and
TDC time are NOT correlated with the same event, and a time window cut must be made. With respect to the drift chambers, a cut on the raw drift time spectrum is made to reduce the background
from multiple TDC hits.\\
\indent Figure \ref{fig:detec_tWin} shows typical examples of the detector time window cuts on an SHMS hodoscope plane,
a drift chamber plane and calorimeter block. The plots show distributions with (red) and without (blue) a 3-hit multiplicity cut.
A narrow peak is clearly distinguishable in all plots with the dashed lines representing the time window cut region. The detector time window cuts were determined for every
PMT channel (or DC plane) of all detectors that were used in the analysis of E12-10-003. See Ref.\cite{cyero_HC_Analysis_steps} for a detailed list of the detector time
windows as well as instructions on how to set the detector time window cuts in the analysis.
\section{Detector Calibrations}
After selecting the right reference times and setting proper detector time window cuts, detector calibrations can be started. Ideally, one would
use specific runs for calibrations in which most of the focal plane (refer to footnote \ref{foot:focalplane}) is illuminated. Sometimes, a magnet de-focused run is used, however, one has to be
careful as some calibrations actually depend on reconstructed quantities at the target, and hence, knowledge of the reconstruction
optics elements. In this case, it is recommended to use single-arm runs over coincidence runs, as the former will be less constrained and occupy a larger
region of the focal plane. 
\subsection{Hodoscopes}
\begin{figure}[H]
  \centering
  \includegraphics[scale=0.41]{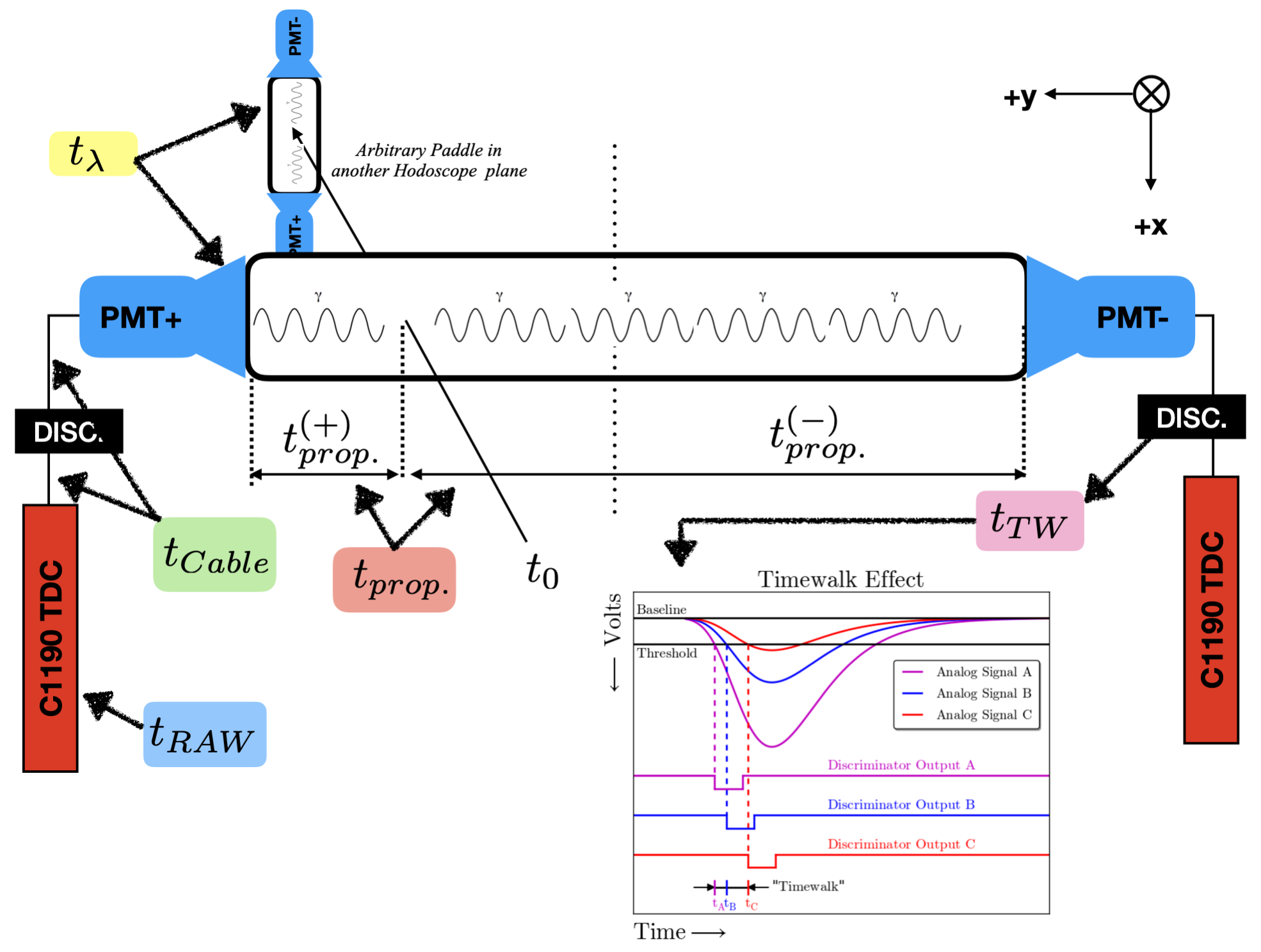}
  \caption{Cartoon of individual scintillator paddles to illustrate the various timing corrections applied. Note: Timewalk Effect illustration reprinted from Ref.\cite{pooser_phdthesis}.}
  \label{fig:Paddle}
\end{figure}

When a particle traverses a hodoscope plane (see Fig. \ref{fig:Paddle}), depending on the trajectory, any paddle (or quartz bar) could in principle be hit.\\
\noindent At this stage, the raw TDC signal has multiple unwanted timing offsets that must be subtracted to obtain the true arrival time of the particle at the hodoscope plane.
The corrected TDC time is then used to determine the correct particle velocity, $\beta = v/c$. The general expression for the corrected TDC time for a hodoscope PMT can
be expressed as:
\begin{equation}
t_{\mathrm{Corr}} = t_{\mathrm{RAW}} - t_{\mathrm{TW}} - t_{\mathrm{Cable}} - t_{\mathrm{prop.}} - t_{\lambda}, 
\end{equation}
where the corrected TDC time represents the particle arrival time at the scintillator paddle (or quartz bar).
The corrections are summarized as follows: 
\begin{itemize}
\item \textbf{Time-Walk Corrections}, $t_{\mathrm{TW}}$: For analog signals arriving at the \textit{Leading Edge Discriminators},
  the logic signal is produced when the signal crosses the discriminator threshold and therefore depends on the signal amplitude (see Fig. \ref{fig:Paddle}).
  The fADCs do not have this disadvantage since they correct for 
  \begin{figure}[H]
    \centering
    \includegraphics[scale=0.52]{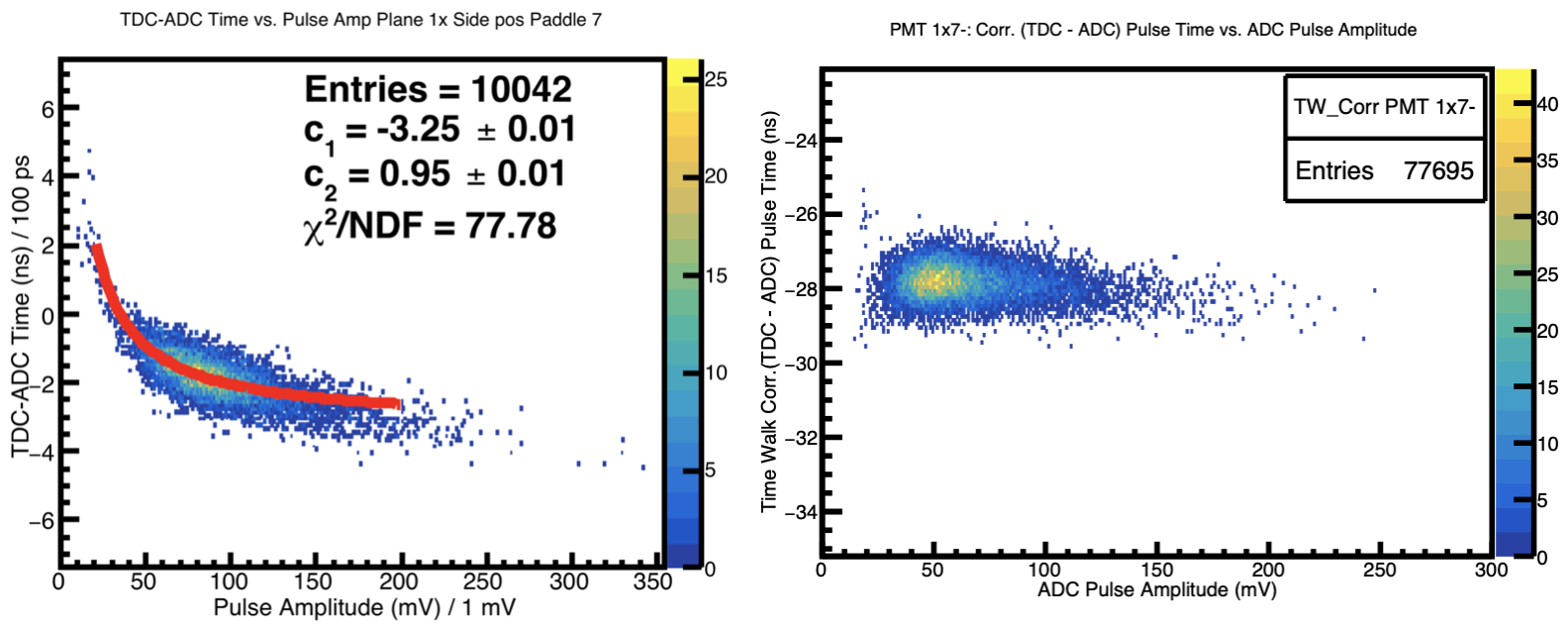}
    \caption{Fit correlation between TDC pulse time and fADC pulse amplitude (left). Time-walk corrected pulse time versus fADC pulse amplitude shows no correlation (right).}
    \label{fig:TWFit}
  \end{figure}
  time-walk internally, and as a result, the fADC pulse time is not correlated with the signal amplitude. The algorithm used by the fADCs
  to effectively remove time-walk effects is similar to that of a constant fraction discriminator (CFD) timing algorithm.
  In the CFD algorithm, the logic signal is generated at a constant fraction of the signal peak height which makes the discrimination
  of the signal independent of the pulse amplitude as illustrated in Fig. 4.7 of Ref.\cite{pooser_phdthesis} or Fig. 7.4 of Ref.\cite{Tech_NuclPhy_Leo}.
  To correct for the TDC time walk, the fADC pulse time is used as a reference by taking the TDC-ADC pulse time difference
  plotted against the fADC amplitude. A model function is fit to this correlation, and the parameters extracted are used to correct the TDC time (see Fig. \ref{fig:TWFit}).
\item \textbf{Cable Time Corrections}, $t_{\mathrm{Cable}}$: This correction takes into account the fact that the analog signal has
  to propagate across signal cables from the PMT all the way into the Counting House electronics rack into the TDC.
  To determine this correction, a correlation between time-walk corrected time and hodoscope paddle track position is fit
  to extract the velocity of propagation across the paddle, and the cable time offset. The propagation velocity is determined from
  \begin{figure}[H]
    \centering
    \includegraphics[scale=0.57]{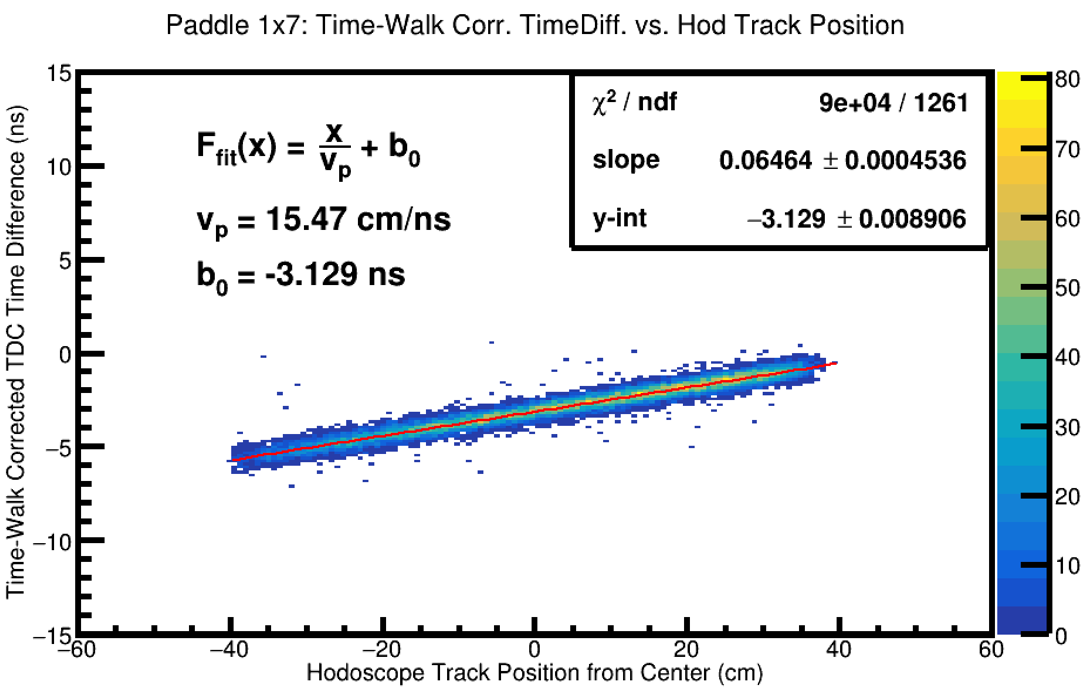}
    \caption{Fit correlation between track position along paddle and time-walk corrected (TDC-ADC) time
      difference used to determined the propagation velocity across the paddle.}
    \label{fig:Vp_fit}
  \end{figure}
  the distance and time of the hit from the center of a paddle. The time is determined by taking half of the time-walk
  corrected TDC time difference between the two ends of a paddle. The \textit{half} is to ensure that if the particle hits the edge
  of the paddle, half of the total propagation time across the entire paddle is taken to obtain the time from the edge to the center.
  The hit distance is determined by extrapolating the distance determined by the drift chambers from tracking. The correlation between
  time and distance is fit to extract the propagation velocity and the cable time difference between the two ends (see Fig. \ref{fig:Vp_fit}).
  The cable time offset parameter is determined for all paddles and the parameter is read by hcana, and added as a correction
  factor to the time-walk corrected TDC time.
\item \textbf{Hodoscope Plane Time Difference Corrections}, $t_{\lambda}$: This correction accounts for any additional time difference (other than the particle propagation
  time to travel across the two paddles) between any two distinct scintillator paddles in different hodoscope planes.
  \begin{figure}[H]
    \centering
    \includegraphics[scale=0.59]{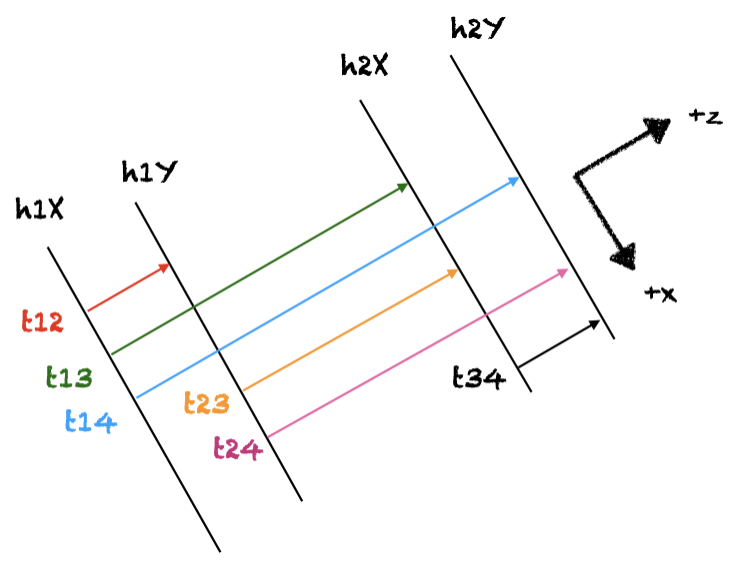}
    \caption{Illustration of all possible time difference combinations that are considered in this correction.}
    \label{fig:hod_planes}
  \end{figure}
  Six possible combinations between
  the four hodoscope planes are considered when correcting for the time difference between any two of their paddles (see Fig. \ref{fig:hod_planes}).
  The combinations of all six possible time differences were expressed as a system of 6 linear equations that were
  solved using the method of Single-Value Decomposition (SVD) to determine the calibration coefficients for each individual PMT.
\end{itemize}
\indent Figure \ref{fig:hodBeta_pm580} shows the representative plots of the calibration results for the 580 MeV/c setting of the E12-10-003 experiment.
The $\beta$ distribution shown uses the tracking information from the drift chambers, whereas the dashed lines use the formula: $\beta = P_{\mathrm{c}}/\sqrt{m^{2} + P_{\mathrm{c}}^{2}}$,
where $P_{\mathrm{c}}$ and $m$ represent the spectrometer central momentum and particle mass, respectively.\\
\begin{figure}[H]
  \centering
  \includegraphics[scale=0.33]{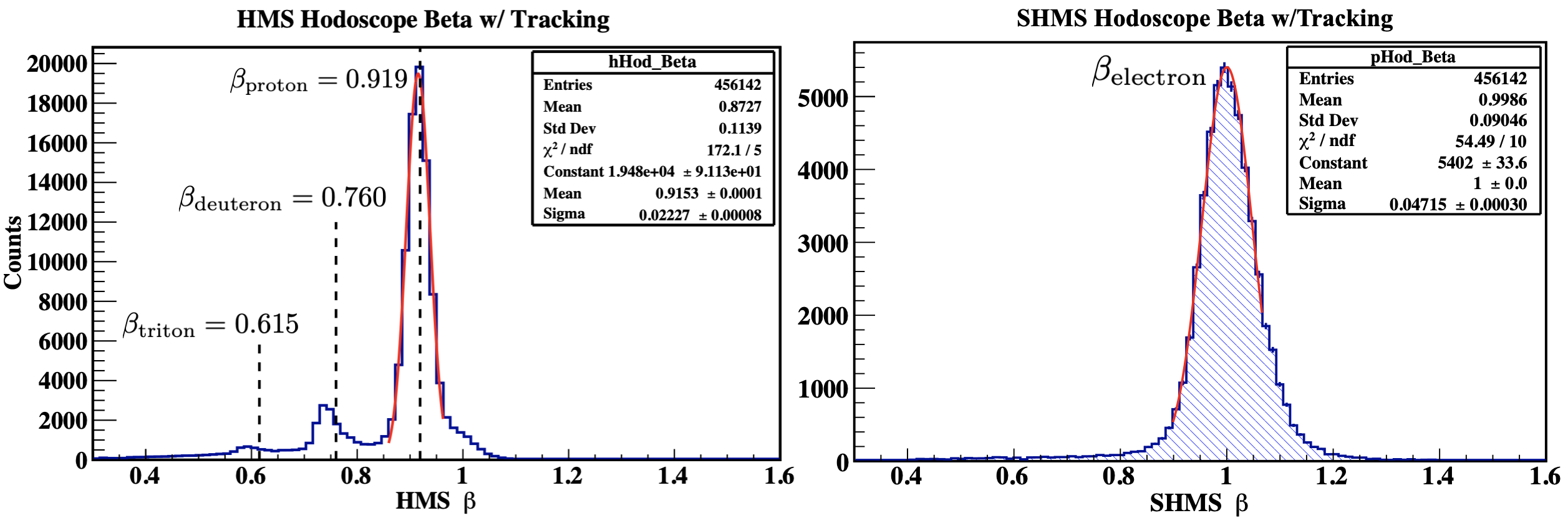}
  \caption{HMS/SHMS hodoscope calibration results for the 580 MeV/c setting of E12-10-003. The histograms are plotted using the drift chamber tracking information to determine $\beta$.}
  \label{fig:hodBeta_pm580}
\end{figure}
\indent The HMS hodoscope $\beta$ distribution shows three distinctive peaks formed from the coincidence with the electrons in the SHMS. The $\beta$ peak corresponding to
the protons are from quasi-elastic scattering off the liquid deuterium target, while the deuteron and triton are produced from knockout reactions, most likely, quasi-elastic
electron scattering off the aluminum walls. The dashed lines were determined based the assumption that the particle mass is either that of a proton ($^{1}$H), deuteron ($^{2}$H) or triton ($^{3}$H) and
momentum acceptance, $\delta = 0 \%$. In the SHMS, the peak at exactly $\beta=1$ clearly represents the electron as its mass is negligible compared to the energy and momentum. See Ref.\cite{cyero_HodoCalib} for a detailed explanation
of the hodoscope calibration procedure.
\subsection{Drift Chambers} \label{sec:DC_Calibration}
When a charged particle traverses the drift chambers, it passes through 12
wire planes, each surrounded by two cathode planes. The wire planes consist of
alternating field and sense wires. The field wires and the cathode planes are kept
at a negative voltage while the sense wires are kept grounded (0 potential). The
potential gradient creates an electric field oriented outwards from the sense wires.
\begin{figure}[H]
  \centering
  \includegraphics[scale=0.35]{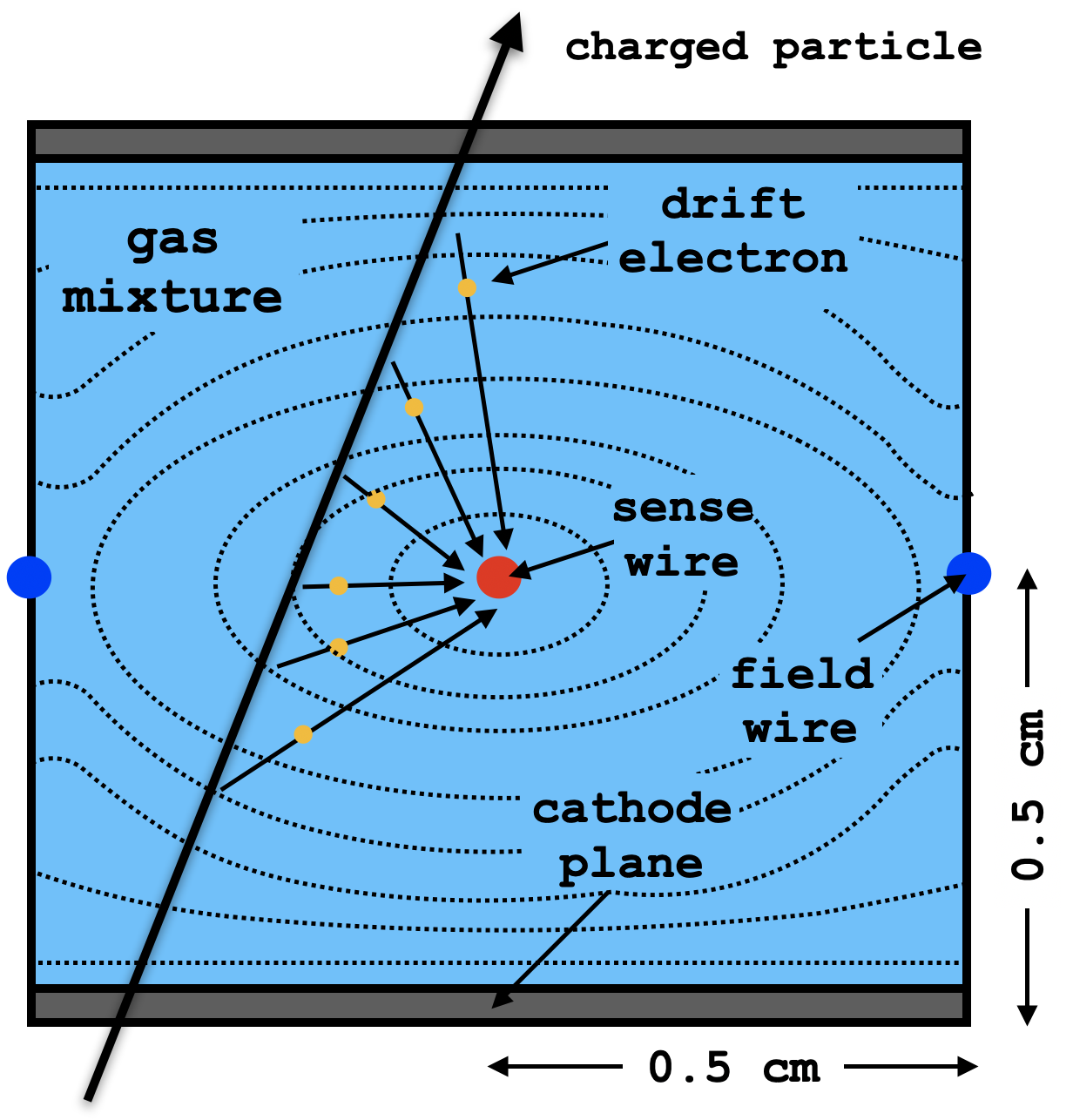}
  \caption{Illustration of a single drift cell (top view) in a drift chamber. The dashed lines represent equipotential
    surfaces where the electric field is perpendicular to the contour. Figure
    adaptation from G. Niculescu.}
  \label{fig:drift_cell}
  \end{figure}
\indent As the charged particle passes through a single drift cell, it ionizes the
gas atoms in the chamber gas mixture, which causes the free electrons from the ionized gas
to drift towards the sense wire producing a measurable current signal. The sense wire
signals are pre-amplified and read out by 16-channel input discriminators which produce
logic signals that are sent to the TDC via 16-channel ribbon cables. The TDC registers
the time when the signal is registered. This time contains the cable delay it would
have taken the signal to propagate across the sense wire, through the ribbon cable and into
the TDC if the particle would have passed through the sense wire itself. The \textit{drift time}
is the time it takes the free electrons to drift towards the sense wire and can be expressed as,
\begin{equation}
  t_{\mathrm{D}} = (t_{\mathrm{meas}} - t_{\mathrm{REF}}) - [\underbrace{(t_{\mathrm{wire}} + t_{\mathrm{cable}})}_{t'} - t_{\mathrm{REF}}]
  \label{eq:4.7}
\end{equation}
where $t_{\mathrm{meas}}$ is the measured time by the TDC, and $t'$ is the time it takes the signal
to propagate across the sense wire, through the cable and into the TDC if the track were to pass
directly through the sense wire. These times are measured relative to a reference time, $t_{\mathrm{REF}}$, used by the
TDC as a common stop. \\
\indent A coarse reconstruction of the track can be carried out with only the knowledge of the wires that fire from a
physics event. Knowing the associated drift times of the wires that were hit, however, allows for a more precise track
reconstruction, as the drift time can be converted to a drift distance from the determination of the time-to-distance maps
resulting from the drift chamber calibration as described below.\\
\indent For a collection of events illuminating all cells in any given wire plane, one obtains a drift time distribution for each sense
wire that can be averaged over an entire group (up to 16 wires in a discriminator card) or over the entire plane to form a
drift time distribution per plane (see Fig. \ref{fig:hdc1x1_time}).\\
\indent Associated with each drift-time spectrum is a quantity called ``$t_{0}$''. The $t_{0}$ corresponds to the location
in the histogram where the ionized particle comes in contact with the wire. If its value is anything other than zero nanoseconds
(0 ns), it is interpreted as the value by which the drift time must be shifted in order to assure that $t_{0}=0$ ns. All
subsequent times in each drift time spectra are measured relative to this time.\\
\begin{figure}[H]
  \centering
  \includegraphics[scale=0.32]{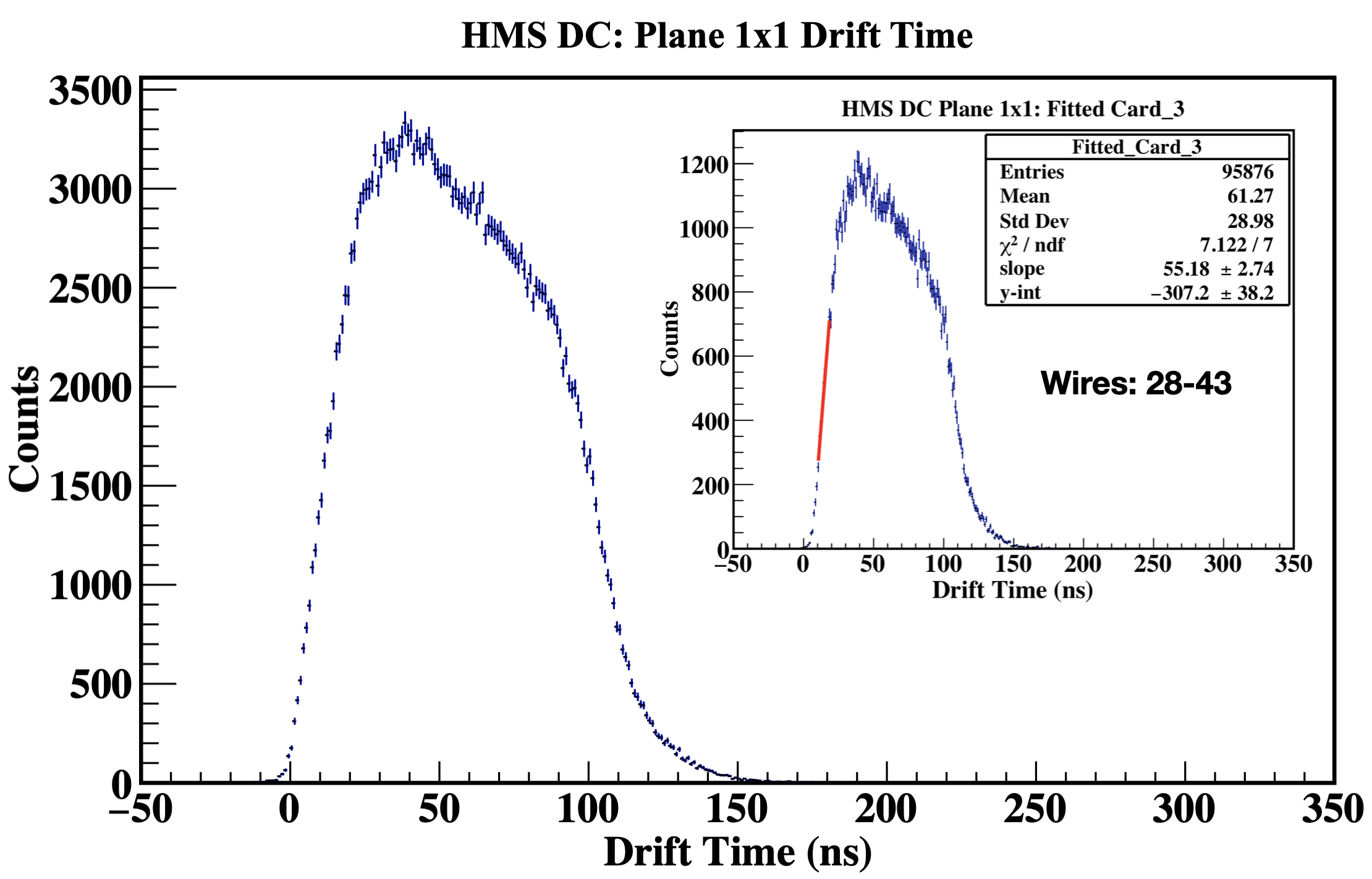}
  \caption{HMS drift time spectrum for plane 1x1. Inset: Fit of the leading edge in a drift time spectrum corresponding to a group of wires from a specific discriminator card of plane 1x1.}
  \label{fig:hdc1x1_time}
\end{figure}
\begin{figure}[H]
  \centering
  \includegraphics[scale=0.32]{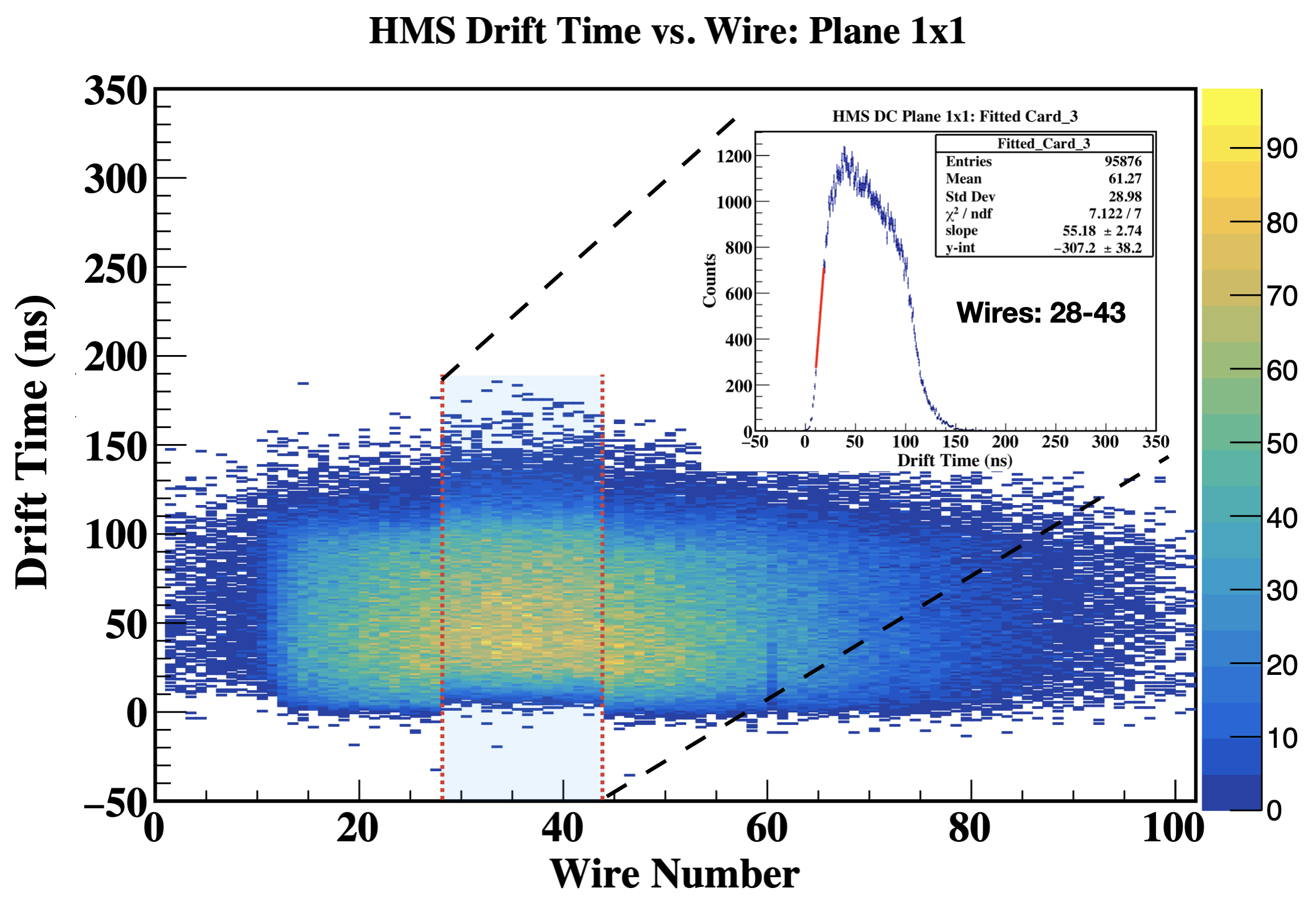}
  \caption{HMS drift times versus wire number for plane 1x1 before $t_{0}$ correction. Inset: Same as in Fig. \ref{fig:hdc1x1_time}.}
  \label{fig:2Dhdc1x1_time_uncorr}
\end{figure}
\begin{figure}[H]
  \centering
  \includegraphics[scale=0.32]{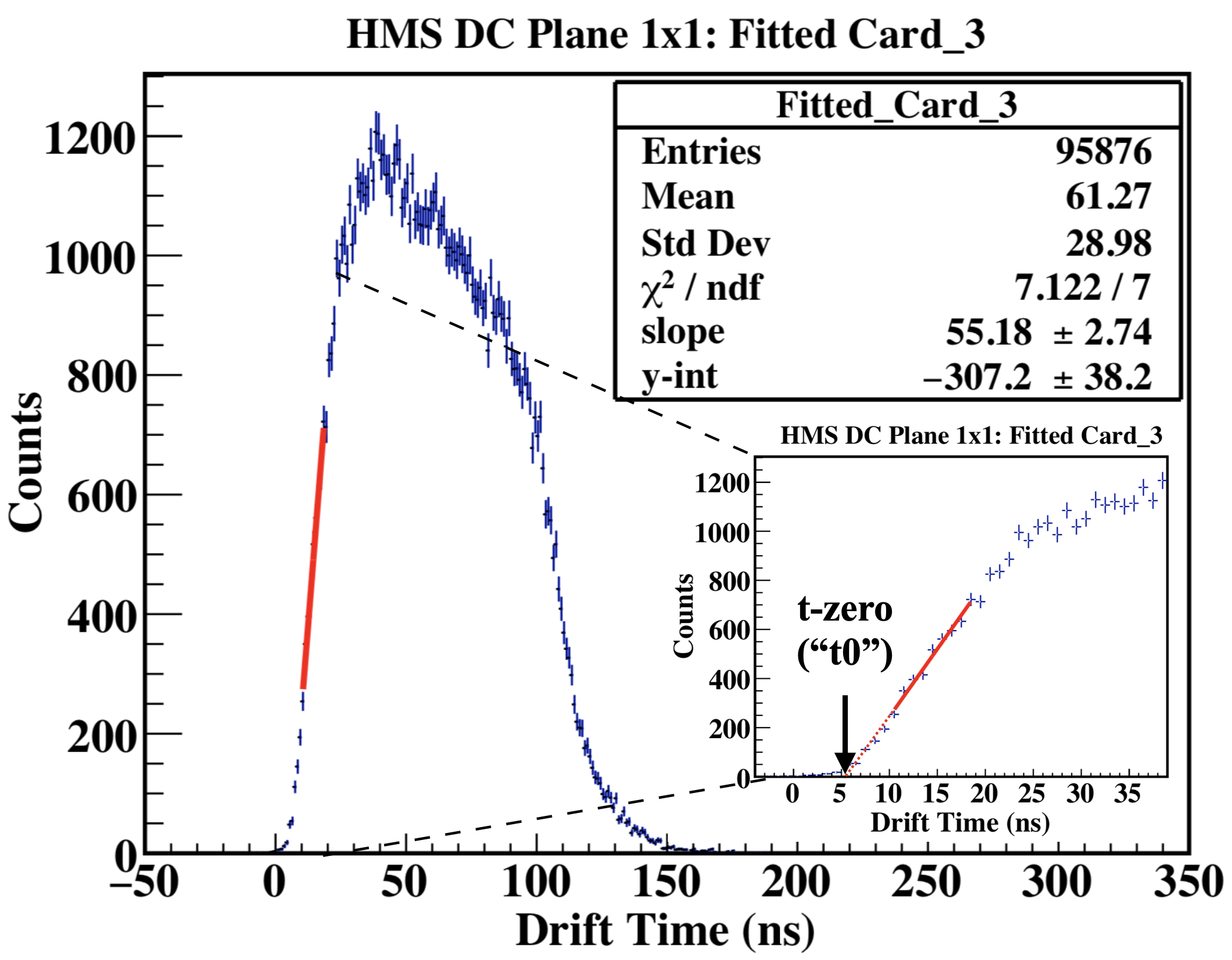}
  \caption{Fit of the leading edge in an HMS drift time spectrum for the wire card $\#$3 of plane 1x1. Inset: Close-up of fit region indicating the $t_{0}$.}
  \label{fig:FIT_hdc1x1_time}
\end{figure}
\begin{figure}[H]
  \centering
  \includegraphics[scale=0.32]{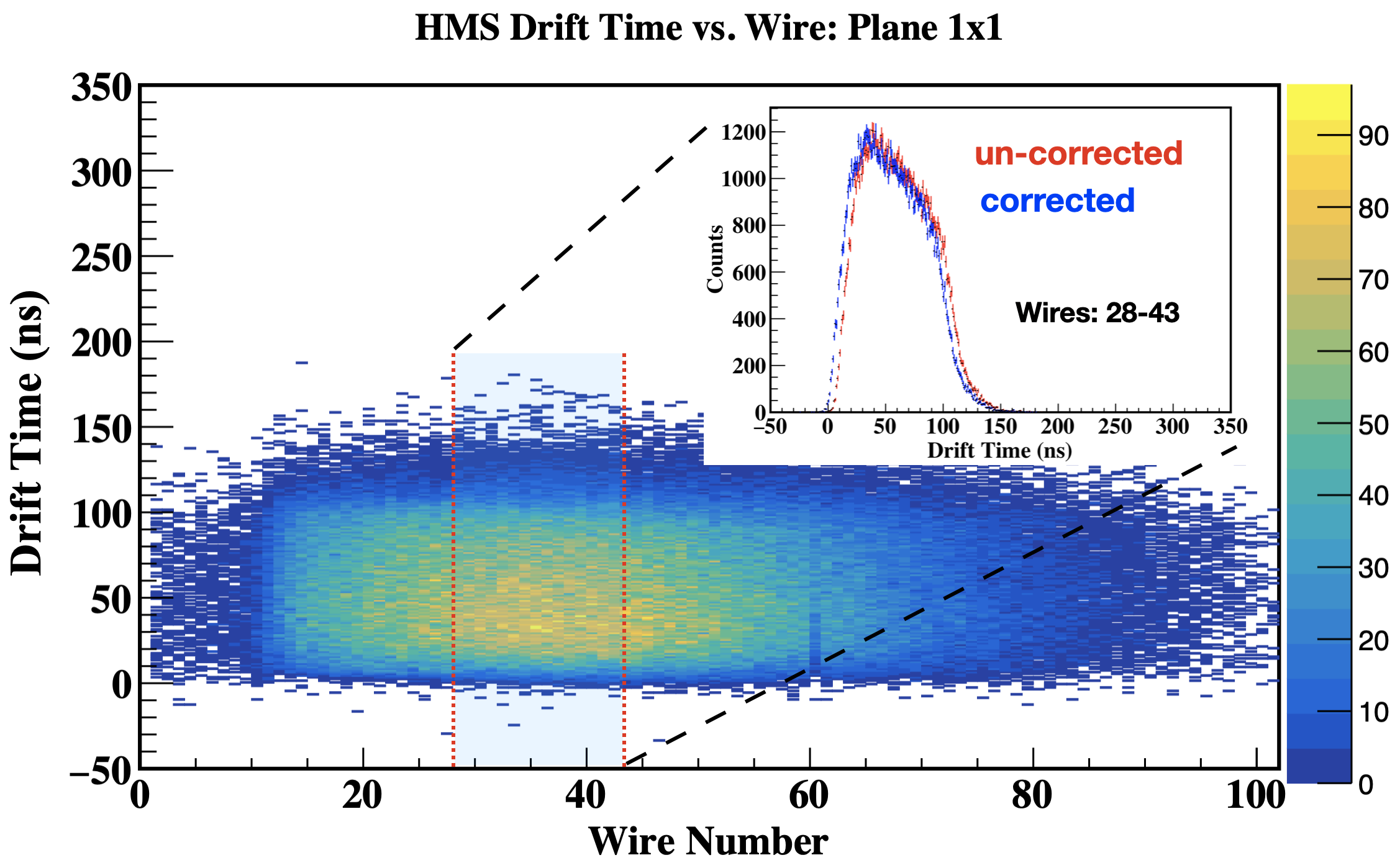}
  \caption{HMS drift times versus wire number for plane 1x1 after $t_{0}$ correction.
    Inset: Drift time for wire group $\#$3 of plane 1x1 before (red) and after (blue) $t_{0}$ correction.}
  \label{fig:2D_hdc1x1_time_corr}
\end{figure}
\indent The $t_{0}$ for each plane is determined by calculating the $t_{0}$ for individual wires in each plane and taking a weighted
average. The $t_{0}$ for individual sense wires (or wire group) is determined by a linear fit of the drift time spectra at
around 20$\%$ of the peak $\pm \Delta t$ for each sense wire (or wire group), where $\Delta t$ is the fit range.
The linear fit is then extrapolated to the horizontal axis (drift time), and this extrapolated value is defined as $t_{0}$ (see Fig. \ref{fig:FIT_hdc1x1_time}).\\
\indent Depending on the calibration method used, the $t_{0}$ correction is applied on a sense wire basis from the individual wire fits,
or on a wire group basis where the same $t_{0}$ correction is applied to all wires of a group. The latter procedure is usually better since
the edge wires have very low statistics that cause the fit to fail, whereas a group of wires will have sufficient statistics for a successful fit. \\
\indent To determine the drift distances from the drift time spectra, it is assumed that the drift distances are uniformly distributed across
the cell. This assumption is based on the fact that a cell is uniformly illuminated with particles, and the ions have an
approximately uniform drift velocity, which implies there should be no preferred drift distance for any ionized charge. Mathematically,
the drift distance is calculated as
\begin{equation}
  d_{\mathrm{drift}}(\tau=T) = \frac{\Delta}{2}\frac{\int^{T\leq t_{\mathrm{max}}}_{t_{0}}F(\tau)d\tau} {\int^{t_{\mathrm{max}}}_{t_{0}}F(\tau)d\tau},
  \label{eq:4.8}
\end{equation}
where $\Delta$ is the cell width and $F(\tau)$ is the drift time distribution integrated from $t_{0}$=0 ns to some arbitrary time $T\leq t_{\mathrm{max}}$
where $t_{\mathrm{max}}$ is the maximum drift time within a cell. In the limiting case of Eq. \ref{eq:4.8}, the drift distance becomes
\begin{equation}
  d_{\mathrm{drift}} = \begin{cases}
    0 \text{ cm}, & \tau = 0 \text{ ns}\\
    0.5 \text{ cm}, & \tau = t_{\mathrm{max}},
  \end{cases}
  \label{eq:4.9}
\end{equation}
which is the expected drift distance at the sense wire ($\tau = 0$ ns) and at the edges of the cell ($\tau=t_{\mathrm{max}}$).\\
\indent Due to the finite resolution of the TDC and other factors involved, the drift times are not determined to infinite precision
and the integral in Eq. \ref{eq:4.8} becomes a sum over a finite bin width,
\begin{equation}
  \int_{\tau}F(\tau)d\tau  \rightarrow \sum_{\mathrm{bin}(\tau)} \underbrace{F(\tau)}_{\text{bin content}} \cdot \underbrace{\Delta \tau}_{\text{bin width}}.
  \label{eq:4.10}
\end{equation}
Re-writing Eq. \ref{eq:4.8} in terms of the finite sums in Eq. \ref{eq:4.10}, one obtains
\begin{equation}
  \frac{\sum\limits^{\text{bin}(t_{0}+T)}_{\text{bin}(t_{0})} F(\tau)\Delta\tau}{\sum\limits^{\text{bin}(t_{0}+t_{\mathrm{max}})}_{\text{bin}(t_{0})} F(\tau)\Delta\tau} \rightarrow \boxed{\frac{1}{N_{\mathrm{tot}}}\sum\limits^{\mathrm{bin}(t_{0}+T)}_{\mathrm{bin}(t_{0})}F(\tau)}
  \label{eq:4.11}
\end{equation}
The ratio in Eq. \ref{eq:4.11} is the lookup value used to convert drift time to distance for an arbitrary drift time bin, $T$.
The numerator represents the sum of all bin contents up to a drift time $T$, and the denominator represents the sum over the
bin contents of all drift times up to $t_{\mathrm{max}}$, in a given plane. The bin width, $\Delta \tau$, is a constant during the sum,
therefore it is cancelled, which simplifies the equation as a ratio of the sum of bin contents (up to some drift time) and the sum over
all bin contents (up to a maximum, $t_{\mathrm{max}}$), $N_{\mathrm{tot}}$.\\
\indent The results of this calibration are per-plane look-up tables that map any given drift time to a drift distance in that plane.
The drift distance for the X-plane of HMS drift chamber 1 is shown in Fig. \ref{fig:hdc1x1_driftDist}. As expected, the drift distances for all planes are
uniformly distributed across the cell width.\\
\begin{figure}
  \centering
  \includegraphics[scale=0.35]{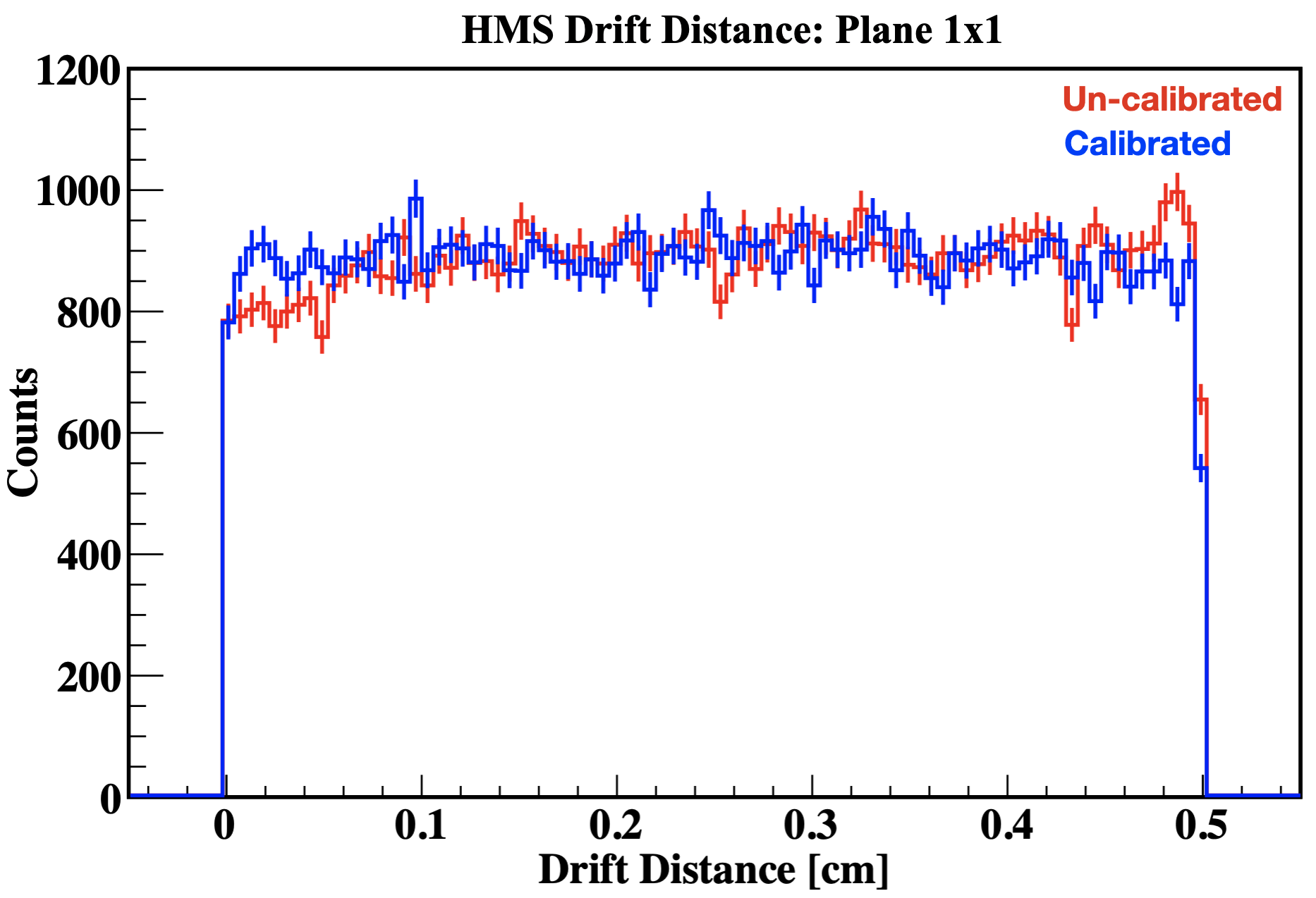}
  \caption{HMS drift distance for plane 1x1 before (red) and after (blue) calibration of the drift maps.}
  \label{fig:hdc1x1_driftDist}
\end{figure}
\begin{figure}
  \centering
  \includegraphics[scale=0.35]{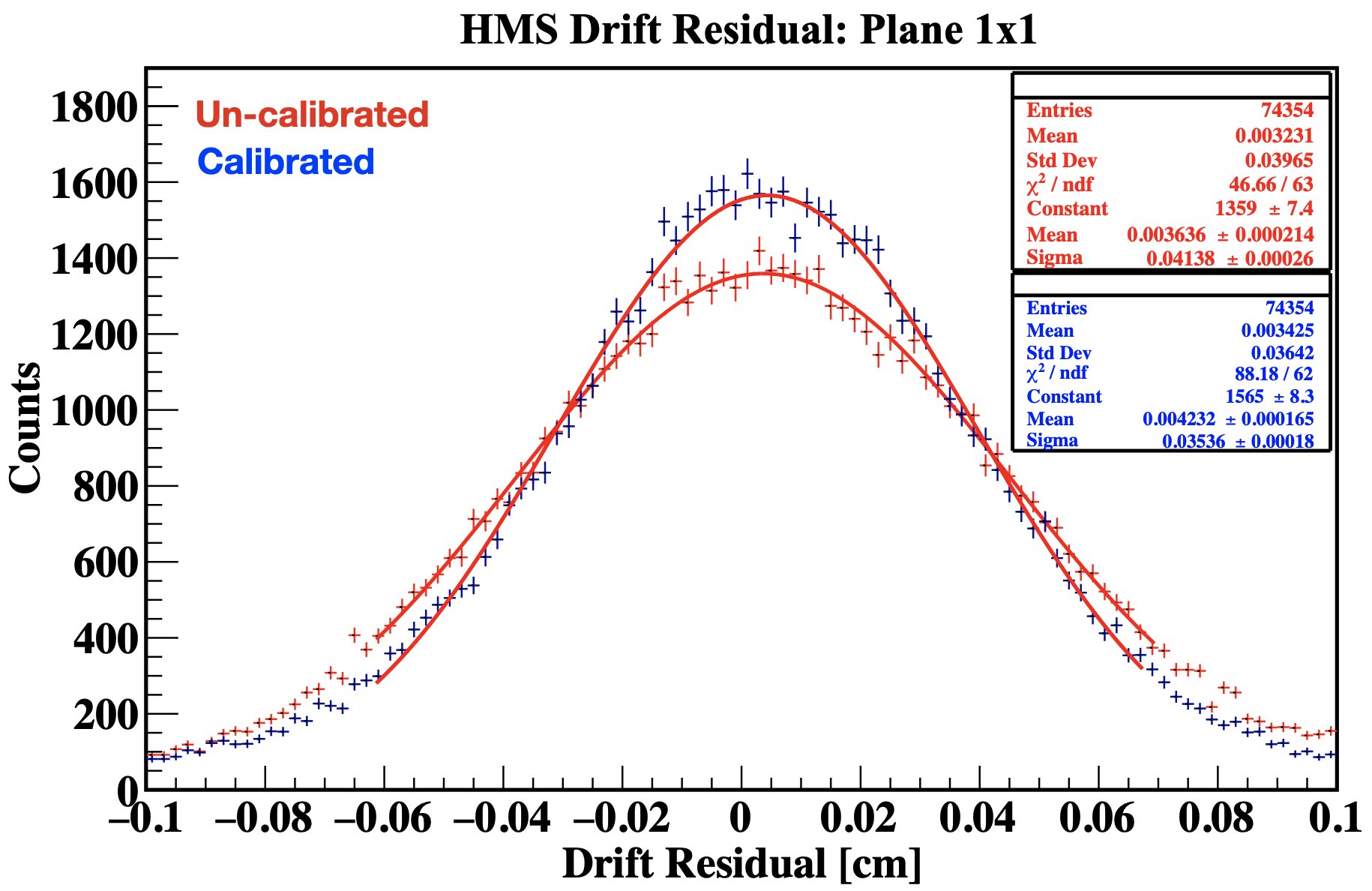}
  \caption{HMS fit drift residuals for plane 1x1 before (red) and after (blue) calibration of the drift maps. The
    standard deviation ($\sigma$) from the fit is representative of the spatial resolution.}
  \label{fig:hdc1x1_driftRes}
\end{figure}
\indent The best way to determine the drift chamber performance is by measuring the spatial resolution, or how well it can
measure the position of particle tracks. This measurement is done through the determination of per-plane residuals. For a particle
traversing at least 4 planes of the chamber, a collection of space-points (X,Y) is measured based on the wires that fired. The space
points are fit with a straight line such as to minimize the chi-square, and obtain a best fit. The line fit is then compared to the
measured space-point from the plane wires that fired, and the difference is called the residual for that plane. The residuals are
calculated on an event by event basis, and should be centered around zero (see Fig. \ref{fig:hdc1x1_driftRes}). \\ 
\indent For the E12-10-003 experiment, typical residuals per plane were found to be on average $\sim$250 $\mu$m for the SHMS and $\sim$350 $\mu$m
for the HMS drift chambers. See Ref.\cite{cyero_DCCalib_steps} for details on the drift chambers calibration procedure.

\subsection{Calorimeters}
\begin{figure}[H]
  \centering
  \includegraphics[scale=0.30]{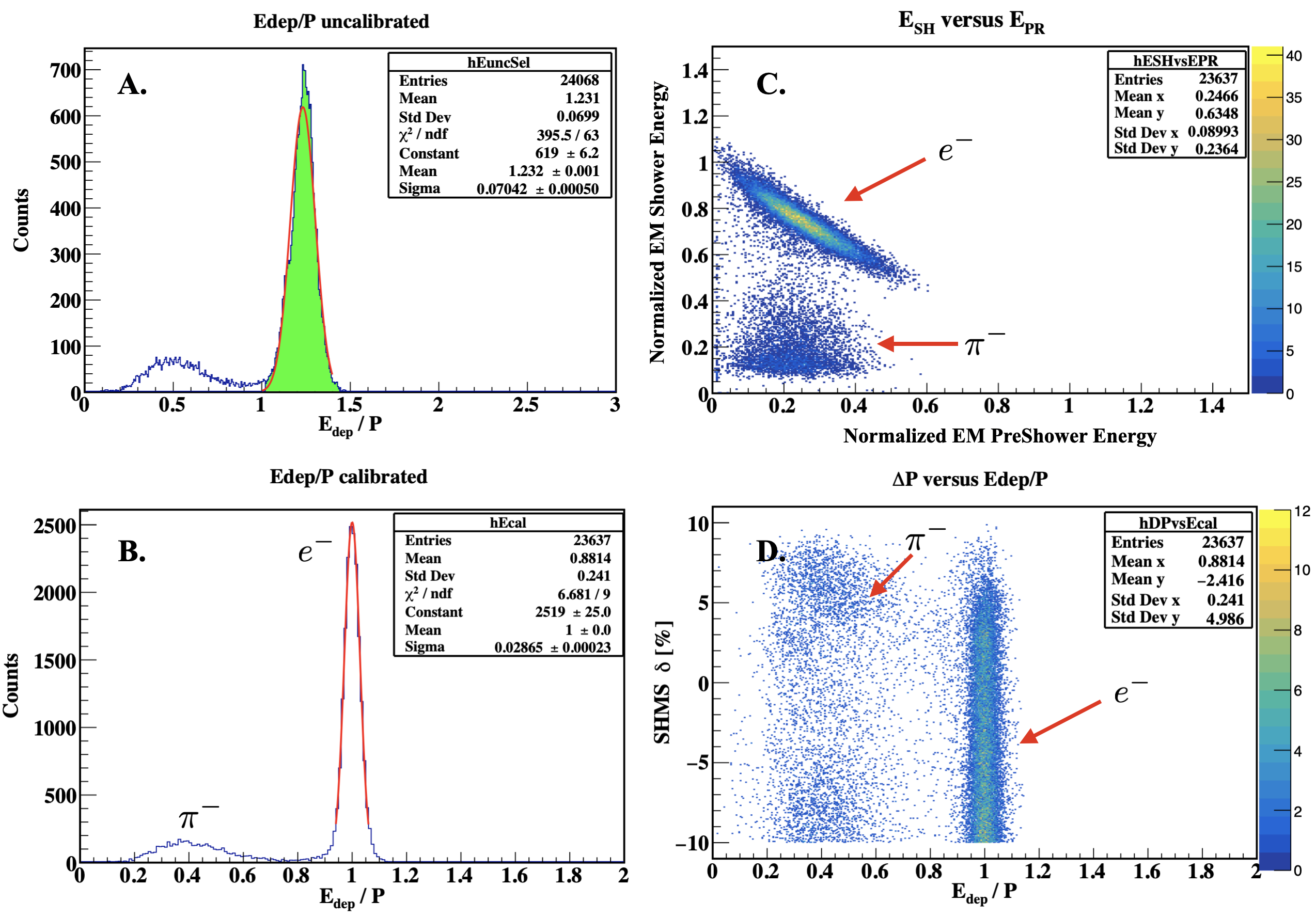}
  \caption{SHMS calorimeter calibration plots for the 580 MeV/c setting of E12-10-003. (A) is before calibration and (B), (C) and (D) are after calibration.}
  \label{fig:pm580_shmsCal}
\end{figure}

The calorimeter in each spectrometer is used primarily for particle identification based on the incident particle track momentum
and the subsequent energy showers detected by the PMTs coupled at the ends of the lead glass blocks. The signals are sent to fADCs
where the signal amplitude is proportional to the fADC channel, which is subsequently converted to the corresponding energy deposited at the
PMT. In order to make the calorimeter trigger efficiency uniform across the calorimeter plane, the output signals were matched by
adjusting the PMTs High Voltage (HV) to make the signal amplitudes as similar as possible\cite{Mkrtchyan_2013}. This resulted in the
PMTs gain being different across the vertical or dispersive direction since particles with higher momentum (lower bend angles)
impact the lower calorimeter blocks and deposit more energy (larger signal amplitude), whereas less energetic particles impact higher
blocks and deposit less energy in the calorimeter. This results in a gain variation across the calorimeter plane that is approximately
equal to the spectrometer momentum acceptance.\\
\indent The purpose of the calibration is thus to correct for the gain variations on a PMT basis across all the calorimeter blocks. The
formula used for the deposited energy in the $i^{\mathrm{th}}$ PMT is estimated to be\cite{Mkrtchyan_2013}:
\begin{equation}
  \epsilon_{i} = c_{i} \cdot (A_{i} - A_{\mathrm{ped},i}) \cdot f(y),
\end{equation}
where $c_{i}$ is a calibration constant, $A_{i}$ is the raw fADC signal, $A_{\mathrm{ped},i}$ is the corresponding fADC pedestal\footnote{\singlespacing The
  pedestal is an electronic offset at the input to the digitization stage.}, and $f(y)$ is a correction factor for the light attenuation across the horizontal hut coordinate, $y$.
The standard calibration algorithm minimizes the variance between the total energy deposited in all channels ($E_{\mathrm{DEP}}=\Sigma{e_{i}}$) relative
to the measured momentum of an incident electron at the face of the calorimeter. The algorithm was developed by Ts. Amatuni in the early 1990s.\\
\indent Figure \ref{fig:pm580_shmsCal} shows the representative calibration plots for the SHMS calorimeter using the combined runs corresponding to the 580 MeV/c setting of E12-10-003.
The upper (A) and lower (B) left plots show the total energy deposition divided by the central spectrometer momenta ($E_{\mathrm{dep}}/P$) before and after calibration, respectively.
Since the electron mass is negligible compared to its momentum, the total energy deposited by the electron is approximately equal to its momentum before entering the calorimeter and therefore
its ratio is unity. The upper right plot (C) shows the correlation between the deposited energy by the electron in the PreShower and Shower normalized by the momentum of the best reconstructed
track. The correlation on this plot shows how the PreShower can be used to augment the electron detection capabilities of the SHMS calorimeter. Finally, the lower right plot (D) shows a
correlation between the SHMS momentum acceptance ($\delta$) and the normalized energy deposited in the calorimeter, which demonstrates that the energy deposited by the electron
is uniform across the entire calorimeter dispersive direction. See Ref.\cite{VTad_CaloCalib} for instructions on how to perform the calorimeter calibration. 
\subsection{Gas/Aerogel \v{C}erenkovs}
The calibration procedures for the threshold gas and aerogel \v{C}erenkovs are very similar. The calibration is based on
identifying the location of the single photo-electron (SPE) peaks relative to the pedestal of
the corresponding PMTs. After pedestal subtraction, each SPE peak is fit with a gaussian. The mean of the gaussian
represents the corrected fADC channel corresponding to the SPE peak, from which a conversion factor between the SPE peak and
fADC channel can be obtained per PMT channel. Each detector calibration procedure employs slightly different methods to identify the SPE peaks
as well as different particle identification requirements. See Ref.\cite{RAmbrose_CerCalib} for more details on the gas \v{C}erenkov detetcor calibrations. For
the aerogel detector calibration, refer to the official Kaon-LT experiment page\cite{KaonLT_official}.
\section{Hall C Coordinate System} \label{sec:hall_coord}
Before discussing the spectrometer optics checks and optimization in the next section (see Section \ref{sec:optics_checks}), the coordinate systems
used to reconstruct the events at the target reaction vertex must be introduced. 
\begin{figure}[H]
  \centering
  \includegraphics[scale=0.4]{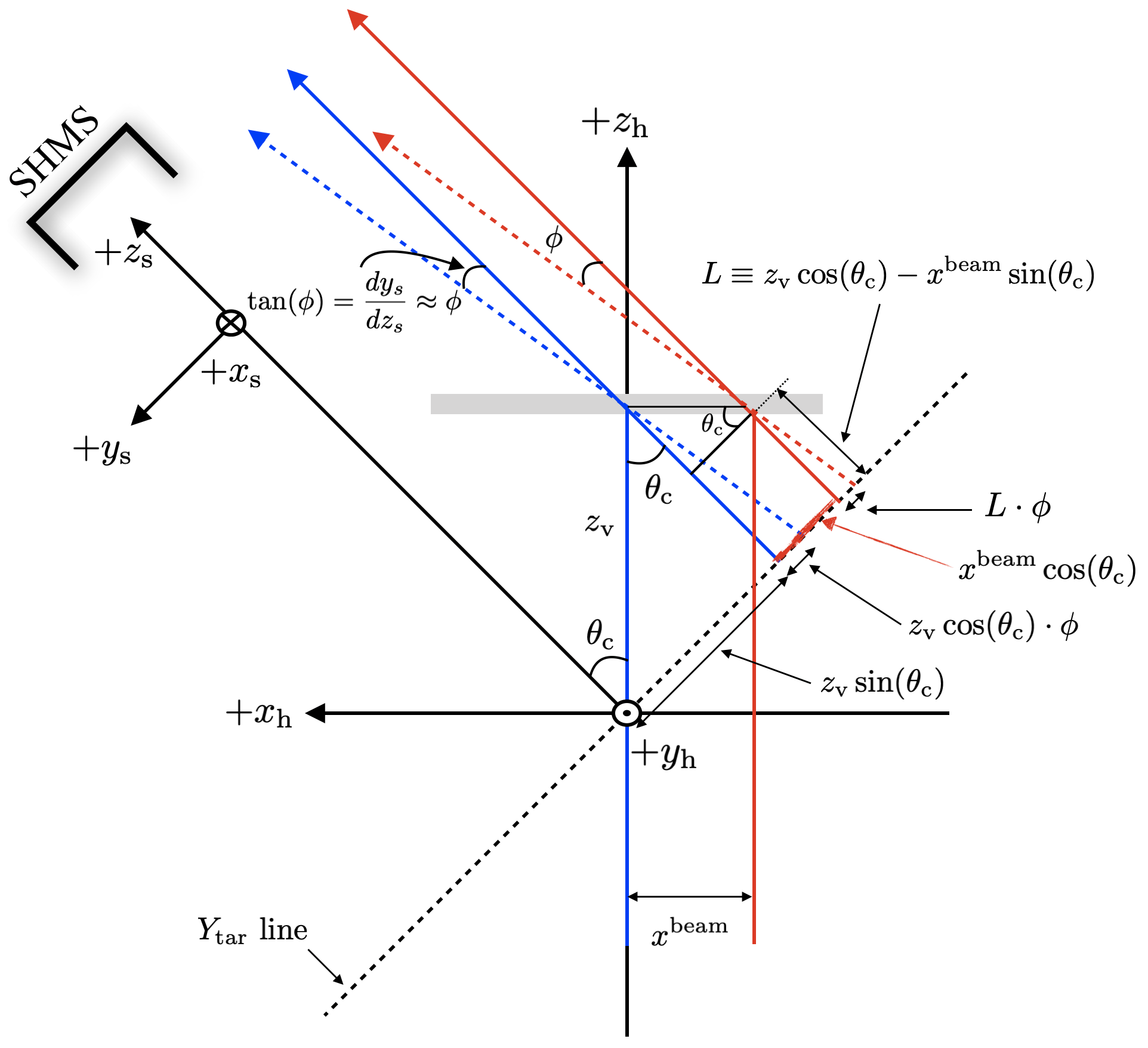}
  \caption{Top view of the spectrometer and hall coordinate systems in Hall C.}
  \label{fig:optics_coord}
\end{figure}
\indent The hall (or vertex) coordinate system is denoted by the subscript ``h'' in Fig. \ref{fig:optics_coord}, where $+z_{\mathrm{h}}$ is parallel to the incident beam direction, $+x_{\mathrm{h}}$ is oriented beam-left
and denotes the horizontal beam position, and $+y_{\mathrm{h}}$ points towards the ceiling and denotes the vertical beam position. These coordinates
are used to describe the beam position at the reaction vertex.\\
\indent The spectrometer coordinate system is denoted by the subscript ``s'' where $+z_{\mathrm{s}}$ is aligned with the spectrometer central ray rotated by the
central spectrometer angle $\theta_{\mathrm{c}}$, $+x_{\mathrm{s}}$ points in the dispersive direction (towards the hall floor), and $+y_{\mathrm{s}}$ is
oriented beam-left in the non-dispersive direction. The detector hut coordinate system is defined by a simple spectrometer coordinate rotation about the
$y_{\mathrm{s}}$-axis to align $+z_{\mathrm{s}}$ with the dipole bend.\\
\indent From Fig. \ref{fig:optics_coord} consider the following case in which the incident electron is aligned with the beam axis (shown in blue).
The electron interacts with the target foil (gray slab) at the reaction vertex $z_{\mathrm{v}}$ and scatters at angle $\theta_{c}$ parallel to the
spectrometer central-ray. The reconstructed event is projected along the dashed (black) line and is denoted as $Y_{\mathrm{tar}}$. In reality, the electron
can also scatter relative to the spectrometer central ray determined by the tangents, $\tan(\phi) = dy_{\mathrm{s}}/dz_{\mathrm{s}}$ (in-plane) and
$\tan(\theta) = dx_{\mathrm{s}}/dz_{\mathrm{s}}$ (out-of-plane). Since the spectrometer
aperture angles are usually very small (see Table \ref{tab:tab3.2}), the tangents can be approximated by $\tan(\phi)\approx\phi$ and $\tan(\theta)\approx\theta$,
which are commonly referred to as  $Y'_{\mathrm{tar}}$ and $X'_{\mathrm{tar}}$, respectively. These derivatives are interpreted as angular distributions of the scattered
particles relative to the spectrometer central-ray. Similar quantities can also be derived in the hut coordinate system using the focal plane variables, ($X_{\mathrm{fp}}, Y_{\mathrm{fp}}$),
to obtain $dX_{\mathrm{fp}}/dZ_{\mathrm{fp}}\approx X'_{\mathrm{fp}}$ and $dY_{\mathrm{fp}}/dZ_{\mathrm{fp}}\approx Y'_{\mathrm{fp}}$.\\
\indent In a more general case, the electron (shown in red) incident on the target is offset by an amount $x^{\mathrm{beam}}$ and scatters parallel to the central ray. As a
result, the reconstructed $Y_{\mathrm{tar}}$ is offset by an amount $x^{\mathrm{beam}}\cos(\theta_{c})$ which is geometrically
equivalent to a spectrometer mispointing along $y_{\mathrm{s}}$. Furthermore, if the electron
scatters at an arbitrary angle $\phi$, the reconstructed $Y_{\mathrm{tar}}$ is further offset by an amount $L\cdot\phi$. Combining all these offsets and adding an
arbitrary $y$-mispointing offset ($y_{\mathrm{mispoint}}$), $Y_{\mathrm{tar}}$ can be expressed in its most general form as
\begin{align}
  Y_{\mathrm{tar}} + y_{\mathrm{mispoint}}\footnotemark  &= z_{\mathrm{v}}\sin(\theta_{\mathrm{c}}) + x^{\mathrm{beam}}\cos(\theta_{\mathrm{c}}) + L\cdot\phi \nonumber \\ 
  &=z_{\mathrm{v}}\sin(\theta_{\mathrm{c}}) + x^{\mathrm{beam}}\cos(\theta_{\mathrm{c}}) + [z_{\mathrm{v}}\cos(\theta_{\mathrm{c}}) - x^{\mathrm{beam}}\sin(\theta_{\mathrm{c}})]Y'_{\mathrm{tar}} \nonumber \\
  &=z_{\mathrm{v}}[\sin(\theta_{\mathrm{c}}) + Y'_{\mathrm{tar}}\cos(\theta_{\mathrm{c}})] + x^{\mathrm{beam}}[\cos(\theta_{\mathrm{c}}) - Y'_{\mathrm{tar}}\sin(\theta_{\mathrm{c}})] \label{eq:4.13}
\end{align}
Alternatively, it is also useful to express Eq. \ref{eq:4.13} in terms of the reaction vertex, 
\begin{equation}
  z_{\mathrm{v}} = \frac{Y_{\mathrm{tar}} + y_{\mathrm{mispoint}} - x^{\mathrm{beam}}[\cos(\theta_{\mathrm{c}}) - Y'_{\mathrm{tar}}\sin(\theta_{\mathrm{c}})]}{\sin(\theta_{\mathrm{c}}) + Y'_{\mathrm{tar}}\cos(\theta_{\mathrm{c}})}.
  \label{eq:4.14}
\end{equation}
\footnotetext{\singlespacing The spectrometer mispointing is defined as a parallel displacement of the spectrometer central ray
  either horizontally ($y$-mispointing) or vertically ($x$-mispointing). The mispointing was determined by survey at various spectrometer angles and a function
  was fit to the $x$- and $y$-mispointing data\cite{holly_HMS_Optics2017,holly_SHMS_Optics2019}, separately.}
\noindent Equation \ref{eq:4.14} is the most general form of the $z$-reaction vertex in terms of measurable quantities. The difference between the $z$-reaction vertex in both spectrometers
was used as an event selection criteria for this experiment (see Section \ref{sec:event_selection}).
\section{Optics Checks and Optimization} \label{sec:optics_checks}
The commissioning of the HMS/SHMS optics took place during the 2017-18 run period and underwent multiple revisions of the reconstruction
matrix elements for both spectrometers during that period \cite{holly_HMS_Optics2017,holly_SHMS_Optics2019}. This section presents the optics optimization checks
and procedures done on the HMS and SHMS for this experiment (E12-10-003) on April 2018. At the time, E12-10-003 also served
as part of the general optics commissioning as during data-taking, it was found that the SHMS Q3 magnet had an unnecessary correction in the matrix elements.
As a result, the data for this experiment is divided into two sections. Only the section after the correction in the SHMS optics was used in the optimization procedure.\\
\indent The problem of optics optimization can be approached in different ways, depending on the circumstances of the experiment. In this particular experiment,
a series of $^{1}$H$(e,e')p$ elastic runs were taken at different configurations such as to cover the entire HMS momentum range corresponding to the $^{2}$H$(e,e'p)n$ reaction kinematics. The
original and corrected $^{1}$H$(e,e')p$ kinematics are summarized below.\\
\begin{table}[ht]
  \begin{tabular}{c c c c c}
    \hline\hline
    Run  & \shortstack{HMS \\ Angle [deg]} & \shortstack{HMS \\ Momentum [GeV/c]} & \shortstack{SHMS \\ Angle [deg]} & \shortstack{SHMS \\ Momentum [GeV/c]} \\
    \hline
    3288 & 37.338 & 2.938 & 12.194 & 8.7 \\
    3371 & 33.545 & 3.48 & 13.93 & 8.7 \\
    3374 & 42.9 & 2.31 & 9.928 & 8.7 \\
    3377 & 47.605 & 1.8899 & 8.495 & 8.7 \\
    \hline
  \end{tabular}
  \label{table:original_heep_kin}
  \caption{Original $^{1}$H$(e,e')p$ elastic kinematics in E12-10-003.}
\end{table}
\begin{table}[H]
  \begin{tabular}{c c c c c}
    \hline\hline
    Run  & \shortstack{HMS \\ Angle [deg]} & \shortstack{HMS \\ Momentum [GeV/c]} & \shortstack{SHMS \\ Angle [deg]} & \shortstack{SHMS \\ Momentum [GeV/c]} \\
    \hline
    3288 & 37.338 & 2.9355 & 12.194 & 8.5342 \\
    3371 & 33.545 & 3.4758 & 13.93 & 8.5342 \\
    3374 & 42.9 & 2.3103 & 9.928 & 8.5342 \\
    3377 & 47.605 & 1.8912 & 8.495 & 8.5342 \\
    \hline
  \end{tabular}
  \label{table:corr_heep_kin}
  \caption{Corrected $^{1}$H$(e,e')p$ elastic kinematics in E12-10-003.}
\end{table}
\begin{table}[H]
  \begin{tabular}{c c c c c}
    \hline\hline
    Spec  & $\delta\theta$[rad] & $\delta\phi$[rad] & $X'_{\mathrm{tar}}$-offset[rad] & $Y'_{\mathrm{tar}}$-offset[rad] \\
    \hline
    HMS & 0.0 & \num{1.521e-3} & \num{2.852e-3} & \num{9.5e-4} \\
    SHMS & 0.0 & 0.0 & 0.0 & 0.0\\ 
    \hline
  \end{tabular}
  \label{table:spec_offsets}
  \caption{Spectrometer offsets determined from $^{1}$H$(e,e')p$ elastic run 3288 in E12-10-003. See Section \ref{sec:spec_off_sec} of this dissertation for more information.}
\end{table}
Since this is a coincidence experiment, the spectrometers are highly correlated, which makes the optics optimization more complicated as changes in
one spectrometer can affect the other spectrometer. Based on the kinematics, it was determined to focus on the HMS first, as the momentum is well below the dipole
saturation ($\sim$5 GeV), and the optics are much better understood from the 6 GeV era.
\subsection{HMS Optics Check}
The procedure to check the HMS Optics involves determining whether a central momentum correction is needed and check that the
HMS momentum fraction $\delta$ is independent of the HMS focal plane variables for constant momenta. That is to say, that there should
not exist a correlation between the momentum fraction and the focal plane variables.
\subsubsection{HMS Central Momentum Correction}\label{sec:hms_optics}
Since the $^{1}$H$(e,e')p$ reaction is used and the HMS is set to detect protons, one can calculate the proton momentum as follows:
\begin{equation}
  P_{\mathrm{calc}} = \frac{2M_{\mathrm{p}}E_{\mathrm{b}}(E_{\mathrm{b}}+M_{\mathrm{p}})\cos(\theta_{\mathrm{p}})}{M^{2}_{\mathrm{p}} + 2M_{\mathrm{p}}E_{\mathrm{b}} + E^{2}_{\mathrm{b}}\sin^{2}(\theta_{\mathrm{p}})},
  \label{eq:4.15}
\end{equation}
where $E_{\mathrm{b}}$ is the initial beam energy and $\theta_{\mathrm{p}}$ is the reconstructed proton angle.
The measured proton momentum, $P_{\mathrm{meas}}$, depends on the $\delta$ from the following definition:
\begin{equation}
  \frac{\delta}{100} = \frac{P_{\mathrm{meas}} - P_{0}}{P_{0}} \rightarrow P_{\mathrm{meas}} = P_{0}(1 + \frac{\delta}{100}),
  \label{eq:4.16}
\end{equation}
where $P_{0}$ is the central momentum of the spectrometer and $\delta$ is the fractional deviation of the particle momentum from the central momentum in $\%$.\\    
\indent From the measured and calculated momentum, the momentum difference is defined as
\begin{equation}
  \Delta P = P_{\mathrm{calc}}-P_{\mathrm{meas}}.
  \label{eq:4.17}
\end{equation}
In Eq. \ref{eq:4.17}, it is assumed that the beam energy and HMS angle are well known, which may not entirely be true, but
serves as the best available approximation. The momentum difference, $\Delta P$, is determined for data and
SIMC independently on an event-by-event basis in terms of ($\theta_{\mathrm{p}}$, $\delta$). It is expected that $\Delta P$ be near zero
in SIMC, as the $\delta$-reconstruction is well described by the TOSCA models, however in data, this may not be the case, as the NMR probe
location in the HMS has changed since the 6 GeV era and the magnetic field for the central momentum may be different from what is expected. 
\begin{figure}[H]
  \centering
  \includegraphics[scale=0.30]{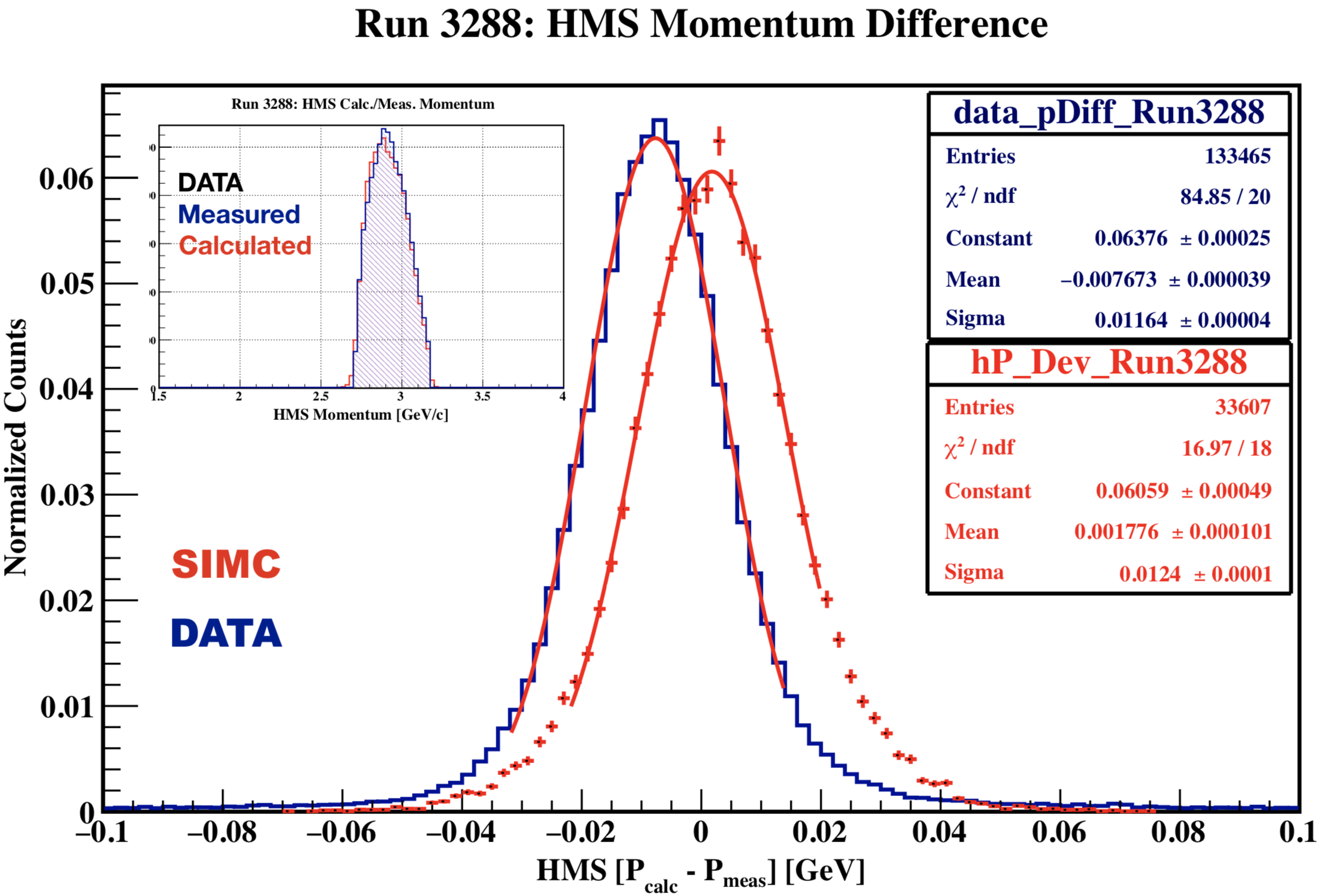}
  \caption{Comparison of HMS momentum difference ($\Delta P$) between data and SIMC. The inset shows the calculated and measured
    HMS momentum distribution for data.}
  \label{fig:hms_Pdiff}
\end{figure}
\indent From the mean of the fit in Fig. \ref{fig:hms_Pdiff}, $\Delta P_{\mathrm{data}}$ is $\sim$8 MeV/c smaller than $\Delta P_{\mathrm{SIMC}}$, or equivalently,
$P_{\mathrm{data}}^{\mathrm{meas}} > P_{\mathrm{data}}^{\mathrm{calc}}$. The data momentum correction factor can be determined as follows: 
\begin{equation}
  f^{\mathrm{HMS}}_{\mathrm{corr}} = 1 - \frac{\Delta P_{\mathrm{SIMC}} - \Delta P_{\mathrm{data}}}{P_{0}},
  \label{eq:4.18}
\end{equation}
and the corrected HMS momentum can then be expressed as
\begin{equation}
  P^{\mathrm{HMS}}_{\mathrm{corr}} = P^{\mathrm{HMS}}_{\mathrm{uncorr}} \cdot f^{\mathrm{HMS}}_{\mathrm{corr}}.
  \label{eq:4.19}
\end{equation}
Figure \ref{fig:hms_Pcorr} shows the difference in the mean of the fit for data and SIMC before and after the HMS momentum
corrections. After correction, the difference between data and SIMC is within $\sim$1 MeV/c for the four elastic runs. It is important to
\begin{figure}[H]
  \centering
  \includegraphics[scale=0.3]{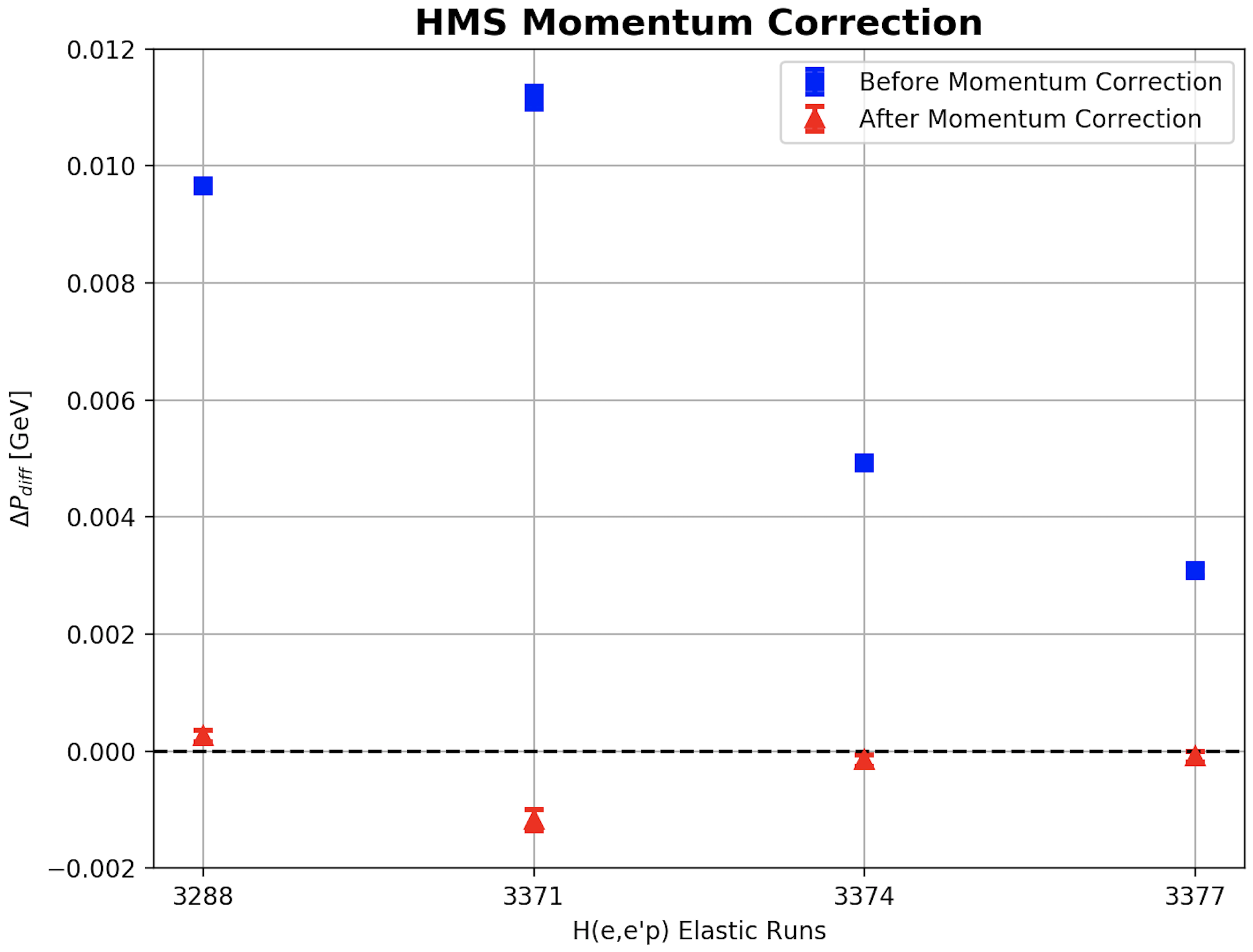}
  \caption{HMS momentum difference, $\Delta P_{\mathrm{diff}}=\Delta P_{\mathrm{SIMC}} - \Delta P_{\mathrm{data}}$, before and after applying the momentum correction to data.}
  \label{fig:hms_Pcorr}
\end{figure}
\begin{figure}[H]
  \centering
  \includegraphics[scale=0.40]{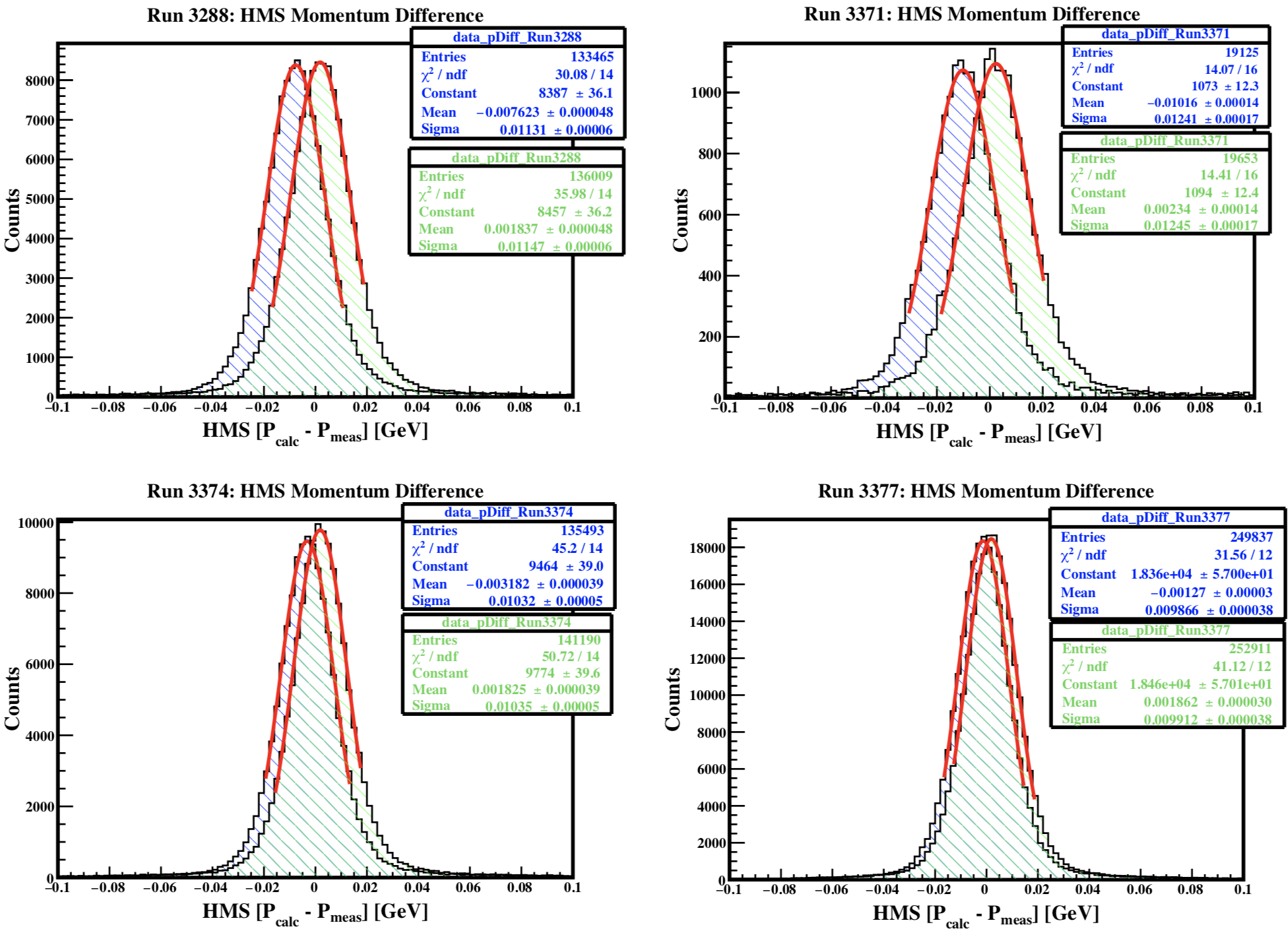}
  \caption{HMS momentum difference for data, before (blue) and after (green) applying the momentum correction ($2^{\mathrm{nd}}$ iteration) to data.}
  \label{fig:hms_dataPdiff}
\end{figure}
\noindent note that the corrections applied are after a second iteration, once the spectrometer offsets were determined. \\
\indent Figure \ref{fig:hms_dataPdiff} shows the difference between the HMS calculated and measured data momentum before and after applying a momentum correction
during the 2nd iteration. The corrected data momentum difference has clearly shifted towards zero, which indicates a successful momentum correction.
\subsubsection{HMS $\delta$ Check}
To check the HMS delta ($\delta$)\footnote{\singlespacing The HMS (or SHMS) $\delta$ is defined as $\delta\equiv\frac{P - P_{0}}{P_{0}}$, where $P$ and $P_{0}$
  are the reconstructed particle momentum ($P$) and central spectrometer momentum ($P_{0}$), respectively.} reconstruction, the HMS fractional momentum is defined as
\begin{equation}
  \Delta P_{\mathrm{frac}} = \frac{P_{\mathrm{calc}} - P_{\mathrm{meas}}}{P_{\mathrm{meas}}}
\end{equation}
and is plotted as a function of the HMS focal plane variables.
\begin{figure}[H]
  \centering
  \includegraphics[scale=0.26]{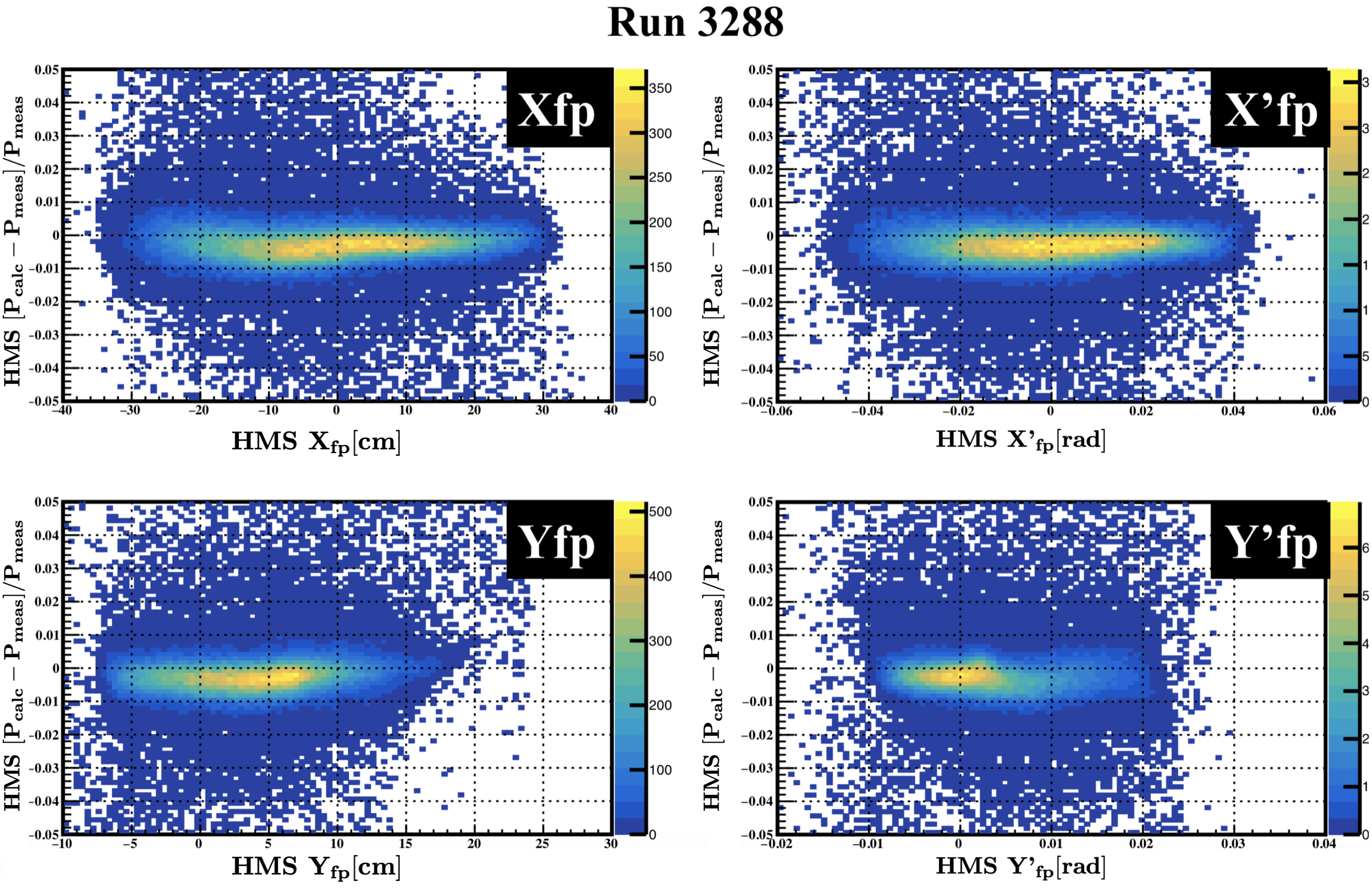}
  \caption{HMS fractional momentum difference vs. focal plane variables for $^{1}$H$(e,e')p$ elastic run 3288.}
  \label{fig:hms_hPdiff_vs_FP_3288}
\end{figure}
\begin{figure}[H]
  \centering
  \includegraphics[scale=0.30]{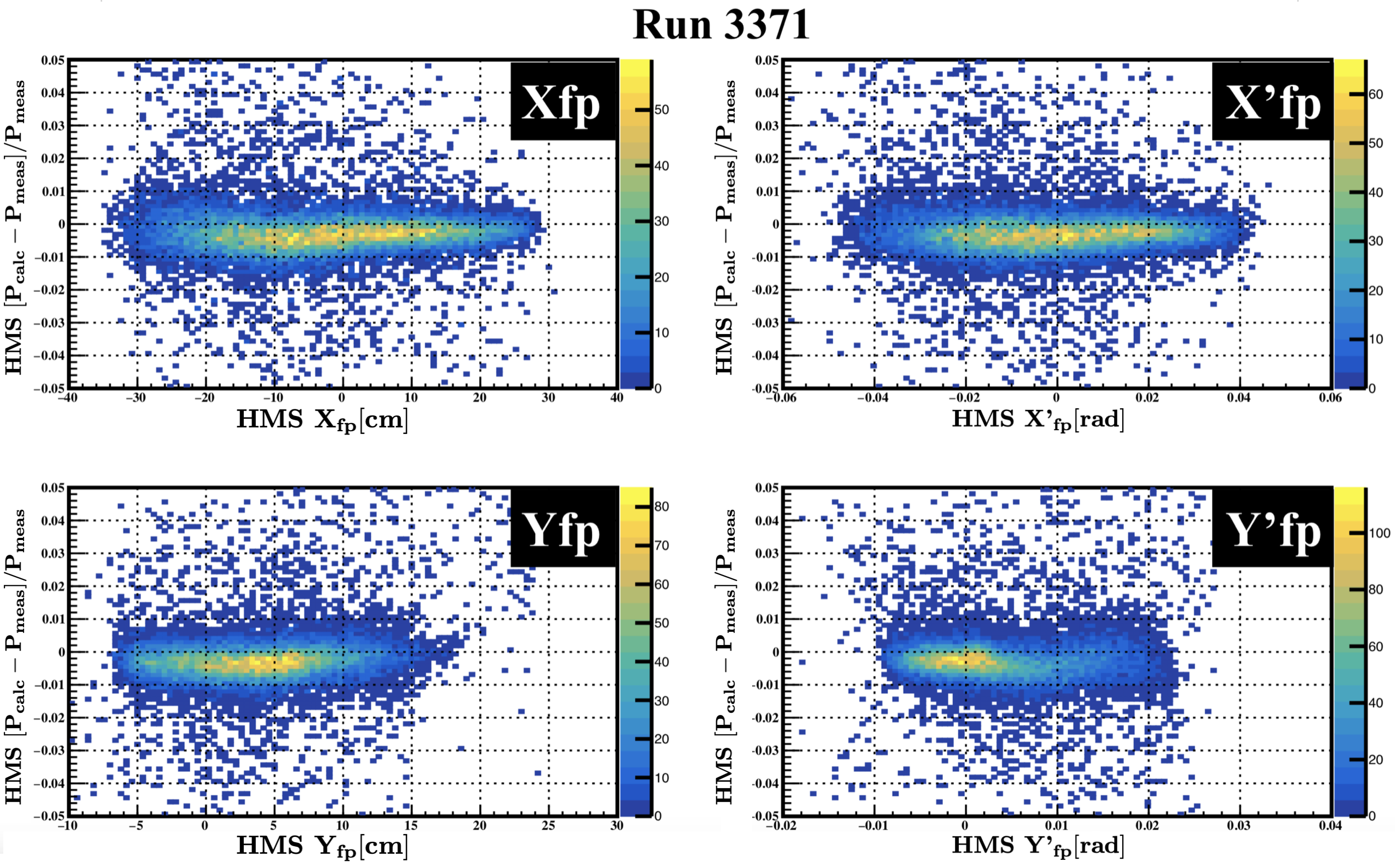}
  \caption{HMS fractional momentum difference vs. focal plane variables for $^{1}$H$(e,e')p$ elastic run 3371.}
  \label{fig:hms_hPdiff_vs_FP_3371}
\end{figure}
\begin{figure}[H]
  \centering
  \includegraphics[scale=0.30]{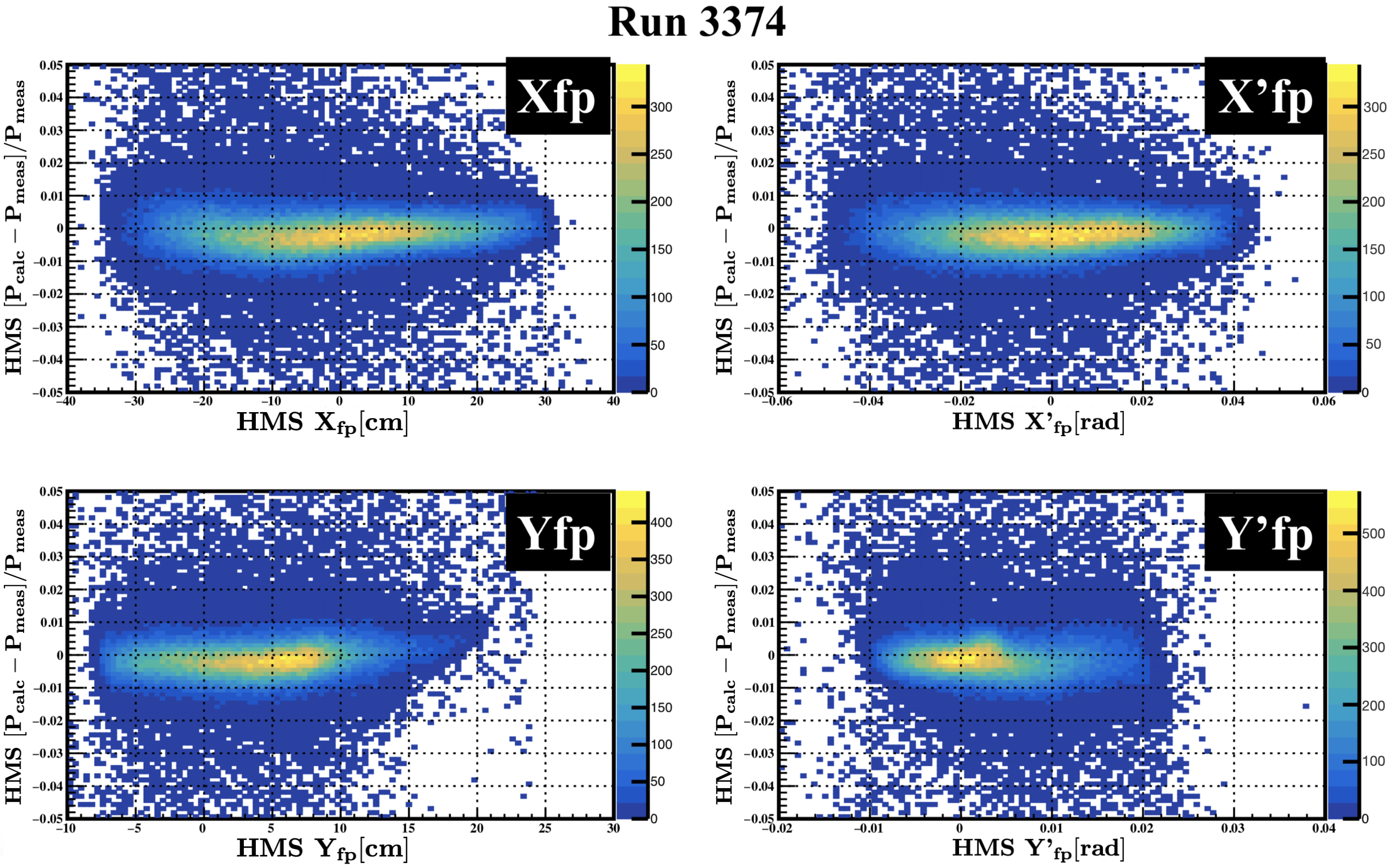}
  \caption{HMS fractional momentum difference vs. focal plane variables for $^{1}$H$(e,e')p$ elastic run 3374.}
  \label{fig:hms_hPdiff_vs_FP_3374}
\end{figure}
\begin{figure}[H]
  \centering
      \includegraphics[scale=0.30]{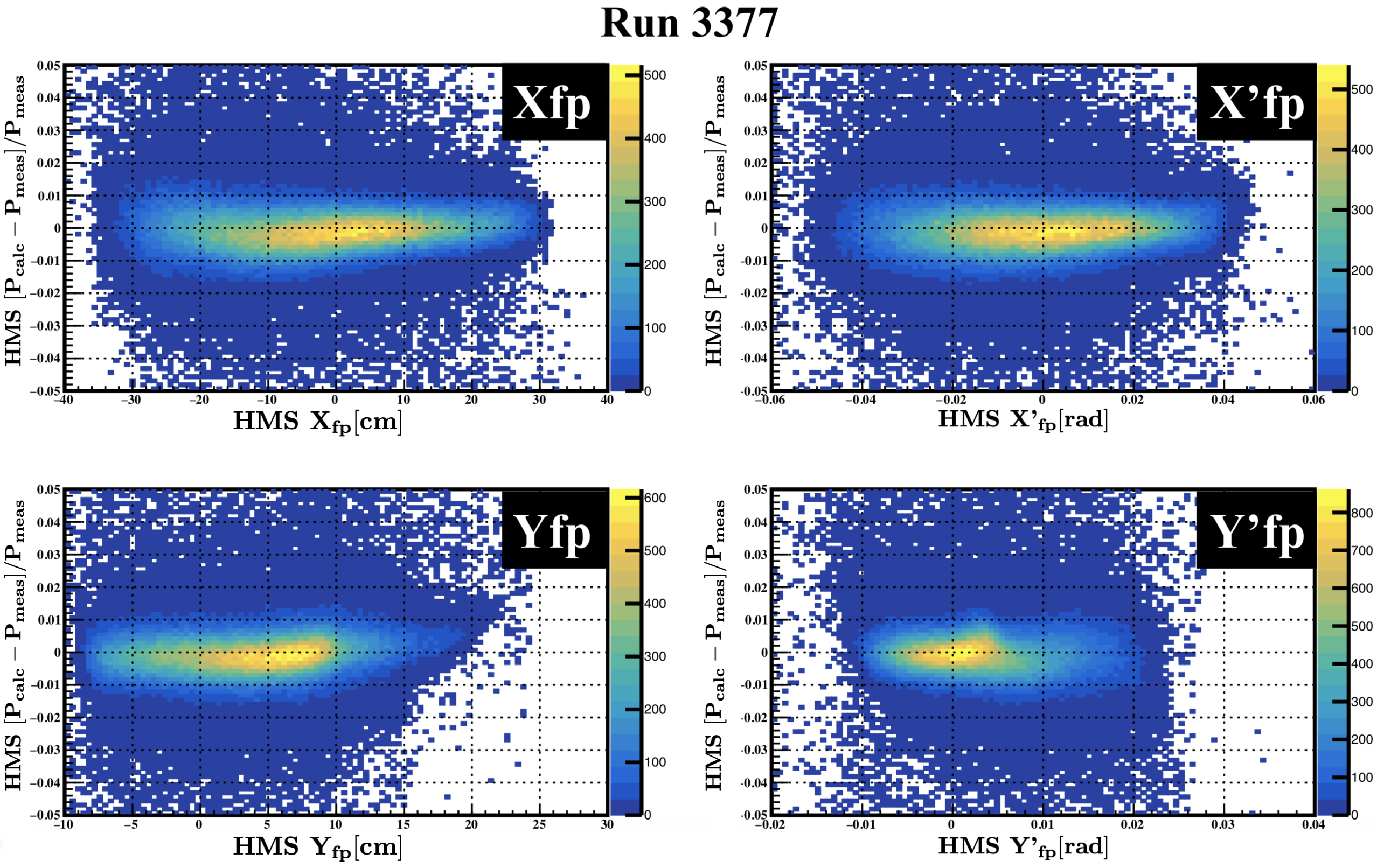}
      \caption{HMS fractional momentum difference vs. focal plane variables for $^{1}$H$(e,e')p$ elastic runs 3377.}
      \label{fig:hms_hPdiff_vs_FP_3377}
\end{figure}
\indent Figures \ref{fig:hms_hPdiff_vs_FP_3288}, \ref{fig:hms_hPdiff_vs_FP_3371}, \ref{fig:hms_hPdiff_vs_FP_3374}, and \ref{fig:hms_hPdiff_vs_FP_3377} show that $\Delta P_{\mathrm{frac}}$
is uncorrelated across each of the HMS focal plane variables, which demonstrates that the $\delta$-reconstruction is already optimized
for the HMS. Now that the HMS optics are well understood, one can move on to the SHMS optics checks.
\subsection{SHMS Optics Check}
Similar to the HMS, the procedure to check the SHMS optics involves determining the central momentum correction
and checking that the reconstructed $\delta$ is uncorrelated with the SHMS focal plane variables. Additional
checks on the $Y_{\mathrm{tar}}, Y'_{\mathrm{tar}}$ and $X'_{\mathrm{tar}}$ reconstruction variables may also be needed
as the SHMS optics optimization is still incomplete.
\subsubsection{SHMS Central Momentum Correction} \label{sec:shms_Pcent_corr}
\begin{figure}[H]
  \centering
  \includegraphics[scale=0.3]{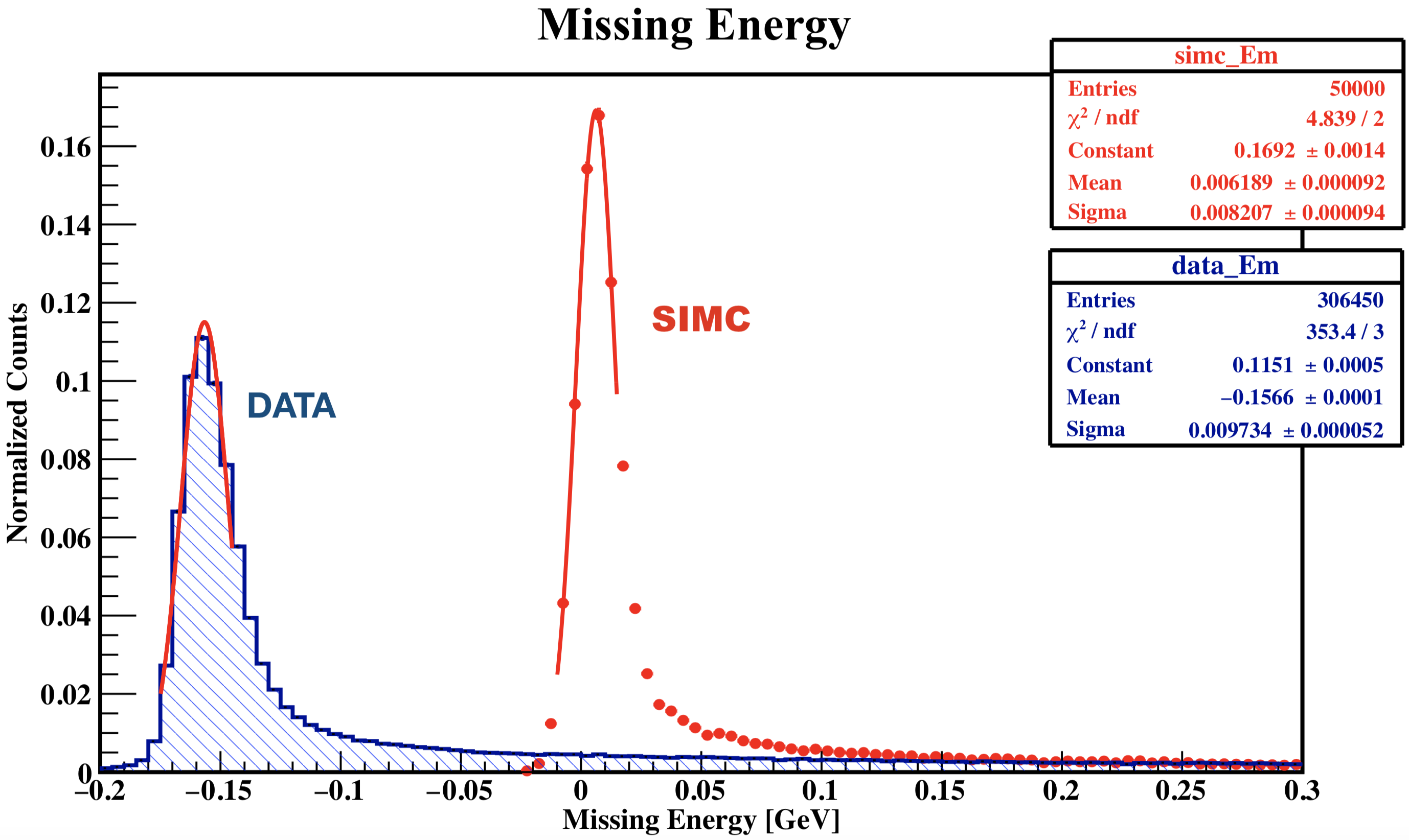}
  \caption{Missing energy spectrum for $^{1}$H$(e,e')p$ elastic run 3288 before central momentum correction.}
  \label{fig:Emiss_UnCorr}
\end{figure} 
To determine the SHMS central momentum correction, one starts with the missing energy definition for elastic scattering on hydrogen, $^{1}$H$(e,e')p$
\begin{equation}
  E_{\mathrm{m}} = (E_{\mathrm{b}} - E') + M_{\mathrm{p}} - E_{\mathrm{p}},
\end{equation}
where $E'$ and $E_{p}$ are the electron and proton final energies, respectively. Since it was assumed that
the beam energy, and the HMS momentum are well known, any deviation from the expected missing
energy is attributed to the electron momentum in the SHMS. The expected location of $E_{\mathrm{m}}$ ideally would
be at zero since $^{1}$H$(e,e')p$ is a completely determined system. However, due to the energy loss and radiative effects,
the peak has a small offset from zero ($\lesssim$10 MeV), which can be simulated. The measured and simulated
$E_{\mathrm{m}}$ are then compared to determine if the SHMS central momentum needs to be corrected.\\
\indent The SHMS central momentum was kept fixed during the entire experiment, which would suggest that the missing
energy offset would be the same for the four elastic runs after the HMS momentum correction. This was found
to be the case due to the fact that the spectrometer offsets have not been determined
at this stage. Alternatively, it was decided to only focus on finding the central momentum correction
for run 3288, as it was the closest kinematic setting to the $^{2}$H$(e,e'p)n$ 80 MeV/c setting. This correction would then
be applied to the remaining elastic runs. \\
\indent Assuming any variation in missing energy between data and SIMC was due to the electron momentum, $E'$,
\begin{equation}
  \frac{\delta E_{\mathrm{m}}}{\delta E'} = -1 \rightarrow \delta E_{\mathrm{m}} = -\delta E', \label{eq:4.22}
\end{equation}
where $\delta E_{\mathrm{m}} = E^{\mathrm{SIMC}}_{\mathrm{m}} - E^{\mathrm{data}}_{\mathrm{m}}$ from the missing energy peak fit.
The electron momentum correction is then
\begin{align}
  E'_{\mathrm{corr}} & =  E'_{\mathrm{uncorr}} + \delta E' \nonumber \\
  & = E'_{\mathrm{uncorr}}(1 - \frac{\delta E_{\mathrm{m}}}{E'_{\mathrm{uncorr}}}),
\end{align}
where the correction factor is defined as
\begin{equation}
  f^{\mathrm{SHMS}}_{\mathrm{corr}} \equiv 1 - \frac{\delta E_{\mathrm{m}}}{E'_{\mathrm{uncorr}}}.
\end{equation}
After correcting the SHMS central momentum, the SHMS $\delta$-reconstruction also needs to be checked as a function
of the SHMS focal plane variables, as it may need to be optimized.    
\subsubsection{SHMS $\delta$ Optimization}
\begin{figure}[H]
  \centering
  \includegraphics[scale=0.34]{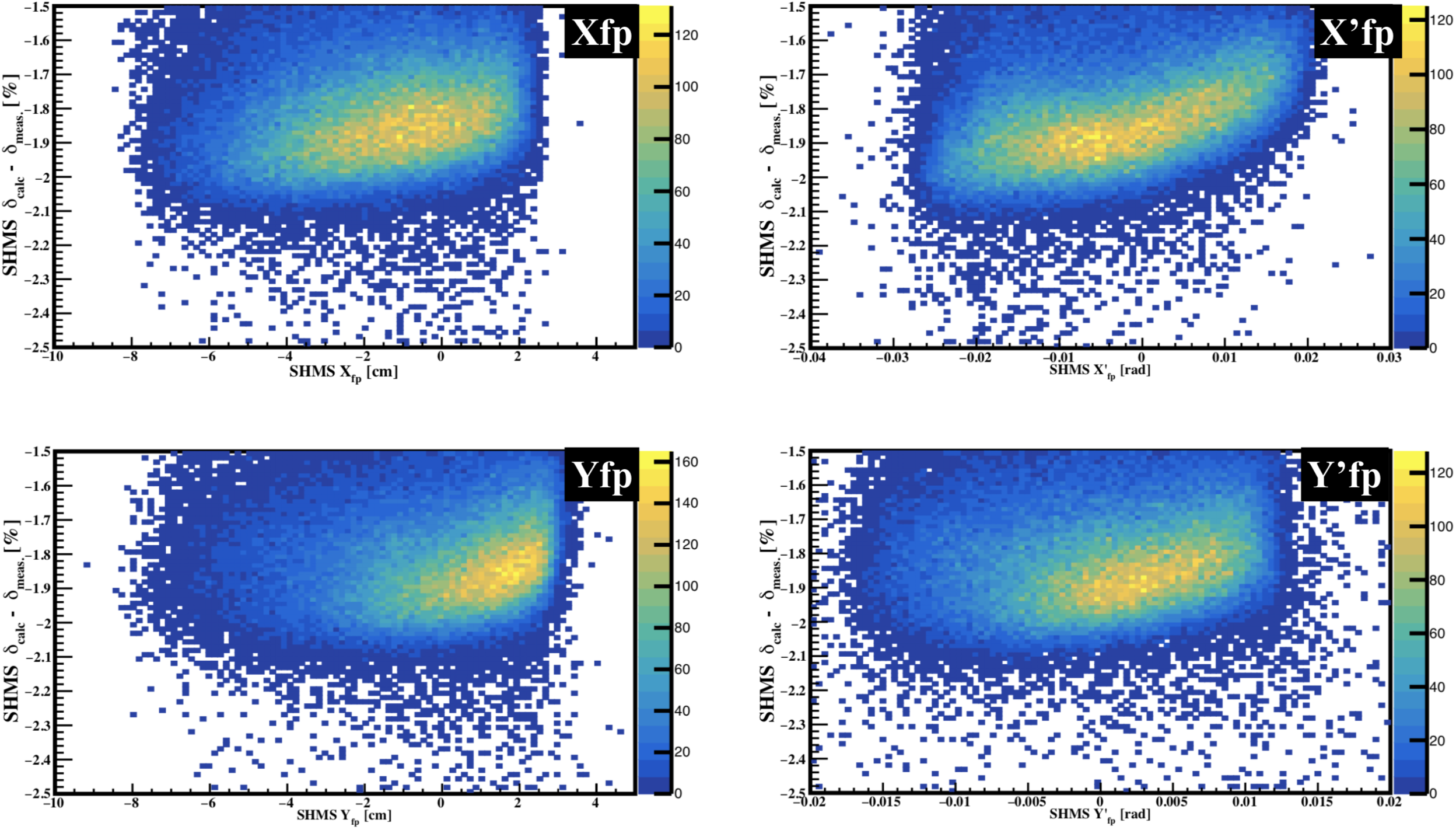}
  \caption{SHMS $(\delta_{\mathrm{calc}}-\delta_{\mathrm{meas}})$ vs. focal plane variables for $^{1}$H$(e,e')p$ elastic run 3288 before $\delta$-optimization.}
  \label{fig:shmsDelta_beforeOptim}
\end{figure}

\begin{figure}[H]
  \centering
  \includegraphics[scale=0.34]{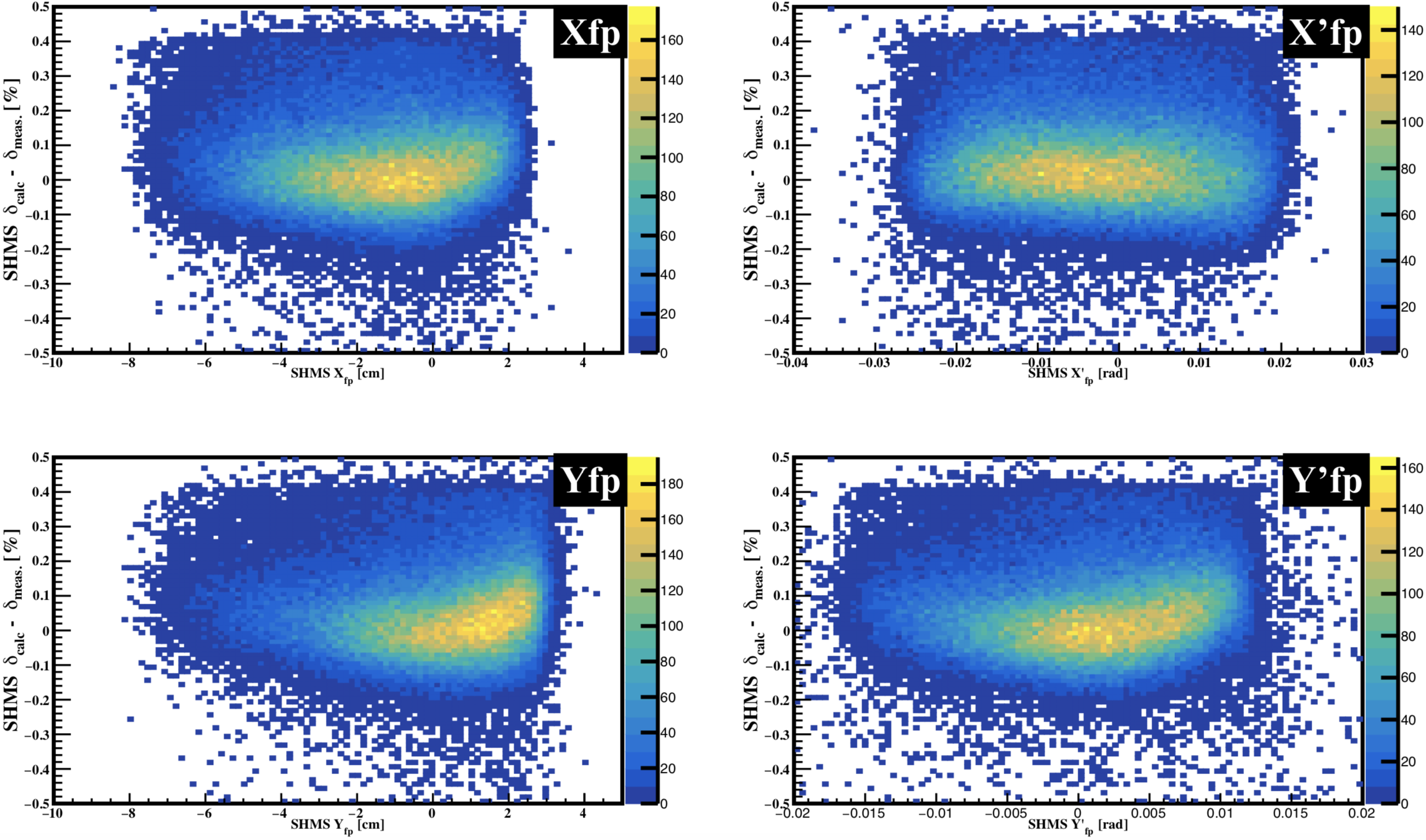}
  \caption{SHMS $(\delta_{\mathrm{calc}}-\delta_{\mathrm{meas}})$ vs. focal plane variables for $^{1}$H$(e,e')p$ elastic run 3288 after $\delta$-optimization.}
  \label{fig:shmsDelta_afterOptim}
\end{figure}

\begin{figure}
  \centering
  \includegraphics[scale=0.30]{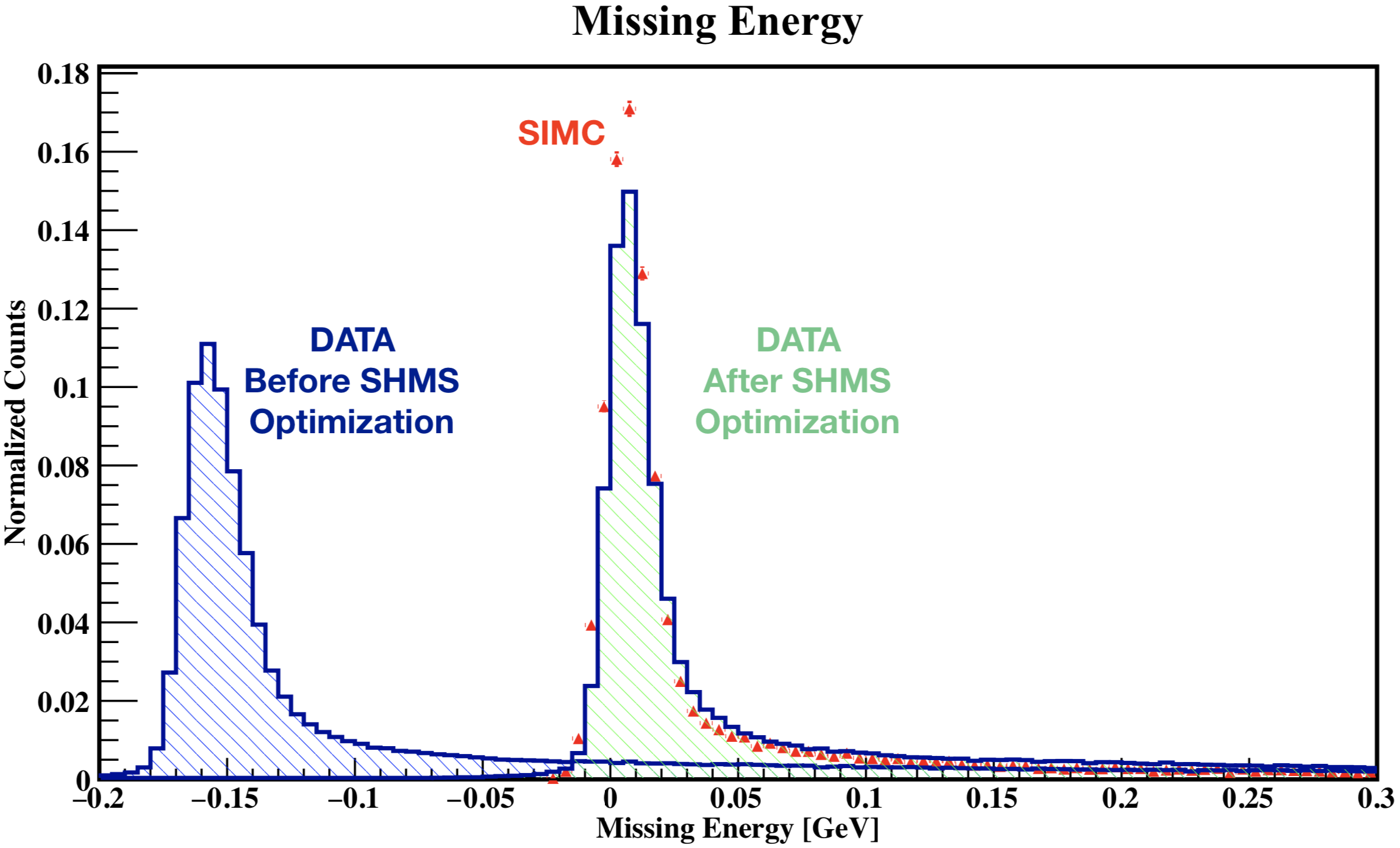}
  \caption{Missing energy spectrum for $^{1}$H$(e,e')p$ elastic run 3288 after central momentum correction and $\delta$-optimization.}
      \label{fig:shmsEm_afterOptim}
\end{figure}

To check the SHMS $\delta$-reconstruction, similar to the HMS, one needs to determine and correct for any 
correlation that might exist between the reconstructed $\delta$ and each of the focal plane variables. In general,
from each of the measured trajectories at the focal plane ($X_{\mathrm{fp}}$, $X'_{\mathrm{fp}}$, $Y_{\mathrm{fp}}$, $Y'_{\mathrm{fp}}$),
one has to reconstruct the measured trajectories at the target, which are characterized by five quantities
($Y_{\mathrm{tar}}$, $X'_{\mathrm{tar}}$, $Y'_{\mathrm{tar}}$, $X_{\mathrm{tar}}$, $\delta$), leading to an underdetermined system.
To overcome this problem, one of the target variables has to be fixed (usually $X_{\mathrm{tar}}$). Once the horizontal target position $X_{\mathrm{tar}}$
has been surveyed, one can express each of the reconstructed target variables as a polynomial expansion of the focal plane variables. For our case,
the $\delta$ component can be expressed as
\begin{equation}
  \delta = \sum_{ijklm} D_{ijklm}x^{i}_{\mathrm{fp}}x'^{j}_{\mathrm{fp}}y^{k}_{\mathrm{fp}}y'^{l}_{\mathrm{fp}}x^{m}_{\mathrm{tar}},
  \label{eq:4.25}
\end{equation}
where $i,j,k,l,m$ are the powers of the focal plane quantities and $D_{ijklm}$ are the \textit{matrix}
coefficients for a particular combination of powers where the target position is typically set to $x^{m}_{\mathrm{tar}}=0$. \\
\indent To optimize the $\delta$-component, we define the calculated $\delta$ as
\begin{equation}
  \delta_{\mathrm{calc}} \equiv \frac{P_{\mathrm{calc}} - P_{0}}{P_{0}},
  \label{eq:4.26}
\end{equation}
where the calculated electron momentum is determined from momentum conservation to be
\begin{equation}
  P_{\mathrm{calc}} = E_{\mathrm{b}} + M_{\mathrm{p}} - E_{\mathrm{p}}. 
  \label{eq:4.27}
\end{equation}
From Eq. \ref{eq:4.27}, the proton energy is $E_{\mathrm{p}} = \sqrt{P_{\mathrm{meas}}^{2} + M_{\mathrm{p}}^{2} }$ and the electron momentum can be approximated by $P_{\mathrm{calc}} \sim E_{\mathrm{calc}}$.
The measured proton momentum, $P_{\mathrm{meas}}$, is the corrected HMS momentum determined in the previous
section. Taking the difference between the calculated and measured momenta,
\begin{equation}
  \chi^{2} \equiv (\delta_{\mathrm{calc}} (E_{\mathrm{b}}, P_{\mathrm{meas}}) - \delta_{\mathrm{meas}} (x_{\mathrm{fp}}, x'_{\mathrm{fp}}, y_{\mathrm{fp}}, y'_{\mathrm{fp}}))^{2}.
  \label{eq:4.28}
\end{equation}
From Eq. \ref{eq:4.28}, the SHMS $\delta$-optimization is now a $\chi^{2}$-minimization problem, where the goal is
to find a set of matrix coefficients,  $D_{ijklm}$, that minimizes the difference between the calculated and measured
$\delta$. For further details, see Ref.\cite{HMSOptics_Jure2017}. 
\indent The $\chi^{2}$-minimization procedure was done simultaneously on the four hydrogen elastic $^{1}$H$(e,e')p$ runs taken during this experiment.
Each of these runs covered a different (also overlapping) region of the SHMS reconstructed $\delta$ with a coverage of $-10 \% <\delta < 12\%$. The advantage
of doing the simultaneous fit on the delta covered by each of these runs is that it allowed the determination of a common table of $D_{ijklm}$ matrix coefficients
as opposed to having done the fits separately for each run, which resulted in matrix correction factors that would have to be applied to each run.\\
\indent Additionally, only the $(x_{\mathrm{fp}}, x'_{\mathrm{fp}})$ focal plane terms were used in the fit since
with elastic events there is a kinematic correlation between the momentum and scattering angle that translates into a correlation
between $x_{\mathrm{fp}}$ and $(y'_{\mathrm{fp}}, y_{\mathrm{fp}})$. So if one fits $(y'_{\mathrm{fp}}, y_{\mathrm{fp}})$ then one can be fitting this kinematic
correlation and not an optics correlation \cite{MKJ_privJuly2019}.\\
\indent The $\delta$ terms that were used in the fit can be expanded from Eq. \ref{eq:4.25} to obtain,
\begin{equation}
  \delta_{\mathrm{meas}} = D_{10000}\cdot x_{\mathrm{fp}} + D_{01000}\cdot x'_{\mathrm{fp}} + D_{11000}\cdot x_{\mathrm{fp}}\cdot x'_{\mathrm{fp}} + D_{20000}\cdot x^{2}_{\mathrm{fp}} + D_{02000}\cdot x'^{2}_{\mathrm{fp}}.
  \label{eq:4.29}
\end{equation}
\noindent The coefficents were optimized for the first and second order $(x_{\mathrm{fp}}, x'_{\mathrm{fp}})$ terms as well as for the cross terms since the correlations
observed were not completely linear as shown in Fig. \ref{fig:shmsDelta_beforeOptim}. After fitting the correlations and determining
the optimum coefficients, these were updated in the SHMS optics parameter file, and the data were re-analyzed.\\
\indent From Fig. \ref{fig:shmsDelta_afterOptim}, there is a noticeable improvement in the $(x_{\mathrm{fp}}, x'_{\mathrm{fp}})$, as the correlations have
been corrected, whereas in the $(y_{\mathrm{fp}}, y'_{\mathrm{fp}})$, the effect is less noticeable as these were not involved in the fit. \\
\indent Figure \ref{fig:shmsEm_afterOptim} shows that after the SHMS central momentum correction and optimization, there
is a clear improvement in the missing energy spectrum. 
\noindent From Fig. \ref{fig:shmsEm_afterOptim}, the improvements observed are:
\begin{itemize}
\item Alignment of data missing energy to SIMC from the central momentum correction
\item Narrower width in data missing energy from $\delta$-optimization of the matrix coefficients.
\end{itemize}
The first bullet point is easy to understand, as the alignment is simply due to a change in the SHMS central momentum.
The second bullet point can be understood from the fact that since the SHMS $\delta_{\mathrm{meas}}$ matrix coefficients
\footnote{\singlespacing The SHMS $\delta$ matrix coefficients are directly associated with the SHMS measured momentum on an event-by-event
  basis, so if these coefficients are optimized, the measured SHMS momentum is optimized, which directly affects where the
  Missing Energy event will be reconstructed.}
have been optimized, an event in the missing energy spectrum that would otherwise be reconstructed far away from the
main peak, is now reconstructed underneath the main peak resulting in an improvement in the resolution as well as in
the recovered events. 
\subsubsection{SHMS $(Y_{\mathrm{tar}}, Y'_{\mathrm{tar}}, X'_{\mathrm{tar}})$ Optimization}
During the E12-10-003 experiment, an optics run with the centered sieve inserted was taken after the optics in SHMS Q3 was fixed. These data were used to
optimize the $(Y_{\mathrm{tar}}, Y'_{\mathrm{tar}}, X'_{\mathrm{tar}})$ components of the reconstruction matrix. The target used consists of three carbon foils positioned at
(-10, 0, 10) cm to mimic the Hall C extended target edges and center. The 3 foils provide events with known and fixed $Y_{\mathrm{tar}}$ positions that are used
to optimize the $Y_{\mathrm{tar}}$ reconstruction whereas the sieve slit provides events with known and fixed sieve holes to optimize the $X'_{\mathrm{tar}}$ and $Y'_{\mathrm{tar}}$
reconstruction. The optimization code used can be found in Ref.\cite{shmsOpt_git_repo}. \\
\indent To check the optics, the SHMS $\delta$ vs. $Y_{\mathrm{tar}}$ was plotted to verify how well the $Y_{\mathrm{tar}}$ has been reconstructed across $\delta$ for each of the
three foils. Below are the plots showing before and after the optimization.
\begin{figure}[H]
  \centering
  \includegraphics[scale=0.33]{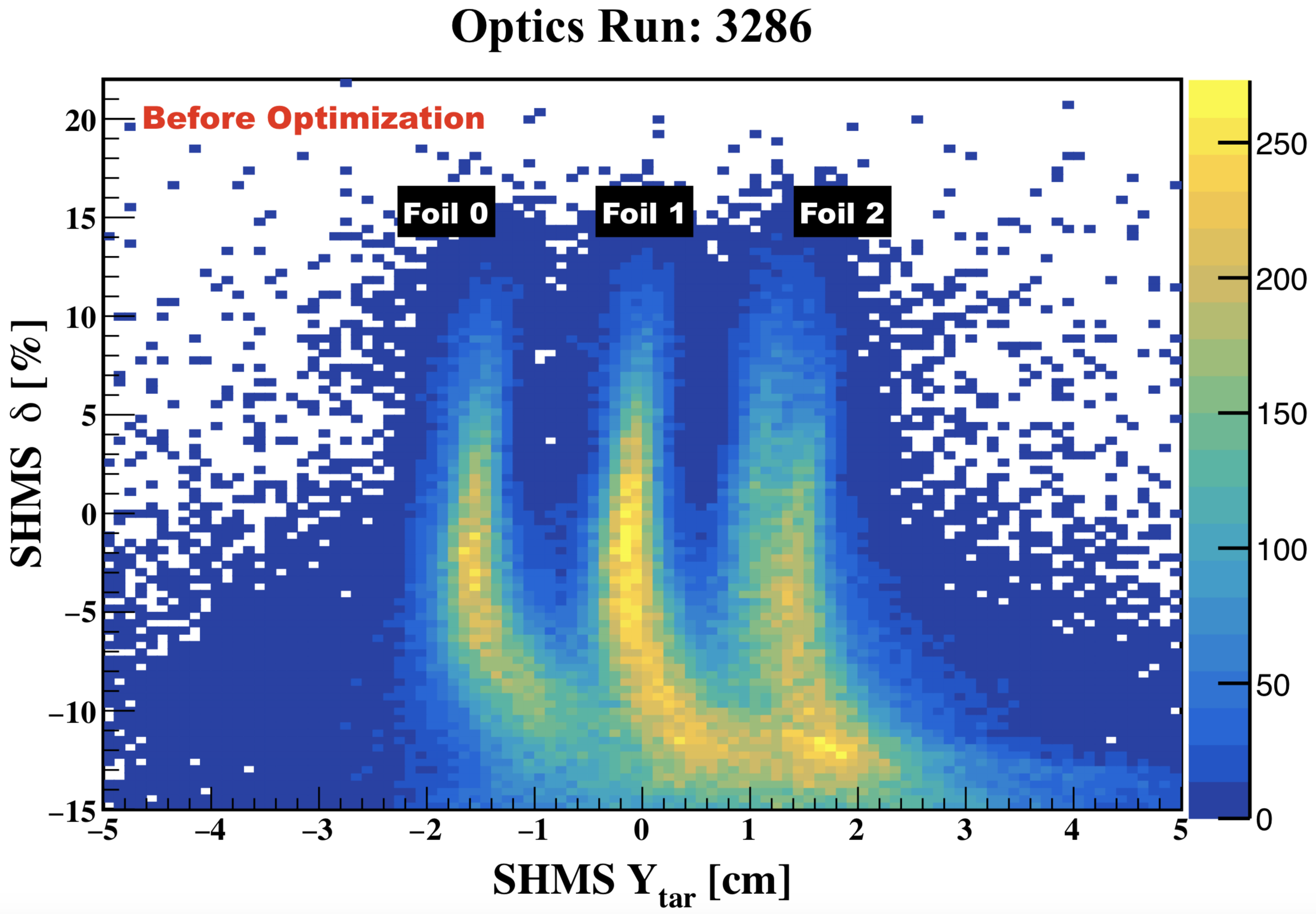}
      \caption{SHMS $\delta$ vs. $Y_{\mathrm{tar}}$ for carbon sieve run 3286 before $Y_{\mathrm{tar}}$-optimization.}
      \label{fig:shmsYtar_beforeOptim}
    \end{figure}
    \begin{figure}[H]
      \centering
      \includegraphics[scale=0.33]{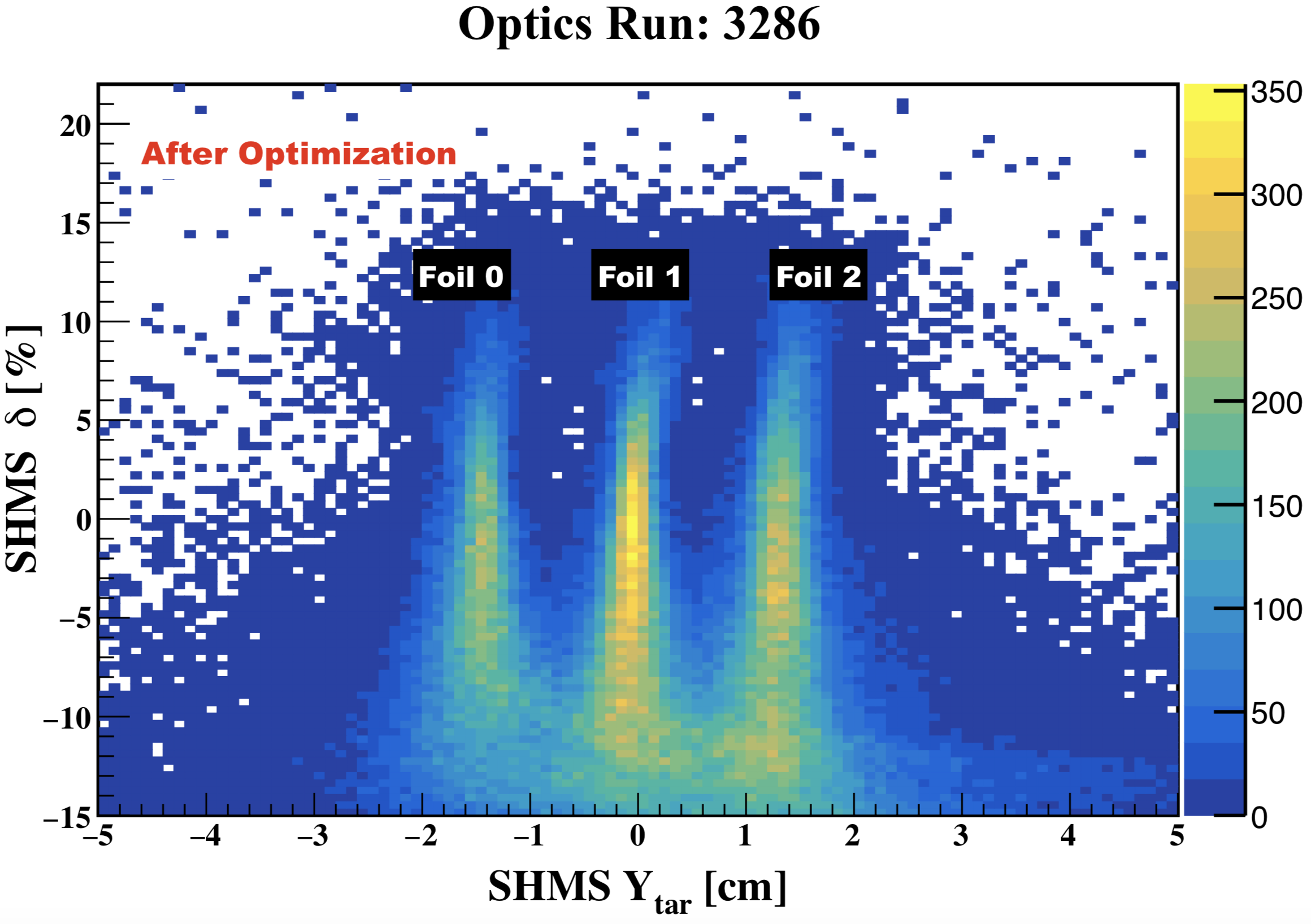}
      \caption{SHMS $\delta$ vs. $Y_{\mathrm{tar}}$ for carbon sieve run 3286 after $Y_{\mathrm{tar}}$-optimization.}
      \label{fig:shmsYtar_afterOptim}
    \end{figure}
    After the optimization, it is clear from Fig. \ref{fig:shmsYtar_afterOptim} that there is almost no
    correlation as compared to before optimization. From the optimized variables $(Y_{\mathrm{tar}}, Y'_{\mathrm{tar}}, X'_{\mathrm{tar}})$, the last two are
    related to the in-plane and out-of-plane angles of the reconstructed particle trajectory relative to the spectrometer central ray as discussed in Section \ref{sec:hall_coord}.
    Removing the correlation in $Y_{\mathrm{tar}}$ improves the determination of the SHMS electron scattering angle, which in turn corrects the location of the invariant mass ($W$) peak as it depends
    on the electron angle.    
    \subsection{Spectrometer Offsets} \label{sec:spec_off_sec}
    The optics optimization was originally done assuming there were no spectrometer offsets. This is not true,
    however, as there were still some small mis-alignments observed in the missing energy spectrum. An extensive study
    of the spectrometer offsets in Hall C has not been performed yet. We have estimated these
    offsets based on observations in $^{1}$H$(e,e')p$ elastic run 3288, as it is closest in kinematics to the deuteron 80 MeV/c setting.
    \subsubsection{Central Angle Offsets}
    The central angle offsets refer to the angular offsets of the spectrometer central ray and can be classified as follows:
    \begin{itemize}
    \item \text{In-plane central angle offset $(\theta_{\mathrm{c}} + \delta\theta^{\mathrm{off}}_{\mathrm{c}})$}, \text{[h(p)\_thetacentral\_offset]}\footnote{\singlespacing The spectrometer central angle offset parameters can be found at\\
      \text{hallc\_replay/PARAM/(S)HMS/GEN/(s)hmsflags.param}}
    \item \text{Out-of-plane central angle offset ($\phi_{\mathrm{c}} + \delta\phi^{\mathrm{off}}_{\mathrm{c}}$)}, \text{[h(p)\_oopcentral\_offset]}
    \end{itemize}
    where the bracketed parameters represent the nomenclature in the analysis software. 
    \textit{In-plane} is parallel to the hall floor, whereas the \textit{out-of-plane} is perpendicular to the hall floor.
    The central angle offsets can be determined from the missing momentum components of $^{1}$H$(e,e')p$ elastic events as these should ideally be centered around
    zero. In the hall coordinate system, the in-plane central angle offsets can be determined by taking the fractional difference between
    the measured (data) and expected (SIMC) X-component of the missing momentum as follows:
    \begin{equation}
      \delta\theta^{\mathrm{off}}_{\mathrm{c}} = \frac{P_{\mathrm{mx}}^{\mathrm{SIMC}} - P_{\mathrm{mx}}^{\mathrm{data}}}{P_{0}}.
    \end{equation}
    The out-of-plane central angle offset can be determined by taking the fractional difference between
    the measured (data) and expected (SIMC) Y-component of the missing momentum as follows:
    \begin{equation}
      \delta\phi^{\mathrm{off}}_{\mathrm{c}} = \frac{P_{\mathrm{my}}^{\mathrm{SIMC}} - P_{\mathrm{my}}^{\mathrm{data}}}{P_{0}}.
    \end{equation}
    \begin{figure}[H]
      \centering
      \includegraphics[scale=0.33]{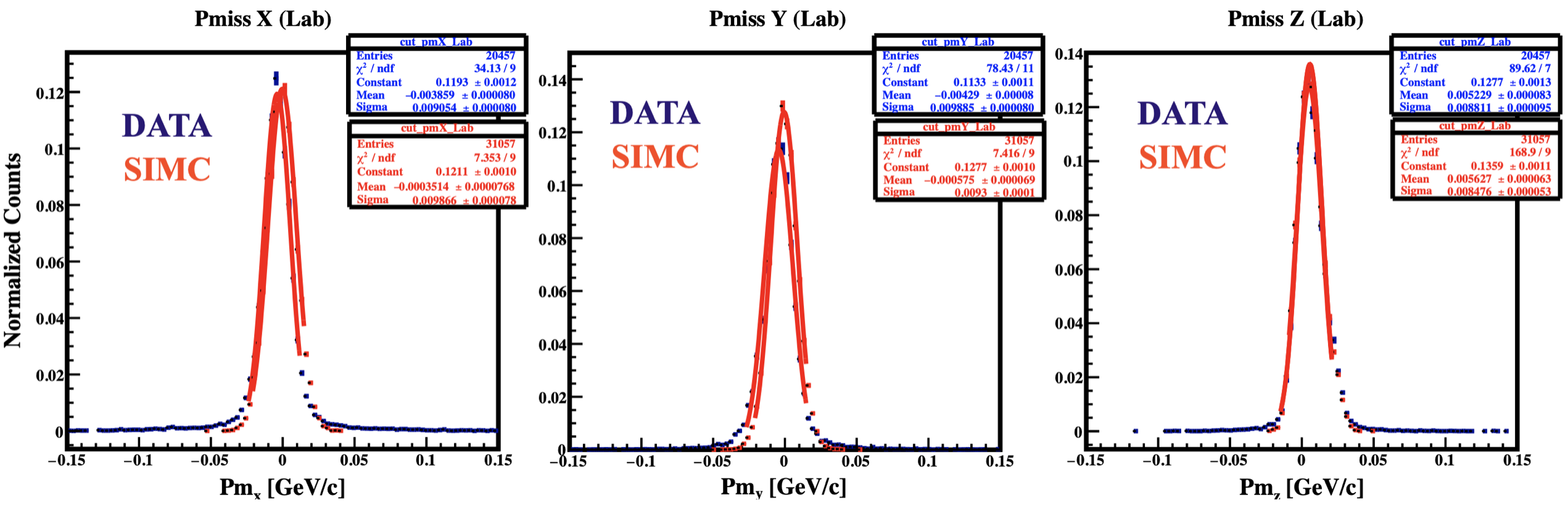}
      \caption{Missing momentum components with no spectrometer offsets applied.}
      \label{fig:PmComp_noOffsets}
    \end{figure}
    \begin{figure}[H]
      \centering
      \includegraphics[scale=0.33]{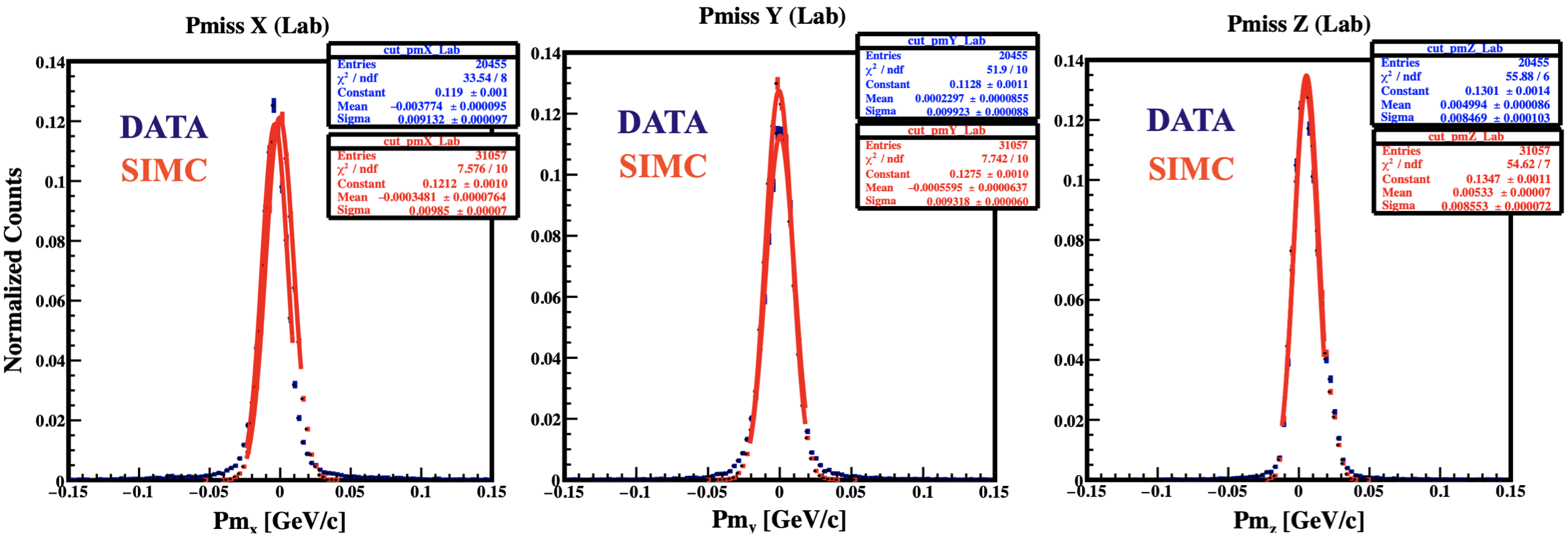}
      \caption{Missing momentum components with out-of-plane central offset applied.}
      \label{fig:PmComp_OopOffset}
    \end{figure}
    \indent After applying the \textit{out-of-plane} offset, the Y-component of the missing momentum agrees with simulation
    as shown in Fig. \ref{fig:PmComp_OopOffset}. With respect to the X-component of the missing momentum, it was decided not
    to apply an \textit{in-plane} angle offset as this would directly impact the location of the invariant mass peak. Alternatively,
    it was decided to apply a relative in-plane angle offset that would align the X-component. The relative angle offsets are discussed
    in the next section.
    \subsubsection{Relative Angle Offsets}
\begin{figure}[H]
      \centering
      \includegraphics[scale=0.49]{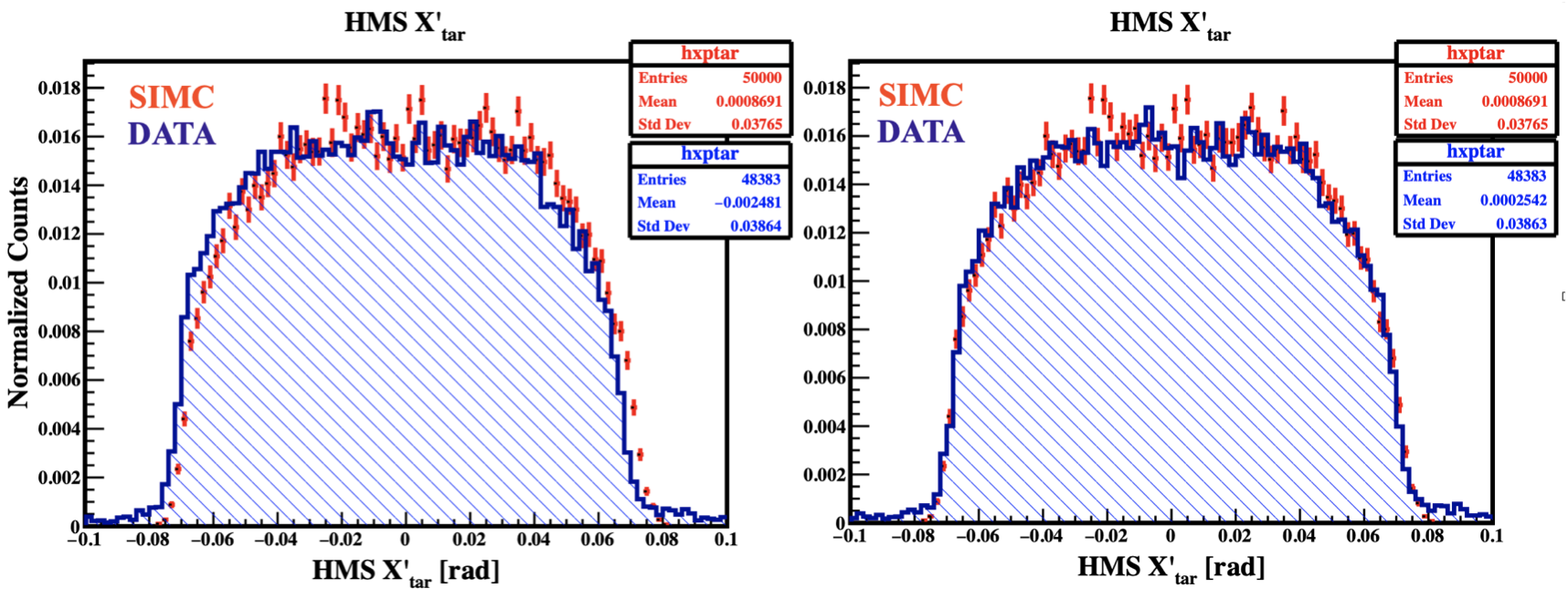}
      \caption{HMS $X'_{\mathrm{tar}}$ for run 3288 before (left) and after (right) applying the offset correction.}
      \label{fig:hxptar_Offset}
    \end{figure}

    The relative angle offsets refer to the angle offset relative to the spectrometer central ray and can be classified as follows:
    \begin{itemize}
    \item \text{In-plane relative angle $(Y'_{\mathrm{tar}} + \delta\theta^{\mathrm{off}})$ offset,} \text{[h(p)theta\_offset]}
    \item \text{Out-of-plane relative angle $(X'_{\mathrm{tar}} + \delta\phi^{\mathrm{off}})$ offset,} \text{[h(p)phi\_offset]}
    \end{itemize}
    The $Y'_{\mathrm{tar}}$ offset is directly related to the spectrometer angle, and therefore has a direct impact on the electron/hadron kinematics, depending on which
    spectrometer is associated with the particle type. In E12-10-003, this offset was determined for the HMS in order to align the X-component of the missing
    momentum as well as to improved the HMS central momentum correction. Recall that in Section \ref{sec:hms_optics} it was assumed that the proton (HMS) angle
    was well known, which is not completely true. \\
    \indent Figure \ref{fig:hxptar_Offset} shows the relative out-of-plane angle distributions for all events within the spectrometer acceptance. The \textit{zero} value in the distribution
    represents events whose trajectory was parallel to the central ray, whereas the events away from the \textit{zero} value represent those events that are at an out-of-plane angle relative
    to the central ray. The $X'_{\mathrm{tar}}$ offset was determined by ``eye'', using the mean of the distribution. \\
    \indent Similarly to the relative out-of-plane angles, the relative in-plane angles in the $Y'_{\mathrm{tar}}$ distribution (not shown) represent angles relative to the central ray, with the
    \textit{zero} value representing particles parallel to the central ray. The $Y'_{\mathrm{tar}}$ offset was determined based on how well the X-component of the missing momentum between data and
    simulation were matched, as well as how well were the HMS momentum from data and simulation matched (see Fig. \ref{fig:hms_Pcorr}). \\
    \indent After determining the spectrometer offsets, a second iteration of the HMS and SHMS Optics check procedure was performed to obtain
    improved results. Finally, the four $^{1}$H$(e,e')p$ elastic runs were used to determine the HMS momentum corrections for the $^{2}$H$(e,e'p)n$ data, to be discussed in the next section.
    \subsection{HMS Momentum Calibration}
\begin{figure}[H]
      \centering
      \includegraphics[scale=0.4]{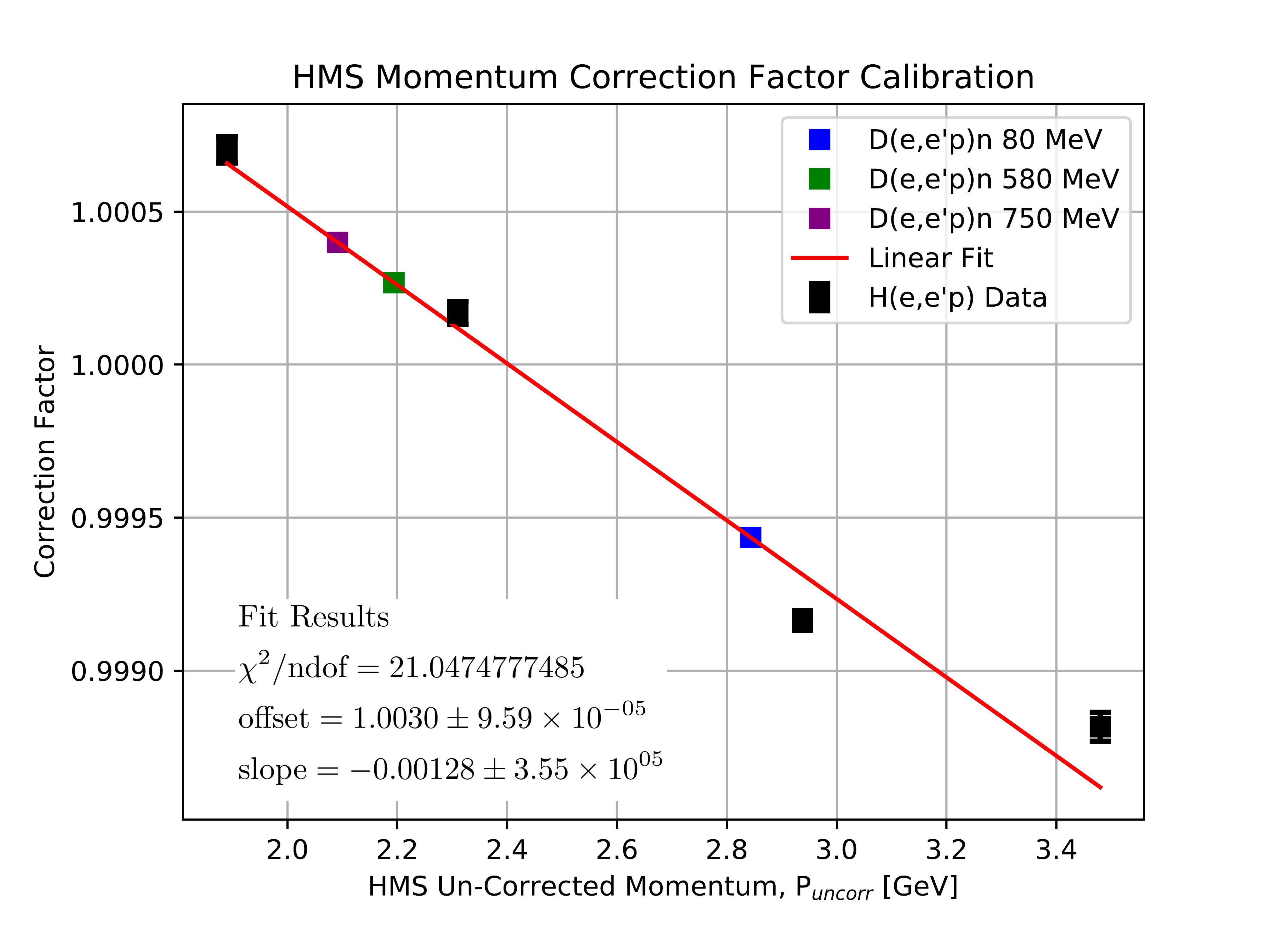}
      \caption{HMS momentum correction for $^{1}$H$(e,e')p$ and $^{2}$H$(e,e'p)n$. }
      \label{fig:hms_Pcorr_fit}
    \end{figure}

    During the E12-10-003 experiment, the four $^{1}$H$(e,e')p$ elastic runs analyzed covered the HMS momentum range such that the $^{2}$H$(e,e'p)n$ measured
    momentum was within the range covered by the elastic data. From this knowledge, one can determine the $^{2}$H$(e,e'p)n$ data momentum correction
    from a simple linear fit of the $^{1}$H$(e,e')p$ data. \\
    \indent From Fig. \ref{fig:hms_Pcorr_fit}, the momentum correction factor is plotted against the original HMS central momentum and the
    four data points are fit with a straight line. Using the line fit, the $^{2}$H$(e,e'p)n$ momentum correction for the three missing
    momentum settings are determined from the $^{2}$H$(e,e'p)n$ original HMS momentum setting.\\
    \indent Tables \ref{tab:original_deep_kin} and \ref{tab:new_deep_kin} summarize the $^{2}$H$(e,e'p)n$
    kinematics before and after the SHMS (see Section \ref{sec:shms_Pcent_corr}) and HMS central momentum
    corrections. Since the SHMS momentum was fixed during the experiment, the single correction factor
    determined from the $^{1}$H$(e,e')p$ data analysis applies for all runs.
    \begin{table}[H]
      \centering
      \scalebox{0.9}{
      \begin{tabular}{c c c c c}
        \hline
        \shortstack{$P_{\mathrm{m}}$ \\ Setting [MeV/c]}  & \shortstack{HMS \\ Angle [deg]} & \shortstack{HMS \\ Momentum [GeV/c]} & \shortstack{SHMS \\ Angle [deg]} & \shortstack{SHMS \\ Momentum [GeV/c]} \\
        \hline\hline
        80 & 38.896 & 2.8438 & 12.194 & 8.7 \\
        580 (set1) & 54.992 & 2.194 & 12.194 & 8.7 \\
        580 (set2) & 55.000 & 2.194 & 12.194 & 8.7 \\
        750 (set1) & 58.391 & 2.091 & 12.194 & 8.7 \\
        750 (set2) & 58.394 & 2.091 & 12.194 & 8.7 \\
        750 (set3) & 58.391 & 2.091 & 12.210 & 8.7 \\
        \hline
      \end{tabular}
      }
      \caption{Original $^{2}$H$(e,e'p)n$ kinematics for E12-10-003.}
      \label{tab:original_deep_kin}
    \end{table}
    \begin{table}[h!]
      \centering
      \scalebox{0.9}{
      \begin{tabular}{c c c c c}
        \hline
        \shortstack{$P_{\mathrm{m}}$ \\ Setting [MeV/c]}  & \shortstack{HMS \\ Angle [deg]} & \shortstack{HMS \\ Momentum [GeV/c]} & \shortstack{SHMS \\ Angle [deg]} & \shortstack{SHMS \\ Momentum [GeV/c]} \\
        \hline\hline
        80 & 38.896 & 2.840 & 12.194 & 8.5342 \\
        580 (set1) & 54.992 & 2.1925 & 12.194 & 8.5342 \\
        580 (set2) & 55.000 & 2.1925 & 12.194 & 8.5342 \\
        750 (set1) & 58.391 & 2.0915 & 12.194 & 8.5342 \\
        750 (set2) & 58.394 & 2.0915 & 12.194 & 8.5342 \\
        750 (set3) & 58.391 & 2.0915 & 12.210 & 8.5342 \\
        \hline
      \end{tabular}
      }
      \caption{Corrected $^{2}$H$(e,e'p)n$ kinematics for E12-10-003.}
      \label{tab:new_deep_kin}
    \end{table}
    \subsection{Spectrometer Acceptance Post-Optimization}
    After optics checks and optimization for each spectrometer, the data and simulated (SIMC) reconstructed variables at the target were compared using $^{1}$H$(e,e')p$ elastics run 3288. The
    ratio of the data-to-simulation was also taken and is plotted below. The data were normalized by the total accumulated charge and corrected for experimental inefficiencies
    that will be discussed in Chapter \ref{chap:chapter5}. In addition, several event selection cuts were applied for both data and simulation (see Section \ref{sec:event_selection}).
    \begin{figure}[H]
      \centering                                                         
      \includegraphics[scale=0.35]{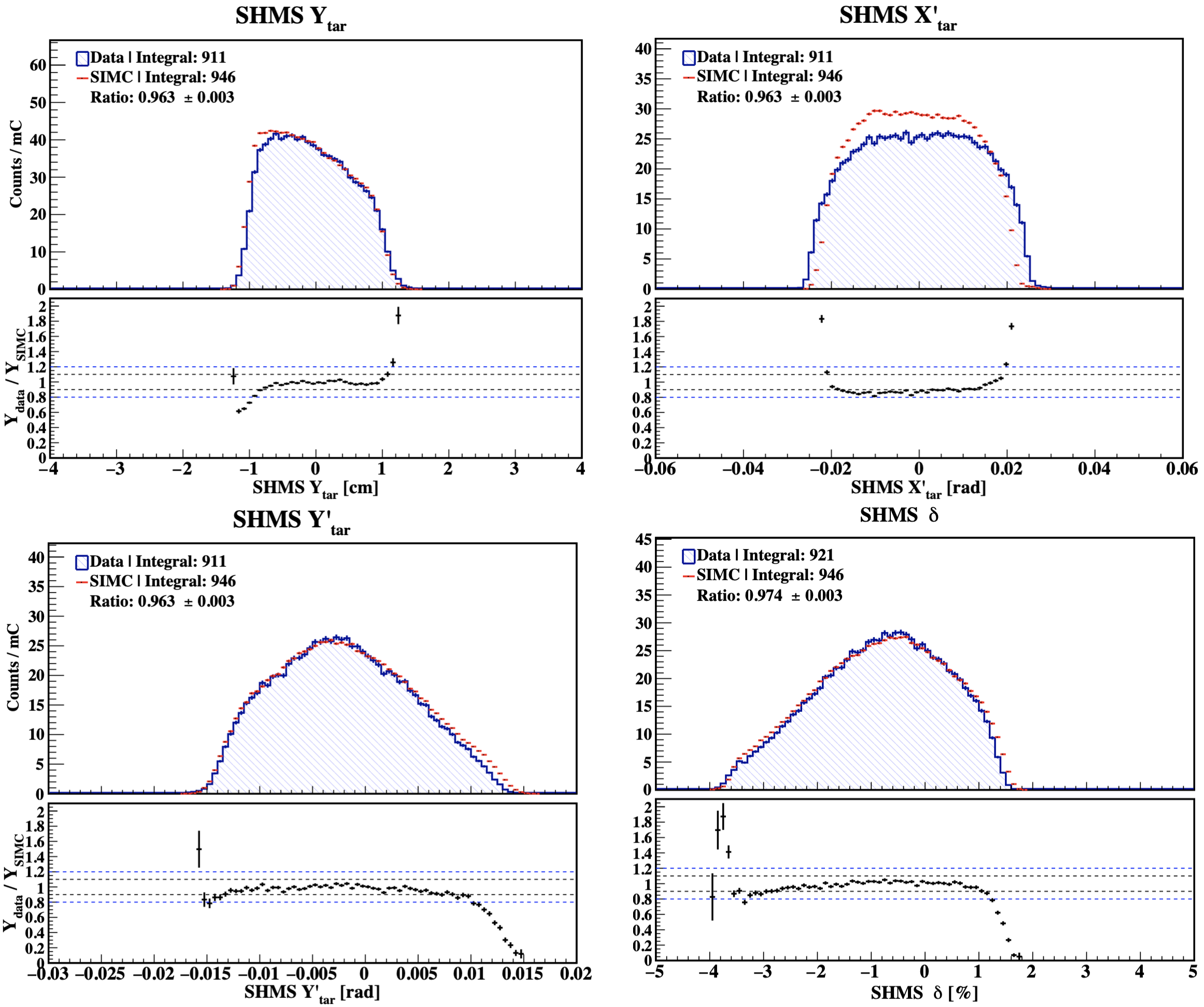}
      \caption{SHMS target reconstruction after optics optimization of $^{1}$H$(e,e')p$ elastic run 3288 for the E12-10-003.}
      \label{fig:shms_accpetance}
    \end{figure}
    \indent Figure \ref{fig:shms_accpetance} shows a generally good agreement betweem data and simulation over a wide range of the spectrometer acceptance. The main
    issues seem to be at the edges of the acceptance where differences beyond 20$\%$ (blue dashed line) can be observed in the ratios. For the most part of the acceptance, the ratios
    indicate a discrepancy of $\sim 10\%$ between data and simulation yields with the exception of $X'_{\mathrm{tar}}$, which seems to have a resolution issue as the simulation appears slighly narrower
    than data. The overall integrated yield over the entire range, however, shows only a discrepancy of $\sim 3-4\%$. The simulation used the proton form factor parametrization of Ref.\cite{PhysRevC.69.022201}
    to generate $^{1}$H$(e,e')p$ events. It is important to keep in mind that the simulation program does not take into account the uncertainties due to the elastic form factors that are used to
    simulate the hydrogen elastic events.
    \begin{figure}[H]
      \centering                                                         
      \includegraphics[scale=0.35]{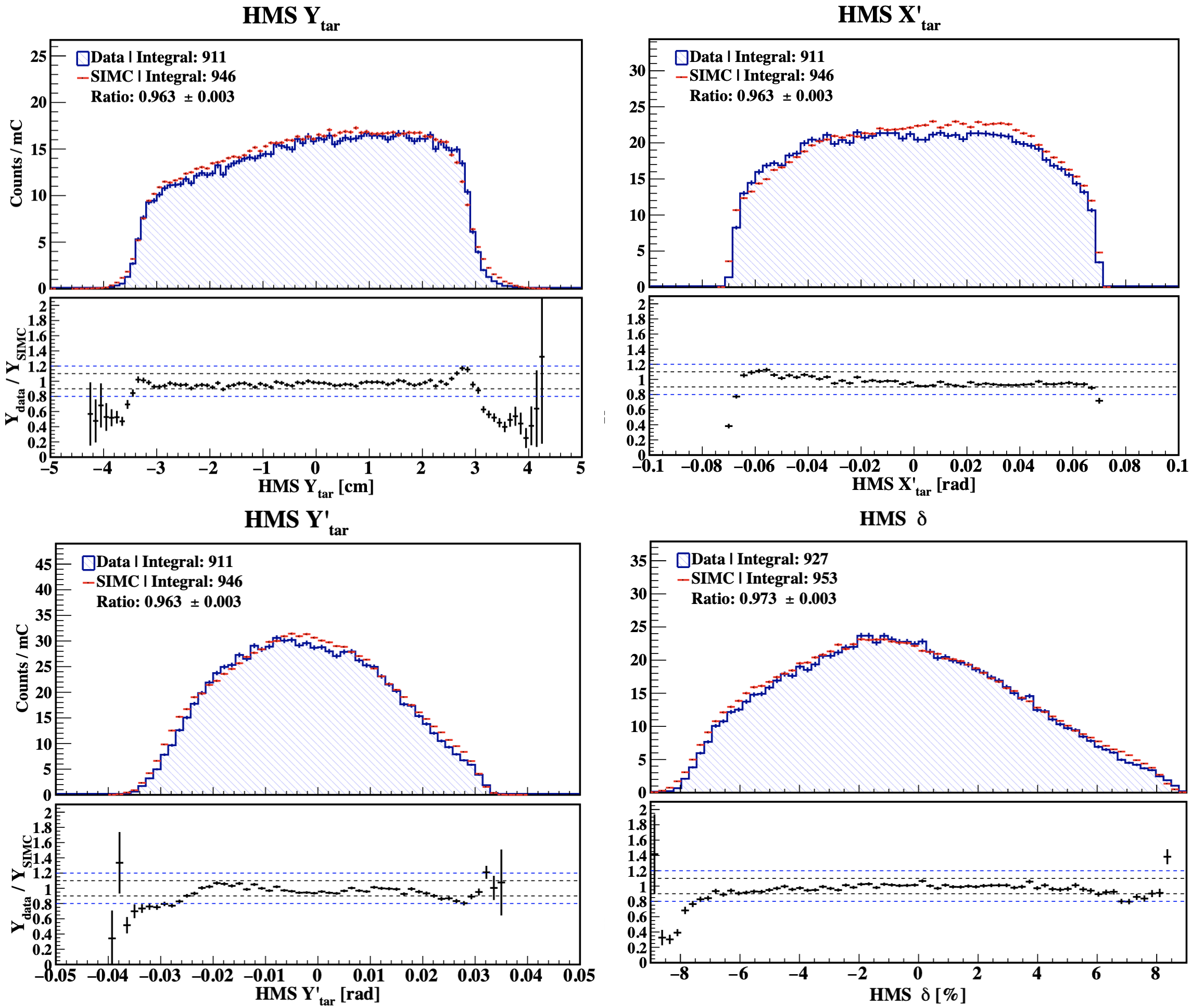}
      \caption{HMS target reconstruction after optics checks of $^{1}$H$(e,e')p$ elastic run 3288 for the E12-10-003.}
      \label{fig:hms_accpetance}
    \end{figure}
    \indent Similarly to the SHMS, the HMS reconstructed variables (see Fig. \ref{fig:hms_accpetance}) show a generally good agreement between data and simulation with discrepancies below $\sim 10\%$ for most of the acceptance range,
    with discrepancies beyond $\sim20\%$ at the edges. \\
    \indent Given that these studies were done using the coincidence elastic $^{1}$H$(e,e')p$ data, determining systematic effects due to our knowledge of the spectrometer acceptances
    is rather complicated given the correlations that exist between both spectrometer arms. Ideally, one would have to look at either single-arm elastic or deep-inelastic (DIS) data
    to carry out a complete spectrometer acceptance systematics study.
    \begin{figure}[H]
      \centering                                                         
      \includegraphics[scale=0.35]{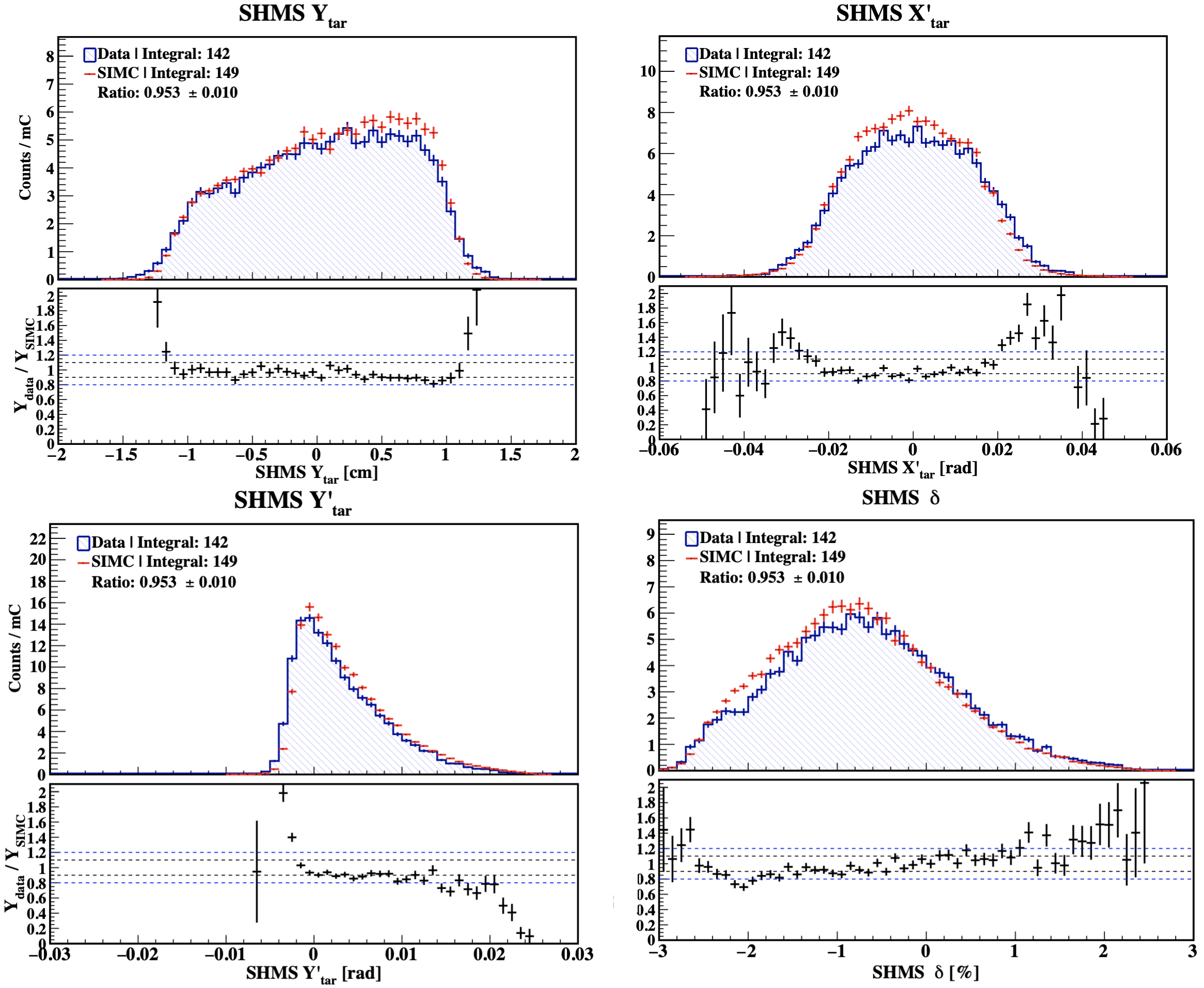}
      \caption{SHMS target reconstruction of $^{2}$H$(e,e'p)n$ run 3289 (80 MeV/c setting) for the E12-10-003.}
      \label{fig:shms_accpetance_3289}
    \end{figure}
    \indent The spectrometer acceptance for the deuteron 80 MeV/c setting are shown in Figs. \ref{fig:shms_accpetance_3289} and \ref{fig:hms_accpetance_3289} since
    the kinematics were very close to that of hydrogen elastics, and were used to check the spectrometer acceptance before looking at the higher momentum settings.
    The data have been normalized by the total charge and corrected for inefficiencies and in addition, have also been integrated over the full range of neutron
    recoil angles ($\theta_{nq}$) for better statistical precision.\\
    \indent Similar to the hydrogen, there is an overall good agreement between data and simulation with up to $\sim 20\%$ difference (blue dashed line) in the yield over
    most of the acceptance range on both spectrometers. The Laget FSI model\cite{LAGET2005} was used in the simulation. 
    \begin{figure}[H]
      \centering                                                         
      \includegraphics[scale=0.35]{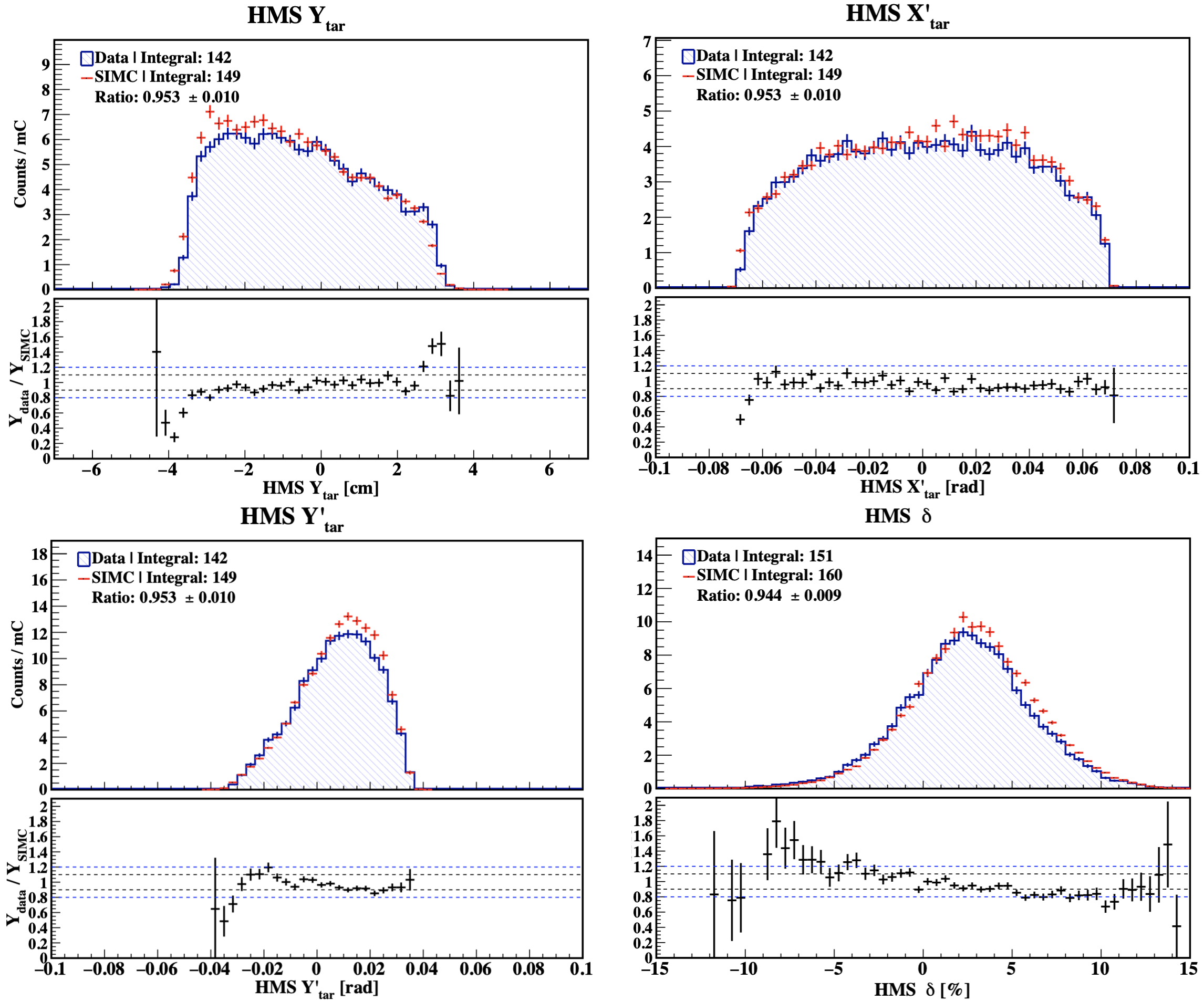}
      \caption{HMS target reconstruction of $^{2}$H$(e,e'p)n$ run 3289 (80 MeV/c setting) for the E12-10-003.}
      \label{fig:hms_accpetance_3289}
    \end{figure}
    \section{Event Selection}\label{sec:event_selection}
    A variety of cuts have been applied during the analysis of E12-10-003 to select true $^{1}$H$(e,e')p$ elastics and $^{2}$H$(e,e'p)n$ events at the reaction vertex.
    Due to the similarity in kinematics between the hydrogen elastic (run 3288) and the deuteron 80 MeV/c setting (run 3289), similar cuts were placed 
    on both hydrogen and deuteron data to select good events. The same cuts were also placed on the simulation for a direct comparison, and ultimately, for
    the determination of the spectrometer phase space from SIMC. An additional kinematic cut was placed on the deuteron data to select events at the highest momentum
    transfers ($Q^{2}$) allowed by the kinematics to further suppress MEC and IC contributions as stated in Section \ref{sec:MEC_IC_theory}. The same cuts placed on the 80 MeV/c setting were also placed on the 580
    and 750 MeV/c settings (not shown below) since the cut ranges were not affected by the change in kinematics from lower to higher missing momenta.
    \subsection{Missing Energy Cut}
    \begin{figure}[H]
      \centering                                                         
      \includegraphics[scale=0.29]{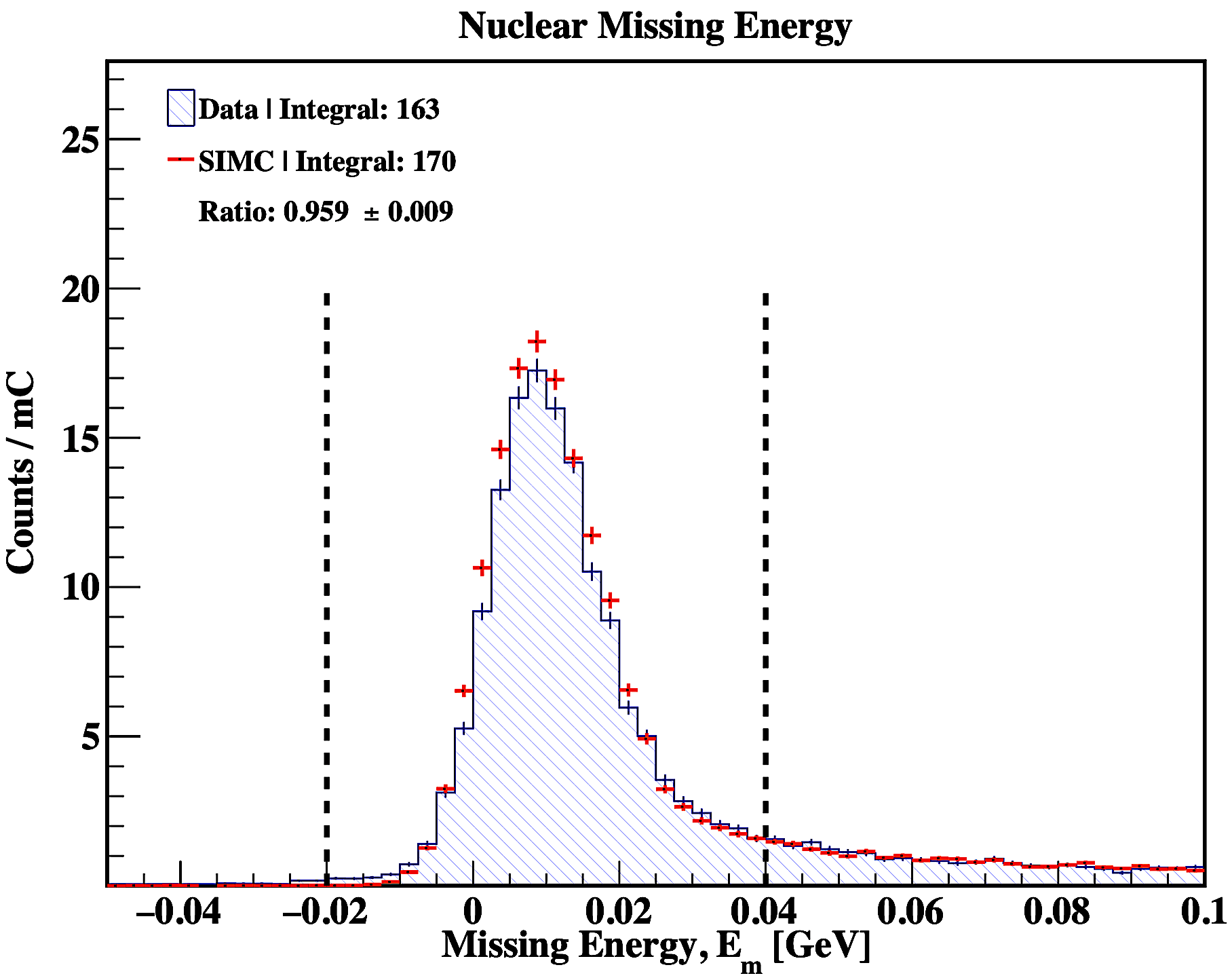}
      \caption{Missing energy cut on the 80 MeV/c setting of E12-10-003. }
      \label{fig:Em_3289}
    \end{figure}
    The primary cut used to select true $^{2}$H$(e,e'p)n$ is a missing energy cut around the deuteron binding energy ($\sim2.2$ MeV) from the formula in Eq. \ref{eq:2.4}
    where the recoiling system is assumed to be a neutron. The peak is not exactly at the deuteron binding energy because energy loss corrections have not been applied to the data nor SIMC.
    \subsection{Momentum Acceptance Cuts}
    \begin{figure}[H]
      \centering                                                         
      \includegraphics[scale=0.34]{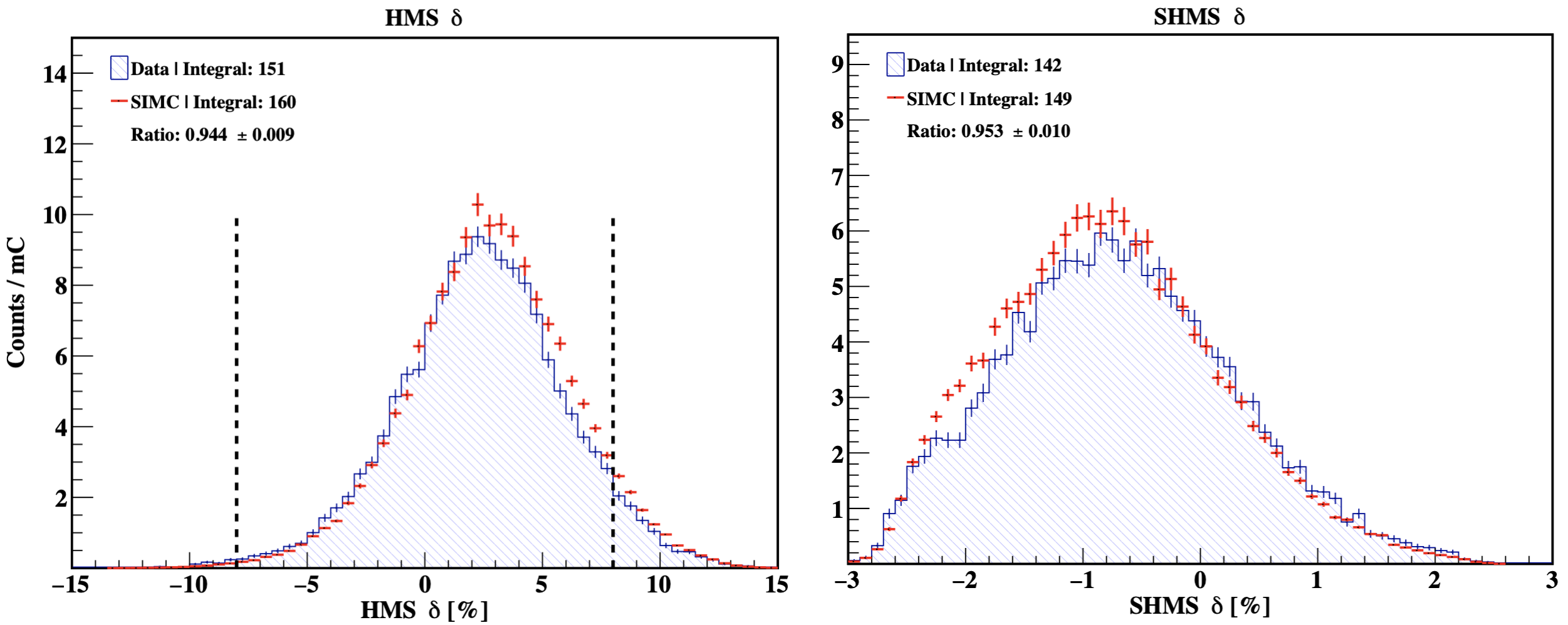}
      \caption{Momentum acceptance cuts on the 80 MeV/c setting of E12-10-003. }
      \label{fig:delta_accpt_3289}
    \end{figure}
    To ensure that events are reconstructed in a momentum acceptance region where the optics reconstruction matrix is reliable,
    a cut is placed on the HMS momentum acceptance in the range $-8\%<\delta_{\mathrm{HMS}}<8\%$, where the reconstruction is well known. Since the two spectrometers are in coincidence,
    there exists a correlation between the momentum acceptances. As a result, the HMS acceptance region at this particular kinematics
    automatically constrains the SHMS acceptance to be in the range $-3\%\lesssim\delta_{\mathrm{SHMS}}\lesssim2\%$. \\
    \subsection{HMS Collimator Cut}
    \begin{figure}[H]
        \centering                                                         
        \subfloat{\includegraphics[scale=0.31]{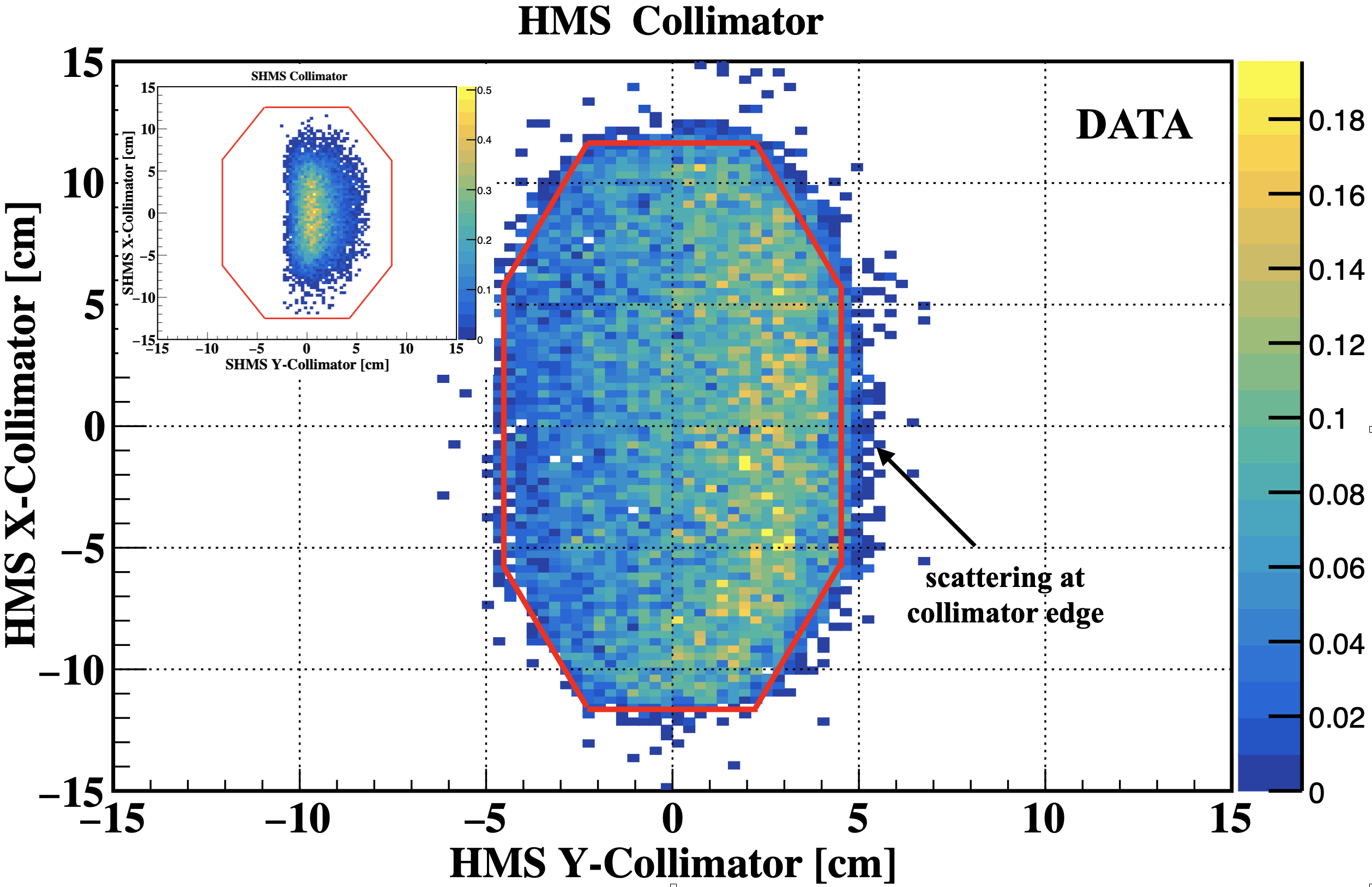}\label{fig:hmsColl_3289_data}}
        \caption{Data HMS collimator cut on the 80 MeV/c setting of E12-10-003. Inset: SHMS collimator geometry and reconstructed events.}
        \subfloat{\includegraphics[scale=0.31]{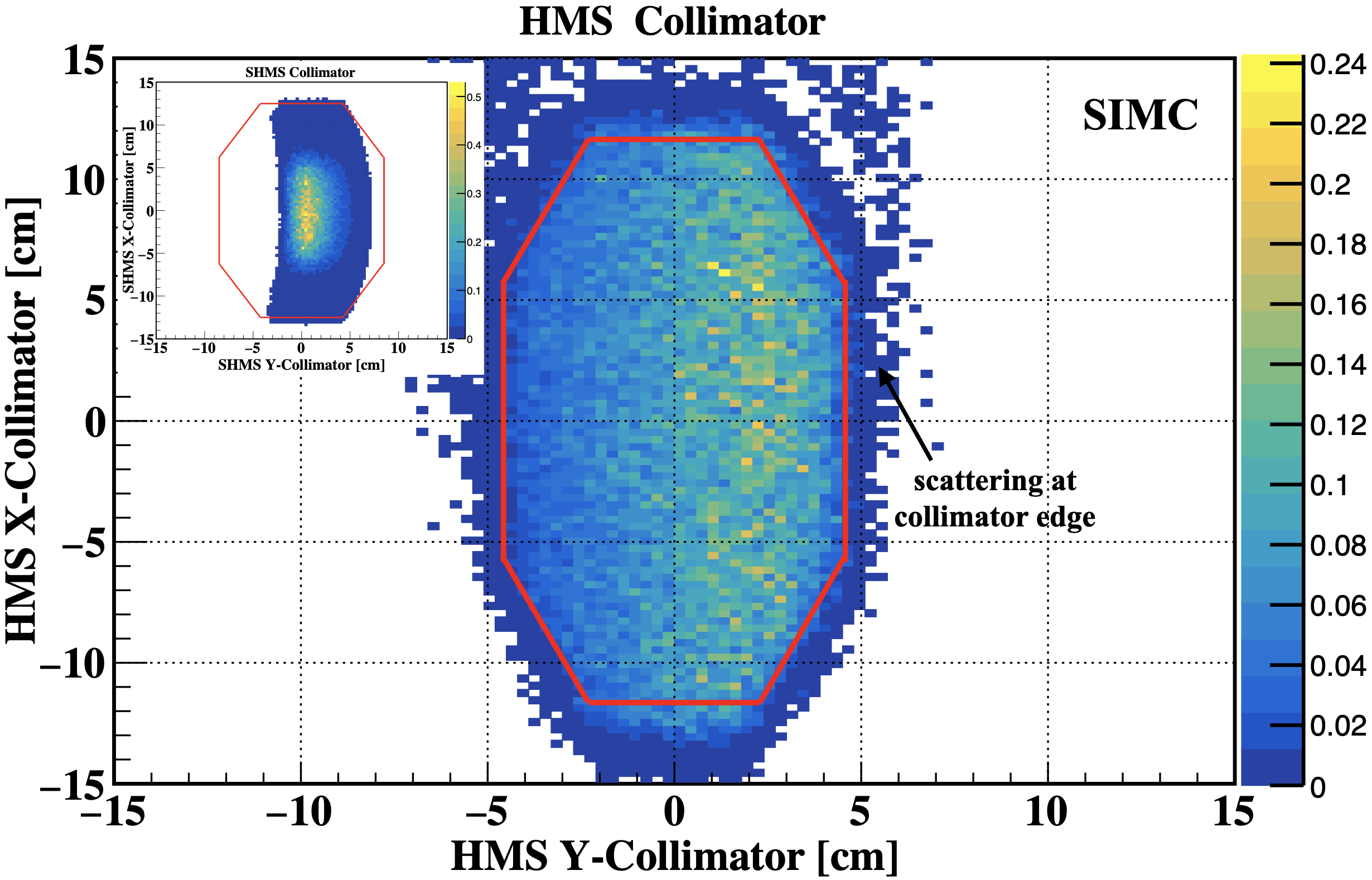}\label{fig:hmsColl_3289_simc}}
        \caption{Simulated HMS collimator cut on the 80 MeV/c setting of E12-10-003. Inset: SHMS collimator geometry and reconstructed events.}
    \end{figure}
    \noindent To make sure that all events that enter the spectrometer pass through the collimator, and not re-scatter at the edges,
    a cut is placed on the HMS collimator. The inset plots represent the SHMS collimator geometry whose events are constrained
    by the HMS collimator entrance. The events shown are projected at the collimator entrance and are functions of
    the reconstructed variables ($Y_{\mathrm{tar}}, X'_{\mathrm{tar}}, Y'_{\mathrm{tar}}, \delta$) and the surveyed collimator position 
    measured from the target center. It was preferred to put a geometrical cut rather than a cut on the reconstructed variables as the latter
    are subject to change when the spectrometer moves whereas the former is a fixed cut that defines the particles that enter the spectrometer.
    \subsection{Reaction $z_{\mathrm{v}}$-Vertex Difference Cut}
    \begin{figure}[h!]
      \centering                                                         
      \includegraphics[scale=0.58]{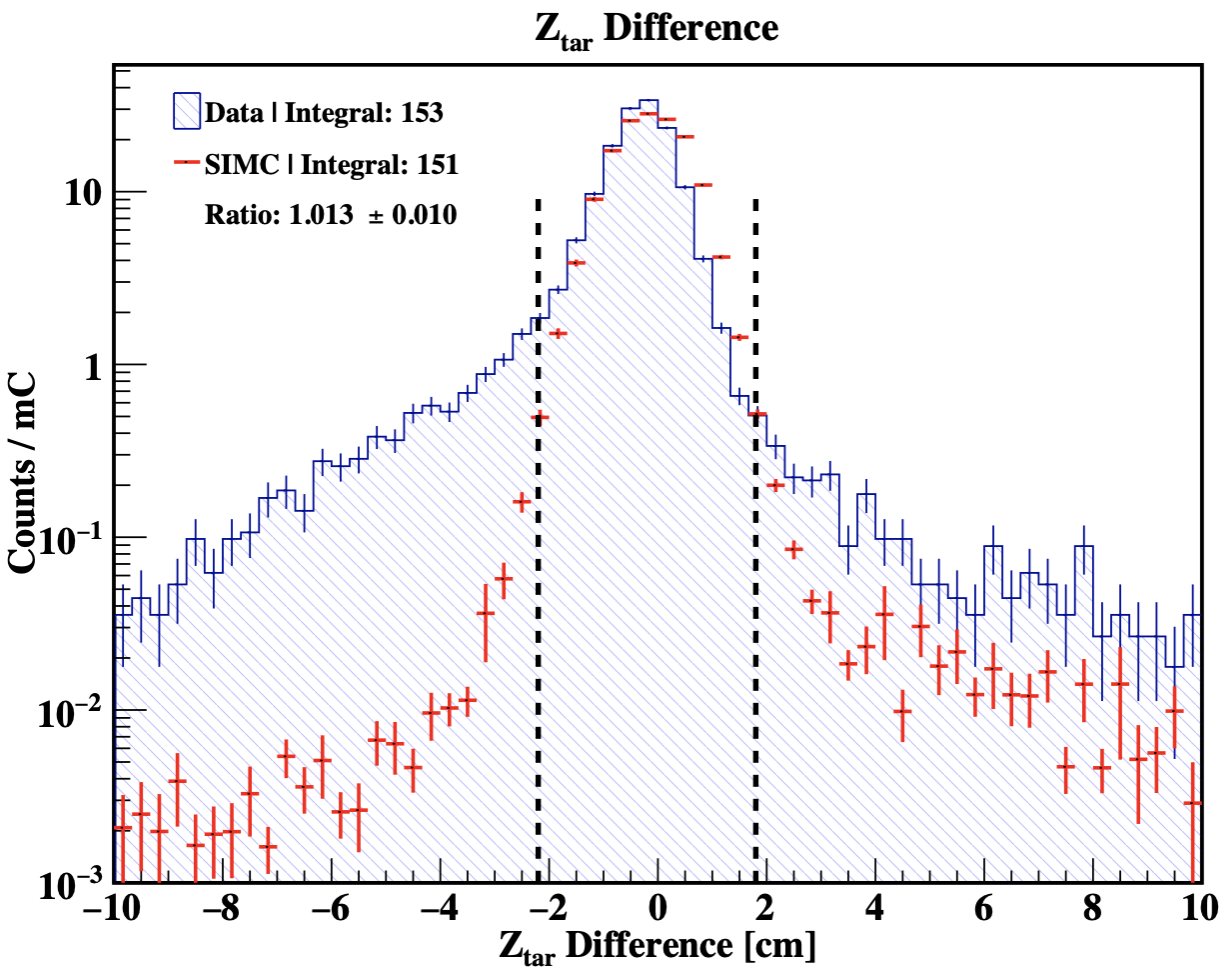}
      \caption{Reaction $z_{\mathrm{v}}$-vertex difference cut on the 80 MeV/c setting of E12-10-003. }
      \label{fig:ZtarDiff_3289}
    \end{figure}
    \noindent To ensure real coincidence events are selected, a cut was made on the difference between HMS and SHMS $z$-reaction vertex (using Eq. \ref{eq:4.14})
    at $\pm$ 2 cm relative to the peak. If the events originated from the same reaction vertex (i.e., true coincidences), the difference should
    peak at zero with a finite resolution width. If the events are uncorrelated (i.e., accidental coincidences), however, the reconstruction along the $z_{\mathrm{v}}$-vertex
    can be significantly different between the two spectrometers, which contributes to the tails of the distribution (see Fig. \ref{fig:ZtarDiff_3289}).
    Additionally, the tails can also arise from a bad $Y_{\mathrm{tar}}$ reconstruction, as the $z_{\mathrm{v}}$-vertex is calculated from this variable.
    \subsection{SHMS Calorimeter Cut}
    \begin{figure}[h!]
      \centering                                                         
      \includegraphics[scale=0.47]{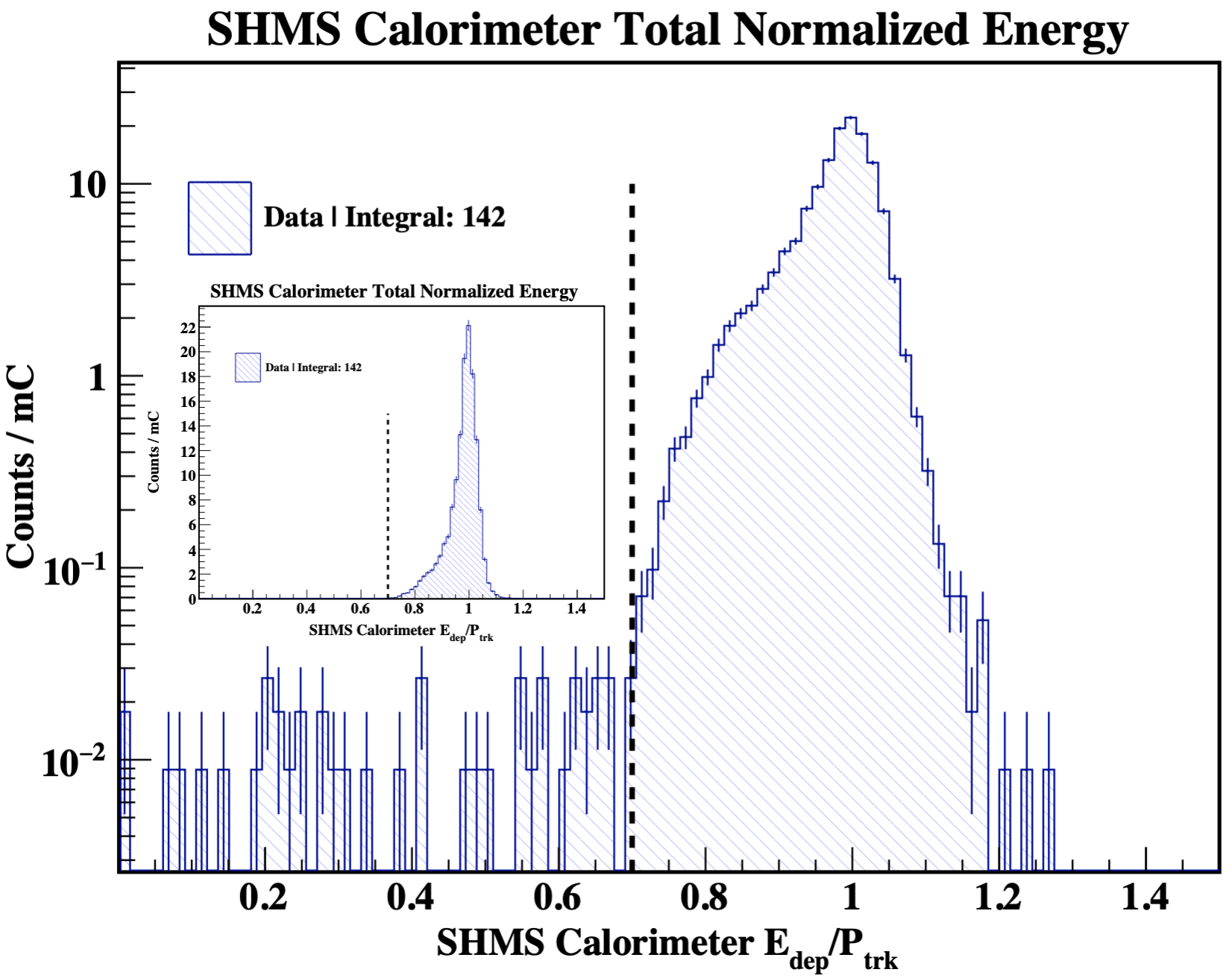}
      \caption{SHMS calorimeter cut on the 80 MeV/c setting of E12-10-003. }
      \label{fig:eCal_3289}
    \end{figure}
    \noindent The SHMS calorimeter was used to separate electrons from background (pions), however, as it is shown in Fig. \ref{fig:eCal_3289}, the
    deposited energy in the calorimeter normalized by the incident particle track shows a very clean distribution with a peak at one indicating
    the detected particles were electrons. The clean electron sample can be attributed to the low accidental trigger rates and low pion background
    in the SHMS to form these coincidences with the protons in the HMS.
    \subsection{Coincidence Time Cut}
    \begin{figure}[h!]
      \centering                                                         
      \includegraphics[scale=0.45]{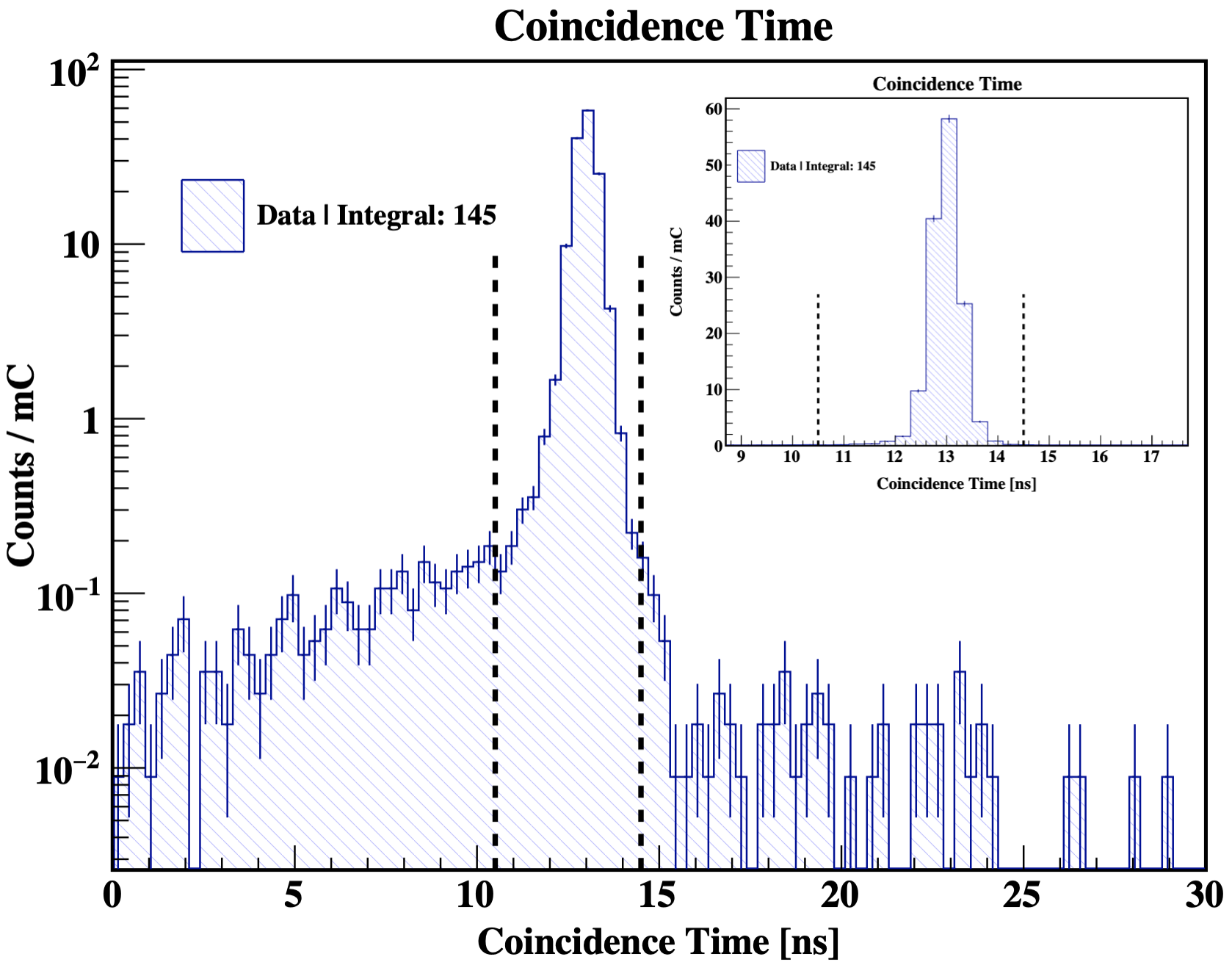}
      \caption{Coincidence time cut on the 80 MeV/c setting of E12-10-003. }
      \label{fig:Ctime_3289}
    \end{figure}
    \noindent To further clean the electron-proton coincidence sample of events, a coincidence cut was made in the range $10.5<t_{\mathrm{coin}}<14.5$ ns.
    Similar to the calorimeter spectrum, the coincidence time spectrum formed between the HMS and SHMS 3/4 triggers is very clean as the beam
    bunch structure is not observed. The out-of-time events at the tails can originate from the radiation in the Hall partially entering the
    detector huts and forming a trigger in coincidence with the other arm.
    \subsection{Four-Momentum Transfer ($Q^{2}$) Cut}
    \begin{figure}[h!]
      \centering                                                         
      \includegraphics[scale=0.50]{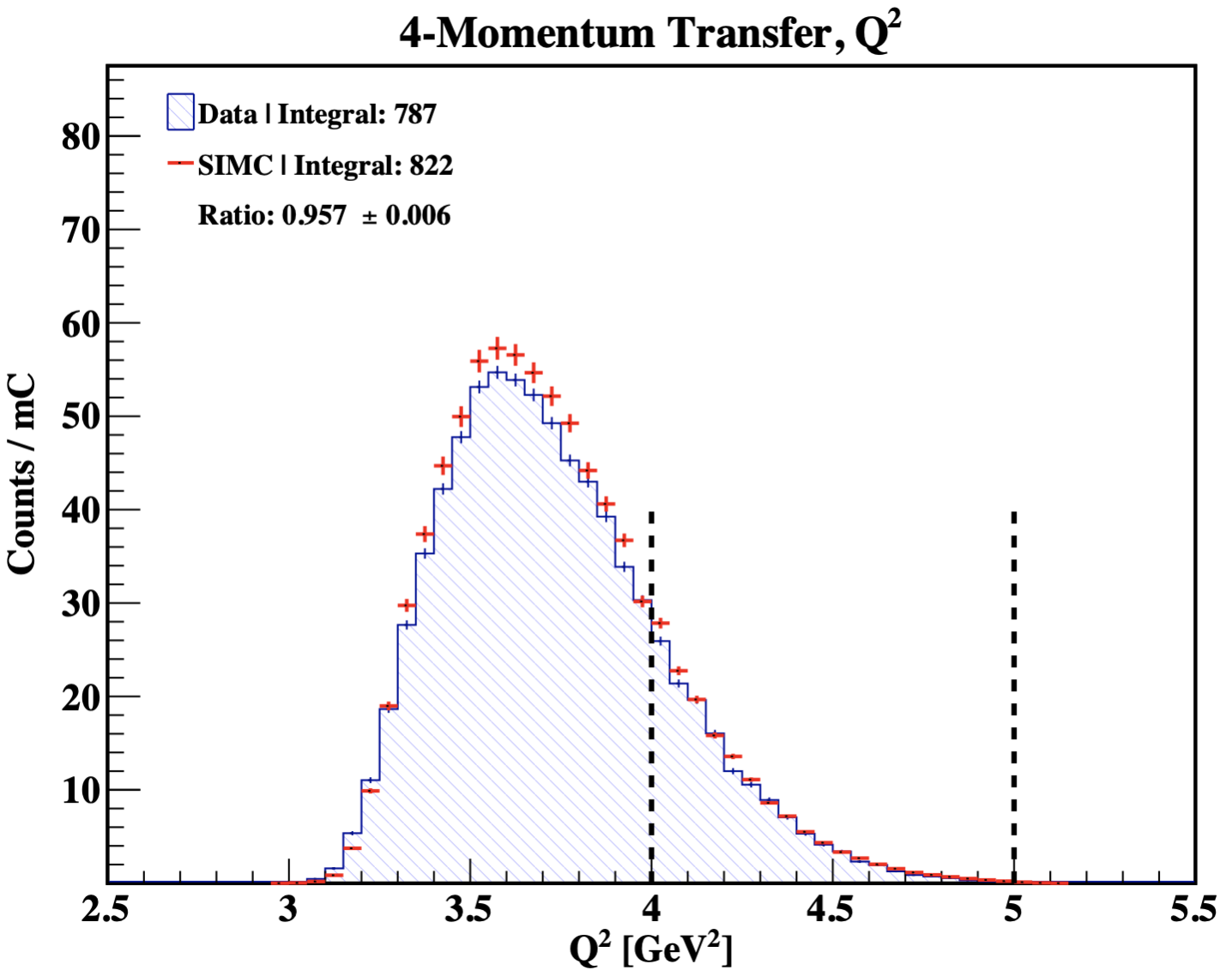}
      \caption{Four-momentum transfer ($Q^{2}$)cut on the 80 MeV/c setting of E12-10-003. }
      \label{fig:Q2_3289}
    \end{figure}
    \noindent A kinematical cut on the 4-momentum transfer is made at $Q^{2} = 4.5\pm0.5$ (GeV/c)$^{2}$ to select events only at the highest possible momentum transfers.
    The previous deuteron experiment in Hall A \cite{PhysRevLett.107.262501} measured the cross sections up to $p_{\mathrm{r}}=$550 MeV/c and $Q^{2}=3.5\pm0.25$ (GeV/c)$^{2}$,
    whereas this experiment seeks to probe the deuteron momentum distributions at extreme kinematics by moving beyond 500 MeV/c recoil momenta
    at the highest $Q^{2}$, where MEC and IC are suppresed. The cross sections for this experiment were also extracted at a lower kinematic
    bin of $Q^{2}=3.5\pm0.5$ for comparison with the Hall A data as well as achieving a higher statistical precision compared to our higher $Q^{2}$ setting.

\chapter{DATA CROSS SECTION EXTRACTION}	\label{chap:chapter5}
In this chapter I discuss how the experimental cross section was determined
for this experiment as well as the various corrections applied to extract the yield. In addition,
I will also describe the studies to determine the systematic uncertainties from various sources on the cross section.
\section{Experimental $^{2}$H$(e,e'p)n$ Cross Section} \label{sec:exp_Xsec}
\begin{figure}[H]
  \centering
  \includegraphics[scale=0.36]{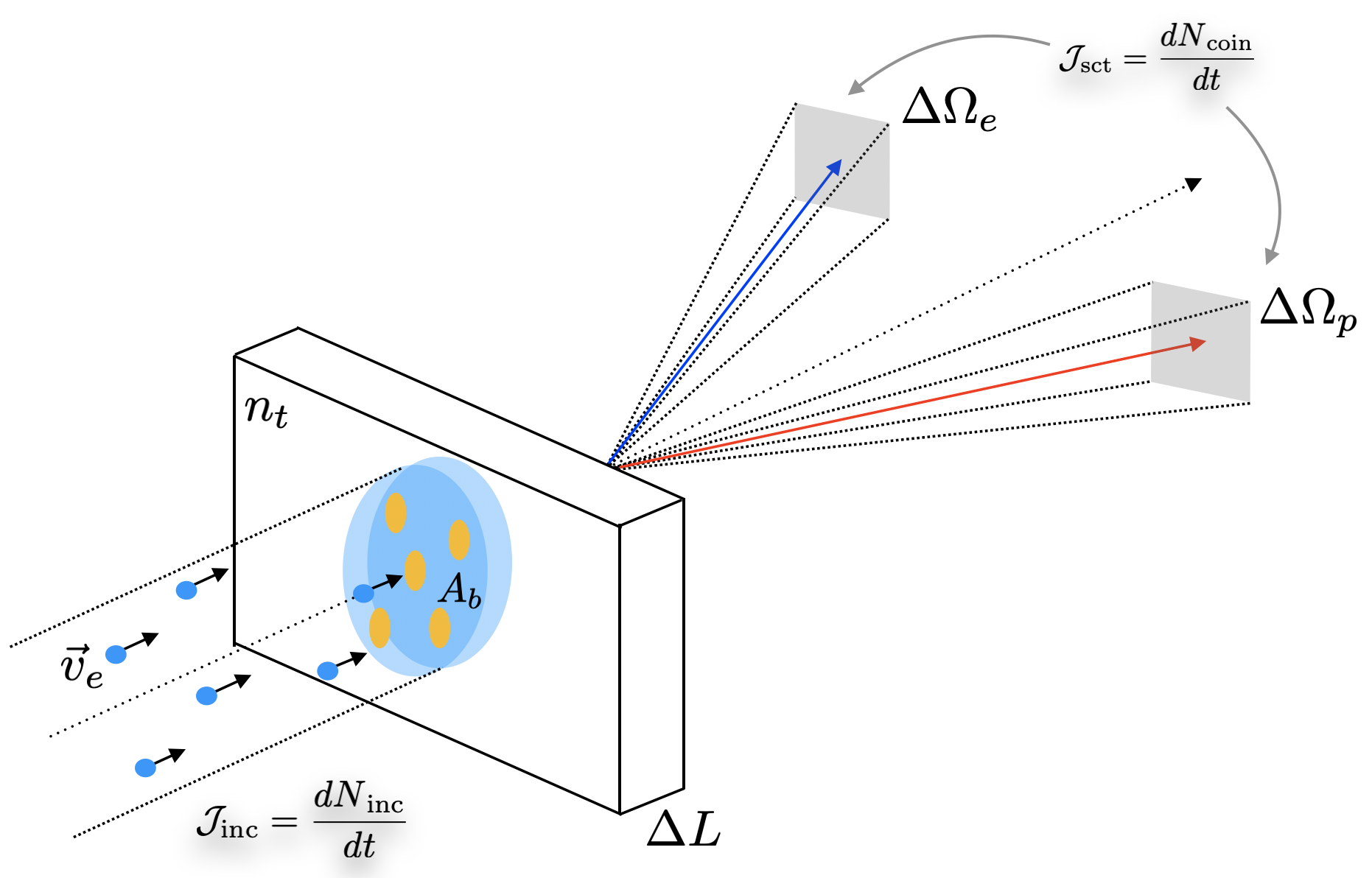}
  \caption{Cartoon representation of a typical coincidence experiment.}
  \label{fig:exp_Xsec}
\end{figure}

The deuteron cross section was introduced in Section \ref{sec:theory_Xsec} from a theoretical approach, however, to make a direct comparison between
theory and experiment, one must also consider how the cross section is determined experimentally. Figure \ref{fig:exp_Xsec} shows a simple cartoon of a
typical coincidence experiment that will be used to derive the experimental cross section.\\
\indent Consider an electron beam with cross-sectional area $A_{b}$[cm$^{2}$] incident on a target of length (thickness) $\Delta L$[cm] and density $n_{t}$([g]/[cm$^{3}$]). The incident beam flux on the
target is then defined as
\begin{equation}
  \mathcal{J}_{\mathrm{inc}} \equiv n_{b} A_{b} v_{b} = \frac{dN_{\mathrm{inc}}}{dt},
  \label{eq:5.1}
\end{equation}
where $n_{b}$ is the beam density in $N_{\mathrm{inc}}/\mathrm{[cm]}^{3}$, $N_{\mathrm{inc}}$ is the total number of incident particles,  and $v_{b}$ is the
electron velocity in $\mathrm{[cm]}/\mathrm{[s]}$. The beam flux is also interpreted as the instantaneous rate of the number of beam particles incident on an target area $A_{b}$ (beam current).
The beam current is ususally measured in milli-Coulomb per second (mC/s), which can also be expressed as the total number of incident electrons per second using the conversion factor: 1 mC/s $\equiv$ 6.2415x10$^{15}$ e$^{-}$/s.
As the electron beam passes through the target, each electron can interact with any target atom within the area $A_{b}$. The total number of atoms that can potentially be scattered by the beam within this area
are determined to be
\begin{equation}
 N_{t} \equiv n_{t} A_{b} \Delta L. 
  \label{eq:5.2}
\end{equation}
In reality, only a certain fraction of the total target atoms that can be scattered will interact with the beam, which is characterized by the probability,
\begin{equation}
  p = N_{t}\frac{\sigma}{A_{b}} = n_{t}\Delta L \sigma,
  \label{eq:5.3}
\end{equation}
where $\sigma$ is defined as the cross section and describes the effective area of interaction for a particular reaction out of a total area $A_{b}$.\\
\indent From the number of interactions between the incident electrons and target atoms, the reaction rate $\mathcal{J}_{\mathrm{sct}}$ of the scattered
particles is determined from the probability that the beam flux interacts with the target atoms, which is defined as
\begin{equation}
  \mathcal{J}_{\mathrm{sct}} \equiv p\mathcal{J}_{\mathrm{inc}} = \mathcal{J}_{\mathrm{inc}}n_{t}\Delta L \sigma.
  \label{eq:5.4}
\end{equation}
From \ref{eq:5.4} a useful quantity to define is the experimental luminosity,
\begin{equation}
  \mathcal{L} \equiv \mathcal{J}_{\mathrm{inc}}n_{t}\Delta L
  \label{eq:5.5}
\end{equation}
in units of [cm$^{-2}$][s$^{-1}$], which describes the total number of interactions that can be produced from beam particles illuminating a specific target area.
From the luminosity, the experimental cross section can be expressed in its simplest form as
\begin{equation}
  \sigma = \frac{\mathcal{J}_{\mathrm{sct}}}{\mathcal{L}},
  \label{eq:5.6}
\end{equation}
where the cross section is a constant of proportionality between the luminosity and reaction rate. \\
\indent To determine the differential cross section in a typical coincidence experiment, consider a general $A(e,e'p)$ reaction where the incident electron knocks out a proton
and the $(A-1)$ recoiling system is undetected. The scattered electron and knocked-out proton are detected in coincidence between each spectrometer within a finite angular
acceptance region limited by either the spectrometer apertures or collimator (if inserted). In reality, the acceptance region (phase space) covered by the electrons
in coincidence with protons is determined by the reaction kinematics, which may be smaller than the full acceptance of the spectrometers. For the specific case of the deuteron
break-up reaction, the phase space was determined via a Monte Carlo simulation (SIMC) by randomly generating a set of values, ($X'_{e,\mathrm{tar}}, Y'_{e,\mathrm{tar}}, X'_{p,\mathrm{tar}}, Y'_{p,\mathrm{tar}}, E'$)\footnote{\singlespacing
  The primed variables represent relative spectrometer angles introduced in Section \ref{sec:hall_coord} and $E'$ is the scattered electron energy (or approximately, its momentum).},
which along with a known beam energy, completely determines the final proton momentum. The volume formed by this hypercube was defined as
\begin{equation}
  V_{\mathrm{PS}} = \frac{N_{\mathrm{acc}}}{N_{\mathrm{gen}}}\Delta\Omega_{e}\Delta\Omega_{p}\Delta E', 
  \label{eq:5.7}
\end{equation}
where $N_{\mathrm{acc}}$ is the number of accepted coincidence events that lie in the hypercube and $N_{\mathrm{gen}}$ is the total number of events generated by the random sampling process.  
The $\Delta\Omega_{(e,p)} = \Delta X'_{(e,p)\mathrm{tar}}\Delta Y'_{(e,p)\mathrm{tar}}$ defines the angular range of the electrons in coincidence with protons and $\Delta E'$ defines the range of the scattered electron momenta.
The differential cross section can then be determined by dividing the total number of detected electron-proton coincidences in the experiment by the Monte Carlo generated phase space at the
same reaction kinematics. By substituting Eqs. \ref{eq:5.5} into \ref{eq:5.6} and dividing by Eq. \ref{eq:5.7} one obtains
\begin{equation}
  \frac{d^{5}\sigma}{d\Omega_{e}d\Omega_{p}dE'} = \frac{\mathcal{J}_{\mathrm{sct}}}{\mathcal{J}_{\mathrm{inc}}n_{t}\Delta L\mathcal{J}_{\mathrm{corr}}(\Delta\Omega_{e,p}\rightarrow d\Omega_{e,p})\frac{N_{\mathrm{acc}}}{N_{\mathrm{gen}}}\Delta\Omega_{e}\Delta\Omega_{p}\Delta E' },
  \label{eq:5.8}
\end{equation}
where $\mathcal{J}_{\mathrm{corr}}(\Delta\Omega_{e,p}\rightarrow d\Omega_{e,p})$ is a Jacobian matrix that is used to convert the solid angles from the spectrometer coordinates to spherical coordinates.
The incident ($\mathcal{J}_{\mathrm{inc}}$) and reaction ($\mathcal{J}_{\mathrm{sct}}$) rates can be integrated over the entire experimental run time to obtain
\begin{align}
  &\int \mathcal{J}_{\mathrm{inc}}dt \equiv \int \frac{dN_{\mathrm{inc}}}{dt} dt = N_{\mathrm{inc}},  \\ 
  &\int \mathcal{J}_{\mathrm{coin}}dt \equiv \int \frac{dN_{\mathrm{coin}}}{dt} dt = N_{\mathrm{coin}},  
\end{align}
where $N_{\mathrm{inc}}$ is the total number of incident electrons, which is normalized to 1 mC in SIMC and $N_{\mathrm{coin}}$ is the true number of detected $^{2}$H$(e,e'p)n$ coincidences,
provided that the event selection cuts defined in Section \ref{sec:event_selection} have been applied. It is important to note that these coincidences, hereafter
referred to as $Y_{\mathrm{uncorr}}$, have not been corrected for detector inefficiencies. Therefore, to obtain the final reaction cross section, the data yield has been
corrected as follows:
\begin{equation}
  Y_{\mathrm{corr}} \equiv \frac{Y_{\mathrm{uncorr}}\cdot f_{\mathrm{rad}}}{Q^{\mathrm{exp}}_{\mathrm{tot}}\cdot\epsilon_{\mathrm{tLT}}\cdot\epsilon_{\mathrm{htrk}}\cdot\epsilon_{\mathrm{etrk}}\cdot\epsilon_{\mathrm{tgt.Boil}}\cdot\epsilon_{\mathrm{pTr}}},
  \label{eq:5.11}
\end{equation}
where  $Q^{\mathrm{exp}}_{\mathrm{tot}}$ is the total experimental accumulated charge from the electron beam at the target, $f_{\mathrm{rad}}$ is the correction due to radiative effects, and the $\epsilon_{i}$'s are corrections
from various experimental inefficiencies described in the following sections. The normalization of the data by the accumulated charge (yield/mC) is necessary for a direct comparison to the SIMC yield.
Substituting Eq. \ref{eq:5.11} into Eq. \ref{eq:5.8}, the final averaged experimental cross section for the $^{2}$H$(e,e'p)n$ can be expressed as
\begin{equation}
  \Big(\frac{d^{5}\bar{\sigma}}{d\Omega_{e}d\Omega_{p}dE'}\Big)_{\mathrm{k}} = \Big(\frac{Y_{\mathrm{corr}}}{n_{t}\Delta L\mathcal{J}_{\mathrm{corr}}(\Delta\Omega_{e,p}\rightarrow d\Omega_{e,p})\frac{N_{\mathrm{acc}}}{N_{\mathrm{gen}}}\Delta\Omega_{e}\Delta\Omega_{p}\Delta E'}\Big)_{\mathrm{k}},
  \label{eq:5.12}
\end{equation}
where the ``average'' refers to the fact that the cross section has been calculated at the center of the k$^{\text{th}}$ kinematic bin in question, where k = $p_{r}, Q^{2}, x_{Bj}, \theta_{nq},$ etc.
In reality, the true cross section must be determined at the averaged kinematics for that bin. See Section \ref{sec:bin_cent_corr} for a detailed discussion of the bin-centering corrections.\\
\indent Similar to the averaged data cross sections, the Laget model cross sections implemented in SIMC were determined using Eq. \ref{eq:5.12}, where the
simulated yield had the same cuts as the data yield, but the efficiencies were all $\epsilon_{i} = 1$ as SIMC does not simulate detector inefficiencies. 
\section{Tracking Efficiencies ($\epsilon_{\mathrm{htrk}}$, $\epsilon_{\mathrm{etrk}}$)}
\begin{figure}[H]
  \centering
  \includegraphics[scale=0.73]{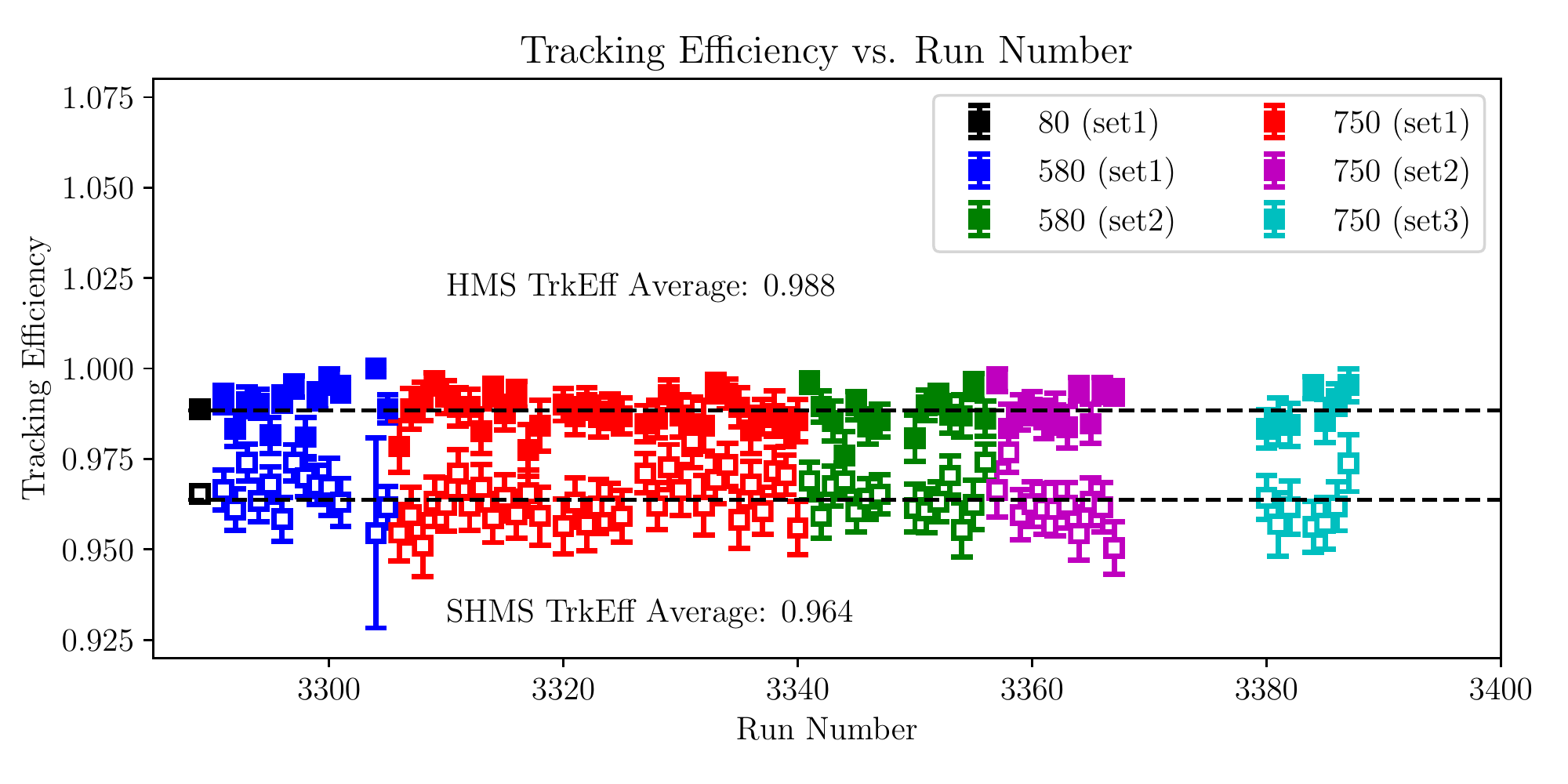}
  \caption{Tracking efficiency of HMS (open) and SHMS (full) for the E12-10-003 experiment.}
  \label{fig:trk_eff}
\end{figure}

\begin{figure}[H]
  \centering
  \includegraphics[scale=0.65]{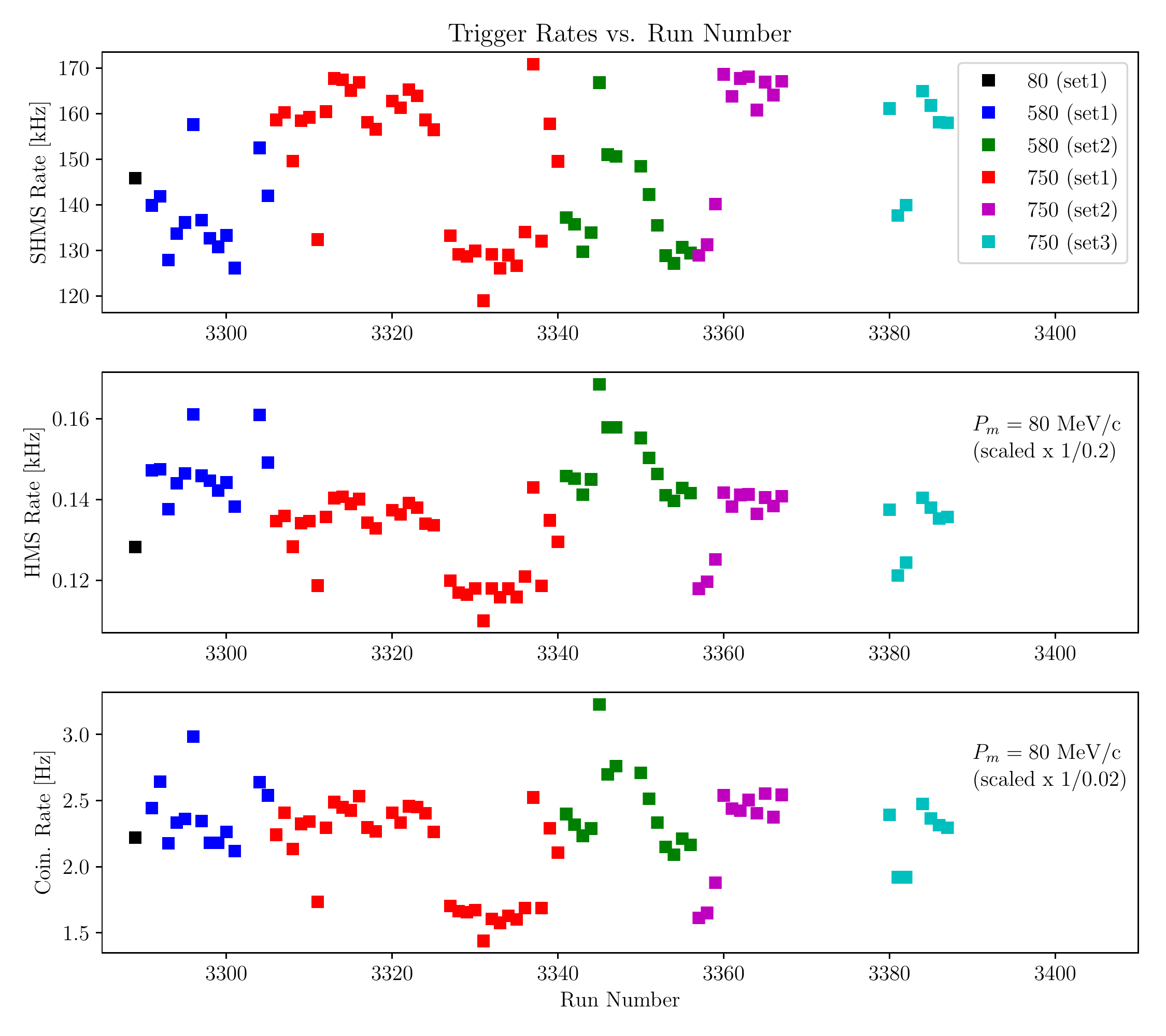}
  \caption{Trigger rates for the SHMS (top), HMS (middle) and coincidence trigger (bottom) during the E12-10-003 experiment.}
  \label{fig:trg_rates}
\end{figure}

To account for the experimental yield loss due to a bad track reconstruction or the selection of the wrong track by the Hall C tracking algorithm,
the tracking efficiencies have been determined. The tracking efficiency per experimental run is generally defined as
\begin{subequations}
\begin{align}
  &\epsilon_{(\mathrm{htrk,etrk})} \equiv \frac{N_{\mathrm{did}}}{N_{\mathrm{should}}},\\
  &\delta\epsilon_{(\mathrm{htrk,etrk})} \equiv \frac{\sqrt{N_{\mathrm{should}}-N_{\mathrm{did}}}}{N_{\mathrm{should}}},
\end{align}
\end{subequations}
where $N_{\mathrm{did}}$ are the number of events for which there was at least one track formed by the drift chambers tracking algorithm given a specific criteria
and $N_{\mathrm{should}}$ are the number of events where at least one track was expected but was not necessarily reconstructed by the algorithm using the same criteria.
For simplicity, we define the logical operator AND as ``$\wedge$''  and the EQUALITY operator as ``$==$'' to be used in the tracking criteria definition.
For the electron arm (SHMS), the criteria for the formation of a track was defined as
\begin{align}
  & N_{\mathrm{should}}\footnotemark \equiv  (N_{\mathrm{goodScinHit}}==1) \wedge (\beta_{\mathrm{notrk}}>0.5) \wedge (\beta_{\mathrm{notrk}}<1.5) \wedge \label{eq:5.14} \\ 
  & (E_{\mathrm{tot.norm}}>0.6) \wedge (N_{\mathrm{NGC,npeSum}}>0.5), \nonumber \\
  & N_{\mathrm{did}} \equiv (N_{\mathrm{should}}) \wedge (N_{\mathrm{DCtrk}}>0).
\end{align}
\footnotetext{\singlespacing As an example the tracking criteria defined in Eq.\ref{eq:5.14} can be read out as: ``the total number of events that \textit{should}
  have passed the tracking criteria require that the total number of good scintillator hits must be exactly 1 AND $\beta_{\mathrm{notrk}}$ be between 0.5 and 1.5 AND the total normalized
calorimeter energy be below 0.6 AND the total number of \v{C}erenkov photoelectrons be below 0.5''}
\noindent For the hadron arm (HMS), a similar set of variables were used and defined as follows:
\begin{align}
  & N_{\mathrm{should}} \equiv  (N_{\mathrm{goodScinHit}}==1) \wedge  (\beta_{\mathrm{notrk}}>0.5) \wedge (\beta_{\mathrm{notrk}}<1.5) \wedge \\ 
  & (E_{\mathrm{tot.norm}}<0.6) \wedge (N_{\mathrm{CER,npeSum}}<0.5), \nonumber \\
  & N_{\mathrm{did}} \equiv (N_{\mathrm{should}}) \wedge (N_{\mathrm{DCtrk}} > 0).
\end{align}
The variables for both spectrometers are defined as
\begin{itemize}
\item  $N_{\mathrm{goodScinHit}}$: Number of good scintillator hits, which can either be 1 or 0. The requirement for one hit is that the candidate track
  passes through a fiducial region in each XY hodoscope plane. The fiducial region is defined by requiring a TDC hit on a certain number
  of scintillator paddles that are adjacent to the paddle the candidate track passed through.
\item $\beta_{\mathrm{notrk}}$: The hodoscope beta ($\beta=v/c$) is calculated \textit{without using tracking information}, as this would make the drift chamber efficiency calculation biased.
\item $E_{\mathrm{tot.norm}}$: The total energy deposited in the calorimeter normalized by the spectrometer central momentum. In the SHMS, a cut is made $>0.6$
  to select electrons, whereas in the HMS, it is chosen to be $<0.6$ to suppress positrons. Due to the low experimental background, both HMS and SHMS spectra
  were very clean after all cuts were applied.
\item $N_{\mathrm{NGCER, CER,npeSum}}$: The total number of photoelectrons from the NGC detector in SHMS or HGC detector on the HMS. A cut is made $>0.5$ for SHMS
  and $<0.5$ for HMS to select the respective particles in each spectrometer. Similar to the calorimeter, the spectra for the \v{C}erenkovs in both spectrometers were
  clean of background sources.
\item $N_{\mathrm{DCtrk}}$: Total number of tracks formed by the tracking algorithm. We require at least one track to increment $N_{\mathrm{did}}$.
\end{itemize}
In the tracking algorithm, it is possible that there may be multiple, but not necessarily real physics tracks that passed the criteria set above. In this case, Hall C uses three distinct methods
to select the \textit{best} track:
\begin{itemize}
\item \textit{scintillator hit method}: Selects the best track as the track closest to the paddle hit in the last scintillator plane (S2Y). In addition,
  it also rejects tracks if they fail certain criteria imposed on the hodoscope calculated $\beta$ and normalized calorimeter energy.
\item \textit{best $\chi^{2}$ method}: Selects the best track as the track fit with the lowest $\chi^{2}$.
\item \textit{pruning method}: Selects among multiple possible tracks, the track with the lowest $\chi^{2}$ after pruning (``cutting'' or ``trimming'') tracks that
  do not meet certain criteria imposed on the reconstructed variables at the target, hodoscope $\beta$, focal plane time and number of PMT hits among others.
  For example, an event with a bad $X'_{\mathrm{tar}}$ reconstruction might have the track with the lowest $\chi^{2}$ (among multiple tracks for a single event), which would have been chosen
  by the \textit{best $\chi^{2}$ method}. The \textit{pruning method}, however, would have pruned this potential ``good track'' candidate, thereby reducing the probability of what would have
  otherwise been considered a ``good track'' by the \textit{best $\chi^{2}$ method}. A judgement needs to be made regarding which are reasonable variables to prune on. 
\end{itemize}
\indent For the E12-10-003, the \textit{best $\chi^{2}$ method} was used to determine the tracking efficiencies. In general, the rates for the SHMS varied
between $\sim$ 120-170 kHz, and in the HMS, between $\sim$110-170 Hz (see Fig. \ref{fig:trg_rates}), which did not present a problem
(multiple real tracks per event) in the determination of the tracking efficiency. In general, the tracking efficiency for both spectrometers
was found to be very stable over the course of the experiment and was measured to be on average, $98.8\%$ for the HMS and $96.4\%$ for the
SHMS (see Fig. \ref{fig:trk_eff}).\\
\section{DAQ Live Time Efficiency ($\epsilon_{\mathrm{tLT}}$)}
\begin{figure}[H]
  \centering
  \includegraphics[scale=0.75]{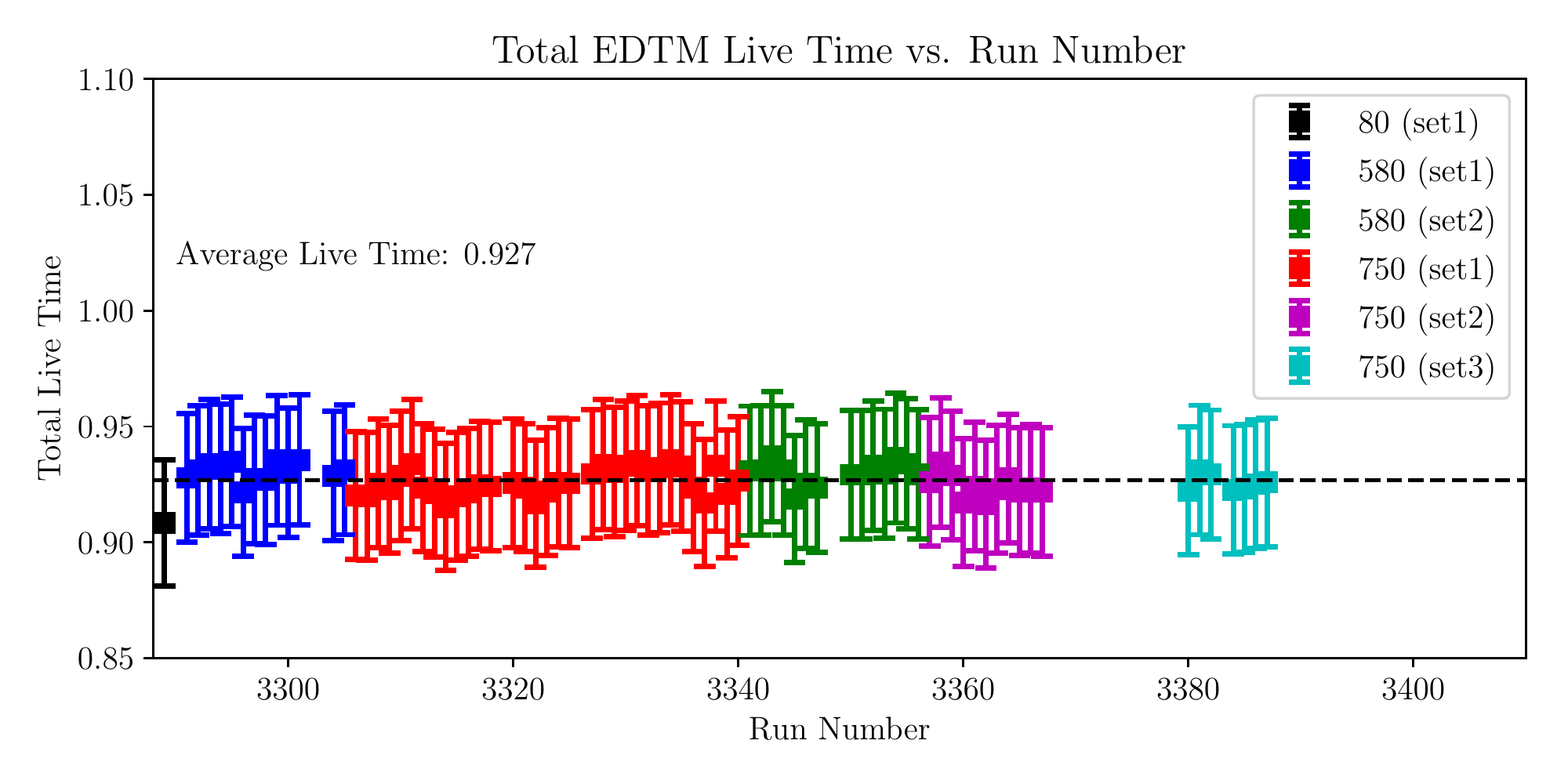}
  \caption{Total EDTM live time determined for the E12-10-003 experiment.}
  \label{fig:edtm_lt}
\end{figure}

Another source of inefficiency in the experimental yield arises from the total dead time of the data acquisition (DAQ) system
which is separated into an electronic and computer deadtime. The electronic deadtime arises from
signal pile-up at the front-end of the electronic modules due to high rates, whereas the computer deadtime
refers to the amount of time the DAQ is unable to accept a pre-trigger.\\
\indent In E12-10-003, the live time calculation was determined using the EDTM system described in Section \ref{subsec:edtm}.
This method determines the electronic and computer live time simultaneously by feeding a clocked EDTM logic signal into the trigger electronics
(mixed with the physics signals) and counting how many of them were accepted by the trigger interface. Using the formula from Eq. \ref{eq:3.24},
the total EDTM live time was determined to be on average $\sim92.7 \%$ (see Fig. \ref{fig:edtm_lt}). 
The uncertainty on the EDTM system is not a straightforward calculation and needs to be given more thought since 1) the numerator and denominator involved in the
calculation (see Eq. \ref{eq:3.24}) are correlated and 2) the accepted number of EDTM logic pulses (numerator) are governed by a random process that can be
understood from the fact that even though these are clocked pulses, they are mixed with the physics pre-triggers, which are random and follow a poissonian distribution.
An educated guess of $\sim3\%$ relative uncertainty on the total live time was made based on a crude estimate of the dominant source of electronic deadtime in the system.
Given that the SHMS S1X hodoscope plane is the dominant source of electronic dead time as it had the highest trigger rates (typically $R\sim1$ MHz) out of the four hodoscope
planes and that the time window of the electronics module (LeCroy 4564 logic module) used to form the S1X pre-trigger was set to $\tau\sim 60$ ns, the electronic deadtime can be approximated to be
$R\tau\sim $0.06 or 6$\%$. Assuming that the computer dead time is negligible given that the coincidence pre-triggers were only $\sim 3$ Hz, we take $\sim6\%$ (or $\pm 3\%$) as the
relative uncertainty in the total live time calculation. Given that this is a statistically dominated experiment ($\sim20-30\%$), a conservative estimate of a few percent
is negligible in comparison.

\section{Target Density Corrections ($\epsilon_{\mathrm{tgt.Boil}}$)} \label{sec:tgt_boil_corr}
\begin{figure}[H]
  \centering
  \includegraphics[scale=0.35]{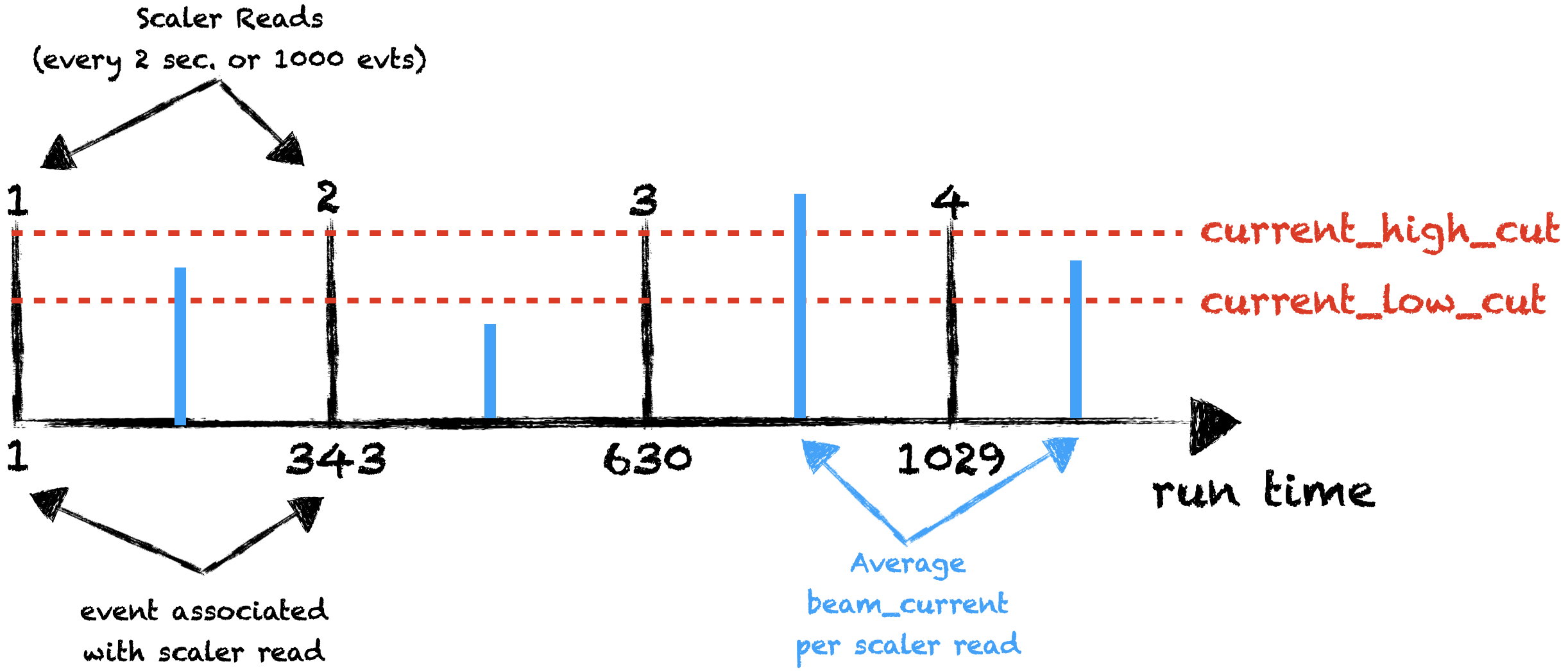}
  \caption{Illustration of scaler and event reads during a typical experimental run.}
  \label{fig:scl_reads}
\end{figure}

The exposure of a cryogenic target to the beam can cause density reductions in the beam path or even localized boiling on the target due to the large amounts of heat deposited by the beam.
To minimize boiling, the beam can be rastered (``smeared out'') by up to 5x5 mm$^{2}$ in area to re-distribute the heat deposited over a larger area on the target (see Section \ref{sec:beam_raster}). 
Even though the rastered beam minimizes the local boiling, there may still be a small reduction in the target density that results in the reduction of the data yield by a few percent. \\
\indent To correct the experimental data yield for target density, a series of dedicated runs (see Table \ref{tab:table5.1}) were taken independently in each spectrometer (single-arm)
at various beam currents. In this section I present the target density study results using the HMS only, however, a similar study using the SHMS should be carried out as a cross-check. \\
\indent The runs in Table \ref{tab:table5.1} were taken at a spectrometer central angle and momentum settings of 25$^{\circ}$ and -4.4 GeV/c (negative polarity), respectively, at a beam energy of $E_{\mathrm{b}}=10.6$ GeV.
At these kinematics, the inclusive electrons detected by the HMS correspond to the inelastic electron scattering off a target nucleon.
The spectrometer was set to negative polarity to detect electrons that scattered from carbon-12, LH$_{2}$ and LD$_{2}$ targets, respectively, at various beam currents.
To select electrons in the HMS, the trigger was set to HMS EL-REAL (see Fig. \ref{fig:HMS_SingleArm_diagram}).\\
\begin{table}[H] 
  \centering
  \scalebox{1.1}{
    \begin{tabular}[t]{ccc}
      \hline
      HMS Run&Target&Beam Current [$\mu$A]\\
      \hline
      \hline
      2093 & Carbon-12  &60\\
      2094 & Carbon-12  &50\\
      2095 & Carbon-12  &35\\
      2075 & LH$_{2}$   &80\\
      2076 & LH$_{2}$   &70\\
      2078 & LH$_{2}$   &10\\
      2080 & LH$_{2}$   &10\\
      2081 & LH$_{2}$   &20\\
      2082 & LH$_{2}$   &35\\
      2083 & LH$_{2}$   &35\\
      2084 & LH$_{2}$   &45\\
      2085 & LH$_{2}$   &55\\
      2073 & LD$_{2}$   &80\\
      2074 & LD$_{2}$   &70\\
      2087 & LD$_{2}$   &55\\
      2088 & LD$_{2}$   &45\\
      2089 & LD$_{2}$   &35\\
      2090 & LD$_{2}$   &20\\
      2091 & LD$_{2}$   &10\\
      \hline
    \end{tabular}
  }
  \caption{Target density reduction studies run list taken on April 02, 2018.}
  \label{tab:table5.1}
\end{table}
\indent The target density reduction analysis consists of determining the charge normalized yield as a function of the beam current to determine the yield loss per unit beam current.
To correctly determine the yield corresponding to a specific beam current, a cut was made to select events that
correspond to a stable beam current period, and not to the beam ramp up or down periods.\\
\newpage
\indent To precisely select events corresponding to a certain beam current threshold, consider the following example illustrated in Fig. \ref{fig:scl_reads}.
In this example, the \textit{scaler reads} (vertical black lines) are registered by the DAQ and dumped into the data-stream at either every 2 seconds or 1000 events.
As the run progresses, the average beam charge and time interval in between scaler reads can be determined from which the average beam current is calculated for each interval.
A beam current \textit{low} and \textit{high} cut can then be set to select only those scaler read intervals for which the average beam current is within the cut limits. In the
data event loop, each event is then compared to the next scaler read and is discarded if it lies within a scaler read interval that did not pass the cut. \\
\indent After selecting only those events that passed the nominal beam current for each run, the charge normalized yields can be determined and plotted
as a function of the average beam current. In this analysis, three separate charge normalized yields were determined, designed to test the computer live time and tracking
efficiency corrections.\\
\indent The yields were defined as
\begin{align}
  & Y_{\mathrm{scl}} = \frac{N_{\mathrm{scl}}}{Q_{\mathrm{tot}}} \label{eq:5.18},\\
  & Y_{\mathrm{no.trk}} = \frac{N_{\mathrm{acc}}}{Q_{\mathrm{tot}}\cdot\epsilon_{\mathrm{cpuLT}}} \label{eq:5.19},\\
  & Y_{\mathrm{trk}} = \frac{N_{\mathrm{acc}}}{Q_{\mathrm{tot}}\cdot\epsilon_{\mathrm{cpuLT}}\cdot\epsilon_{\mathrm{htrk}}}, \label{eq:5.20} 
\end{align}
where Eq. \ref{eq:5.18} is the yield calculated from the total number of HMS pre-trigger scaler counts normalized by the total charge, Eq. \ref{eq:5.19} defines the
charge normalized yield (using accepted HMS triggers) corrected for computer live time but does not use tracking information in the event selection criteria, and Eq. \ref{eq:5.20}
defines a charge normalized yield that uses tracking information in the event selection criteria and is therefore also corrected for the tracking efficiency. The associated
histograms used to determine the counts for each of these yield calculations are shown below.
\begin{figure}[H]
  \centering
  \includegraphics[scale=0.45]{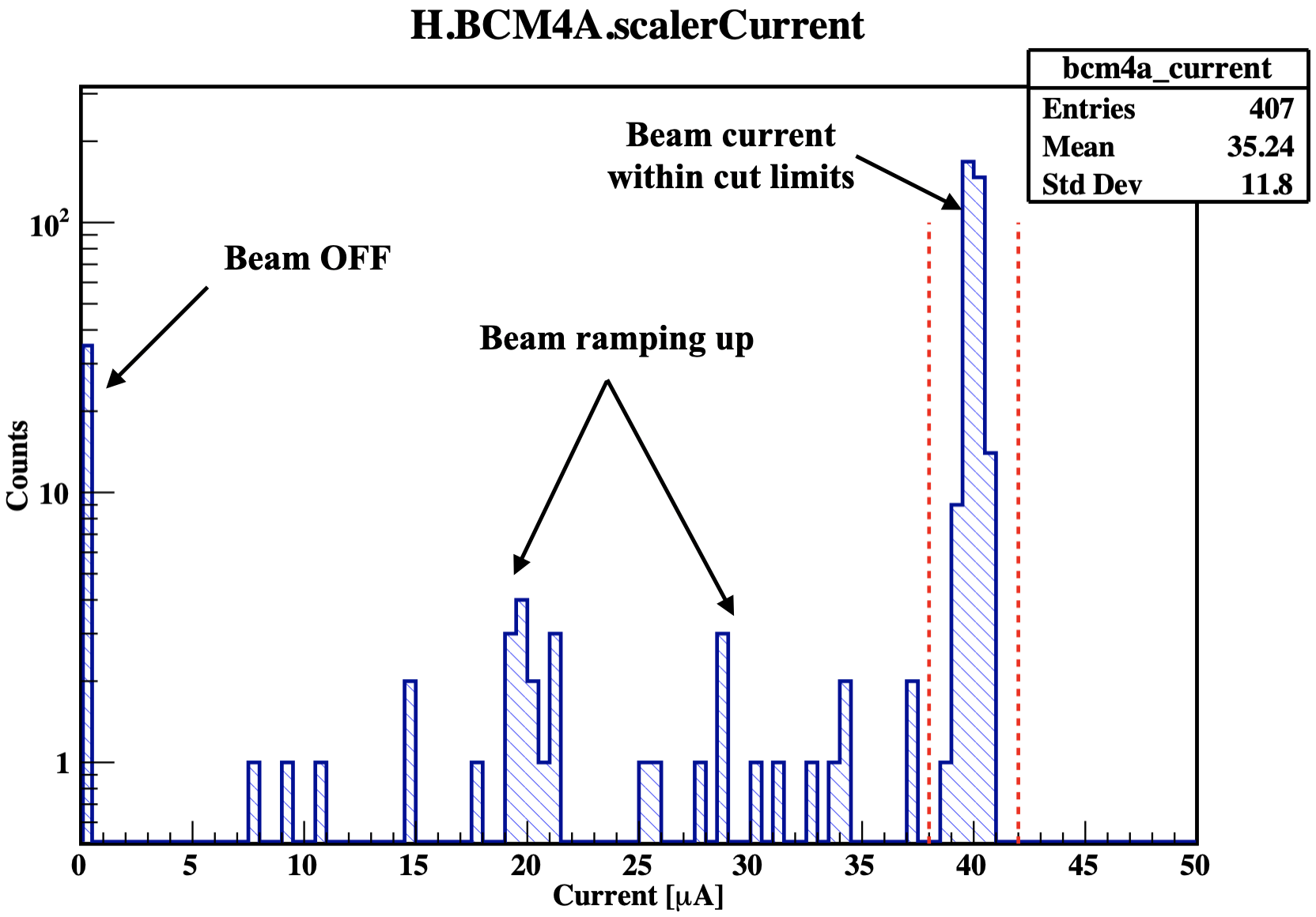}
  \caption{Example of a BCM scaler current cut used to determine the yield.}
  \label{fig:scl_yield}
\end{figure}
\begin{figure}[H]
  \centering
  \includegraphics[scale=0.45]{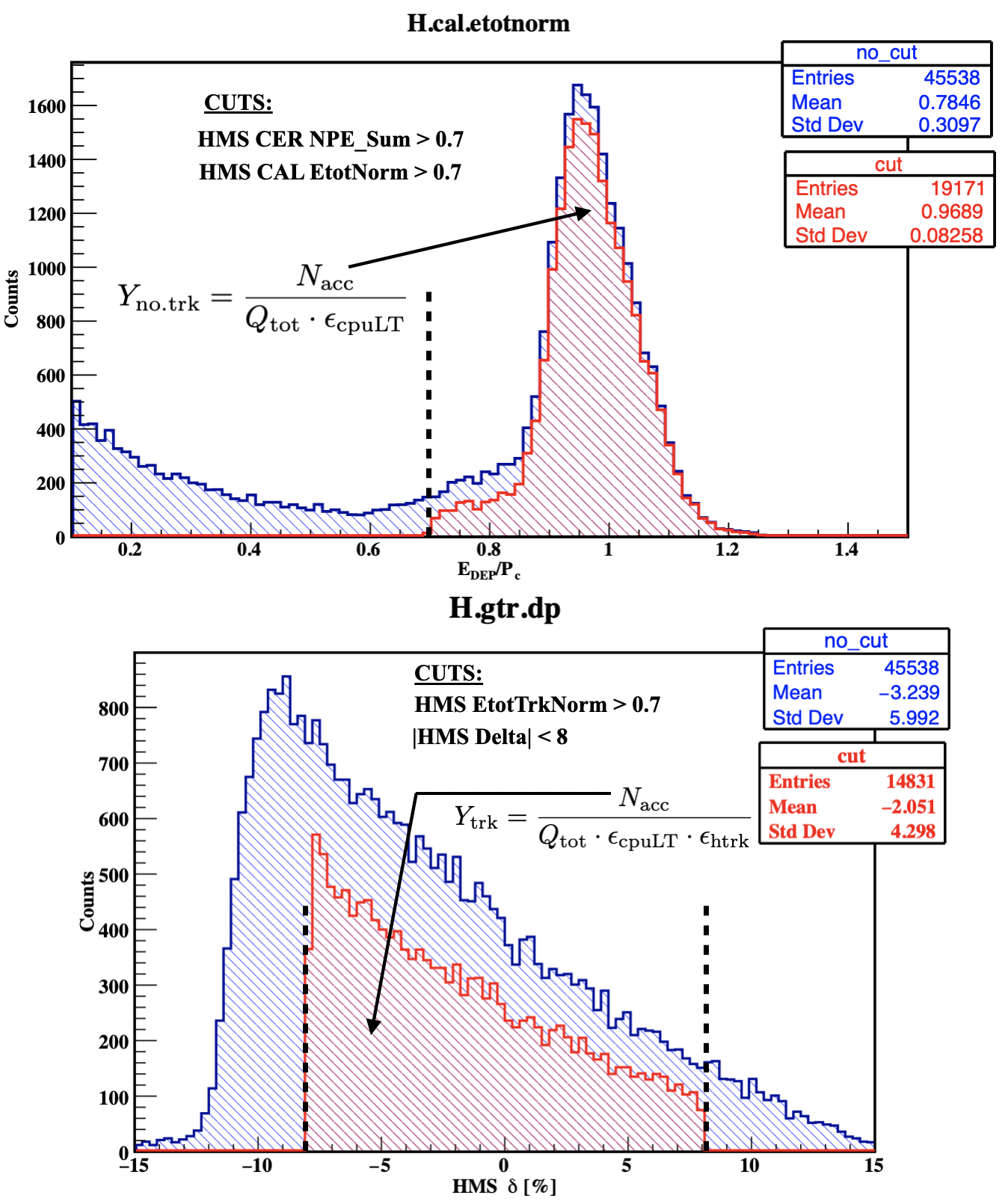}
  \caption{Example of the histograms used to determine the non-tracking (top) and tracking (bottom) yields. Top: $x$-axis shows the total deposited energy in the calorimeter normalized by the
  central spectrometer momentum, $E_{\text{DEP}}/P_{\text{c}}$. Bottom: $x$-axis shows the HMS momentum acceptance, $\delta$, in percent.}
  \label{fig:ntrk_trk_yields}
\end{figure}
Figure \ref{fig:bcm4a_saturation} shows the charge normalized yields using BCM4A (red) and BCM4B (black) beam current cuts in the event selection process.
This study was done to check the behavior of both BCMs at very high currents. As can be seen from the normalized yields using the LH$_{2}$ and LD$_{2}$ targets,
above $\sim75$ $\mu$A, the BCM4A yield is significantly lower than the BCM4B yield indicating that BCM4A begins to saturate above this current (see Ref.\cite{bcm_studies_2018}).\\
\begin{figure}[H]
  \centering
  \includegraphics[scale=0.6]{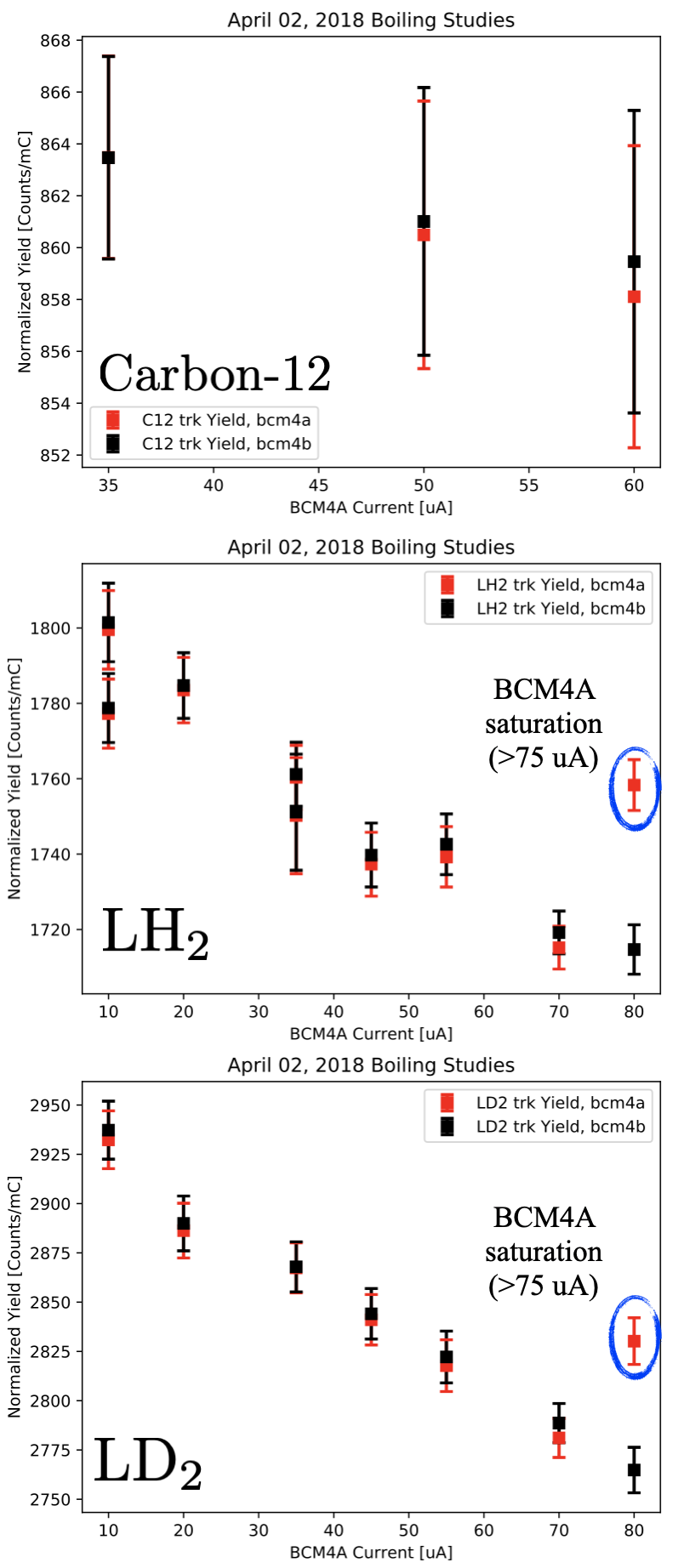}
  \caption{Normalized tracking yields using BCM4A (red) and BCM4B (black) beam current cuts on carbon-12 (top), LH$_{2}$ (middle) and LD$_{2}$ (bottom) targets.}
  \label{fig:bcm4a_saturation}
\end{figure}
\noindent Due to the saturation of BCM4A at beam currents $>75$ $\mu$A, it was decided to use BCM4B for the remaining target density studies. \\
\begin{figure}[H]
  \centering
  \includegraphics[scale=0.45]{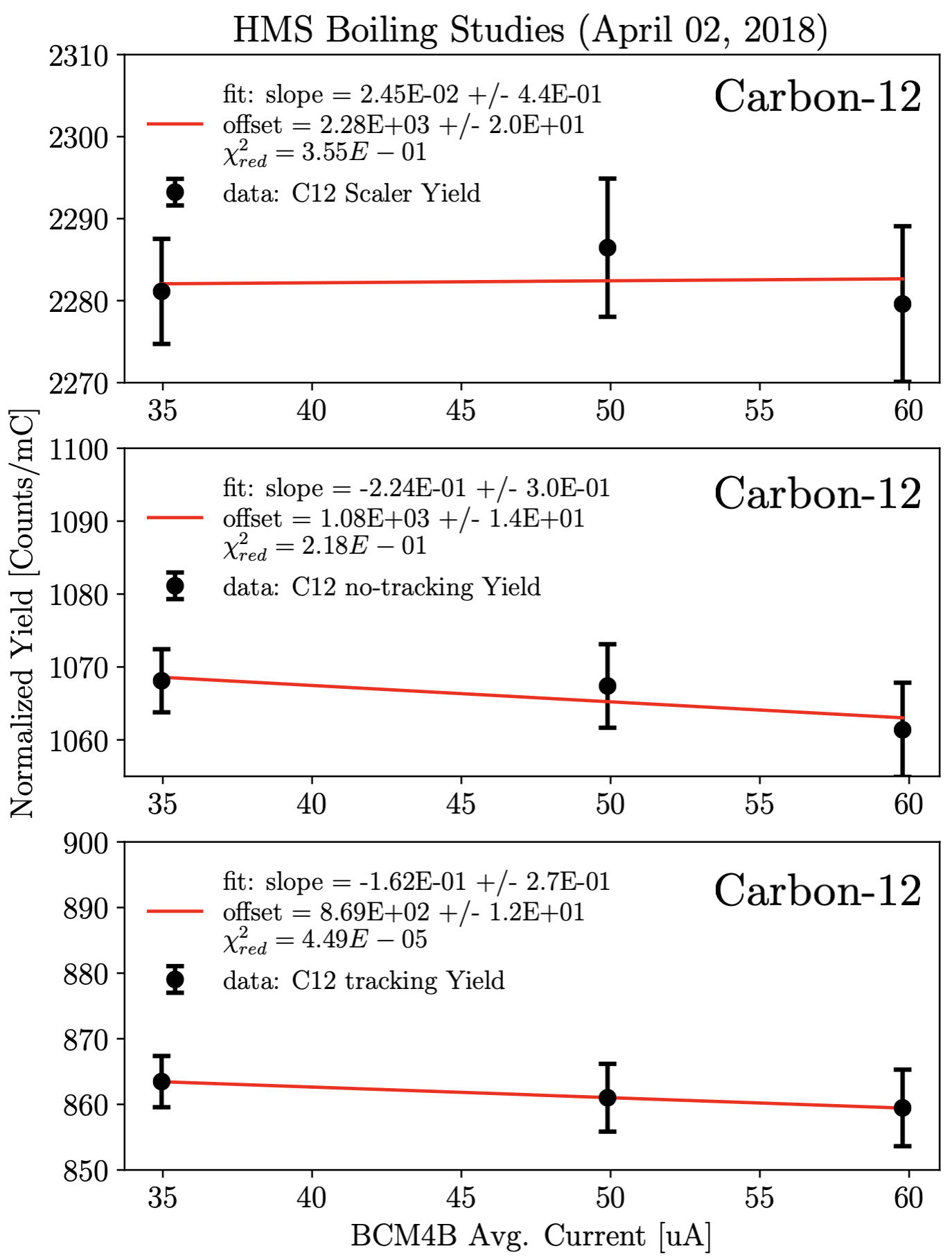}
  \caption{Linear fit of the charge normalized yields for carbon-12.}
  \label{fig:C12_fit}
\end{figure}
\indent The normalized yields for each of the targets were determined from Eqs. \ref{eq:5.18}, \ref{eq:5.19} and \ref{eq:5.20} and fit as a function of the BCM4B average beam current using the fit function
\begin{equation}
  Y_{\mathrm{norm}} = m\cdot I_{\mathrm{beam}} + Y_{0},
\end{equation}
where $Y_{\mathrm{norm}}$ is the charge normalized yield ($y$-axis), $I_{\mathrm{beam}}$ is the average beam current ($x$-axis), and ($m$, $Y_{0}$) are the slope and $y$-intercept parameters, respectively.
The fit results are shown in Figs. \ref{fig:C12_fit}, \ref{fig:LH2_fit} and \ref{fig:LD2_fit}.   
\begin{figure}[H]
  \centering
  \includegraphics[scale=0.45]{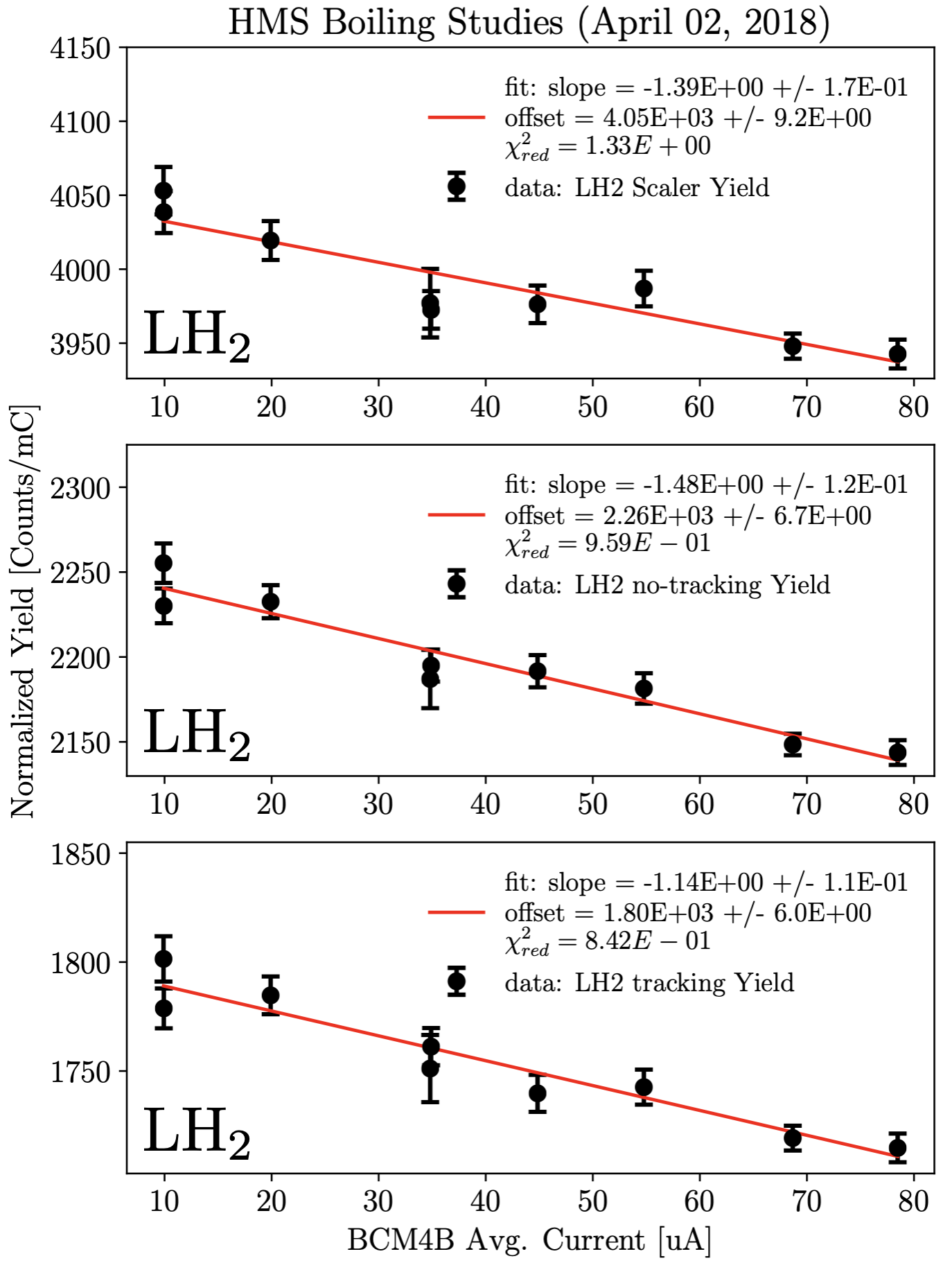}
  \caption{Linear fit of the charge normalized yields for LH$_{2}$.}
  \label{fig:LH2_fit}
\end{figure}
A gradual improvement is observed in the $\chi^{2}$ fit from the top to bottom panels for each of the targets, indicating that the yield calculated using Eq. \ref{eq:5.20} gives the best fit
results. To apply the target density corrections, one needs to determine the fractional yield loss per $\mu$A. This is done by normalizing the yield to the $y$-intercept, $Y_{0}$, which corresponds
to the yield at 0 $\mu$A to obtain a relative yield. From the relative yield and the slope, the fractional yield loss at any beam current within the fit range can be calculated. This correction
is then applied to the experimental yield.
\begin{figure}[H]
  \centering
  \includegraphics[scale=0.44]{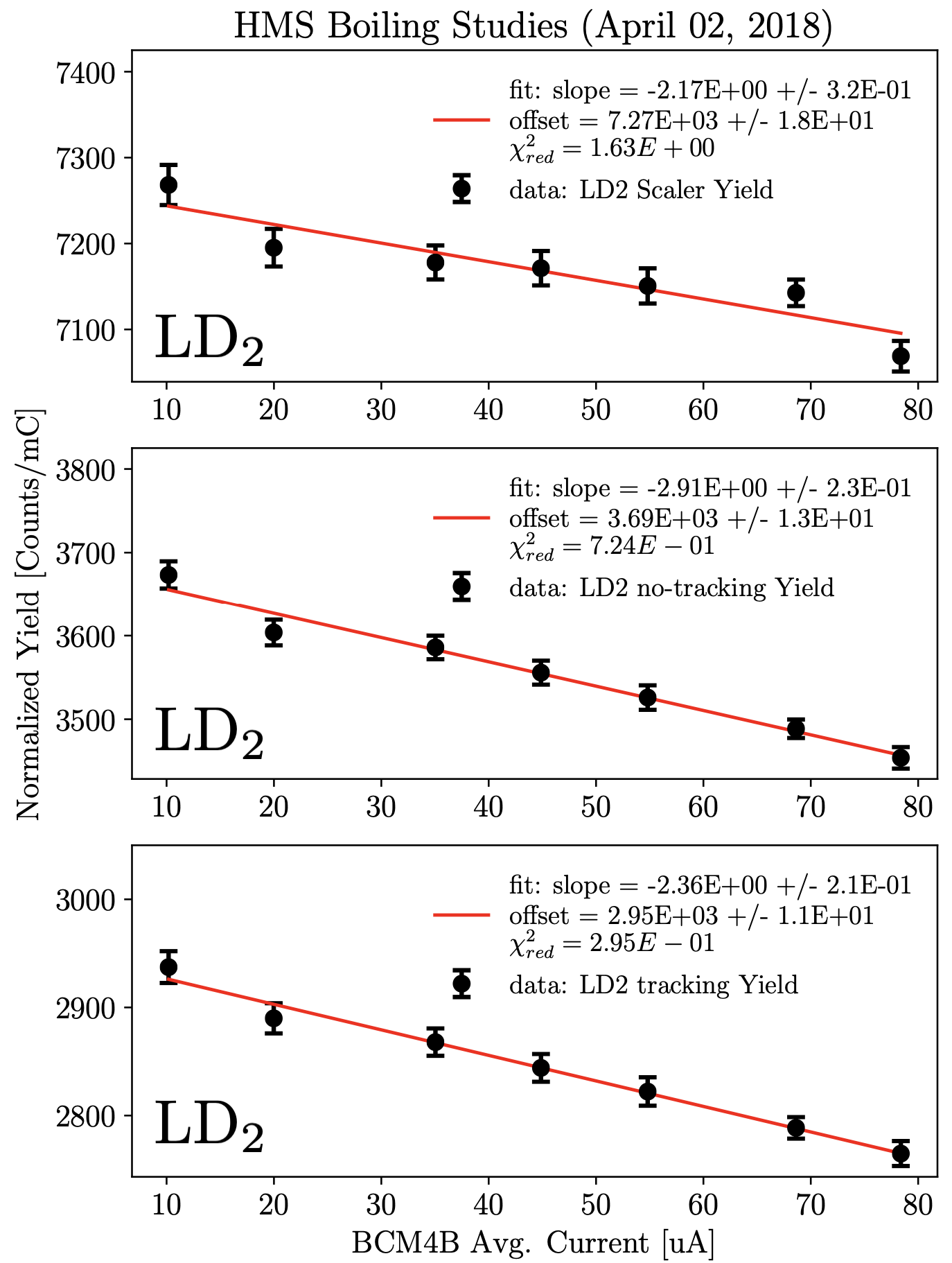}
  \caption{Linear fit of the charge normalized yields for LD$_{2}$.}
  \label{fig:LD2_fit}
\end{figure}
Figure \ref{fig:tgt_boil_results} shows the fit results after normalizing the fit function by $Y_{0}$.
From the final fit results of the HMS target density studies, carbon-12 has a reduction in the yield of $\sim 0.18\%/\mu$A, which shows almost non-existent density changes
as is expected from a solid target. Liquid hydrogen and deuterium, however, show a significant reduction in the yield corresponding to $\sim0.063\%/\mu$A and $\sim0.080\%/\mu$A,
respectively.
\begin{figure}[H]
  \centering
  \includegraphics[scale=0.35]{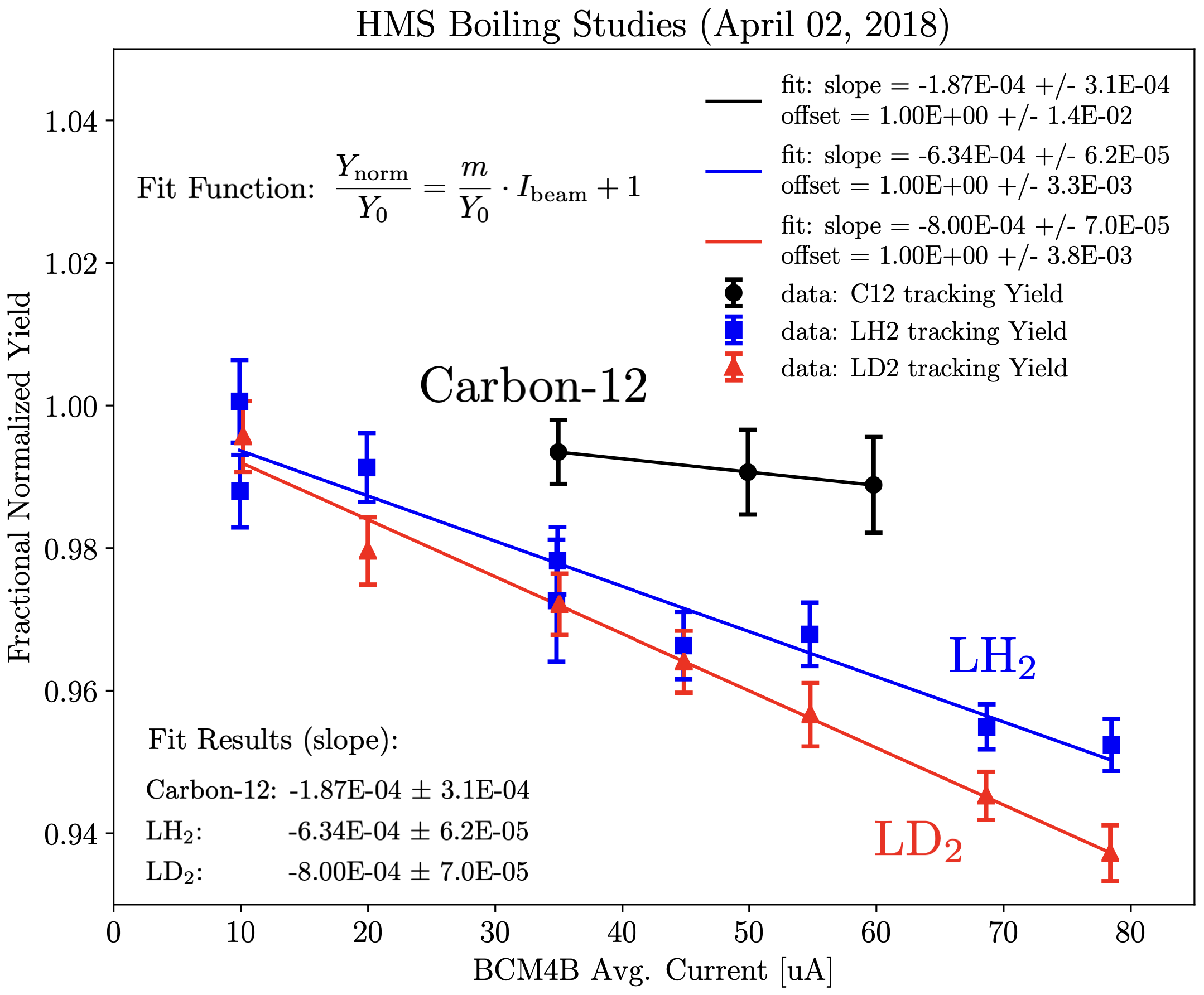}
  \caption{Linear fit function normalized to the $y$-intercept, $Y_{0}$ for each of the three targets.}
  \label{fig:tgt_boil_results}
\end{figure}
\begin{table}[ht!] 
  \centering
  \scalebox{1.}{
    \begin{tabular}[t]{ccc}
      \hline
      Target&Slope, $m_{0}\equiv m/Y_{0}$&Slope Error, $\delta(m_{0})$\\
      \hline
      \hline
      Carbon-12&-1.87x10$^{-4}$&3.1x10$^{-4}$\\
      LH$_{2}$&-6.34x10$^{-4}$&6.2x10$^{-5}$\\
      LD$_{2}$&-8.00x10$^{-4}$&7.0x10$^{-5}$\\      
      \hline
    \end{tabular}
  }
  \caption{Target boiling (or density reduction) studies fit results normalized to the $y$-intercept.}
  \label{tab:table5.2}
\end{table}
\indent The target density correction was applied on a run by run basis using the formula,
\begin{subequations}
  \begin{align}
    &\epsilon_{\mathrm{tgt.Boil}} = 1 - m_{0}\cdot I_{\mathrm{avg}}, \\
    &\delta\epsilon_{\mathrm{tgt.Boil}} = \sqrt{I^{2}_{\mathrm{avg}}\delta^{2}_{m_{0}} + m^{2}_{0}\delta^{2}_{I_{\mathrm{avg}}}},
  \end{align}
\end{subequations}
where $m_{0}$ is the slope of the corresponding target and $I_{\mathrm{avg}}$ is the averaged BCM4A beam current over the entire run, provided the beam current cuts have been applied.
It is important to note that the beam current cut on the data analysis was less strict ($>10$ $\mu$A) than in the target density studies to avoid cutting out possible coincidences during the
ramping up periods of the beam.\\
\begin{figure}[H]
  \centering
  \includegraphics[scale=0.66]{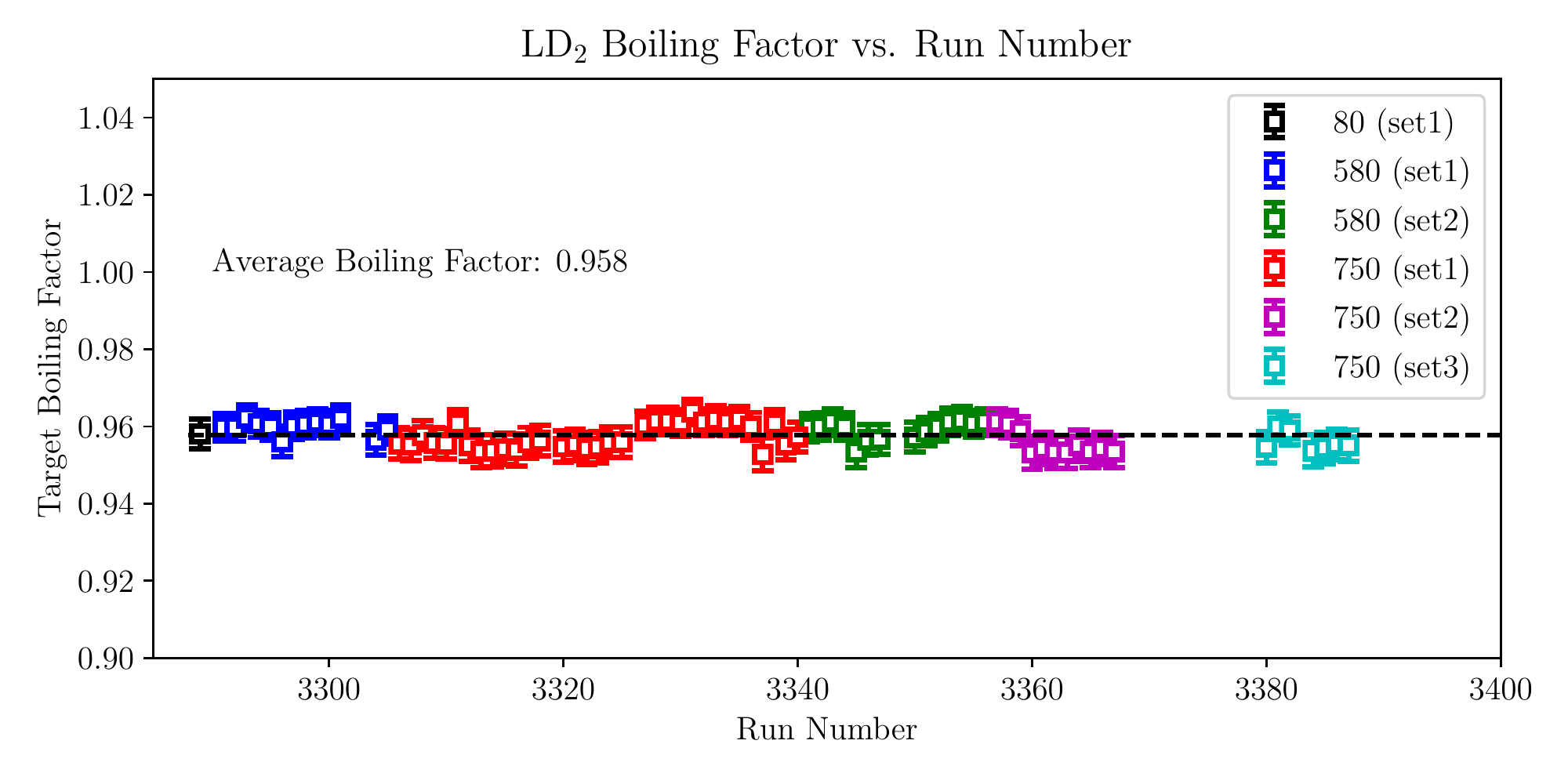}
  \caption{Target density correction of deuterium determined for the E12-10-003 experiment.}
  \label{fig:tgt_boil_vs_run}
\end{figure}
Figure \ref{fig:tgt_boil_vs_run} shows the target density correction factor for all data sets of the E12-10-003 experiment. The correction factor seems stable over all runs,
which is a result of the beam currents during this experiment were stable and within a small range of $\sim45-60$ $\mu$A as shown in Fig. \ref{fig:beam_current}.
\begin{figure}[H]
  \centering
  \includegraphics[scale=0.66]{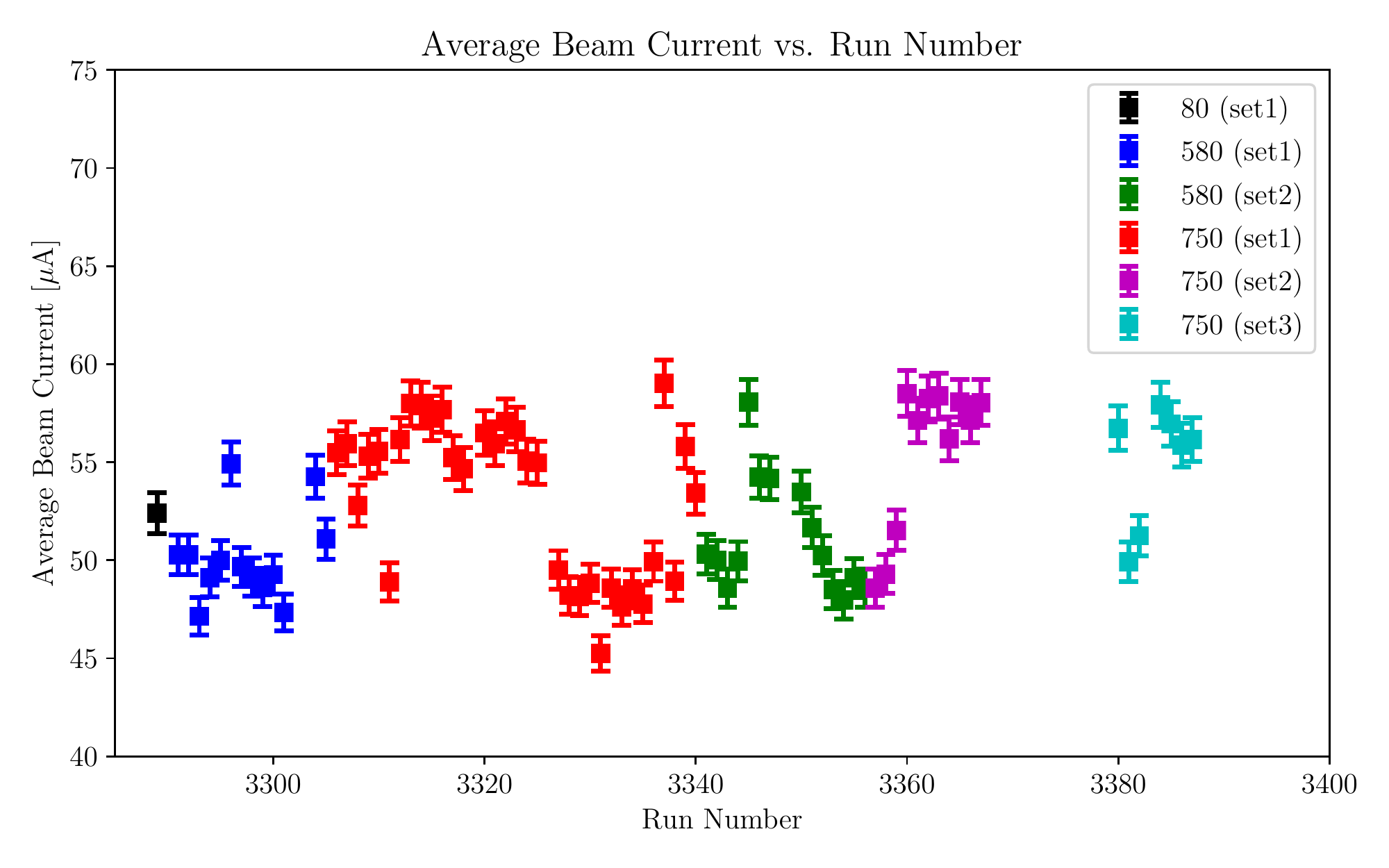}
  \caption{Averaged beam currents (BCM4A) determined for the E12-10-003 experiment.}
  \label{fig:beam_current}
\end{figure}
\section{Proton Absorption Corrections ($\epsilon_{\mathrm{pTr}}$)}
\begin{figure}[H]
  \centering
  \includegraphics[scale=0.32]{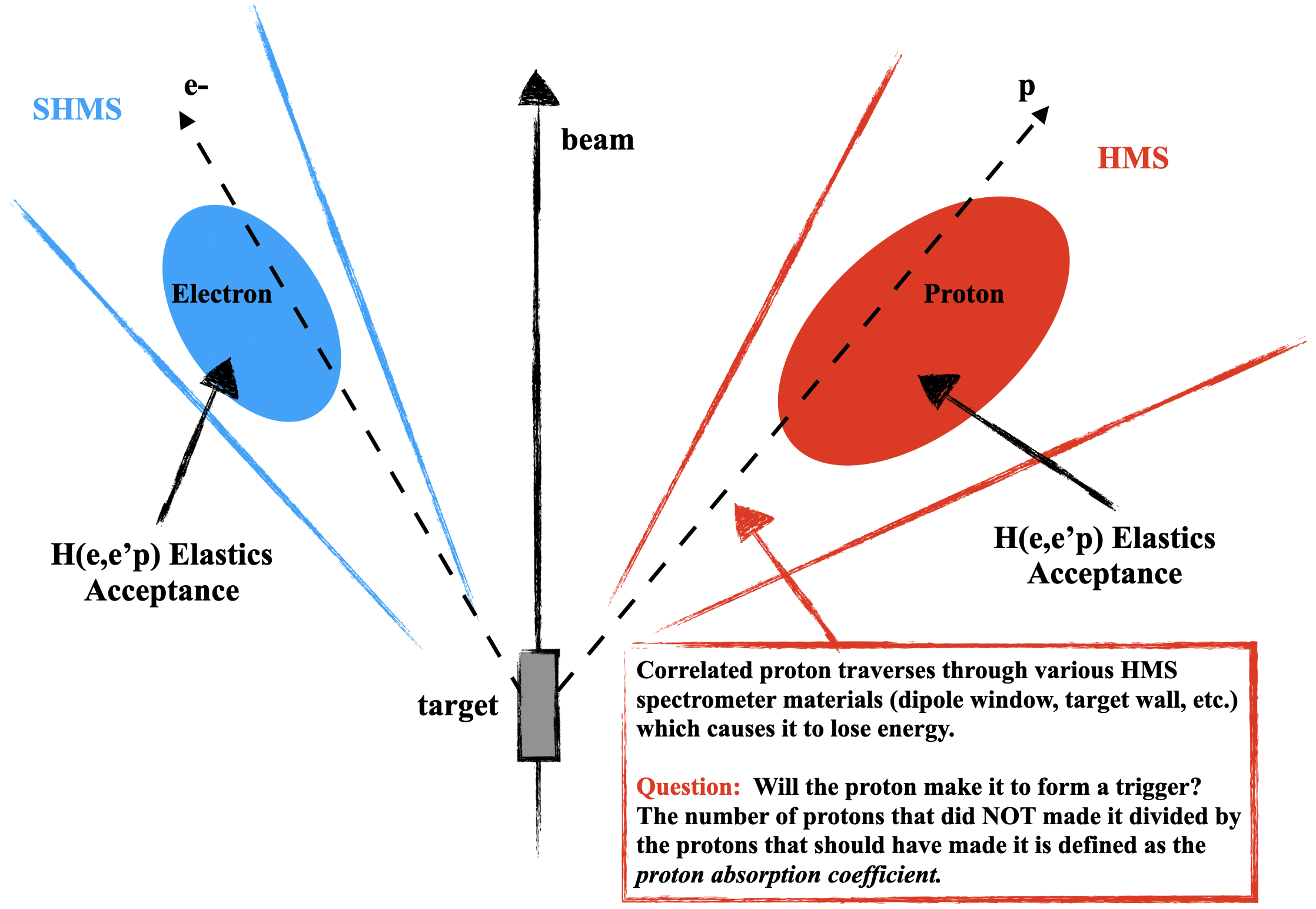}
  \caption{Cartoon to illustrate how the proton absorption coefficient is determined experimentally by selecting the $ep$ elastics acceptance region.}
  \label{fig:pAbs_concept}
\end{figure}

In general $A(e,e'p)$ coincidence experiments there is a small ($\sim$ few percent) probability that the knocked-out proton never makes it to the required detectors to form a coincidence trigger
with the electron. This is due to the fact that as the proton leaves the target, traverses various windows (e.g., the target cell window, the spectrometer entrance/exit windows) and enters the
detectors, it can lose energy and/or outscatter and so fails to form a trigger or pass cuts.\\
\indent The proton absorption coefficient can be determined in theory from the knowledge of the material thickness and interaction length, which depends on the
cross section of the interaction process. In the E12-10-003 experiment, the proton absorption coefficient was determined by taking several
dedicated hydrogen elastic runs (see Table \ref{tab:table3.1}) using either only a single-arm or a coincidence trigger at the same kinematics.
The general steps taken were:
\begin{itemize}
\item Use the $^{1}$H$(e,e')p$ coincidence runs to determine the spectrometer acceptance region for both HMS/SHMS corresponding to $ep$ elastics events (see Fig. \ref{fig:pAbs_concept}).
\item Use the $^{1}$H$(e,e')p$ SHMS single-arm runs to determine the number of electrons that lie within SHMS the acceptance region corresponding to SHMS acceptance for $ep$ elastics determined in the first step.
\end{itemize}
\indent By selecting the SHMS acceptance region (single-arm) corresponding to the $ep$ elastics, one is selecting the total number of electrons that \textit{should} have
been in coincidence (during the coincidence runs) with the proton in the HMS, but because the proton was absorbed by some material, never made it to form a trigger. The
number of electrons that \textit{did} pass the requirements to be in coincidence with the protons were determined by imposing additional restrictions (``cuts'') on the HMS variables.
It should be noted that even though only the SHMS singles triggers were used for data readout, since these runs were taken with the DAQ in coincidence mode,
the HMS detectors are also readout, which enables one to put cuts on the HMS related variables.
\subsubsection{Coincidence $^{1}$H$(e,e')p$ Acceptance Selection}
\begin{figure}[H]
  \centering
  \includegraphics[scale=0.4]{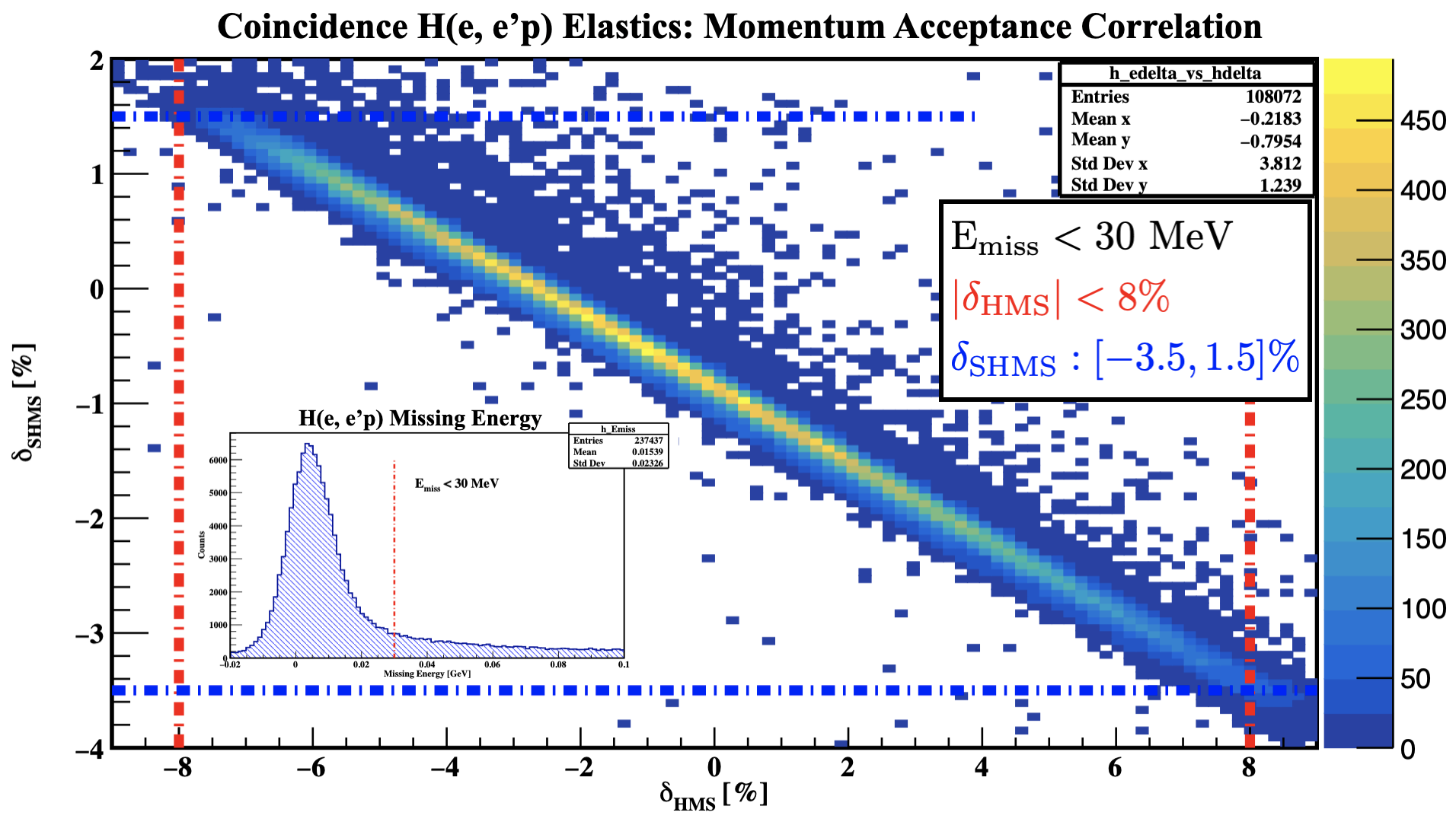}
  \caption{Momentum acceptance correlation between SHMS and HMS for coincidence $^{1}$H$(e,e')p$ data run 3248. Inset: Missing energy spectrum cut below 30 MeV to select true elastic events.}
  \label{fig:coin_delta_correlation}
\end{figure}

To determine the acceptance region in the SHMS corresponding to elastic events, the momentum acceptance correlation between the two spectrometers is plotted in Fig. \ref{fig:coin_delta_correlation}
and shows that the SHMS momentum acceptance corresponding to electrons from $ep$ elastics is completely determined by the HMS momentum acceptance for protons. \\
\begin{figure}[H]
  \centering
  \includegraphics[scale=0.35]{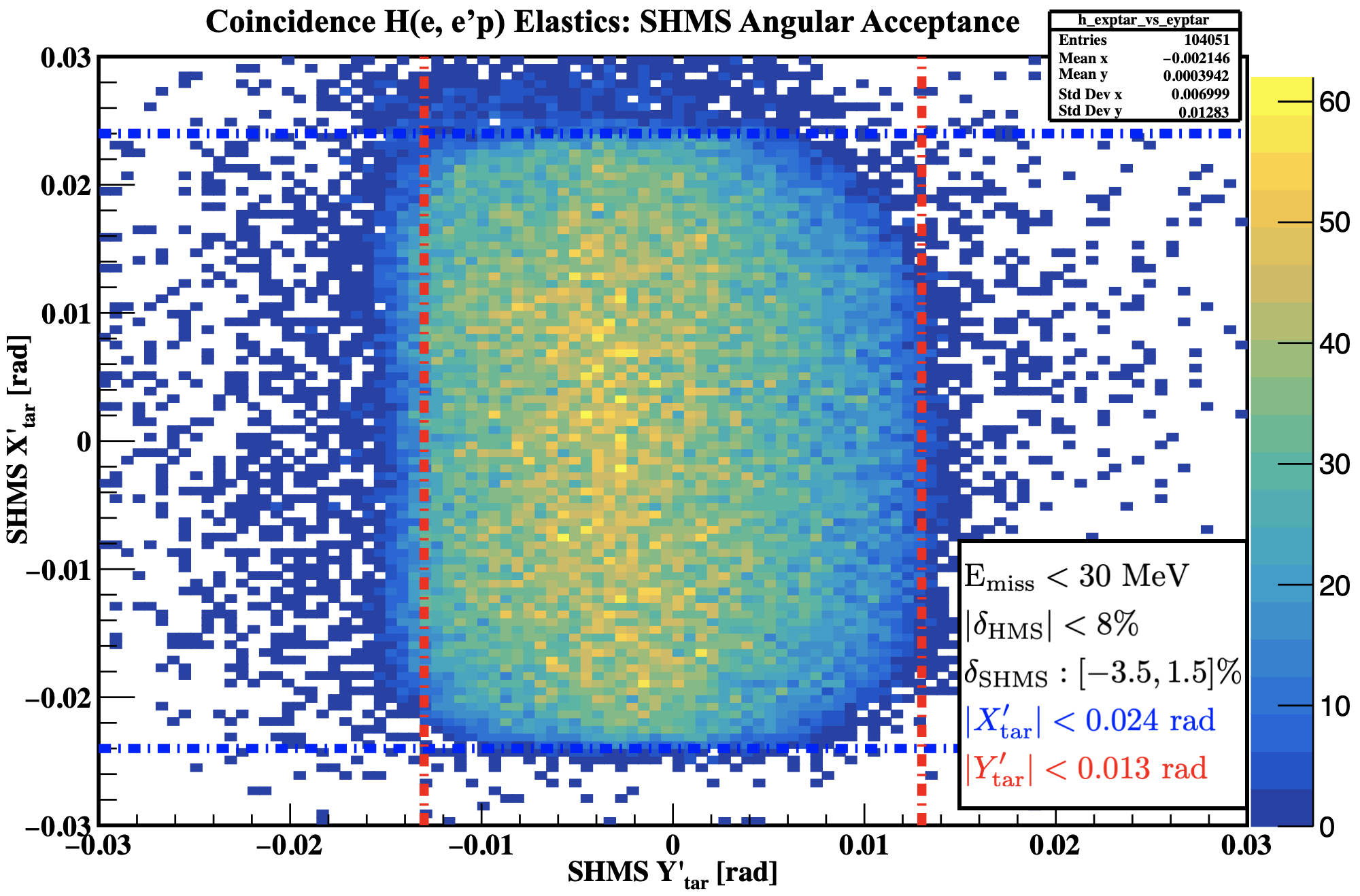}
  \caption{SHMS electron angular acceptance for coincidence $^{1}$H$(e,e')p$ data run 3248.}
  \label{fig:coin_eXptar_eYptar}
\end{figure}
\begin{figure}[H]
  \centering
  \includegraphics[scale=0.35]{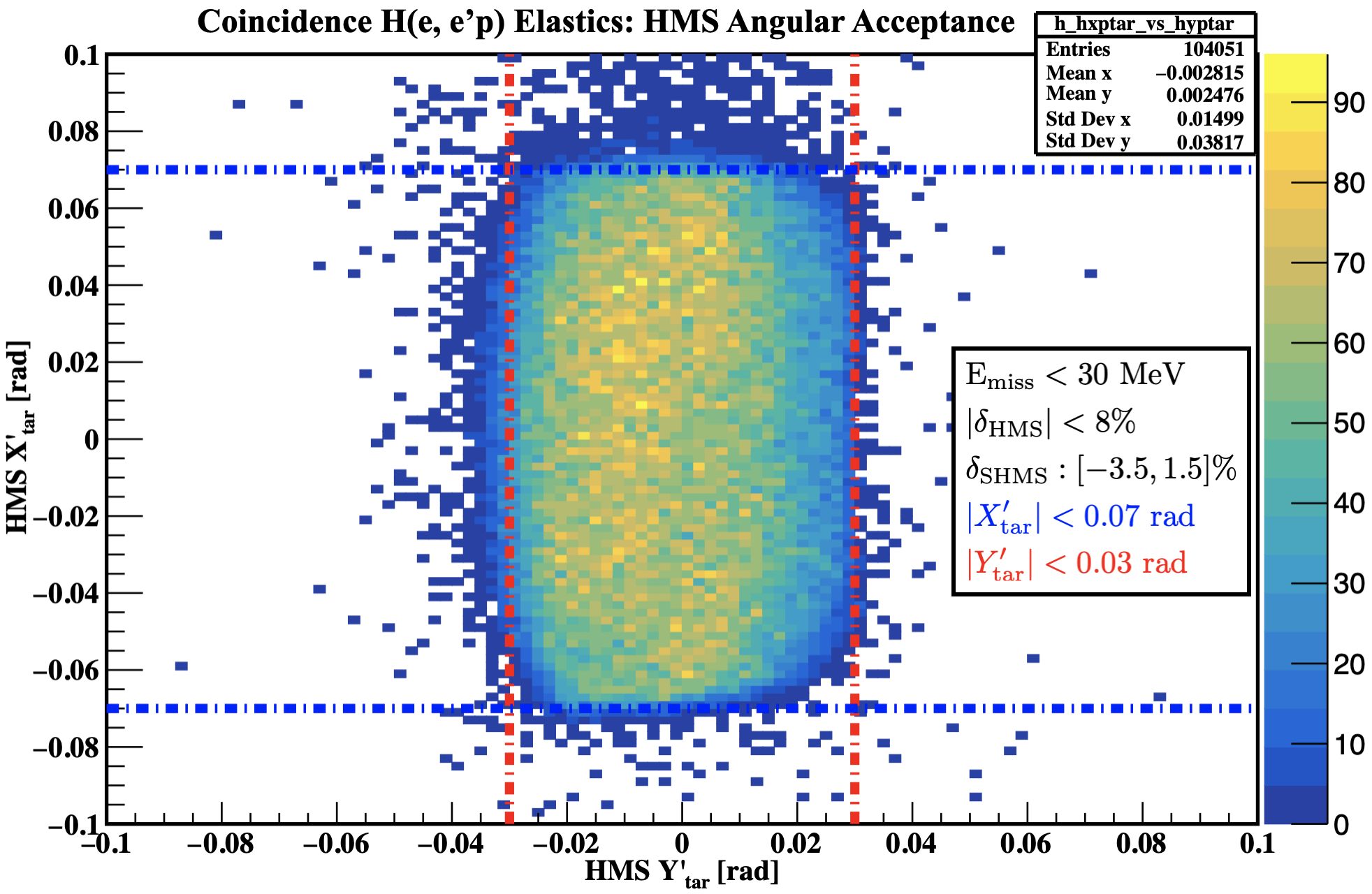}
  \caption{HMS proton angular acceptance for coincidence $^{1}$H$(e,e')p$ data run 3248.}
  \label{fig:coin_hXptar_hYptar}
\end{figure}
Figure \ref{fig:coin_eXptar_eYptar} shows the SHMS angular acceptance where the $\delta$ momentum acceptance and missing energy cuts determined from Fig. \ref{fig:coin_delta_correlation}
have been applied. Similarly, for the HMS angular acceptance shown in Fig. \ref{fig:coin_hXptar_hYptar}, the same cuts on missing energy and momentum acceptance  have been applied.
The dashed color lines in both Figs. \ref{fig:coin_eXptar_eYptar} and \ref{fig:coin_hXptar_hYptar} define the angular acceptance cuts for $ep$ elastics events. The SHMS acceptance cuts
determined in this section will be applied to the SHMS singles $ep$ elastics data in the next section.
\subsubsection{SHMS Singles $^{1}$H$(e,e')p$ Acceptance Selection}
\begin{figure}[H]
  \centering
  \includegraphics[scale=0.40]{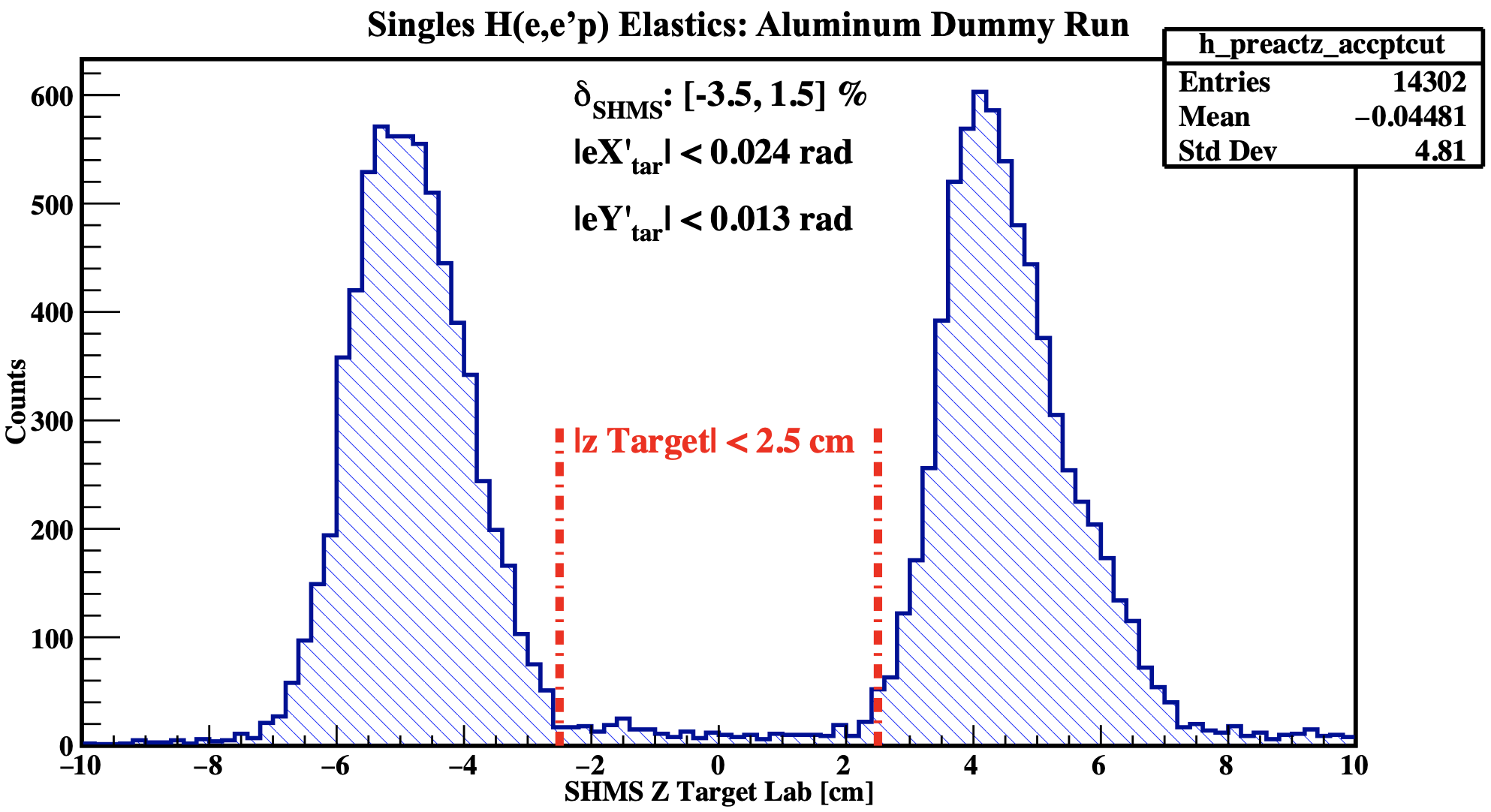}
  \caption{Reaction $z$-vertex cut determination from aluminum dummy run 3254 taken with SHMS singles trigger.}
  \label{fig:singles_AlDummy}
\end{figure}

To help suppress quasi-elastic electrons which scattered from the target cell walls,
an aluminum dummy run was taken at the same kinematics as the hydrogen elastics data. Figure \ref{fig:singles_AlDummy}
shows the reconstructed events along the $z$-vertex with the two peaks representing the target cell end caps. A $z$-vertex cut
of $\pm2.5$ cm on the dummy target was determined to remove the target windows from the good elastic events.\\
\begin{figure}[H]
  \centering
  \includegraphics[scale=0.42]{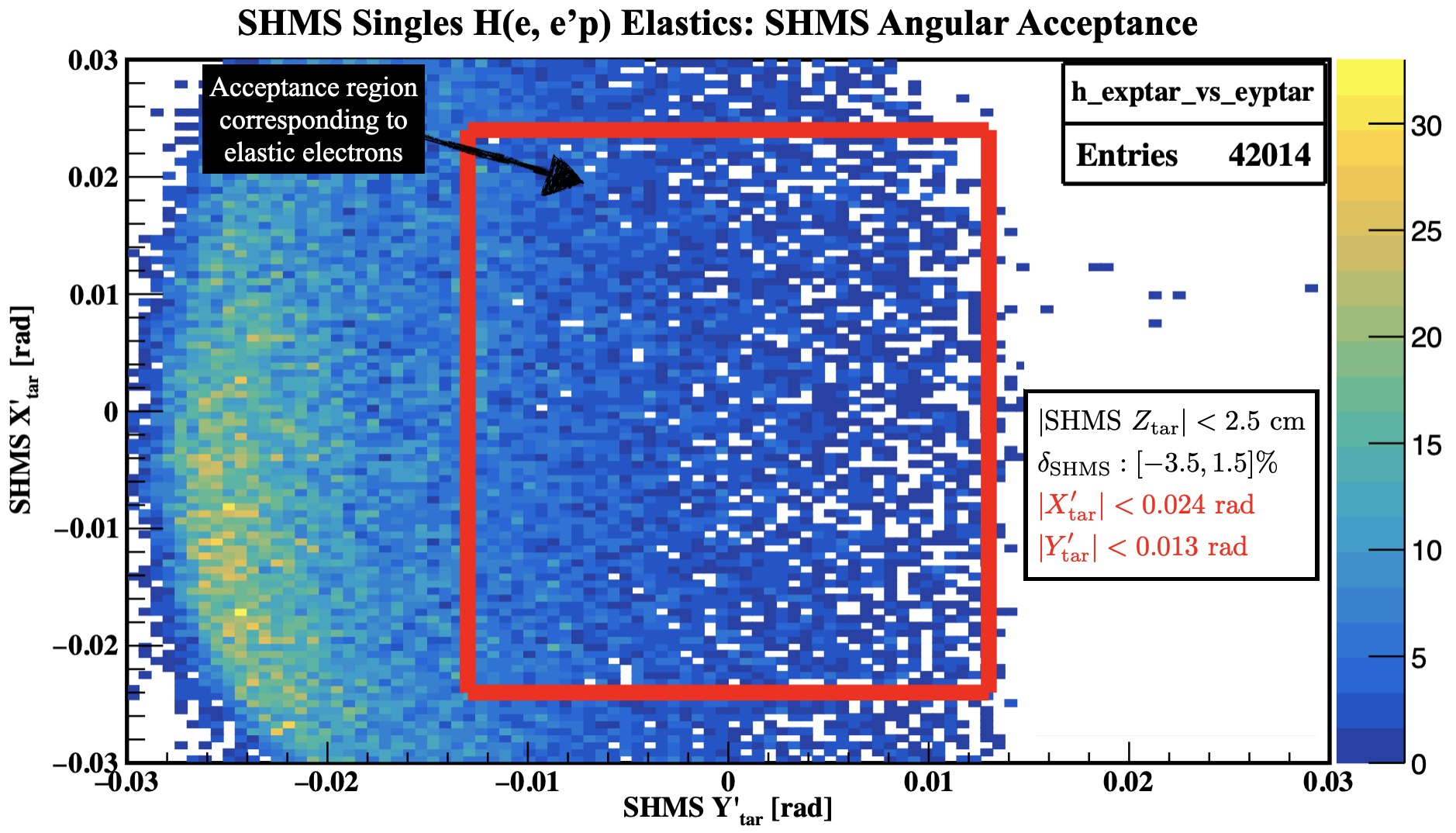}
  \caption{SHMS angular acceptance from electron singles run 3259. The red square is the elastic acceptance region determined from coincidence elastics data at the same kinematics.}
  \label{fig:singles_eXptar_eYptar}
\end{figure}
\begin{figure}[H]
  \centering
  \includegraphics[scale=0.35]{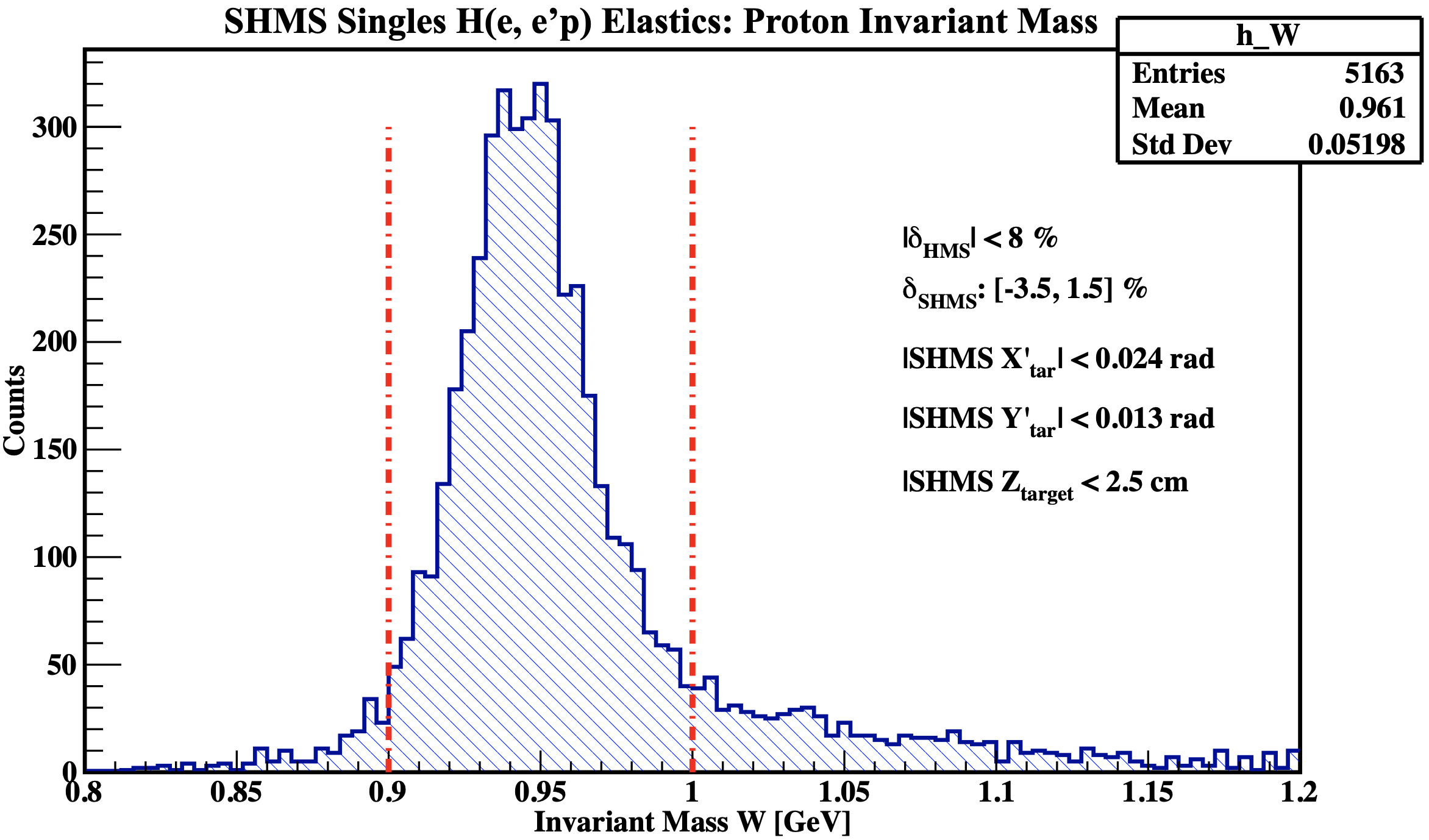}
  \caption{Proton invariant mass ($W$) determined from SHMS electron singles run 3259.}
  \label{fig:singles_Wcut}
\end{figure}
\indent After doing the dummy target analysis, we now focus on the analysis of the single-arm $^{1}$H$(e,e')p$ elastic runs. Figure \ref{fig:singles_eXptar_eYptar} shows the
SHMS angular acceptance from electron singles. The boxed (red) region corresponds to the angular acceptance
from $ep$ elastics determined from the coincidence data. A cut was also placed on the SHMS momentum acceptance determined from the elastic coincidence data
as well as the $z$-target cut mentioned above.\\
\indent Finally, a cut on the proton invariant mass $W$ (reconstructed from $^{1}$H$(e,e')p$ electron singles) has been placed between 0.9 and 1.0 GeV/c to ensure electrons that
correspond to true $ep$ elastics are selected (see Fig. \ref{fig:singles_Wcut}).\\
\indent After determining the cuts to select electrons singles that truly originated from hydrogen elastic scattering, we define the proton transmission coefficient as
\begin{subequations}
  \begin{align}
  &\epsilon_{\mathrm{pTr}} \equiv \frac{e^{-}_{\mathrm{did}}/\epsilon_{\mathrm{htrk}}}{e^{-}_{\mathrm{should}}}, \label{eq:5.23a}\\
  &\delta\epsilon_{\mathrm{pTr}} \equiv \frac{\sqrt{e^{-}_{\mathrm{should}}-(e^{-}_{\mathrm{did}}/\epsilon_{\mathrm{htrk}})}}{e^{-}_{\mathrm{should}}}, \label{eq:5.23b}
  \end{align}
\end{subequations}
where $e^{-}_{\mathrm{should}}$ is the number of electrons that passed the above-mentioned cuts for which the
correlated proton should have been detected in the HMS and $e^{-}_{\mathrm{did}}$ is a superset of (or contains) $e^{-}_{\mathrm{should}}$ with the additional requirement that the HMS reconstructed
events are within the well defined HMS momentum acceptance ($|\delta|<8\%$) and that there was an HMS 3/4 hodoscope trigger (hTRIG1 TDC $>$0). The $e^{-}_{\mathrm{did}}$ has also been corrected for
HMS tracking efficiency $\epsilon_{\mathrm{htrk}}$ since HMS tracking-related variables have been used in the determination of $e^{-}_{\mathrm{did}}$. Using these definitions, the number of electrons singles that
\textit{should} have and \textit{did} pass the cuts can be expressed as
\begin{align}
  &e^{-}_{\mathrm{should}} &\equiv (\delta_{\mathrm{SHMS}}) \wedge (\Delta \Omega_{\mathrm{SHMS}}) \wedge (\Delta Z_{\mathrm{tar}}) \wedge (\Delta \text{W}), \label{eq:5.24} \\
  &e^{-}_{\mathrm{did}} &\equiv (e^{-}_{\mathrm{should}}) \wedge (\delta_{\mathrm{HMS}}) \wedge (\text{hTRIG 3/4}), \label{eq:5.25}
\end{align}
where $\wedge$ represent the logical AND operator, and the variables in parentheses represent the applied cuts ($\Delta \Omega_{\mathrm{SHMS}}\rightarrow$ angular acceptance cuts).
Using these cuts, the ratio of $e^{-}_{\mathrm{did}}/e^{-}_{\mathrm{should}}$ using Eqs. \ref{eq:5.24} and \ref{eq:5.25} was taken for the $X'_{\mathrm{tar}}$ and $Y'_{\mathrm{tar}}$ to determine the variations in the proton transmission
factor across the acceptance of the SHMS. The plots are shown in Figs. \ref{fig:xptar_ratio} and \ref{fig:yptar_ratio}, respectively.
\begin{figure}[H]
  \centering
  \includegraphics[scale=0.34]{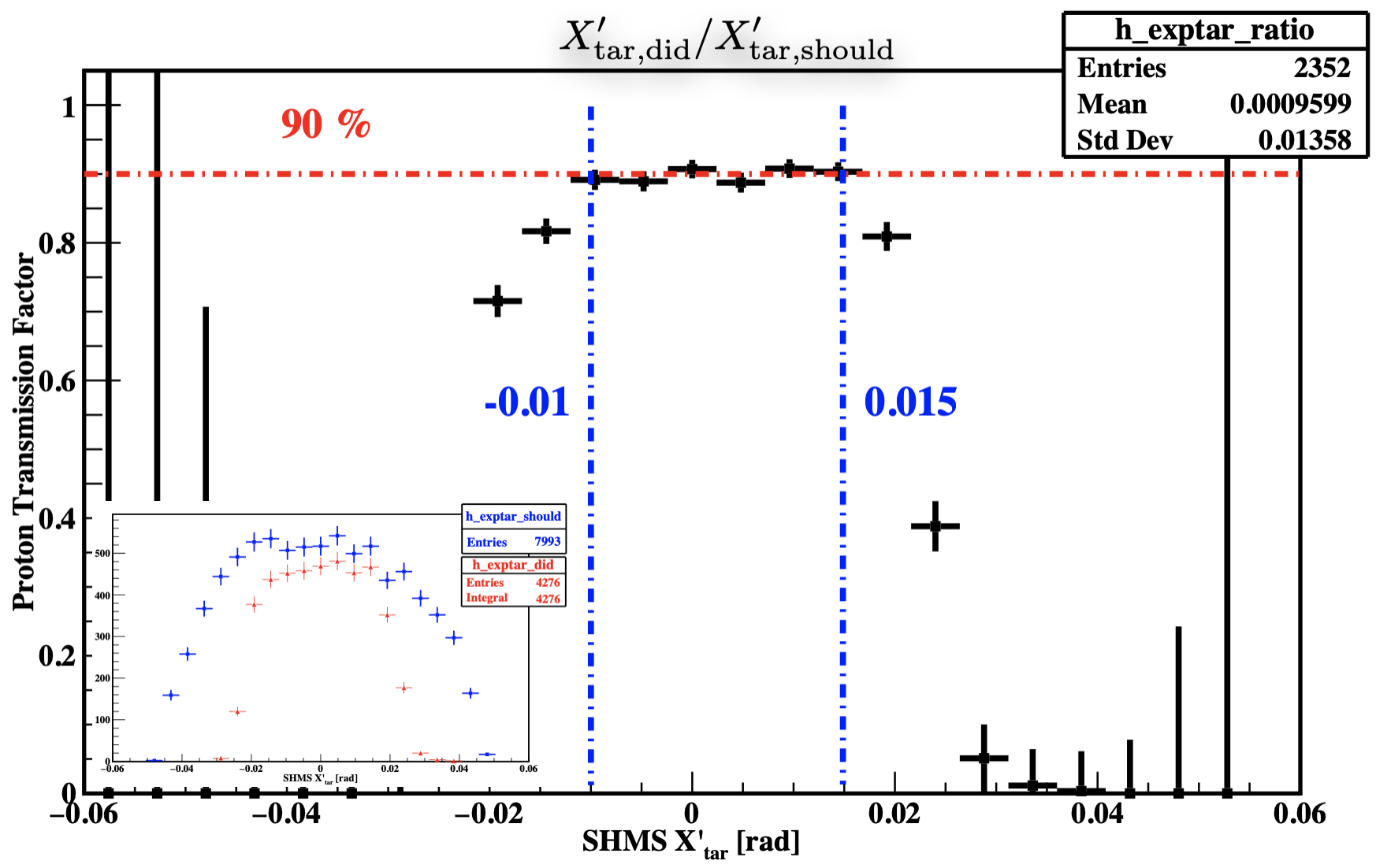}
  \caption{Ratio of $X'_{\mathrm{tar}}$ from SHMS electron singles run 3259.
    Inset: $X'_{\mathrm{tar}}$ histograms before taking the ratio, where did is in red and should is in blue.}
  \label{fig:xptar_ratio}
\end{figure}                                                                                                                         
\begin{figure}[H]
  \centering
  \includegraphics[scale=0.34]{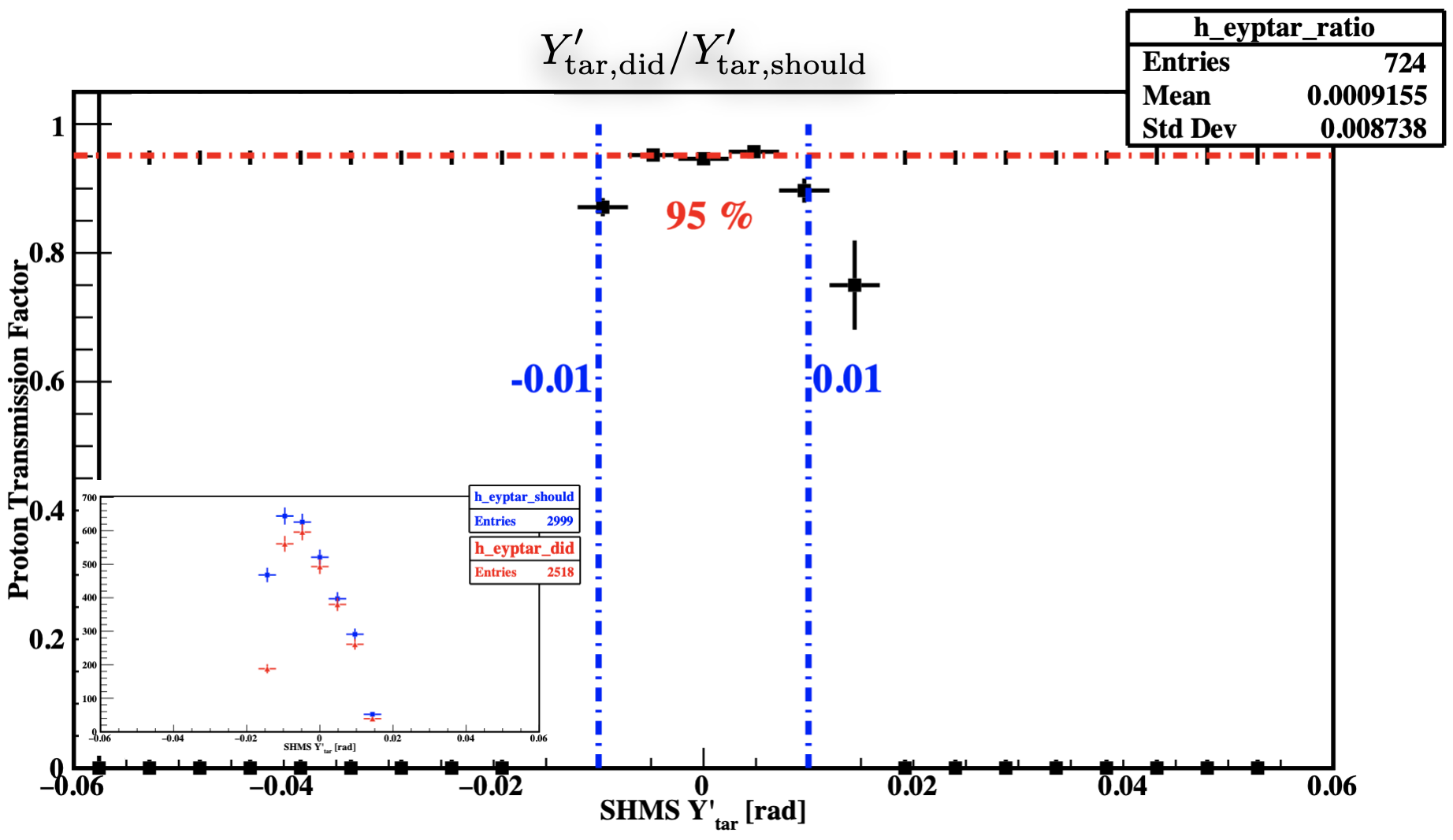}
  \caption{Ratio of $Y'_{\mathrm{tar}}$ from SHMS electron singles run 3259.
    Inset: $Y'_{\mathrm{tar}}$ histograms before taking the ratio, where did is in red and should is in blue.}
  \label{fig:yptar_ratio}
\end{figure}
\indent From these ratios, the proton transmission factor is $\sim90\%$ for the $X'_{\mathrm{tar}}$ ratio and $\sim95\%$ for the $Y'_{\mathrm{tar}}$ near the center (Note: these early plots were not corrected for the tracking efficiency). At the
edges, however, the ratio drops rapidly which indicates that the angular acceptance cuts need to be tightened. This is presumably due to the HMS momentum acceptance rolling off at large $X'_{\mathrm{tar}}$. The dashed blue lines indicate the region where the new
SHMS angular acceptance cuts will be placed.\\
\indent In addition to the tighter acceptance cuts, a cut on the total energy deposited in the SHMS calorimeter normalized by the central momentum was made to eliminate any
possible pion background. The $e^{-}_{\mathrm{did}}$ also needed to be corrected for the HMS tracking efficiency as this correction factor does not cancel in
the ratio of Eq. \ref{eq:5.23a}. The tracking efficiency for the analyzed run was determined to be 99.07$\%$ ($<1\%$ correction).\\
\begin{figure}[H]
  \centering
  \includegraphics[scale=0.32]{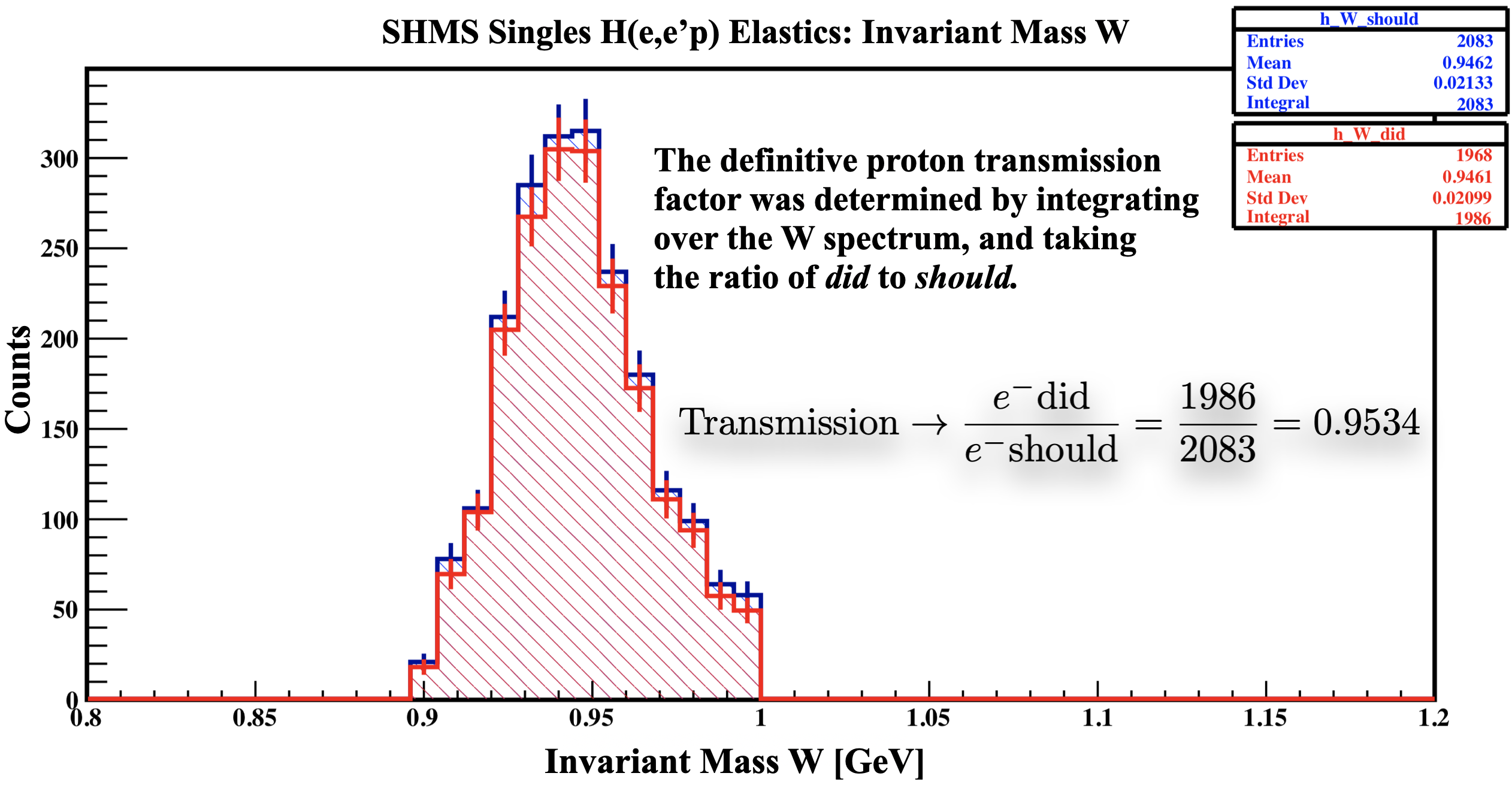}
  \caption{Invariant mass $W$ determined from SHMS elastic electron singles run 3259.}
  \label{fig:invariant_W}
\end{figure}
\indent The final proton transmission factor was determined by taking the ratio of the integrated invariant mass histograms reconstructed from
SHMS electron singles (see Fig. \ref{fig:invariant_W}). The $W_{\mathrm{did}}$ and $W_{\mathrm{should}}$ have their respective cuts,
$e^{-}_{\mathrm{did}}$ and $e^{-}_{\mathrm{should}}$, as described above where the tighter acceptacne cuts, calorimeter cut and tracking efficiency
corrections have been included. \\
\indent From Eqs. \ref{eq:5.23a}, \ref{eq:5.23b} and the number of electrons from Fig. \ref{fig:invariant_W}, the final proton transmission coefficient
and its uncertainty were determined to be: $\epsilon_{\mathrm{pTr}} = 0.9534 \pm 0.0047$. This result is interpreted
as the fraction of the protons that made it to the detector hut to form the trigger.
\section{Charge Normalization ($Q_{\mathrm{tot}}$)}
To make a direct comparison between data and SIMC and determine the experimental cross section, the data must be normalized by the total experimental charge.
The charge normalization for E12-10-003 was done on a data-set basis.
That is, for each of the missing momentum sets, the runs were combined and scaled by the corresponding accumulated charge. The precise determination of the
uncertainty in the charge is dependent on the BCM calibration, which is a work in progress. For this experiment, a conservative estimate of the relative
uncertainty on the BCM4A charge was determined to be $dQ/Q = 0.02$ (or 2$\%$)\cite{DMack_privNov2019}. 
\section{Hydrogen Normalization Check}
The $^{1}$H$(e,e')p$  is the ideal reaction to study spectrometer acceptances, as well as determining spectrometer/kinematical offsets and misalignments. This is possible due to the wide variety of data
that exists at different kinematics, which has enabled the determination of the electric ($G_{\mathrm{Ep}}$) and magnetic ($G_{\mathrm{Mp}}$) form factors very precisely over a wide kinematic range\cite{YE20188}. 
\begin{figure}[H]
  \centering
  \includegraphics[scale=0.38]{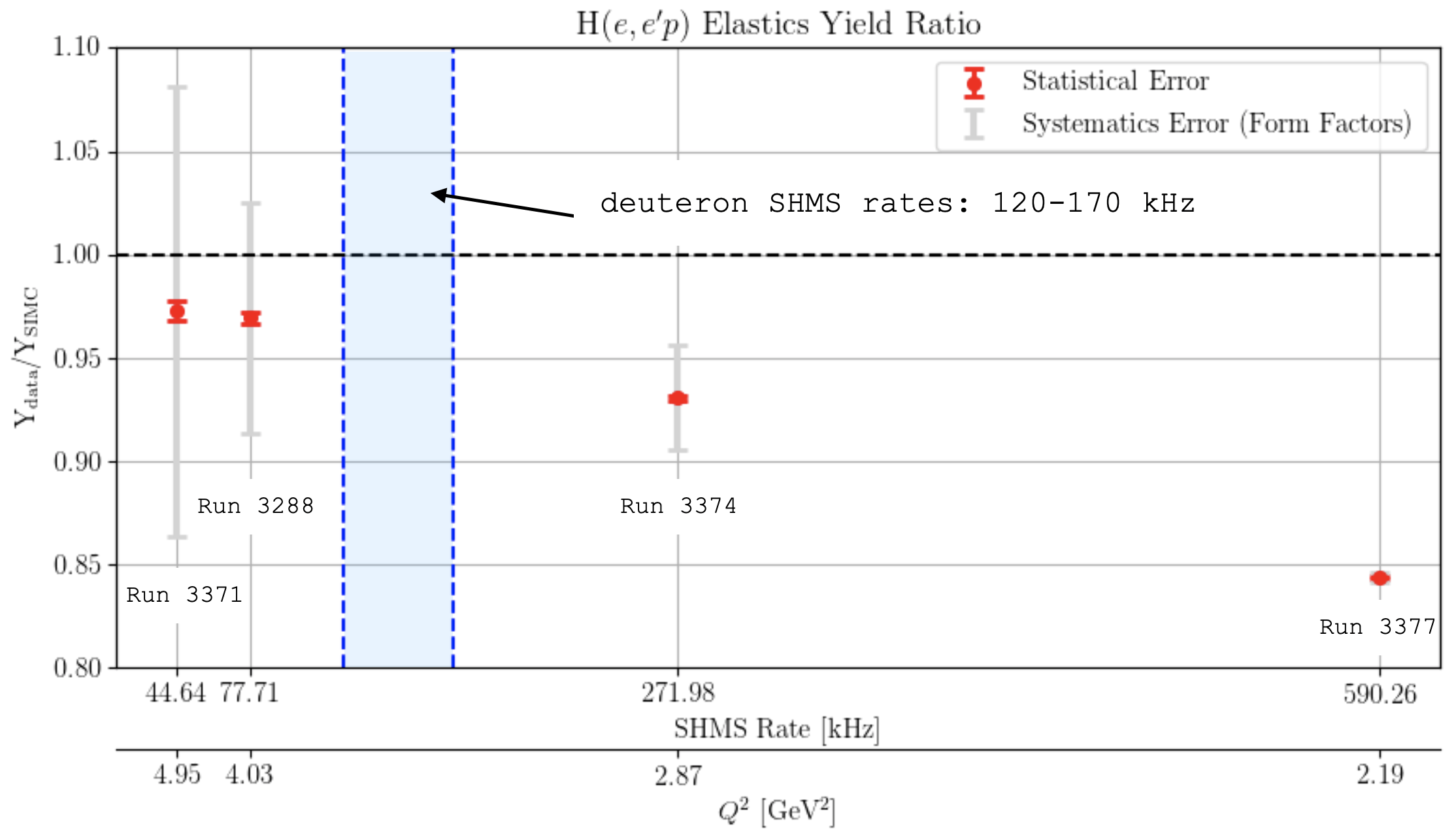}
  \caption{Fully corrected experimental data to SIMC yield ratio for four $^{1}$H$(e,e')p$ elastic data.}
  \label{fig:heep_ratio}
\end{figure}
\indent In the E12-10-003, four $^{1}$H$(e,e')p$ elastic data points taken at kinematics that covered a significant part of the SHMS acceptance ($\delta_{\mathrm{SHMS}}: -10 \text{ to } 12 \%$) were mainly
used for optics optimization (see Section \ref{sec:optics_checks}). In addition, the hydrogen data were also used to check how well they would agree with the calculated cross sections using the
form factor parametrization from Ref.\cite{PhysRevC.69.022201}. This was achieved by taking the ratio of the fully corrected data yield to the SIMC yield, which is equivalent
to the ratio of the cross sections (see Fig. \ref{fig:heep_ratio}). The yields were determined by integrating the invariant mass $W$ in the range [0.85, 1.05] GeV using similar event
selection cuts as described in Section \ref{sec:event_selection}. There is a significant drop in the ratio with increasing SHMS trigger rates, which is currently not well
understood, but likely due to unresolved issues with the tracking efficiency and/or the electronic deadtime at very high rates. However, with respect to the first two data points, the ratio is very close to one. \\
\indent We have decided to focus on the second data point (run 3288), as it is the closest in kinematics to the 80 MeV/c deuteron data. From this data point, the $^{1}$H$(e,e')p$ elastic cross section has been
measured with a normalization systematic error of about 3$\%$ at a $Q^{2}$ region where the absolute cross section is known at the $\sim$6$\%$ level (shaded gray error) determined from the uncertainties in the form factors
(Supplemental Materials of Ref.\cite{YE20188}). We quote the precise ratio of this data point to be:
\begin{align}
  R &\equiv \frac{Y_{\mathrm{data}}}{Y_{\mathrm{SIMC}}} \pm \delta_{\mathrm{stats}} \pm \delta^{(\mathrm{norm})}_{\mathrm{syst}} \pm \delta^{(G_{\mathrm{Ep}},G_{\mathrm{Mp}})}_{\mathrm{syst}} \nonumber \\
  &= 0.969 \pm 0.0027 \pm 0.031 \pm 0.0561
\end{align}
This supports our estimate of the normalization systematic error for the 80 MeV/c setting, $\sim0.036$ (see Table \ref{tab:table5.4}), but it does not tell us the normalization systematic error better than the $5.6\%$ error from the
form factors. For this reason, we have decided not to normalize the deuteron data to the $ep$ elastics run 3288.
\section{Radiative Corrections ($f_{\mathrm{rad}}$)} \label{sec:section_rad_corr}
Radiative effects contribute significantly to the determination of the experimental cross sections. These effects refer to the process by which the electron interacts with
the electric field of a nearby nucleus causing the electron to change its velocity and emit either real or virtual photons known as \textit{bremsstrahlung} radiation. This
process can be further divided into either external or internal bremsstrahlung radiation. In \textit{external bremsstrahlung}, the electron interacts with the electric field
of a nucleus other than the nucleus involved in the scattering process, whereas in \textit{internal bremsstrahlung}, it interacts with the electric field of the same nucleus it scatters off. 
Furthermore, the electron can radiate before and (or) after the (hard) scattering process. As a result, the reaction kinematics and the cross section can be modified significantly and the events might not
necessarily be reconstructed at the vertex, but at some other kinematics. This process becomes evident in the so-called radiative tails present in some of the reconstructed histograms, for example,
the reconstructed missing energy in Fig. \ref{fig:Em_3289} or the invariant mass in Fig. \ref{fig:singles_Wcut}.\\
\begin{figure}[H]
  \centering
  \includegraphics[scale=0.45]{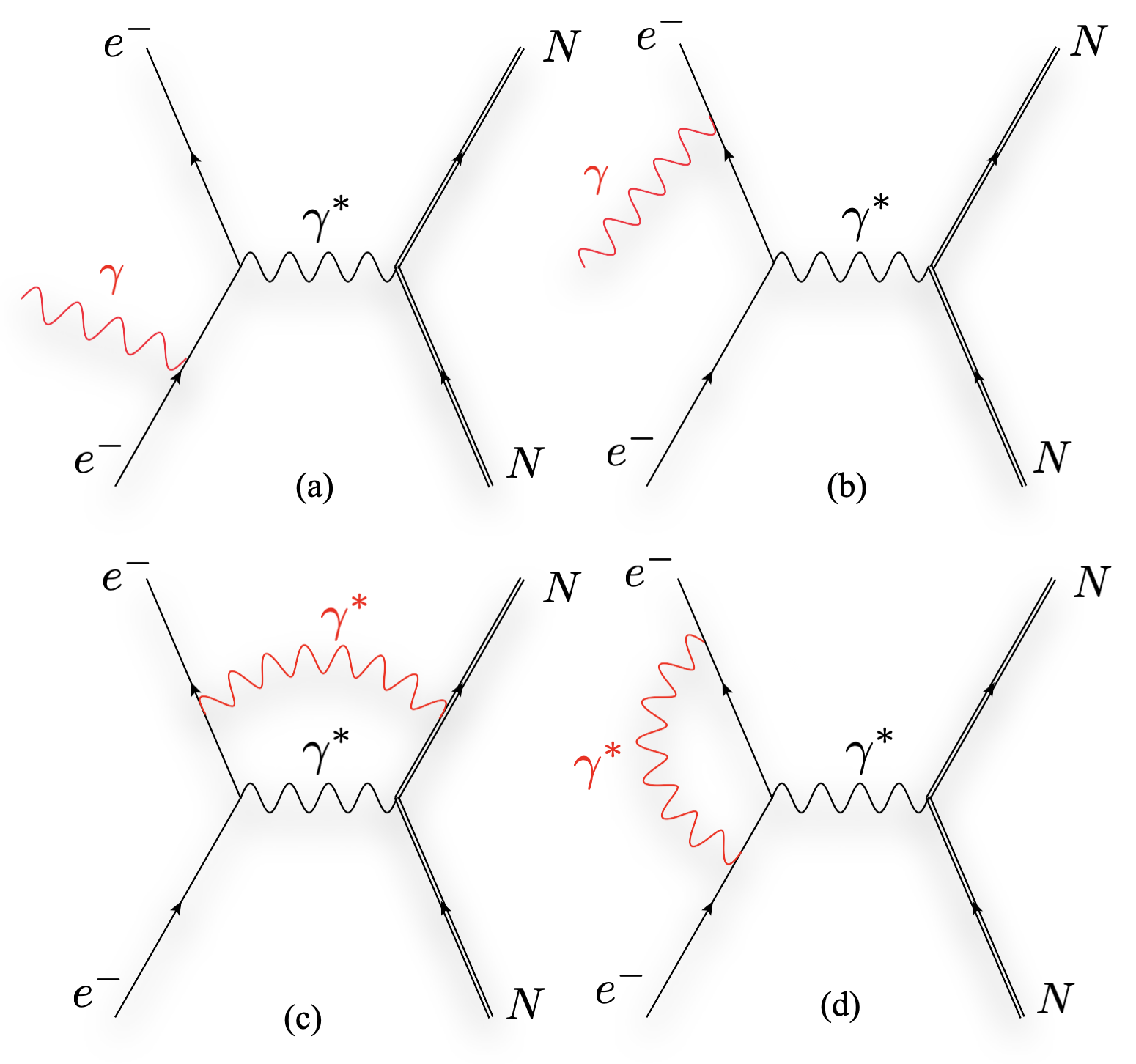}
  \caption{Examples of internal bremsstrahlung photons (red) for a general $N(e,e')N$. In the top diagrams, the electron emits a real bremsstrahlung photon (a) before and (b) after scattering off a nucleus $N$.
  In the bottom diagrams, the electron exchanges a virtual bremsstrahlung photon with (c) the nucleus $N$ and (d) between its own initial and final state.}
  \label{fig:eN_rad_diag}
\end{figure}
\indent Theoretical models do not account for these radiative effects in their calculations of the cross sections, therefore, the experimental data must be corrected before a comparison with theory
can be made. The theoretical calculations for radiative effects were first carried out by Schwinger\cite{PhysRev.76.790} and later improved by Mo and Tsai \cite{RevModPhys.41.205}. 
The simulation program SIMC uses the radiative correction formulas for coincidence $(e,e'p)$ reactions calculated using the Mo and Tsai formulation by R. Ent \textit{et al.} in Ref.\cite{PhysRevC.64.054610}.
A brief report describing how radiative effects are simulated in SIMC can be found in Ref.\cite{RadCorr_dutta}. Figure \ref{fig:eN_rad_diag} shows a Feynman diagram representation of a typical radiative
process involving internal bremsstrahlung.\\
\indent In the E12-10-003 experiment, the radiative corrections were applied by multiplying by the ratio of non-radiative to radiative SIMC yields,\\
\begin{equation}
  f_{\mathrm{rad,k}} = \frac{Y_{\mathrm{norad}}}{Y_{\mathrm{rad}}},
\end{equation}
where k is an arbitrary kinematic bin that is defined based on the choice of kinematic for binning the cross sections.
For convenience, the radiative and non-radiative yield histograms were binned in k=($p_{\mathrm{r}}, \theta_{nq}$) bins and a ratio was taken and multiplied
by the experimental yield per data set, also binned in k=($p_{\mathrm{r}}, \theta_{nq}$). As an illustrative example, Fig. \ref{fig:2d_radCorr} shows
a 2D histogram of the ratio between the non-radiative and radiative SIMC yields. The inset shows a vertical projection along $p_{\mathrm{r}}$ for the bin
$\theta_{nq}=35\pm5^{\circ}$, where the vertical axis of the inset corresponds to radiative correction factor, $f_{\mathrm{rad}}$.\\
\indent Figure \ref{fig:pm_radcorr} shows the experimental data yield binned in missing momentum before and after applying the radiative correction factor of the inset plot of Fig. \ref{fig:2d_radCorr}.
After radiative corrections, it is clear from Fig. \ref{fig:pm_radcorr} that there is a significant increase in the number of coincidence counts per mC for each bin in $p_{r}$. These ``recovered'' counts
are interpreted as the number of coincidences that should have been detected if it were not for the radiative effects modifying the reaction kinematics at the vertex, as shown in Fig. \ref{fig:eN_rad_diag}.
\begin{figure}[H]
  \centering
  \includegraphics[scale=0.31]{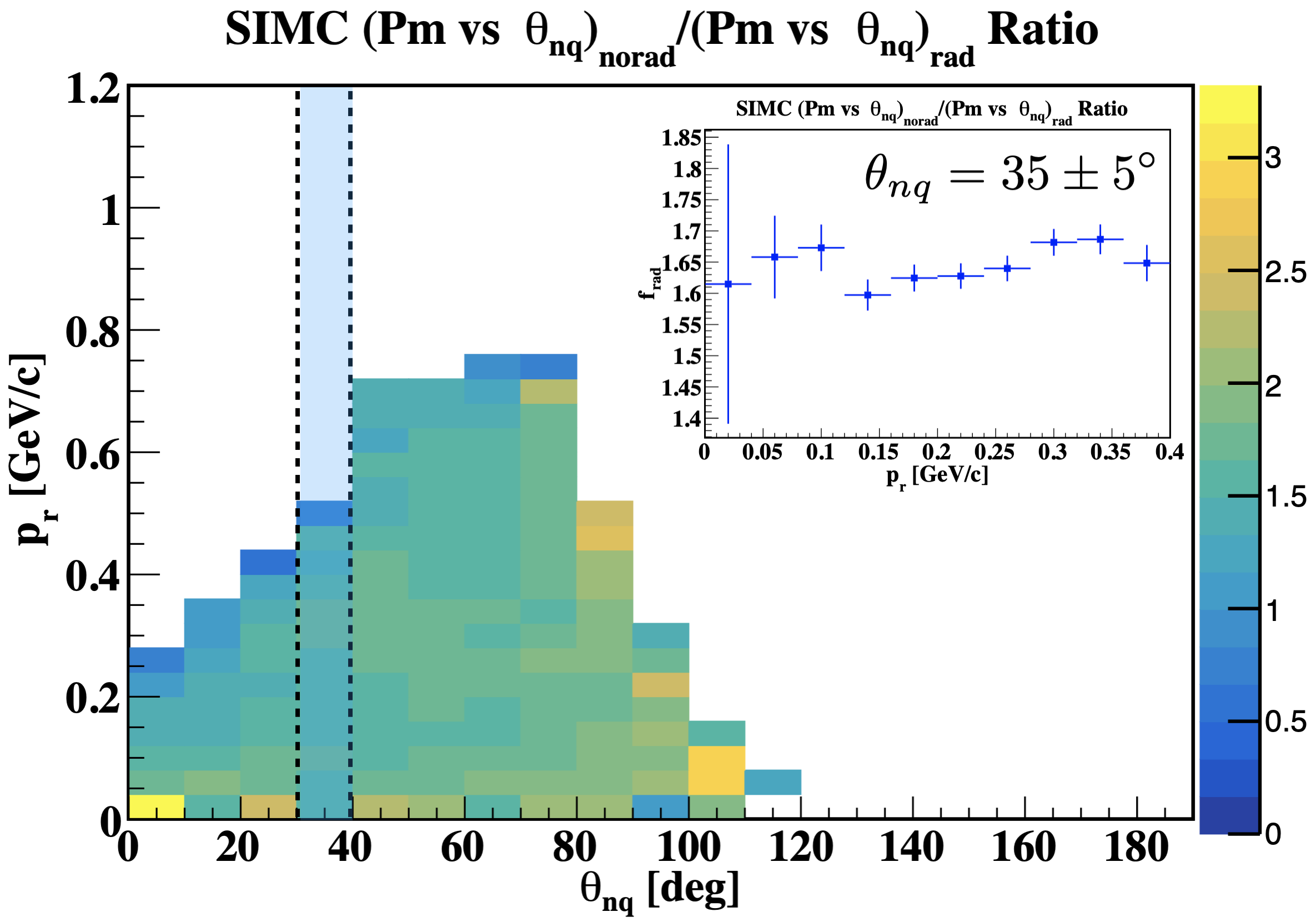}
  \caption{2D histogram of radiative correction factor, $f_{\mathrm{rad}}$, binned in $(p_{\mathrm{r}},\theta_{nq})$ for the 80 MeV/c setting of E12-10-003. Inset: Y-projection of $\theta_{nq}$ between 30 and 40 degrees.}
  \label{fig:2d_radCorr}
\end{figure}
\begin{figure}[H]
  \centering
  \includegraphics[scale=0.31]{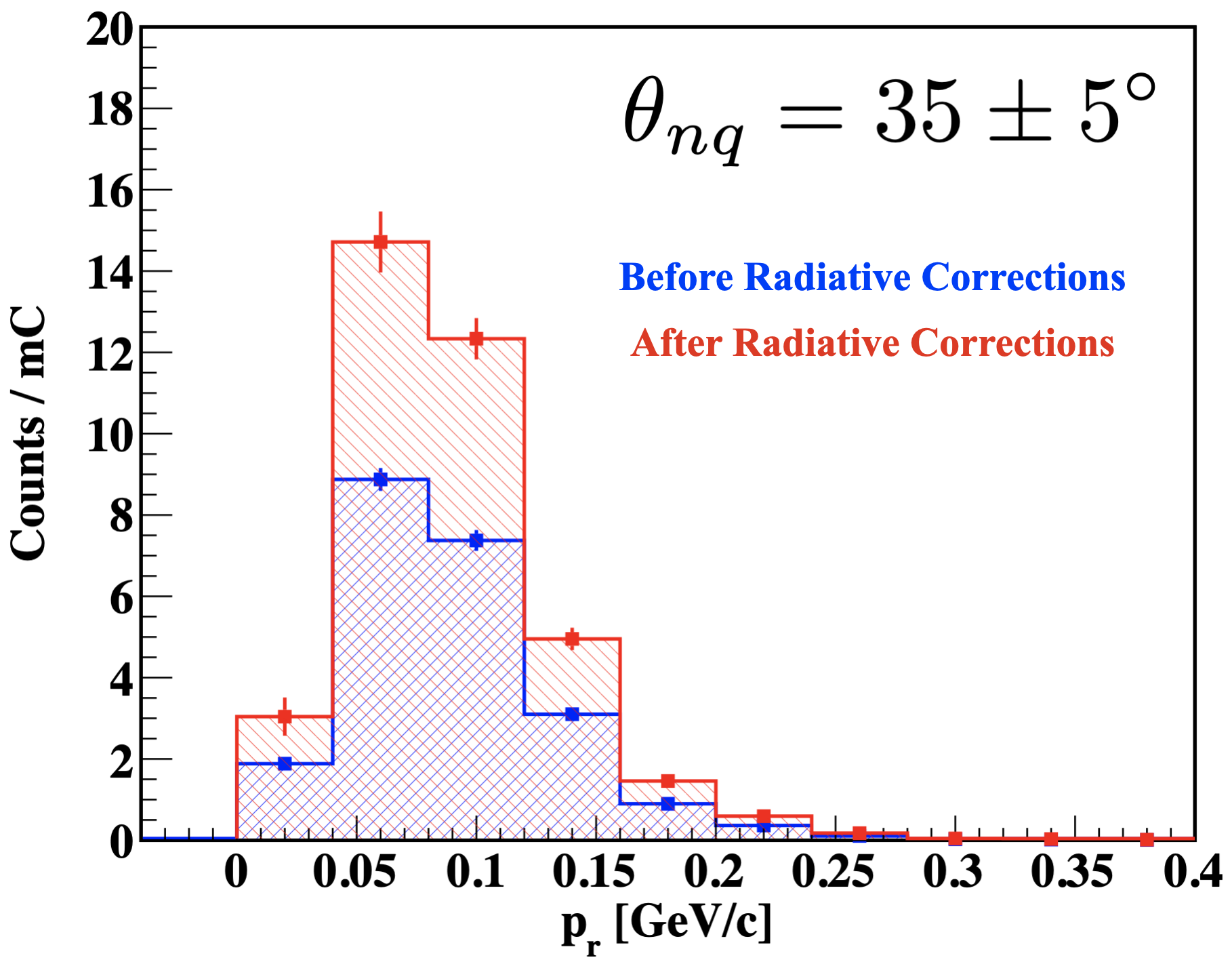}
  \caption{Missing momentum yield for $\theta_{nq}=35\pm5^{\circ}$ before and after radiative corrections for the 80 MeV/c setting of E12-10-003.}
  \label{fig:pm_radcorr}
\end{figure}
\section{Bin Centering Corrections ($f_{\mathrm{bc}}$)} \label{sec:bin_cent_corr}
\begin{figure}[H]
  \centering
  \includegraphics[scale=0.35]{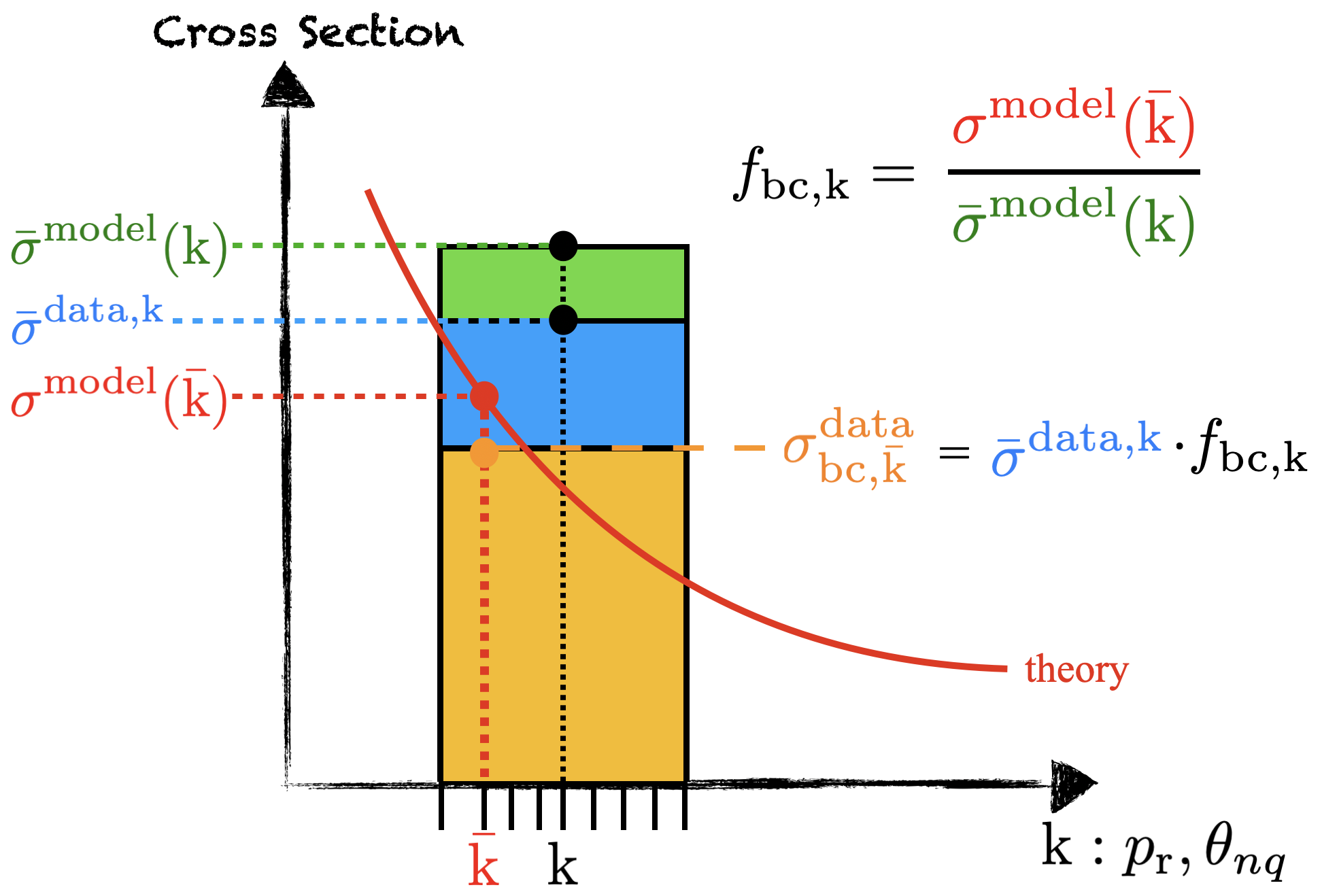}
  \caption{Cartoon illustrating bin-centering calculation for this experiment.}
  \label{fig:bin_centering}
\end{figure}

\begin{figure}[H]
  \centering
  \includegraphics[scale=0.65]{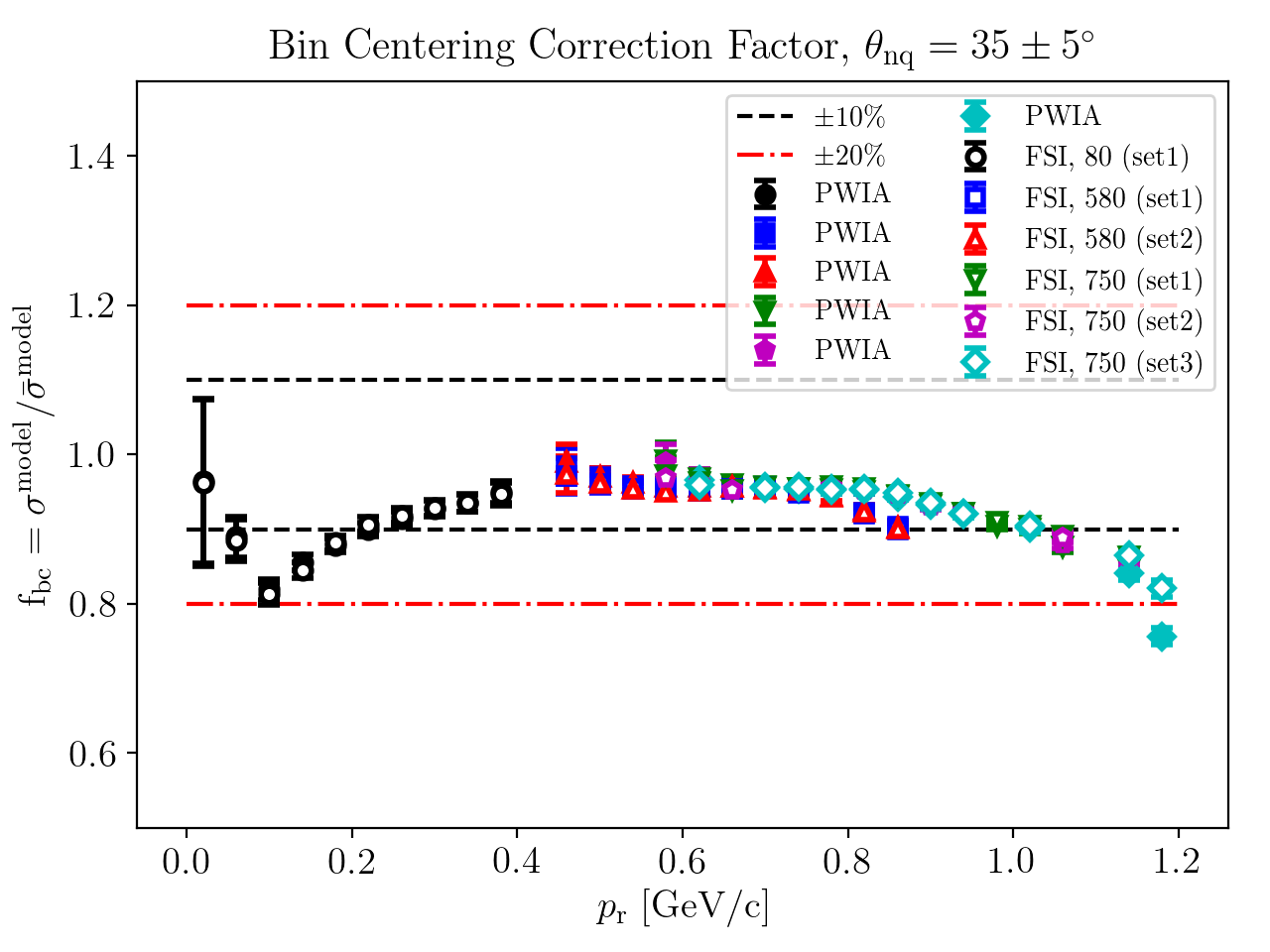}
  \caption{Bin-centering correction factor at $\theta_{nq}=35\pm5^{\circ}$ for each $p_{\mathrm{r}}$ setting of E12-10-003.
    The dashed reference lines indicate $\pm10\%$ (black) or $\pm20\%$ (red) deviation from unity.
    The correction factor was calculated by taking the ratio of cross sections (see Eq. \ref{eq:5.28}) either within the
    PWIA (full data points) or by including FSI (empty data points) for each dataset (see Section \ref{sec:exp_overview}).
    The theoretical cross section calculations were by J.M. Laget\cite{LAGET2005} using the Paris potential\cite{Paris_NN_Lacombe1980}.}
  \label{fig:bin_cent_35deg}
\end{figure}

\begin{figure}[H]
  \centering
  \includegraphics[scale=0.65]{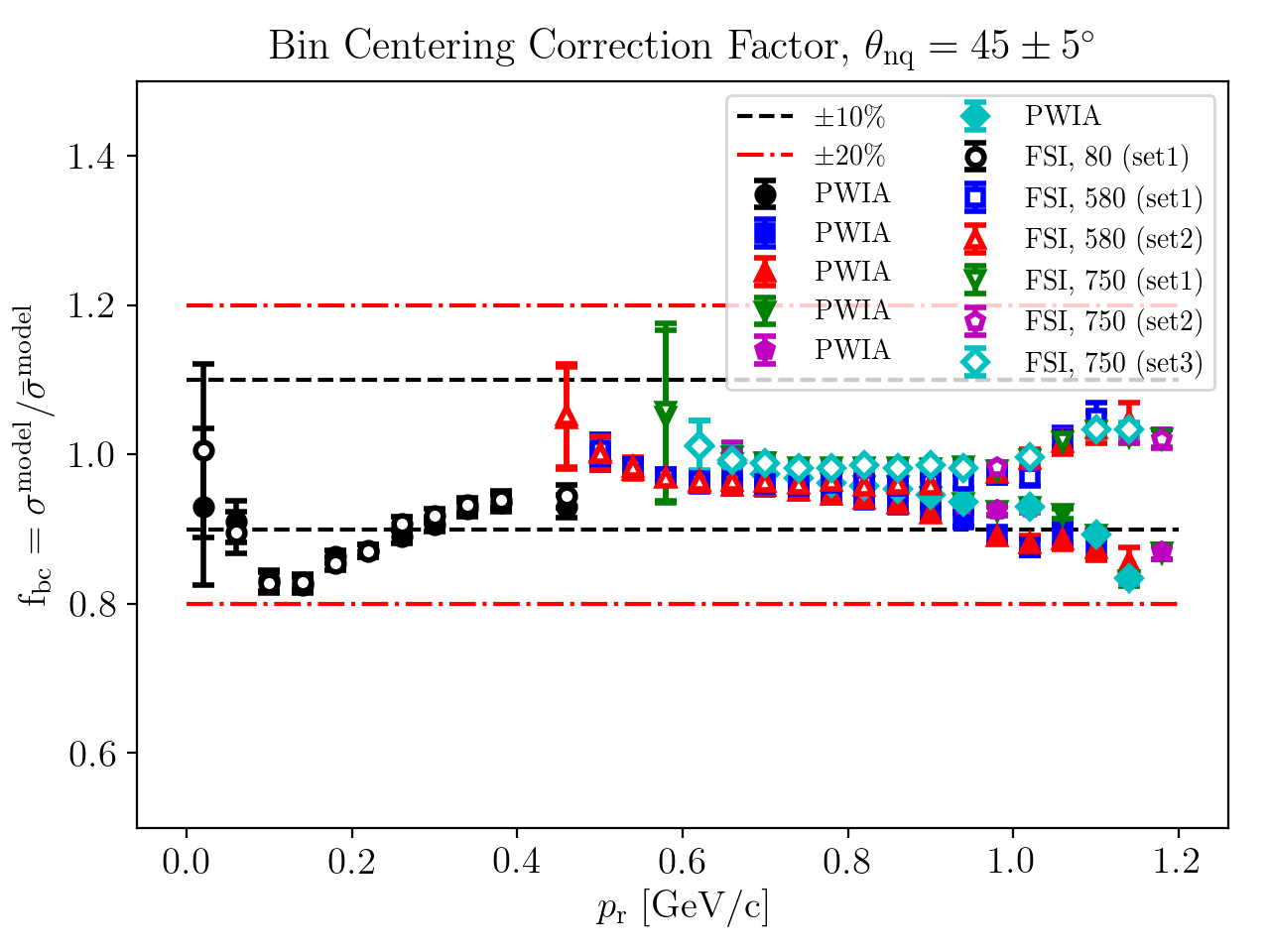}
  \caption{Bin-centering correction factor at $\theta_{nq}=45\pm5^{\circ}$ for each $p_{\mathrm{r}}$ setting of E12-10-003. See Fig. \ref{fig:bin_cent_35deg} for definition of lines and data points.}
  \label{fig:bin_cent_45deg}
\end{figure}

As mentioned in Section \ref{sec:exp_Xsec}, the experimental cross sections are averaged over a kinematic bin k, which was re-defined as k=($p_{\mathrm{r}}, \theta_{nq}$) in Section \ref{sec:section_rad_corr}.
In reality,  $(p_{\mathrm{r}}, \theta_{nq})$ has a sub-range of kinematics over which the theoretical cross section can vary rapidly so the question arises as to
which kinematic value within this bin should be associated with the experimental cross section.\\
\indent We define the bin-centering correction factor for the k=($p_{\mathrm{r}}, \theta_{nq}$) kinematic bin as
\begin{equation}
  f_{\mathrm{bc,k}} \equiv \frac{\sigma^{\mathrm{model}}(\bar{\mathrm{k}})}{\bar{\sigma}^{\mathrm{model}}(\mathrm{k})},
  \label{eq:5.28}
\end{equation}
where the numerator is the theoretical cross section calculated at the averaged kinematics (calculation is external to SIMC) and the denominator is the theoretical cross section determined from SIMC,
averaged over the kinematic bin k=($p_{\mathrm{r}}, \theta_{nq}$). The Laget FSI model was used for the theoretical cross section calculations. See Section \ref{sec:syst_effects_studies}
for systematic studies using the PWIA and FSI Laget models.\\
\indent The bin-center corrected data cross sections at each kinematic bin were determined by multiplying the average experimental cross section by the correction factor
at each kinematic bin as follows:
\begin{equation}
  \sigma^{\mathrm{data}}_{\mathrm{bc,\bar{k}}} \equiv \bar{\sigma}^{\mathrm{data,k}} \cdot f_{\mathrm{bc,k}},
\end{equation}
where $\sigma^{\mathrm{data}}_{\mathrm{bc,\bar{k}}}$ is the bin-centered corrected data cross section evaluated at the averaged kinematics ($\bar{\mathrm{k}}$) over bin k=($p_{\mathrm{r}}, \theta_{nq}$).\\
\indent Figure \ref{fig:bin_centering} shows an illustrative example of how the bin-centering corrections were done for this experiment. The blue
and green bins represent the bin content of the average data and model cross sections, and the theoretical curve represents the same model
cross section evaluated at the averaged kinematics, denoted by a bar. The orange bin represents the bin content of the data cross sections after
applying the bin-centering corrections. The advantage of performing the bin-centering corrections this way is that one
can avoid time-consuming Monte Carlo calculations when comparing the experimental data to different theoretical models.  \\
\indent The averaged kinematics on this experiment were determined from SIMC at the reaction vertex for every ($p_{\mathrm{r}}, \theta_{nq}$) bin since the
vertex quantities are corrected for energy loss at the target. The average for a kinematic quantity $X$ for the kinematic bin k=($p_{\mathrm{r}}, \theta_{nq}$)
was calculated as
\begin{equation}
  \bar{X}_{\mathrm{k}} = \frac{(\sum\limits_{i} X_{i}w_{i})_{\mathrm{k}}}{(\sum\limits_{i}w_{i})_{\mathrm{k}}},
\end{equation}
where $X_{i}$ is the value of the kinematic quantity $X$ for the event $i$ and the model cross section was used as weight ($w_{i}$).\\
\indent Figures \ref{fig:bin_cent_35deg} and \ref{fig:bin_cent_45deg} show the bin centering correction versus $p_{\mathrm{r}}\pm$20 MeV/c bins at $\theta_{nq}$=35 and 45$\pm5^{\circ}$ bins
using either the PWIA or FSI models from Ref.\cite{LAGET2005}. \\
At the low missing momentum setting (80 MeV/c), the corrections are relatively large (ignoring the lowest missing momenta at $\sim20$ MeV/c) with correction factors between 10-20$\%$ for recoil momenta up to $p_{\mathrm{r}}\sim240$ MeV/c whereas at higher
recoil momenta up to $\sim1$ GeV/c, the correction factors are typically within 10$\%$. The larger bin-centering corrections at lower missing momenta reflect the rapid fall-off of the deuteron wavefunction with
increasing $p_{\mathrm{r}}$, whereas the smaller corrections at higher missing momenta are representative of a smaller (less steep) fall-off as will be shown in the reduced cross sections presented in Chapter \ref{chap:chapter6}.

\section{Systematic Uncertainty Studies} \label{sec:syst_effects_studies}
\begin{figure}[H]
  \centering
  \includegraphics[scale=0.59]{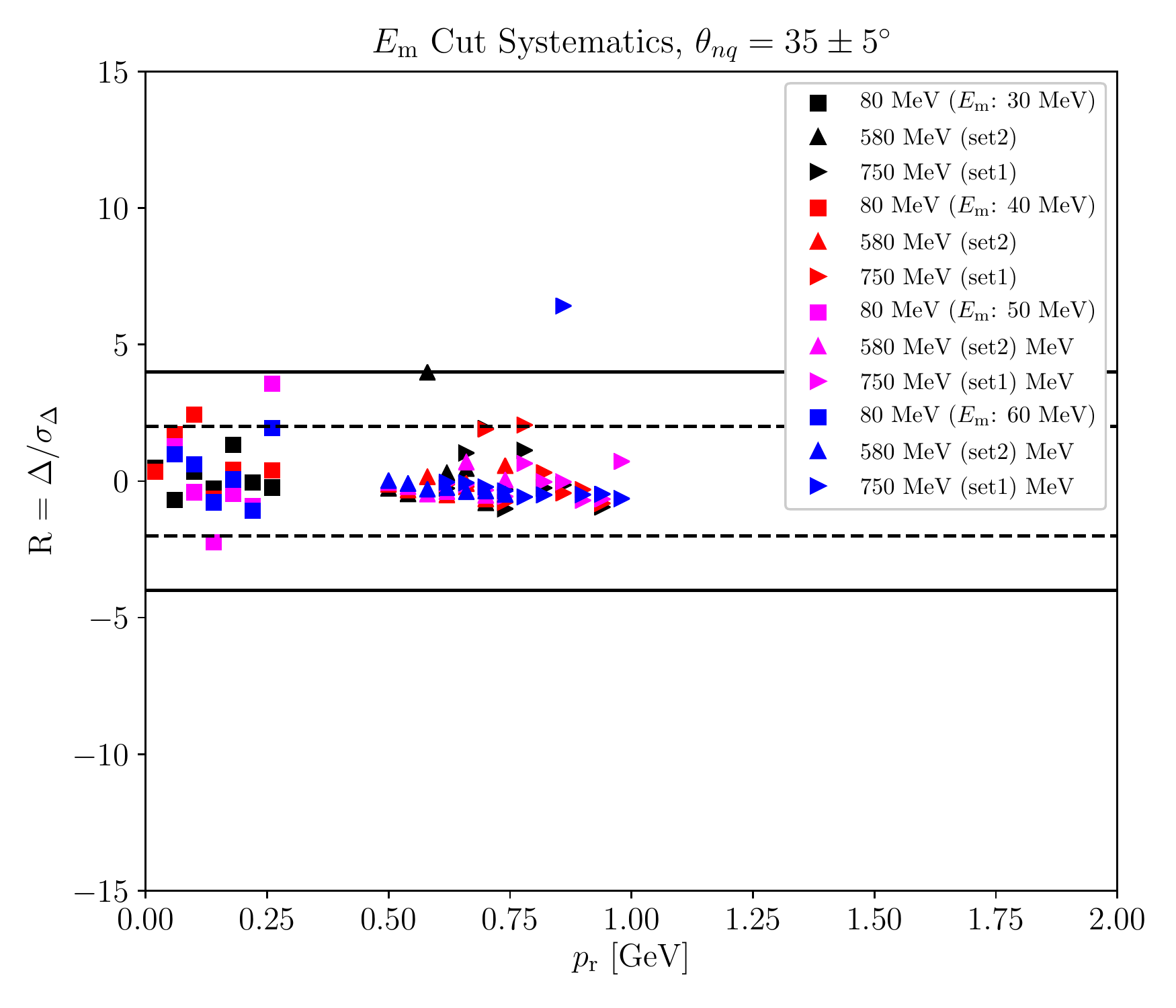}
  \caption{Systematic effects of the missing energy cut. The inner dashed and outer solid lines represent the $\Delta=\pm2\sigma_{\Delta}$ and
    $\pm4\sigma_{\Delta}$ boundaries, respectively.}
  \label{fig:Em_cut_study}
\end{figure}

\begin{figure}[H]
  \centering
  \includegraphics[scale=0.60]{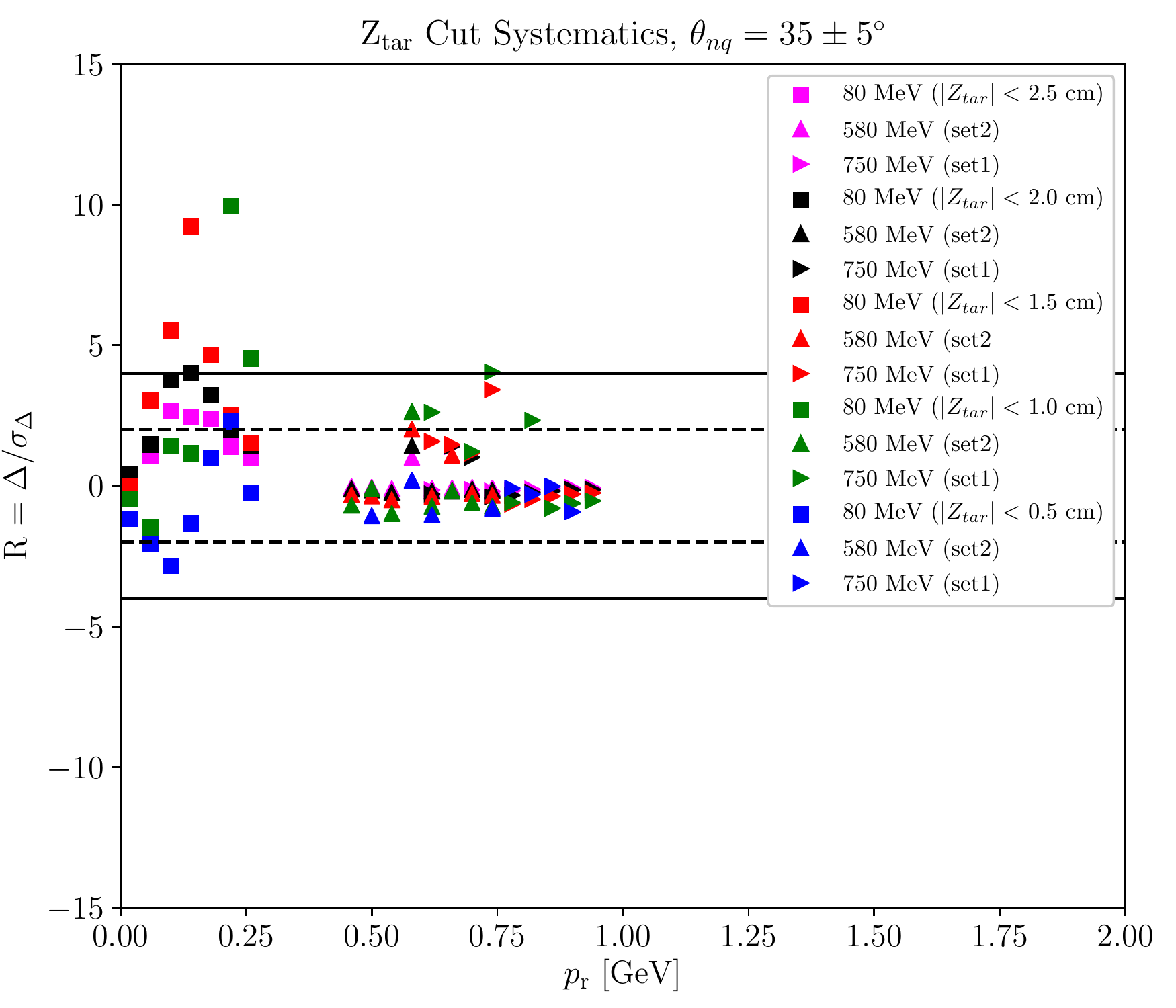}
  \caption{Systematic effects of the $Z_{\mathrm{tar}}$ difference cut. The lines are described in Fig. \ref{fig:Em_cut_study}.}
  \label{fig:Ztar_cut_study}
\end{figure}

\begin{figure}[H]
  \centering
  \includegraphics[scale=0.60]{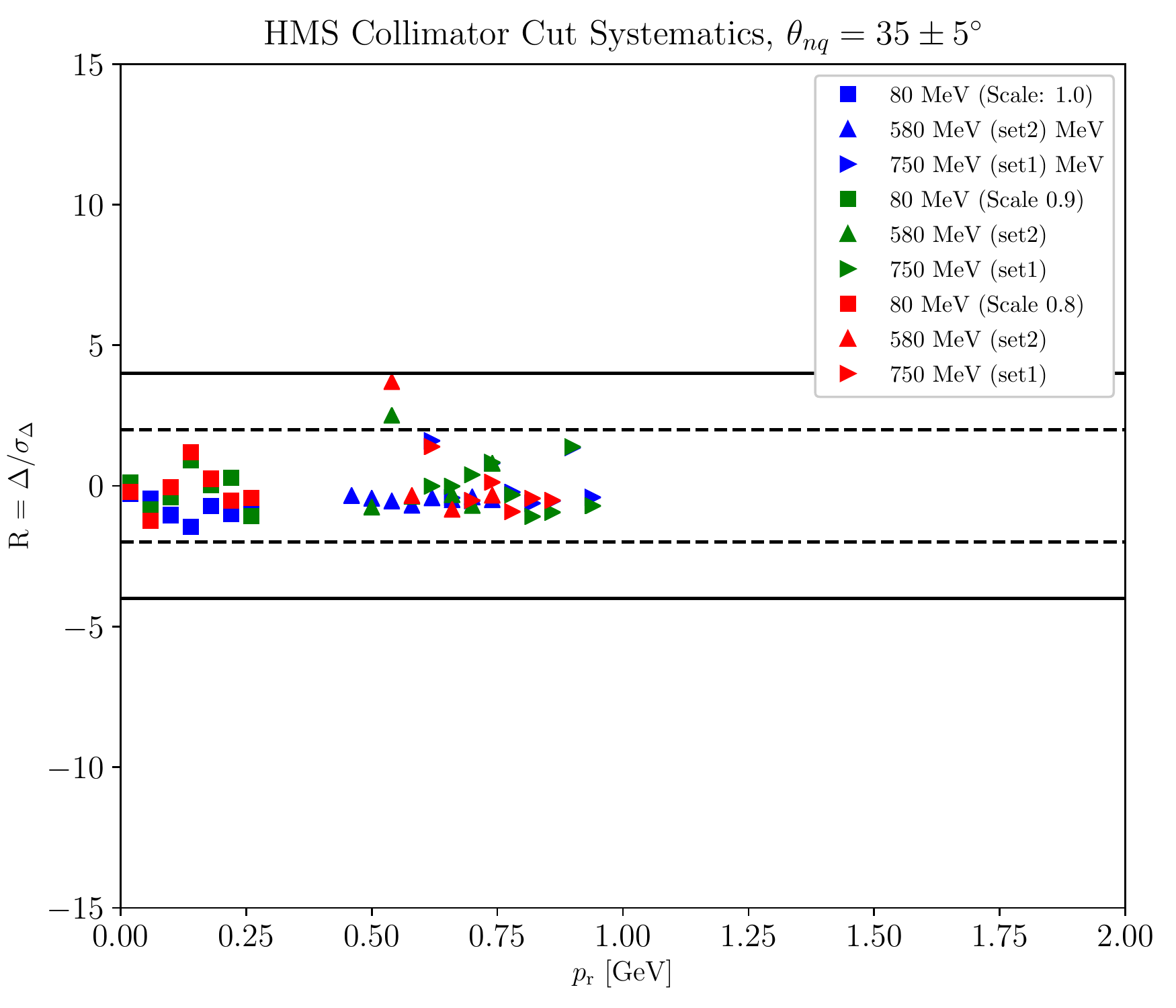}
  \caption{Systematic effects of the HMS collimator cut. The lines are described in Fig. \ref{fig:Em_cut_study}.}
  \label{fig:hcoll_cut_study}
\end{figure}

\begin{figure}[H]
  \centering
  \includegraphics[scale=0.60]{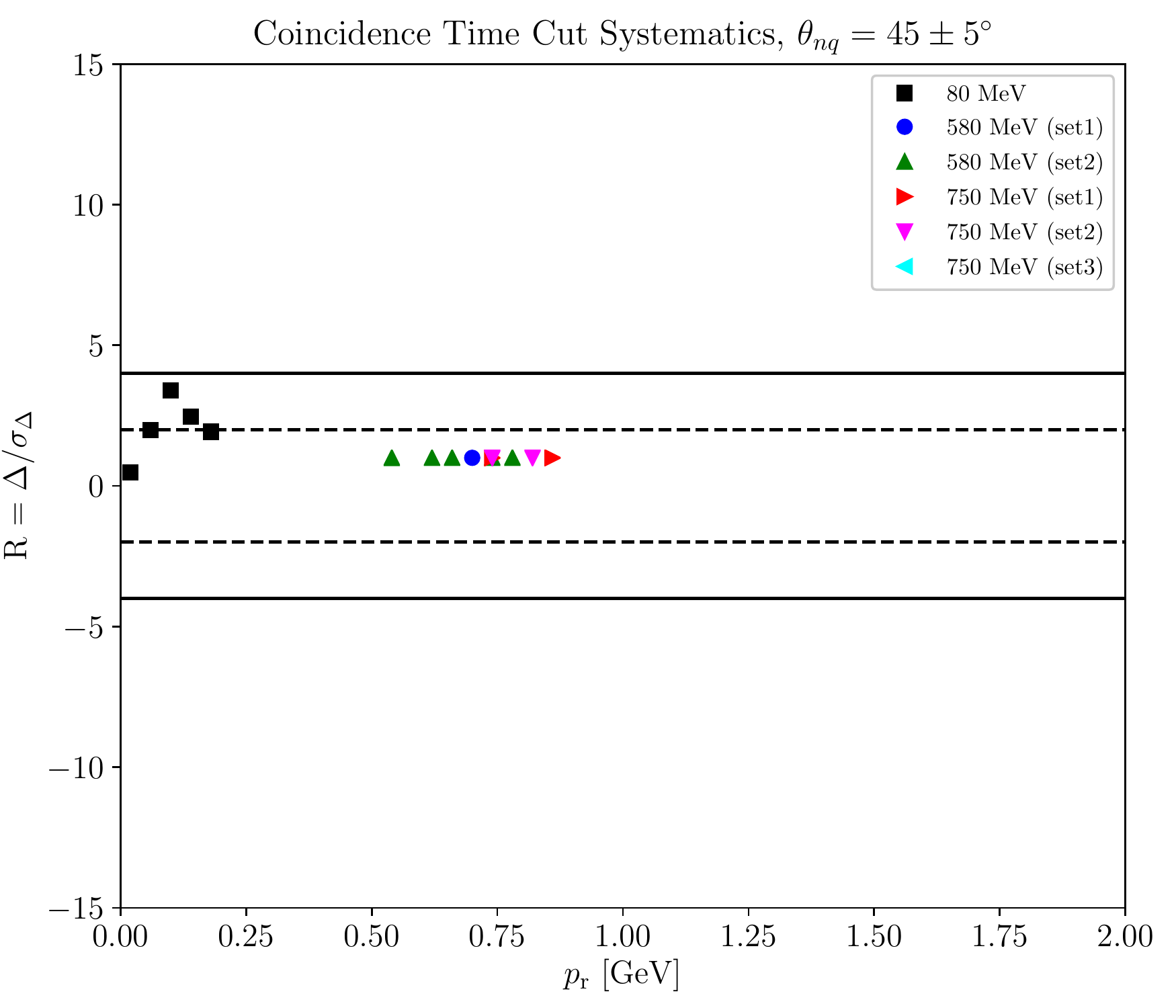}
  \caption{Systematic effects of the coincidence time cut. The lines are described in Fig. \ref{fig:Em_cut_study}.}
  \label{fig:ctime_cut_study}
\end{figure}

\begin{figure}[H]
  \centering
  \includegraphics[scale=0.60]{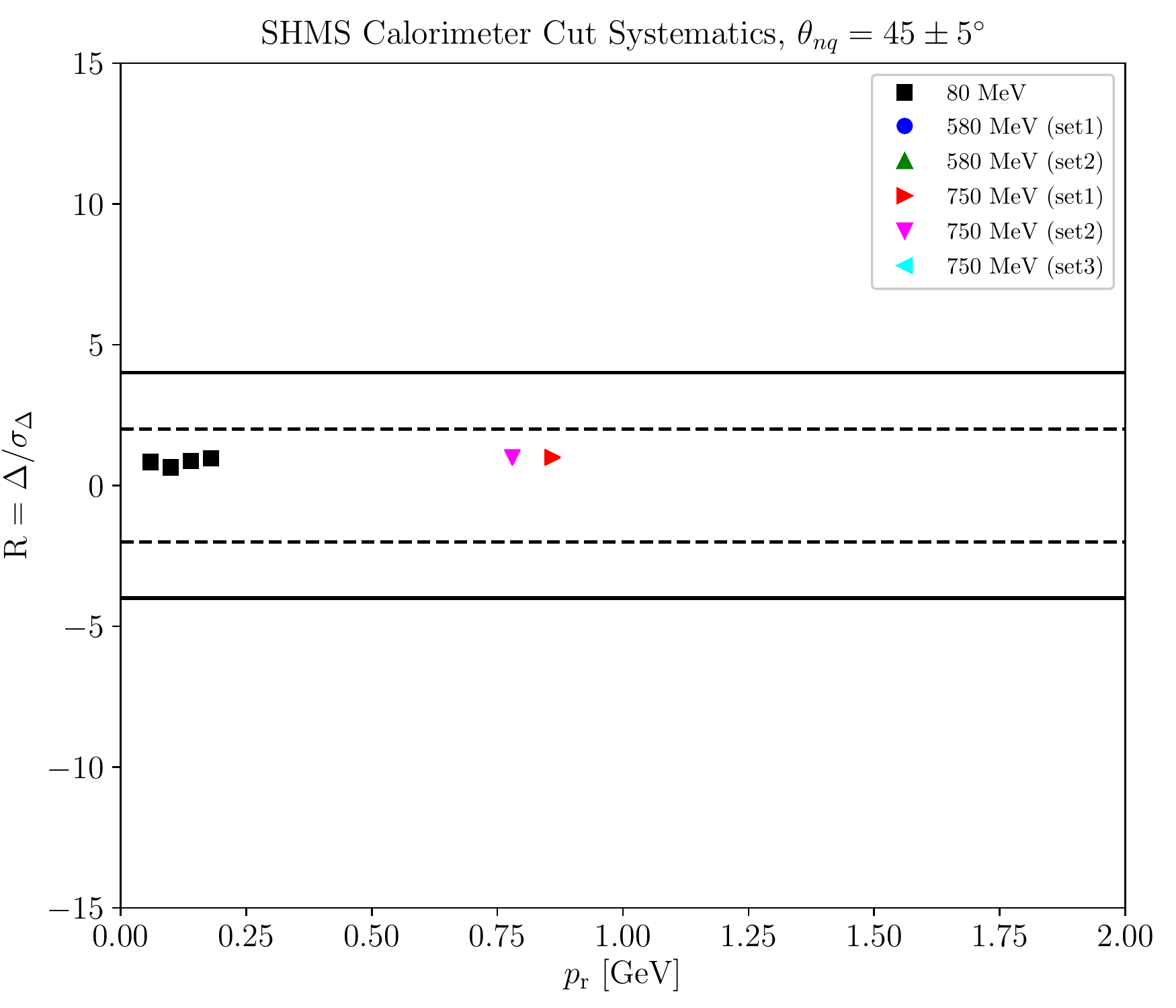}
  \caption{Systematic effects of the SHMS calorimeter cut. The lines are described in Fig. \ref{fig:Em_cut_study}.}
  \label{fig:shms_cal_cut_study}
\end{figure}

A study of the sensitivity of the bin-center corrected experimental cross sections to variations in
the event selection cuts described in Section \ref{sec:event_selection} was carried out.
Comparing two data subsets one needs to take into account the fact that part of the data are correlated.
To determine if the variation in each of the cuts contributes
to a systematic effect and whether this contribution is significant enough to be considered as a systematic error, we use the approach by R. Barlow described in Ref.\cite{barlow2002systematic}.\\
\indent Consider a cross section measurement done two different ways (i.e., apply different cuts). Let the measurements and their statistical uncertainties be:
($\sigma^{\mathrm{exp}}_{\mathrm{bc,1}}\pm\delta\sigma^{\mathrm{exp}}_{\mathrm{bc,1}}$) and ($\sigma^{\mathrm{exp}}_{\mathrm{bc,2}}\pm\delta\sigma^{\mathrm{exp}}_{\mathrm{bc,2}}$) where one
of the measurements is a subset of the other. The difference and its associated uncertainty can be expressed as,
\begin{subequations}
  \begin{align}
    &\Delta \equiv \sigma^{\mathrm{exp}}_{\mathrm{bc,1}} - \sigma^{\mathrm{exp}}_{\mathrm{bc,2}}, \\
    &\sigma^{2}_{\Delta} \equiv (\delta\sigma^{\mathrm{exp}}_{\mathrm{bc,1}})^{2} - (\delta\sigma^{\mathrm{exp}}_{\mathrm{bc,2}})^{2},
  \end{align}
\end{subequations}
where the error of the difference between the two measurements is found by taking the difference of their variance. As demonstrated in Ref.\cite{barlow2002systematic}, this
error accounts for the possible correlation between the two measurements. By taking the ratio
\begin{equation}
  R_{\mathrm{Barlow}} \equiv \frac{\Delta}{\sigma_{\Delta}},
\end{equation}
a criterion imposed on $R_{\mathrm{Barlow}}$ determines whether the difference is significant enough to be considered as a systematic error or sufficiently small that it may be
ignored. This criterion requires knowledge of the correlation between the subsets, but in general, as suggested in Ref.\cite{barlow2017}: if $R_{\mathrm{Barlow}} < 2$ (or $\Delta <2\sigma_{\Delta}$)
the test passes and if $R_{\mathrm{Barlow}} > 4$ (or $\Delta >4\sigma_{\Delta}$), the test fails and the discrepancy must be added as a systematic error. For $2<R_{\mathrm{Barlow}}<4$, a judgement must be made.\\
\indent In E12-10-003, for each of the event selection cuts, the difference between the data cross section corresponding to the largest cut and each of the subset cuts was taken and
divided by the corresponding difference in their variances to the half power to determine the ratio for each ($p_{\mathrm{r}}, \theta_{nq}$) bin. As an example, the results for a single $\theta_{nq}$
bin on each of the event selection cuts are shown in Figs. \ref{fig:Em_cut_study}-\ref{fig:shms_cal_cut_study}.
\noindent The Barlow ratio shown in these plots are used mainly as a check that the deviations in the cross section due to applied cuts are within a
reasonable boundary ($<4\sigma_{\Delta}$) such that the systematic effects can be neglected. Some important points to keep in mind on the interpretation of these systematic studies results are:
\begin{itemize}
\item As can be seen from some of these plots, for example, in Fig. \ref{fig:ctime_cut_study} or \ref{fig:shms_cal_cut_study},
the data points are scarce, which indicates that there was no difference in the measured cross sections giving a ratio of zero.
This is understood from the fact that this experiment was very clean of any background sources, and so taking the difference in cross sections
with and without these cuts does not affect the final result at all. 
\item For the missing energy cut, we do expect the ratio to change, as we are changing how much of the radiative tail we include in the cut,
so a different number of counts is expected. The ratio was found to be more spread out in the 80 MeV/c setting, however, it
was mostly within the boundaries at the higher missing momentum settings. This could be attributed to the low number of statistics in the larger settings and hence
larger variances (and statistical fluctuations).
\item Some of the ratios might be slightly negative. So one asks, how it can be that when a subset cut cross section is subtracted
from the cross section of the total set, we end up with negative? Which could only mean that the subset cross section is larger.
One possible explanation is that this is due to the radiative correction factor becoming smaller with larger cuts, which means
that a very tight cut (subset) can have a larger cross section due to a larger radiative correction factor. Another possibility is that
a smaller cut reduces the acceptance of the spectrometer.
\item If the variances of the two measurements are almost the same, this can give very large values of the ratio, as the denominator
is the difference in the variances. 
\end{itemize}
\begin{figure}[H]
  \centering
  \includegraphics[scale=0.57]{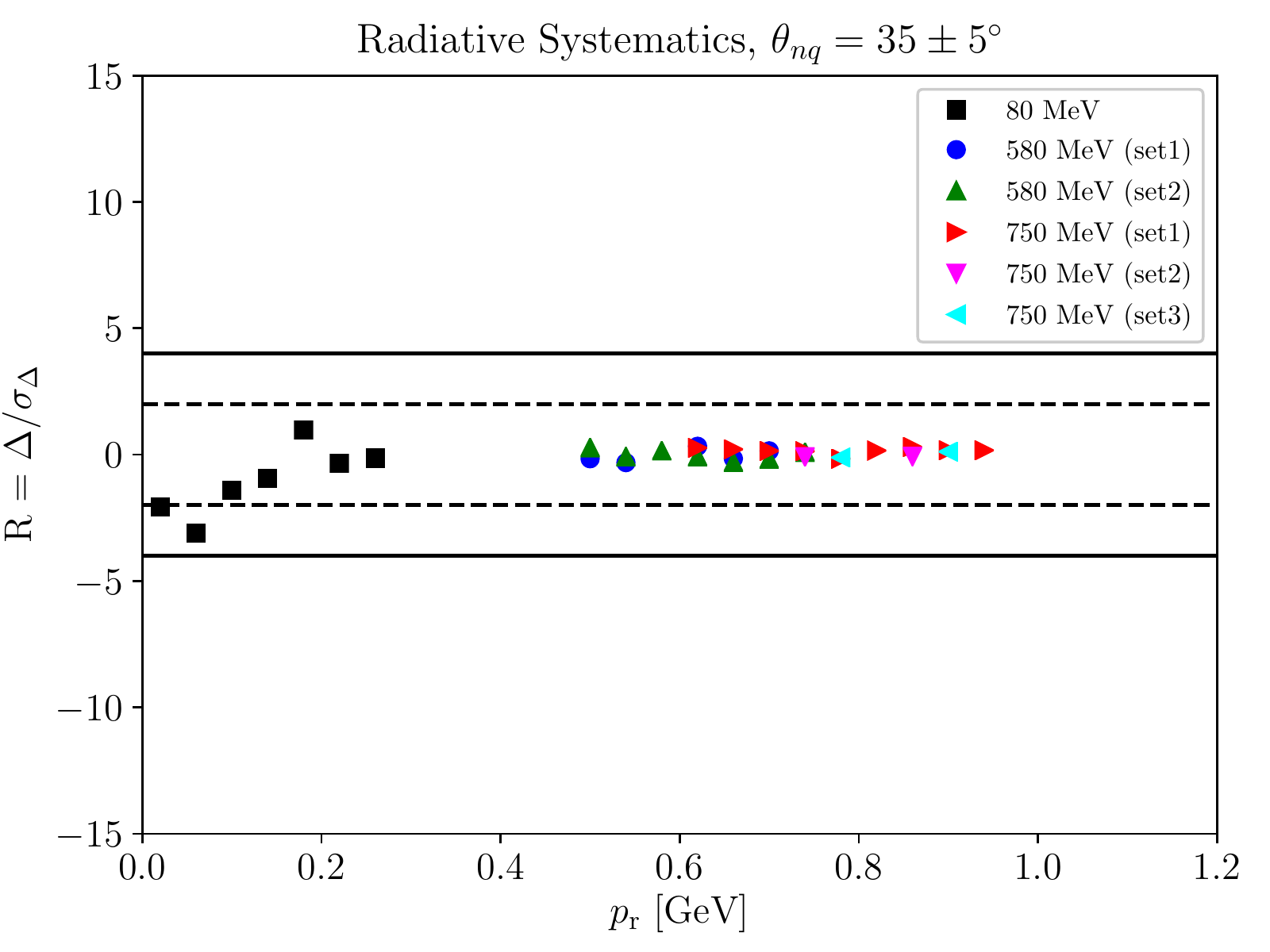}
  \caption{Systematic effects of the model dependency of radiative corrections. The lines are described in Fig. \ref{fig:Em_cut_study}.}
  \label{fig:rad_corr_study}
\end{figure}
\begin{figure}[H]
  \centering
  \includegraphics[scale=0.57]{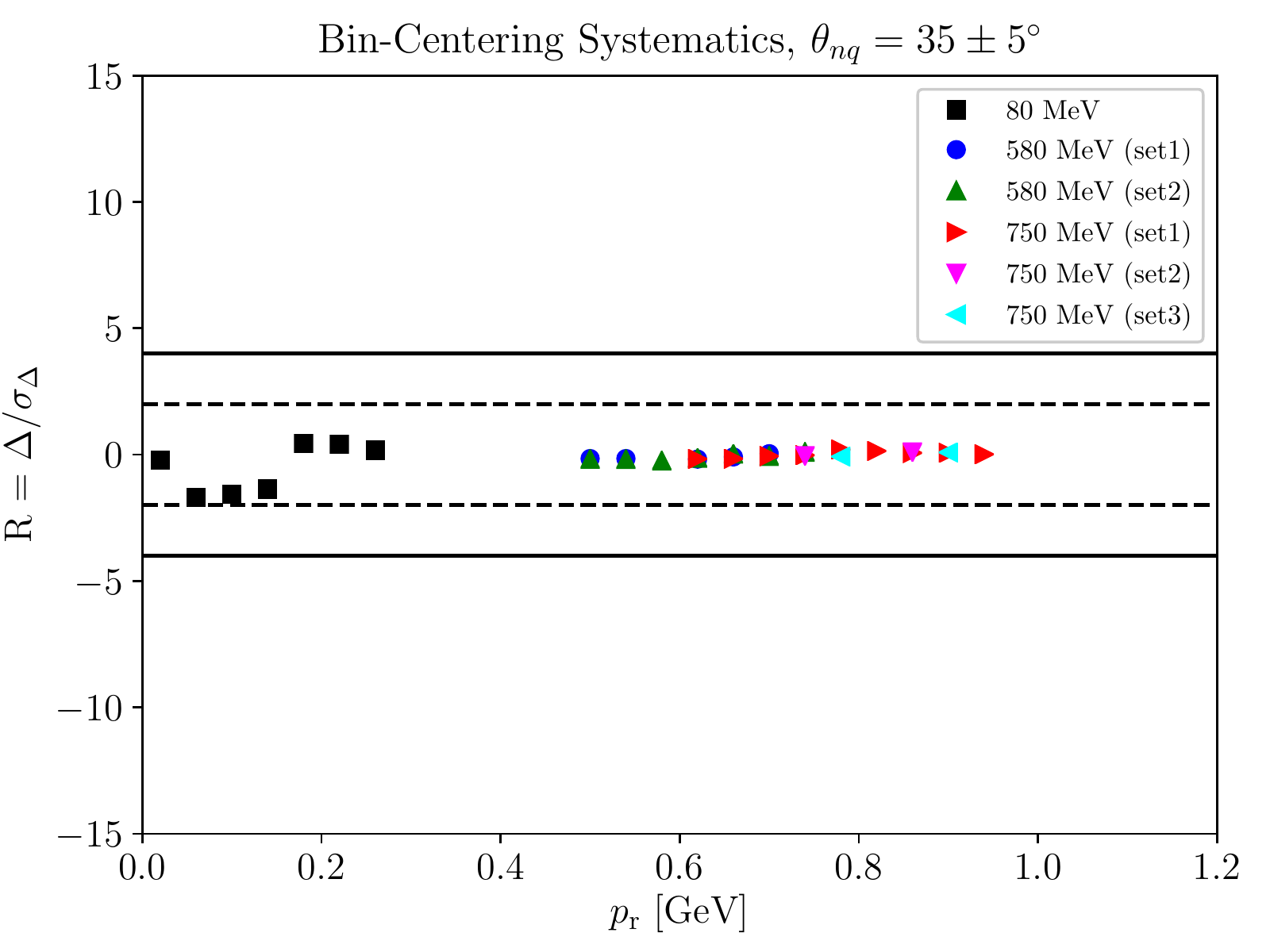}
  \caption{Systematic effects of the model dependency of bin-centering corrections. The lines are described in Fig. \ref{fig:Em_cut_study}.}
  \label{fig:bc_corr_study}
\end{figure}
\indent An additional study was carried out to determine the magnitude of the systematic effects due to the model dependency of the radiative
and bin-centering corrections applied to the experimental data. In this case, the Barlow ratio was calculated from the difference between the experimental cross
sections using the Laget PWIA and FSI models for both radiative and bin-centering corrections. In other words, the experimental cross sections were radiation and
bin-center corrected using both the Laget PWIA and FSI models and the difference between these models was compared. Figures \ref{fig:rad_corr_study} and \ref{fig:bc_corr_study}
are shown for a single bin in $\theta_{nq}$ and demonstrate that the effects of model dependency on the correction factors are negligible on the measured cross sections.
\section{Normalization Systematics}
Starting from the fully corrected experimental cross section,
\begin{equation}
  \sigma^{\mathrm{exp}} = \sigma^{\mathrm{exp}}_{\mathrm{uncorr}} \cdot f_{1} \cdot f_{2} . . . \cdot f_{n}, \label{eq:5.33}
\end{equation}
where $\sigma^{\mathrm{exp}}_{\mathrm{uncorr}}$ is the uncorrected data yield ($Y_{\mathrm{uncorr}}$) divided by the SIMC phase space, and $f_{n}'s$ are the correction factors,
which are re-defined for covenience as
\begin{align}
  f_{1} &= \frac{1}{\epsilon_{\mathrm{tgt.Boil}}}   \hspace{1.5cm} \text{target density factor} \label{eq:5.34} \\
  f_{2} &= \frac{1}{\epsilon_{\mathrm{pTr}}}       \hspace{2cm} \text{proton transmission factor} \label{eq:5.35}  \\
  f_{3} &= \frac{1}{\epsilon_{\mathrm{etrk}}}      \hspace{2cm} \text{electron tracking efficiency} \label{eq:5.36} \\
  f_{4} &= \frac{1}{\epsilon_{\mathrm{htrk}}}      \hspace{2cm} \text{hadron (proton) tracking efficiency} \label{eq:5.37} \\
  f_{5} &= \frac{1}{\epsilon_{\mathrm{tLT}}}       \hspace{2.1cm} \text{total live time} \label{eq:5.38} \\
  f_{6} &= \frac{1}{Q^{\mathrm{exp}}_{\mathrm{tot}}} \hspace{2cm} \text{total accumulated charge} \label{eq:5.39} 
\end{align}
\begin{align}
  f_{7} &= f_{\mathrm{rad}}                        \hspace{2.2cm} \text{radiative correction factor} \label{eq:5.40} \\
  f_{8} &= f_{\mathrm{bc}}                         \hspace{2.3cm} \text{bin-centering correction factor} \label{eq:5.41} 
\end{align}
The systematic uncertainty on the cross section due to the uncertainty in each of these correction factors are added in quadrature as 
\begin{align}
  (d\sigma^{\mathrm{exp}})^{2} = \sum_{i=1}^{8} \Big( \frac{\partial\sigma^{\mathrm{exp}}}{\partial f_{i}}\Big)^{2} df_{i}^{2}. \label{eq:5.42}
\end{align}
From Eq. \ref{eq:5.33}, the derivative with respect to factor $f_{i}$ is
\begin{equation}
  \frac{\partial\sigma^{\mathrm{exp}}}{\partial f_{i}} = \frac{\sigma^{\mathrm{exp}}}{f_{i}}. \label{eq:5.43}
\end{equation}
Substituting Eq. \ref{eq:5.43} into Eq. \ref{eq:5.42}, one obtains
\begin{align}
  (d\sigma^{\mathrm{exp}})^{2} = (\sigma^{\mathrm{exp}})^{2}\sum_{i=1}^{8} \Big( \frac{df_{i}}{f_{i}}\Big)^{2} \label{eq:5.44}
\end{align}
From Eq. \ref{eq:5.44}, one can see that the overall relative error on the cross section due to the normalization uncertainties is simply the
quadrature sum of the relative errors of each normalization factor. For the radiative and bin-centering correction factors, the relative error
is zero as determined from the systematics studies in the previous section.\\
\indent The normalization uncertainties on the cross sections were determined per data set by averaging each normalization factor
and its associated uncertainty over all runs of said set. The average was taken as there were no significant fluctuations in the factors over all runs
for each data set. The results from each data set were added in quadrature for overlapping missing momentum bins once the reduced cross sections were combined
for all sets. For the relative normalization uncertainties that did not vary over the entire experiment, that is, the uncertainties associated with the proton
transmission factor, the total live time and charge, these were added in quadrature as an overall factor to the final error (see Tables \ref{tab:table5.3} and \ref{tab:table5.4}).
\begin{table}[H] 
  \centering
  \scalebox{1.0}{
    \begin{tabular}{c c c c c c c}
      \hline
      \shortstack{$p_{\mathrm{r}}$ [MeV/c]} & $\epsilon_{\mathrm{htrk}}$ & $\epsilon_{\mathrm{etrk}}$ & $\epsilon_{\mathrm{tgt.Boil}}$ & $\epsilon_{\mathrm{pTr}}$ & $\epsilon_{\mathrm{tLT}}$ & $Q^{\mathrm{exp}}_{\mathrm{tot}}$[mC] \\
      \hline
      \hline
      80         & 0.989 & 0.965 & 0.958 & 0.953 & 0.908 & 142.140 \\
      580(set1)  & 0.990 & 0.965 & 0.960 & 0.953 & 0.929 & 1686.83 \\
      580(set2)  & 0.987 & 0.964 & 0.959 & 0.953 & 0.929 & 1931.77 \\
      750(set1)  & 0.988 & 0.964 & 0.957 & 0.953 & 0.924 & 5329.49 \\
      750(set2)  & 0.989 & 0.962 & 0.956 & 0.953 & 0.923 & 1894.01 \\
      750(set3)  & 0.989 & 0.962 & 0.956 & 0.953 & 0.924 & 1083.70 \\
      \hline
    \end{tabular}
  }
  \caption{Summary of the averaged normalization correction factors (or efficiencies) in fractional form and the total accumulated charge per data set.}
  \label{tab:table5.3}
\end{table}
\begin{table}[H] 
  \centering
  \scalebox{0.87}{
    \begin{tabular}{c c c c c c c}
      \hline
      $p_{\mathrm{r}}$ [MeV/c] & $\delta\epsilon_{\mathrm{htrk}}/\epsilon_{\mathrm{htrk}}$ & $\delta\epsilon_{\mathrm{etrk}}/\epsilon_{\mathrm{etrk}}$ & $\delta\epsilon_{\mathrm{tgt.Boil}}/\epsilon_{\mathrm{tgt.Boil}}$ & $\delta\epsilon_{\mathrm{pTr}}/\epsilon_{\mathrm{pTr}}$ & $\delta\epsilon_{\mathrm{tLT}}/\epsilon_{\mathrm{tLT}}$ & $\delta dQ^{\mathrm{exp}}_{\mathrm{tot}}/Q^{\mathrm{exp}}_{\mathrm{tot}}$ \\
      \hline
      \hline
      80 & 0.0344 & 0.0413 & 0.3948 & 0.4951 & 3.0 & 2.0 \\
      580(set1) & 0.3999 & 0.7586 & 0.3766 & 0.4951 & 3.0 & 2.0 \\
      580(set2) & 0.4786 & 0.6041 & 0.3842 & 0.4951 & 3.0 & 2.0 \\
      750(set1) & 0.5329 & 0.7155 & 0.4013 & 0.4951 & 3.0 & 2.0 \\
      750(set2) & 0.4719 & 0.7089 & 0.4196 & 0.4951 & 3.0 & 2.0 \\
      750(set3) & 0.5127 & 0.7584 & 0.4150 & 0.4951 & 3.0 & 2.0 \\
      \hline
    \end{tabular}
  }
  \caption{Summary of relative systematic error on the measured cross sections due to the normalization factors (units are in percent).}
  \label{tab:table5.4}
\end{table}
\section{Kinematical Systematics} \label{sec:kin_systematics}
\begin{figure}[H]
  \centering
  \includegraphics[scale=0.70]{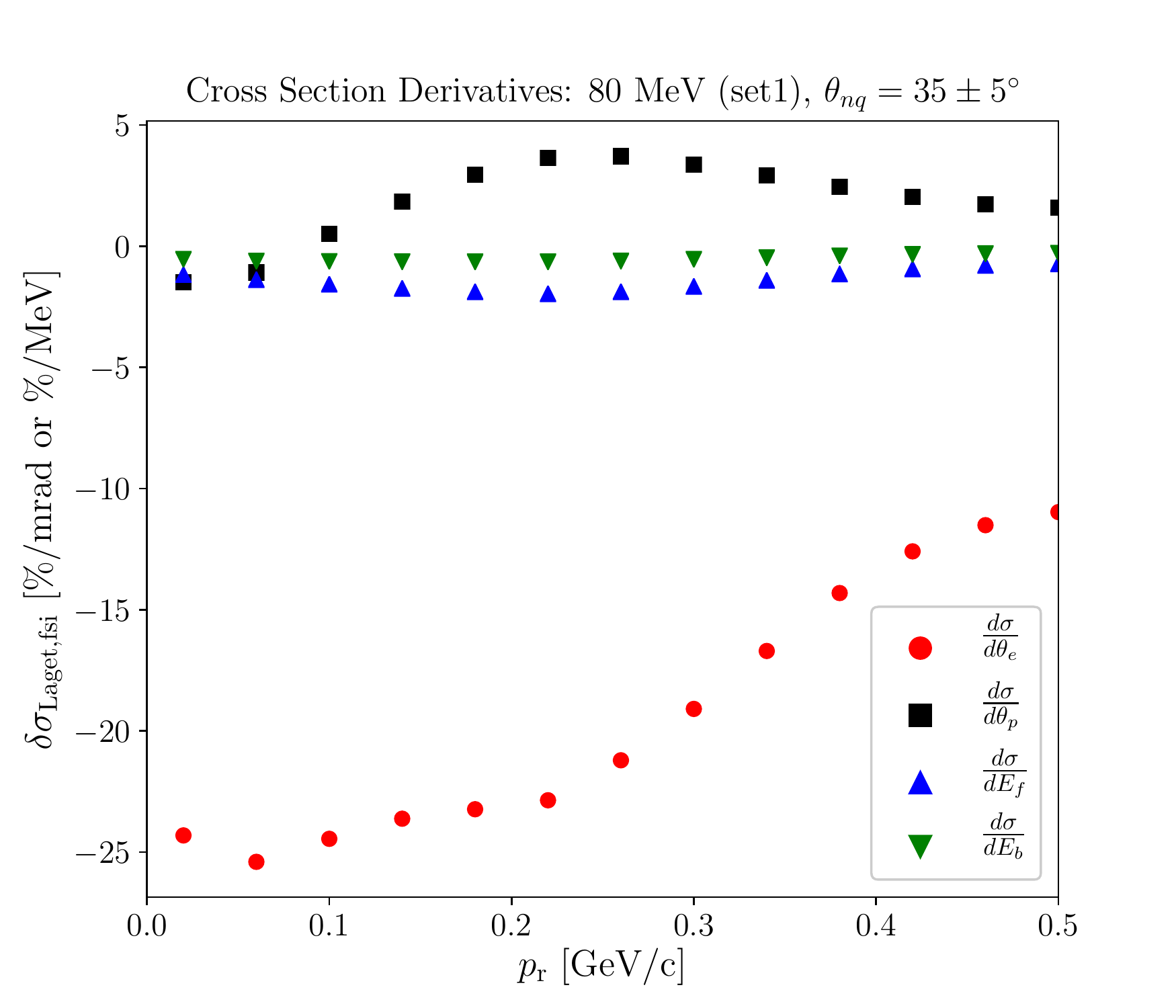}
  \caption{Laget FSI model cross section derivatives at $\theta_{nq}=35\pm5^{\circ}$ for the 80 MeV/c setting.}
  \label{fig:kin_derv_pm80}
\end{figure}

\begin{figure}[H]
  \centering
  \includegraphics[scale=0.53]{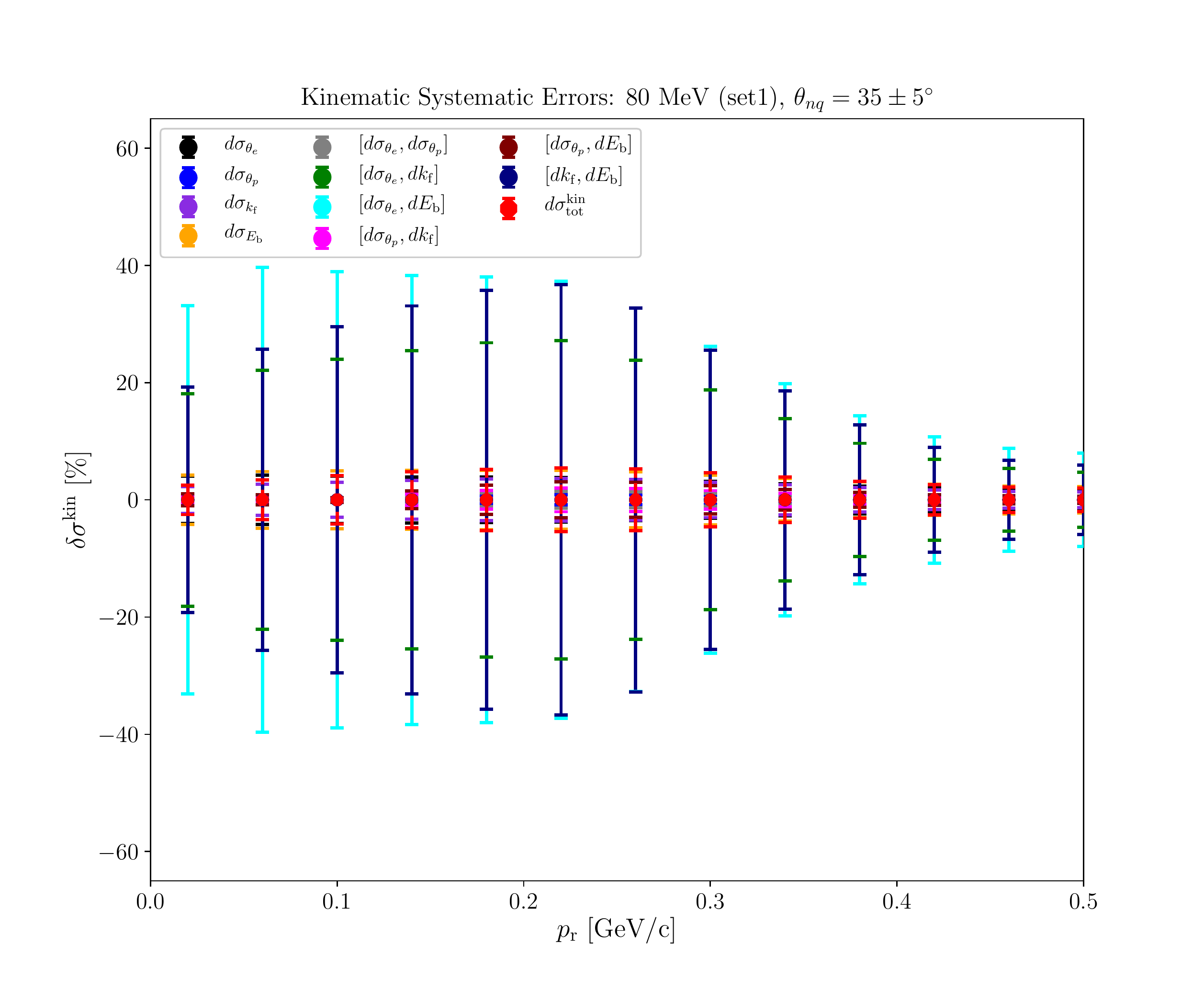}
  \caption{Kinematic systematics relative error contributions to the experimental cross sections at $\theta_{nq}=35\pm5^{\circ}$ for the 80 MeV/c setting. The overlall relative error on the cross section is shown in red.}
  \label{fig:kin_syst_Xsec_pm80}
\end{figure}

The determination of the experimental cross sections depends on the spectrometer kinematics. In particular for this coincidence experiment, the determination of
the beam energy ($E_{\mathrm{b}}$), final electron angle and momentum ($\theta_{e}, k_{\mathrm{f}}$) and either the proton angle ($\theta_{p}$) or momentum ($p_{\mathrm{f}}$) completely determines
the deuteron reaction kinematics. Therefore, how well can we measure these quantities (kinematic uncertainties) determines how well can we measure the experimental cross sections.\\
\indent Ideally, the kinematic uncertainties can be determined by taking a series of dedicated $^{1}$H$(e,e')p$ elastic singles for each spectrometer arm at a wide range of kinematics as well
as beam energies. The data can then be simultaneously fit to determine the kinematical offsets as well as the kinematic uncertainties. In reality, this is very difficult to do
due to the availability of the beam as well as the required time for these runs to take place in a very busy experimental schedule at Jefferson Lab.\\
\indent In E12-10-003, we had a limited and usable hydrogen elastic data set (runs 3288, 3371 and 3374) taken in coincidence, which we used to simultaneously determine the
kinematical uncertainties using the procedure described in Ref.\cite{cyero_specKinUnc}. This method basically consisted of a general $\chi^{2}$-minimization procedure using a matrix approach
that enables one to extract the variance-covariance matrix that contain the uncorrelated (diagonal) as well as the correlated (off-diagonal) kinematical uncertainties. Of the
several models described, we used the results from MODEL 2 of Ref.\cite{cyero_specKinUnc}, which simultaneously fit the elastic data to determine the kinematical as well as the correlated uncertainties
on ($E_{\mathrm{b}}, k_{\mathrm{f}}, \theta_{e}, \theta_{p}$). Tables \ref{tab:table5.5} and \ref{tab:table5.6} summarize the kinematical and correlated uncertainties on each of the variables.\\
\begin{table}[H] 
  \centering
  \scalebox{1.0}{
    \begin{tabular}{l l}
      \hline
      \multicolumn{2}{c}{Kinematic Uncertainties}\\ 
      \hline\hline
      $\delta\theta_{e}$ & 0.1659 [mr]  \\
      $\delta\theta_{p}$  & 0.2369 [mr]  \\
      $\delta k_{\mathrm{f}}$/$k_{\mathrm{f}}$        & 9.132$\times$10$^{-4}$  \\
      $\delta E_{\mathrm{b}}$/$E_{\mathrm{b}}$        & 7.498$\times$10$^{-4}$  \\
      \hline
    \end{tabular}
  }
  \caption{Kinematic uncertainties corresponding to the diagonal elements of the correlation matrix.}
  \label{tab:table5.5}
\end{table}
\begin{table}[H] 
  \centering
  \scalebox{1.0}{
    \begin{tabular}{l l}
      \hline
      \multicolumn{2}{c}{Correlated Kinematic Uncertainties} \\ 
      \hline\hline
      cov($E_{\mathrm{b}}$, $k_{\mathrm{f}}$) & 6.838$\times$ 10$^{-7}$ \\
      cov($E_{\mathrm{b}}$, $\theta_{e}$) & -1.213$\times$ 10$^{-7}$ [rad]\\
      cov($E_{\mathrm{b}}$, $\theta_{p}$) & -6.267$\times$ 10$^{-8}$ [rad] \\
      cov($k_{\mathrm{f}}$, $\theta_{e}$) & -1.488$\times$ 10$^{-7}$ [rad]\\
      cov($k_{\mathrm{f}}$, $\theta_{p}$) & -7.014$\times$ 10$^{-8}$ [rad]\\
      cov($\theta_{e}$,$\theta_{p}$) & 8.432$\times$ 10$^{-9}$ [rad]$^{2}$ \\
      \hline
    \end{tabular}
  }
  \caption{Kinematic uncertainties corresponding to the off-diagonal elements of the correlation matrix.}
  \label{tab:table5.6}
\end{table}
\indent As a side note, recall from the beam energy measurements made in Section \ref{sec:Eb_meas}, that the relative error in the beam energy was determined to
be $\delta E_{\mathrm{b}}/E_{\mathrm{b}} =  4.64 \times 10^{-4} $ as compared to $\delta E_{\mathrm{b}}/E_{\mathrm{b}} =  7.498 \times 10^{-04}$ determined from Ref.\cite{cyero_specKinUnc}.
These two measurements are very close to each other and on the same order of magnitude at the $\sim 10^{-4}$ level. Given the small effect that variations in the beam energy (by itself, and not correlated)
has on the deuteron cross sections, we decided to use the more conservative and slighly larger value determined from Ref.\cite{cyero_specKinUnc}.\\
\indent To determine the systematic effects of the kinematical uncertainties on the measured cross sections, the derivatives of the cross section with respect to each of the kinematic
variables are needed. For this experiment, a table of the cross section derivatives has already been calculated using the Laget FSI model. These derivatves have been calculated
using the averaged kinematics determined per data set as input. Using the standard error propagation formula, the table of cross section derivatives and the kinematical uncertainties,
the full kinematic uncertainty contribution to the experimental cross section can be expressed as
\begin{align}
  (\delta\sigma^{\mathrm{exp}}_{\mathrm{kin}})^{2} &= \Big(\frac{d\sigma}{d\theta_{e}}\delta\theta_{e}\Big)^{2} + \Big(\frac{d\sigma}{d\theta_{p}}\delta\theta_{p}\Big)^{2}
  + \Big(\frac{d\sigma}{dk_{\mathrm{f}}}\frac{\delta k_{\mathrm{f}}}{k_{\mathrm{f}}}k_{\mathrm{f}}\Big)^{2} + \Big(\frac{d\sigma}{dE_{\mathrm{b}}}\frac{\delta E_{\mathrm{b}}}{E_{\mathrm{b}}}E_{\mathrm{b}}\Big)^{2} \nonumber  \\
  &+ 2\frac{d\sigma}{dE_{\mathrm{b}}}\frac{d\sigma}{dk_{\mathrm{f}}}\text{cov($E_{\mathrm{b}}, k_{\mathrm{f}}$)} + 2\frac{d\sigma}{dE_{\mathrm{b}}}\frac{d\sigma}{d\theta_{e}}\text{cov($E_{\mathrm{b}}, \theta_{e}$)} + 2\frac{d\sigma}{dE_{\mathrm{b}}}\frac{d\sigma}{d\theta_{p}}\text{cov($E_{\mathrm{b}}, \theta_{p}$)}\nonumber \\
  &+ 2\frac{d\sigma}{dk_{\mathrm{f}}}\frac{d\sigma}{d\theta_{e}}\text{cov($k_{\mathrm{f}}, \theta_{e}$)} + 2\frac{d\sigma}{k_{\mathrm{f}}}\frac{d\sigma}{d\theta_{p}}\text{cov($k_{\mathrm{f}}, \theta_{p}$)} + 2\frac{d\sigma}{d\theta_{e}}\frac{d\sigma}{d\theta_{p}}\text{cov($\theta_{e}, \theta_{p}$)}. \label{eq:5.45}
\end{align}
\indent By including the covariance errors from Eq. \ref{eq:5.45}, the overall kinematics uncertainty can actually be reduced in the case where any two
variables are anti-correlated (``-'' covariance sign) as it is the case from most of the variables in Table \ref{tab:table5.6}. \\
\indent As an example, Fig. \ref{fig:kin_derv_pm80} shows the model cross section
derivatives wirh respect to each kinematic variable and Fig. \ref{fig:kin_syst_Xsec_pm80} shows the systematic contribution to the measured cross section due to the kinematics
as well as the correlated uncertaintites for $\theta_{nq}=35\pm5^{\circ}$.
\indent In Fig. \ref{fig:kin_derv_pm80}, the electron scattering angle uncertainty ($\delta\theta_{e}$) clearly has the largest effect on the cross sections with a $\sim20-25\%$/mrad
effect for up to $p_{\mathrm{r}}\sim 300$ MeV/c, whereas the other kinematics have a $\lesssim 5\%$ effect per mrad (or MeV). This large effect on the electron angle is attributed to the Mott cross
section, $\sigma_{\mathrm{Mott}}$, which has a dependence of $\sigma_{\mathrm{Mott}}\propto\frac{1}{\sin^{4}(\theta_{e})}$ on the electron angle. Therefore, it is crucial that we determine the SHMS (electron arm)
angle to much better than 1 mrad of uncertainty. The variations on the higher momentum settings are still dominated by the electron angle but to a much lower extent of $\sim 10\%$/mrad
(not shown). \\
\indent Figure \ref{fig:kin_syst_Xsec_pm80} shows the contributions from each of the terms on Eq. \ref{eq:5.45} on the total relative kinematic systematic error. It is clear
that the main contributors to the systematic error are the correlated errors, cov($E_{\mathrm{b}},\theta_{e}$) in cyan, cov($E_{\mathrm{b}}, k_{\mathrm{f}}$) in navy blue, and cov($k_{\mathrm{f}}, \theta_{e}$) in green.
Two out of these three major contributors, however, are anti-correlated so when the errors are added in quadrature, the final result is actually smaller, as can be seen by the total
kinematic systematic error (color red). \\
\indent The kinematic systematic errors, such as that shown on Fig. \ref{fig:kin_syst_Xsec_pm80}, were determined poin-to-point in $(p_{r}, \theta_{nq})$ bins for each missing momentum setting, and
added in quadrature for overlapping $p_{\mathrm{r}}$ bins. The overall systematic uncertainty in the final cross section was determined by the quadrature sum of the normalization and kinematic systematic uncertainties.
This result was then added in quadrature to the statistical uncertainty (20-30$\%$ on average) to obtain the final uncertainty in the cross section. See Appendices \hyperref[appendix:appA1]{A1} and \hyperref[appendix:appA2]{A2}
for a summary of the relative statistical and systematic uncertainties for each $(p_{r},\theta_{nq})$ bin at $Q^{2}=3.5\pm0.5$ and $Q^{2}=4.5\pm0.5$ (GeV/c)$^{2}$, respectively.

\chapter{RESULTS AND DISCUSSION}  \label{chap:chapter6}
In this chapter I will discuss how the experimental reduced cross
sections were extracted. Finally, I will present the final experimental
reduced cross sections compared to various theoretical models as well as their
cross section ratios and discuss the implications of these results. 
\section{Extraction of the $^{2}$H$(e,e'p)n$ Reduced Cross Sections}\label{sec:6.1}
From the theoretical cross sections introduced in Eq. \ref{eq:2.11} under the PWIA assumption, we define the experimental (or theoretical) reduced
cross section evaluated at a kinematic bin k=($p_{\mathrm{r}}, \theta_{nq}$) as
\begin{equation}
  (\sigma^{\mathrm{exp, th}}_{\mathrm{red}})_{\mathrm{k}} \equiv \frac{(\sigma^{\mathrm{exp,th}})_{\mathrm{k}}}{(E_{\mathrm{f}}p_{\mathrm{f}}f_{\mathrm{rec}}\sigma_{\mathrm{cc1}})_{\mathrm{k}}}, \label{eq:6.1}
\end{equation}
where $\sigma^{\mathrm{exp,th}}$ is the fully corrected data (or theoretical) cross section, ($E_{\mathrm{f}}$, $p_{\mathrm{f}}$) are the final proton energy and momentum, respectivelty, $f_{\mathrm{rec}}$
is a recoil factor introduced in Eq. \ref{eq:2.12}, and $\sigma_{\mathrm{cc1}}=\sigma_{\mathrm{cc1}}(G_{\mathrm{E}_{p}}, G_{\mathrm{M}_{p}})$ is one of two variations of the deForest\cite{DEFOREST1983}
off-shell cross section that describes the scattering between an electron and a loosely bound (``free'') proton. \\
\indent By dividing by the kinematical and recoil factors as well as the deForest cross section, which depends on the proton elastic form factors, most of the kinematical
dependencies found on the cross sections are cancelled leaving only a dependence on the neutron recoil momentum, $p_{\mathrm{r}}$. Therefore,
the reduced cross sections are closely related to the genuine momentum distributions of the nucleons provided the kinematics have
been chosen such that the PWIA is dominant. In analogy, this is also observed in $^{1}$H$(e,e')p$, where the elastic cross sections are divided by $\sigma_{\mathrm{Mott}}$ to extract the proton elastic form factors
which describe the internal structure of the proton.\\
\indent It is important to note that each of the kinematic variables in the denominator of Eq. \ref{eq:6.1}, similar to the fully corrected cross sections ($\sigma^{\mathrm{exp,th}}$) determined in Section \ref{sec:bin_cent_corr},
were also determined at the averaged kinematic setting for each kinematic bin in ($p_{\mathrm{r}}$, $\theta_{nq}$). Finally, the reduced cross sections calculated for each missing momentum data set were combined for overlapping bins in ($p_{\mathrm{r}},\theta_{nq}$)
to gain better statistical precision. 
\section{$^{2}$H$(e,e'p)n$ Momentum Distributions}
\begin{figure}[H]
  \centering
  \includegraphics[scale=0.7]{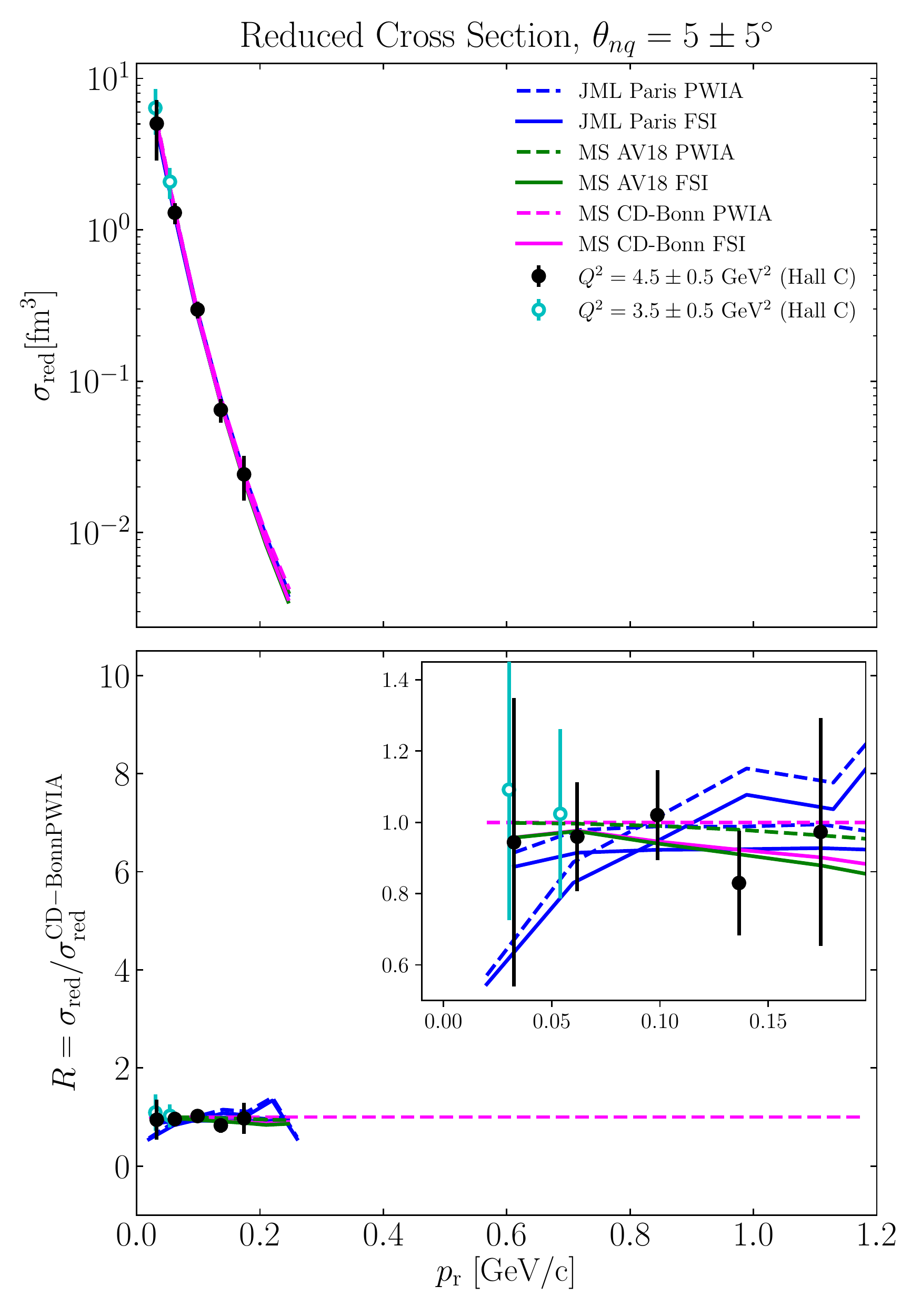}
  \caption{$^{2}$H$(e,e'p)n$ reduced cross sections at $\theta_{nq}=5\pm5^{\circ}$. Top panel: The blue lines represent the theoretical calculations by J.M. Laget\cite{LAGET2005} using the
    Paris potential\cite{Paris_NN_Lacombe1980} denoted by JML and the green/magenta lines are calculations from M. Sargsian\cite{PhysRevC.82.014612} using either the AV18 (green)\cite{AV18_1995} or CD-Bonn (magenta)\cite{CDBonn_NN_Machleidt2001}
    potentials denoted by MS. The dashed lines are calculations within the PWIA and the solid lines are calculations including FSI. Bottom panel: The dashed reference (magenta)
    line refers to MS CD-Bonn PWIA calculation (or momentum distribution) by which the data and all models are divided. Inset (bottom panel): Close-up plot of the reduced cross section ratio shown in the bottom panel. }
  \label{fig:redXsec_5deg}
\end{figure}

\begin{figure}[H]
  \centering
  \includegraphics[scale=0.8]{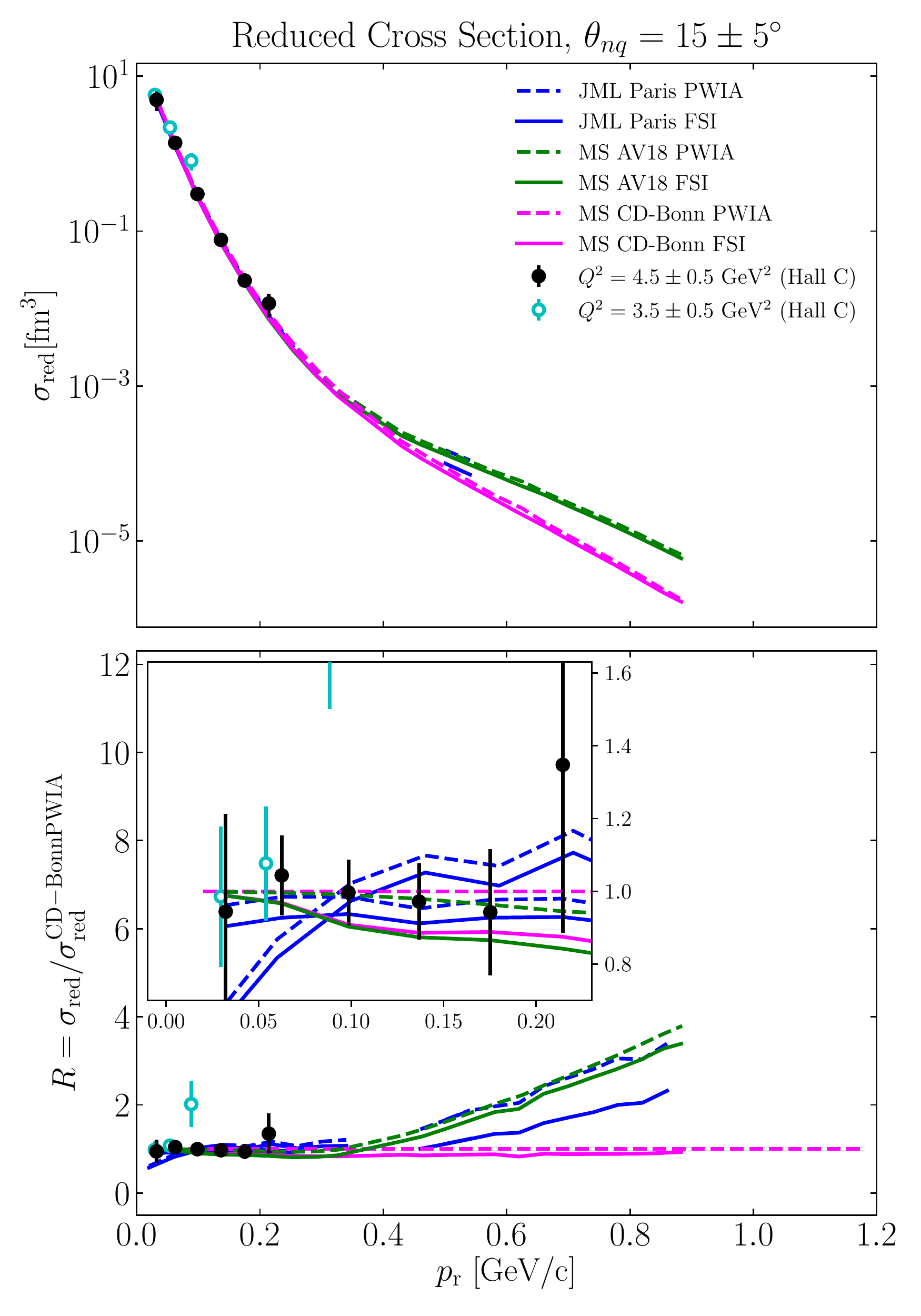}
  \caption{$^{2}$H$(e,e'p)n$ reduced cross sections at $\theta_{nq}=15\pm5^{\circ}$. The lines are described in Fig. \ref{fig:redXsec_5deg}.}
  \label{fig:redXsec_15deg}
\end{figure}

\begin{figure}[H]
  \centering
  \includegraphics[scale=0.8]{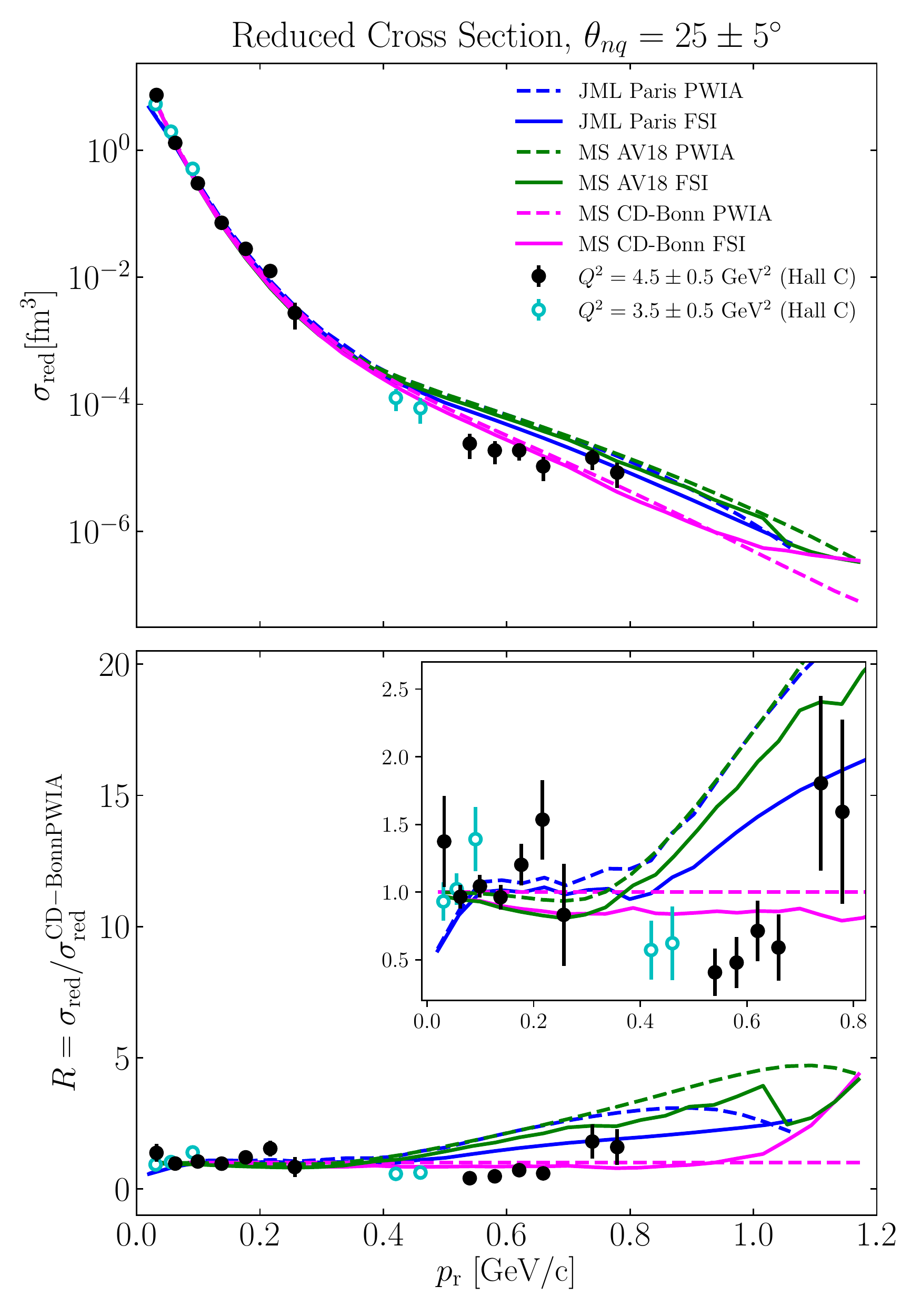}
  \caption{$^{2}$H$(e,e'p)n$ reduced cross sections at $\theta_{nq}=25\pm5^{\circ}$. The lines are described in Fig. \ref{fig:redXsec_5deg}.}
  \label{fig:redXsec_25deg}
\end{figure}

\begin{figure}[H]
  \centering
  \includegraphics[scale=0.8]{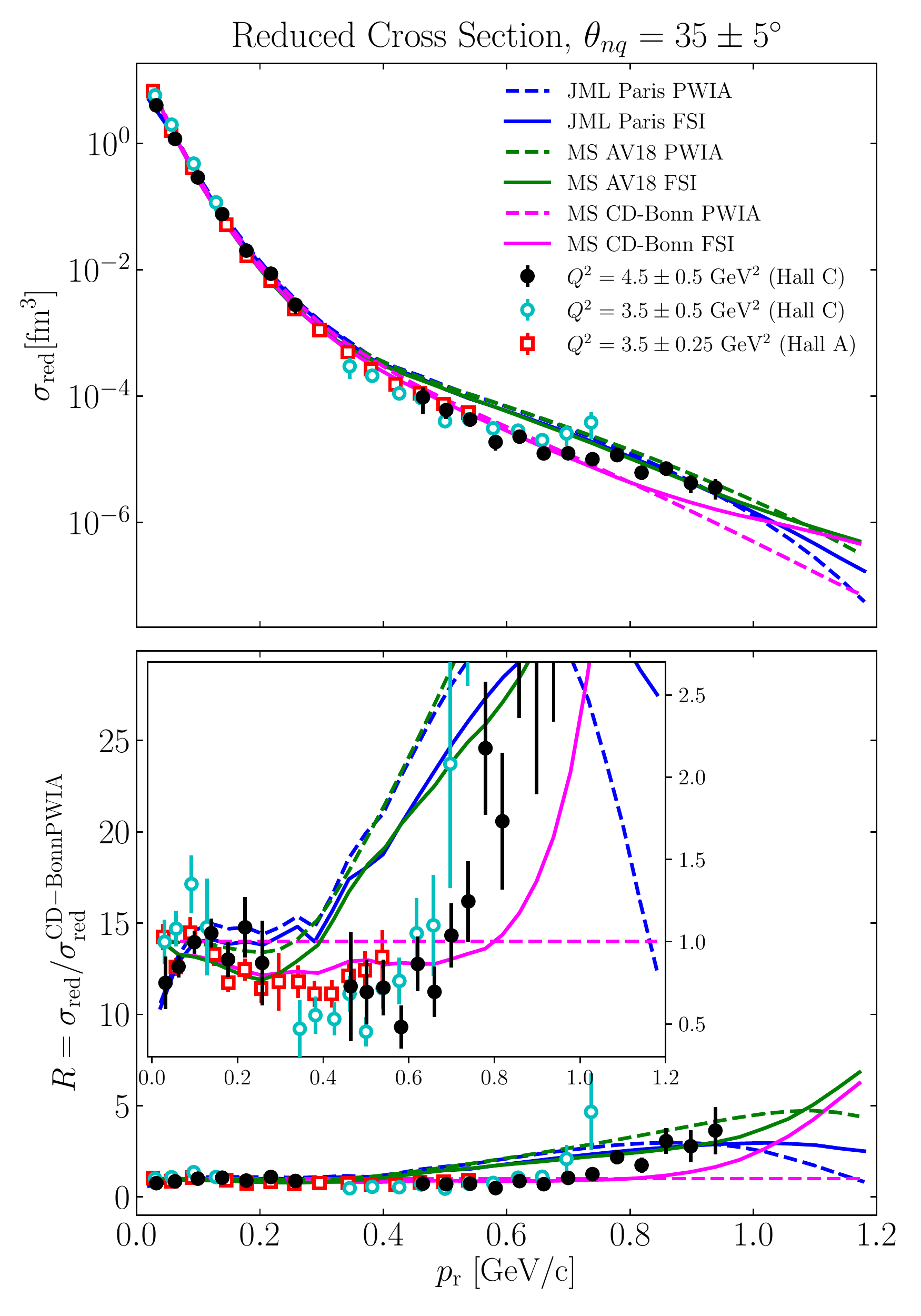}
  \caption{$^{2}$H$(e,e'p)n$ reduced cross sections at $\theta_{nq}=35\pm5^{\circ}$. The lines are described in Fig. \ref{fig:redXsec_5deg}.}
  \label{fig:redXsec_35deg}
\end{figure}

\begin{figure}[H]
  \centering
  \includegraphics[scale=0.8]{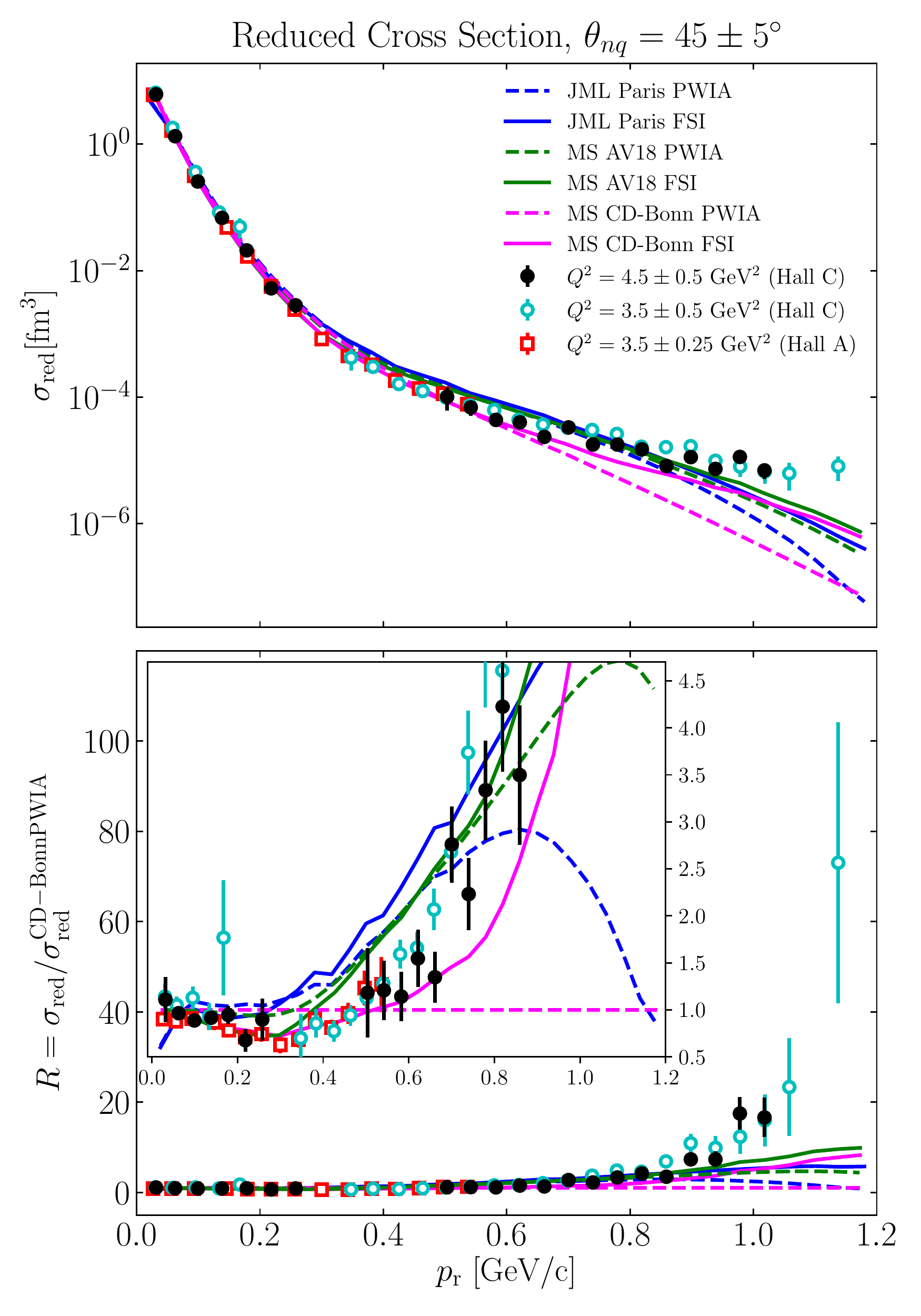}
  \caption{$^{2}$H$(e,e'p)n$ reduced cross sections at $\theta_{nq}=45\pm5^{\circ}$. The lines are described in Fig. \ref{fig:redXsec_5deg}.}
  \label{fig:redXsec_45deg}
\end{figure}

\begin{figure}[H]
  \centering
  \includegraphics[scale=0.8]{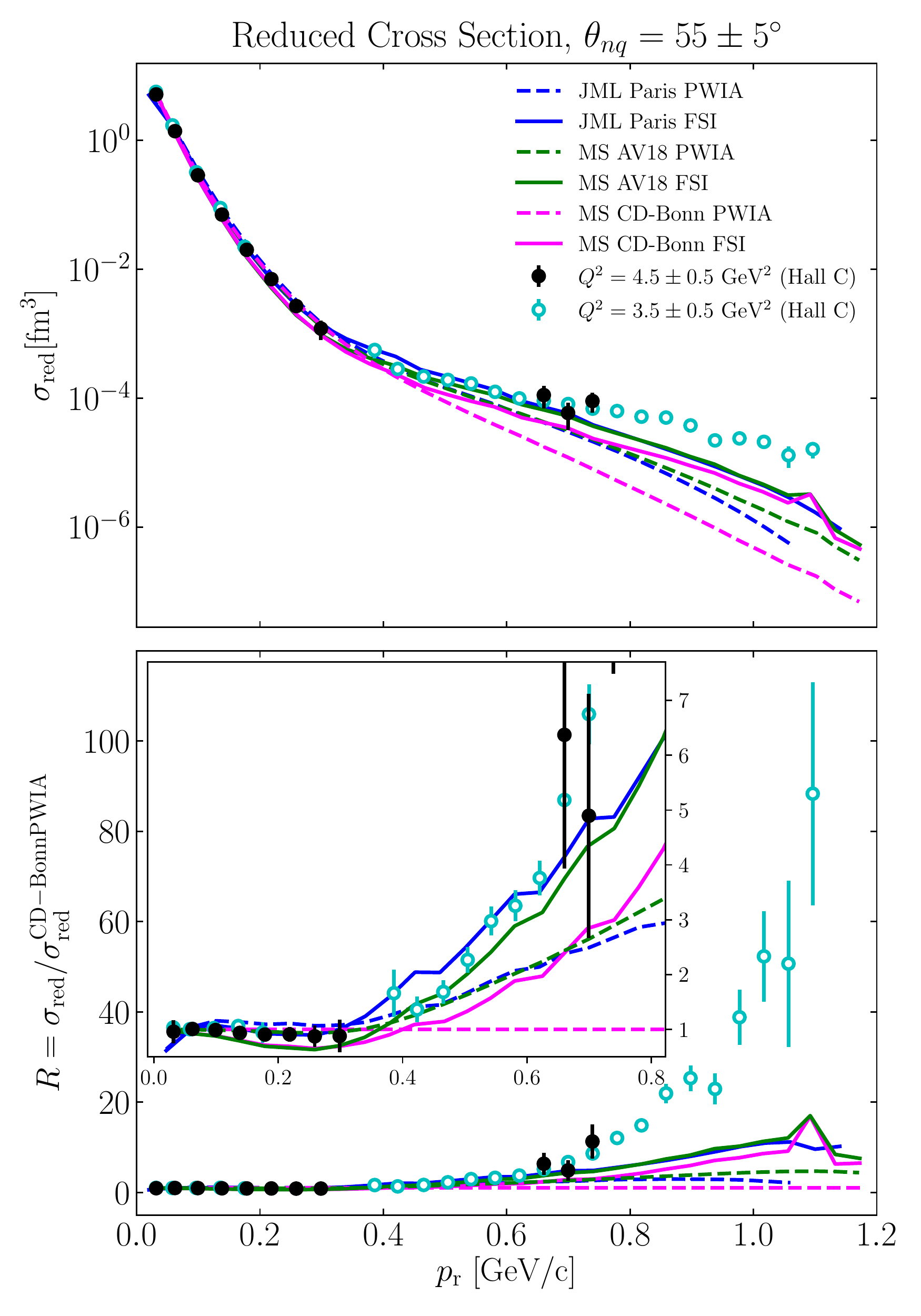}
  \caption{$^{2}$H$(e,e'p)n$ reduced cross sections at $\theta_{nq}=55\pm5^{\circ}$. The lines are described in Fig. \ref{fig:redXsec_5deg}.}
  \label{fig:redXsec_55deg}
\end{figure}

\begin{figure}[H]
  \centering
  \includegraphics[scale=0.8]{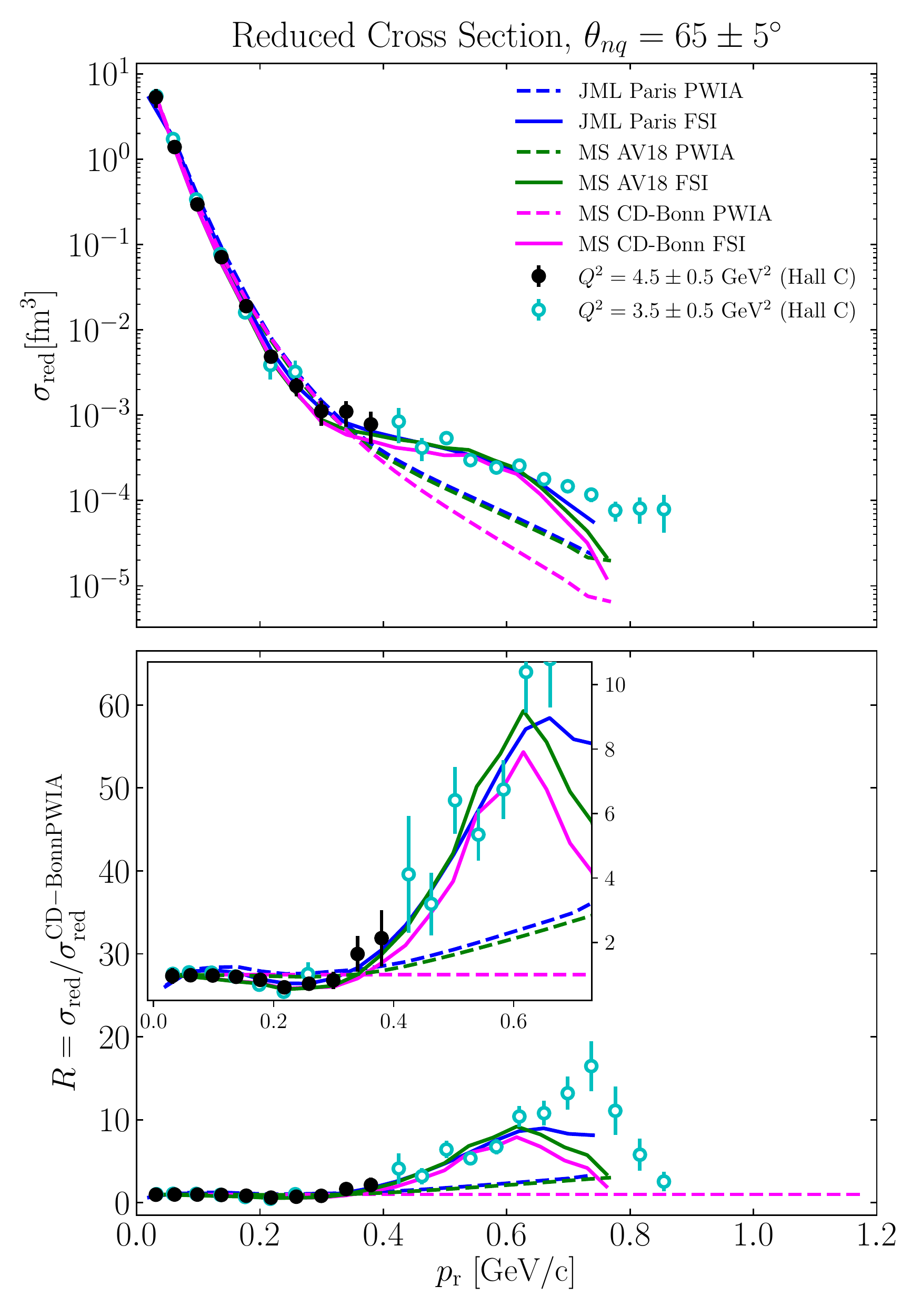}
  \caption{$^{2}$H$(e,e'p)n$ reduced cross sections at $\theta_{nq}=65\pm5^{\circ}$. The lines are described in Fig. \ref{fig:redXsec_5deg}.}
  \label{fig:redXsec_65deg}
\end{figure}

\begin{figure}[H]
  \centering
  \includegraphics[scale=0.8]{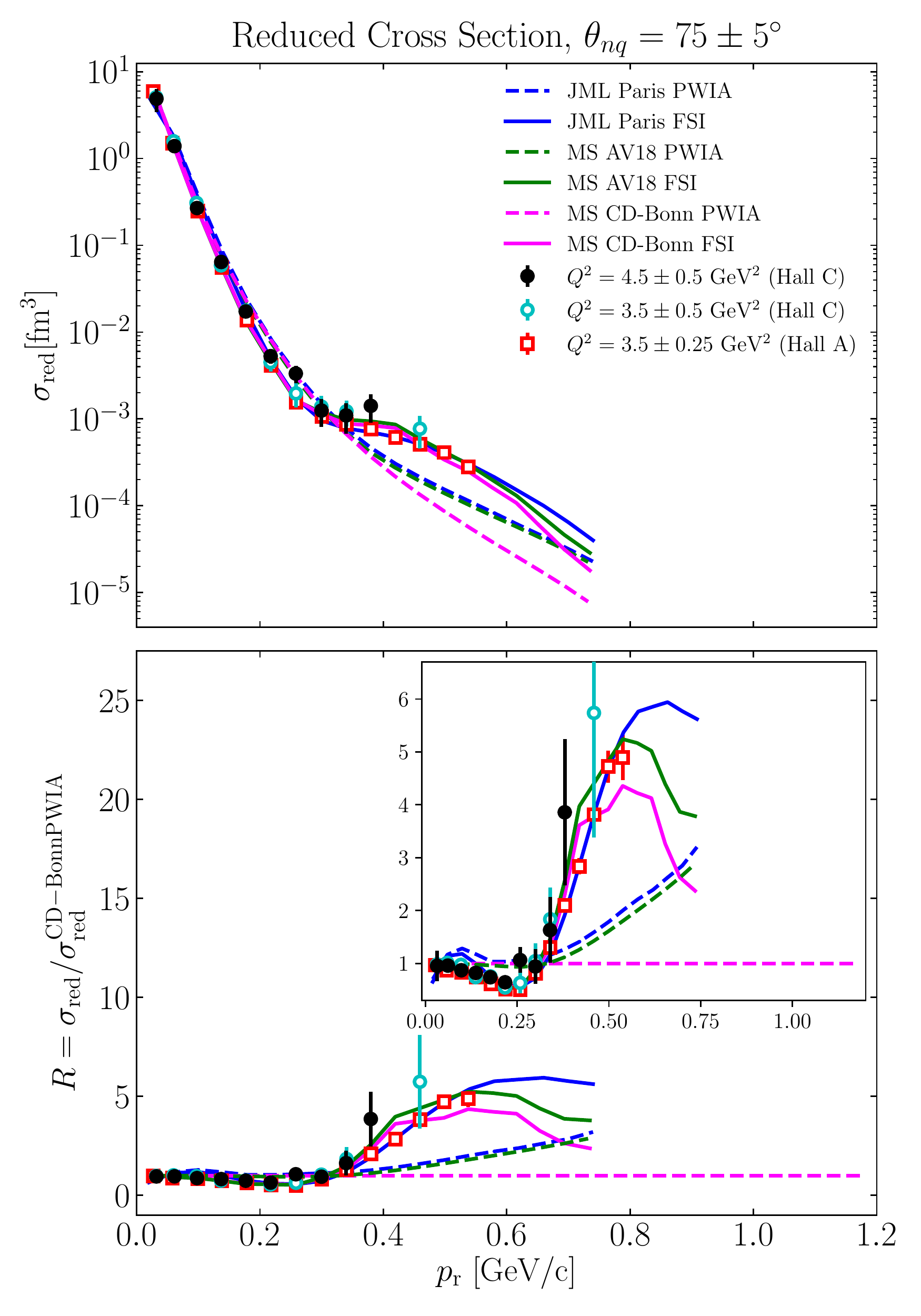}
  \caption{$^{2}$H$(e,e'p)n$ reduced cross sections at $\theta_{nq}=75\pm5^{\circ}$. The lines are described in Fig. \ref{fig:redXsec_5deg}.}
  \label{fig:redXsec_75deg}
\end{figure}

\begin{figure}[H]
  \centering
  \includegraphics[scale=0.8]{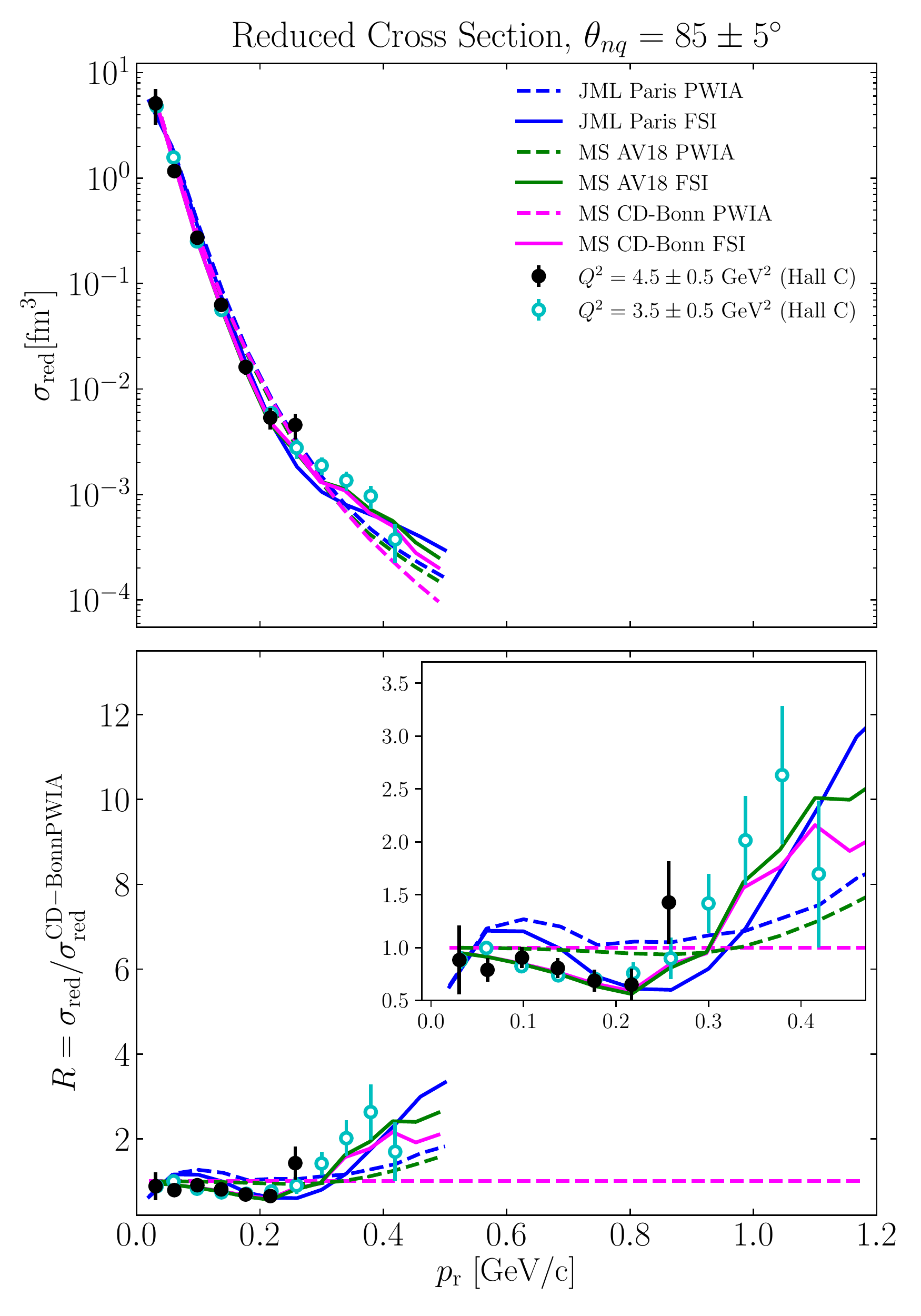}
  \caption{$^{2}$H$(e,e'p)n$ reduced cross sections at $\theta_{nq}=85\pm5^{\circ}$. The lines are described in Fig. \ref{fig:redXsec_5deg}.}
  \label{fig:redXsec_85deg}
\end{figure}

\begin{figure}[H]
  \centering
  \includegraphics[scale=0.8]{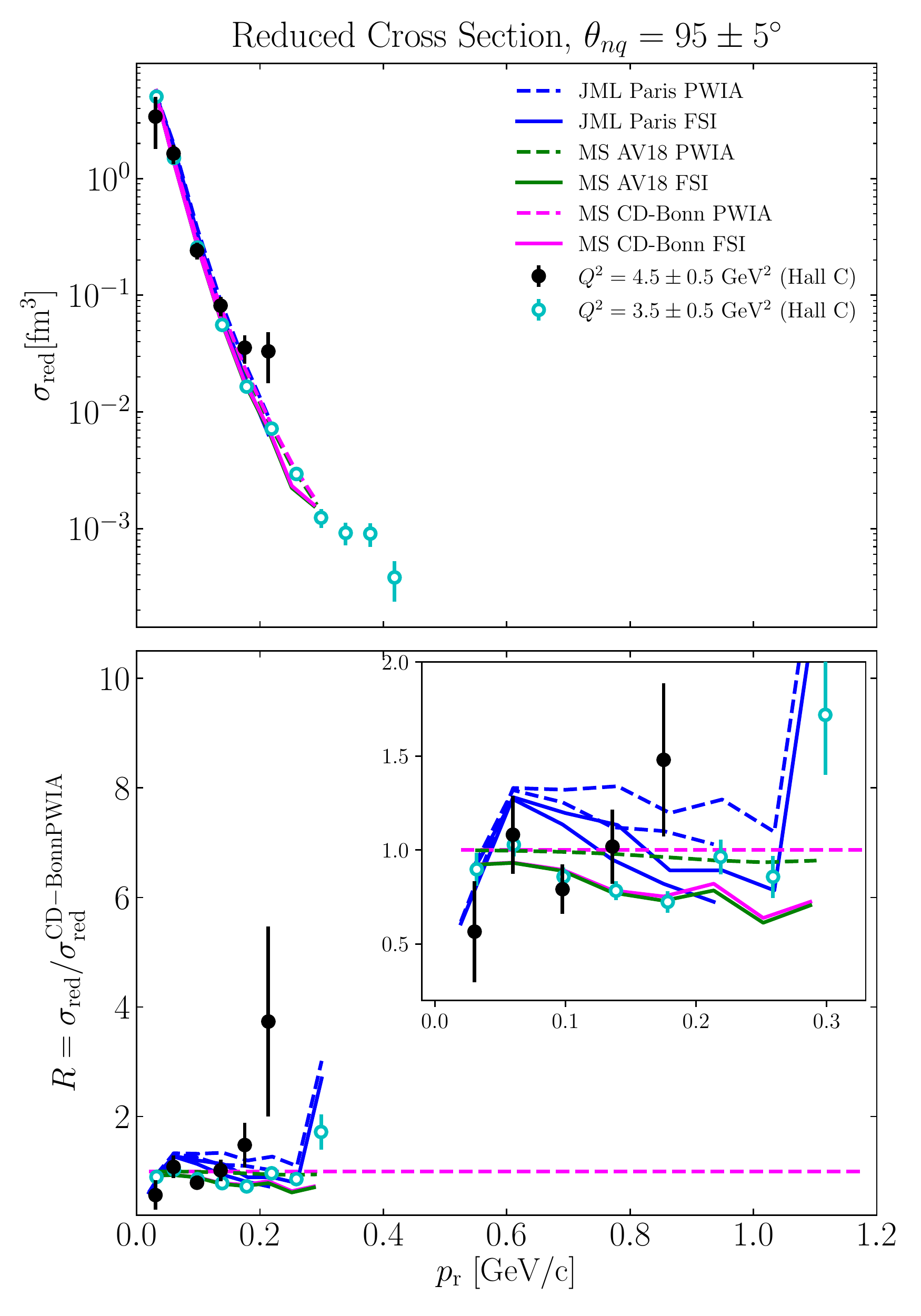}
  \caption{$^{2}$H$(e,e'p)n$ reduced cross sections at $\theta_{nq}=95\pm5^{\circ}$. The lines are described in Fig. \ref{fig:redXsec_5deg}.}
  \label{fig:redXsec_95deg}
\end{figure}

The experimental and theoretical reduced cross sections have been combined for overlapping bins in $p_{\mathrm{r}}$ for each data set and are shown in Figs. \ref{fig:redXsec_5deg}-\ref{fig:redXsec_95deg}.
The ratio of the experimental and theoretical reduced cross sections to the CD-Bonn PWIA model has also been evaluated where the inset plot shows a more detailed (close-up)
representation of the ratios taken. In addition to the results presented at $Q^{2} = 4.5\pm0.5$ (GeV/c)$^{2}$, reduced cross sections have also been determined at $Q^{2} = 3.5\pm0.5$ (GeV/c)$^{2}$ for
comparison. For $\theta_{nq} = 35^{\circ}, 45^{\circ}$ and $75^{\circ}$ we have plotted the reduced cross sections from the previous deuteron break-up experiment
performed in Hall A\cite{PhysRevLett.107.262501} at $Q^{2} = 3.5 \pm 0.25$ (GeV/c)$^{2}$. \\
\indent The error on the cross sections have been determined by adding the statistical and systematic errors in quadrature. See Appendices \hyperref[appendix:appA1]{A1} and \hyperref[appendix:appA2]{A2} 
for the tabulated reduced cross sections and the associated statistical and systematic errors for each $(p_{\mathrm{r}}, \theta_{nq})$ bin at $Q^{2}=3.5\pm0.5$ and $Q^{2}=4.5\pm0.5$ GeV$^{2}$, respectively. \\
\section{Discussion of Results}
Even though the focus of this experiment was to extract the $^{2}$H$(e,e'p)n$ reduced cross sections at $35^{\circ}<\theta_{nq}<45^{\circ}$ where the FSI are significantly reduced,
it has been possible to determine the reduced cross sections at a different range of neutron recoil angles, $5^{\circ}<\theta_{nq}<25^{\circ}$ and $55^{\circ}<\theta_{nq}<95^{\circ}$, for comparison.
In general, there is an overall
good agreement between the Halls A and C data ($\theta_{nq}=35^{\circ},45^{\circ}$ and $75^{\circ}$) even though they were taken at different $Q^{2}$ kinematics. This can be understood
from the fact that the kinematical dependencies of the cross sections have been divided out in the reduced cross sections, as mentioned in Section \ref{sec:6.1}. The good agreement
between both experiments at lower $p_{\mathrm{r}}$ gives us confidence in the measurements made at the higher missing momentum settings for which no previous data exist. \\
\indent For all recoil angles shown in Figs. \ref{fig:redXsec_5deg}-\ref{fig:redXsec_95deg}, at recoil momenta $p_{r}\leq250$ MeV/c, the reduced cross sections are well reproduced by
all models when FSI are included. The agreement at $p_{r}\leq250$ MeV/c can be understood from the fact that this region corresponds to the long-range part of the $NN$ potential where
the One Pion Exchange Potential (OPEP) is well known and common to all modern potentials. \\
\indent Beyond $p_{r}\sim250$ MeV/c at $\theta_{nq}=35^{\circ}$ and $45^{\circ}$,  the JML Paris and MS AV18
models significantly differ from the MS CD-Bonn calculation. In this region, the JML Paris and MS AV18 cross sections are dominated by the PWIA and within good agreement of each other up to $p_{r}\sim700$ MeV/c. The MS CD-Bonn based cross sections, in contrast, are generally smaller than those calculated with the JML Paris or MS AV18 wave function in this region.
In addition, for $\theta_{nq}=35^{\circ}$, the MS CD-Bonn cross sections are dominated by the PWIA up to $p_{\mathrm{r}}\sim800$ MeV/c while for $\theta_{nq}=45^{\circ}$,
FSI start to contribute already above $p_{\mathrm{r}}\sim600$ MeV/c.\\
\indent At $\theta_{nq} = 75^{\circ}$ and $p_{\mathrm{r}}>180$ MeV/c, FSI become the dominant contribution to the cross sections for all models that exhibit a similar behavior
(smaller fall-off) that eliminates any possibility of extracting the momentum distributions.\\
\indent The difference between the deuteron wave functions with CD-Bonn, Paris and AV18 potentials\cite{Misak_privMay2020} is 
how the $NN$ potential is modeled based on the empirical $NN$ scattering data.
The CD-Bonn model is based on the One-Boson-Exchange approach in which the 
nucleon-meson-meson couplings are constrained to describe the $NN$ scattering phase shifts
extracted from the data. The interaction potential represents the static limit of 
this potential. The Paris and AV18 models are phenomenological in which the 
Yukawa type interaction is introduced and parameters are fit to describe the 
same $NN$ scattering phase-shifts. The major difference between CD-Bonn and Paris/AV18 
potentials is that the former predicts a much softer repulsive interaction at short distance that 
results in a smaller high momentum component in the deuteron wave function in momentum space.
The effect of these local approximations on the $NN$ potential are shown in Fig. 2 of Ref. \cite{CDBonn_NN_Machleidt2001}. \\
\indent To quantify the discrepancy observed between data and theory at higher missing momenta for $\theta_{nq}=35^{\circ}$ and $45^{\circ}$, the ratio of the experimental and theoretical reduced cross sections to the
deuteron momentum distribution calculated using the CD-Bonn potential is shown in the lower subplot of Figs. \ref{fig:redXsec_5deg}-\ref{fig:redXsec_95deg}.
For $\theta_{nq}=35^{\circ}$ and $45^{\circ}$, the data are best described by the MS CD-Bonn FSI calculation for recoil momenta up to $p_{\mathrm{r}}\sim 700$ MeV/c and $\sim 600$ MeV/c, respectively, with
a ratio of $R\sim 0.5-1$ as compared to $R\sim2-4$ at $\theta_{nq}=75^{\circ}$ which indicates a significant reduction in FSI at forward $\theta_{nq}$ angles.
Furthermore, the agreement between the Halls A and C data supports the Hall A approach of selecting a kinematic region where recoil angles are small and FSI are reduced.\\
\newpage
\indent At larger recoil momenta, where the ratio is $R>1$ and increasing, for $\theta_{nq}=35^{\circ}$, FSI start to dominate for missing momenta typically above 800 MeV/c for the MS CD-Bonn
calculation, while the other models predict still relatively small FSI below 900 MeV/c. At $\theta_{nq}=45^{\circ}$, the FSI dominance starts earlier for all models above 800 MeV/c and for the
MS CD-Bonn based calculation, above 600 MeV/c. \\
\indent Overall, it is interesting to note that none of the calculations can reproduce the measured $p_{\mathrm{r}}$ dependence above 600 MeV/c in a region where FSI are still relatively
small ($<30\%$). This behavior of the data is new and additional data in this kinematic region are necessary to improve the statistics.\\
\indent At $\theta_{nq}=75^{\circ}$, FSI are small below $p_{\mathrm{r}}\sim 180$ MeV/c, but do not exactly cancel the PWIA/FSI interference term in the scattering amplitude, which results
in a small dip in this region in agreement with the data. At $p_{\mathrm{r}}>300$ MeV/c, the data were statistically limited as our focus was on the smaller recoil angles. The Hall A
data, however, show a reasonable agreement with the FSI from all models, which gives us confidence in our understanding of FSI at smaller recoil angles.
\section{Conclusion}
This experiment extended the previous Hall A cross section measurements\cite{PhysRevLett.107.262501} on the $^{2}$H$(e,e'p)n$ reaction to very high neutron recoil momenta ($p_{\mathrm{r}}>500$ MeV/c) at
kinematic settings where FSI were predicted to be small and the cross section was dominated by the PWIA and sensitive to the short range part of the deuteron wave function. The experimental
and theoretical reduced cross sections were extracted and found to be in good agreement with the Hall A data. Furthermore, the MS CD-Bonn model was found to be significantly different
than the JML Paris or MS AV18 models and was able to partially describe the data over a larger range in $p_{\mathrm{r}}$. At higher missing momenta, however, all models were unable to
describe the missing momentum dependence of the data. Additional measurements of the $^{2}$H$(e,e'p)n$ would be required at a wider range in central missing momentum, as stated in the
original proposal\cite{e12_10_003_proposal}, to reduce the statistical uncertainties in this very high missing momentum region ($p_{\mathrm{r}}>500$ MeV/c) and to better understand the large
deviations observed between the different models and data. 


\bibliographystyle{unsrt}
\bibliography{ms}

\cleardoublepage
\appendix

\setcounter{table}{0}
\renewcommand{\thetable}{A\arabic{table}}

\renewcommand{\thechapter}{} 
\phantomsection
\addcontentsline{toc}{chapter}{APPENDICES}
\chapter*{\normalfont APPENDICES}

\renewcommand{\thechapter}{\Alph{chapter}} 

\setcounter{secnumdepth}{0}

\section*{Appendix A: Reduced Cross Section Data Table} \phantomsection\label{appendix:appA}
The $^{2}$H$(e,e'p)n$ reduced cross sections are tabulated for fixed $Q^{2}$ and $\theta_{nq}$ bins.
The $p_{\mathrm{r,bin}}$ represents the neutron recoil (missing) momentum central bin with a bin width of $\pm0.02$ GeV/c
and the $p_{\mathrm{r,avg}}$ represents the recoil momentum averaged over each  $p_{\mathrm{r,bin}}$. The uncertainties in the
reduced cross sections ($\sigma_{\mathrm{red}}$) are expressed as relative statistical ($\delta\sigma_{\mathrm{stat}}$), normalization ($\delta\sigma_{\mathrm{norm}}$),
kinematic ($\delta\sigma_{\mathrm{kin}}$) and systematic ($\delta\sigma_{\mathrm{syst}}$) and total ($\delta\sigma_{\mathrm{tot}}$) where
$\delta\sigma^{2}_{\mathrm{sys}} = \delta\sigma^{2}_{\mathrm{norm}} + \delta\sigma^{2}_{\mathrm{kin}}$ and
$\delta\sigma^{2}_{\mathrm{tot}} = \delta\sigma^{2}_{\mathrm{stat}} + \delta\sigma^{2}_{\mathrm{syst}}$.\\
\subsection*{A1. Reduced Cross Sections at $Q^{2}=3.5\pm0.5$ \textnormal{(GeV/c)$^{2}$}} \phantomsection\label{appendix:appA1}
\begin{table}[H]
\captionsetup{singlelinecheck = false, justification=justified}
\caption{$\theta_{nq}=5\pm5^{\circ}$ at $Q^{2}=3.5\pm0.5$ (GeV/c)$^{2}$}
 \scalebox{1.0}{
\begin{tabular}{l l l l l l l l}
\hline
\hline
\shortstack{$p_{\mathrm{r,bin}}$\\$\mathrm{[GeV/c]}$} &  \shortstack{$p_{\mathrm{r,avg}}$\\$\mathrm{[GeV/c]}$} & \shortstack{$\sigma_{\mathrm{red}}$\\$[\mathrm{fm}^{3}]$} & \shortstack{$\delta\sigma_{\mathrm{stats}}$\\$[\%]$} & \shortstack{$\delta\sigma_{\mathrm{norm}}$\\$[\%]$} & \shortstack{$\delta\sigma_{\mathrm{kin}}$\\$[\%]$} & \shortstack{$\delta\sigma_{\mathrm{syst}}$\\$[\%]$} & \shortstack{$\delta\sigma_{\mathrm{tot}}$\\$[\%]$} \\
\hline
0.020    &       0.03027   &    6.37735E+00      &        33.23        &        3.66       &          3.40        &         5.00        &         33.61 \\               
0.060    &       0.05398   &    2.07666E+00      &        22.38        &        3.66       &          5.18        &         6.34        &         23.26 \\
\hline
\end{tabular}
}
\label{tab:redXsec_table_Q2_3to4_5deg}
\end{table}
\begin{table}[H]
\captionsetup{singlelinecheck = false, justification=justified}
\caption{$\theta_{nq}=15\pm5^{\circ}$ at $Q^{2}=3.5\pm0.5$ (GeV/c)$^{2}$}
 \scalebox{1.0}{
\begin{tabular}{l l l l l l l l}
\hline
\hline
\shortstack{$p_{\mathrm{r,bin}}$\\$\mathrm{[GeV/c]}$} &  \shortstack{$p_{\mathrm{r,avg}}$\\$\mathrm{[GeV/c]}$} & \shortstack{$\sigma_{\mathrm{red}}$\\$[\mathrm{fm}^{3}]$} & \shortstack{$\delta\sigma_{\mathrm{stats}}$\\$[\%]$} & \shortstack{$\delta\sigma_{\mathrm{norm}}$\\$[\%]$} & \shortstack{$\delta\sigma_{\mathrm{kin}}$\\$[\%]$} & \shortstack{$\delta\sigma_{\mathrm{syst}}$\\$[\%]$} & \shortstack{$\delta\sigma_{\mathrm{tot}}$\\$[\%]$} \\
\hline
0.020     &      0.02958   &    5.72716E+00      &        18.91       &         3.66        &         3.61        &         5.14        &         19.59  \\              
0.060     &      0.05397   &    2.17760E+00      &        13.01       &         3.66        &         5.46        &         6.57        &         14.57  \\              
0.100     &      0.08841   &    8.09013E-01      &        24.49       &         3.66        &         6.83        &         7.75        &         25.69  \\
\hline
\end{tabular}
}
\label{tab:redXsec_table_Q2_3to4_15deg}
\end{table}
\begin{table}[H]
\captionsetup{singlelinecheck = false, justification=justified}
\caption{$\theta_{nq}=25\pm5^{\circ}$ at $Q^{2}=3.5\pm0.5$ (GeV/c)$^{2}$}
 \scalebox{1.0}{
\begin{tabular}{l l l l l l l l}
\hline
\hline
\shortstack{$p_{\mathrm{r,bin}}$\\$\mathrm{[GeV/c]}$} &  \shortstack{$p_{\mathrm{r,avg}}$\\$\mathrm{[GeV/c]}$} & \shortstack{$\sigma_{\mathrm{red}}$\\$[\mathrm{fm}^{3}]$} & \shortstack{$\delta\sigma_{\mathrm{stats}}$\\$[\%]$} & \shortstack{$\delta\sigma_{\mathrm{norm}}$\\$[\%]$} & \shortstack{$\delta\sigma_{\mathrm{kin}}$\\$[\%]$} & \shortstack{$\delta\sigma_{\mathrm{syst}}$\\$[\%]$} & \shortstack{$\delta\sigma_{\mathrm{tot}}$\\$[\%]$} \\
\hline
0.020    &       0.03069   &    5.29384E+00      &        14.29        &        3.66        &         3.41       &          5.00       &          15.14 \\               
0.060    &       0.05562   &    1.95390E+00      &        9.28         &        3.66        &         5.70       &          6.78       &          11.49 \\               
0.100    &       0.09083   &    5.07866E-01      &        14.93        &        3.66        &         7.36       &          8.22       &          17.04 \\
0.420    &       0.42023   &    1.26426E-04      &        37.87        &        3.86        &         0.43       &          3.88       &          38.07 \\               
0.460    &       0.46019   &    8.69872E-05      &        43.33        &        3.98        &         0.55       &          4.02       &          43.51 \\
\hline
\end{tabular}
}
\label{tab:redXsec_table_Q2_3to4_25deg}
\end{table}
\begin{table}[H]
\captionsetup{singlelinecheck = false, justification=justified}
\caption{$\theta_{nq}=35\pm5^{\circ}$ at $Q^{2}=3.5\pm0.5$ (GeV/c)$^{2}$}
 \scalebox{1.0}{
\begin{tabular}{l l l l l l l l}
\hline
\hline
\shortstack{$p_{\mathrm{r,bin}}$\\$\mathrm{[GeV/c]}$} &  \shortstack{$p_{\mathrm{r,avg}}$\\$\mathrm{[GeV/c]}$} & \shortstack{$\sigma_{\mathrm{red}}$\\$[\mathrm{fm}^{3}]$} & \shortstack{$\delta\sigma_{\mathrm{stats}}$\\$[\%]$} & \shortstack{$\delta\sigma_{\mathrm{norm}}$\\$[\%]$} & \shortstack{$\delta\sigma_{\mathrm{kin}}$\\$[\%]$} & \shortstack{$\delta\sigma_{\mathrm{syst}}$\\$[\%]$} & \shortstack{$\delta\sigma_{\mathrm{tot}}$\\$[\%]$} \\
\hline
0.020     &      0.02981   &    5.75153E+00      &        12.70        &        3.66       &          2.77        &         4.59         &        13.51   \\             
0.060     &      0.05656   &    1.97889E+00      &        7.55         &        3.66       &          5.60        &         6.69         &        10.09   \\             
0.100     &      0.09245   &    4.77584E-01      &        9.78         &        3.66       &          7.48        &         8.33         &        12.85   \\             
0.140     &      0.12870   &    1.16331E-01      &        25.84        &        3.66       &          7.18        &         8.06         &        27.06   \\             
0.340     &      0.34540   &    2.95606E-04      &        36.50        &        3.86       &          1.07        &         4.00         &        36.72   \\             
0.380     &      0.38192   &    2.09371E-04      &        20.20        &        3.86       &          0.93        &         3.97         &        20.58   \\             
0.420     &      0.42582   &    1.09994E-04      &        18.63        &        4.09       &          1.22        &         4.27         &        19.11   \\
\hline
\end{tabular}
}
\label{tab:redXsec_table_Q2_3to4_35deg}
\end{table}
\begin{table}[H]
\captionsetup{singlelinecheck = false, justification=justified}
\caption{$\theta_{nq}=45\pm5^{\circ}$ at $Q^{2}=3.5\pm0.5$ (GeV/c)$^{2}$}
 \scalebox{1.0}{
\begin{tabular}{l l l l l l l l}
\hline
\hline
\shortstack{$p_{\mathrm{r,bin}}$\\$\mathrm{[GeV/c]}$} &  \shortstack{$p_{\mathrm{r,avg}}$\\$\mathrm{[GeV/c]}$} & \shortstack{$\sigma_{\mathrm{red}}$\\$[\mathrm{fm}^{3}]$} & \shortstack{$\delta\sigma_{\mathrm{stats}}$\\$[\%]$} & \shortstack{$\delta\sigma_{\mathrm{norm}}$\\$[\%]$} & \shortstack{$\delta\sigma_{\mathrm{kin}}$\\$[\%]$} & \shortstack{$\delta\sigma_{\mathrm{syst}}$\\$[\%]$} & \shortstack{$\delta\sigma_{\mathrm{tot}}$\\$[\%]$} \\
\hline
0.020     &      0.03068   &    6.49330E+00      &        11.19       &         3.66         &        2.46       &          4.41       &          12.03    \\            
0.060     &      0.05804   &    1.80699E+00      &        6.22        &         3.66         &        5.23       &          6.39       &          8.91     \\            
0.100     &      0.09531   &    3.64016E-01      &        7.30        &         3.66         &        7.08       &          7.97       &          10.80    \\            
0.140     &      0.13365   &    8.33640E-02      &        13.32       &         3.66         &        6.86       &          7.77       &          15.43    \\            
0.180     &      0.16738   &    4.94696E-02      &        33.93       &         3.66         &        6.25       &          7.24       &          34.70    \\            
0.340     &      0.34768   &    4.20104E-04      &        36.82       &         3.86         &        1.72       &          4.22       &          37.06    \\            
0.380     &      0.38332   &    3.01566E-04      &        17.06       &         3.86         &        1.50       &          4.14       &          17.56    \\
\hline
\end{tabular}
}
\label{tab:redXsec_table_Q2_3to4_45deg}
\end{table}
\begin{table}[H]
\captionsetup{singlelinecheck = false, justification=justified}
\caption{$\theta_{nq}=55\pm5^{\circ}$ at $Q^{2}=3.5\pm0.5$ (GeV/c)$^{2}$}
 \scalebox{1.0}{
\begin{tabular}{l l l l l l l l}
\hline
\hline
\shortstack{$p_{\mathrm{r,bin}}$\\$\mathrm{[GeV/c]}$} &  \shortstack{$p_{\mathrm{r,avg}}$\\$\mathrm{[GeV/c]}$} & \shortstack{$\sigma_{\mathrm{red}}$\\$[\mathrm{fm}^{3}]$} & \shortstack{$\delta\sigma_{\mathrm{stats}}$\\$[\%]$} & \shortstack{$\delta\sigma_{\mathrm{norm}}$\\$[\%]$} & \shortstack{$\delta\sigma_{\mathrm{kin}}$\\$[\%]$} & \shortstack{$\delta\sigma_{\mathrm{syst}}$\\$[\%]$} & \shortstack{$\delta\sigma_{\mathrm{tot}}$\\$[\%]$} \\
\hline
0.020    &       0.03135   &    5.68088E+00      &        10.16      &          3.66       &          1.76       &          4.06       &          10.94    \\             
0.060    &       0.05799   &    1.71593E+00      &        5.42       &          3.66       &          4.45       &          5.76       &          7.91     \\            
0.100    &       0.09654   &    3.21520E-01      &        5.85       &          3.66       &          5.97       &          7.01       &          9.13     \\            
0.140    &       0.13567   &    8.93298E-02      &        8.62       &          3.66       &          5.88       &          6.92       &          11.06    \\            
0.180    &       0.17490   &    2.23853E-02      &        18.36      &          3.66       &          5.14       &          6.31       &          19.41    \\
0.380    &       0.38595   &    5.59781E-04      &        25.63      &          3.86       &          1.96       &          4.33       &          25.99    \\            
0.420    &       0.42326   &    2.84214E-04      &        16.63      &          3.86       &          1.77       &          4.24       &          17.16    \\
\hline   
\end{tabular}
}
\label{tab:redXsec_table_Q2_3to4_55deg}
\end{table}
\begin{table}[H]
\captionsetup{singlelinecheck = false, justification=justified}
\caption{$\theta_{nq}=65\pm5^{\circ}$ at $Q^{2}=3.5\pm0.5$ (GeV/c)$^{2}$}
 \scalebox{1.0}{
\begin{tabular}{l l l l l l l l}
\hline
\hline
\shortstack{$p_{\mathrm{r,bin}}$\\$\mathrm{[GeV/c]}$} &  \shortstack{$p_{\mathrm{r,avg}}$\\$\mathrm{[GeV/c]}$} & \shortstack{$\sigma_{\mathrm{red}}$\\$[\mathrm{fm}^{3}]$} & \shortstack{$\delta\sigma_{\mathrm{stats}}$\\$[\%]$} & \shortstack{$\delta\sigma_{\mathrm{norm}}$\\$[\%]$} & \shortstack{$\delta\sigma_{\mathrm{kin}}$\\$[\%]$} & \shortstack{$\delta\sigma_{\mathrm{syst}}$\\$[\%]$} & \shortstack{$\delta\sigma_{\mathrm{tot}}$\\$[\%]$} \\
\hline
0.020     &      0.03201   &    5.46553E+00      &        9.55        &         3.66        &         1.27        &         3.87        &         10.30     \\           
0.060     &      0.05898   &    1.72017E+00      &        4.87        &         3.66        &         2.83        &         4.63        &         6.72      \\           
0.100     &      0.09658   &    3.37215E-01      &        4.86        &         3.66        &         4.24        &         5.60        &         7.42      \\           
0.140     &      0.13579   &    7.71106E-02      &        7.00        &         3.66        &         4.18        &         5.56        &         8.94      \\           
0.180     &      0.17593   &    1.60043E-02      &        13.98       &         3.66        &         3.31        &         4.94        &         14.83     \\           
0.220     &      0.21683   &    3.86961E-03      &        31.70       &         3.66        &         2.54        &         4.46        &         32.02     \\           
0.260     &      0.25715   &    3.21213E-03      &        35.77       &         3.66        &         1.97        &         4.16        &         36.01     \\
\hline    
\end{tabular}
}
\label{tab:redXsec_table_Q2_3to4_65deg}
\end{table}
\begin{table}[H]
\captionsetup{singlelinecheck = false, justification=justified}
\caption{$\theta_{nq}=75\pm5^{\circ}$ at $Q^{2}=3.5\pm0.5$ (GeV/c)$^{2}$}
 \scalebox{1.0}{
\begin{tabular}{l l l l l l l l}
\hline
\hline
\shortstack{$p_{\mathrm{r,bin}}$\\$\mathrm{[GeV/c]}$} &  \shortstack{$p_{\mathrm{r,avg}}$\\$\mathrm{[GeV/c]}$} & \shortstack{$\sigma_{\mathrm{red}}$\\$[\mathrm{fm}^{3}]$} & \shortstack{$\delta\sigma_{\mathrm{stats}}$\\$[\%]$} & \shortstack{$\delta\sigma_{\mathrm{norm}}$\\$[\%]$} & \shortstack{$\delta\sigma_{\mathrm{kin}}$\\$[\%]$} & \shortstack{$\delta\sigma_{\mathrm{syst}}$\\$[\%]$} & \shortstack{$\delta\sigma_{\mathrm{tot}}$\\$[\%]$} \\
\hline
0.020     &      0.03171   &    5.06893E+00     &         9.13       &          3.66       &          0.76       &          3.74      &           9.86    \\             
0.060     &      0.05974   &    1.57799E+00     &         4.50       &          3.66       &          1.54       &          3.97      &           6.01    \\             
0.100     &      0.09702   &    3.07899E-01     &         4.23       &          3.66       &          1.77       &          4.07      &           5.87    \\             
0.140     &      0.13679   &    5.87380E-02     &         5.97       &          3.66       &          1.65       &          4.02      &           7.20    \\             
0.180     &      0.17715   &    1.75816E-02     &         10.67      &          3.66       &          1.10       &          3.82      &           11.34   \\             
0.220     &      0.21708   &    4.50003E-03     &         21.85      &          3.66       &          0.82       &          3.75      &           22.17   \\             
0.260     &      0.25832   &    1.96842E-03     &         30.79      &          3.66       &          0.55       &          3.70      &           31.01   \\
\hline
\end{tabular}
}
\label{tab:redXsec_table_Q2_3to4_75deg}
\end{table}
\begin{table}[H]
\captionsetup{singlelinecheck = false, justification=justified}
\caption{$\theta_{nq}=85\pm5^{\circ}$ at $Q^{2}=3.5\pm0.5$ (GeV/c)$^{2}$}
 \scalebox{1.0}{
\begin{tabular}{l l l l l l l l}
\hline
\hline
\shortstack{$p_{\mathrm{r,bin}}$\\$\mathrm{[GeV/c]}$} &  \shortstack{$p_{\mathrm{r,avg}}$\\$\mathrm{[GeV/c]}$} & \shortstack{$\sigma_{\mathrm{red}}$\\$[\mathrm{fm}^{3}]$} & \shortstack{$\delta\sigma_{\mathrm{stats}}$\\$[\%]$} & \shortstack{$\delta\sigma_{\mathrm{norm}}$\\$[\%]$} & \shortstack{$\delta\sigma_{\mathrm{kin}}$\\$[\%]$} & \shortstack{$\delta\sigma_{\mathrm{syst}}$\\$[\%]$} & \shortstack{$\delta\sigma_{\mathrm{tot}}$\\$[\%]$} \\
\hline
0.020    &       0.03231   &    4.79539E+00     &         8.87        &         3.66       &          0.84       &          3.76       &          9.63   \\              
0.060    &       0.05981   &    1.57054E+00     &         4.30        &         3.66       &          0.53       &          3.70       &          5.67   \\              
0.100    &       0.09790   &    2.51174E-01     &         3.88        &         3.66       &          0.88       &          3.76       &          5.41   \\              
0.140    &       0.13759   &    5.62479E-02     &         4.99        &         3.66       &          1.15       &          3.84       &          6.29   \\              
0.180    &       0.17779   &    1.61925E-02     &         7.74        &         3.66       &          1.24       &          3.86       &          8.65   \\              
0.220    &       0.21862   &    5.88967E-03     &         12.69       &         3.66       &          1.28       &          3.88       &          13.27  \\              
0.260    &       0.25921   &    2.77418E-03     &         21.35       &         3.66       &          1.30       &          3.88       &          21.70  \\
\hline
\end{tabular}
}
\label{tab:redXsec_table_Q2_3to4_85deg}
\end{table}
\begin{table}[H]
\captionsetup{singlelinecheck = false, justification=justified}
\caption{$\theta_{nq}=95\pm5^{\circ}$ at $Q^{2}=3.5\pm0.5$ (GeV/c)$^{2}$}
 \scalebox{1.0}{
\begin{tabular}{l l l l l l l l}
\hline
\hline
\shortstack{$p_{\mathrm{r,bin}}$\\$\mathrm{[GeV/c]}$} &  \shortstack{$p_{\mathrm{r,avg}}$\\$\mathrm{[GeV/c]}$} & \shortstack{$\sigma_{\mathrm{red}}$\\$[\mathrm{fm}^{3}]$} & \shortstack{$\delta\sigma_{\mathrm{stats}}$\\$[\%]$} & \shortstack{$\delta\sigma_{\mathrm{norm}}$\\$[\%]$} & \shortstack{$\delta\sigma_{\mathrm{kin}}$\\$[\%]$} & \shortstack{$\delta\sigma_{\mathrm{syst}}$\\$[\%]$} & \shortstack{$\delta\sigma_{\mathrm{tot}}$\\$[\%]$} \\
\hline
0.020    &       0.03213   &    5.03786E+00      &        8.99       &          3.66       &          1.11        &         3.83       &          9.77   \\              
0.060    &       0.06053   &    1.50083E+00      &        4.27       &          3.66       &          1.98        &         4.16       &          5.96   \\              
0.100    &       0.09861   &    2.56864E-01      &        3.63       &          3.66       &          2.57        &         4.48       &          5.77   \\              
0.140    &       0.13869   &    5.57178E-02      &        4.38       &          3.66       &          3.07        &         4.78       &          6.48   \\              
0.180    &       0.17838   &    1.64757E-02      &        6.26       &          3.66       &          3.08        &         4.79       &          7.88   \\              
0.220    &       0.21883   &    7.18822E-03      &        8.31       &          3.66       &          3.04        &         4.76       &          9.58   \\              
0.260    &       0.25901   &    2.93957E-03      &        12.22      &          3.66       &          2.77        &         4.59       &          13.06  \\
\hline
\end{tabular}
}
\label{tab:redXsec_table_Q2_3to4_95deg}
\end{table}

\subsection*{A2. Reduced Cross Sections at $Q^{2}=4.5\pm0.5$ \textnormal{(GeV/c)$^{2}$}} \phantomsection\label{appendix:appA2}
\begin{table}[H]
\captionsetup{singlelinecheck = false, justification=justified}
\caption{$\theta_{nq}=5\pm5^{\circ}$ at $Q^{2}=4.5\pm0.5$ (GeV/c)$^{2}$}
\begin{tabular}{l l l l l l l l}
\hline
\hline
\shortstack{$p_{\mathrm{r,bin}}$\\$\mathrm{[GeV/c]}$} &  \shortstack{$p_{\mathrm{r,avg}}$\\$\mathrm{[GeV/c]}$} & \shortstack{$\sigma_{\mathrm{red}}$\\$[\mathrm{fm}^{3}]$} & \shortstack{$\delta\sigma_{\mathrm{stats}}$\\$[\%]$} & \shortstack{$\delta\sigma_{\mathrm{norm}}$\\$[\%]$} & \shortstack{$\delta\sigma_{\mathrm{kin}}$\\$[\%]$} & \shortstack{$\delta\sigma_{\mathrm{syst}}$\\$[\%]$} & \shortstack{$\delta\sigma_{\mathrm{tot}}$\\$[\%]$} \\
\hline
0.020     &      0.03259   &    5.03296E+00      &        42.52       &         3.66        &         4.31       &          5.65       &          42.90  \\              
0.060     &      0.06183   &    1.29738E+00      &        14.47       &         3.66        &         5.46       &          6.58       &          15.89  \\              
0.100     &      0.09878   &    2.97336E-01      &        10.63       &         3.66        &         5.28       &          6.42       &          12.42  \\              
0.140     &      0.13646   &    6.46574E-02      &        16.60       &         3.66        &         5.02       &          6.21       &          17.73  \\              
0.180     &      0.17414   &    2.42630E-02      &        32.29       &         3.66        &         4.87       &          6.10       &          32.86  \\             
\hline
\end{tabular}
\label{tab:redXsec_table_Q2_4to5_5deg}
\end{table}
\begin{table}[H]
\captionsetup{singlelinecheck = false, justification=justified}
\caption{$\theta_{nq}=15\pm5^{\circ}$ at $Q^{2}=4.5\pm0.5$ (GeV/c)$^{2}$}
\begin{tabular}{l l l l l l l l}
\hline
\hline
\shortstack{$p_{\mathrm{r,bin}}$\\$\mathrm{[GeV/c]}$} &  \shortstack{$p_{\mathrm{r,avg}}$\\$\mathrm{[GeV/c]}$} & \shortstack{$\sigma_{\mathrm{red}}$\\$[\mathrm{fm}^{3}]$} & \shortstack{$\delta\sigma_{\mathrm{stats}}$\\$[\%]$} & \shortstack{$\delta\sigma_{\mathrm{norm}}$\\$[\%]$} & \shortstack{$\delta\sigma_{\mathrm{kin}}$\\$[\%]$} & \shortstack{$\delta\sigma_{\mathrm{syst}}$\\$[\%]$} & \shortstack{$\delta\sigma_{\mathrm{tot}}$\\$[\%]$} \\
\hline
0.020     &      0.03223   &    4.95610E+00      &        27.92       &         3.66        &         4.08        &         5.48       &          28.46   \\             
0.060     &      0.06248   &    1.37967E+00      &        8.42        &         3.66        &         5.29        &         6.43       &          10.60   \\             
0.100     &      0.09853   &    3.01980E-01      &        6.37        &         3.66        &         5.21        &         6.36       &          9.00    \\             
0.140     &      0.13669   &    7.71973E-02      &        8.79        &         3.66        &         5.18        &         6.35       &          10.84   \\             
0.180     &      0.17523   &    2.30324E-02      &        17.23       &         3.66        &         5.34        &         6.47       &          18.41   \\             
0.220     &      0.21438   &    1.16646E-02      &        33.63       &         3.66        &         5.59        &         6.68       &          34.29   \\          
\hline
\end{tabular}
\label{tab:redXsec_table_Q2_4to5_15deg}
\end{table}
\begin{table}[H]
\captionsetup{singlelinecheck = false, justification=justified}
\caption{$\theta_{nq}=25\pm5^{\circ}$ at $Q^{2}=4.5\pm0.5$ (GeV/c)$^{2}$}
\begin{tabular}{l l l l l l l l}
\hline
\hline
\shortstack{$p_{\mathrm{r,bin}}$\\$\mathrm{[GeV/c]}$} &  \shortstack{$p_{\mathrm{r,avg}}$\\$\mathrm{[GeV/c]}$} & \shortstack{$\sigma_{\mathrm{red}}$\\$[\mathrm{fm}^{3}]$} & \shortstack{$\delta\sigma_{\mathrm{stats}}$\\$[\%]$} & \shortstack{$\delta\sigma_{\mathrm{norm}}$\\$[\%]$} & \shortstack{$\delta\sigma_{\mathrm{kin}}$\\$[\%]$} & \shortstack{$\delta\sigma_{\mathrm{syst}}$\\$[\%]$} & \shortstack{$\delta\sigma_{\mathrm{tot}}$\\$[\%]$} \\
\hline
0.020     &      0.03215   &    7.38449E+00      &        24.01       &         3.66       &          3.15        &         4.83        &         24.49   \\             
0.060     &      0.06256   &    1.30063E+00      &        6.82        &         3.66       &          4.61        &         5.89        &         9.01    \\             
0.100     &      0.09921   &    3.00250E-01      &        5.06        &         3.66       &          4.88        &         6.10        &         7.93    \\             
0.140     &      0.13785   &    7.18342E-02      &        6.82        &         3.66       &          5.21        &         6.36        &         9.33    \\            
0.180     &      0.17685   &    2.79423E-02      &        10.91       &         3.66       &          5.59        &         6.68        &         12.79   \\             
0.220     &      0.21664   &    1.24979E-02      &        17.84       &         3.66       &          5.81        &         6.86        &         19.12   \\             
0.260     &      0.25638   &    2.73724E-03      &        44.80       &         3.66       &          5.63        &         6.71        &         45.30   \\
\hline
\end{tabular}
\label{tab:redXsec_table_Q2_4to5_25deg}
\end{table}
\begin{table}[H]
\captionsetup{singlelinecheck = false, justification=justified}
\caption{$\theta_{nq}=35\pm5^{\circ}$ at $Q^{2}=4.5\pm0.5$ (GeV/c)$^{2}$}
\begin{tabular}{l l l l l l l l}
\hline
\hline
\shortstack{$p_{\mathrm{r,bin}}$\\$\mathrm{[GeV/c]}$} &  \shortstack{$p_{\mathrm{r,avg}}$\\$\mathrm{[GeV/c]}$} & \shortstack{$\sigma_{\mathrm{red}}$\\$[\mathrm{fm}^{3}]$} & \shortstack{$\delta\sigma_{\mathrm{stats}}$\\$[\%]$} & \shortstack{$\delta\sigma_{\mathrm{norm}}$\\$[\%]$} & \shortstack{$\delta\sigma_{\mathrm{kin}}$\\$[\%]$} & \shortstack{$\delta\sigma_{\mathrm{syst}}$\\$[\%]$} & \shortstack{$\delta\sigma_{\mathrm{tot}}$\\$[\%]$} \\
\hline
0.020     &      0.03181   &    4.00473E+00      &        20.91       &         3.66        &         2.52        &         4.45       &          21.38     \\           
0.060     &      0.06217   &    1.18093E+00      &        6.32        &         3.66        &         3.36        &         4.97       &          8.04      \\           
0.100     &      0.09914   &    2.88897E-01      &        4.60        &         3.66        &         4.09        &         5.49       &          7.17      \\           
0.140     &      0.13875   &    7.58109E-02      &        5.77        &         3.66        &         4.76        &         6.00       &          8.33      \\           
0.180     &      0.17780   &    2.00561E-02      &        10.12       &         3.66        &         5.22        &         6.38       &          11.97     \\           
0.220     &      0.21766   &    8.63454E-03      &        15.72       &         3.66        &         5.46        &         6.57       &          17.04     \\           
0.260     &      0.25754   &    2.78087E-03      &        28.92       &         3.66        &         5.25        &         6.40       &          29.62     \\
\hline
\end{tabular}
\label{tab:redXsec_table_Q2_4to5_35deg}
\end{table}
\begin{table}[H]
\captionsetup{singlelinecheck = false, justification=justified}
\caption{$\theta_{nq}=45\pm5^{\circ}$ at $Q^{2}=4.5\pm0.5$ (GeV/c)$^{2}$}
\begin{tabular}{l l l l l l l l}
\hline
\hline
\shortstack{$p_{\mathrm{r,bin}}$\\$\mathrm{[GeV/c]}$} &  \shortstack{$p_{\mathrm{r,avg}}$\\$\mathrm{[GeV/c]}$} & \shortstack{$\sigma_{\mathrm{red}}$\\$[\mathrm{fm}^{3}]$} & \shortstack{$\delta\sigma_{\mathrm{stats}}$\\$[\%]$} & \shortstack{$\delta\sigma_{\mathrm{norm}}$\\$[\%]$} & \shortstack{$\delta\sigma_{\mathrm{kin}}$\\$[\%]$} & \shortstack{$\delta\sigma_{\mathrm{syst}}$\\$[\%]$} & \shortstack{$\delta\sigma_{\mathrm{tot}}$\\$[\%]$} \\
\hline
0.020    &       0.03147   &    6.12525E+00       &       21.33       &         3.66        &         2.06        &         4.20        &         21.74      \\          
0.060    &       0.06241   &    1.33303E+00       &       6.20        &         3.66        &         1.87        &         4.11        &         7.44       \\          
0.100    &       0.09921   &    2.56214E-01       &       4.58        &         3.66        &         2.79        &         4.60        &         6.49       \\          
0.140    &       0.13844   &    6.81144E-02       &       5.64        &         3.66        &         3.74        &         5.23        &         7.69       \\          
0.180    &       0.17838   &    2.08912E-02       &       8.57        &         3.66        &         4.24        &         5.61        &         10.24      \\          
0.220    &       0.21825   &    5.24976E-03       &       16.75       &         3.66        &         4.36        &         5.69        &         17.69      \\          
0.260    &       0.25782   &    2.83066E-03       &       24.31       &         3.66        &         4.10        &         5.50        &         24.93      \\
\hline
\end{tabular}
\label{tab:redXsec_table_Q2_4to5_45deg}
\end{table}
\begin{table}[H]
\captionsetup{singlelinecheck = false, justification=justified}
\caption{$\theta_{nq}=55\pm5^{\circ}$ at $Q^{2}=4.5\pm0.5$ (GeV/c)$^{2}$}
\begin{tabular}{l l l l l l l l}
\hline
\hline
\shortstack{$p_{\mathrm{r,bin}}$\\$\mathrm{[GeV/c]}$} &  \shortstack{$p_{\mathrm{r,avg}}$\\$\mathrm{[GeV/c]}$} & \shortstack{$\sigma_{\mathrm{red}}$\\$[\mathrm{fm}^{3}]$} & \shortstack{$\delta\sigma_{\mathrm{stats}}$\\$[\%]$} & \shortstack{$\delta\sigma_{\mathrm{norm}}$\\$[\%]$} & \shortstack{$\delta\sigma_{\mathrm{kin}}$\\$[\%]$} & \shortstack{$\delta\sigma_{\mathrm{syst}}$\\$[\%]$} & \shortstack{$\delta\sigma_{\mathrm{tot}}$\\$[\%]$} \\
\hline
0.020    &       0.03172   &    5.18815E+00      &        21.36       &         3.66       &          0.65        &         3.72        &         21.68    \\            
0.060    &       0.06218   &    1.39568E+00      &        6.48        &         3.66       &          0.77        &         3.74        &         7.48     \\            
0.100    &       0.09928   &    2.88239E-01      &        4.72        &         3.66       &          0.58        &         3.71        &         6.00     \\            
0.140    &       0.13833   &    7.06831E-02      &        5.86        &         3.66       &          1.67        &         4.03        &         7.11     \\            
0.180    &       0.17855   &    2.00666E-02      &        9.24        &         3.66       &          2.35        &         4.35        &         10.21    \\            
0.220    &       0.21843   &    7.04340E-03      &        14.24       &         3.66       &          2.59        &         4.49        &         14.93    \\            
0.260    &       0.25868   &    2.68690E-03      &        21.99       &         3.66       &          2.40        &         4.38        &         22.42    \\
\hline
\end{tabular}
\label{tab:redXsec_table_Q2_4to5_55deg}
\end{table}
\begin{table}[H]
\captionsetup{singlelinecheck = false, justification=justified}
\caption{$\theta_{nq}=65\pm5^{\circ}$ at $Q^{2}=4.5\pm0.5$ (GeV/c)$^{2}$}
\begin{tabular}{l l l l l l l l}
\hline
\hline
\shortstack{$p_{\mathrm{r,bin}}$\\$\mathrm{[GeV/c]}$} &  \shortstack{$p_{\mathrm{r,avg}}$\\$\mathrm{[GeV/c]}$} & \shortstack{$\sigma_{\mathrm{red}}$\\$[\mathrm{fm}^{3}]$} & \shortstack{$\delta\sigma_{\mathrm{stats}}$\\$[\%]$} & \shortstack{$\delta\sigma_{\mathrm{norm}}$\\$[\%]$} & \shortstack{$\delta\sigma_{\mathrm{kin}}$\\$[\%]$} & \shortstack{$\delta\sigma_{\mathrm{syst}}$\\$[\%]$} & \shortstack{$\delta\sigma_{\mathrm{tot}}$\\$[\%]$} \\
\hline
0.020    &       0.03116   &    5.29860E+00      &        25.05       &         3.66        &         2.00        &         4.17        &         25.40   \\             
0.060    &       0.06157   &    1.38526E+00      &        7.47        &         3.66        &         2.78        &         4.60        &         8.77    \\             
0.100    &       0.09847   &    2.94809E-01      &        5.23        &         3.66        &         2.33        &         4.34        &         6.79    \\             
0.140    &       0.13752   &    7.09719E-02      &        6.39        &         3.66        &         1.25        &         3.87        &         7.47    \\             
0.180    &       0.17772   &    1.89440E-02      &        9.85        &         3.66        &         0.38        &         3.68        &         10.52   \\             
0.220    &       0.21785   &    4.84834E-03      &        17.33       &         3.66        &         0.46        &         3.69        &         17.72   \\             
0.260    &       0.25887   &    2.20160E-03      &        24.19       &         3.66        &         0.67        &         3.72        &         24.48   \\
\hline
\end{tabular}
\label{tab:redXsec_table_Q2_4to5_65deg}
\end{table}
\begin{table}[H]
\captionsetup{singlelinecheck = false, justification=justified}
\caption{$\theta_{nq}=75\pm5^{\circ}$ at $Q^{2}=4.5\pm0.5$ (GeV/c)$^{2}$}
\begin{tabular}{l l l l l l l l}
\hline
\hline
\shortstack{$p_{\mathrm{r,bin}}$\\$\mathrm{[GeV/c]}$} &  \shortstack{$p_{\mathrm{r,avg}}$\\$\mathrm{[GeV/c]}$} & \shortstack{$\sigma_{\mathrm{red}}$\\$[\mathrm{fm}^{3}]$} & \shortstack{$\delta\sigma_{\mathrm{stats}}$\\$[\%]$} & \shortstack{$\delta\sigma_{\mathrm{norm}}$\\$[\%]$} & \shortstack{$\delta\sigma_{\mathrm{kin}}$\\$[\%]$} & \shortstack{$\delta\sigma_{\mathrm{syst}}$\\$[\%]$} & \shortstack{$\delta\sigma_{\mathrm{tot}}$\\$[\%]$} \\
\hline
0.020     &      0.03205   &    4.86919E+00      &        29.30        &        3.66        &         4.72        &         5.98       &          29.90  \\              
0.060     &      0.06131   &    1.38732E+00      &        8.83         &        3.66        &         5.60        &         6.69       &          11.08  \\              
0.100     &      0.09788   &    2.68460E-01      &        6.34         &        3.66        &         4.58        &         5.87       &          8.64   \\              
0.140     &      0.13715   &    6.42418E-02      &        7.43         &        3.66        &         3.55        &         5.10       &          9.01   \\              
0.180     &      0.17692   &    1.72753E-02      &        10.83        &        3.66        &         2.58        &         4.48       &          11.72  \\              
0.220     &      0.21724   &    5.28407E-03      &        17.61        &        3.66        &         1.75        &         4.06       &          18.07  \\              
0.260     &      0.25809   &    3.33159E-03      &        22.42        &        3.66        &         1.23        &         3.86       &          22.75  \\
\hline
\end{tabular}
\label{tab:redXsec_table_Q2_4to5_75deg}
\end{table}
\begin{table}[H]
\captionsetup{singlelinecheck = false, justification=justified}
\caption{$\theta_{nq}=85\pm5^{\circ}$ at $Q^{2}=4.5\pm0.5$ (GeV/c)$^{2}$}
\begin{tabular}{l l l l l l l l}
\hline
\hline
\shortstack{$p_{\mathrm{r,bin}}$\\$\mathrm{[GeV/c]}$} &  \shortstack{$p_{\mathrm{r,avg}}$\\$\mathrm{[GeV/c]}$} & \shortstack{$\sigma_{\mathrm{red}}$\\$[\mathrm{fm}^{3}]$} & \shortstack{$\delta\sigma_{\mathrm{stats}}$\\$[\%]$} & \shortstack{$\delta\sigma_{\mathrm{norm}}$\\$[\%]$} & \shortstack{$\delta\sigma_{\mathrm{kin}}$\\$[\%]$} & \shortstack{$\delta\sigma_{\mathrm{syst}}$\\$[\%]$} & \shortstack{$\delta\sigma_{\mathrm{tot}}$\\$[\%]$} \\
\hline
0.020     &      0.03071   &    5.11718E+00      &        36.38       &         3.66        &         6.22       &          7.22       &          37.09    \\            
0.060     &      0.06108   &    1.16687E+00      &        11.57       &         3.66        &         7.25       &          8.12       &          14.13    \\            
0.100     &      0.09839   &    2.72231E-01      &        8.45        &         3.66        &         6.04       &          7.06       &          11.01    \\            
0.140     &      0.13708   &    6.28358E-02      &        9.87        &         3.66        &         5.20       &          6.36       &          11.74    \\            
0.180     &      0.17669   &    1.61772E-02      &        14.42       &         3.66        &         4.32       &          5.66       &          15.49    \\            
0.220     &      0.21679   &    5.34681E-03      &        22.55       &         3.66        &         3.65       &          5.17       &          23.14    \\            
0.260     &      0.25716   &    4.56447E-03      &        26.89       &         3.66        &         2.92       &          4.68       &          27.30    \\                                                     
\hline
\end{tabular}
\label{tab:redXsec_table_Q2_4to5_85deg}
\end{table}
\begin{table}[H]
\captionsetup{singlelinecheck = false, justification=justified}
\caption{$\theta_{nq}=95\pm5^{\circ}$ at $Q^{2}=4.5\pm0.5$ (GeV/c)$^{2}$}
\begin{tabular}{l l l l l l l l}
\hline
\hline
\shortstack{$p_{\mathrm{r,bin}}$\\$\mathrm{[GeV/c]}$} &  \shortstack{$p_{\mathrm{r,avg}}$\\$\mathrm{[GeV/c]}$} & \shortstack{$\sigma_{\mathrm{red}}$\\$[\mathrm{fm}^{3}]$} & \shortstack{$\delta\sigma_{\mathrm{stats}}$\\$[\%]$} & \shortstack{$\delta\sigma_{\mathrm{norm}}$\\$[\%]$} & \shortstack{$\delta\sigma_{\mathrm{kin}}$\\$[\%]$} & \shortstack{$\delta\sigma_{\mathrm{syst}}$\\$[\%]$} & \shortstack{$\delta\sigma_{\mathrm{tot}}$\\$[\%]$} \\
\hline
0.020     &      0.03048   &    3.39458E+00      &        46.80       &         3.66        &         6.23        &         7.23        &         47.35   \\             
0.060     &      0.05984   &    1.63840E+00      &        16.85       &         3.66        &         8.27        &         9.05        &         19.12   \\             
0.100     &      0.09783   &    2.41467E-01      &        14.49       &         3.66        &         7.19        &         8.07        &         16.58   \\             
0.140     &      0.13606   &    8.15315E-02      &        17.95       &         3.66        &         6.17        &         7.18        &         19.33   \\             
0.180     &      0.17524   &    3.55247E-02      &        26.77       &         3.66        &         5.23        &         6.38        &         27.52   \\             
0.220     &      0.21358   &    3.30502E-02      &        46.09       &         3.66        &         4.55        &         5.84        &         46.46   \\                                            
\hline
\end{tabular}
\label{tab:redXsec_table_Q2_4to5_95deg}
\end{table}

\newpage

\setcounter{equation}{0}
\renewcommand{\theequation}{B.\arabic{equation}}

\setcounter{figure}{0}
\renewcommand{\thefigure}{B.\arabic{figure}}

\section*{Appendix B: The Hall C Fast Raster} \phantomsection\label{appendix:appB}
\begin{figure}[H]
\centering
\includegraphics[scale=0.3]{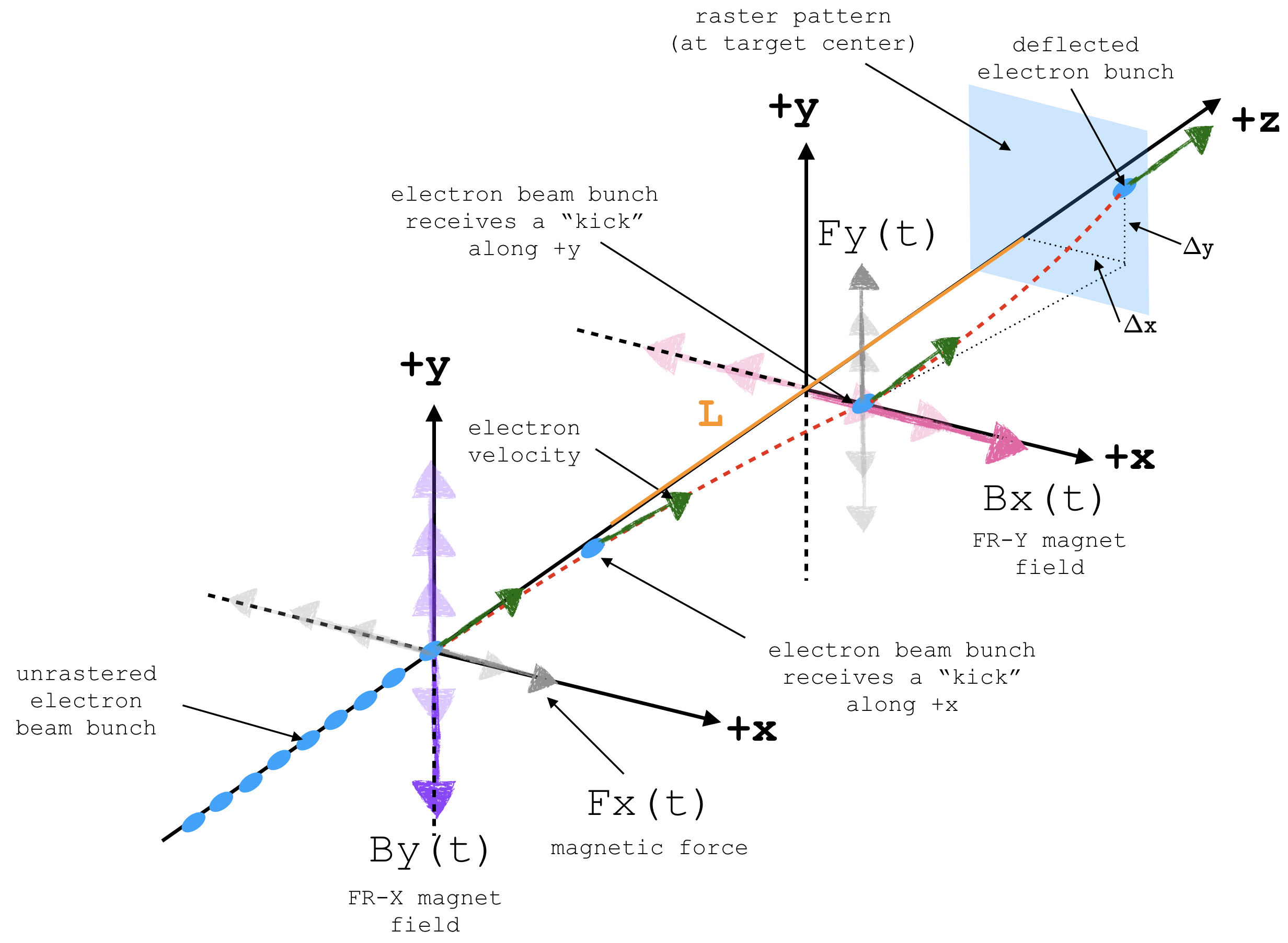}
\caption{Cartoon showing how the fast-raster system works. The beam bunches feel a kick (time-varying magnetic force) along 
the $(x,y)$ coordinates due to a time-varying magnetic field and form a rectangular pattern at the target.}
\label{fig:figB1}
\end{figure}
\indent The Hall C fast raster (FR) system consists of a set of X and Y air-core magnets. A linear\footnote{\singlespacing The first linear raster in Hall C has been
operating since August 2002. The advantage of the linear (triangular) waveform over the original sinusoidal waveform is the higher linear velocity and suppression of the turning time at the 
crests and troughs of the oscillating magnetic field. This suppression significantly reduces the dwelling time of the field at the edges of the raster and contributes to less 
beam energy being deposited at the boundaries. As a result, the overheating generated in the cryogenic target due to the edges of the raster is reduced. For a detailed description 
of the linear raster, see Ref.\cite{Raster_Yan2005}.}
(triangular) waveform generator provides a $\sim25$ kHz\cite{Raster_Yan2005} time-varying current to the FR magnets coils which produces an oscillating magnetic field transverse
to the beam axis. As the electron beam bunches pass by the FR-X and FR-Y magnets, they are deflected by a magnetic force (``kick'') along the $x$ and $y$ axis, respectively.
The ``kick'' to coordinate ($\Delta x$,$\Delta y$) at the target effectively smears out the beam in a rectangular raster pattern (See Fig. \ref{fig:figB1}). \\
\indent The deflection angle of the electron beam bunch along $(x,y)$\footnote{\singlespacing The deflection equations of the FR-X and FR-Y magnets have the same form and are denoted by the $(x,y)$ subscripts to differentiate between the magnets.} 
-axes at the target center is given by 
\begin{equation}
  \theta_{\Delta x,y}=\Delta x,y / L_{\mathrm{FR}},
  \label{eq:B1}
\end{equation}
where the length from the center of the target to the average distance between FR-X and FR-Y magnets is  $L_{\mathrm{FR}}=13.56$ m.
To estimate the deflection of the beam at the target, substitute Eq. \ref{eq:B1} into the equation of motion for a charged particle 
under a magnetic field (Eq. \ref{eq:3.3}) to obtain
\begin{equation}
  p_{e} = C_{k}\frac{\int B_{y,x}(t)d\ell}{\Delta x,y / L_{\mathrm{FR}}}.
  \label{eq:B2}
\end{equation}
From the FR-magnet specifications in Table 1 of Ref.\cite{Raster_Yan2005}, the field integral is given by
\begin{equation}
  \int B_{y,x}d\ell = 8.1 \times 10^{-5} \frac{\mathrm{[T][m]}}{\mathrm{[A]}} \times I_{x,y}.
  \label{eq:B3}
\end{equation}
Substituting Eq. \ref{eq:B3} in Eq. \ref{eq:B2} and solving for $\Delta x$ and $\Delta y$,
\begin{equation}
  \Delta x,y = 0.329 \times \frac{I_{x,y}}{p_{e}},
  \label{eq:B4}
\end{equation}
where $\Delta x,y$ is the electron beam bunch deflection in [mm] at the target center,  $I_{x,y}$ is the FR-magnet current in [A] and $p_{e}$ 
is the electron momentum in [GeV/c]. Since each FR-magnet has
two coils and each coil gives equal deflection, both coils will give double deflection, therefore $I_{\mathrm{coil}_{x,y}} = I_{x,y}/2$.
Using this expression, Eq. \ref{eq:B4} can be expressed as
\begin{equation}
  \Delta x,y = 0.329 \times  \frac{2I_{\mathrm{coil}_{x,y}}}{p_{e}},
  \label{eq:B5}
\end{equation}
and the raster dimension along the $(x,y)$-axes are defined as twice the deflection angle,
\begin{equation}
  R_{x,y} \equiv 2\Delta x,y = 1.316 \times  \frac{I_{\mathrm{coil}_{x,y}}}{p_{e}}.
  \label{eq:B6}
\end{equation}
From Table 2 of Ref.\cite{Raster_Yan2005}, the maximum operating current of each FR-magnet is $I_{\mathrm{peak}}\approx100$ A, and assuming
that running the power supply at 80$\%$ of the voltage is best for the long-term lifetime of the power supply, $I_{\mathrm{peak}}\approx80$ A\cite{MKJ_privFeb2020}.
The maximum operating current per coil is then given by $I_{\mathrm{coil,max}_{x,y}} = I_{\mathrm{peak}_{x,y}} / 2 = 40$ A.
Using these approximations and Eq. \ref{eq:B6}, the maximum raster dimension is restricted to
\begin{align}
  R_{(x,y),\mathrm{max}} = \frac{52.64}{p_{e}}.
  \label{eq:B7}
\end{align}
Therefore, for a 5-pass beam ($p_{e}=$10.6 GeV/c)
the maximum possible raster size is 
\begin{equation}
  R_{x,\mathrm{max}}\times R_{y,\mathrm{max}}= 5\times5 \text{ mm}^{2}.
  \label{eq:B8}
\end{equation}
From these results and Eq. \ref{eq:B6}, the maximum deflection is $\Delta x_{\mathrm{max}},y_{\mathrm{max}} = R_{(x,y)_{\mathrm{max}}}/2 = 2.5$ mm.
Starting from Eq. \ref{eq:B1}, the maximum deflection angle at both ends (-,+) of a target of length $L_{\mathrm{tgt}}$ is given by
\begin{align}
&\theta^{\mathrm{max}}_{\Delta x_{\mp}} = \frac{\Delta x_{\mathrm{max}}}{L_{\mathrm{FR}} \mp \frac{L_{\mathrm{tgt}}}{2}}, \label{eq:B9} \\
&\theta^{\mathrm{max}}_{\Delta y_{\mp}} = \frac{\Delta y_{\mathrm{max}}}{L_{\mathrm{FR}} \mp \frac{L_{\mathrm{tgt}}}{2}}. \label{eq:B10}
\end{align}
Inserting the numerical values $\Delta x,y = 2.5$ mm, $L_{\mathrm{FR}} = 13560$ mm, and $L_{\mathrm{tgt}} = 100$ mm (10-cm long target) in Eqs.\ref{eq:B9} and \ref{eq:B10}, 
\begin{align}
\theta^{\mathrm{max}}_{\Delta x_{\mp}}, \theta^{\mathrm{max}}_{\Delta y_{\mp}} = (1.85 \times 10^{-4}, 1.83 \times 10^{-4}) \text{ rad}. \label{eq:B11} 
\end{align}
From Eq. \ref{eq:B11} the maximum deflection angle at both ends of the target along the $(x,y)$ axes is negligible for the maximum possible raster size, therefore,
the raster is approximately uniform across the target length. This experiment used a raster size of $2\times2$ mm$^{2}$ across a 10-cm long target at
5-pass, therefore, the deflection angles are expected to be even smaller than in Eq. \ref{eq:B11}. \\
\indent To detect and process the raster signals, the fast-raster system is equipped with 2 current probes and a field pickup probe. The current probes are used for precise measurements
of the magnet current amplitude and direct current (DC) offset detection and the field pickup probe is used to detect the variations in the ramping magnetic field.   
The signals are sent to a beam raster monitor in the hall that compares the setting and readback parameters to create a fast shutdown detection for the machine safety
operation\cite{Raster_Yan2005}. The signal from the field pickup probe is sent to an ADC module in the Hall C Counting House and 
the raw ADC signals are then further processed by the analysis software.\\

\begin{vita}
\begin{center}
CARLOS YERO PEREZ \\[4ex]
\end{center}

\noindent
\begin{tabular}{ll}
                Born, Santiago de Cuba, Cuba \\[2ex]

2010 \hspace{2in} & A.A., Biology \\
	        & Miami-Dade College \\
		& Miami, Florida \\[2ex]

2014 \hspace{2in} & B.Sc., Physics \\
	        & Florida International University \\
		& Miami, Florida \\[2ex]
              
2014--2015      & Graduate Research Assistant \\
                & Florida International University\\
                & Miami, Florida \\[2ex]

2015--2016      & Graduate Teaching Assistant \\
                & Florida International University\\
                & Miami, Florida \\[2ex]

2016--2018      & Graduate Research Assistant  \\
		& Florida International University \\
                & Miami, Florida \\[3ex]

2018 \hspace{2in} & M.Sc. in Physics  \\
		  & Florida International University \\
                  & Miami, Florida \\[3ex]

2018--2020        &Ph.D Candidate in Physics \\
                  & Florida International University \\
                  & Miami, Florida \\[3ex]
\end{tabular}

\noindent
PUBLICATIONS AND PRESENTATIONS \\[2ex]
Pooser, E., Barbosa, F., Boeglin, W., Hutton, C., Ito, M.M., Kamel, M., Khetarpal, P., Llodra, A., Sandoval, N., Taylor, S., Whitlatch, T., Worthington, S., Yero, C. and Zihlmann, B.
\href{https://doi.org/10.1016/j.nima.2019.02.029}{The GlueX Start Counter Detector}. \textit{Nucl.Instrum.Meth. A927}:330-342, May 2019.  \\
\\
\noindent Yero, C. \textit{et al.} (2020). \textit{Probing the Deuteron at Very Large Internal Momenta}. Manuscript in preparation for Physical Review Letters (PRL).\\  
\newpage
\noindent Yero, C. (2019, March). \textit{Deuteron Electro-Disintegration Experiment at Hall C}. Workshop on Quantitative Challenges in SRC and EMC Research,
Boston, Massachusetts.\\
\\
\noindent Yero, C. (2019, October). \textit{First Cross Section Results of $D(e,e'p)n$ at Very High Recoil Momenta}. Division of Nuclear Physics (DNP) Conference,
Arlington, Virginia.\\

\noindent

\end{vita}

\end{document}